%% file: main.tex
\documentclass[final]{cvpr}
\usepackage{times}
\usepackage{epsfig}
\usepackage{graphicx}
\usepackage{amsmath}
\usepackage{amssymb}
\usepackage{soul}
\usepackage{multirow}
\usepackage{booktabs}
\usepackage{wrapfig}
\usepackage{footnote}
\usepackage{subcaption}

\usepackage{algorithmicx}
\usepackage{algorithm}
\usepackage{algpseudocode}

\usepackage[pagebackref=true,breaklinks=true,colorlinks,bookmarks=false]{hyperref}

\makeatletter
\def\blfootnote{\xdef\@thefnmark{}\@footnotetext}
\makeatother

\def\cvprPaperID{10371} 
\def\confYear{CVPR 2021}

\input{macros}

\begin{document}

\title{Inferring CAD Modeling Sequences Using Zone Graphs}

\author{\normalsize Xianghao Xu\\
\normalsize Brown University\\

\and
\normalsize Wenzhe Peng \\
\normalsize MIT\\

\and
\normalsize Chin-Yi Cheng\\
\normalsize Autodesk Research\\

\and
\normalsize Karl D.D. Willis\\
\normalsize Autodesk Research\\

\and
\normalsize Daniel Ritchie\\
\normalsize Brown University\\
}

\maketitle

\begin{abstract}
\input{00-abstract}
\end{abstract}

\blfootnote{This work was done partially during Xianghao Xu and Wenzhe Peng's internship at Autodesk Research}


\input{01-intro}

\input{02-related}

\input{03-method}

\input{04-results}

\input{05-conclusion}

\input{06-acknowledgments}

{\small
\bibliographystyle{ieee_fullname}
\bibliography{main}
}

\newpage

\input{supp}

\end{document}

%% file: macros.tex
\newcommand{\denselist}{\itemsep -0.3em \parsep=0pt\partopsep 0pt\vspace{-\topsep}}

%% file: 00-abstract.tex
In computer-aided design (CAD), the ability to ``reverse engineer'' the modeling steps used to create 3D shapes is a long-sought-after goal. This process can be decomposed into two sub-problems: converting an input mesh or point cloud into a boundary representation (or B-rep), and then inferring modeling operations which construct this B-rep. In this paper, we present a new system for solving the second sub-problem. Central to our approach is a new geometric representation: the zone graph. Zones are the set of solid regions formed by extending all B-Rep faces and partitioning space with them; a zone graph has these zones as its nodes, with edges denoting geometric adjacencies between them. Zone graphs allow us to tractably work with industry-standard CAD operations, unlike prior work using CSG with parametric primitives. We focus on CAD programs consisting of sketch + extrude + Boolean operations, which are common in CAD practice. We phrase our problem as search in the space of such extrusions permitted by the zone graph, and we train a graph neural network to score potential extrusions in order to accelerate the search. We show that our approach outperforms an existing CSG inference baseline in terms of geometric reconstruction accuracy and reconstruction time, while also creating more plausible modeling sequences.

%% file: 01-intro.tex
\section{Introduction}
\label{sec:intro}

\begin{figure}[t!]
    \centering
    \includegraphics[width=\columnwidth]{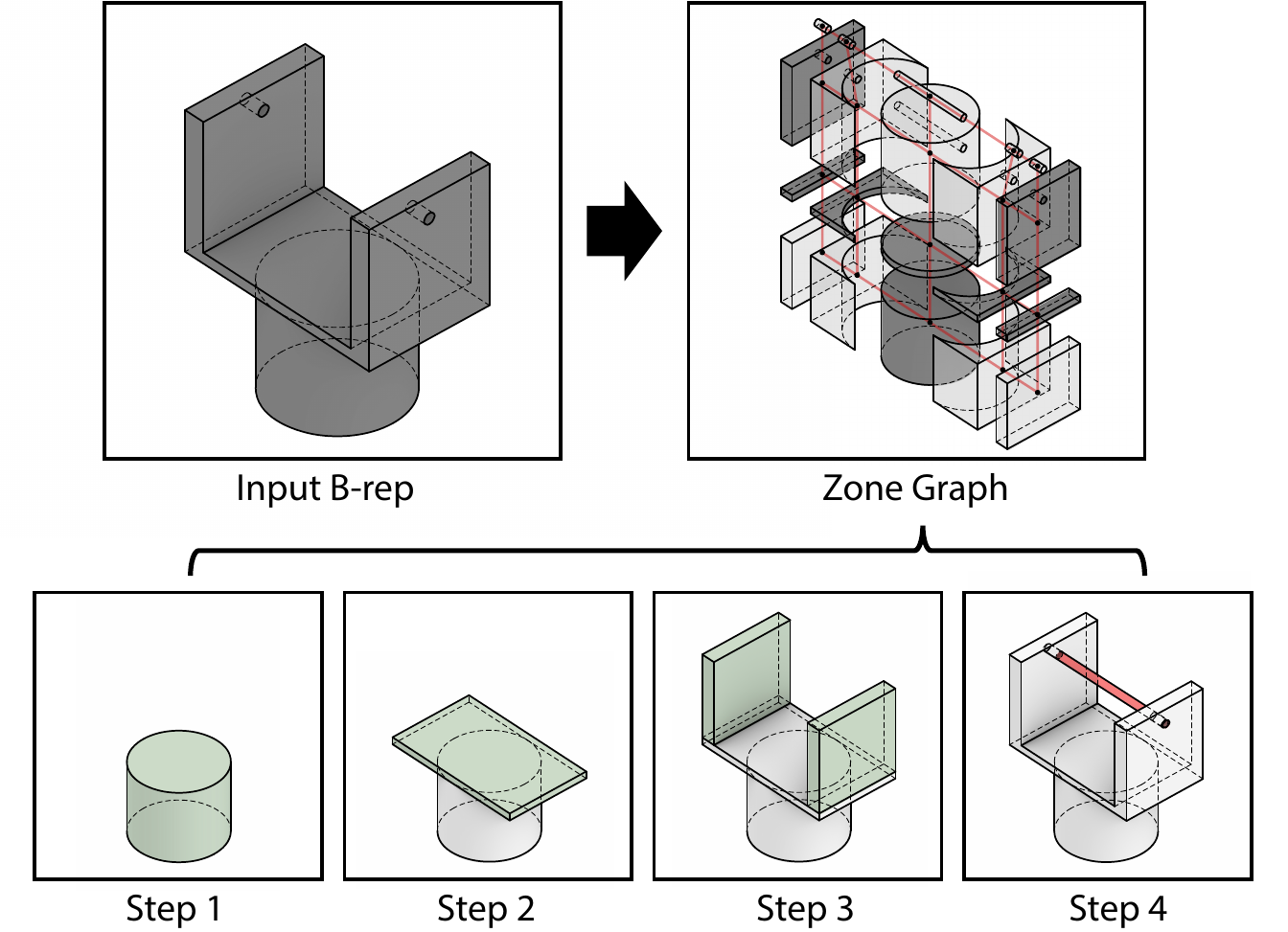}
    \caption{A solid shape (top-left) can be decomposed into a graph of \emph{zones} which make up its interior volume and surrounding space (top-right).
    This representation permits efficient search for modeling sequences that reconstruct the shape, using industry standard modeling operations (bottom).
    Light red lines in the zone graph indicate connections between zone nodes.
    In the bottom modeling sequence, green denotes sketch + extrude + union and red denotes sketch + extrude + difference.}
    \label{figure:teaser}
\end{figure}

Many real-world 3D objects begin their existence as parametric CAD programs.
If one could recover such programs for everyday objects, it would enable powerful editing and re-purposing abilities, with many applications in mechanical and industrial design.
As such, it's unsurprising that this problem has become a popular research topic in the graphics, vision, and machine learning communities, with multiple recent papers examining how to infer CAD-like programs for an input shape~\cite{sharma2017csgnet,du2018inversecsg,zhou2020unsupervised,ellis2019write,kania2020ucsg}.

One property shared by these works is their use of constructive solid geometry (CSG) as a modeling language.
In CSG, shapes are formed by combining primitive solids (e.g. spheres, cylinders, boxes) with Boolean union, intersection, and difference.
Its small library of parametric primitive shapes makes it appealing for CAD program inference, as this reduces the search space of possible programs.

Unfortunately, parametric primitive CSG is not the modeling language used by CAD practitioners today.
Instead, modern CAD workflows use \emph{feature-based modeling}, in which a solid object is created by iteratively adding features such as holes, slots, or bosses~\cite{hoffmann1989geometric}.
Feature creation is typically performed via operations on surfaces, as these are intuitive for users to reason about: for instance, creating a slot in the surface of an object by sketching the profile of the slot and then specifying how deep it goes.
Throughout this process, the object's geometry is stored as a \emph{boundary representation}, or B-rep, which is a watertight assembly of surface patches which enclose the object's solid volume~\cite{weiler1986}.
Feature-based modeling with B-reps supports Boolean operations between solids, making it strictly more expressive than parametric primitive CSG.
This expressiveness comes at a cost for program inference, though, as the space of feature-based modeling programs is much larger.

How can we enable tractable inference of feature-based CAD programs?
We first note another common property of prior work: solving CAD program inference ``in one shot,'' going directly from input unstructured geometry to a CAD program.
However, this problem actually decomposes into two sub-problems: (1) convert the input geometry into a B-rep, (2) infer a CAD program which generates that B-rep.
The first sub-problem is of huge importance to the CAD industry, where it is known as \emph{reverse engineering}.
Many semi-automatic commercial tools exist~\cite{Geomagic,SpaceClaim} and recent work has begun to leverage learning based approaches~\cite{sharma2020parsenet, wang2020pie}.
The second problem is supported by all major CAD software using a rule-based approach, e.g.~\cite{inventorFR, featureWorks}.
However, such systems often require active input from a human designer, can fail to automatically infer earlier steps of longer modeling sequences, and struggle to generalize beyond pre-defined rules.

In this paper, we show that assuming a B-rep as input allows us to tractably attack the problem of automatically inferring industry-standard, feature-based CAD programs.
Specifically, it allows us to use a new geometry representation that enables tractable program search.
Extending all surface faces of the input B-rep into infinite surfaces and then partitioning space with those surfaces results in an arrangement of solid zones.
The spatial connectivity of these zones forms a data structure which we call the \emph{zone graph} (Figure~\ref{figure:teaser}, top-right).
Searching for a CAD program that reconstructs the input shape then becomes a search for a sequence of modeling operations that fill in all the zones which are inside the input B-rep (Figure~\ref{figure:teaser}, bottom).
This perspective reduces the search space of possible programs from an infinite set (a space of programs with many continuous parameters) to a finite set (the set of operation sequences that fill particular zones).

In this paper, we focus on modeling via sketch + extrude + Boolean operations, which are commonly-used in CAD workflows to create new solid masses and cut holes and slots out of existing ones.

In addition, we \emph{learn} how to guide the search using a large dataset of CAD modeling sequences.
We train a graph neural network that takes the zone graph as input and predicts scores for different candidate modeling operations, allowing search to focus on higher-scoring options first.
This allows our approach to infer modeling sequences which use meaningful design patterns without active guidance from a human designer.

In experiments on real-world CAD shapes, our zone graph method outperforms a recent state-of-the-art CSG inference method: it achieves better reconstruction accuracy by inferring more plausible programs that use industry-standard CAD operations.

In summary, the contributions of this paper are:
\begin{itemize}
\denselist
    \item The zone graph representation of B-rep solids, which reduces CAD program search space to a finite size.
    \item A search algorithm for inferring CAD modeling sequences from zone graphs.
    \item A graph neural network for learning to score candidate CAD operations during program search.
\end{itemize}

%% file: 02-related.tex
\section{Related Work}
\label{sec:related}

\paragraph{Space Partitioning } 
The zone graph is a partition of space formally known as an arrangement of surfaces~\cite{ArrangmentsChapter}.
Use of spatial partitioning data structures is common in computer graphics to accelerate spatial queries, particularly in ray tracing~\cite{AccelerationStructureSurvey}.
A more recent body of work has explored space partitioning for geometry reconstruction tasks. 
Polyfit~\cite{nan2017polyfit} performs surface reconstruction from point clouds by extracting and intersecting planar primitives; other recent reconstruction works use a similar method~\cite{fang2020connect, bauchet2020kinetic}.
Learning-based methods have also begun to use space partitioning representations~\cite{chen2020bsp,deng2020cvxnet}.
BSP-Net~\cite{chen2020bsp} uses binary space partitioning to build up a constructive solid geometry (CSG) tree for compact mesh generation. 
The zone graph representation differs from prior work as it builds up an incidence graph of an arrangement of parametric surfaces and considers curved surfaces in addition to planar surfaces.
We also apply zone graphs to a different reconstruction task: finding a sequence of modeling operations to reproduce a shape.

\paragraph{CAD Reconstruction} 
CAD reconstruction involves recovering a sequence of CAD modeling operations from meshes, point clouds, or B-rep models. Such sequences are critical for preserving editability of CAD models, enabling downstream edits, such as model simplification for simulation, or adjusting tolerances for manufacturing. Although CAD reconstruction has been the subject of significant research~\cite{shah2001discourse}, it remains a challenging problem due to the diverse ways that CAD models are constructed. 
Commercial CAD software uses rule-based feature recognition, often with user assistance, to detect and remove features such as holes, pockets, and fillets before re-applying them parametrically~\cite{inventorFR}. This strategy can recover some modeling operations but may fail to completely rebuild the parametric modeling history from the first step. We focus on a fully automatic approach that can recover the entire construction sequence in a manner consistent with human designs.

CAD reconstruction can also be framed as a program synthesis problem.
InverseCSG \cite{du2018inversecsg} is one example; it uses a constraint-based program synthesizer to find CSG programs whose output is consistent with the input geometry.
More recently, learning-based approaches have been developed.
CSGNet~\cite{sharma2017csgnet} leverages a neural network to infer a sequence of CSG operations on simple primitives such as spheres and cuboids.
Other works in this area~\cite{ellis2019write, tian2018learning, kania2020ucsg} also utilize simple primitives with CSG operations.
However, professional mechanical design tools use a different paradigm, first creating 2D engineering sketches then lifting them to 3D using operations such as extrude, revolve, and sweep.

CAD reconstruction is a \emph{visual program induction} problem~\cite{EG2020STAR}.
One common approach for such problems is \emph{neurally-guided search}, in which a neural network guides a search algorithm by prioritizing search options~\cite{ritchie2016neurally,ellis2019write,DreamCoder}.
We also leverage neurally-guided search, developing novel search proposal and ranking algorithms for zone graphs.

%% file: 03-method.tex
\newcommand{\brep}{\ensuremath{\mathcal{B}}}
\newcommand{\op}{\ensuremath{\mathbf{o}}}
\newcommand{\zonegraph}{\ensuremath{\mathcal{G}}}
\newcommand{\zones}{\ensuremath{\mathcal{Z}}}
\newcommand{\edges}{\ensuremath{\mathcal{E}}}
\newcommand{\zone}{\ensuremath{Z}}
\newcommand{\intzone}{\ensuremath{\zone^\bullet}}
\newcommand{\extzone}{\ensuremath{\zone^\circ}}
\newcommand{\canvas}{\ensuremath{\mathcal{C}}}
\newcommand{\target}{\ensuremath{\mathcal{T}}}
\newcommand{\extrusion}{\ensuremath{\mathcal{X}}}
\newcommand{\sketch}{\ensuremath{\mathcal{S}}}

\section{Task \& Approach}
\label{sec:approach}

Given a 3D shape specified as a B-rep $\brep$, our goal is to find a sequence of modeling operations $[\op_1, \op_2, \ldots , \op_n]$ which reproduce $\brep$.
\begin{wrapfigure}{r}{0.25\textwidth}
    \includegraphics[width=0.25\textwidth]{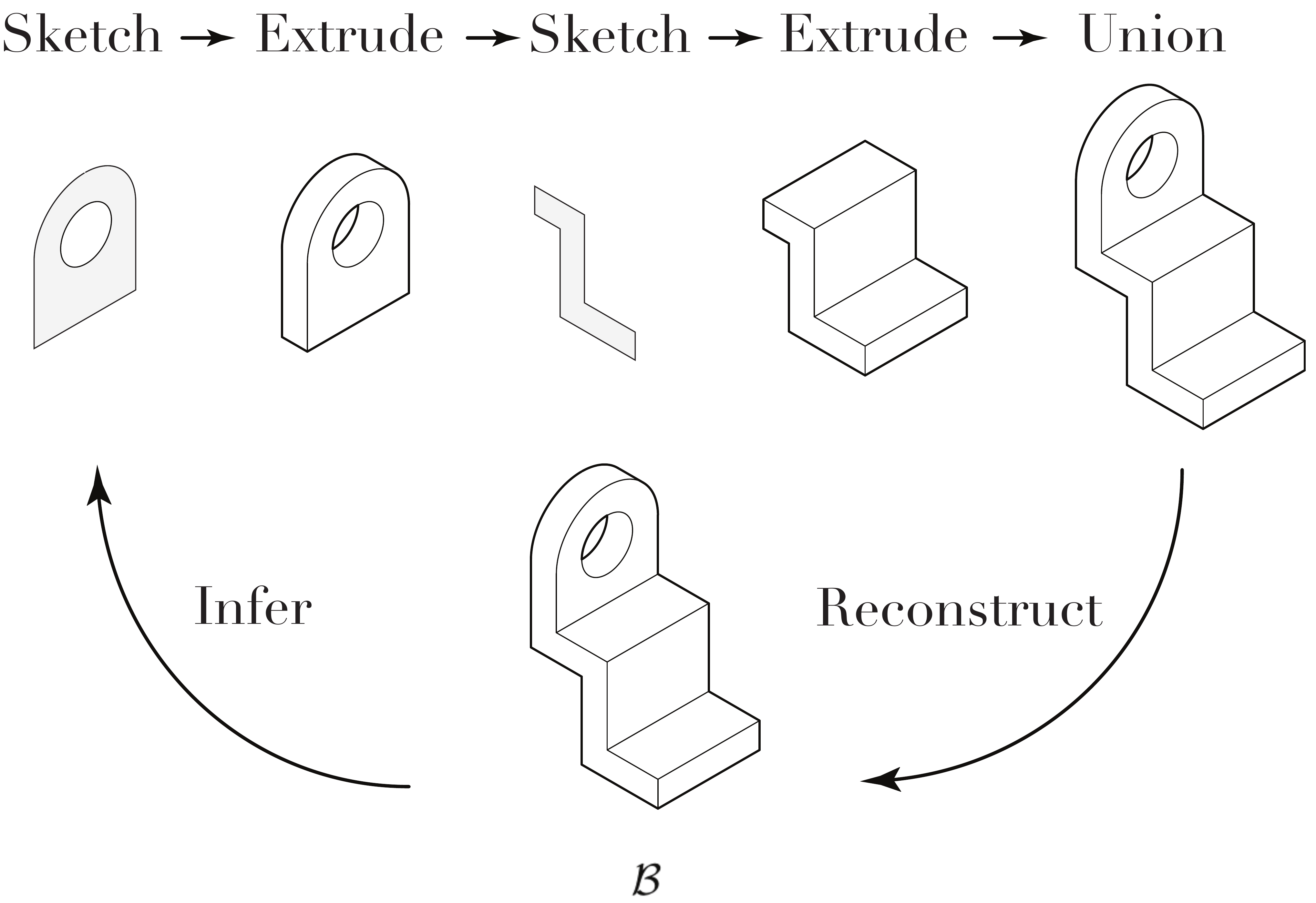}
\end{wrapfigure}
We focus on sketch + extrude + Boolean operations, in which the user (a) sketches one or more closed profile curves on a plane (b) extrudes this sketch to form one or more solid regions, and (c) applies the extrusion to the current partial B-rep by union or difference (see inset).

To solve this problem, we first turn the input B-rep $\brep$ into a zone graph $\zonegraph$ (Section~\ref{sec:zonegraph}).
We then search for a sequence of operations that reproduces this zone graph by enumerating and evaluating sketch + extrude + Boolean operations that are consistent with the zone graph (Section~\ref{sec:search}).
As there may be a large number of such operations, we use a learned guidance network to prioritize ones that are similar to those seen in a large dataset of example CAD programs (Section~\ref{sec:network}). We leverage the recently-released Fusion 360 Gallery reconstruction dataset that provides ground-truth, human-designed CAD programs created with sketch and extrude modeling operations~\cite{willis2020fusion}.

\section{The Zone Graph Representation}
\label{sec:zonegraph}

CAD modeling operates via the addition and removal of solid volumes to create 3D shapes, using intuitive feature-based operations.
To search for CAD programs that can reconstruct a given target shape, one could explore the space of all parameterizations of all such modeling operations.
This is a huge search space.
Instead, we note that the target shape contains strong ``clues'' as to the operations used to create it.
For example, suppose that a hole was created in a shape by subtracting some solid volume.
While that volume is not present in the target, the interior faces of the hole it created reveal its shape.
Our key idea is to formalize this notion of such ``hidden volumes'' and then restrict search to consider only the modeling operations that produce them or target shape interior volumes.

\begin{figure}[t!]
    \centering
    \includegraphics[width=\columnwidth]{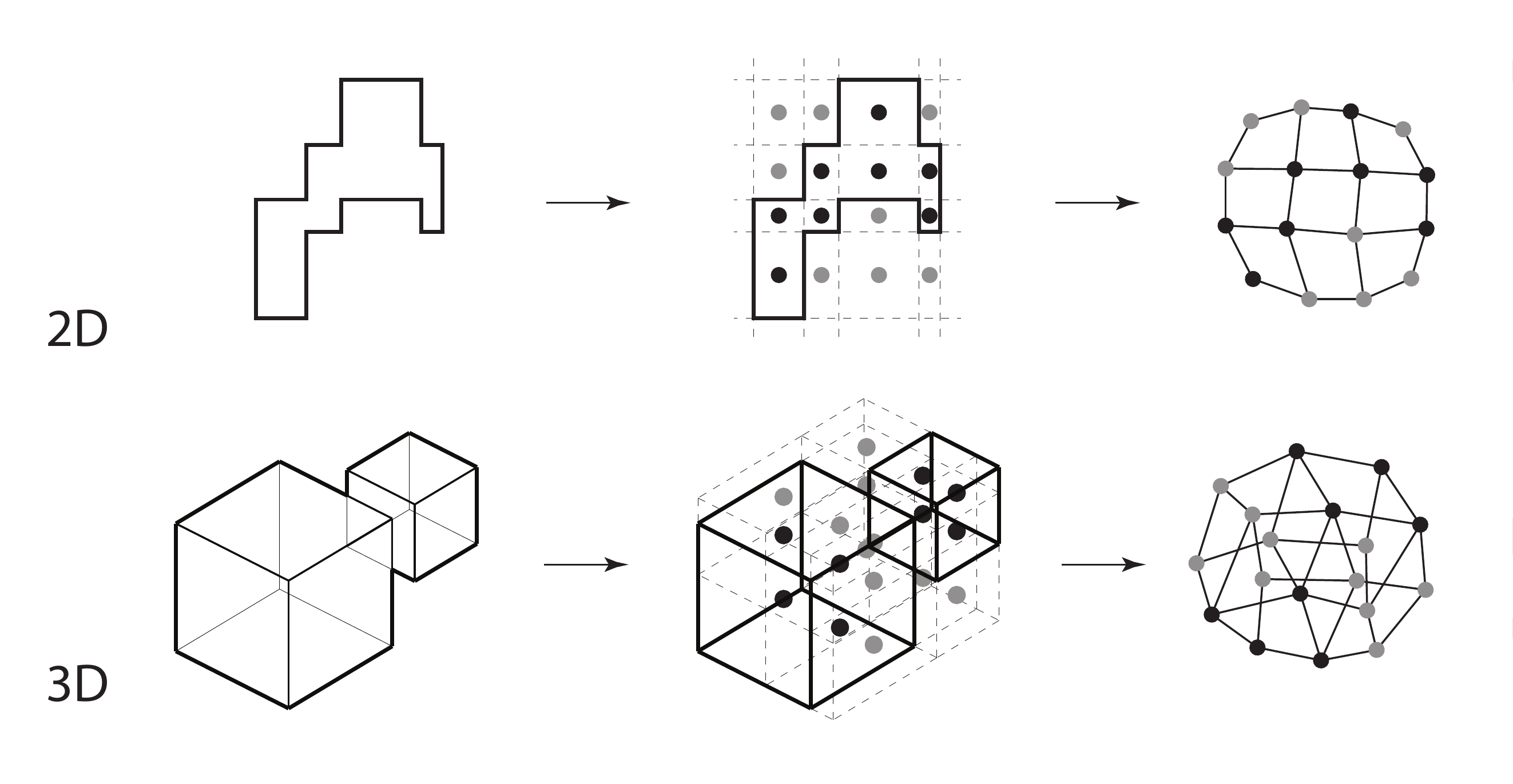}
    \caption{
    \emph{(Left)} Input 2D and 3D shapes.
    \emph{(Middle)} Shape decomposition into solid zones.
    \emph{(Right)} The zone graph.
    }
    \label{figure:representation}
\end{figure}

Our fist step is to construct from each input B-rep $\brep$ a spatial data structure $\zonegraph$ that we call a \emph{zone graph} (Figure~\ref{figure:representation}).
While we focus on 3D shapes in this paper, zone graphs are also well-defined in 2D; Figure~\ref{figure:representation} includes a 2D example for illustrative purposes.
A zone graph in 3D (2D) is a graph $\zonegraph = (\zones, \edges)$ whose nodes $\zones = \{ \zone_1, \zone_2, \ldots , \zone_n \}$ represent solid regions (areas) called \emph{zones} and whose edges $\edges$ represent surface patches (curves) connecting those regions.
A zone graph has the property that the union of all zones is the axis-aligned bounding box (AABB) of the input B-rep.

In Figure~\ref{figure:representation}, zones are colored black if they are \emph{interior zones} $\intzone$ that fill inside the volume enclosed by $\brep$, and gray if they are \emph{exterior zones} $\extzone$.

\paragraph{Face Extension} 
Zone graph construction begins by extending faces of $\brep$ into infinite surfaces that partition space:

\begin{itemize}
\denselist
    \item \emph{Planar faces} are extended into an infinite plane.
    \item \emph{Generalized cylindrical faces} are extended along the face's direction of zero curvature.
    \item \emph{Spherical faces} are extended into a complete sphere.
    \item \emph{Free form faces} are not extended, as there is no known parametric form by which to extend them.
\denselist
\end{itemize}
After extension, all faces are clipped by $\text{AABB}(\brep)$.

\paragraph{Simplification}
The number of zones increases superlinearly with $\brep$'s face count, leading to a larger search space.

It helps to skip face extensions that are not likely to be useful for extrusion-based modeling.
We search $\brep$ for ``face loops,'' sets of faces where (a) the faces form a cycle, (b) the edges shared by each consecutive pair of faces are parallel (we call this direction the \emph{extrusion direction}), and (c) each face has exactly 4 edges in its outer wire.
Faces in a loop are only extended along their extrusion direction.
See the supplement for details.

\noindent
\newline
Figure~\ref{figure:zone_graph_stats} visualizes some statistics for zone graphs extracted from the Fusion 360 Gallery reconstruction dataset~\cite{willis2020fusion}.
On the left, we plot zone count against face count.

On the right, we plot zone count against the time required to construct the zone graph (in seconds).
Build time increases linearly with zone count, with most zone graphs taking under a minute.
The build tests were run on an Intel Core i5-8259U processor using OpenCASCADE's general fuse algorithm (GFA) for solid partitioning via FreeCAD.

\begin{figure}[t!]
    \centering
    \setlength{\tabcolsep}{1pt}
    \begin{tabular}{cc}
        \includegraphics[width=0.49\linewidth]{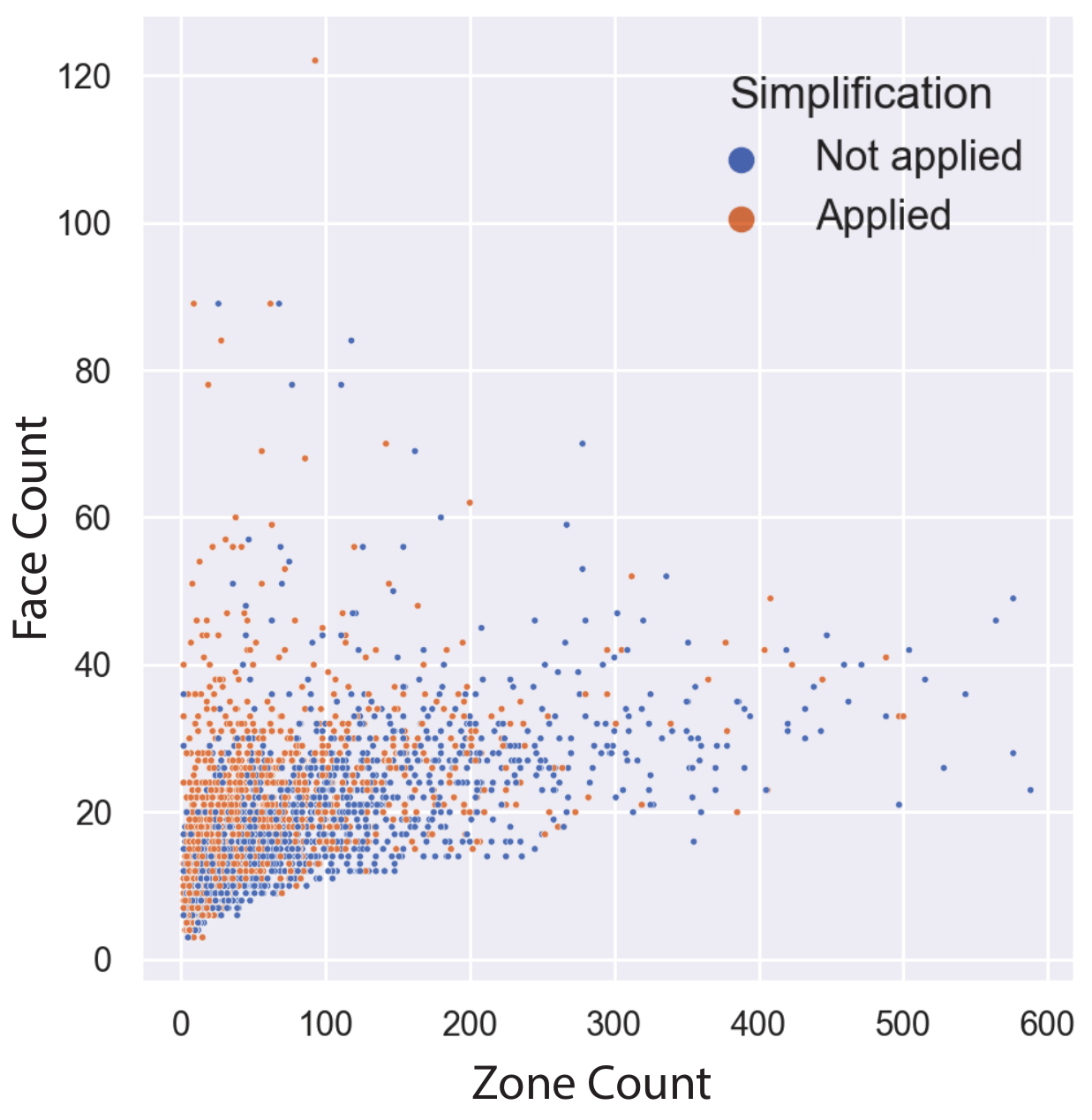} &
        \includegraphics[width=0.49\linewidth]{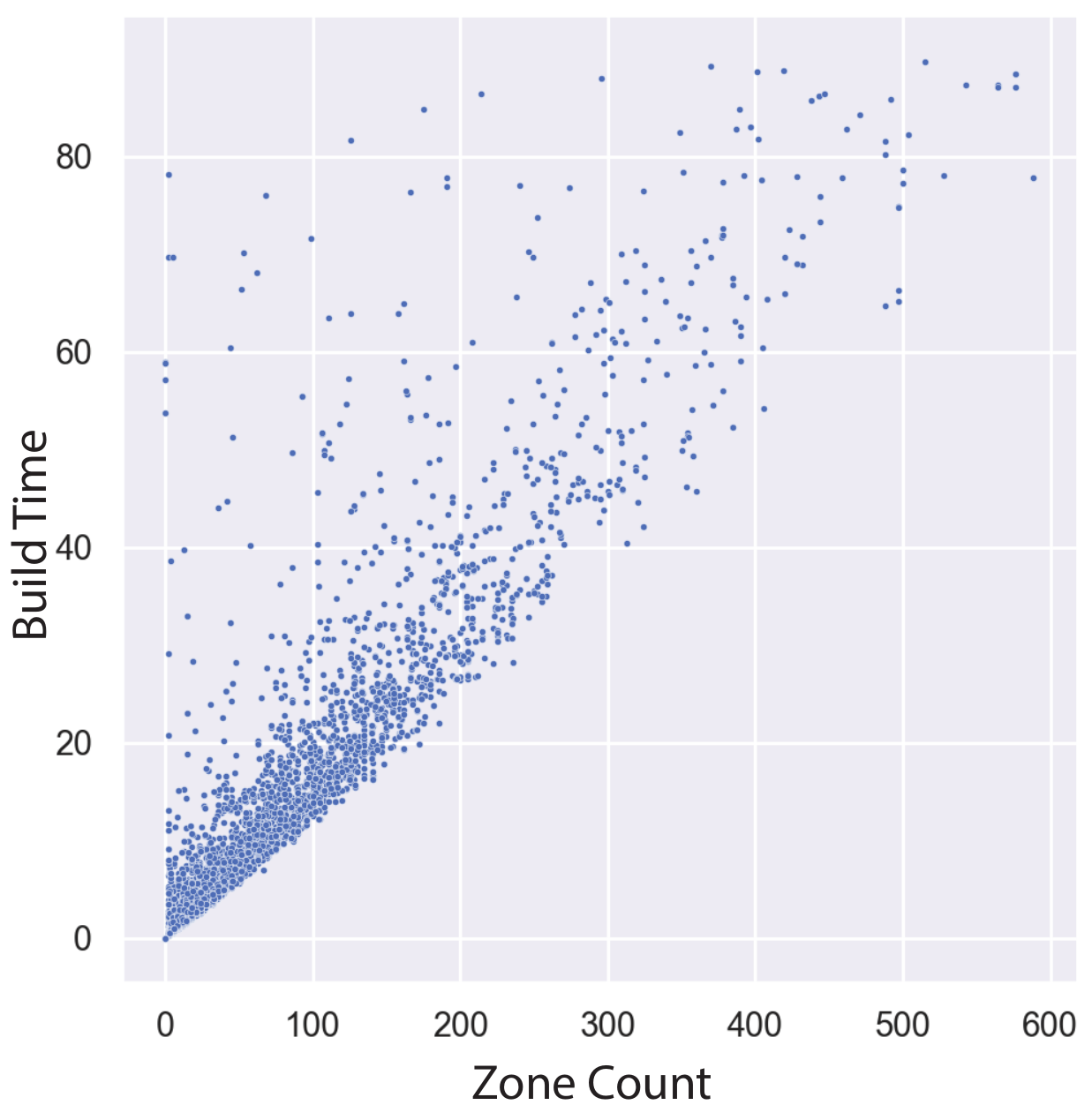}
    \end{tabular}
    \caption{
    Zone graph construction statistics.
    \emph{(Left)} Zone count vs. B-rep face count. 
    \emph{(Right)} Zone count vs. zone graph construction time (in seconds).
    }
    \label{figure:zone_graph_stats}
\end{figure}

Our method can reconstruct 6900/8625 models in the dataset. Failure cases are due to (1) erroneous output from GFA caused by numerical robustness issues (such issues are well known~\cite{weiler1998robust, kettner2008classroom} and the subject of ongoing research) and (2) unsupported operations such as tapered extrude and revolve.
Of these 6900 models, the zone graph can represent the ground truth modeling sequence for 5175 of them.
Failure cases are due to (1) sequences not captured by our extrusion proposals (Section~\ref{sec:proposals}) and (2) sequences not expressible by the zone graph because some operations do not leave any trace in the target shape (see Section~\ref{sec:conclusion} for an example).
See the supplemental material for more details, including ablation studies on zone graph simplification.

\section{Searching for Modeling Sequences}
\label{sec:search}

Given a zone graph $\zonegraph$ for an input B-rep $\brep$, our goal is to search for a sequence of CAD modeling operations $[\op_1, \op_2, \ldots , \op_n]$ that satisfies two properties:
\begin{enumerate}
\denselist
    \item Executing this sequence produces the input shape $\brep$, i.e. $\text{exec}([\op_1, \op_2, \ldots , \op_n]) = \bigcup \intzone_i$
    \item Each step produces a shape which is a combination of zones in $\zonegraph$, i.e. $\forall j,  \exists {\zones' \subset \zones} \text{\ s.t.\ } \text{exec}([\op_1, \ldots , \op_j]) = \bigcup {(\zone_i \in \zones')}$.
\denselist
\end{enumerate}
Our search algorithm satisfies property 2 by construction: it only considers modeling operations whose output coincides with available zones.
If $\brep$ is expressible as a sequence of sketch + extrude + boolean operations, and search time is unbounded, our algorithm also satisfies property 1.
Otherwise, property 1 is only approximately satisfied.

\begin{figure*}[t!]
    \centering
    \includegraphics[width=\linewidth]{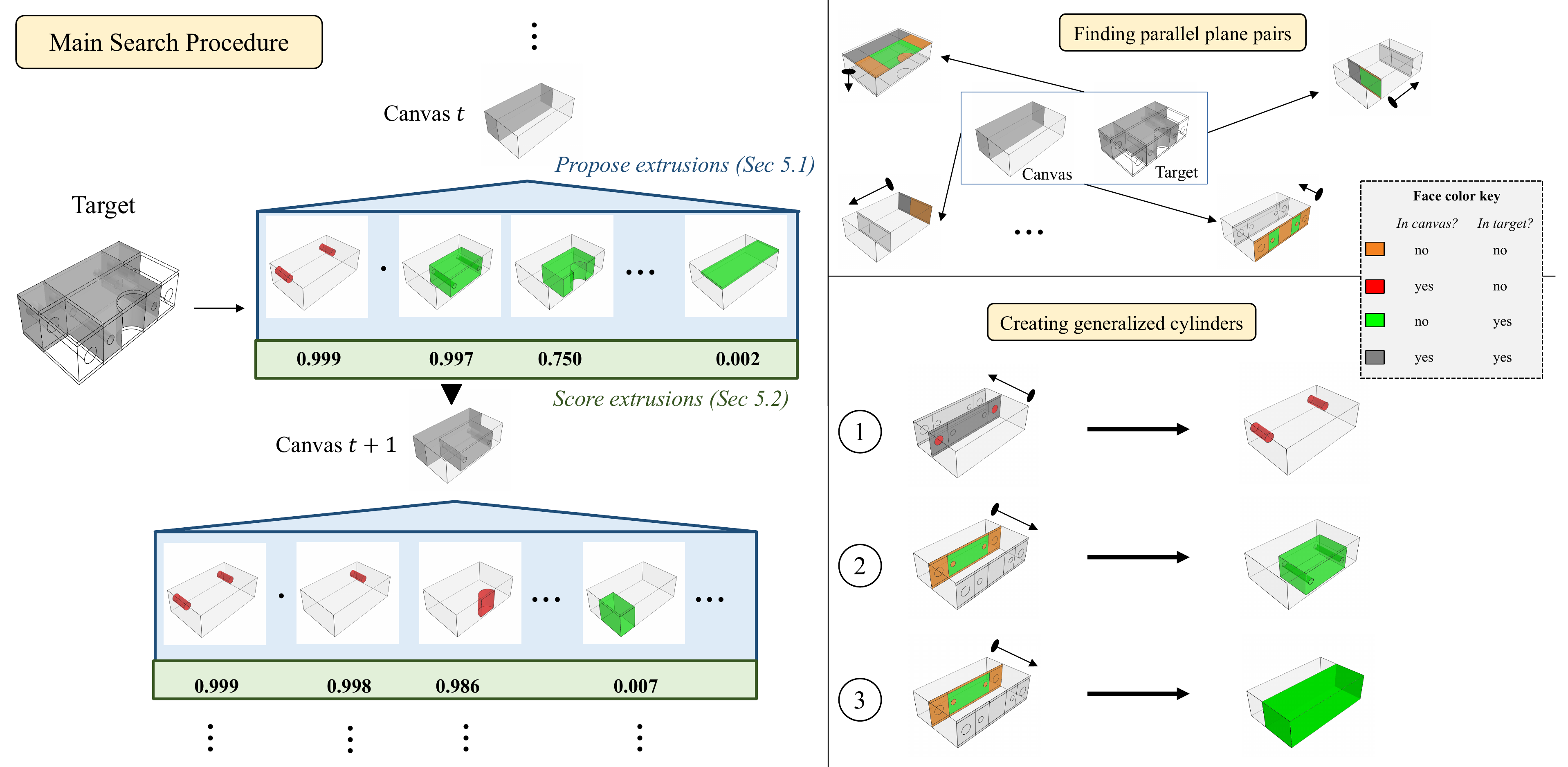}
    \caption{
    Searching for modeling sequences that reconstruct a Target shape.
    \emph{(Left)} Search maintains a canvas (the partial shape constructed thus far), proposes possible next modeling operations, scores those operations, and then selects one of them to further explore based on its score.
    \emph{(Right)} Candidate extrude operations are found by first finding all pairs of parallel planes and identifying which face groups could be used as the starting sketch (top-right) and then extruding this sketch to create generalized cylinders (bottom-right).
    }
    \label{figure:search_process}
\end{figure*}

Figure~\ref{figure:search_process} shows an overview of our search algorithm.
The input B-rep $\brep$ is denoted as the \emph{target}.
The algorithm maintains a \emph{canvas}, which contains the shape constructed by the modeling operations chosen by search thus far and is initially empty.
At each step, the algorithm enumerates all combinations of zones which could be produced via a valid sketch + extrude operation (Section~\ref{sec:proposals}); these are the valid next steps for search to consider.
The algorithm then scores how likely each of these proposals is to lead to a correct reconstruction of the target (Section~\ref{sec:network}).
It retains the top $k$ highest-scoring proposals and initially chooses the top 1 to explore next (i.e. best-first search).
We use $k=5$ unless otherwise specified.
If the search reaches a terminal state (i.e. no valid extrusion proposals), but the canvas does not match the target, it backtracks to the previous step and considers the next extrusion in the top $k$ set.
Search terminates when the canvas matches the target, or when a computation time budget has been exhausted (in the latter case, the canvas with the highest reconstruction IoU is returned).

\subsection{Generating Candidate Modeling Operations}
\label{sec:proposals}

Figure~\ref{figure:search_process} right illustrates our process for identifying candidate modeling operations.
It begins by finding pairs of parallel planes---the start and end planes for an extrusion (Figure~\ref{figure:search_process} top-right).
We define the \emph{extrusion direction} $d$ as the vector from the start plane to the end plane.
Next, we identify the \emph{starting sketch} $\sketch$, a set of faces on the start plane.
Considering all possible such sets is intractable.
Fortunately, the zone that a face is adjacent to along $d$ suggests the operations it might be used in:
\begin{enumerate}
\denselist
    \item Faces $d$-adjacent to zones in the canvas $\canvas$ but not in the target $\target$ could start an extrude + subtract (Figure~\ref{figure:search_process} bottom-right, 1).
    \item Faces $d$-adjacent to zones in $\target$ but not in $\canvas$ could start an extrude + union (Figure~\ref{figure:search_process} bottom-right, 2).
    \item Faces $d$-adjacent to zones in $\target$ or empty zones could start an extrude + union (Figure~\ref{figure:search_process} bottom-right, 3).
\denselist
\end{enumerate}
Candidate starting sketches $\sketch$ are the connected components of each group above, plus the union of these from each group (e.g. Figure~\ref{figure:search_process} bottom-right, 1).
The supplemental material describes other candidate enumeration strategies.

Each candidate sketch is then extruded along $d$ to create a generalized cylinder; the zones which fall within this cylinder form the proposed extrusion volume $\extrusion$.
If $\extrusion \subset \canvas$, it is marked as a subtraction.
If $\extrusion \cup \canvas = \varnothing$, it is marked as a union.
Otherwise, we create proposals of both types.
Computation of these proposals is memorized, as proposals frequently re-occur at different search iterations.
To prevent search cycles, we discard a proposed extrusion if it is the inverse of one performed earlier in search (e.g. re-adding a volume that was previously subtracted).

While we have focused on extrude, a similar procedure could be applied to construct proposals for other operations.
For example, fillet, chamfer, and taper all affect the zone graph in well-defined ways; it is possible to identify combinations of zones which can result from such operations.

\subsection{Ranking Candidate Modeling Operations}
\label{sec:network}

Given a set of proposed modeling operations, our search algorithm must decide which ones to prioritize in its best-first exploration.
We propose to \emph{learn} what operations are best from a dataset of CAD modeling sequences.
We train a neural network that takes as input the zone graph $\zonegraph$, the current canvas $\canvas \subset \zones$ (i.e. which zones are ``filled''), the target shape $\target = \{ \intzone_i \}$ (i.e. the set of interior zones), and a modeling operation $\op = (\extrusion \subset \zones, t \in \{ \cup, - \})$, where $\extrusion$ is a generalized cylinder extrusion and $t$ denotes the type of operation (union or difference).
The network's task is to predict $p(\op | \zonegraph, \canvas, \target)$, how likely $\op$ is to be the next modeling operation, based on the patterns it has seen in its training set.

\begin{figure}[t!]
    \centering
    \includegraphics[width=\linewidth]{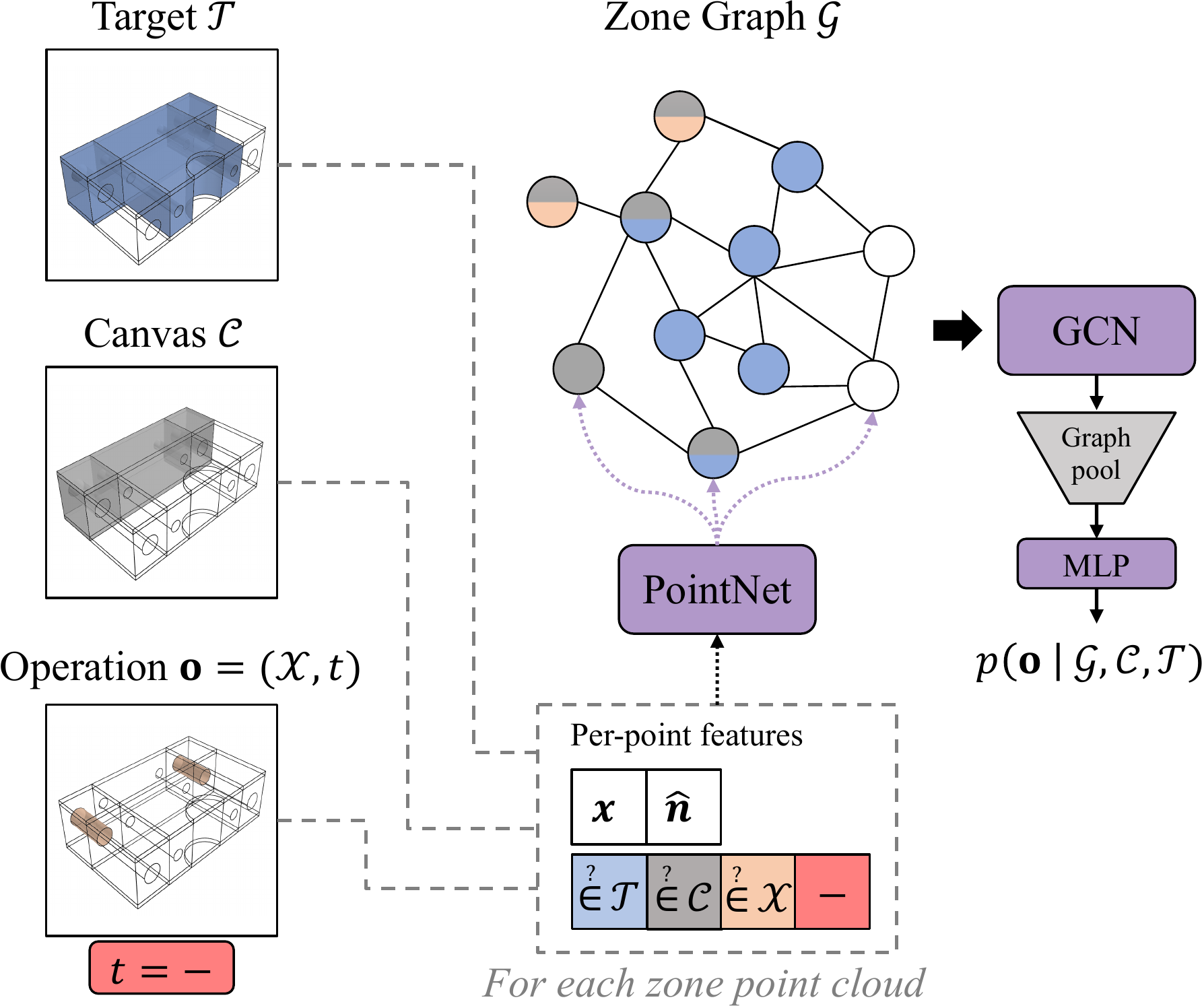}
    \caption{
    Architecture of our search proposal scoring network.
    Each zone is represented as a point cloud with positions $\mathbf{x}$ and normals $\mathbf{\hat{n}}$; labels for whether the zone is part of the target shape $\target$, current canvas $\canvas$, or proposed extrusion $\extrusion$; and the type $t$ of the proposed extrusion.
    Zones are encoded via a PointNet~\cite{qi2017pointnet} to form the input node feature vectors for a graph convolution network (GCN).
    The output node vectors from the GCN are pooled and fed through an MLP to produce the final network output.
    }
    \label{figure:network_arch}
\end{figure}

\paragraph{Network architecture}
Figure~\ref{figure:network_arch} shows our network architecture, which is a message passing graph convolutional network (GCN)~\cite{NeuralMessagePassing}.
The node features are derived from the geometry of the zone it represents and the information in the target shape $\target$, canvas $\canvas$, and proposed extrusion $\op$.
Each zone is represented as a point cloud with per-point: positions $\mathbf{x}$, normals $\mathbf{\hat{n}}$; binary labels indicating whether the zone is part of $\target$, $\canvas$, and/or $\extrusion$; the type $t$ of the proposed extrusion.
These point clouds are encoded using a PointNet~\cite{qi2017pointnet} to produce the GCN initial node vectors.
After 3 rounds of message passing, node vectors are aggregated via global max pooling and fed into a 3-layer MLP to produce the final output probability.
The supplemental material contains ablations on the number of message passing rounds and the influence of point cloud features.

\paragraph{Training}
To create training data, we use the zone graphs constructed from modeling sequences in Section~\ref{sec:zonegraph}.
We use 3000 sequences for training and hold out 440 for testing.
Each step in each sequence is one training datum.
One could treat the ``ground truth'' (GT) extrusion used at this step as a positive example, all others as negative examples, and minimize a binary cross entropy loss.
This approach yields poor performance: there are often multiple approaches to construct a shape, and an approach that is not taken for one shape in the dataset may be taken for another.
Treating all non-GT extrusions as negative examples thus confuses the network.
We instead use ternary labels: positive, negative, and \emph{neutral}.
Positive labels are assigned to GT extrusions.
Non-GT extrusions are labeled as neutral or negative via Monte Carlo tree search.
For each extrusion, we perform $N$ random modeling sequence completions starting with that extrusion ($N =$ the number of remaining steps in the ground truth sequence) and record the percentage $p$ of these completions which yield a match to the target shape.
If any extrusion has $p=0$, that extrusion is labeled as a negative example.
Otherwise, if all extrusions have $p>0$, the one with the smallest $p$ is labeled as negative.
All other extrusions are labeled as neutral.
We then minimize a focal loss \cite{8237586} using only positive and negative examples.

\paragraph{Heuristic Ranking}
We also considered whether heuristics could work in our setting.
The best heuristic we found is one that executes a candidate extrusion and, like IoU, penalizes the filled zones that the resulting canvas and target do not have in common: $\frac{|\zones| - (|\target \cup \canvas| - |\target \cap \canvas|)}{|\zones|}$. 
Ties are broken using volumetric IoU between canvas and target.
This performs considerably better than random but not as well as our network; the next section provides more detail.

%% file: 04-results.tex
\section{Results \& Evaluation}

To evaluate our method's performance, we are interested in three questions:
\begin{enumerate}
\denselist
    \item How consistent are our method's output sequences with sequences it was trained on?
    \item How well does our method reconstruct new input shapes, given a particular compute budget?
    \item How desirable are our method's output sequences, as judged by CAD designers?
\denselist
\end{enumerate}
For (1), we examine how our network scores the modeling steps used in its training data (Section~\ref{sec:results_ranking}).
For (2), we quantify our method's tradeoff between computation budget and reconstruction accuracy (Section~\ref{sec:results_ablation}) and compare to a recent CSG inference system (Section~\ref{sec:results_comparison}).
For (3), we conduct a perceptual study (Section~\ref{sec:results_perceptual}).

\subsection{Search Proposal Ranking}
\label{sec:results_ranking}

\begin{figure}[t!]
    \centering
    \includegraphics[width=0.99\linewidth]{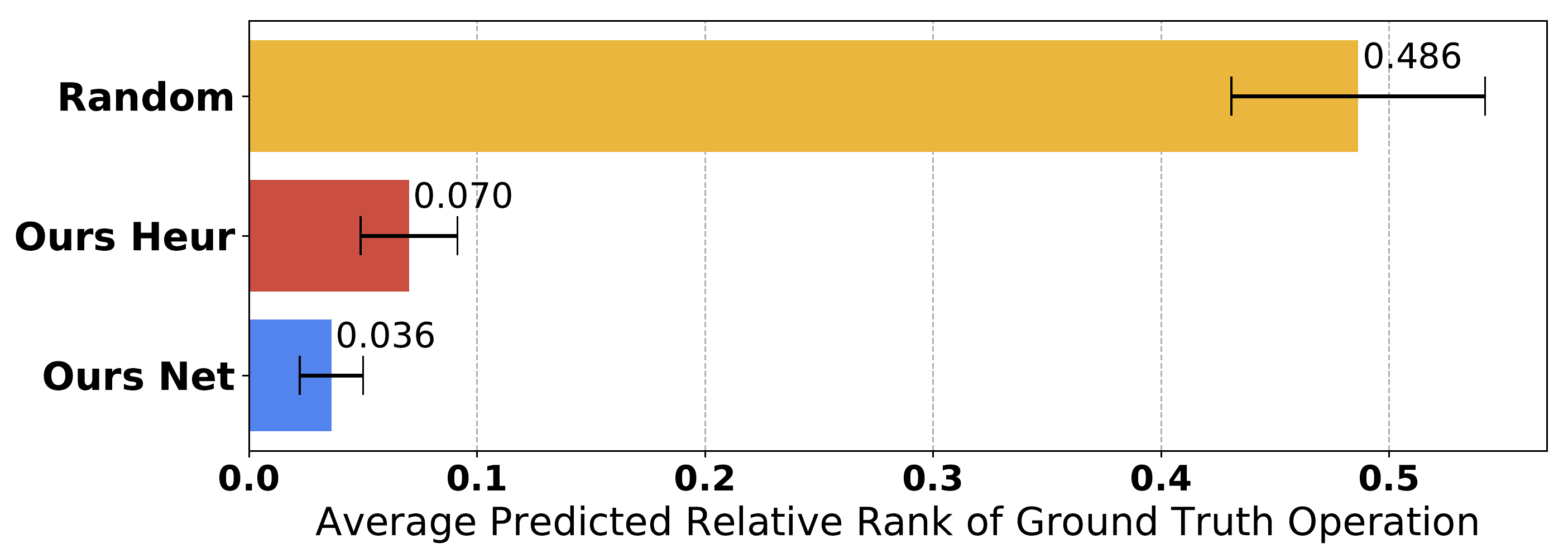}
    \caption{
    Comparing how different methods rank the ground truth extrusions used in modeling sequences from the Fusion 360 dataset.
    The x axis is $\frac{\text{predicted rank of ground truth}}{\text{number of candidate extrusions}}$ averaged over all steps in all test sequences.
    Lower is better.
    90\% confidence intervals are shown as error bars.
    \emph{Ours Net} performs about 2x better than \emph{Ours Heur}.
    }
    \label{figure:ranking_performance}
\end{figure}

\begin{figure*}[t!]
    \centering
    \setlength{\tabcolsep}{1pt}
    \begin{tabular}{c cr cccccc}
        Target & & & \multicolumn{6}{c}{Inferred Sequence}
        \\
        \raisebox{2.9em}{\multirow{3}{*}{\includegraphics[width=0.28\linewidth]{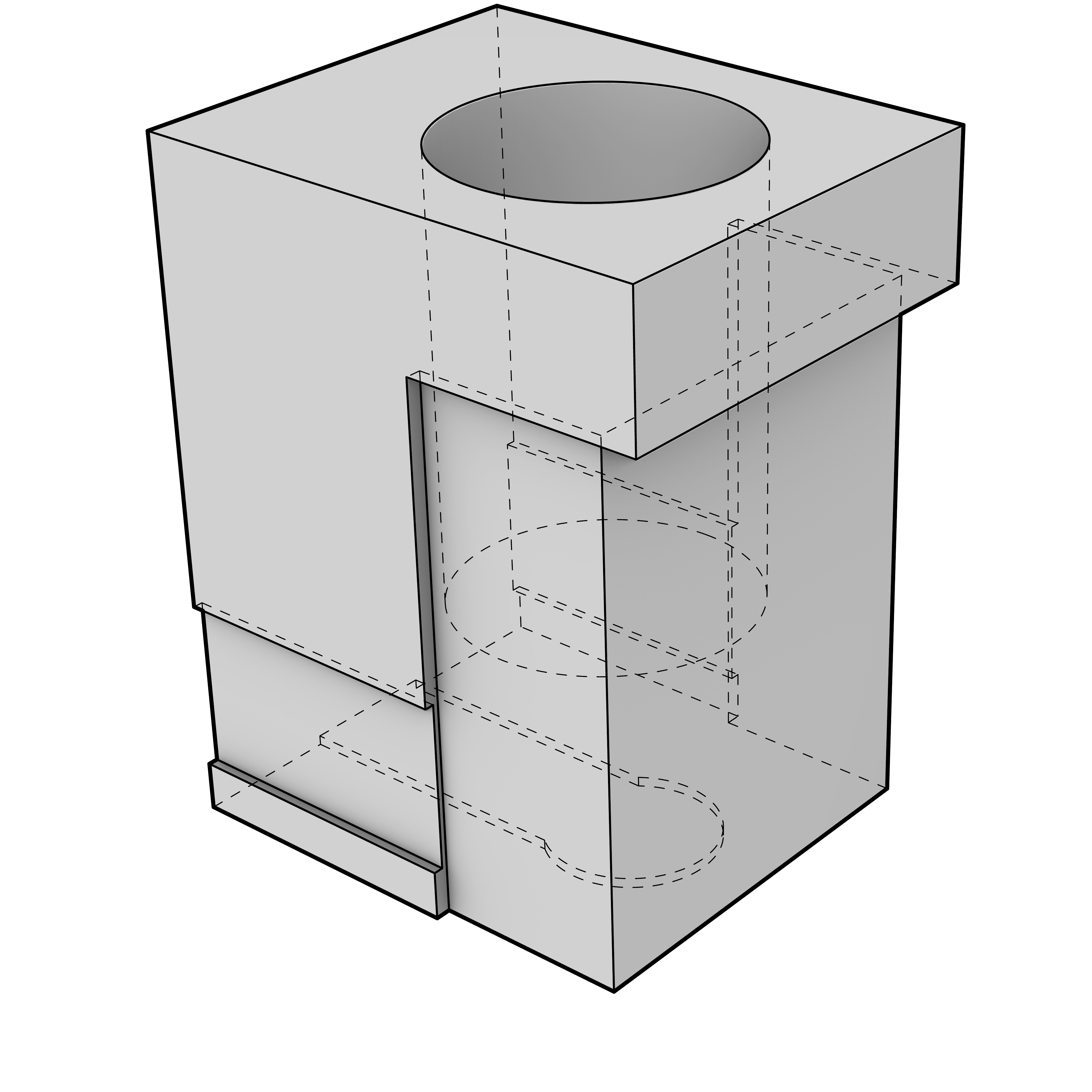}}} &
        \qquad & \raisebox{1.2em}{Random} &
        \includegraphics[width=0.08\linewidth]{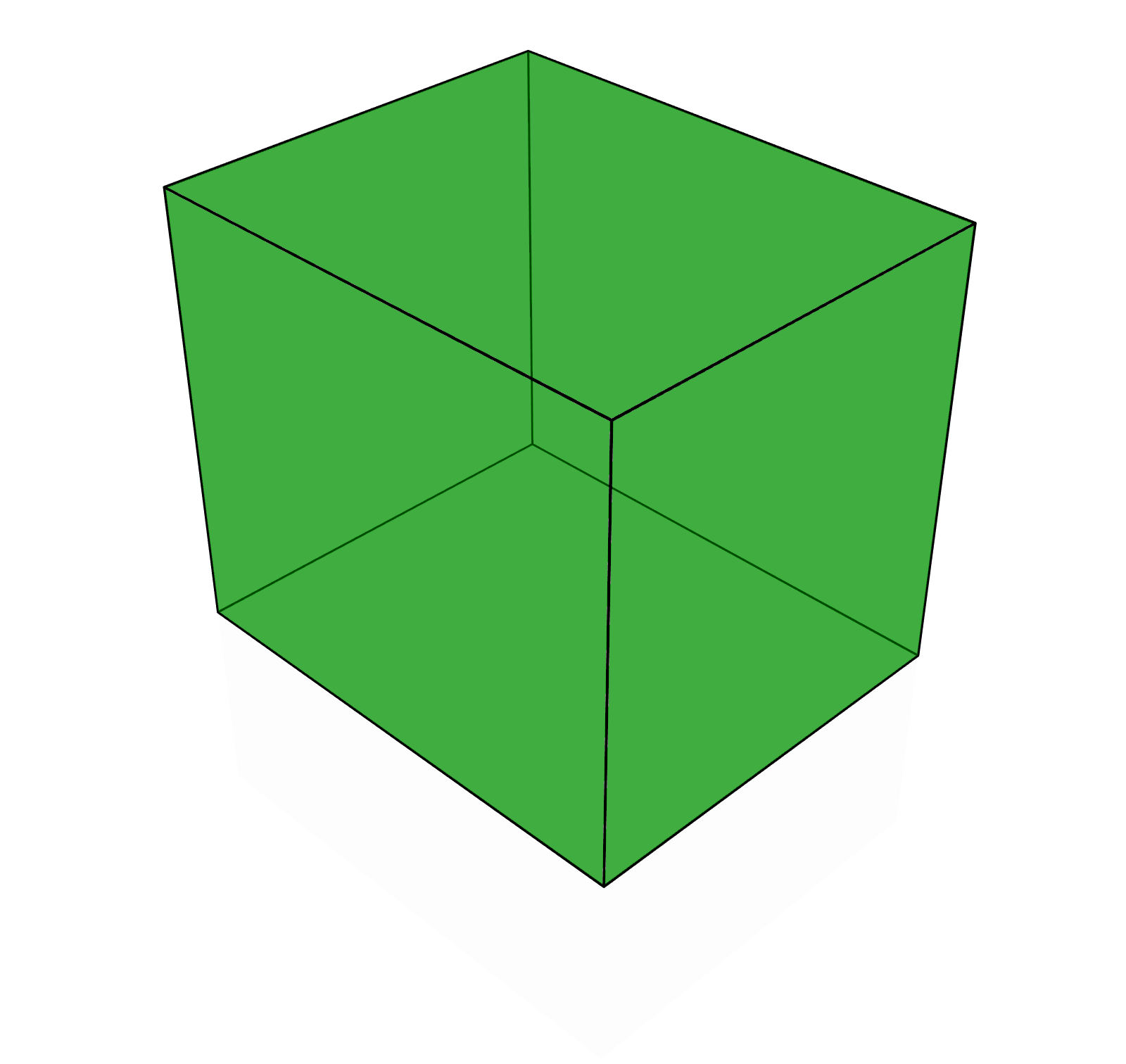} &
        \includegraphics[width=0.08\linewidth]{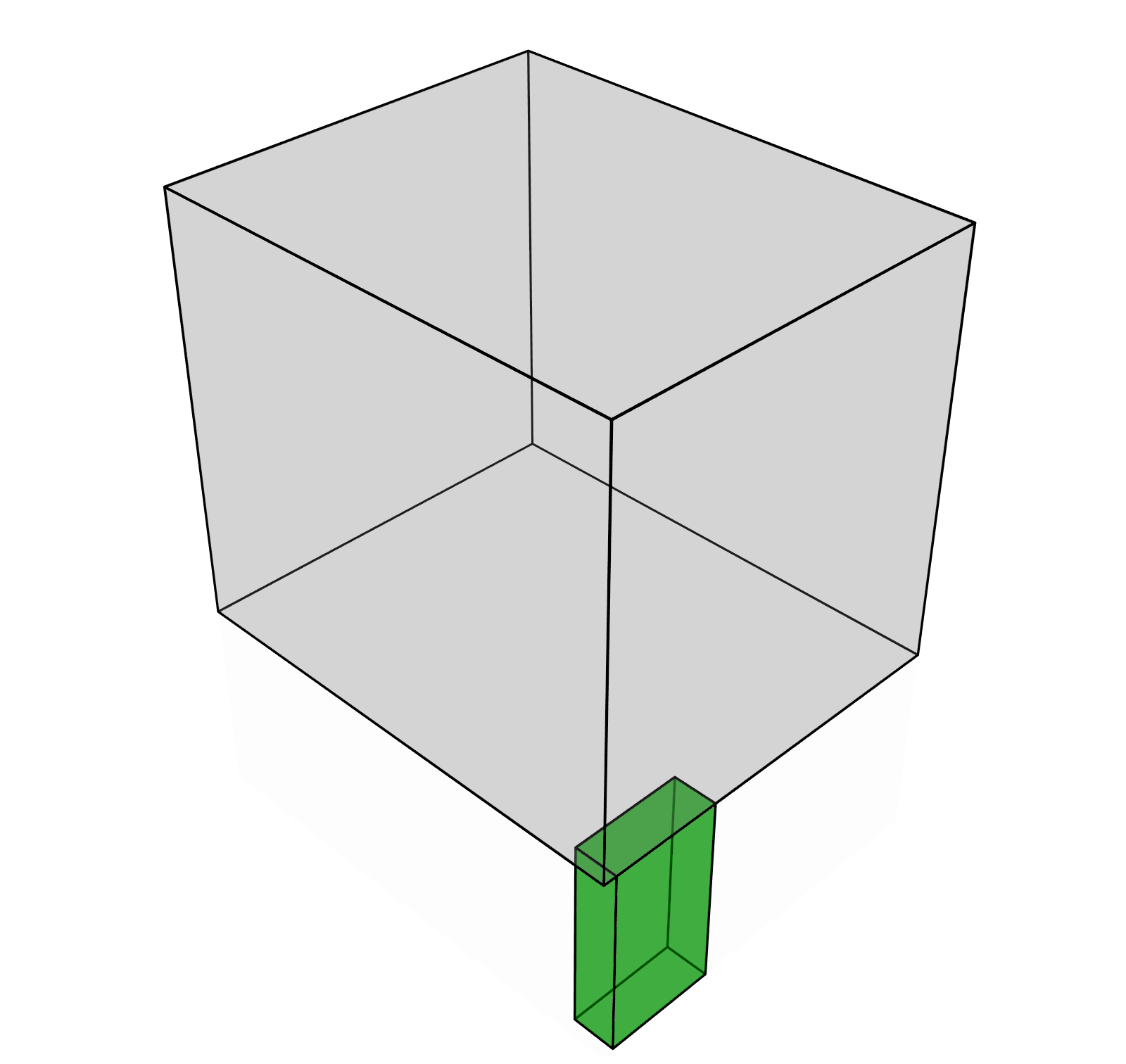} &
        \includegraphics[width=0.08\linewidth]{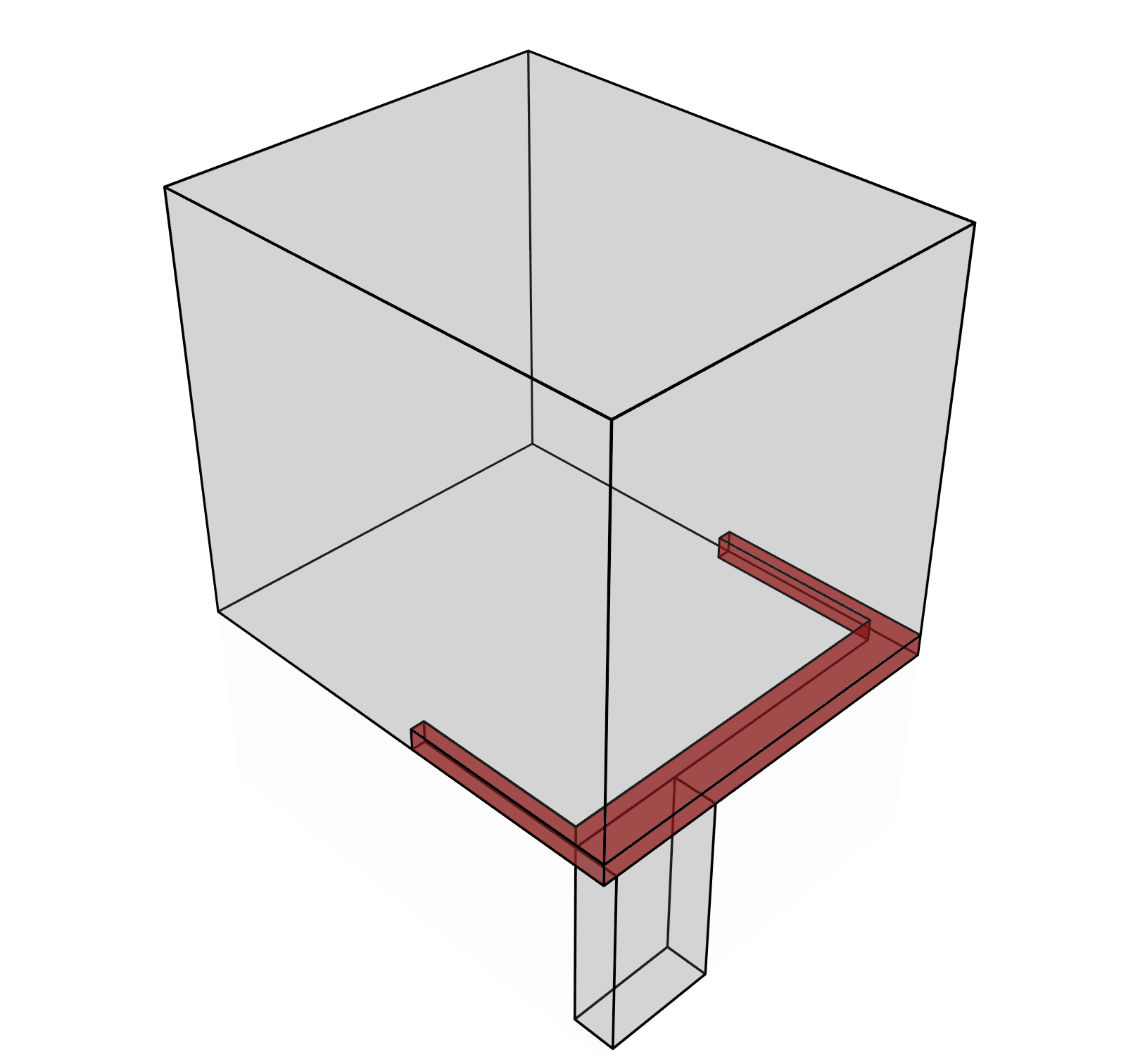} &
        \includegraphics[width=0.08\linewidth]{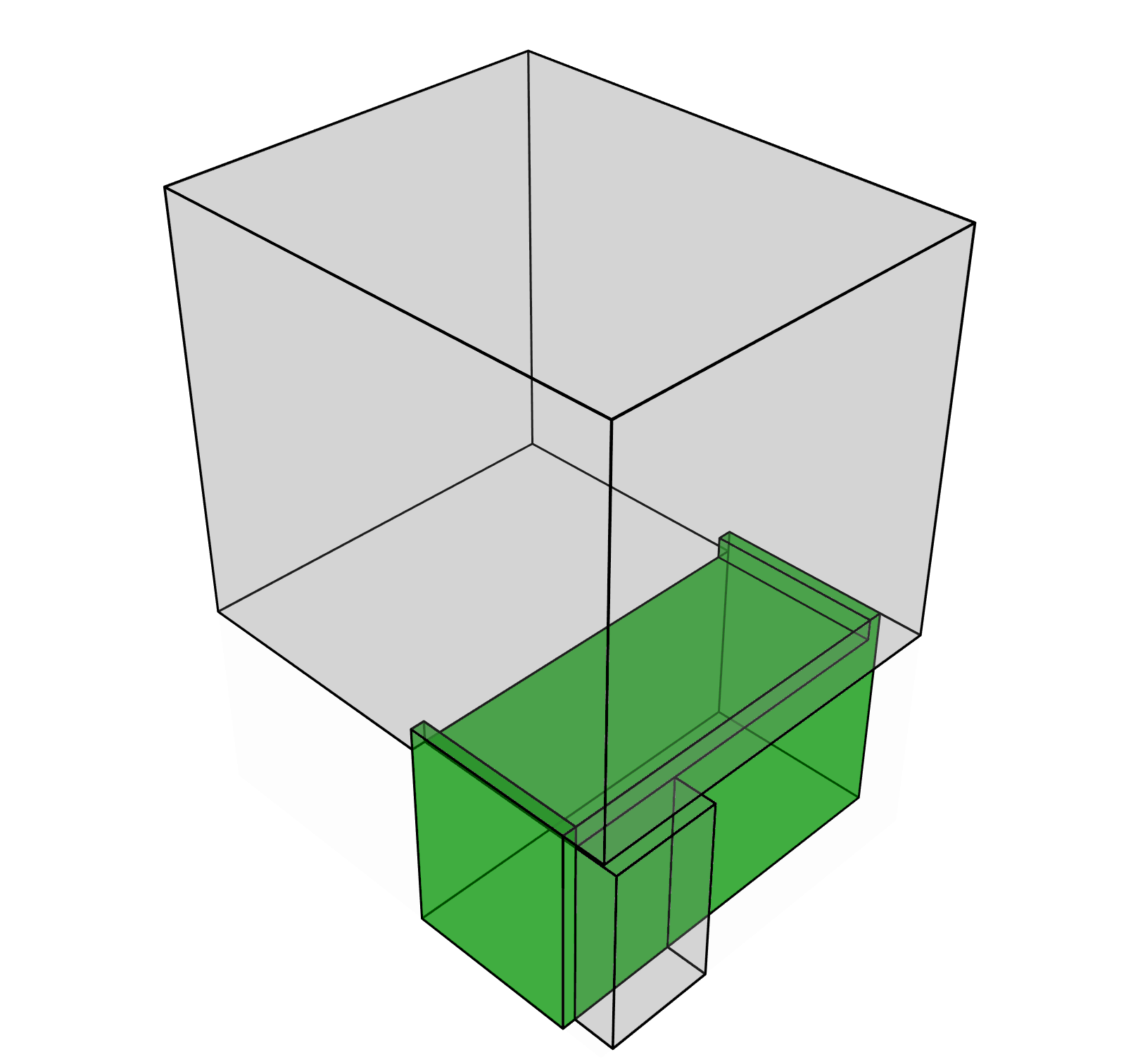} &
        \includegraphics[width=0.08\linewidth]{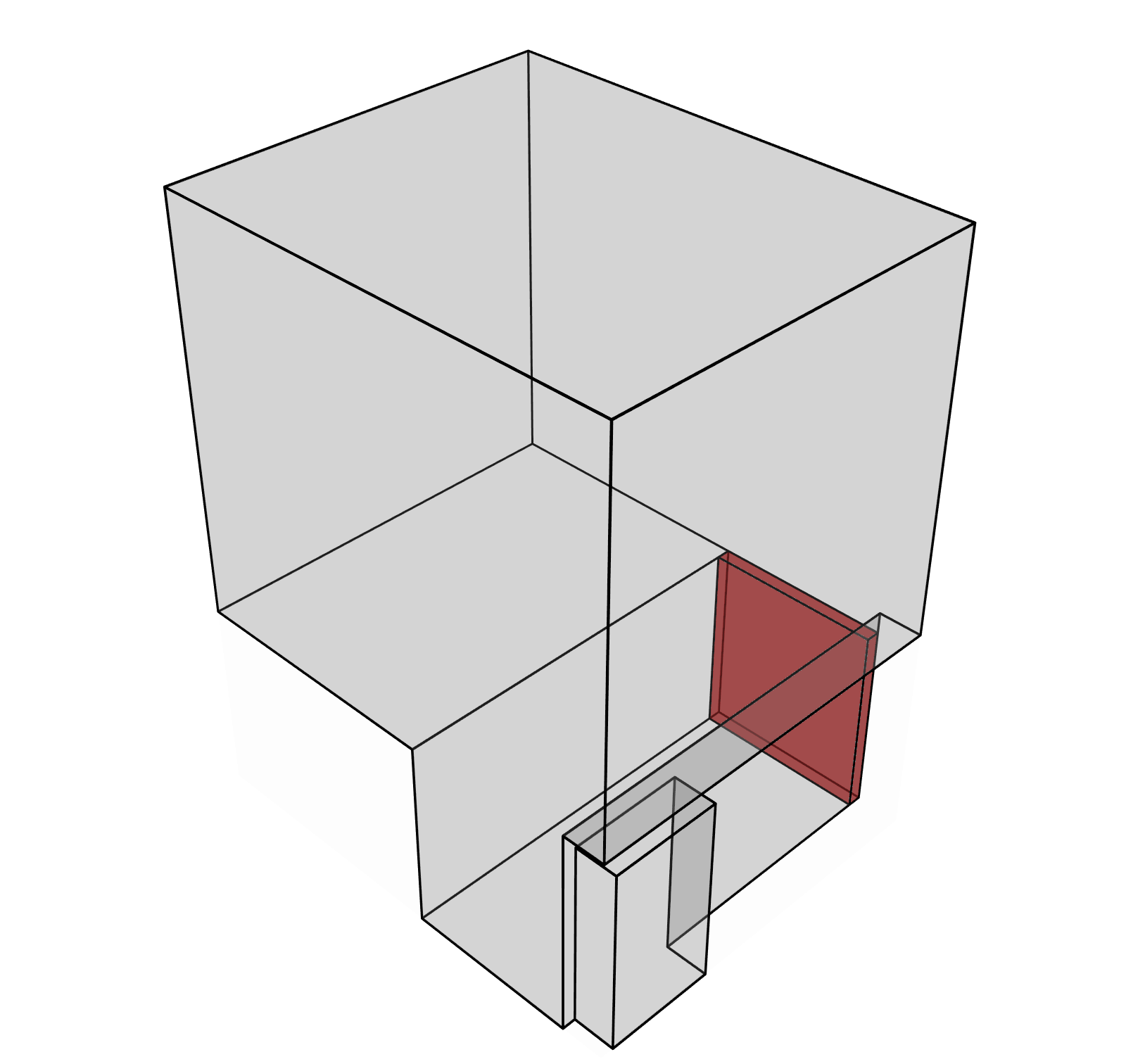} &
        \includegraphics[width=0.08\linewidth]{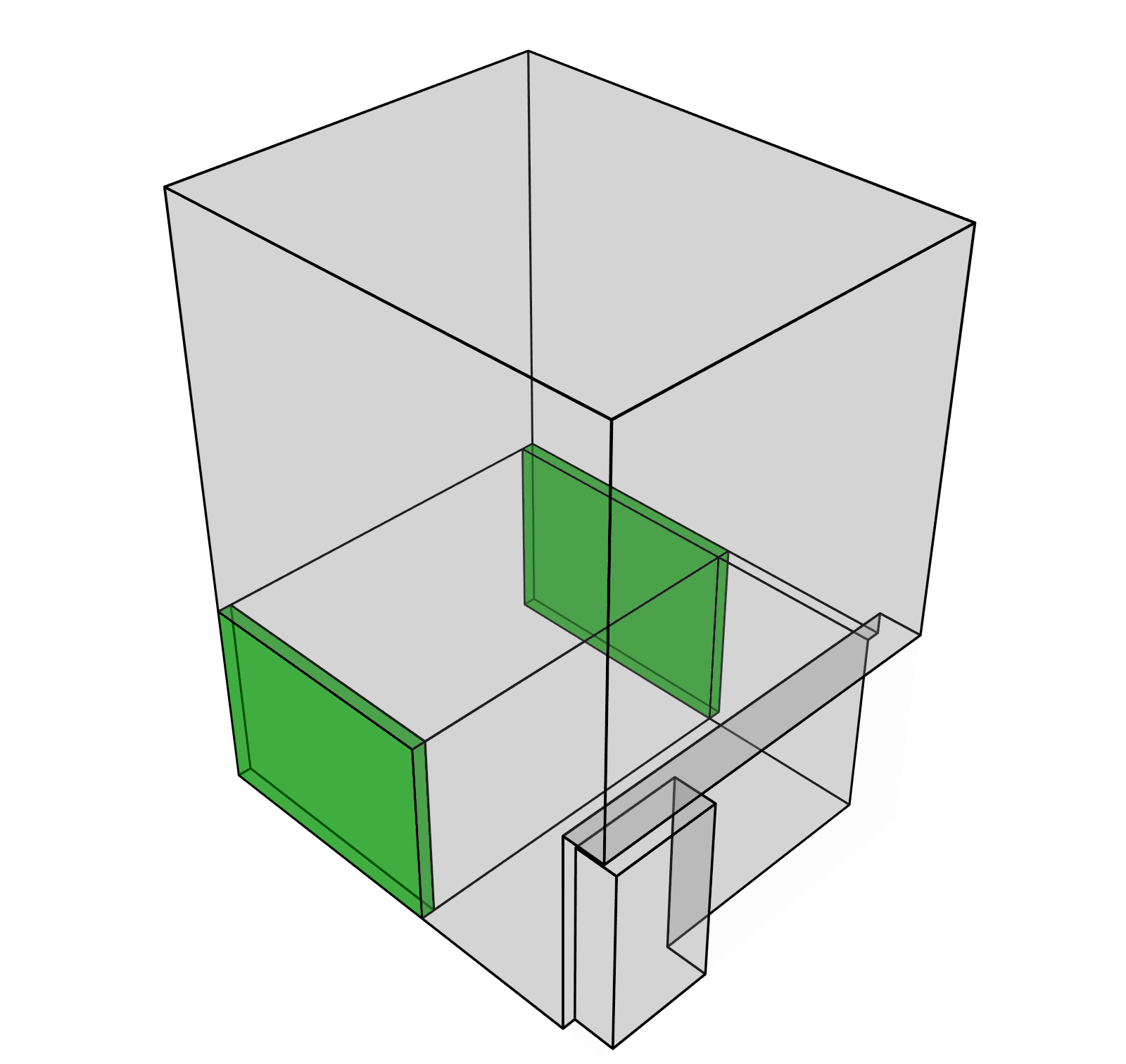}
        \\
        & \qquad & \raisebox{1.2em}{Ours Heur} & 
        \includegraphics[width=0.08\linewidth]{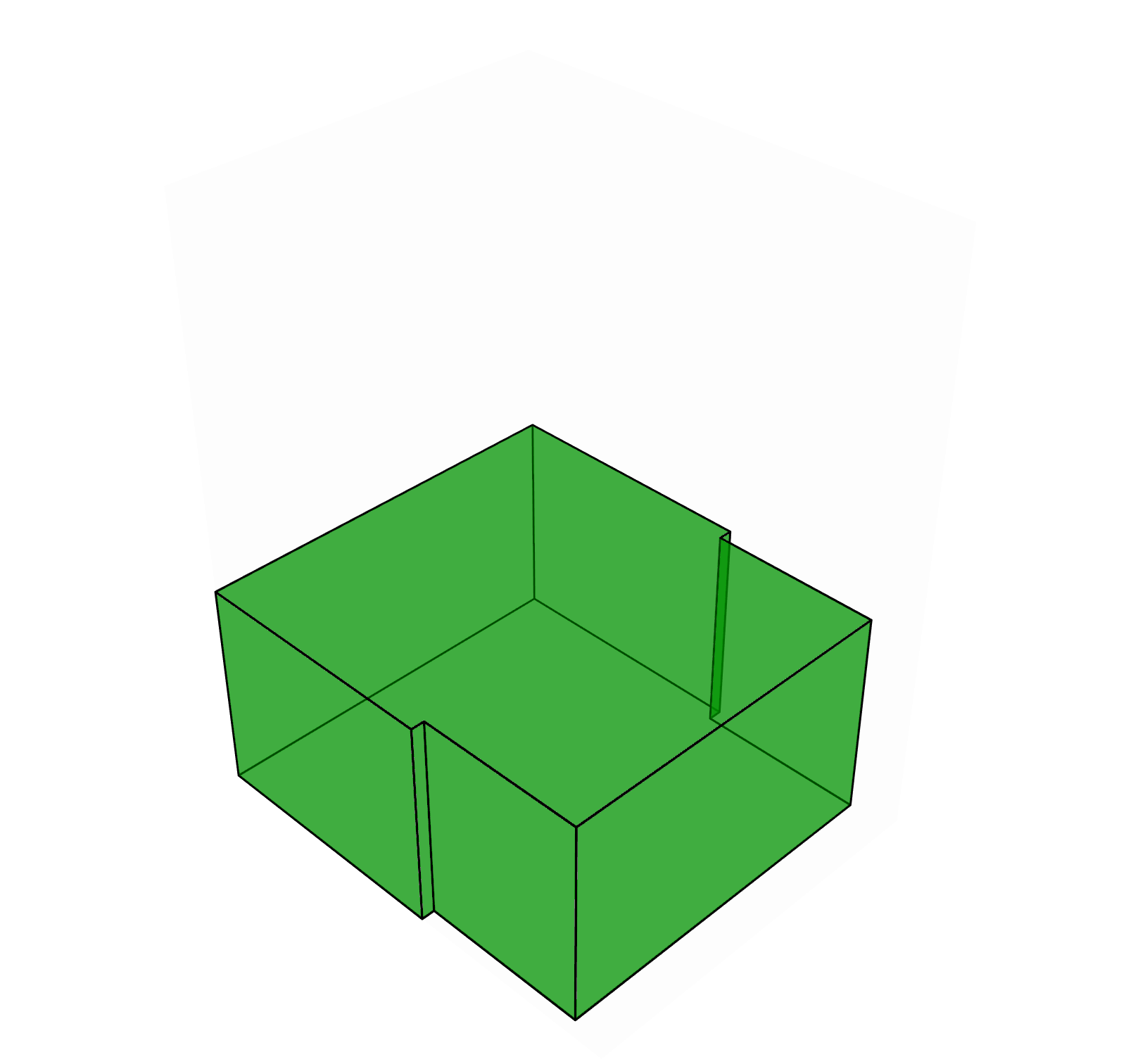} &
        \includegraphics[width=0.08\linewidth]{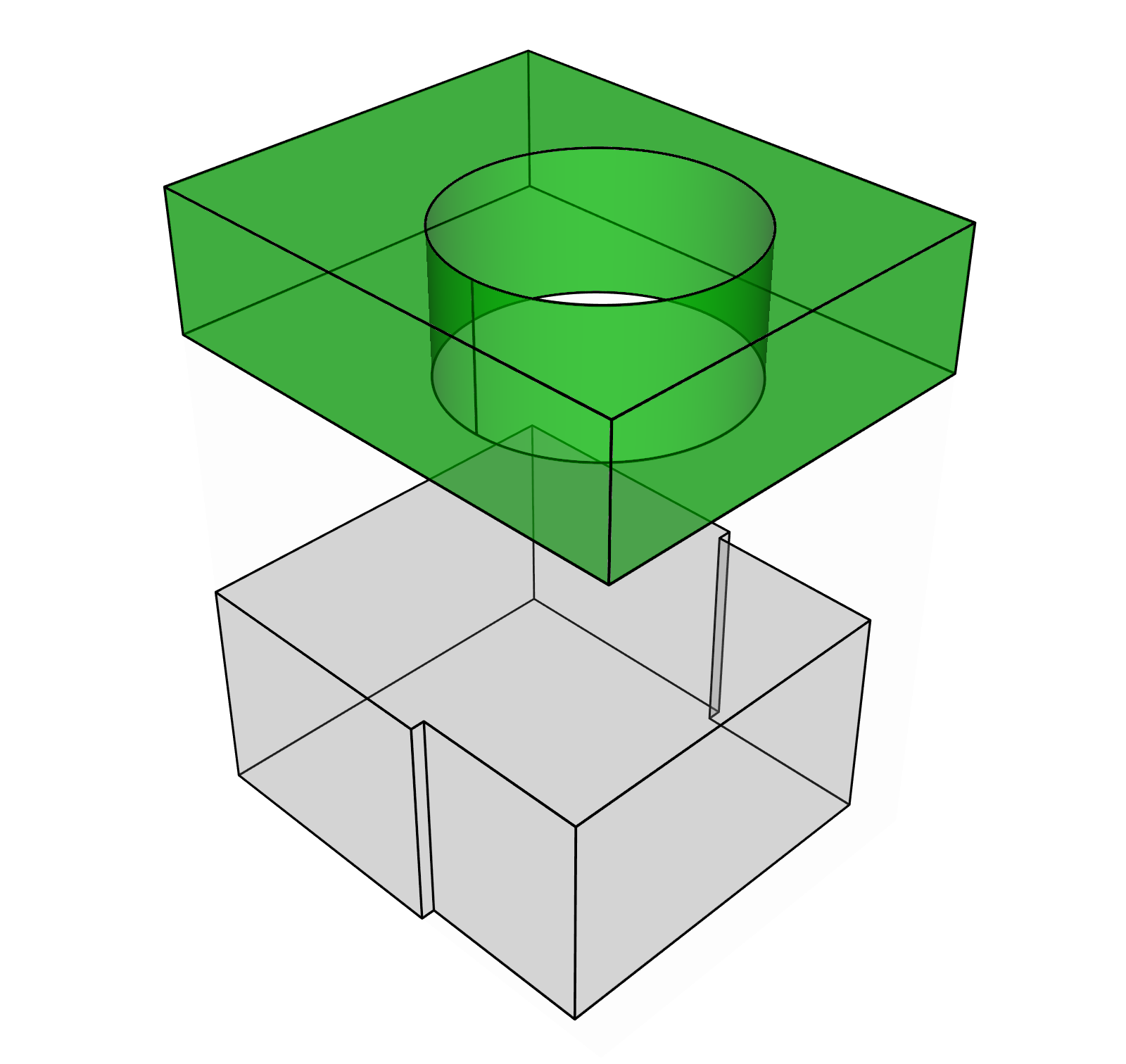} &
        \includegraphics[width=0.08\linewidth]{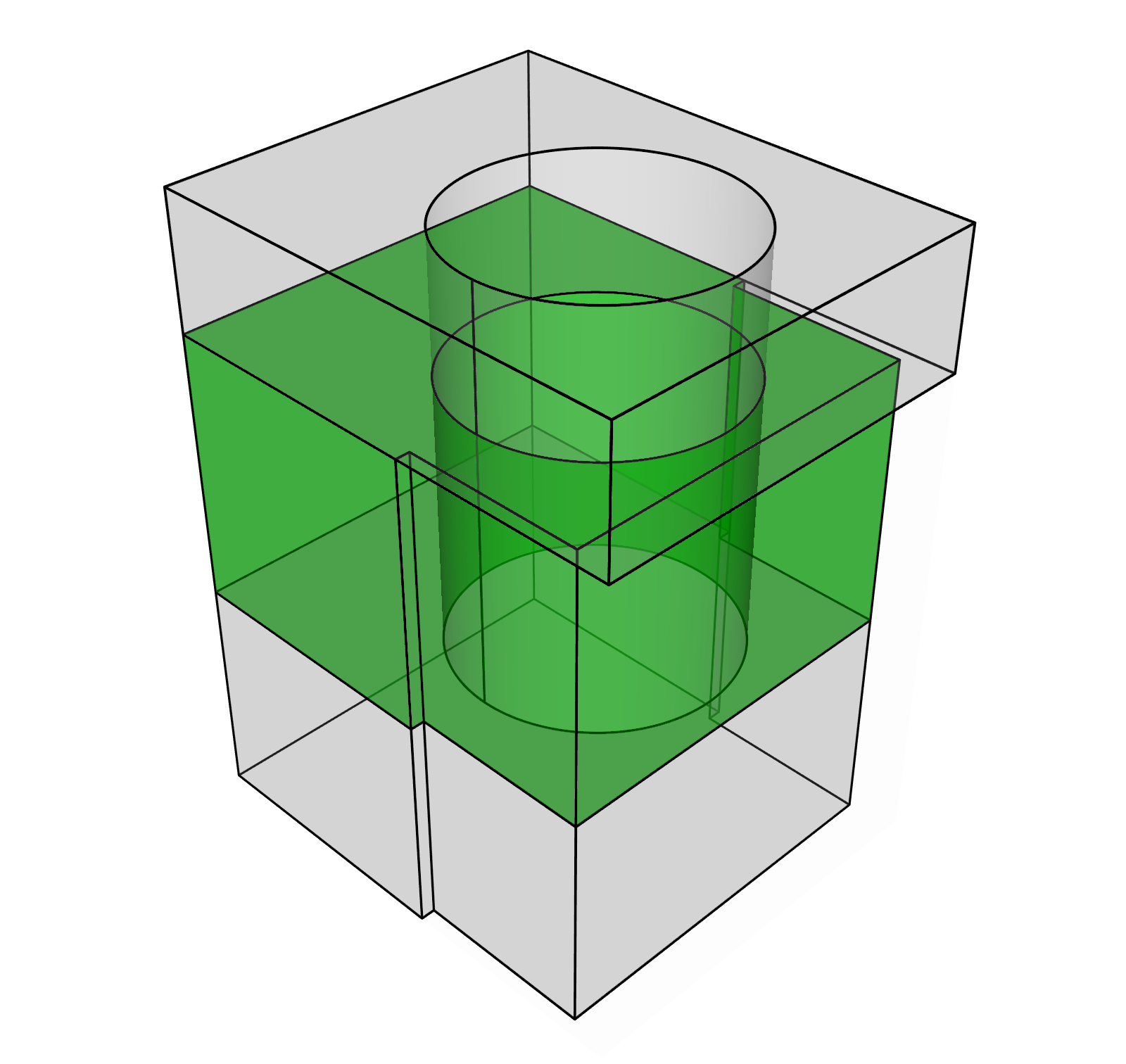} &
        \includegraphics[width=0.08\linewidth]{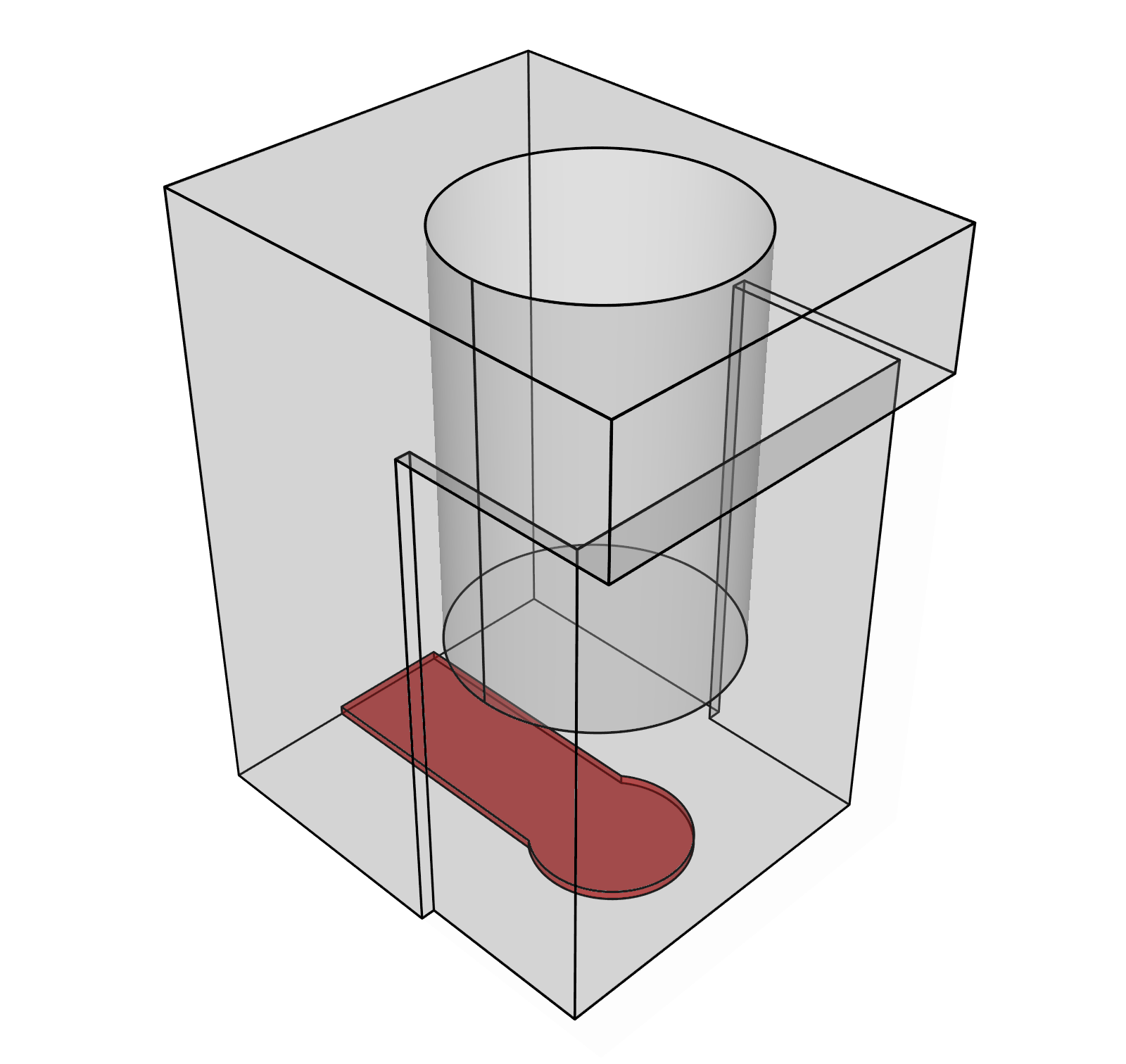} &
        \includegraphics[width=0.08\linewidth]{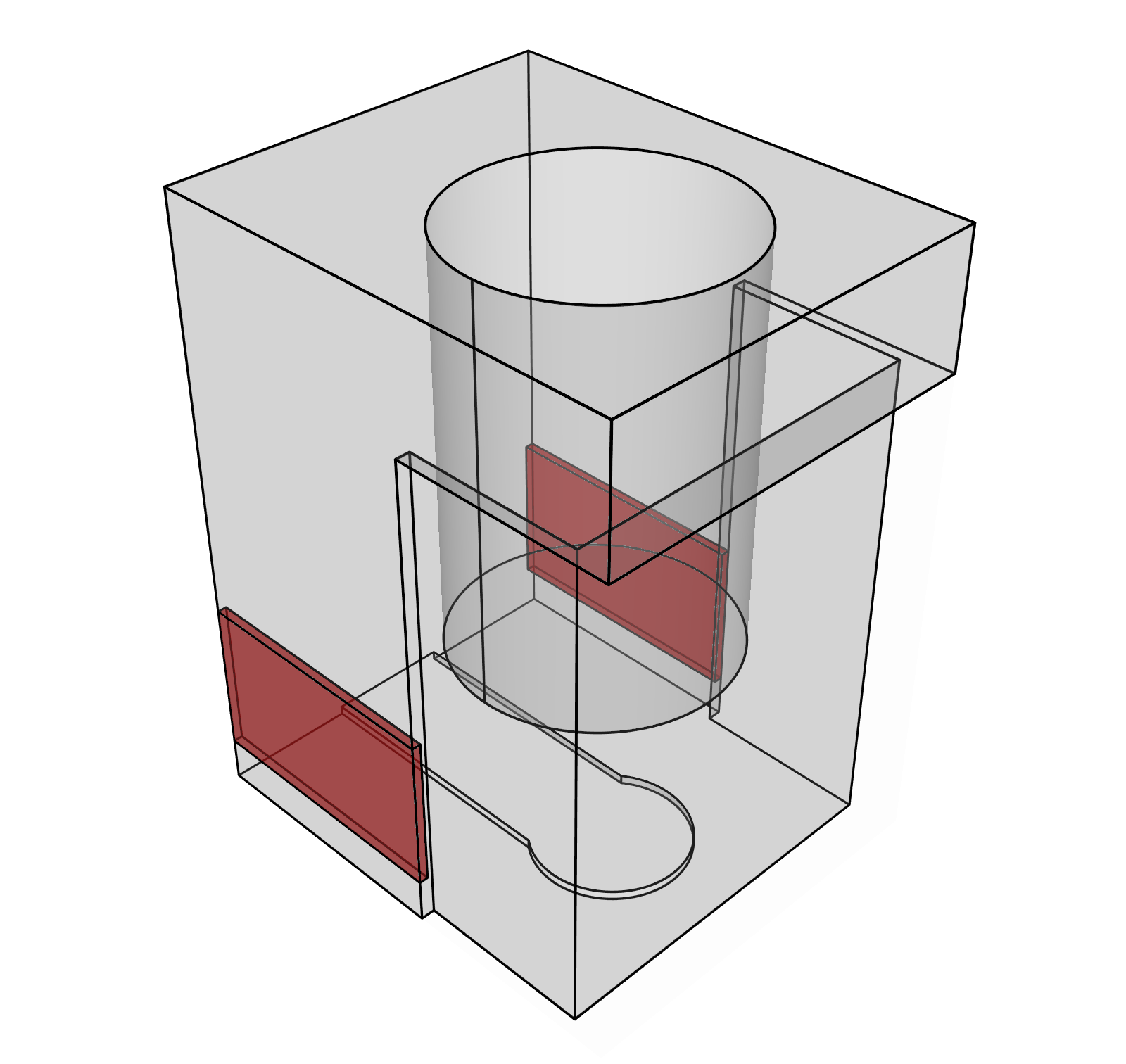} &
        \\
        & \qquad & \raisebox{1.2em}{Ours Net} & 
        \includegraphics[width=0.08\linewidth]{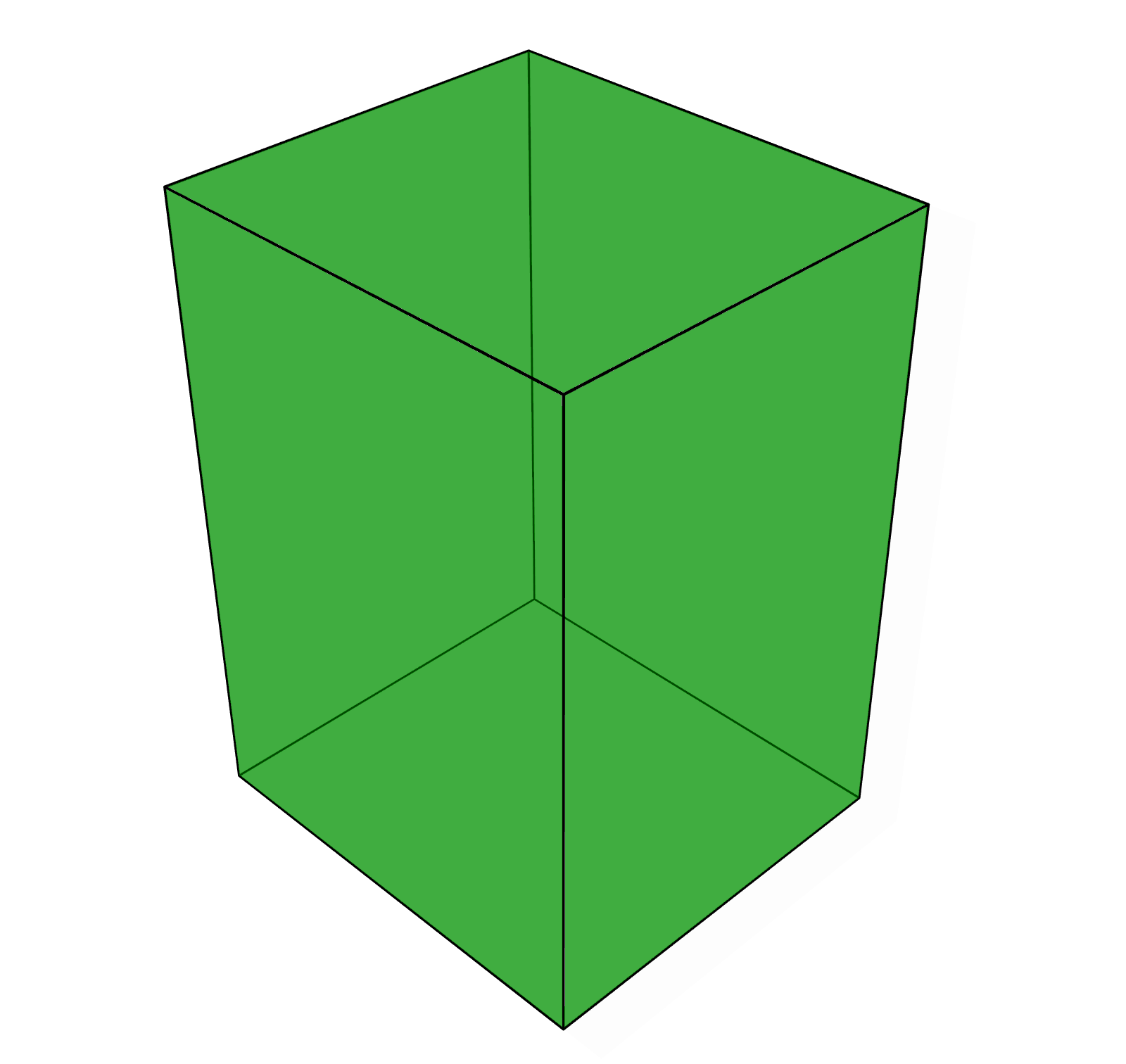} &
        \includegraphics[width=0.08\linewidth]{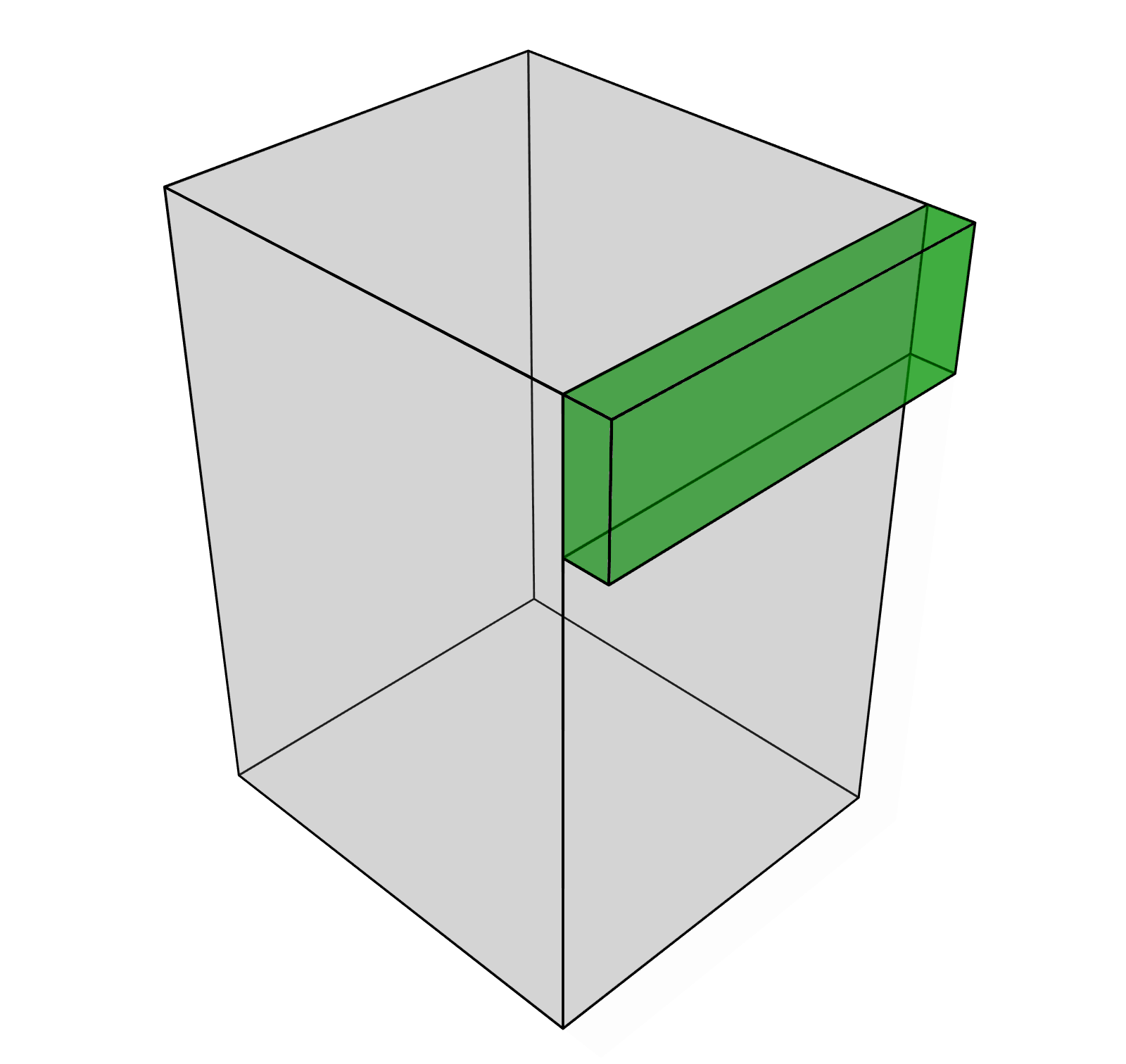} &
        \includegraphics[width=0.08\linewidth]{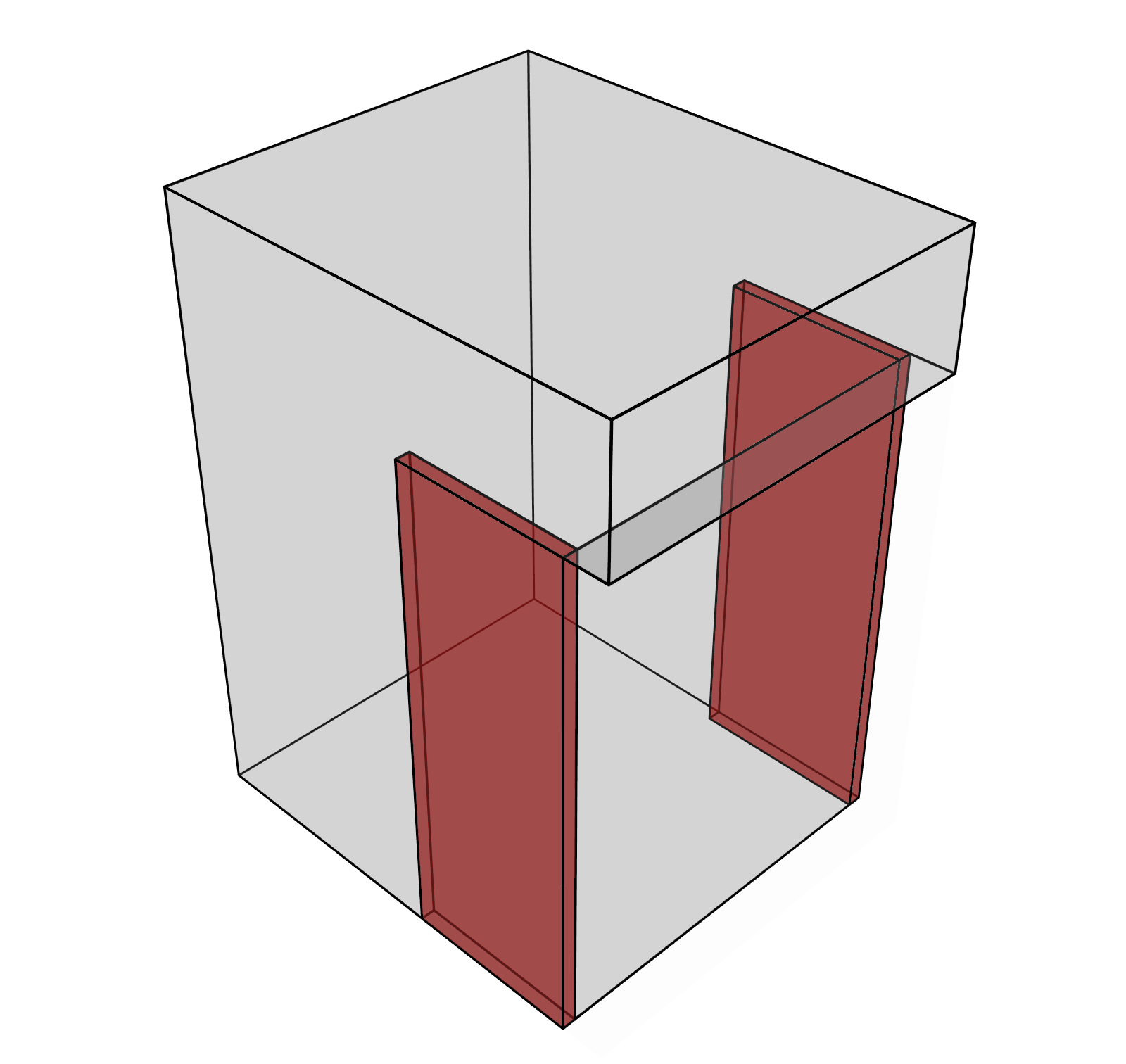} &
        \includegraphics[width=0.08\linewidth]{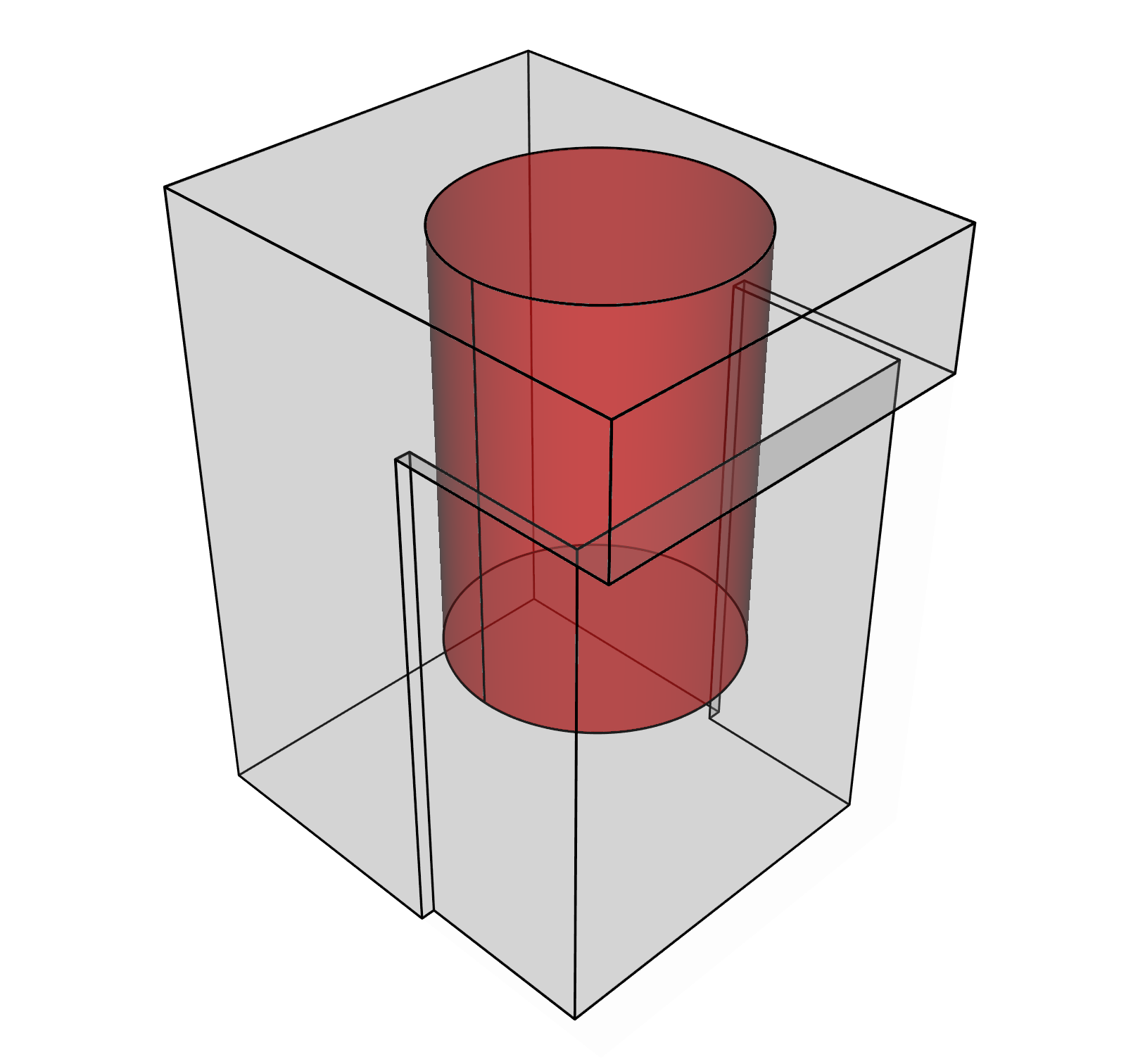} &
        \includegraphics[width=0.08\linewidth]{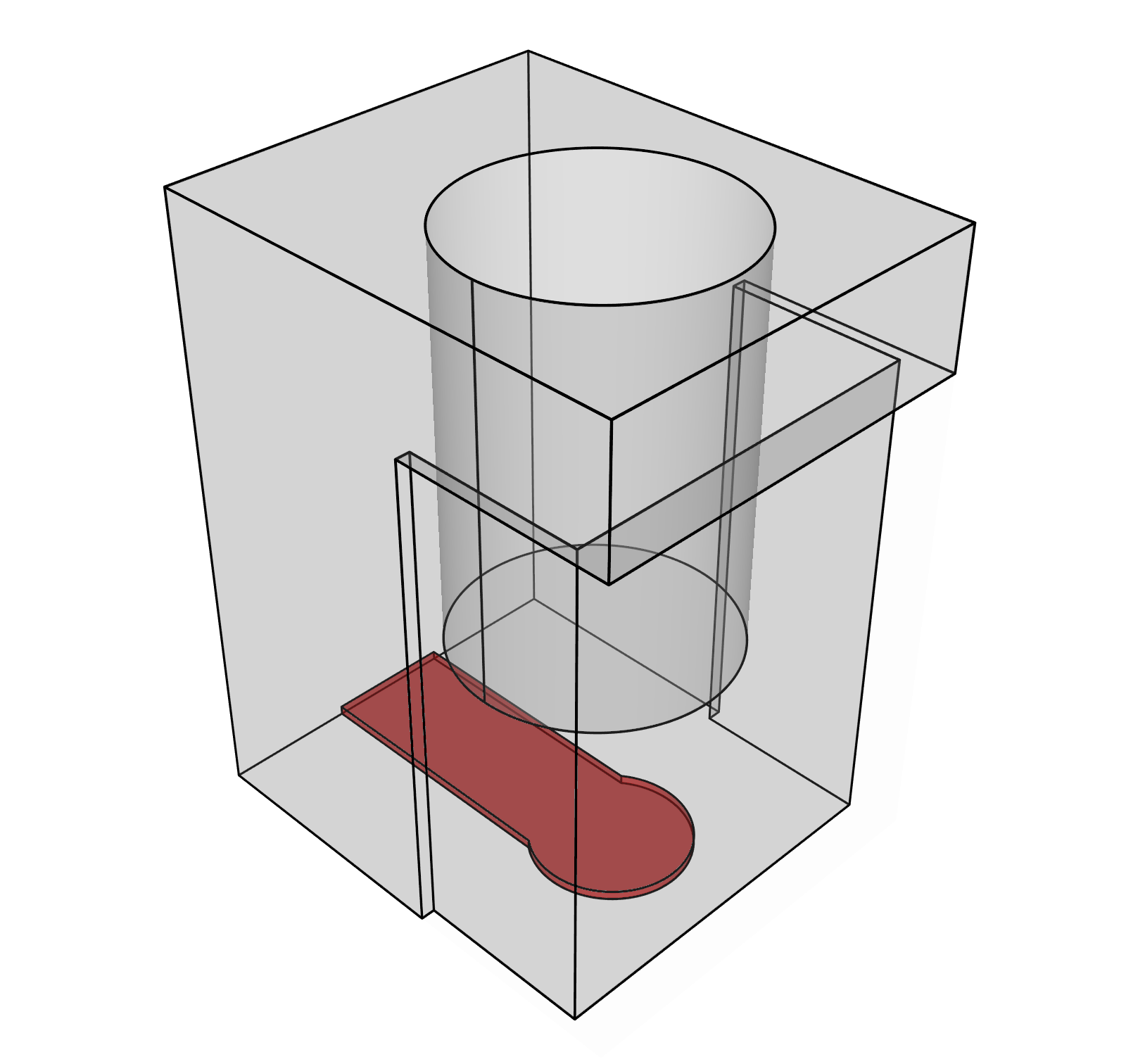} &
        \includegraphics[width=0.08\linewidth]{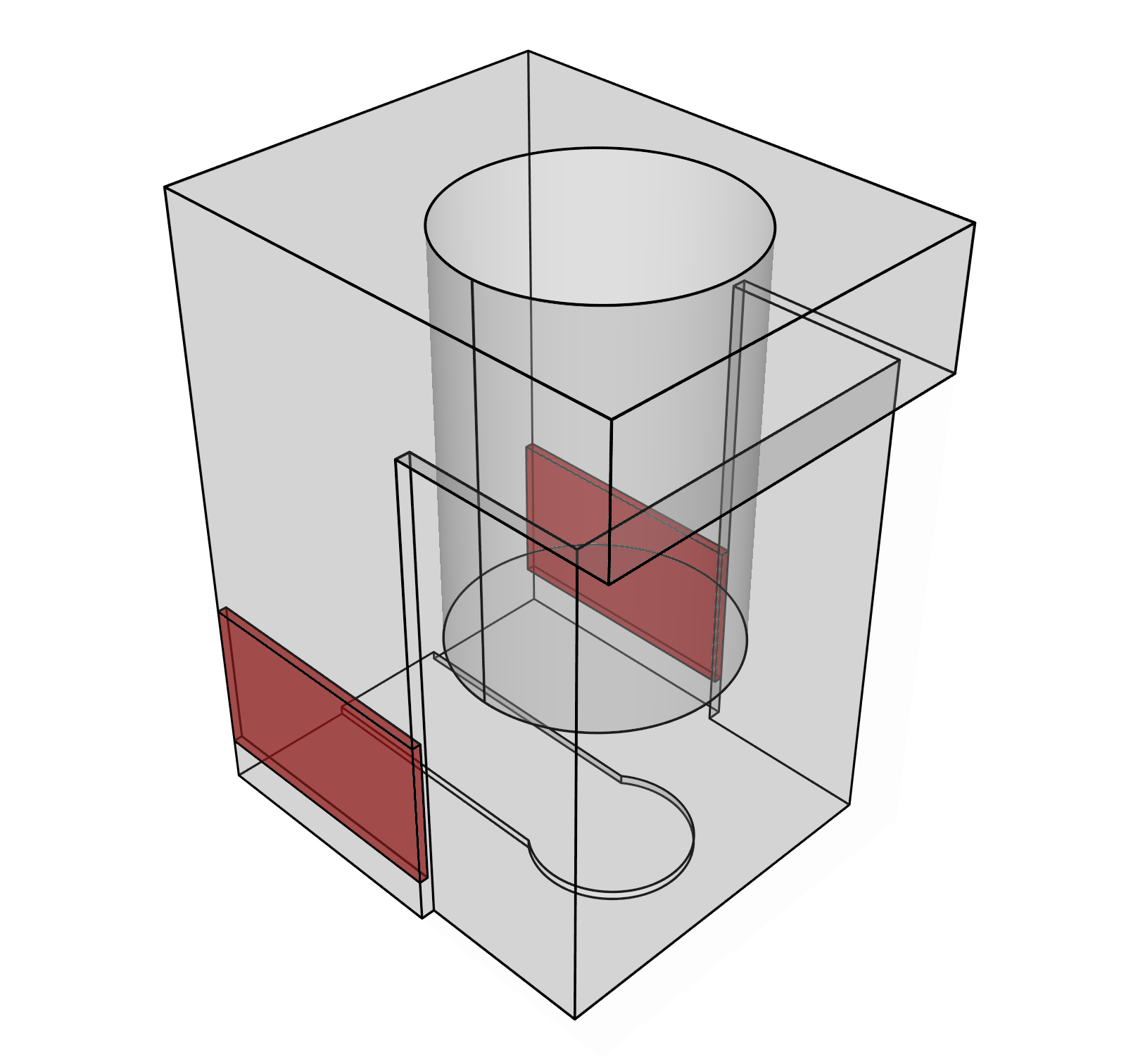}
    \end{tabular}
    \vspace{0.2em}
    \caption{Modeling sequences inferred under different search guidance. Green: addition; Red: subtraction; Grey: current.
    }
    \label{figure:example_sequences}
\end{figure*}

If our network has learned the design patterns within its training data, then for each step in a dataset sequence, when candidate extrusions are ordered by predicted score, the ``ground truth'' extrusion should be ranked near the top.

Figure~\ref{figure:ranking_performance} shows the average relative rank of the ground truth extrusion computed by our network and other baselines.
The evaluation was conducted on the held-out test sequences from the Fusion 360 Gallery reconstruction dataset~\cite{willis2020fusion}.
We compare our network (\textit{Our Net}) against the heuristic (\textit{Our Heur}), as well as a method which picks a random extrusion (\textit{Random}).
Our network and the heuristic both significantly outperform the random baseline, and the network also dominates the heuristic by a factor of two.
Search guided by our network should thus mimic the design patterns in the dataset.
Ablation studies are shown in the supplemental material.

Figure~\ref{figure:example_sequences} shows modeling sequences inferred under different guidance (random, heuristic, network) for the same input shape.
While the heuristic produces the shortest sequence, the network produces a more intuitive one.

\subsection{Reconstruction Performance Ablation Study}
\label{sec:results_ablation}

We next evaluate how well the modeling sequences inferred by our method reconstruct the input shape (via volumetric IoU).
As stated in Section~\ref{sec:search}, if a sketch + extrude + boolean sequence that reconstructs a shape exists, our method is guaranteed to find it given unbounded time; here we are interested in our method's performance under finite time constraints.
Timings were collected on a machine with a GeForce RTX 2080 Ti GPU and an Intel i9-9900K CPU.

\begin{figure}[t!]
    \centering
    \includegraphics[width=0.99\linewidth]{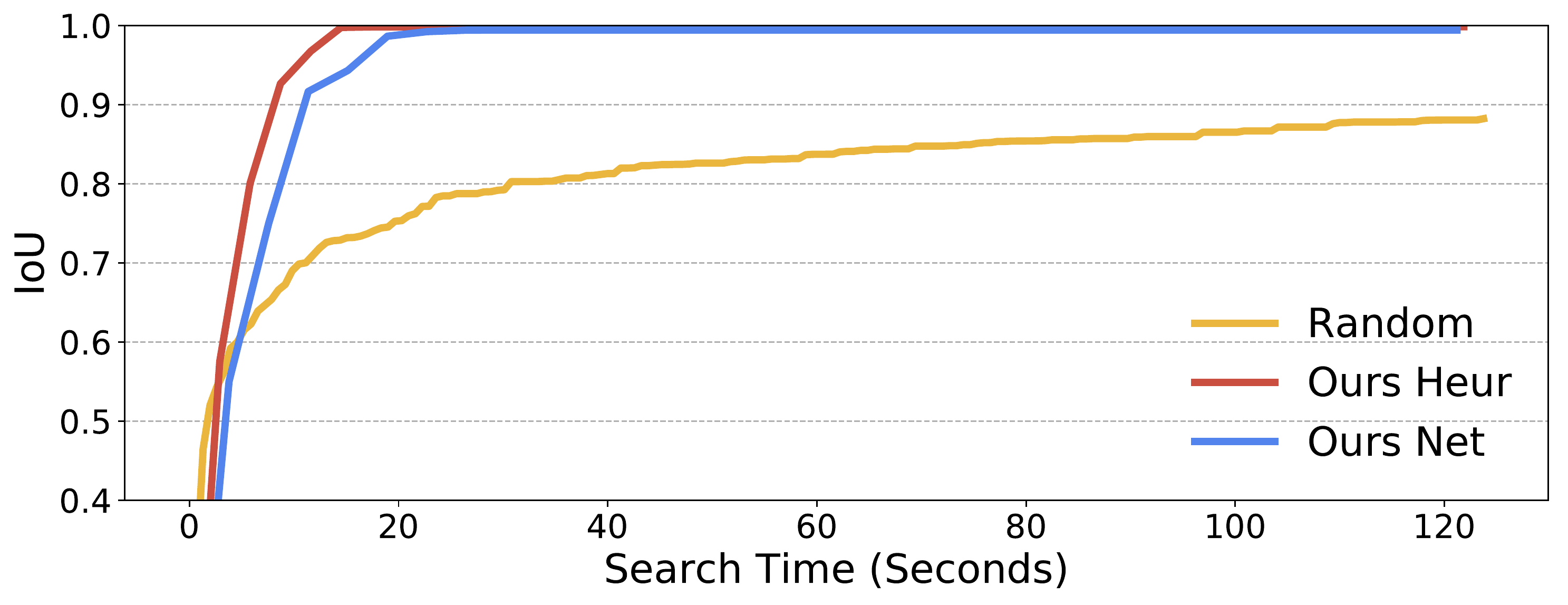}
    \caption{Reconstruction accuracy of the outputs of inferred programs vs. the time used to infer them.
    }
    \label{figure:recon_vs_time}
\end{figure}

Figure~\ref{figure:recon_vs_time} compares the average IoU of inferred programs on our held-out shapes (with a ground truth modeling sequence length of at least 3) using random, heuristic, and network guidance.
Zone graph construction time is excluded, as this is a fixed cost incurred by all methods.
Both network and heuristic guidance converge to 100\% reconstruction accuracy after $\sim 20$ seconds, whereas random guidance does not converge.
Heuristic guidance is faster than the network because (a) it does not incur the cost of network evaluation, and (b) by construction it tries to use the fewest possible modeling operations.
However, shorter sequences are not always better (e.g. Figure~\ref{figure:example_sequences}).
Ablation studies for different search widths $k$ are shown in the supplemental material.

\subsection{Comparison to InverseCSG}
\label{sec:results_comparison}

We next compare to InverseCSG~\cite{du2018inversecsg}, a recent system for inferring CSG programs from complex input shapes.
We evaluate on InverseCSG's test set of 50 3D shapes.
As B-rep files are not provided, we hired freelance CAD designers to reproduce them using Autodesk Fusion 360. Our method successfully build zone graphs for 33 of these, which we use as our test set. From the test set, 27 shapes are expressible using sketch + extrude + boolean operations, which our method can reconstruct exactly.
By contrast, InverseCSG uses parametric primitives that may introduce error when the primitive set is insufficiently expressive.
As these shapes are more complex than those in prior experiments, for our method we set the search width $k=15$ and decrease it by 0.5 for each search depth increment (giving more search options to important early steps in the modeling process).
Finally, we note that InverseCSG takes a triangle mesh as input to solve the CAD reconstruction problem in one shot, whereas we focus on the second sub-problem of inferring a modeling sequence from a B-rep.

\begin{table}[t!]
    \centering
    \footnotesize
    \begin{tabular}{l cc cc}
        \toprule
       \textbf{Method} & \multicolumn{2}{c}{\textbf{All Models}} & 
       \multicolumn{2}{c}{\textbf{Sketch + Extrude Models}} \\
       \midrule
       & \textbf{Error (\%)}   & 
       \textbf{Time (s)}  & 
       \textbf{Error (\%)}             & 
       \textbf{Time (s)}            
       \\
    Ours Net    & 0.72               & 242       & 0.23                    & 242             \\
    Ours Heur   & \textbf{0.50}               & \textbf{197}       & \textbf{0.15}                    & \textbf{197}             \\
    InverseCSG & 0.88               & 900       & 0.79                    & 900 \\
    \bottomrule
    \end{tabular}

     \caption{
     Reconstruction results comparing InverseCSG and our method using heuristic (Ours Heur) or network guided (Ours Net) search. We report the median reconstruction error computed using IoU and the median search time in seconds for all models in the InverseCSG test set and the subset of sketch + extrude models. Lower values are better.
     }
    \label{tab:reconstruction_stat}
\end{table}

Table~\ref{tab:reconstruction_stat} shows the results of this experiment. Our method achieves a lower median reconstruction error (computed using IoU) and a shorter median search time in all cases. Our method reconstructs 13 of the 33 models exactly, with a reconstruction error of less than 0.01\%, in comparison to 1 exact reconstruction by InverseCSG. This result indicates that our representation can more accurately represent typical CAD models when compared to parametric primitive CSG. We provide per model reconstruction results and additional details in the supplemental material.

\begin{figure}
    \centering
    \setlength{\tabcolsep}{1pt}
    \begin{tabular}{ccc}
        Target & InverseCSG & Ours Net
        \\
        \includegraphics[width=0.3\linewidth]{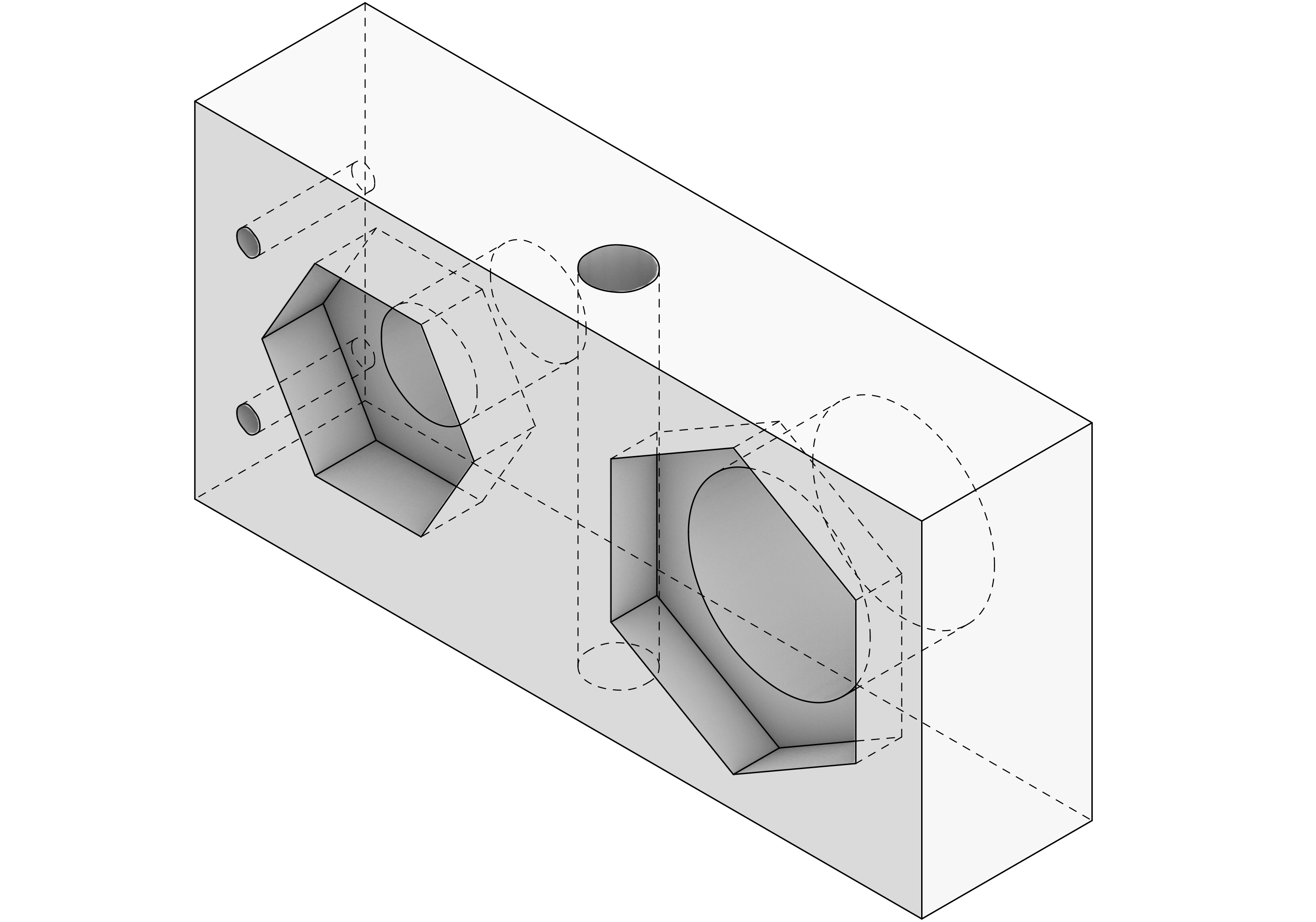} &
        \includegraphics[width=0.3\linewidth]{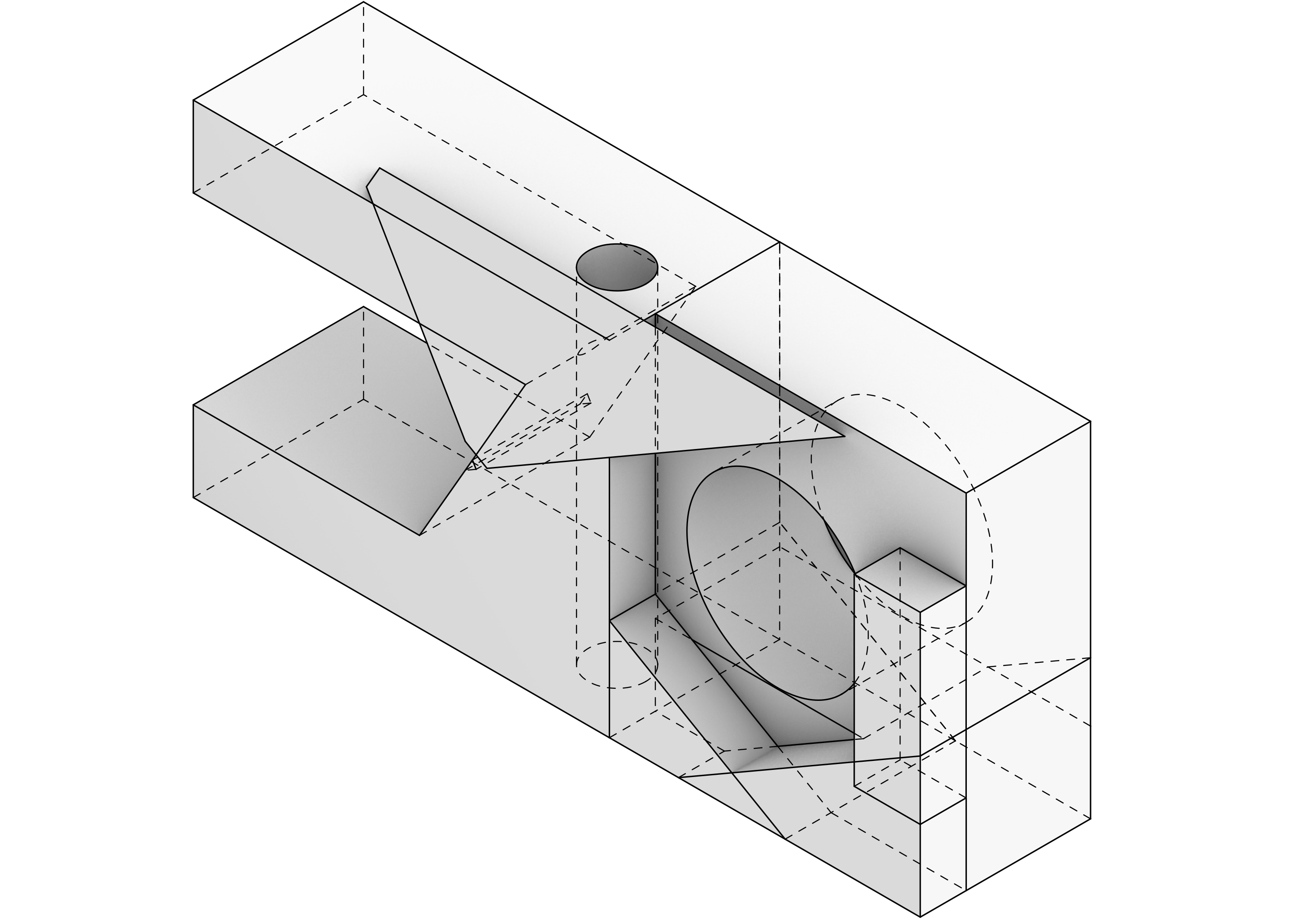} &
        \includegraphics[width=0.3\linewidth]{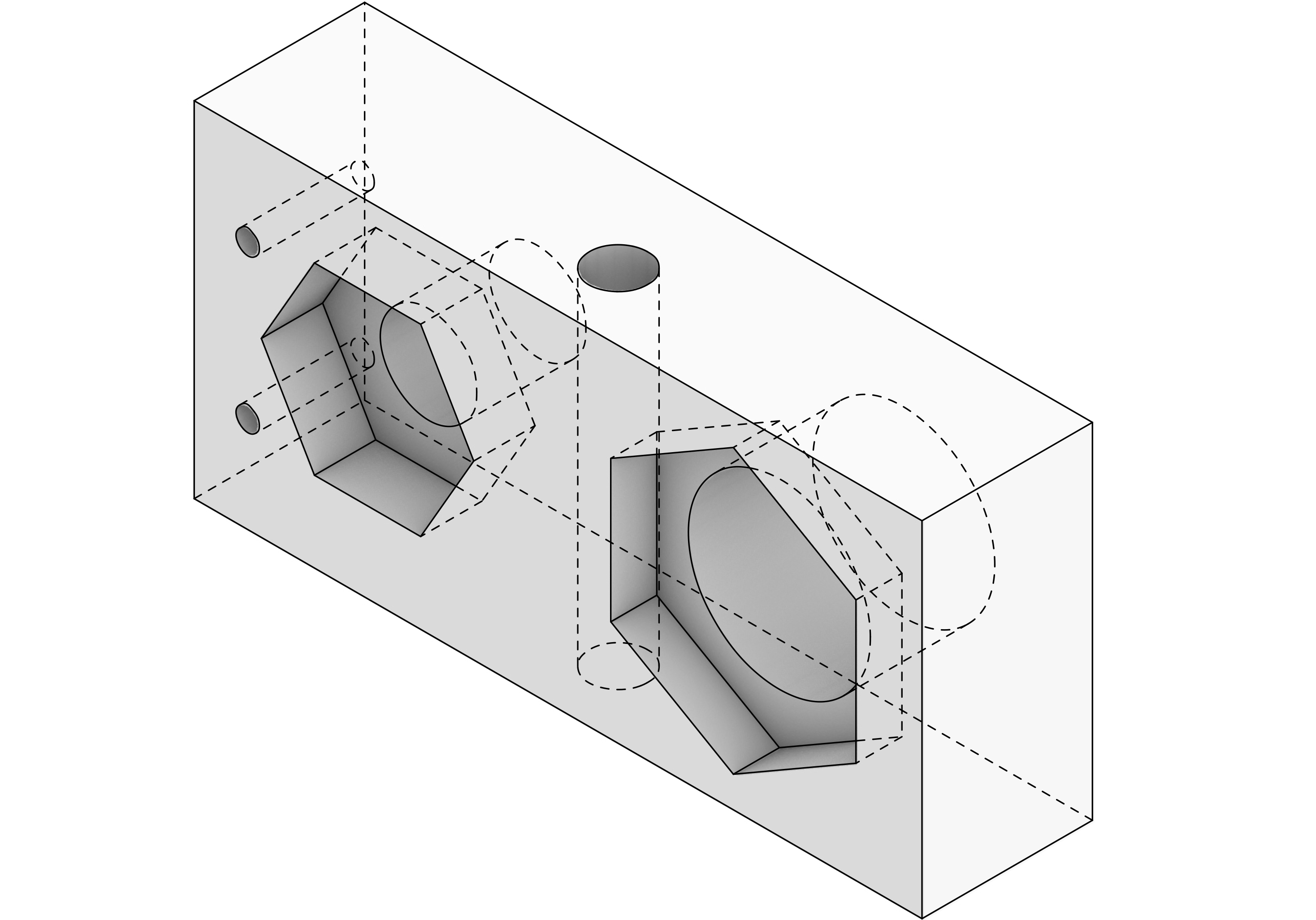}
        \\
        \includegraphics[width=0.3\linewidth]{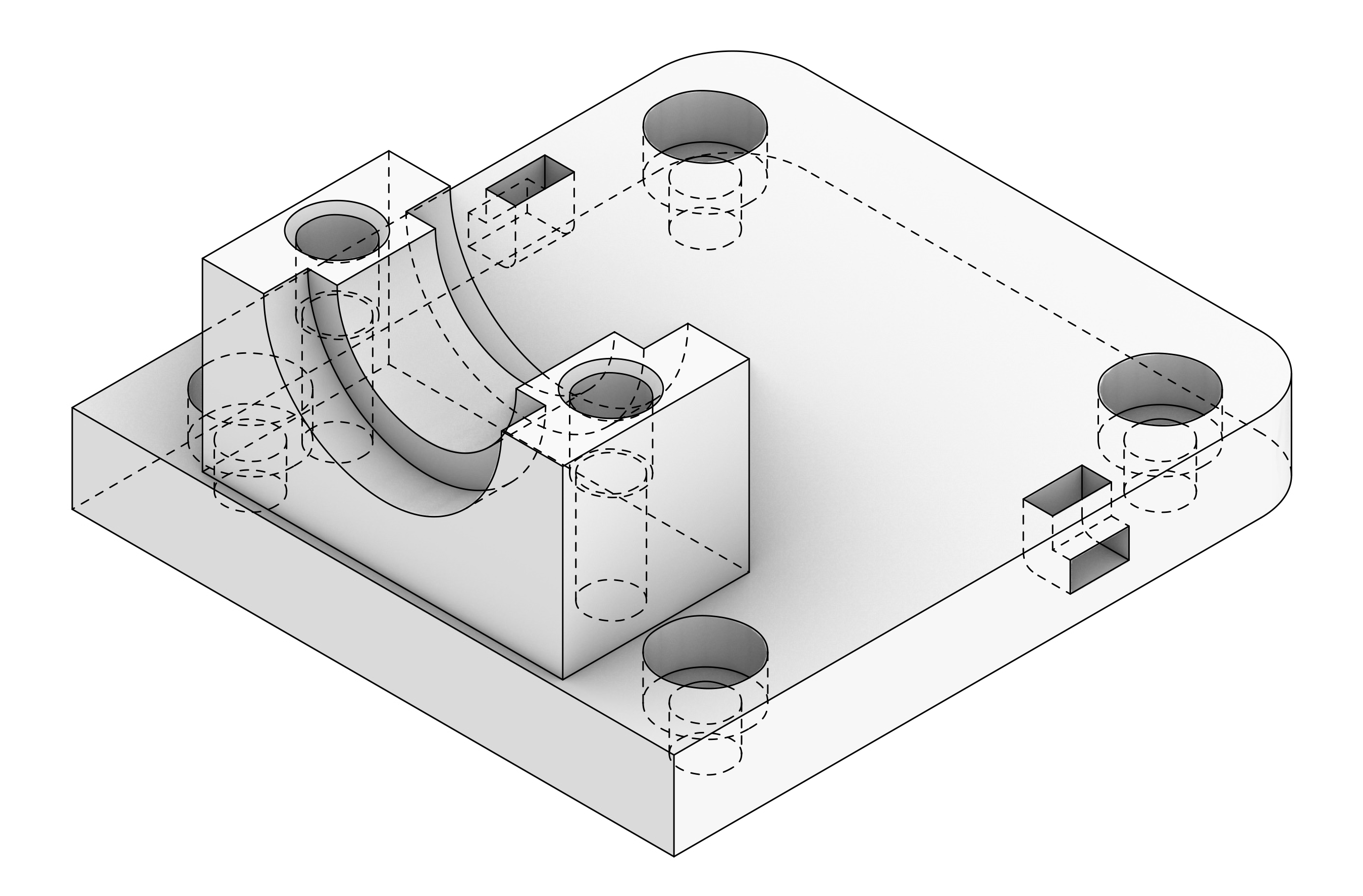} &
        \includegraphics[width=0.3\linewidth]{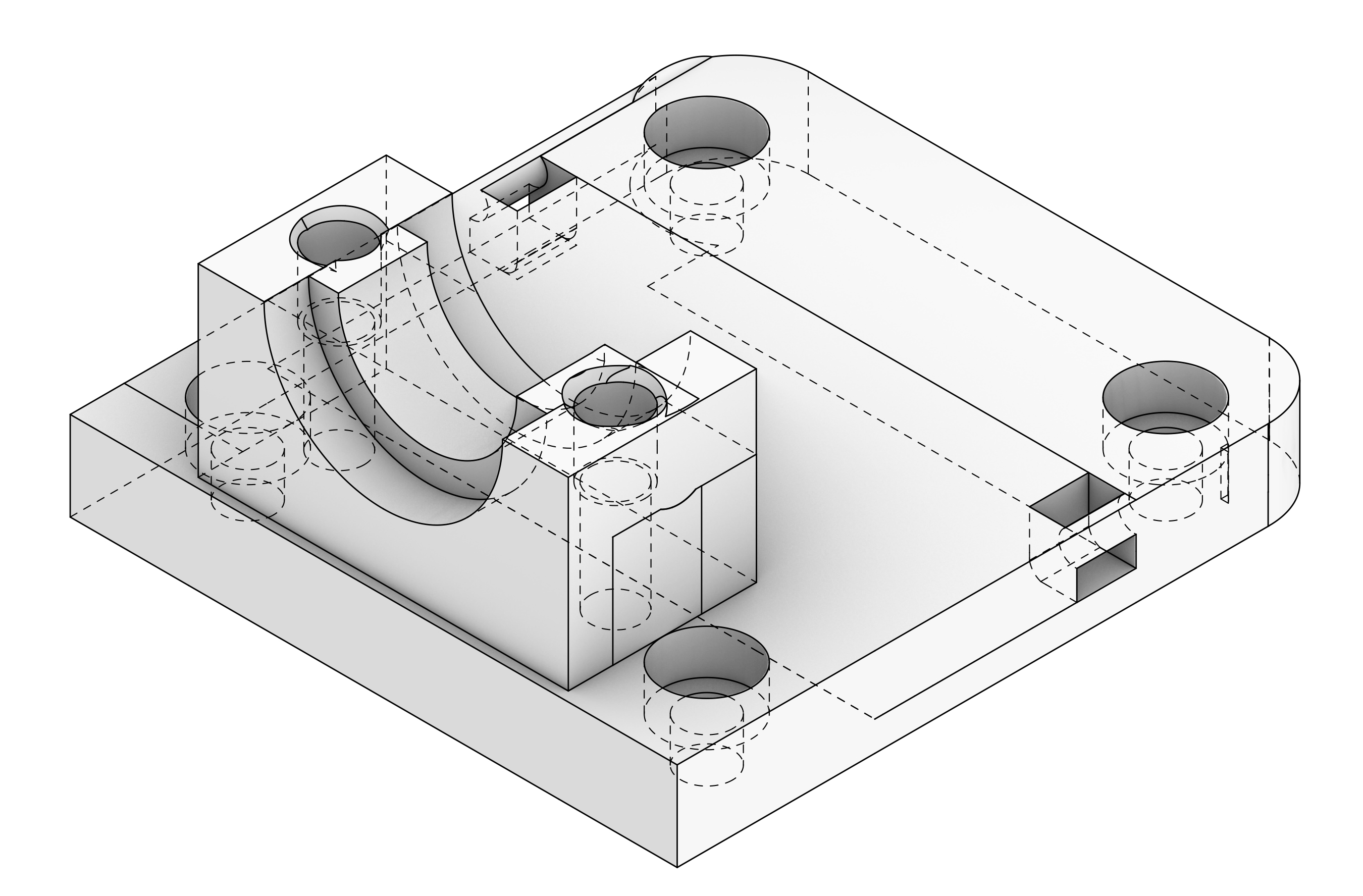} &
        \includegraphics[width=0.3\linewidth]{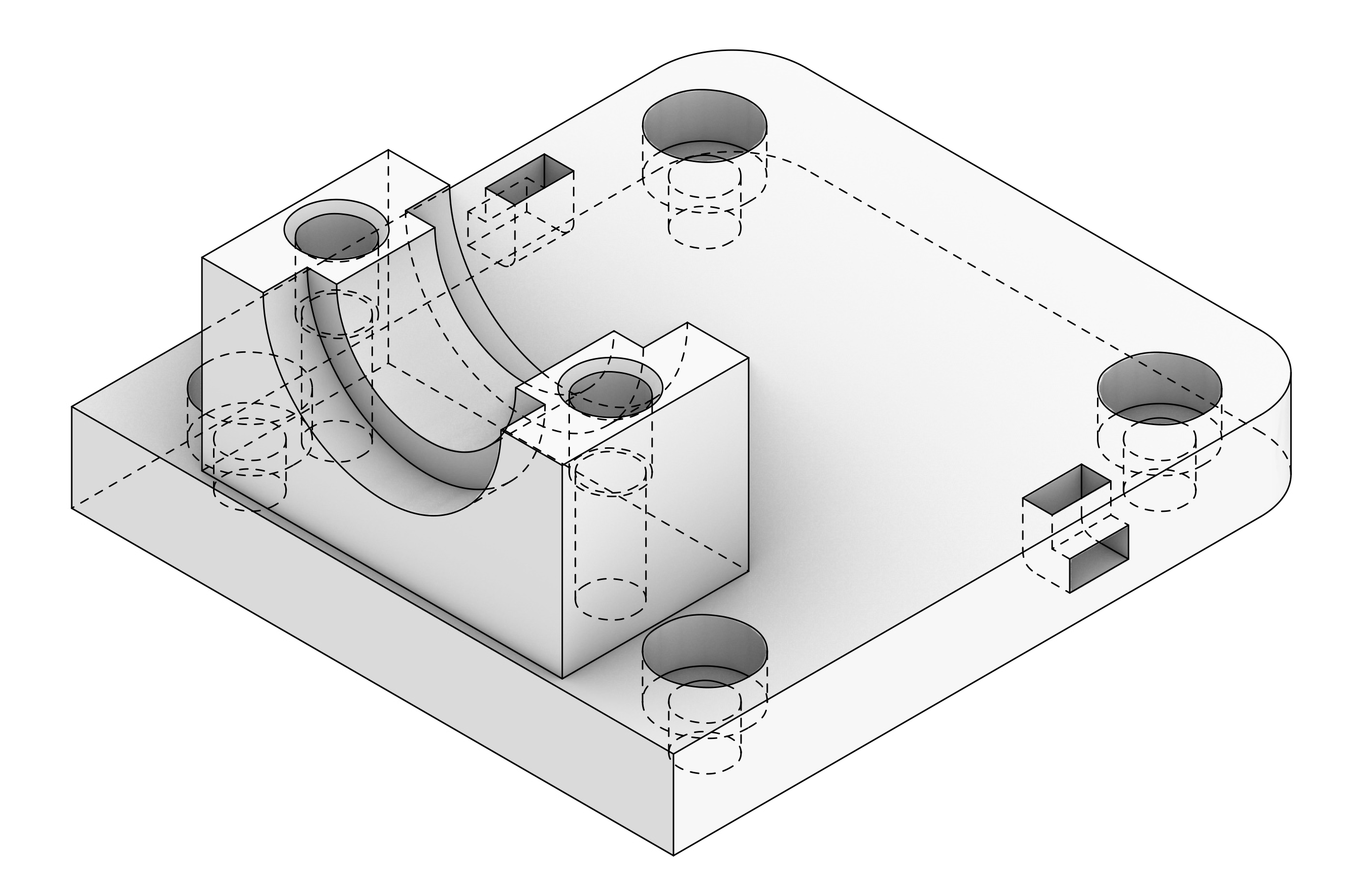}
        \\
        \includegraphics[width=0.3\linewidth]{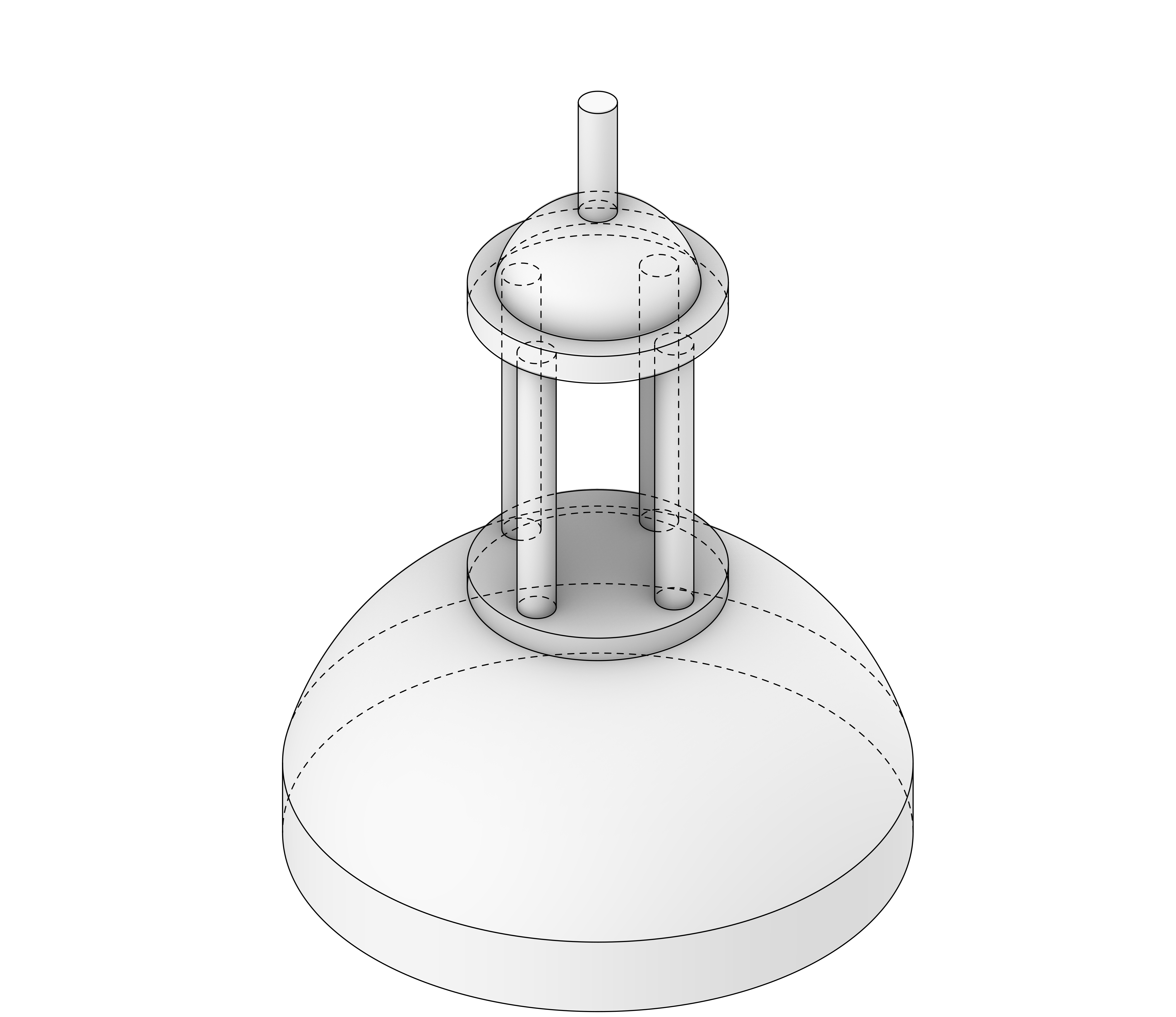} &
        \includegraphics[width=0.3\linewidth]{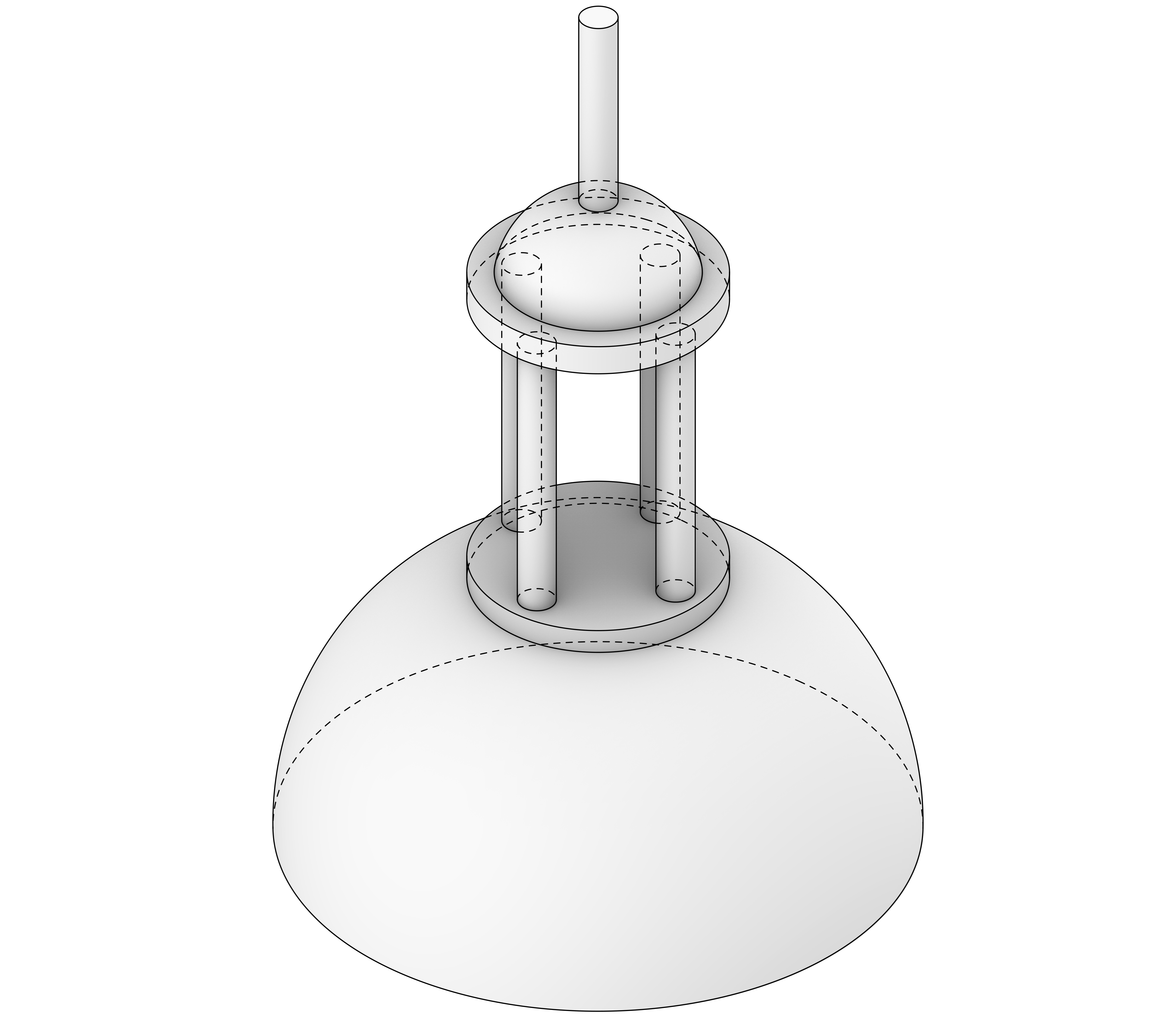} &
        \includegraphics[width=0.3\linewidth]{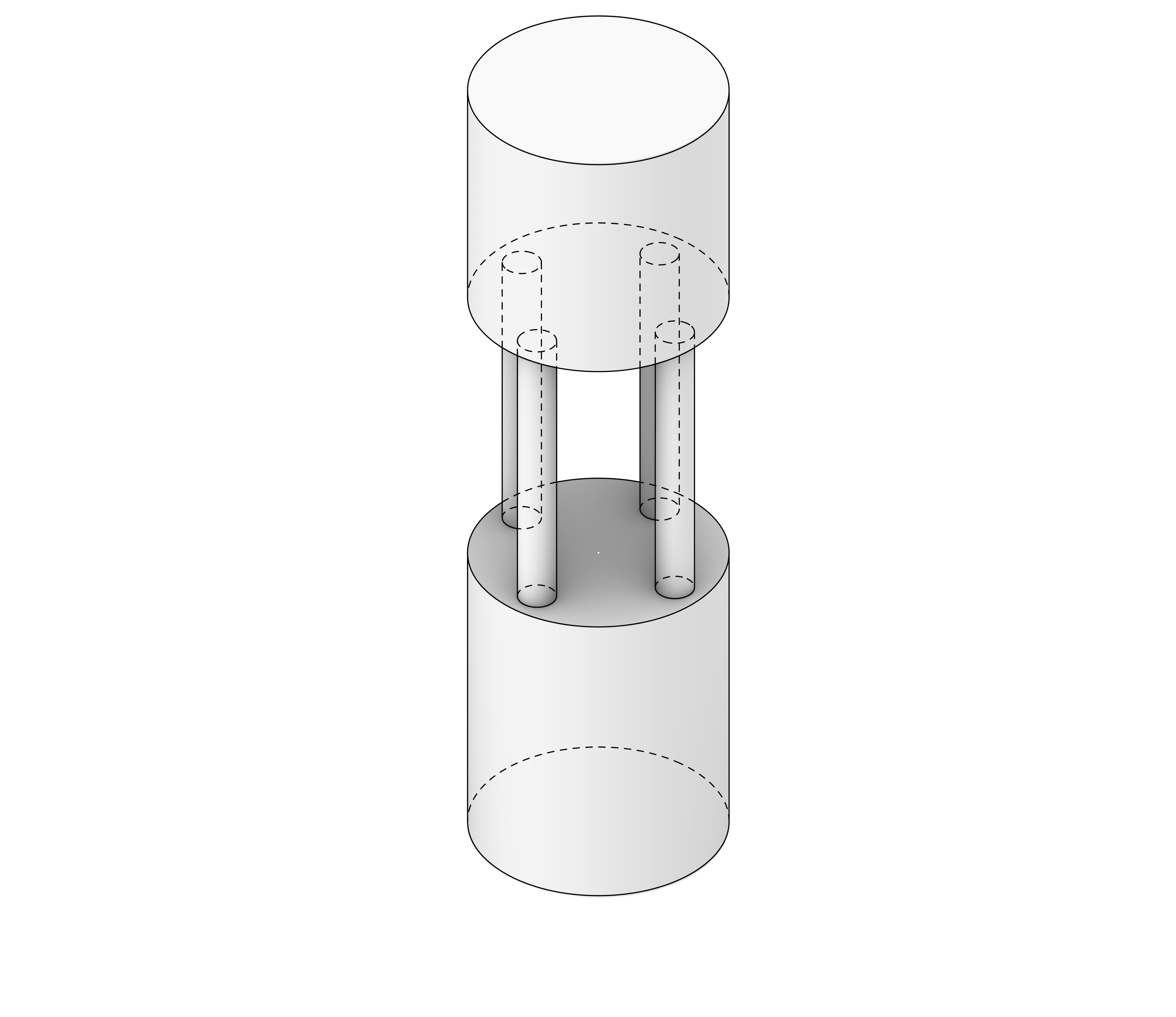}
    \end{tabular}
    \caption{
    Qualitative comparison of the output of our model's inferred programs vs. those of InverseCSG.
    }
    \label{figure:inversecsg_qualitative}
\end{figure}

Figure~\ref{figure:inversecsg_qualitative} shows reconstructions by our method and InverseCSG.
The first row shows a case where our method perfectly reconstructs the shape but InverseCSG incurs some approximation error due to its use of primitives.
The second row shows a case where both methods perform well.
The third row shows a failure case for our method (a revolve operation is needed to express this shape).

\subsection{Expert Perceptual Evaluation}
\label{sec:results_perceptual}

Finally, we evaluate inferred programs via a perceptual study.
We recruited 16 experienced CAD users from the fields of design (5) and engineering (11).
Each participant is asked to describe the type of modeling practice they engaged in more often: reconstructing existing shapes, or exploratory design.
Each participant then performs 45 comparisons of two modeling sequences shown in random order.
In each comparison, the participant is shown a target shape along with two modeling sequences for it and asked to (a) select the one most similar to how they would have constructed that shape, or (b) indicate that they cannot decide.
Comparisons were between (1) Ours Net vs. InverseCSG on the InverseCSG test set, (2) Ours Net vs. Ground Truth from the Fusion 360 Gallery test set, (3) Our Net vs. Ours Heur on the Fusion 360 Gallery test set.

\begin{table}[t!]
    \centering
    \footnotesize
    \setlength{\tabcolsep}{5pt}
    \begin{tabular}{lccc}
        \toprule
        & \multicolumn{3}{c}{\rule[1.5pt]{7em}{0.3pt} \textbf{\% Chosen} \rule[1.5pt]{7em}{0.3pt}}
        \\
        \textbf{Ours Net vs.} & Overall & By ``explorers'' & By ``reconstructers''
        \\
        \midrule
        Ground Truth & 34.1 & 33.3 & 34.0
        \\
        InverseCSG & 84.7 & 76.7 & 86.4
        \\
        Ours Heur & 41.0 & 52.5 & 37.9
        \\
        \bottomrule
    \end{tabular}
    \caption{Results of a perceptual study in which CAD users compared modeling sequences inferred by our method to those produced by InverseCSG.}
    \label{tab:comparison_exprt}
\end{table}

Table~\ref{tab:comparison_exprt} shows the results of this experiment.
Participants strongly preferred Our Net sequences to those of InverseCSG.
Our Net sequences were also judged as better than the ground truth about a third of the time.
Participants who focus on exploratory design preferred Our Net sequences vs. Our Heur ones at a significantly higher rate than participants focused on reconstruction.
This indicates that the sequences found by the learned guidance better support exploratory modification, whereas the heuristic sequences may be better suited to direct reconstruction.

%% file: 05-conclusion.tex
\section{Conclusion}
\label{sec:conclusion}

In this paper, we presented a new representation for CAD reconstruction, the \emph{zone graph}.
We showed how it reduces the search space of CAD modeling sequences that reconstruct a shape to finite size, and we presented an algorithm for searching the space of sketch + extrude + Boolean modeling sequences.
We also introduced a graph neural network that learns which search paths to explore first based on design patterns in a dataset of CAD modeling sequences.
Our experiments showed that our approach reconstructs a large percentage of input shapes and does so with more desirable modeling sequences than InverseCSG.

Some modeling sequences are not recoverable from the zone graph of the output shape,
\begin{wrapfigure}{r}{0.4\linewidth}
    \centering
    \includegraphics[width=\linewidth]{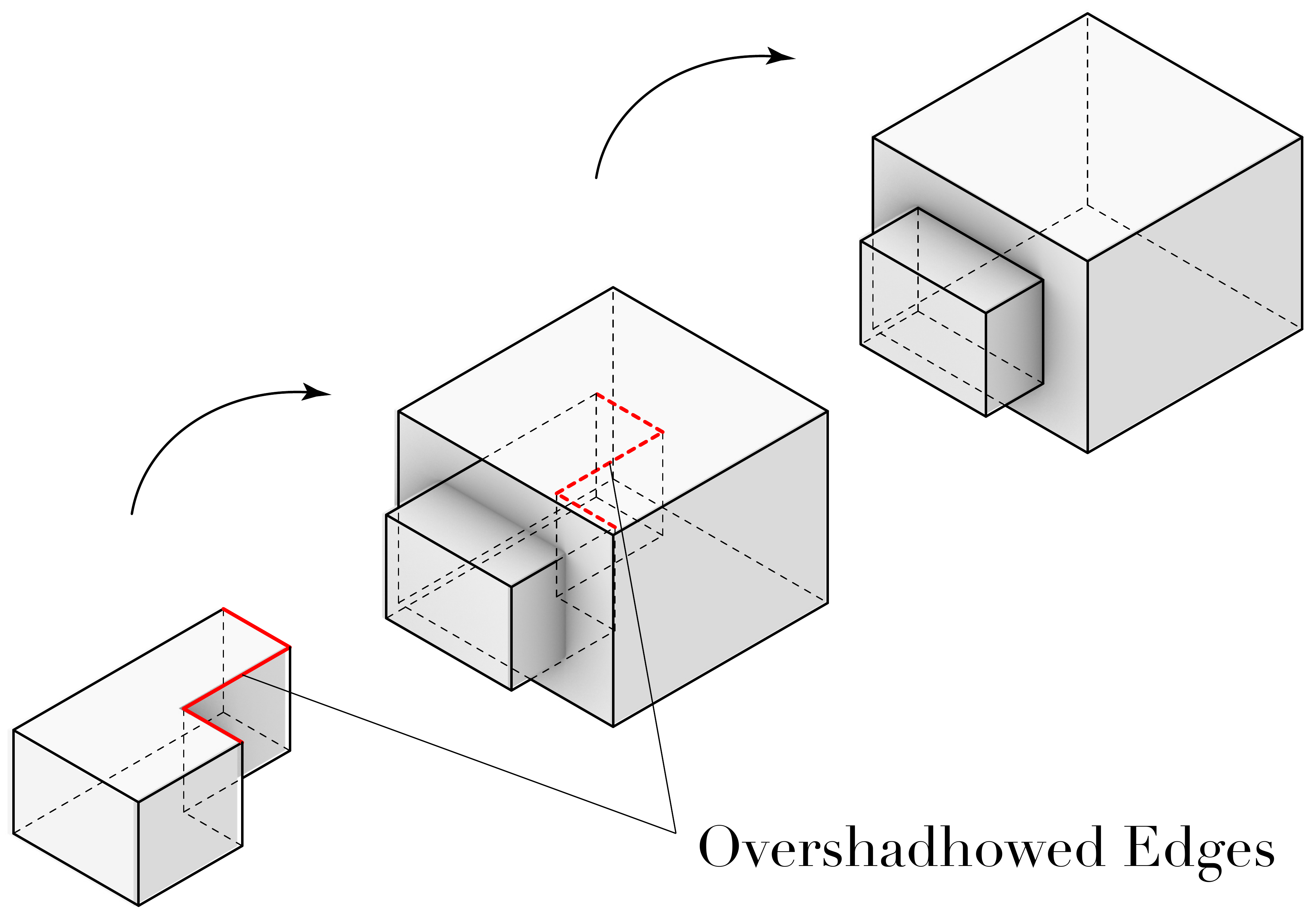}
\end{wrapfigure}
as they may include one or more steps which leaves no trace of itself in the output B-rep (see inset).
We can still infer alternative (and potentially equally-good) reconstruction sequences for such shapes, however.

Finally, we used a large CAD dataset to train our proposal scoring network.
Aside from the Fusion 360 Gallery reconstruction dataset, such data is not widely available.
It is important to investigate weakly-supervised approaches that require only a small amount of human input about what constitutes a meaningful design pattern.

%% file: 06-acknowledgments.tex
\section*{Acknowledgments}
We would like to thank Justin Solomon for pointing us toward the literature on arrangements of surfaces within computational geometry.
Thanks also to the Autodesk designers who participated in the study.
Daniel Ritchie is an advisor to Geopipe and owns equity in the company. Geopipe is a start-up that is developing 3D technology to build immersive virtual copies of the real world with applications in various fields, including games and architecture.

%% file: supp.tex


\clearpage
\setcounter{section}{0}
\renewcommand\thesection{\Alph{section}}
\renewcommand\thesubsection{\thesection.\arabic{subsection}}
\section{Supplementary Material}

\algrenewcommand\textproc{}
\algblock{Input}{EndInput}
\algnotext{EndInput}
\algblock{Output}{EndOutput}
\algnotext{EndOutput}
\newcommand{\Desc}[2]{\State \makebox[2em][l]{#1}#2}

\subsection{Simplification by face loop}
\begin{figure*}[h]
    \centering
    \includegraphics[width=\linewidth]{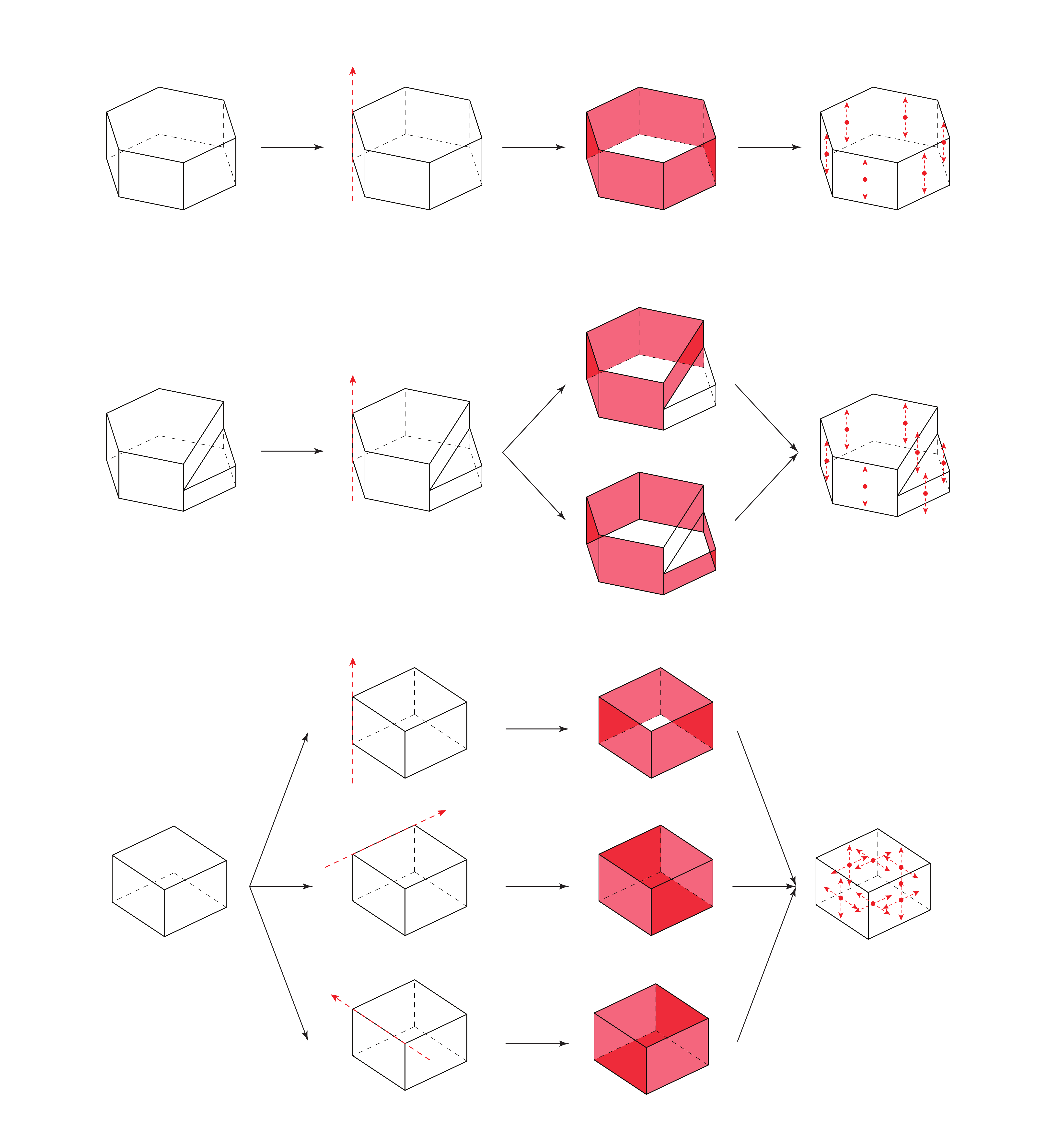}
    \caption{
    Zone graph simplification by finding \textit{face loops}. Column 1: Target shapes; Column 2: Extrusion directions marked with red arrows; Column 3: Found \textit{face loops}, highlighted in red; Column 4: Face extension directions determined by the extrusion directions associated with their \textit{face loops}. Note that when multiple extrusion directions are found for a face, it will be extended in all found directions.
    }
    \label{figure:face_loop}
\end{figure*}

Please see Figure~\ref{figure:face_loop} for a demonstration of finding \textit{face loops} in the target geometry for zone graph simplification.

\begin{figure}[h!]
\includegraphics[width=0.99\linewidth]{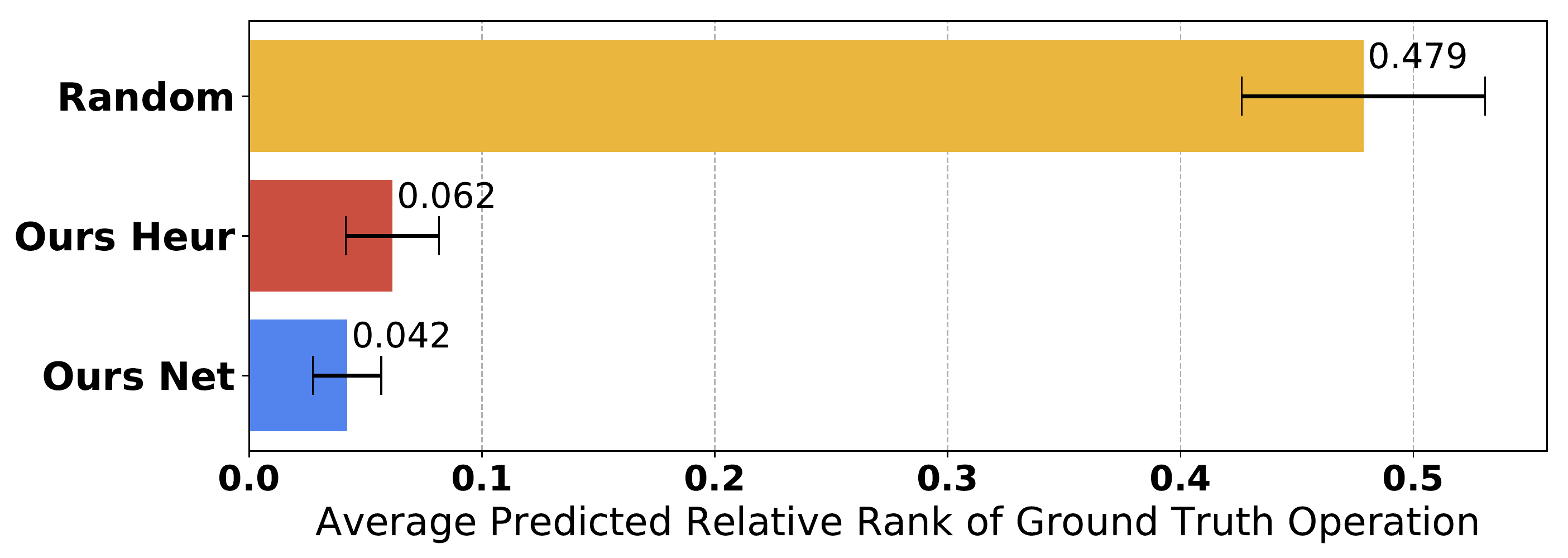}
    \caption{
    Comparing how different methods rank the ground truth extrusions used in modeling sequences from the Fusion 360 Gallery dataset without zone graph simplification.
    }
    \label{figure:rank_sim_comp}
\end{figure}

\subsection{Search}

Please see Algorithm ~\ref{alg:extrusion} and Algorithm ~\ref{alg:search} and for extrusion formulation and search details.

\begin{algorithm}
\begin{algorithmic}[1]
\Input
\Desc{Input zone graph $zg$}
\EndInput
\Output
\Desc{Proposed extrusion list $exts$}
\EndOutput
\Procedure {getExtrusions}{$zg$}
\State $exts \gets [ ]$
\For {$(sp, ep)$ in $plane\_pairs$} \algorithmiccomment{iterating parallel plane pairs}
\State $v \gets sp.nor * dis_{sp \rightarrow ep}$ \algorithmiccomment{compute extrusion vector}
\State $cgs$ $\gets$ $getGroups(sp.faces, sp.cur\_graph)$
\State $tgs$ $\gets$ $getGroups(sp.faces, sp.tgt\_graph)$
\State $igs$ $\gets$ $getGroups(sp.faces, sp.idle\_graph)$
\State $cyds \gets genCylinders(cgs, tgs, igs, v)$ \algorithmiccomment{grouping faces into sketches and generating extrusions}
\For {$cyd$ in $cyds$}
\State $e = Extrusion()$
\State $e.zones \gets findZones(cyd, zg)$\algorithmiccomment{finding and labeling inside zones}
\State $e.bool \gets boolType(e, zg)$
\State $exts.add(e)$
\EndFor
\EndFor
\State\Return $exts$
\EndProcedure
\caption{
    Extrusion Formulation 
    }
\label{alg:extrusion}
\end{algorithmic}
\end{algorithm}

\begin{algorithm}
\begin{algorithmic}[1]
\Input
\Desc{Input zone graph $zg$}
\EndInput
\Output
\Desc{Reconstruction sequence $seq$}
\EndOutput
\Procedure {search}{$zg, seq=[ ]$}
\If {$is\_target(zg)$}
\State\Return $True$
\EndIf
\If {$terminate()$}
\State\Return $False$
\EndIf
\State $exts = getExtrusions(zg)$
\State $exts\_ranked = rankExtrusions(exts)$
\For {$e$ in $exts\_ranked$}
\State $zg = zg.update(e)$
\State $ret \gets search(zg, seq)$ \algorithmiccomment{recursive search}
\If {$ret = True$}
\State $seq.add(e)$
\State\Return $True$
\EndIf
\EndFor
\EndProcedure
\caption{
    Search 
    }
\label{alg:search}
\end{algorithmic}
\end{algorithm}

\begin{figure}[h!]
\includegraphics[width=0.99\linewidth]{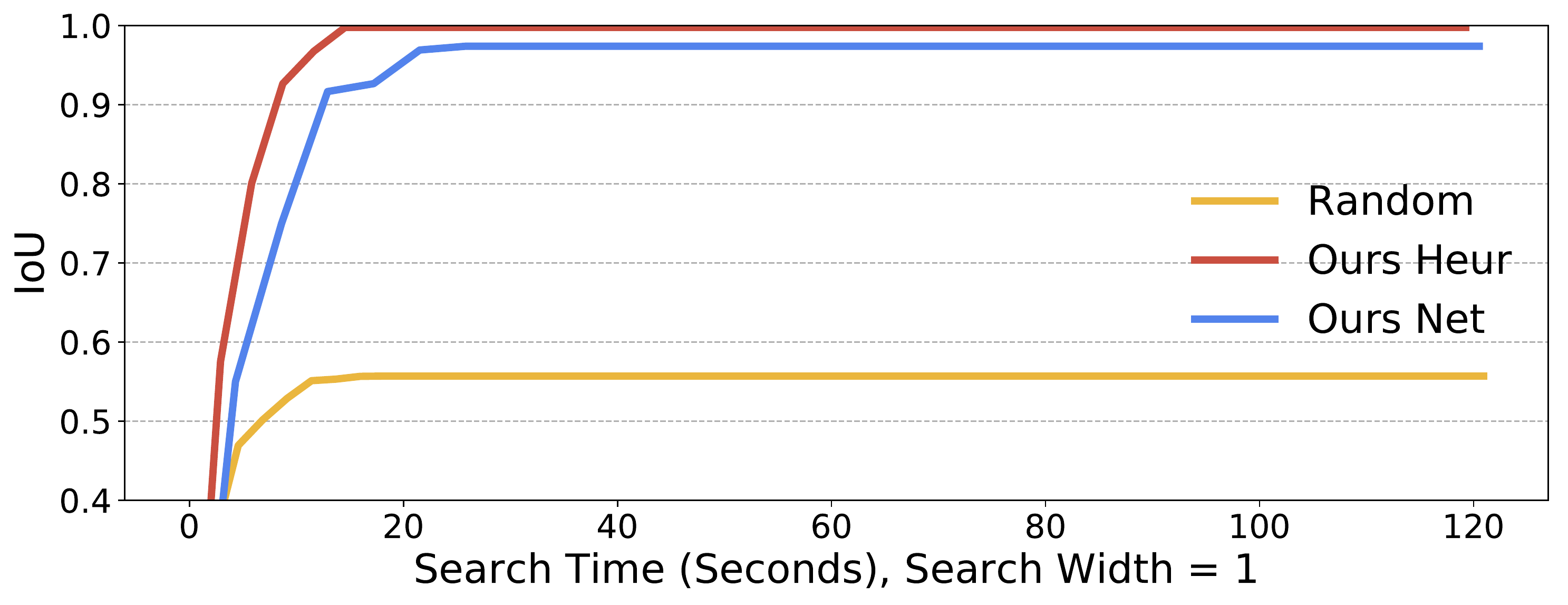}
    \caption{
    Reconstruction accuracy of the outputs of inferred programs vs. the time used to infer them. Search bandwidth = 1.
    }
    \label{figure:recon_k_10}
\end{figure}

\begin{figure}[h!]
\includegraphics[width=0.99\linewidth]{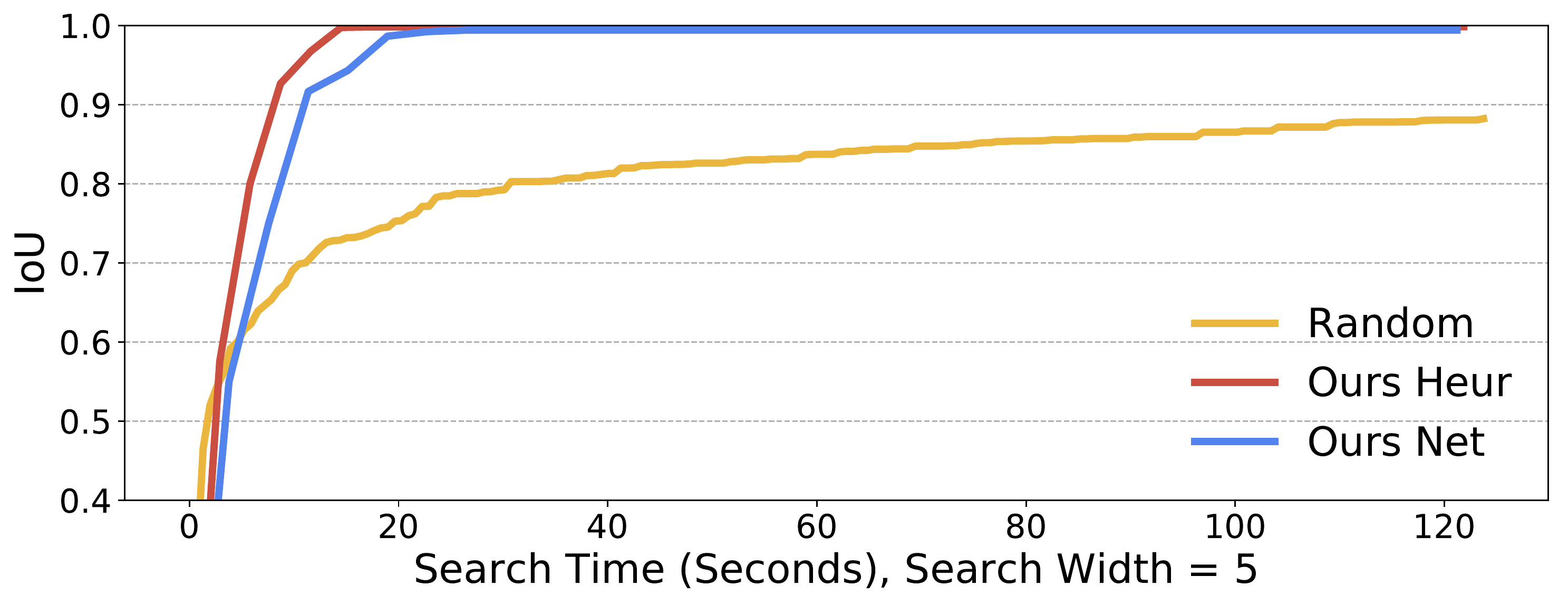}
    \caption{
    Reconstruction accuracy of the outputs of inferred programs vs. the time used to infer them. Search bandwidth = 5.
    }
    \label{figure:recon_k_5}
\end{figure}

\begin{figure}[h!]
\includegraphics[width=0.99\linewidth]{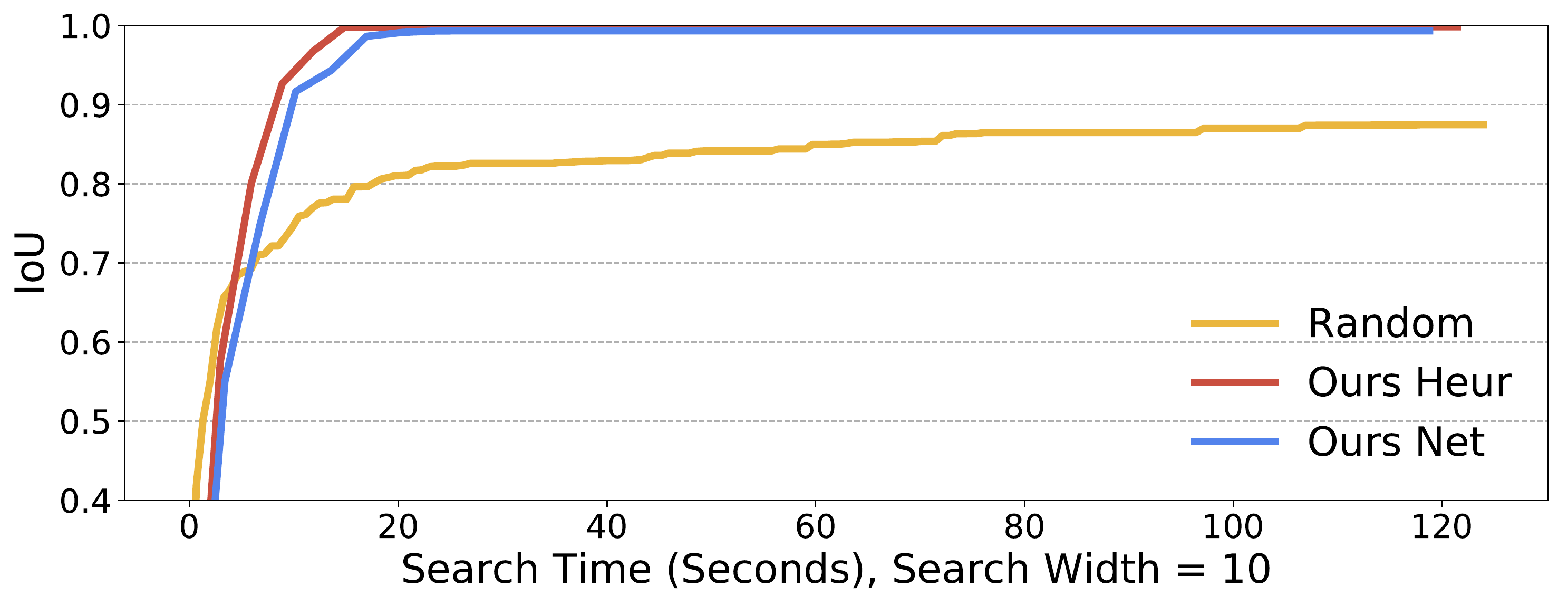}
    \caption{
    Reconstruction accuracy of the outputs of inferred programs vs. the time used to infer them. Search bandwidth = 10.
    }
    \label{figure:recon_k_10}
\end{figure}

\subsection{Network Architecture}

Please see Table~\ref{tab:pointnet}, \ref{tab:msg}, \ref{tab:gcn} and \ref{tab:mlp} for network architecture details.

\begin{figure}[h!]
\includegraphics[width=0.99\linewidth]{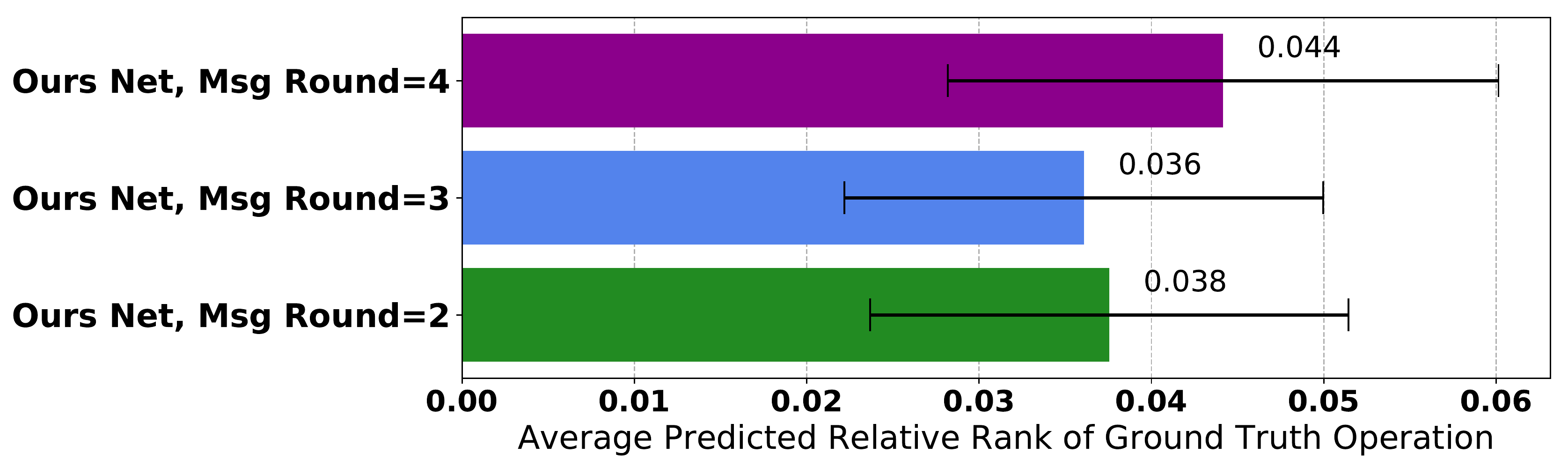}
    \caption{
    Comparing how different message passing round in GCN affects the average relative extrusion ranking.
    }
    \label{figure:rank_msg_comp}
\end{figure}

\begin{figure}[h!]
\includegraphics[width=0.99\linewidth]{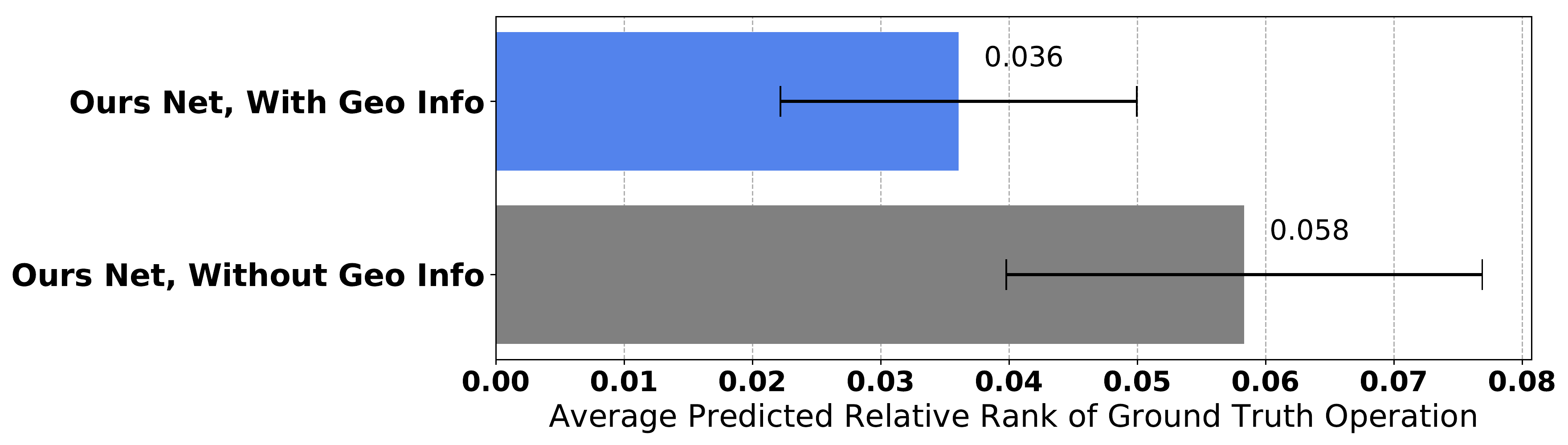}
\caption{
    Comparing how geometry information (with and without point cloud) in the GCN affects the average relative extrusion ranking.
    }
    \label{figure:rank_geo_comp}
\end{figure}

Please see Figures ~\ref{figure:rank_msg_comp}, \ref{figure:rank_geo_comp} for some ablation experiments with different model architectures. 

\begin{table}[t!]
    \centering
    \small
    \begin{tabular}{@{}c@{}}
        \toprule
        \textbf{PointNet}
        \\
        \midrule
        \textbf{Conv1d} $\left(10, 64, 1 \right)$\\
        \textbf{Batchnorm1d}\\
        \textbf{LeakyRelu}\\
        \textbf{Conv1d} $\left(64, 128, 1 \right)$\\
        \textbf{Batchnorm1d}\\
        \textbf{LeakyRelu}\\
        \textbf{Conv1d} $\left(128, 128, 1 \right)$\\
        \textbf{MaxPool}\\
        \textbf{FC}$\left( 128 \times 128 \right)$\\
        \textbf{Batchnorm1d}\\
        \textbf{LeakyRelu}\\
        \textbf{FC}$\left( 128 \times 128 \right)$\\
        \bottomrule
    \end{tabular}
    \vspace{0.1em}
    \caption{
    Detailed architecture of the PointNet we used in the project 
    }
    \label{tab:pointnet}
    \vspace{0.1mm}
\end{table}

\begin{table}[t!]
    \centering
    \small

    \begin{tabular}{@{}c@{}}
        \toprule
        \textbf{Msg}
        \\
        \midrule
        \textbf{FC}$\left( 128 \times 128 \right)$\\
        \textbf{Batchnorm1d}\\
        \textbf{LeakyRelu}\\
        \textbf{FC}$\left( 128 \times 128 \right)$\\
        \textbf{Batchnorm1d}\\
        \textbf{LeakyRelu}\\
        \textbf{FC}$\left( 128 \times 128 \right)$\\
        \bottomrule
    \end{tabular}
    \vspace{0.1em}
    \caption{
    Detailed architecture of the Message Passing Layers we used in the project
    }
    \label{tab:msg}
    \vspace{3mm}
\end{table}

\begin{table}[t!]
    \centering
    \small
    \begin{tabular}{@{}c@{}}
        \toprule
        \textbf{GCN}
        \\
        \midrule
        \textbf{Msg}\\
        \textbf{Msg}\\
        \textbf{Msg}\\
        \textbf{MaxPool}\\
        \bottomrule
    \end{tabular}
    \vspace{0.1em}
    \caption{
    Detailed architecture of the Graph Convolutional Network we used in the project
    }
    \label{tab:gcn}
    \vspace{3mm}
\end{table}

\begin{table}[t!]
    \centering
    \small
    \begin{tabular}{@{}c@{}}
        \toprule
        \textbf{MLP}
        \\
        \midrule
        \textbf{FC}$\left( 128 \times 128 \right)$\\
        \textbf{Batchnorm1d}\\
        \textbf{LeakyRelu}\\
        \textbf{FC}$\left( 128 \times 128 \right)$\\
        \textbf{Batchnorm1d}\\
        \textbf{LeakyRelu}\\
        \textbf{FC}$\left( 128 \times 2 \right)$\\
        \textbf{Softmax}\\
        \bottomrule
    \end{tabular}
    \vspace{0.1em}
    \caption{
    Detailed architecture of the Multi-Layer Perceptron Network we used in the project 
    }
    \label{tab:mlp}
    \vspace{3mm}
\end{table}

\subsection{Fusion 360 Gallery Success/Fail Case Summary}

Tables ~\ref{tab:dataset1}, ~\ref{tab:dataset2}, ~\ref{tab:dataset3} detail  the  percentage  of  Fusion 360 Gallery shapes  that  our method  can/cannot  reconstruct,  breaking  down  successes and  failures  into  subcategories.   Overall,  our  method  can reconstruct  80\%  of  the  shapes in  the  dataset.

We also show the effect of different strategies for constructing extrusion proposals.
As described in Section 5.1 of the main paper, we consider only individual connected components or the union of all connected components in a face group as candidate extrusions.
Here, we introduce a generalization of this scheme based on the idea of proposal \emph{levels}.
For a proposal level of $k$, within each face group, all subsets with size 1, 2 ... k and subsets with size N-k, N-k+1 ... N will be used as candidate extrusions (the scheme presented in Section 5.1 of the main paper uses $k=1$).
Larger $k$ leads to more potential extrusions, which increases the percentage of the ground-truth modeling sequences from the dataset which can be covered, at the cost of a larger search space (and therefore more computation time).
As show in Tables~\ref{tab:dataset1} and~\ref{tab:dataset2}, by increasing extrusion proposal level $k$ from 1 to 3, more ground truth extrusion sequences (from 60\% to 63\%) are captured by our proposals.

\begin{table}[t!]
    \centering
    \scriptsize
    \begin{tabular}{@{}rl@{}}
        \toprule
        \textbf{\% of data} & \textbf{Description}
        \\
    \midrule
\textbf{80\%} & \textbf{Can reconstruct target}\\
60\% & with GT sequence\\
8\% & with different sequence (GT not captured by our proposals)\\
12\% & with different sequence (GT overshadowed \& unrecoverable)\\
\midrule
\textbf{20\%} & \textbf{Cannot reconstruct target}\\
16\% & Insufficiently complete zone graph / GFA error\\
2\% & Unsupported operations (e.g. tapered extrude, revolve)\\
2\% & Crash/Hang/Timeout\\
    \bottomrule
    \end{tabular}
     \caption{
     Percentage of Fusion360 shapes our method(With Zone Graph simplification, Proposal level = 1) can/cannot reconstruct (and why)
     }
    \label{tab:dataset1}
\end{table}

\begin{figure*}
    \includegraphics[width=\linewidth]{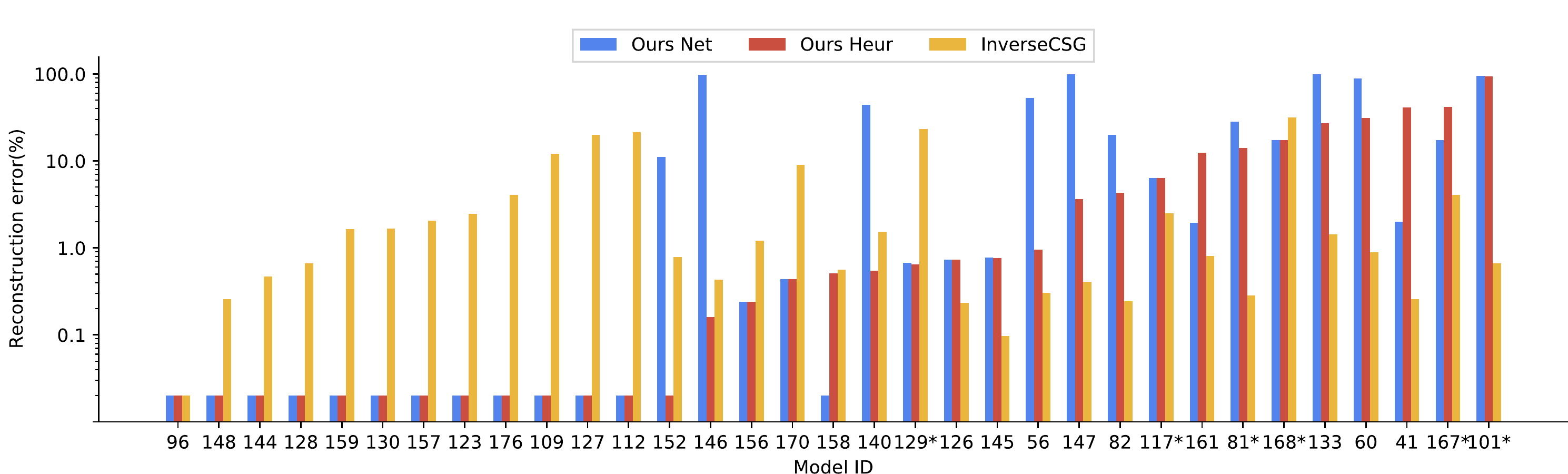}
    \includegraphics[width=\linewidth]{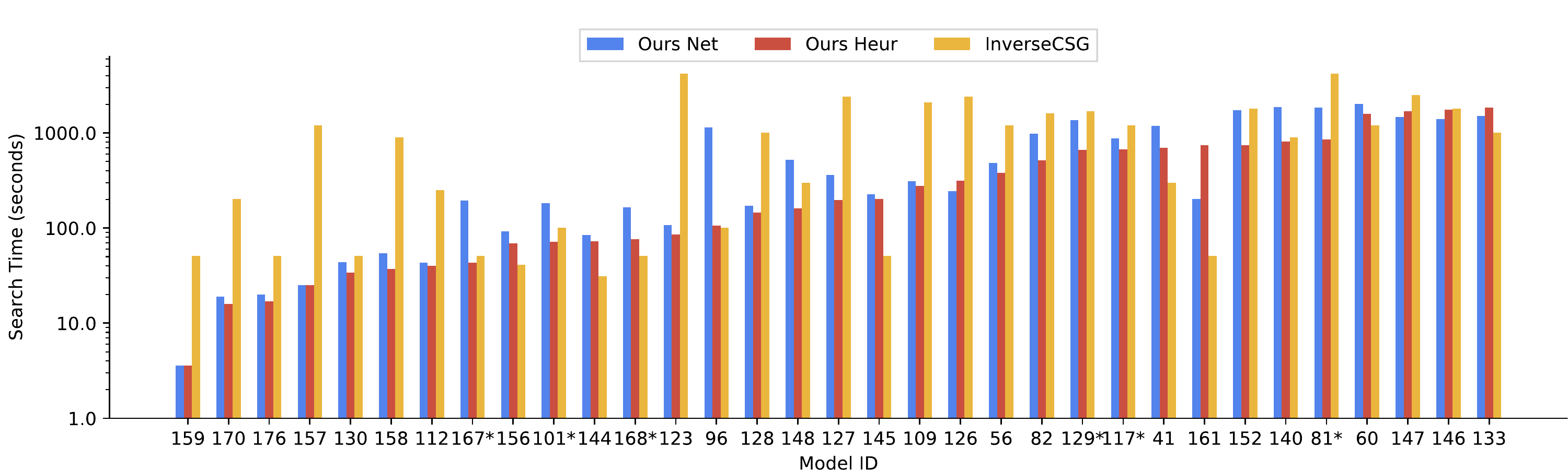}
    \caption{
        Per model comparison between InverseCSG and our method using heuristic (Ours Heur) or network guided (Ours Net) search. Results are show for 33 of the 50 models in the InverseCSG test set. Models IDs with an * contain operations other than sketch + extrude. Top: Reconstruction error computed using IoU. We use the reconstructed models released by the InverseCSG authors to evaluate. Bottom: Search time in seconds. To compare we use the official times from the InverseCSG paper.
    }
    \label{figure:comparison_histogram}
\end{figure*}

\begin{table}[t!]
    \centering
    \scriptsize
    \begin{tabular}{@{}rl@{}}
        \toprule
        \textbf{\% of data} & \textbf{Description}
        \\
    \midrule
\textbf{80\%} & \textbf{Can reconstruct target}\\
63\% & with GT sequence\\
5\% & with different sequence (GT not captured by our proposals)\\
12\% & with different sequence (GT overshadowed \& unrecoverable)\\
\midrule
\textbf{20\%} & \textbf{Cannot reconstruct target}\\
16\% & Insufficiently complete zone graph / GFA error\\
2\% & Unsupported operations (e.g. tapered extrude, revolve)\\
2\% & Crash/Hang/Timeout\\
    \bottomrule
    \end{tabular}
     \caption{
     Percentage of Fusion 360 Gallery shapes our method (with zone graph simplification, proposal level = 3) can/cannot reconstruct (and why)
     }
    \label{tab:dataset2}
\end{table}

\begin{table}[t!]
    \centering
    \scriptsize
    \begin{tabular}{@{}rl@{}}
        \toprule
        \textbf{\% of data} & \textbf{Description}
        \\
    \midrule
\textbf{82\%} & \textbf{Can reconstruct target}\\
62\% & with GT sequence\\
8\% & with different sequence (GT not captured by our proposals)\\
12\% & with different sequence (GT overshadowed \& unrecoverable)\\
\midrule
\textbf{18\%} & \textbf{Cannot reconstruct target}\\
14\% & Insufficiently complete zone graph / GFA error\\
2\% & Unsupported operations (e.g. tapered extrude, revolve)\\
2\% & Crash/Hang/Timeout\\
    \bottomrule
    \end{tabular}
     \caption{
     Percentage of Fusion 360 Gallery shapes our method (without zone graph simplification, proposal level = 1) can/cannot reconstruct (and why)
     }
    \label{tab:dataset3}
\end{table}

\subsection{Qualitative Results}
\subsubsection{Ours vs. random and heuristics}

\begin{figure*}[h!]
    \centering
    \setlength{\tabcolsep}{1pt}
    \begin{tabular}{cccccccc}
        \multicolumn{2}{c}{Target} & & & &
        \\
        \multicolumn{2}{c}{\includegraphics[width=0.25\linewidth]{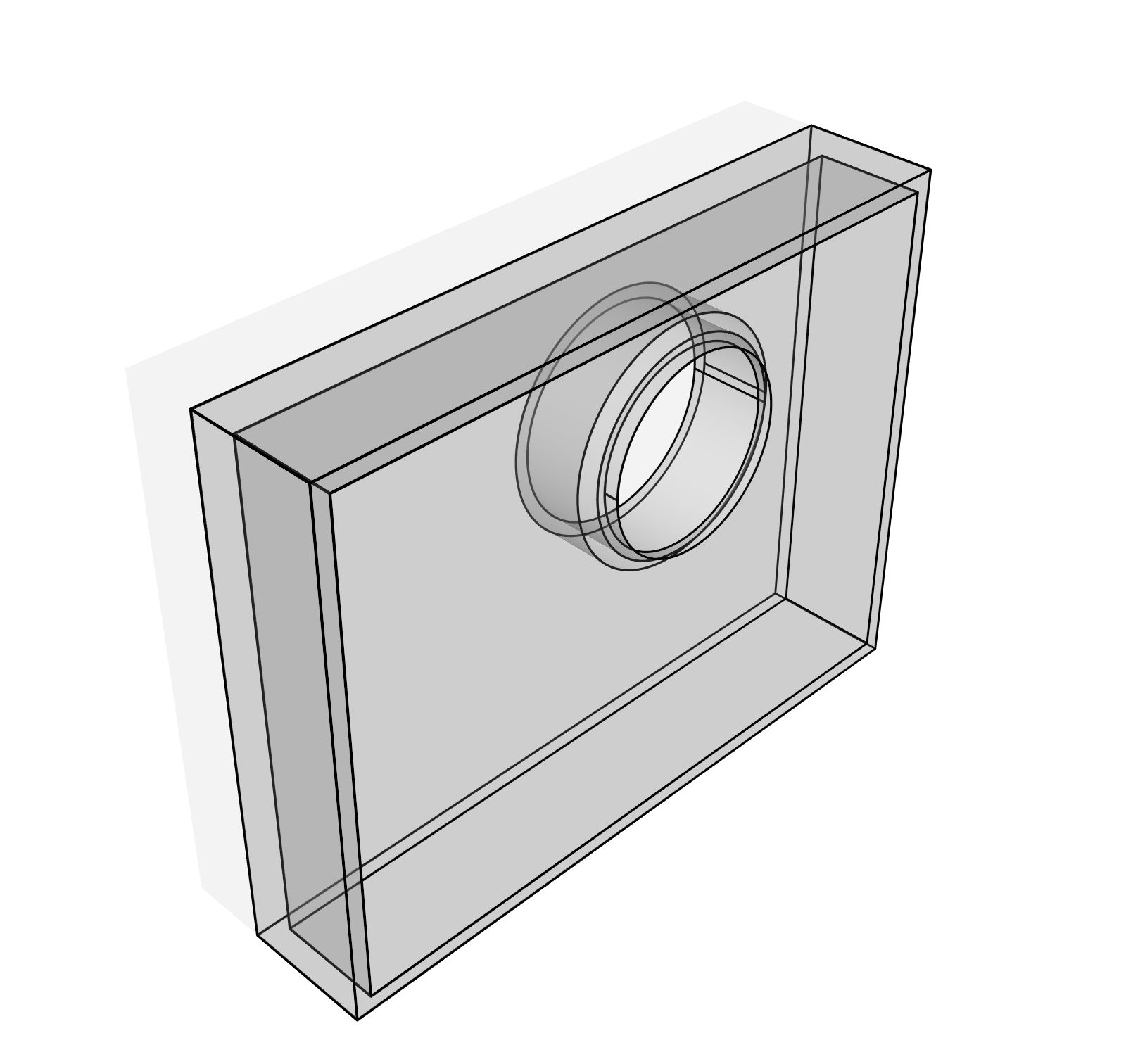}} & & & &
        \\
        \multicolumn{1}{l}{Random} & & & & &
        \\
        \includegraphics[width=0.124\linewidth]{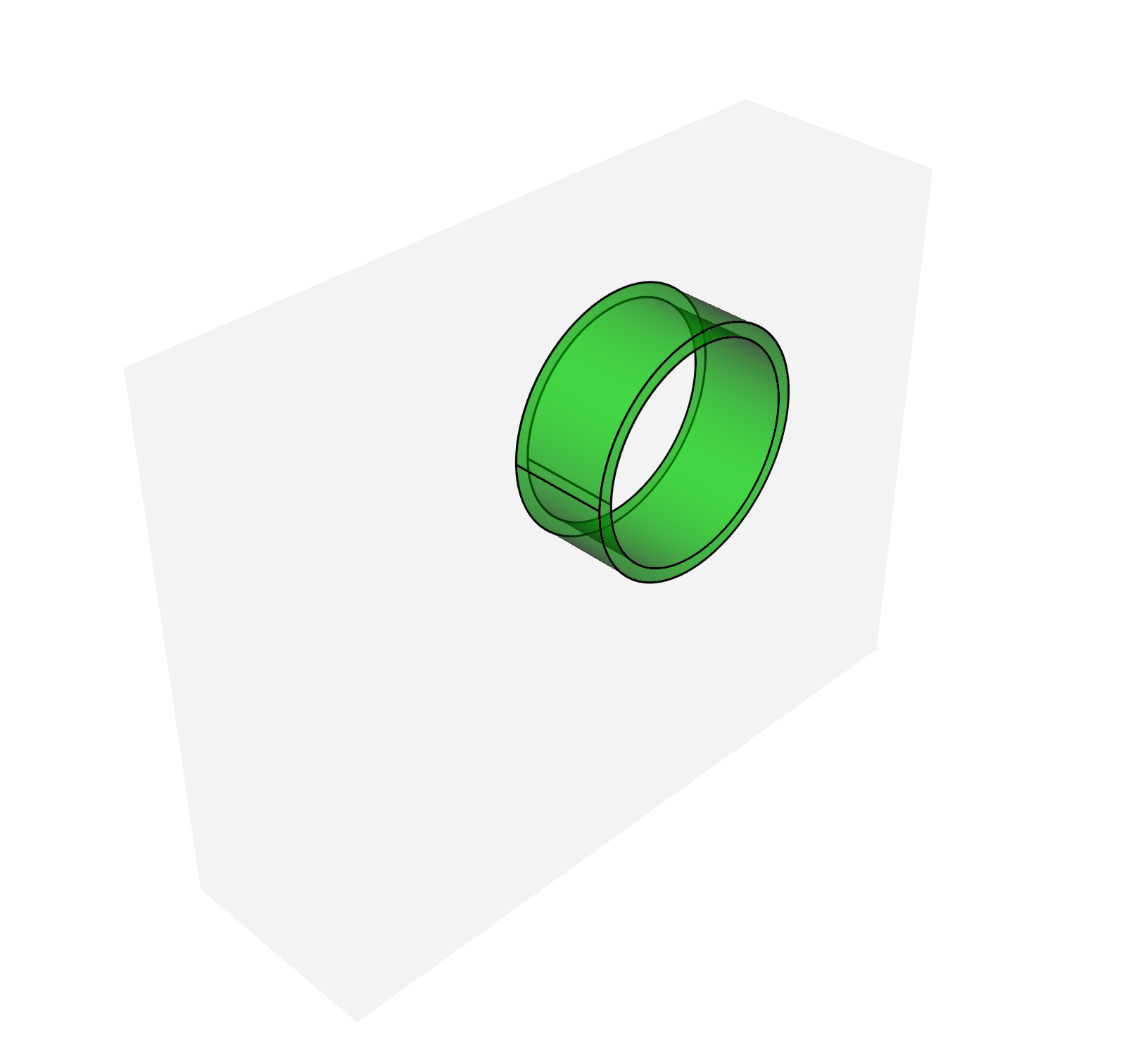} &
        \includegraphics[width=0.124\linewidth]{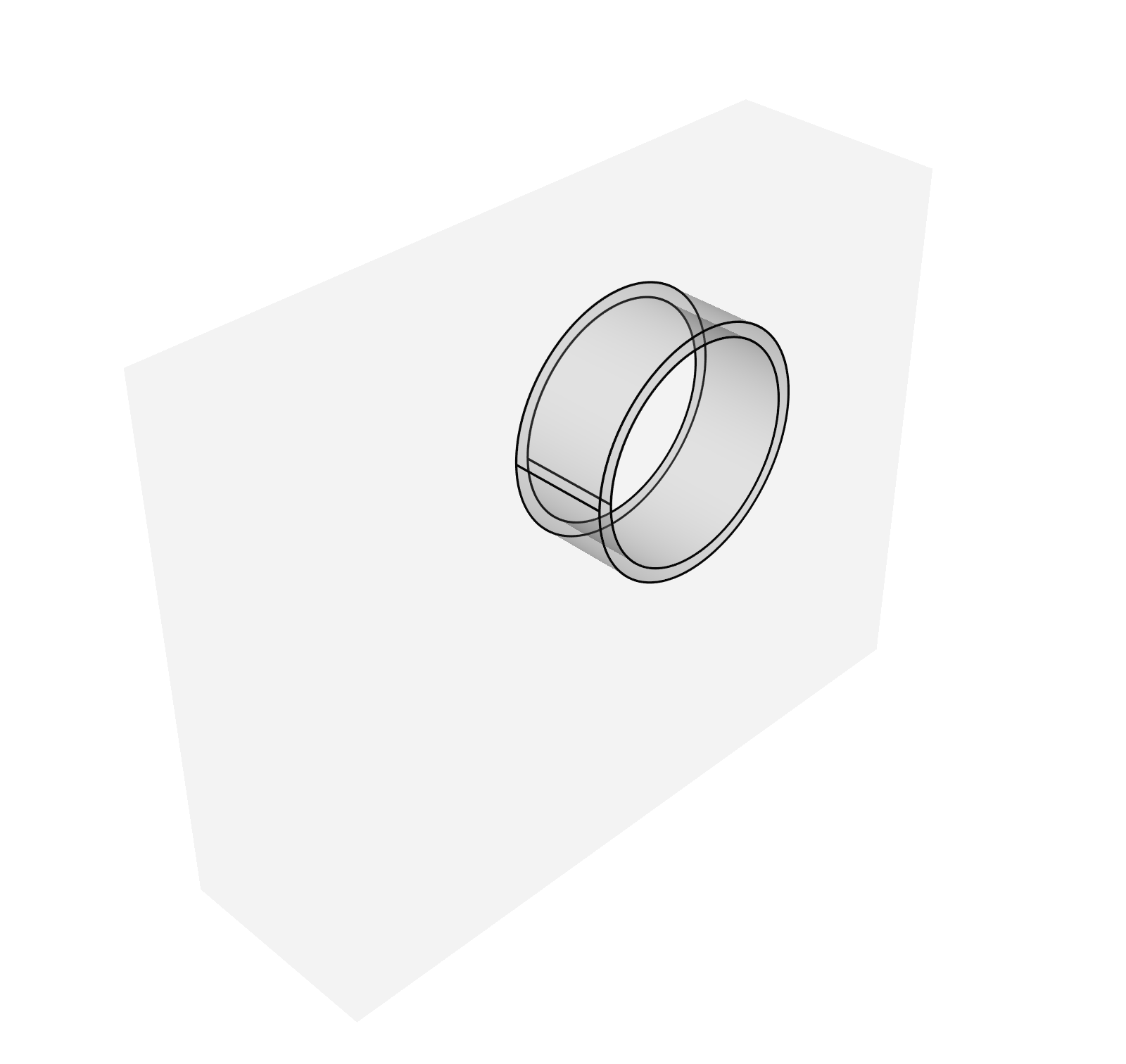} &
        \includegraphics[width=0.124\linewidth]{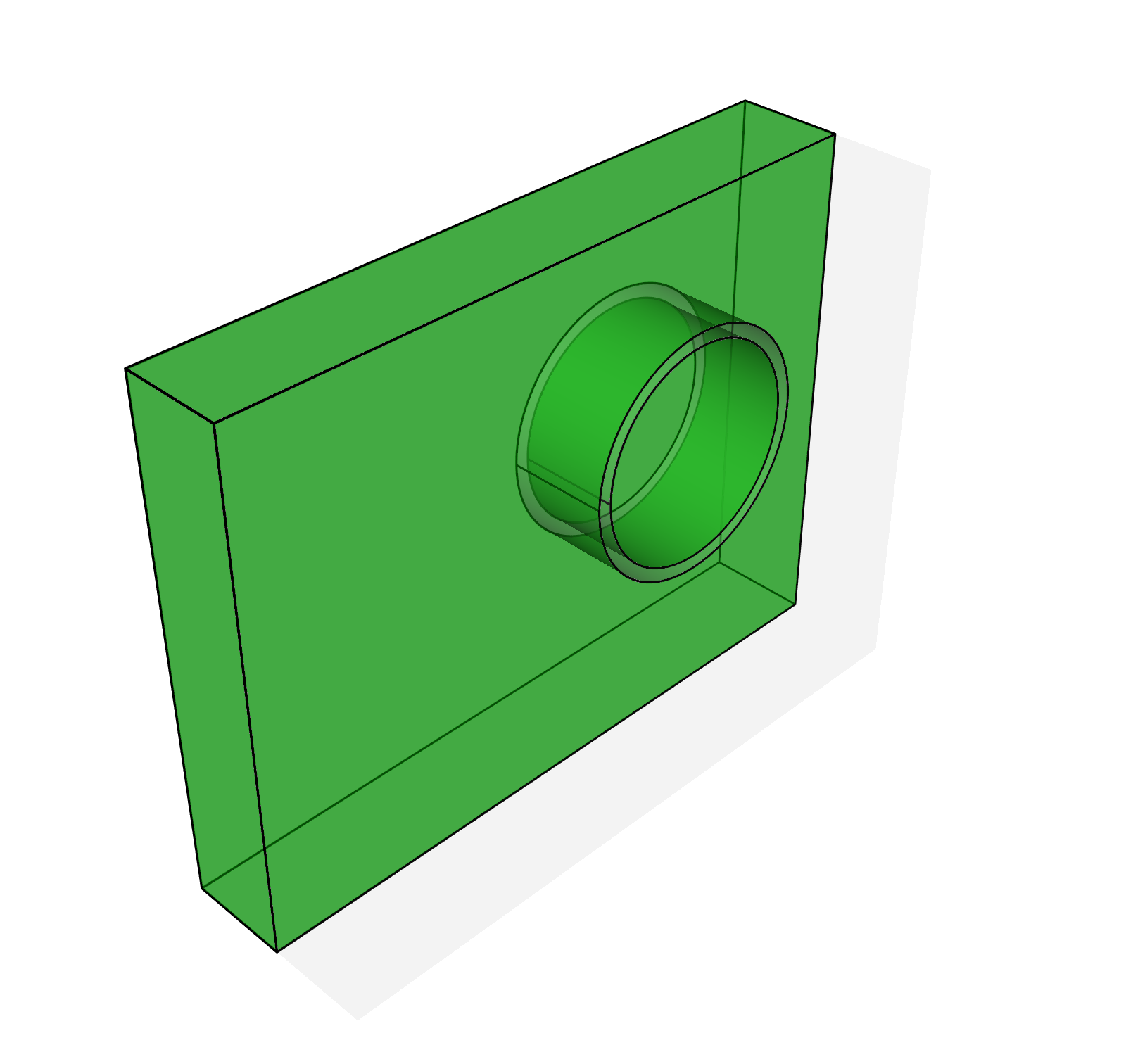} &
        \includegraphics[width=0.124\linewidth]{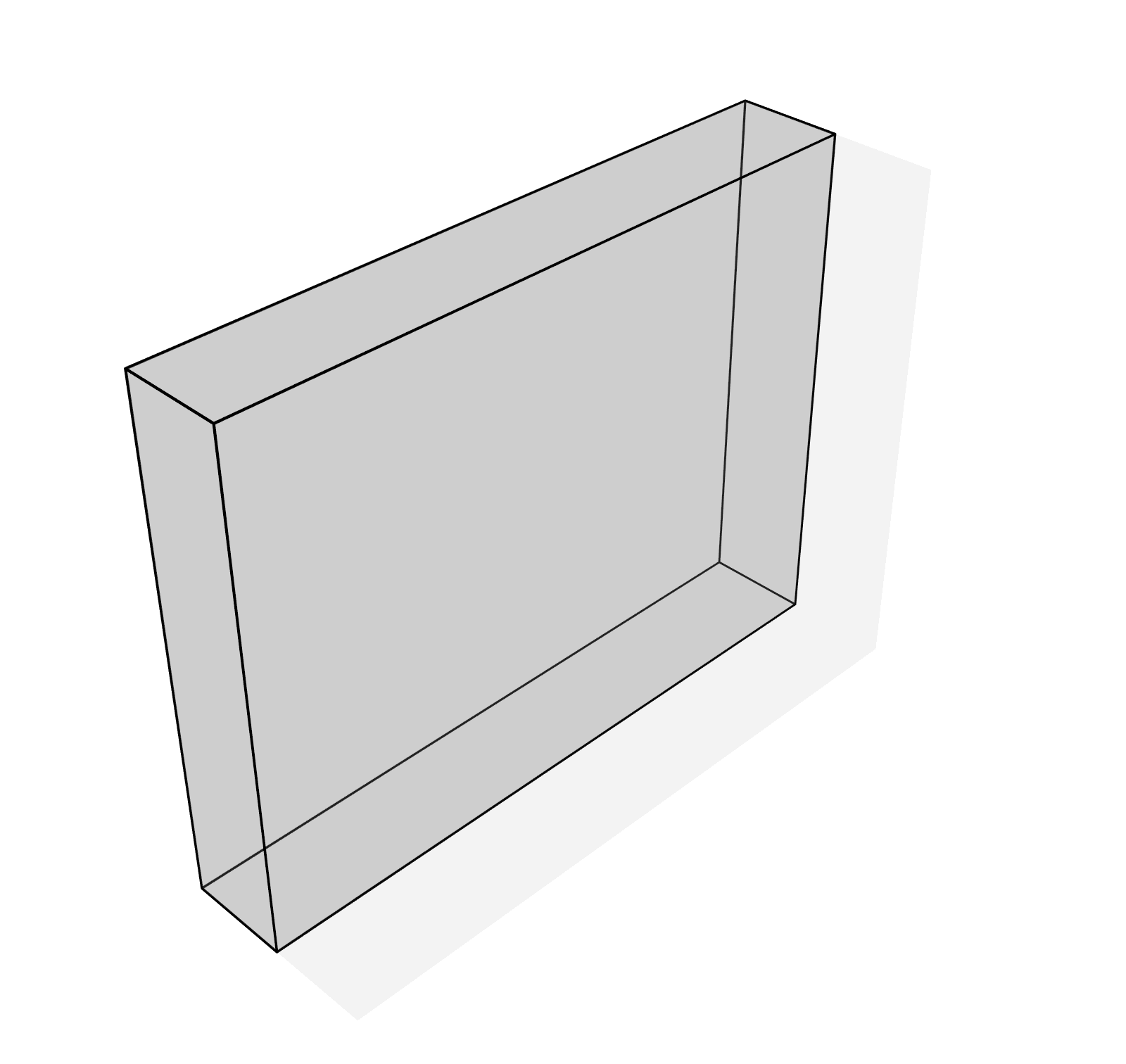} &
        \includegraphics[width=0.124\linewidth]{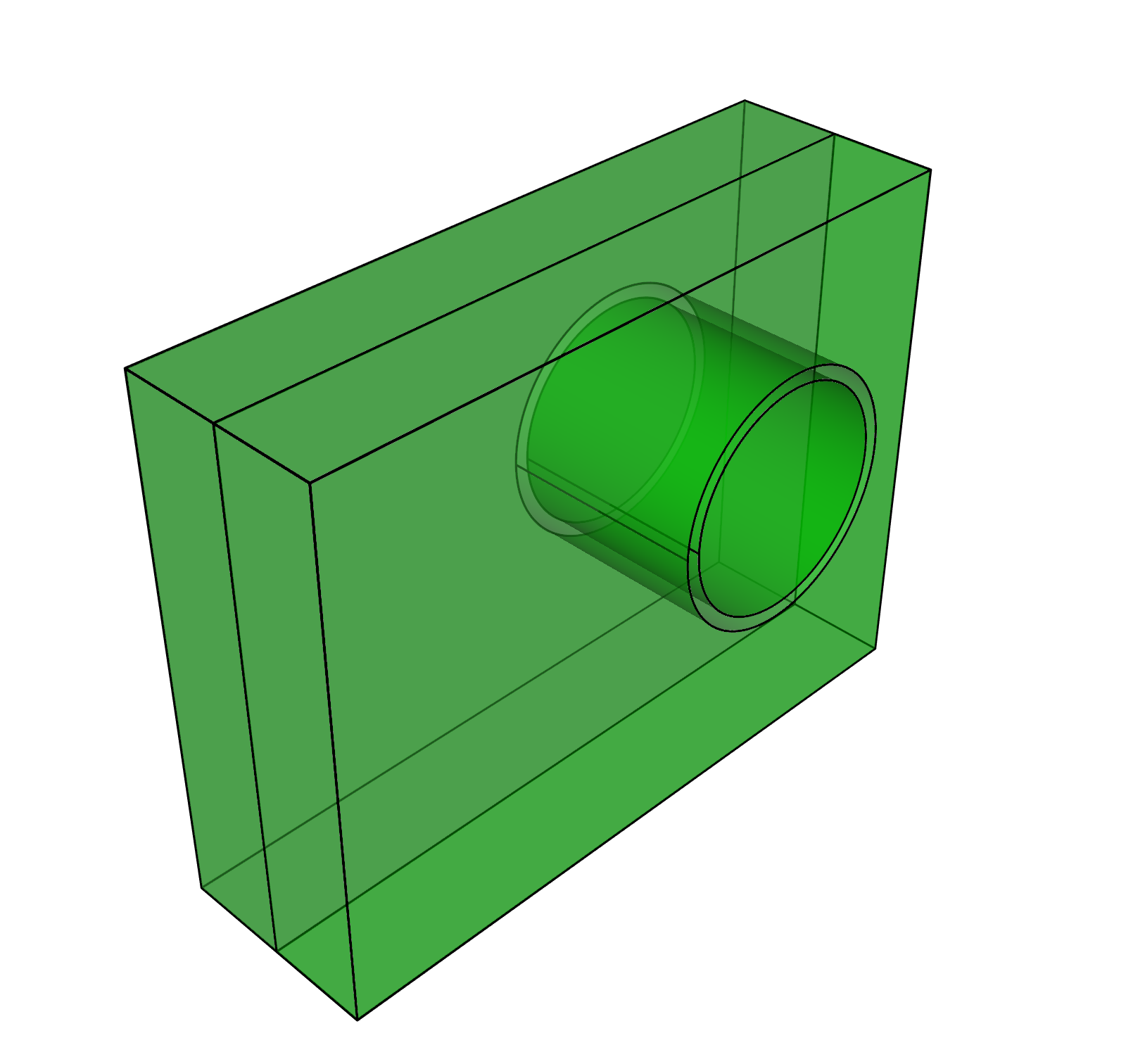} &
        \includegraphics[width=0.124\linewidth]{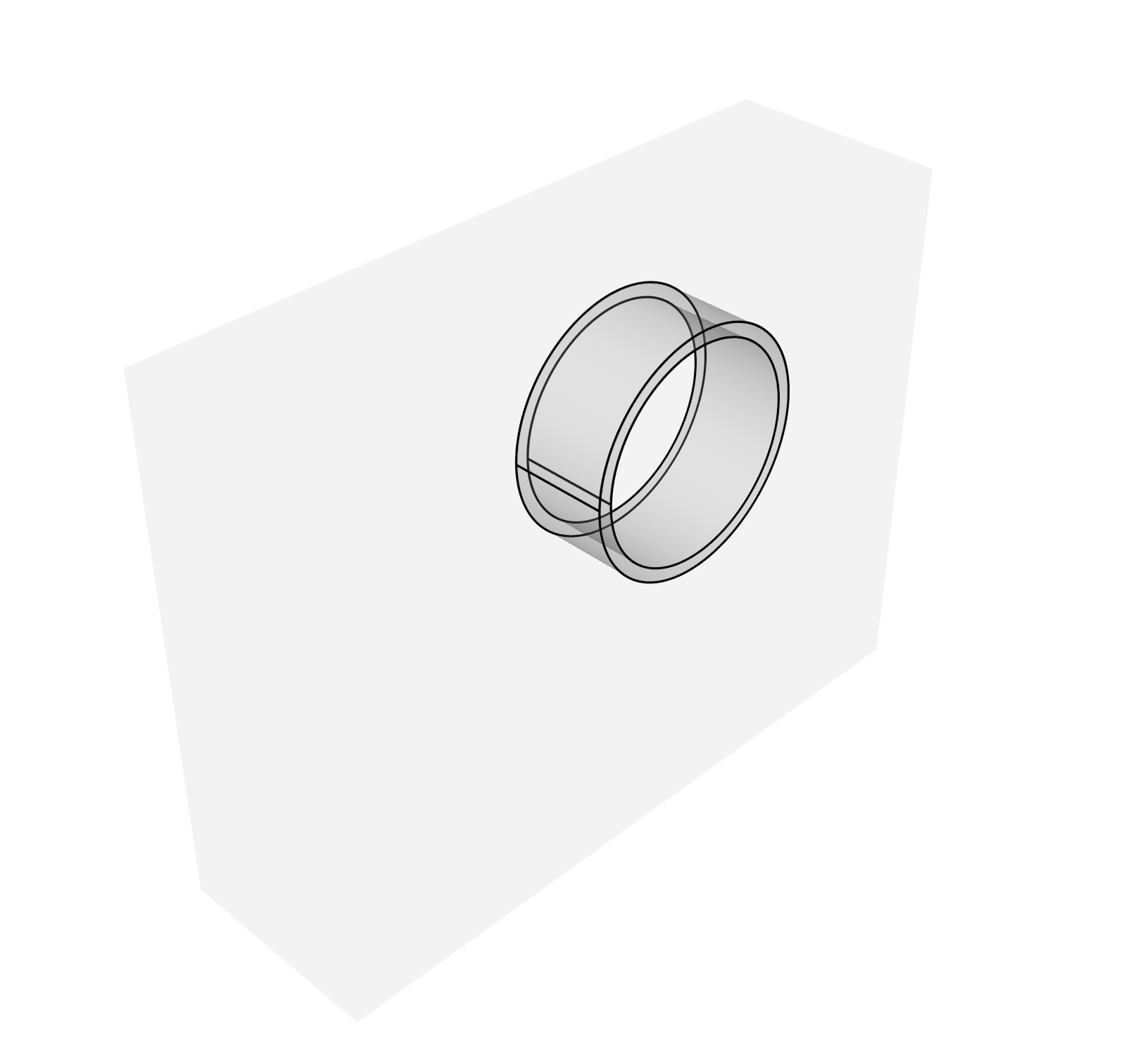} &
        \includegraphics[width=0.124\linewidth]{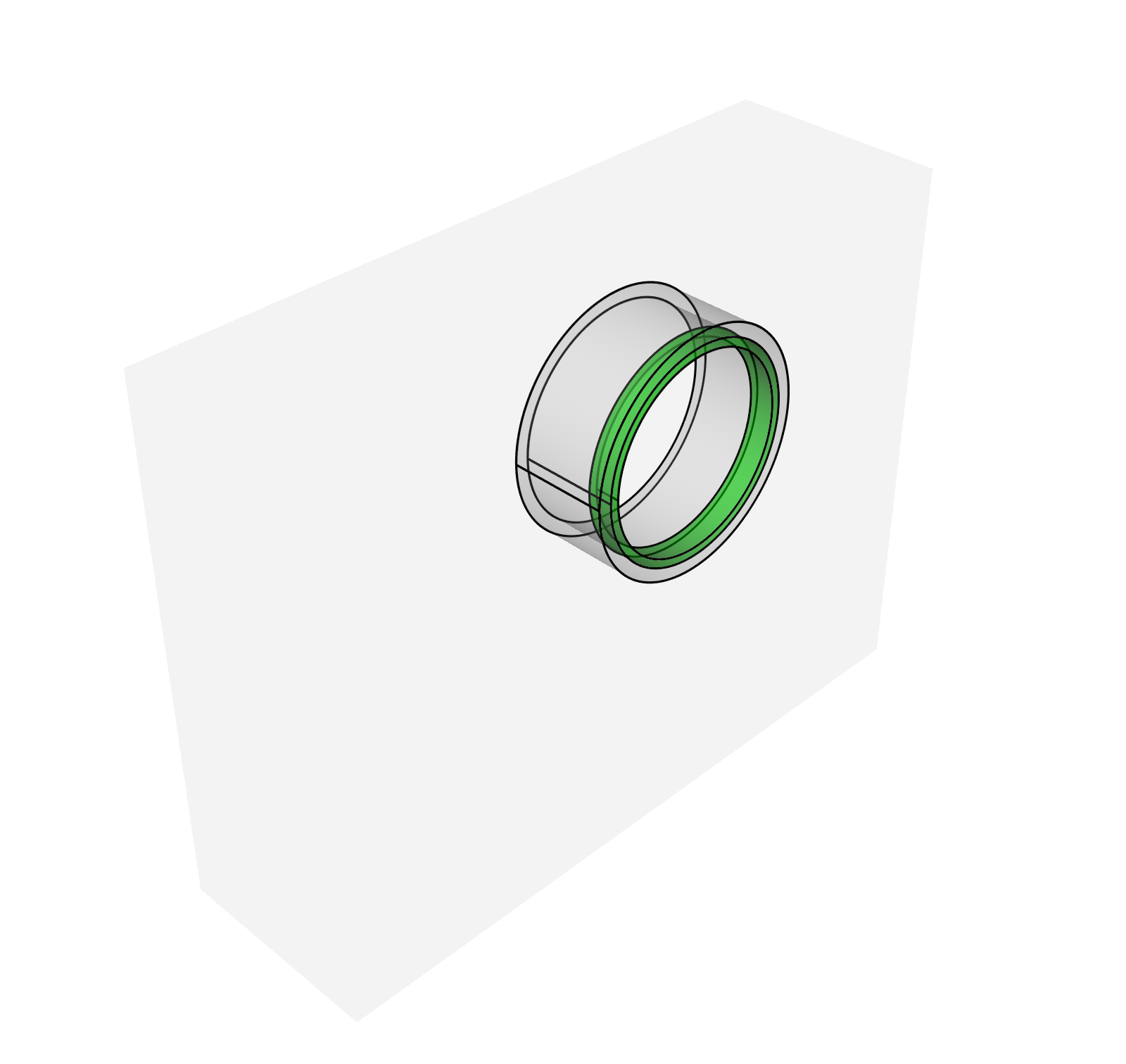} &
        \includegraphics[width=0.124\linewidth]{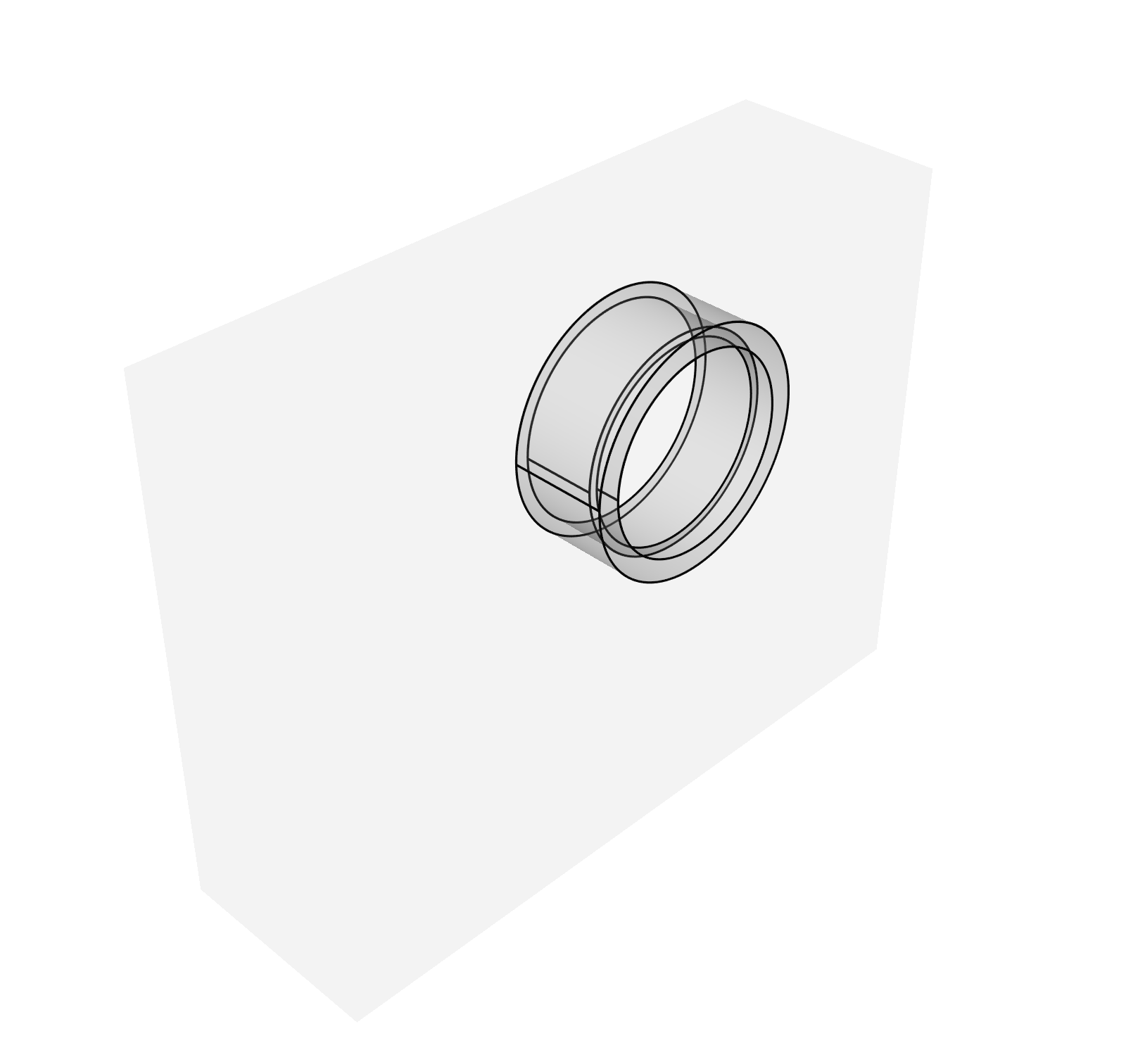}
        \\
        \multicolumn{1}{l}{Ours Heur} & & & & &
        \\
        \includegraphics[width=0.124\linewidth]{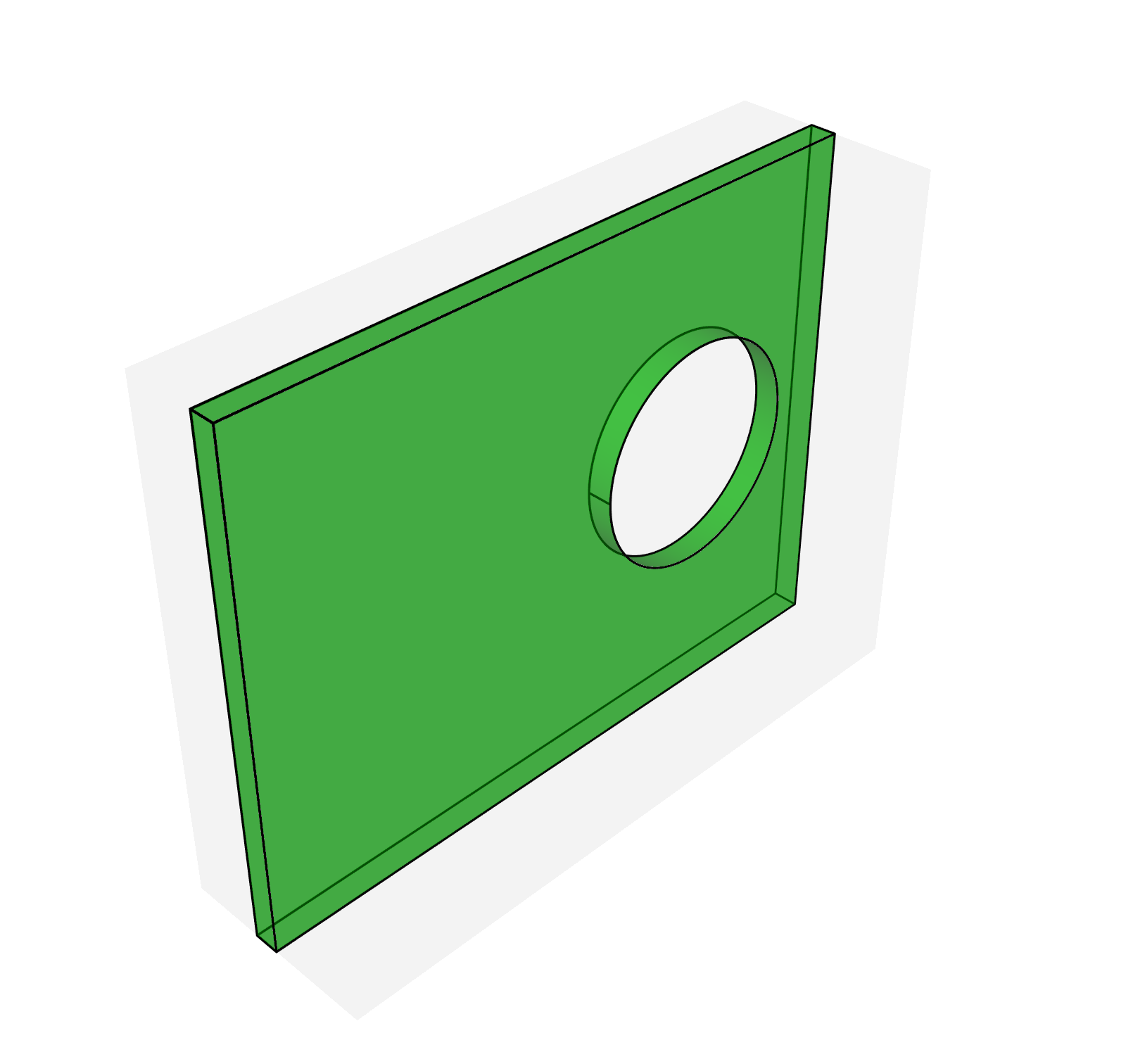} &
        \includegraphics[width=0.124\linewidth]{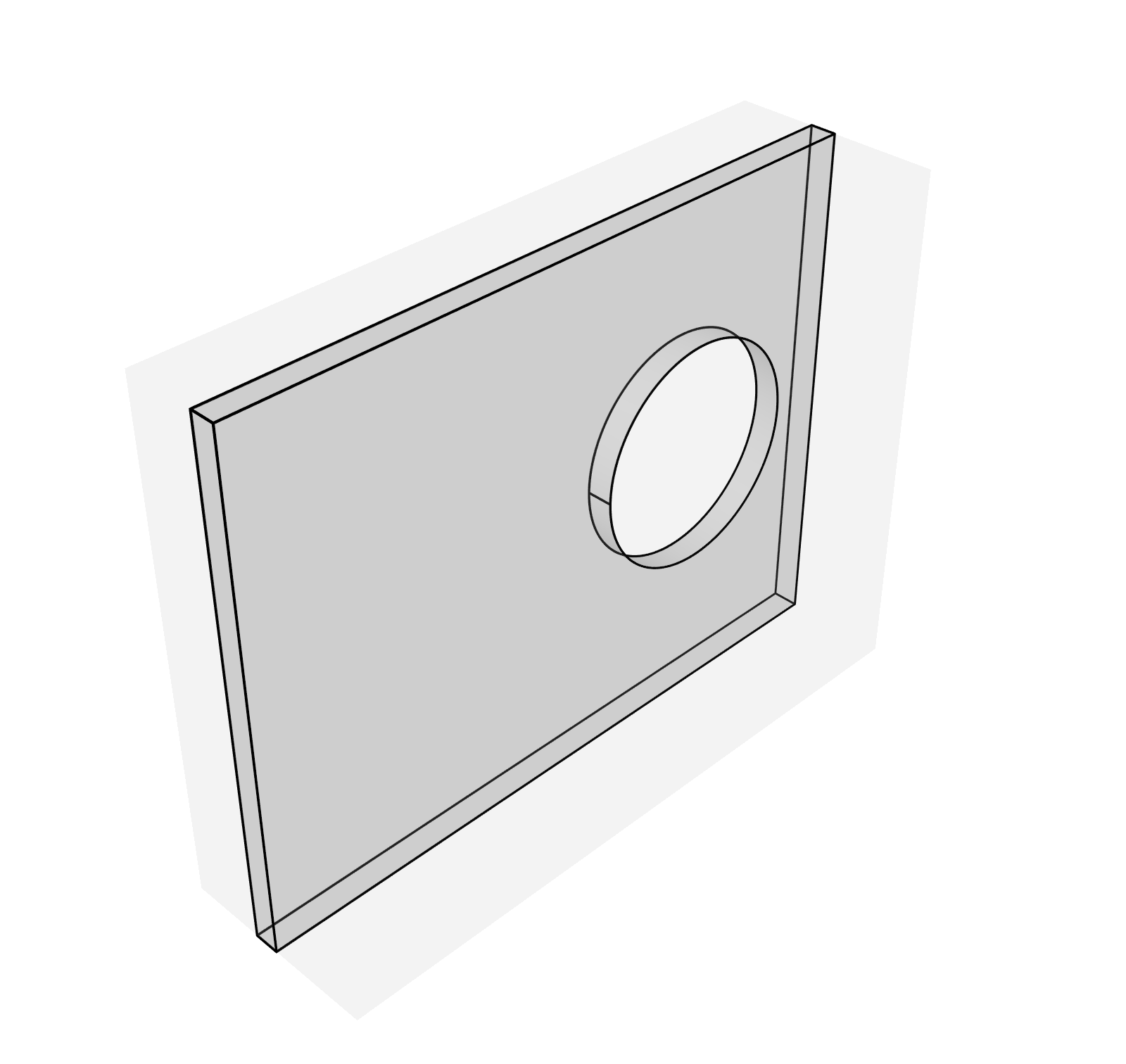} &
        \includegraphics[width=0.124\linewidth]{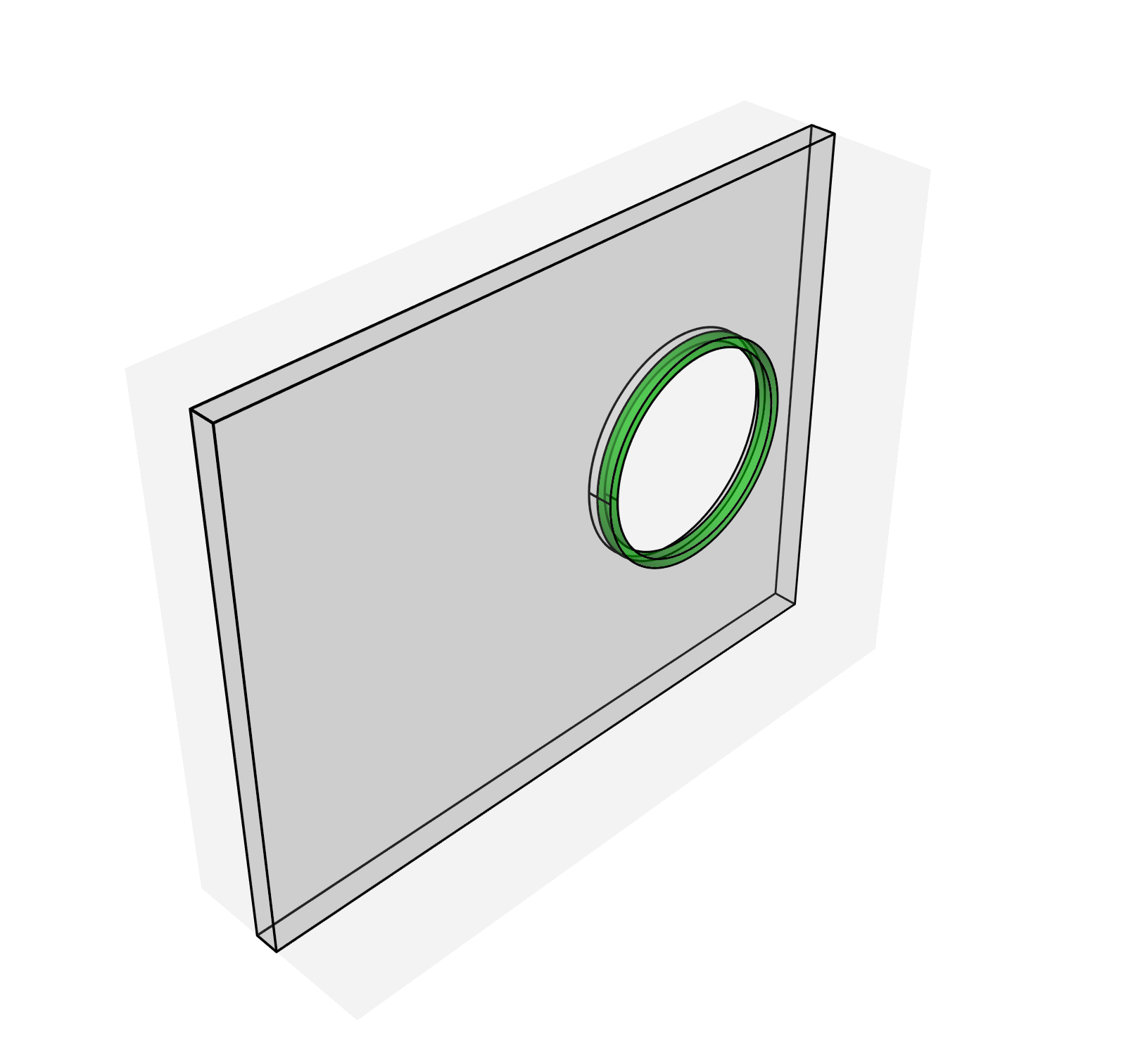} &
        \includegraphics[width=0.124\linewidth]{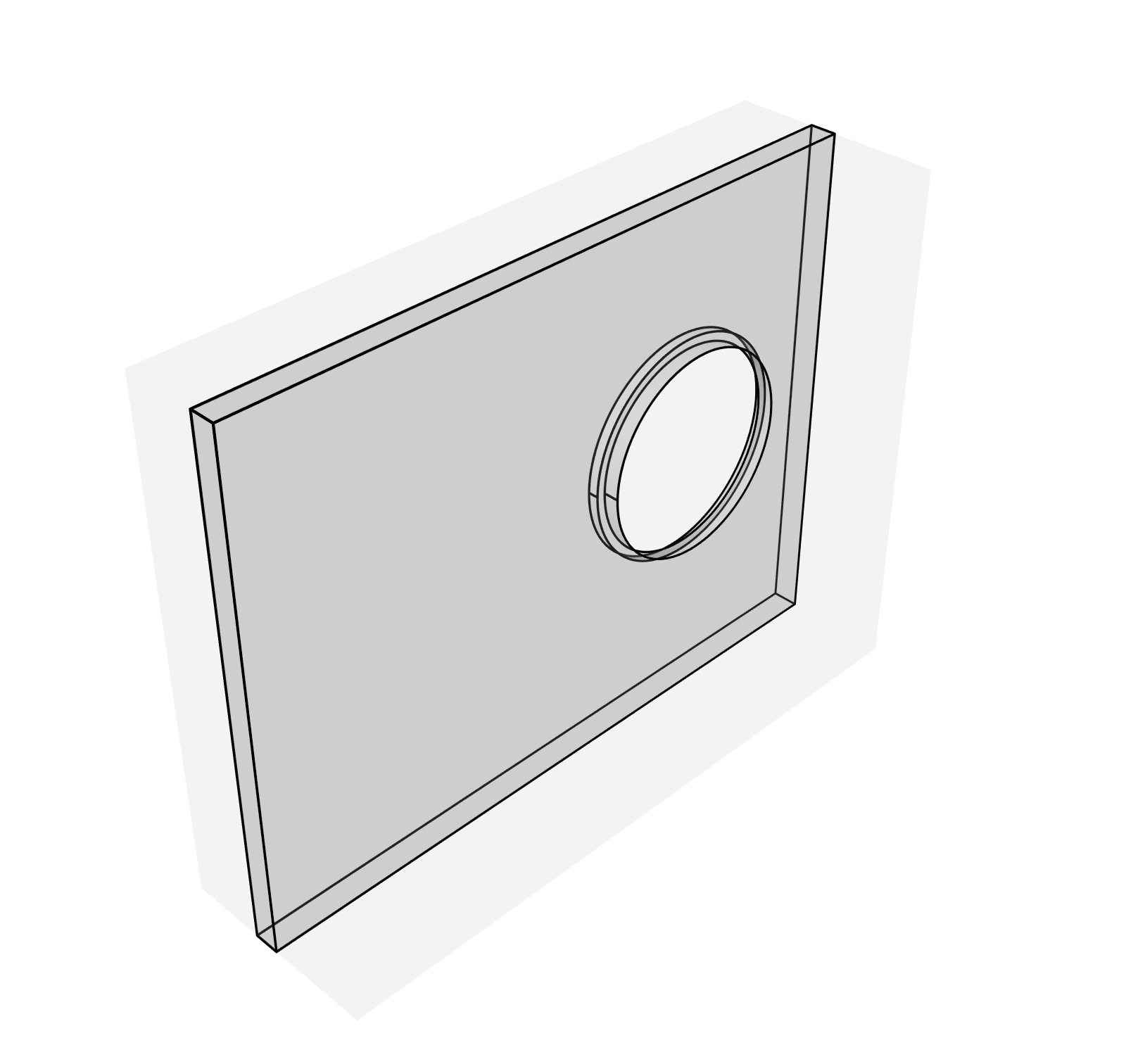} &
        \includegraphics[width=0.124\linewidth]{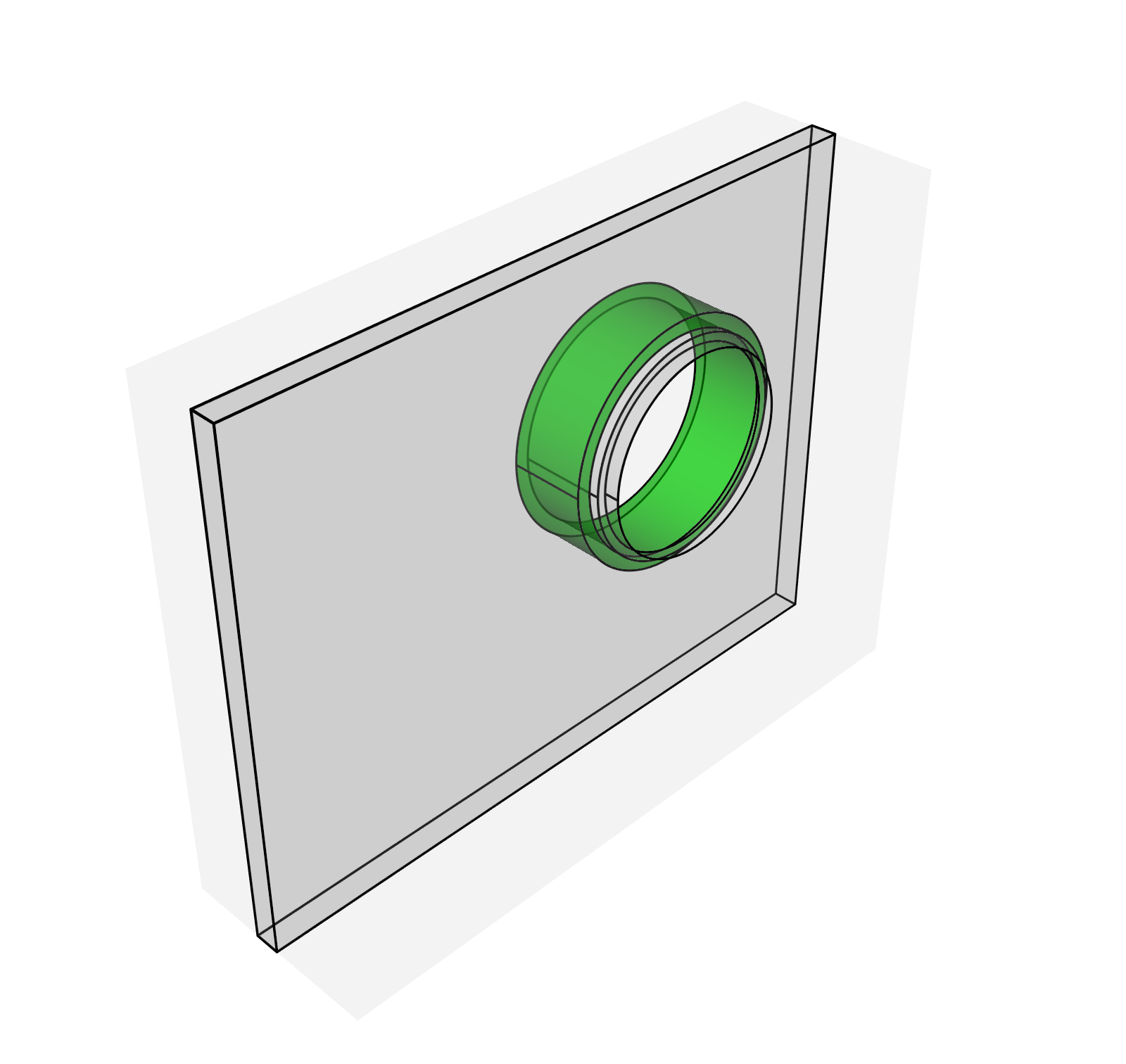} &
        \includegraphics[width=0.124\linewidth]{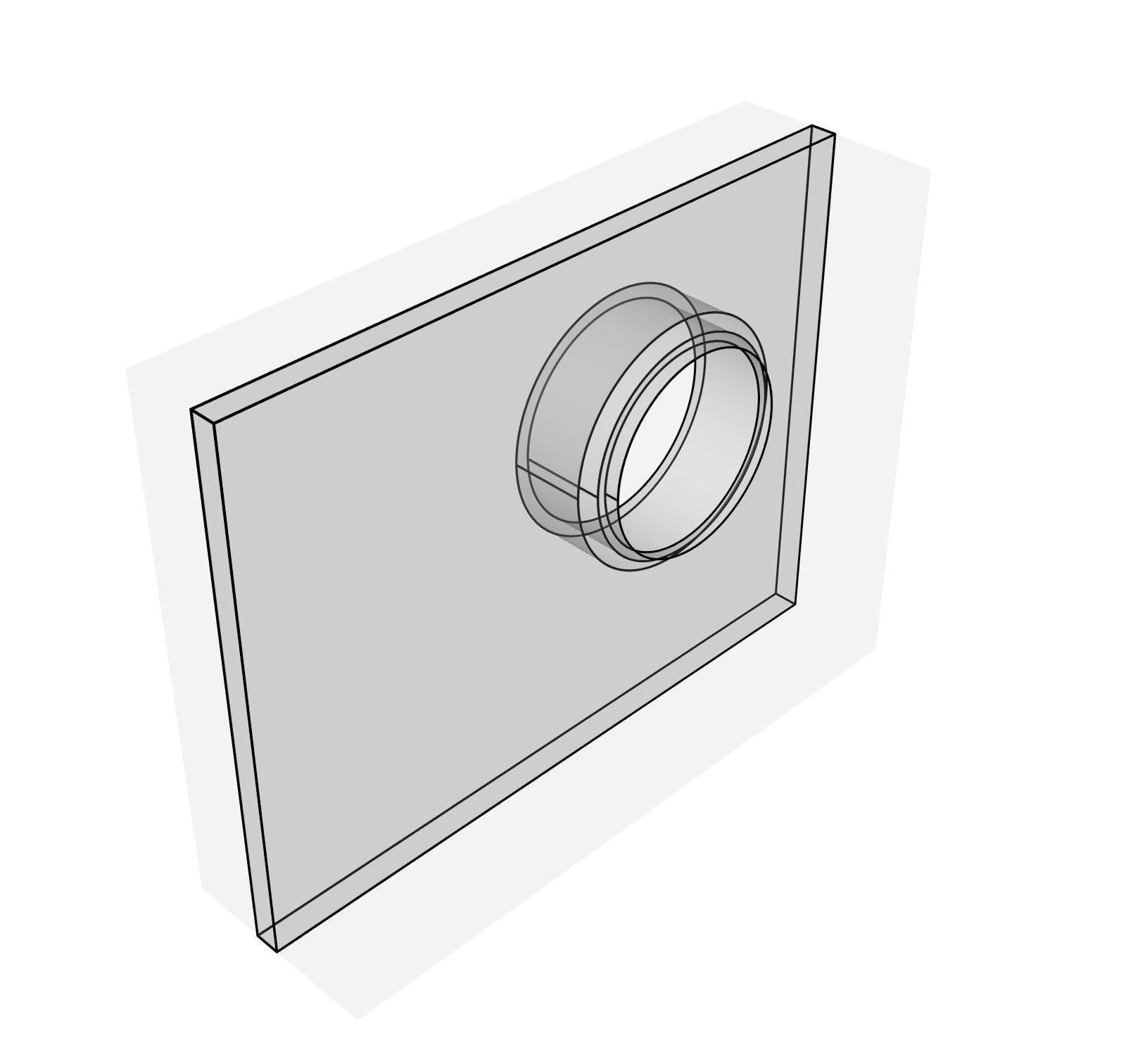} &
        \includegraphics[width=0.124\linewidth]{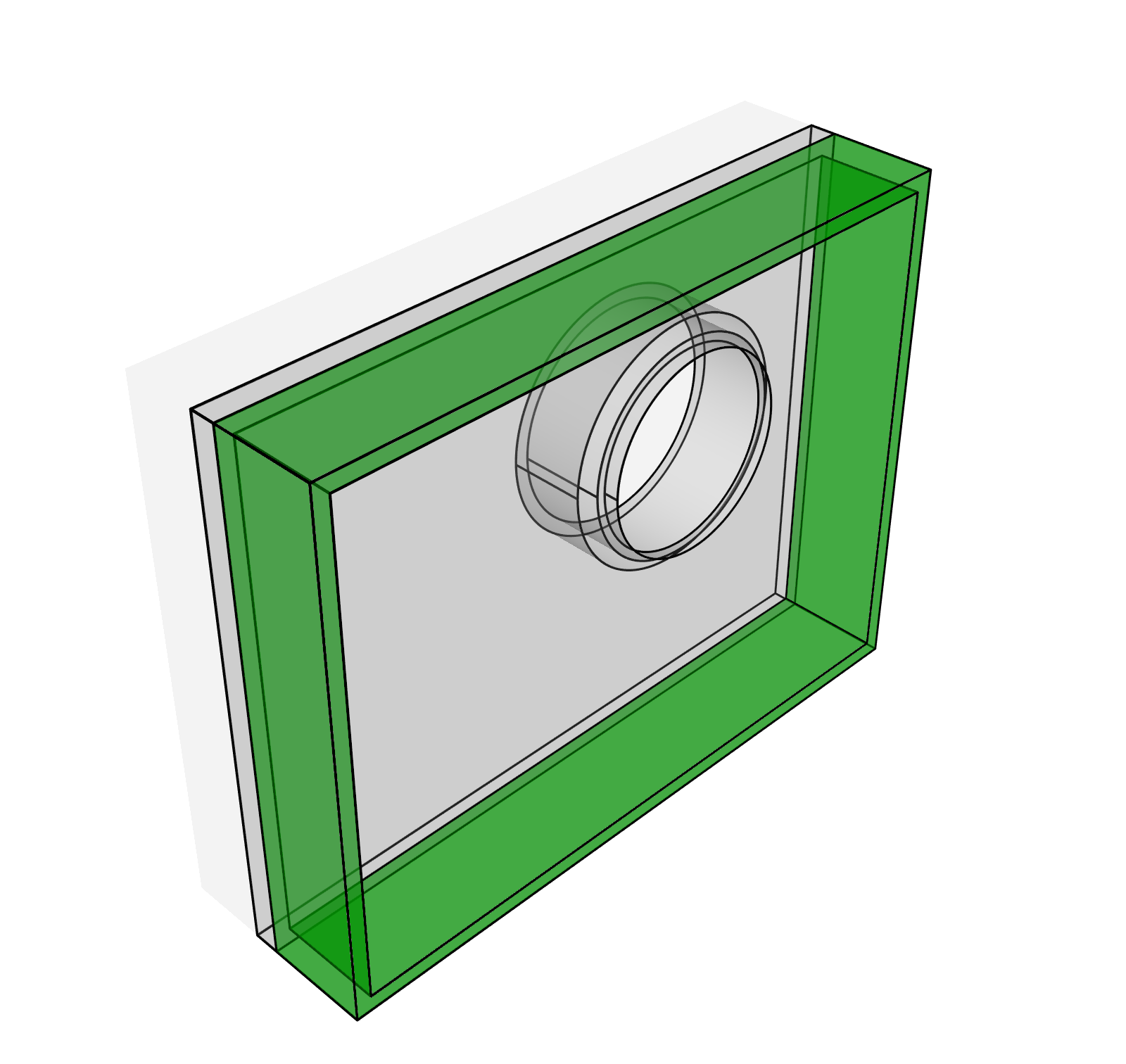} &
        \includegraphics[width=0.124\linewidth]{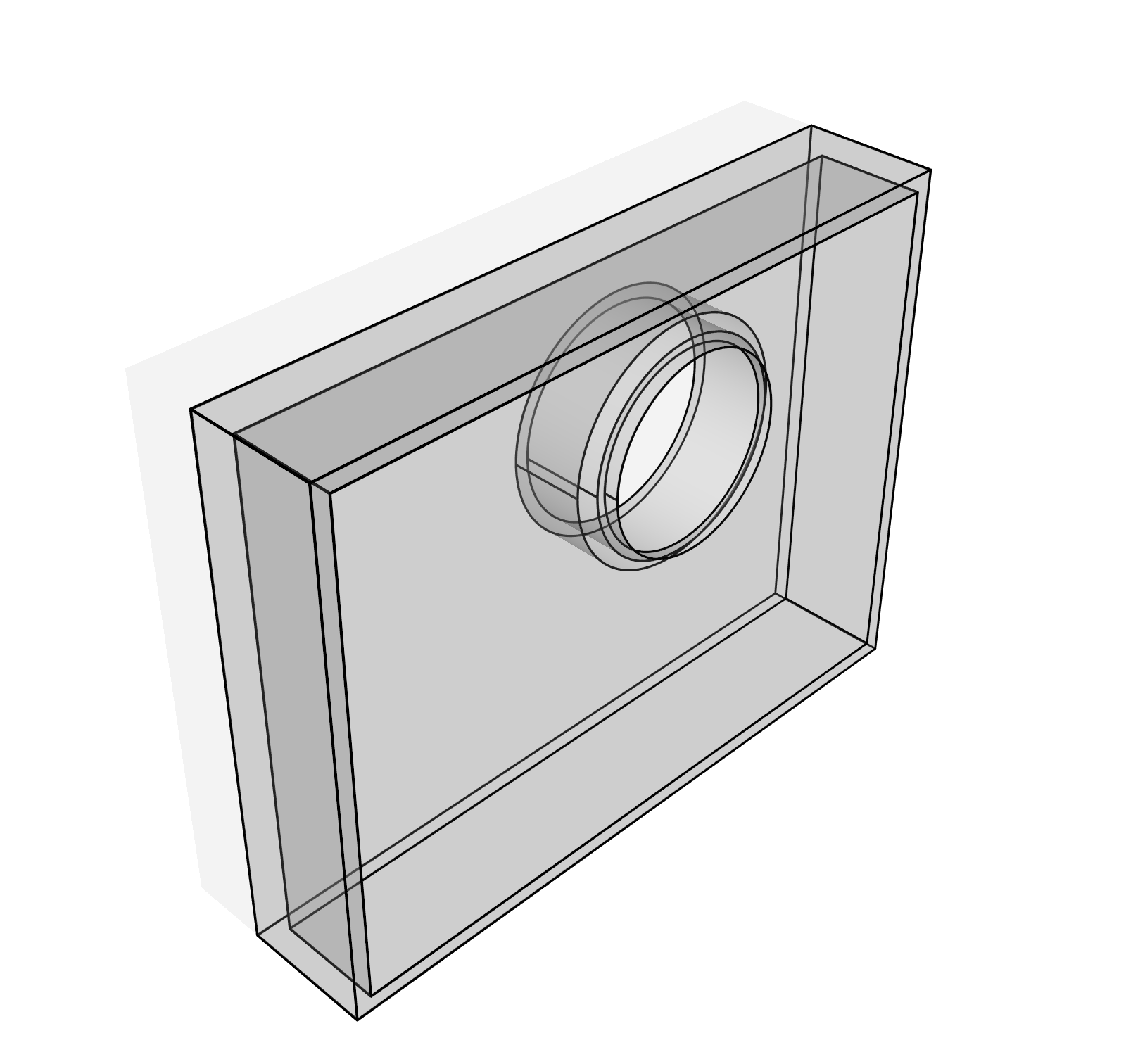}
        \\
        \multicolumn{1}{l}{Ours Net} & & & & &
        \\
        \includegraphics[width=0.124\linewidth]{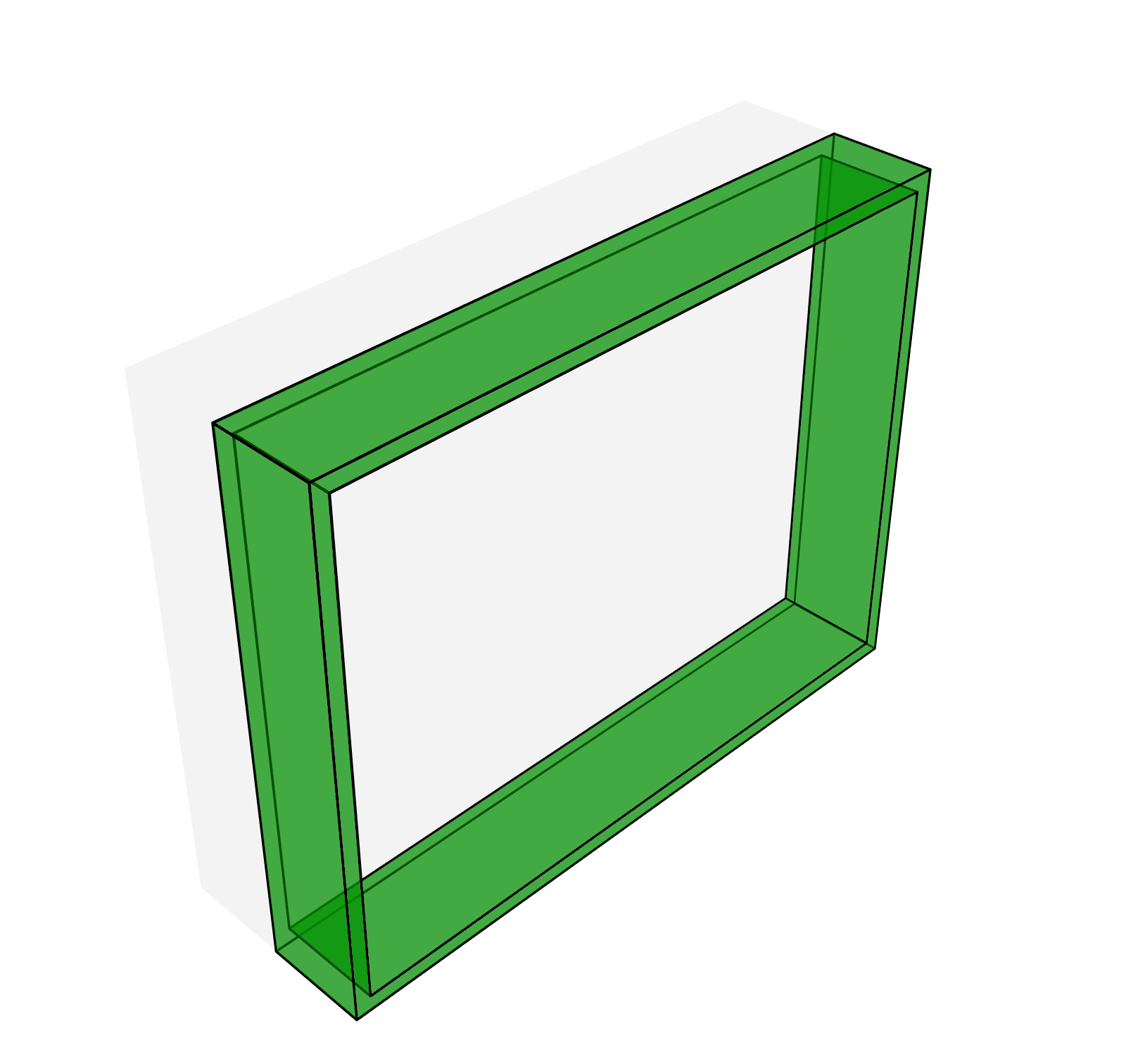} &
        \includegraphics[width=0.124\linewidth]{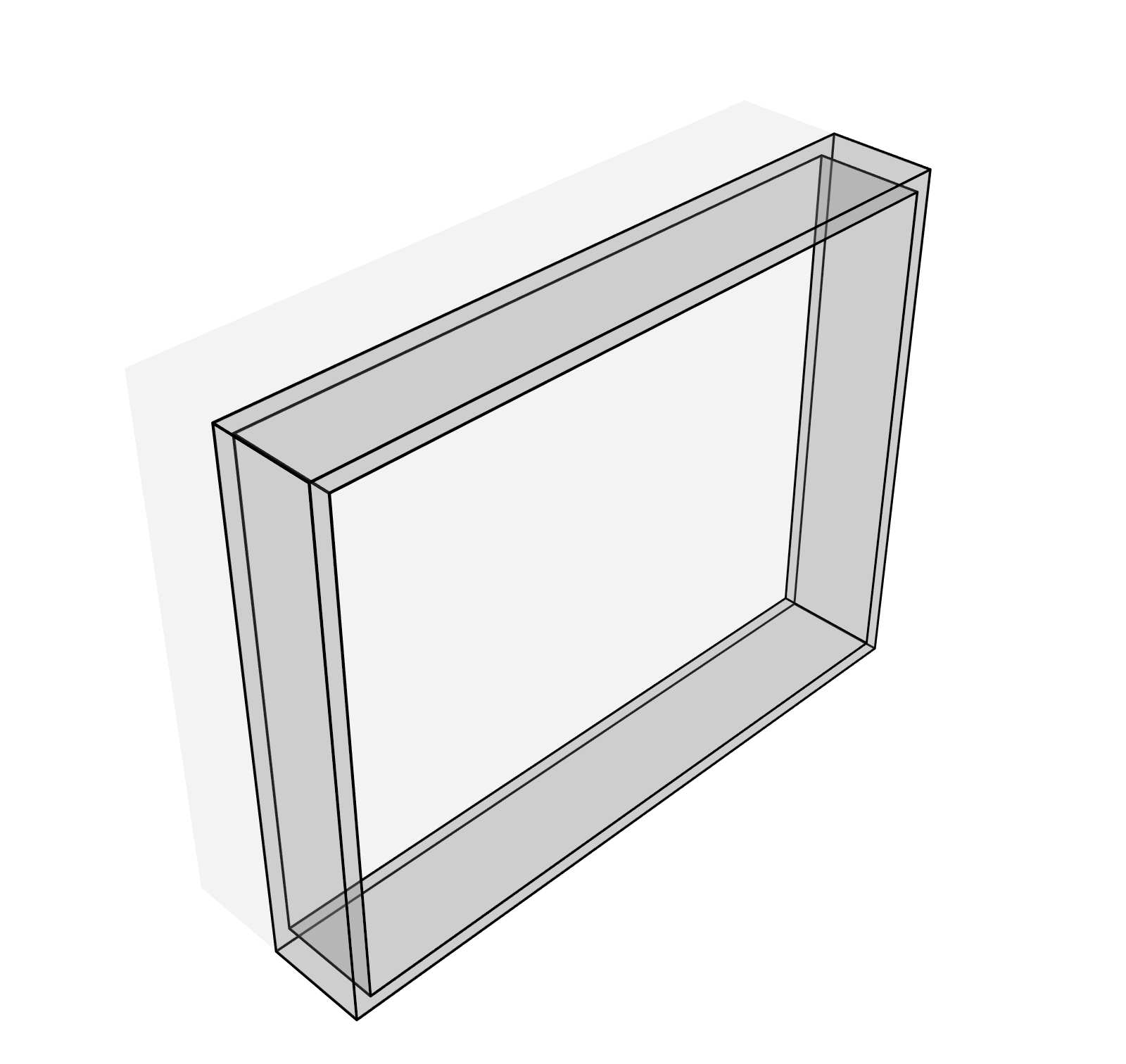} &
        \includegraphics[width=0.124\linewidth]{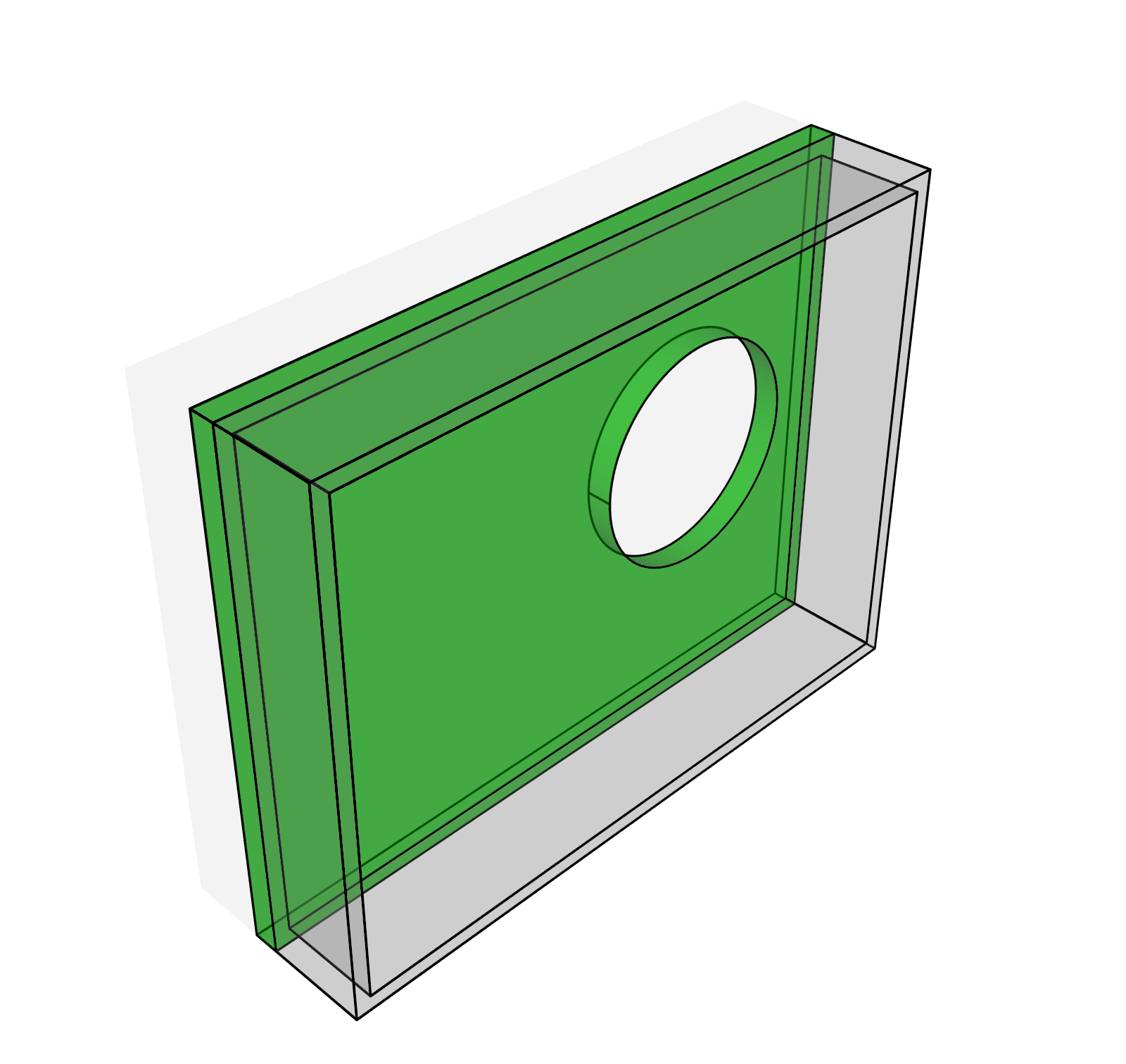} &
        \includegraphics[width=0.124\linewidth]{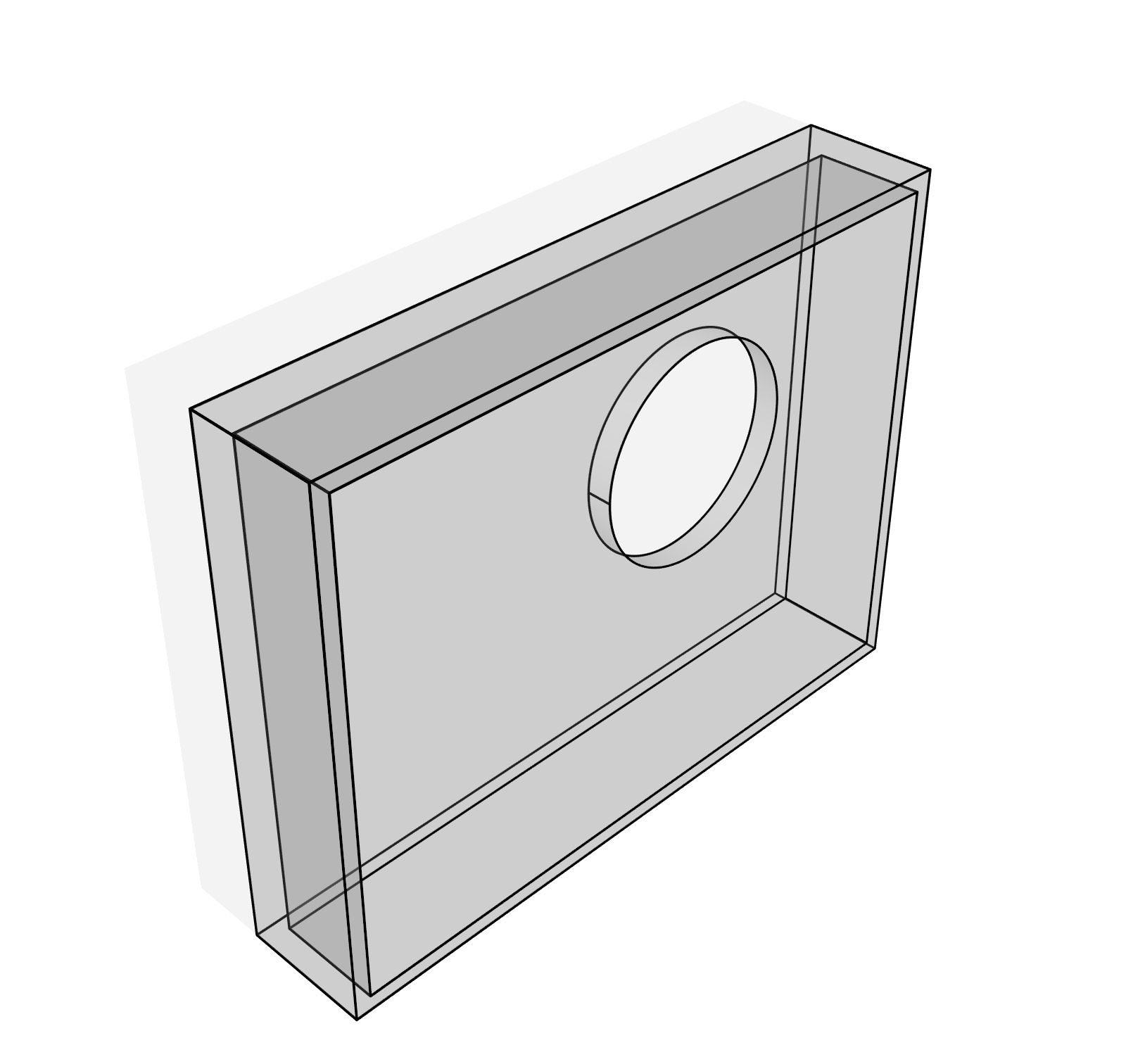} &
        \includegraphics[width=0.124\linewidth]{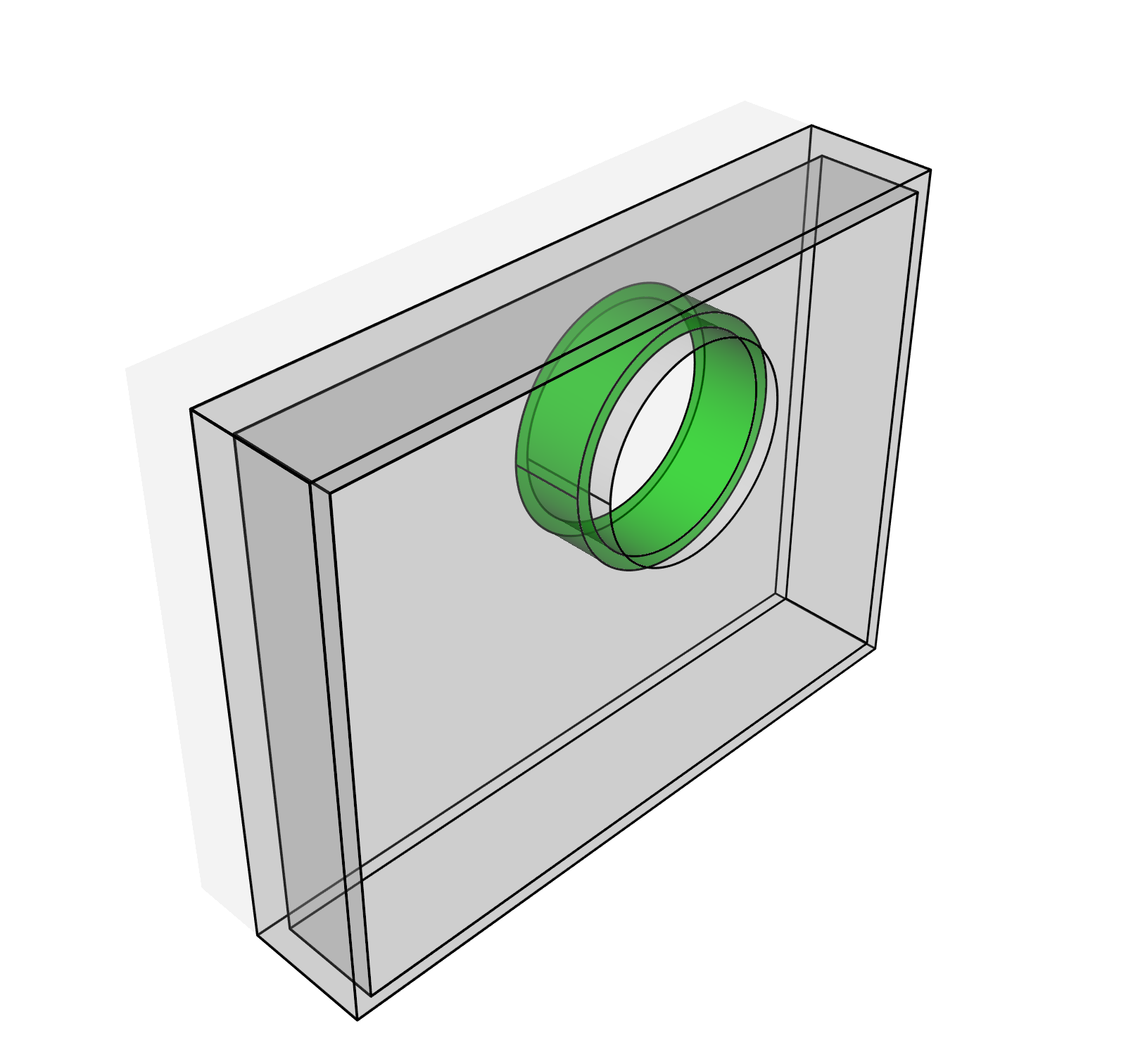} &
        \includegraphics[width=0.124\linewidth]{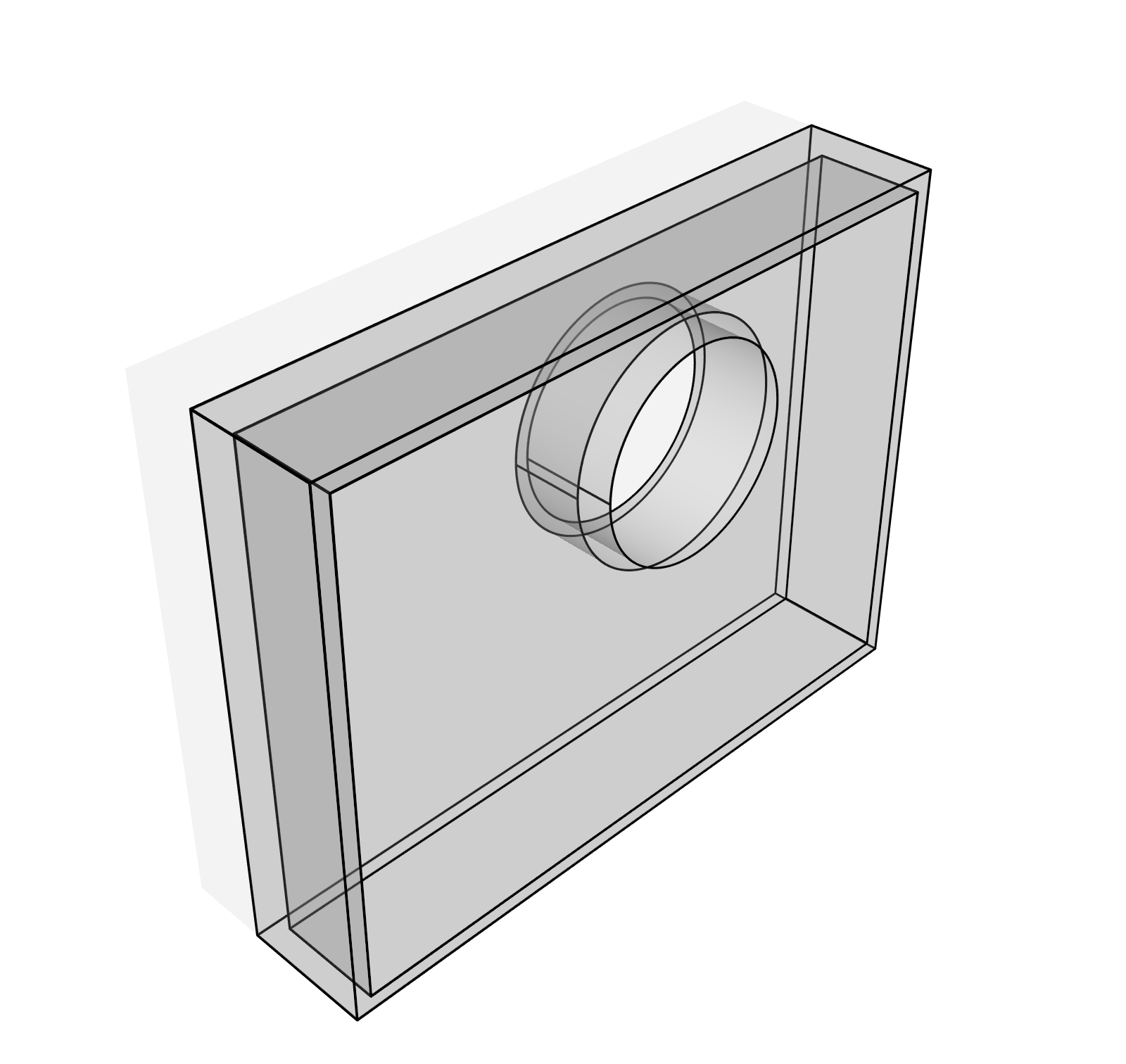} &
        \includegraphics[width=0.124\linewidth]{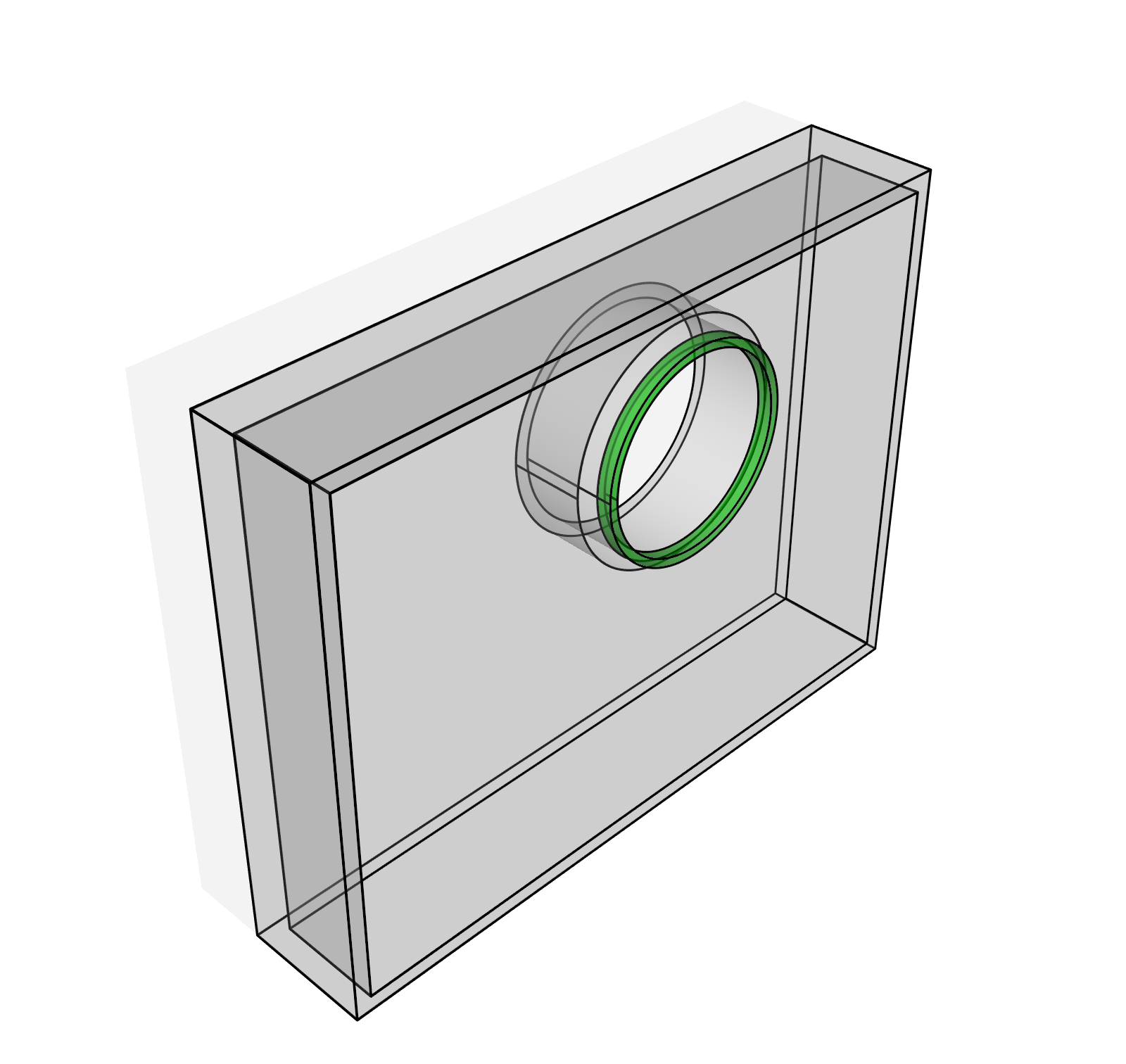} &
        \includegraphics[width=0.124\linewidth]{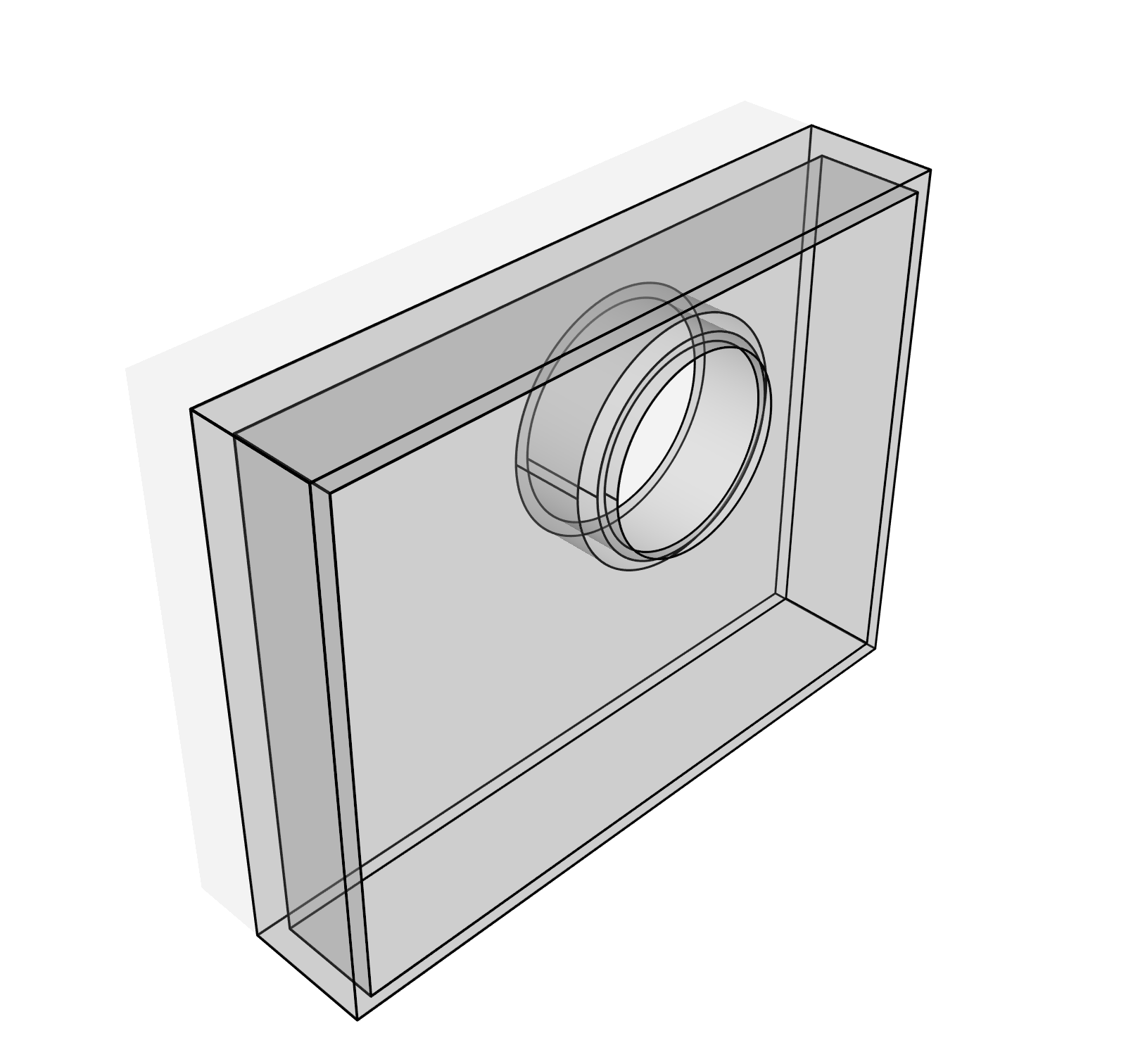}
        \\
    \end{tabular}
    \caption{
    Qualitative comparison of the output of our model's inferred programs (Ours Net) vs. those of Random and Ours Heur. Green: addition, Red: subtraction, Grey: current. (Case 1)
    }
    \label{figure:qualitative_comparision1}
\end{figure*}

\begin{figure*}[h!]
    \centering
    \setlength{\tabcolsep}{1pt}
    \begin{tabular}{cccccccc}
        \multicolumn{2}{c}{Target} & & & &
        \\
        \multicolumn{2}{c}{\includegraphics[width=0.25\linewidth]{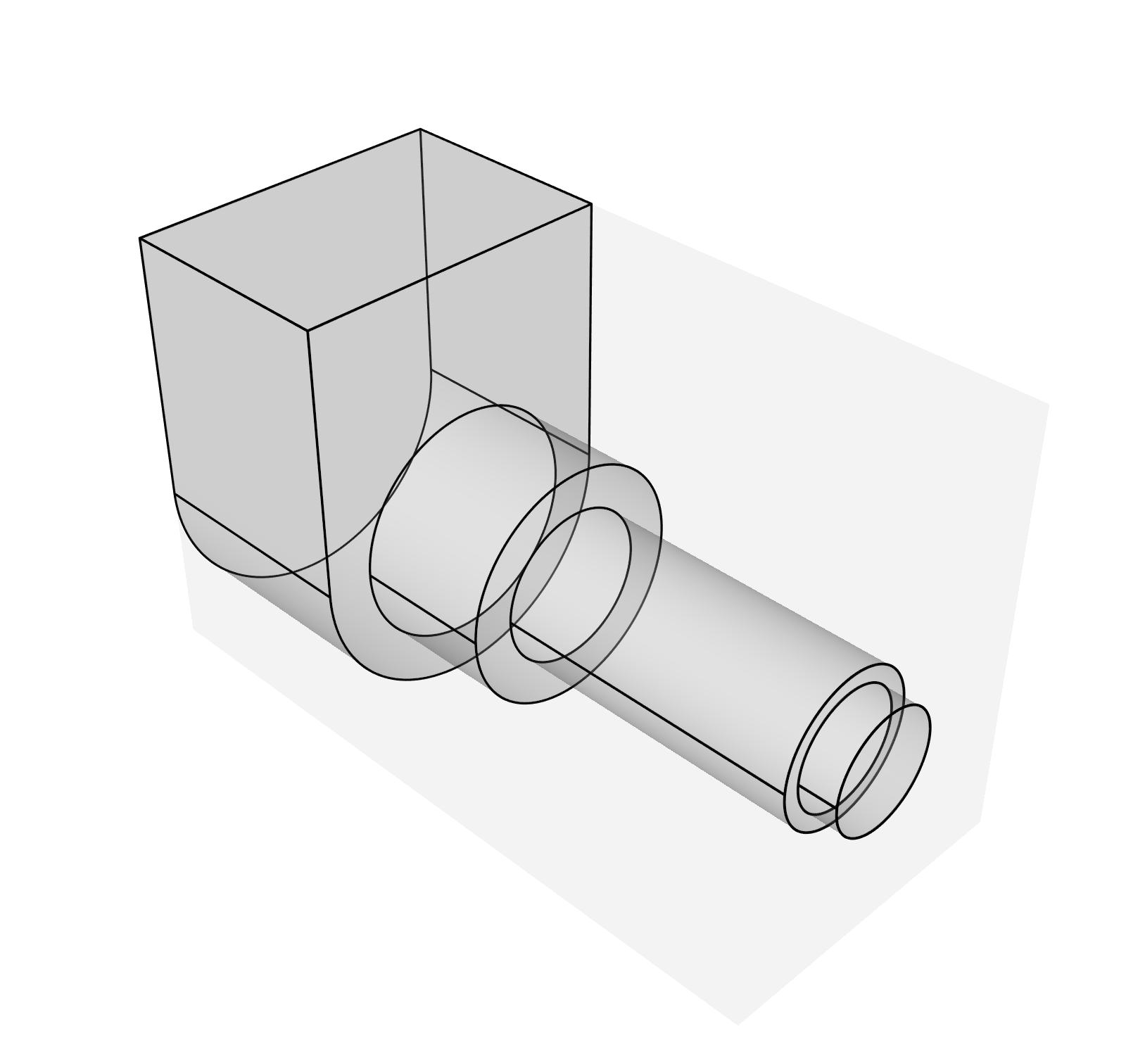}} & & & &
        \\
        \multicolumn{1}{l}{Random} & & & & &
        \\
        \includegraphics[width=0.124\linewidth]{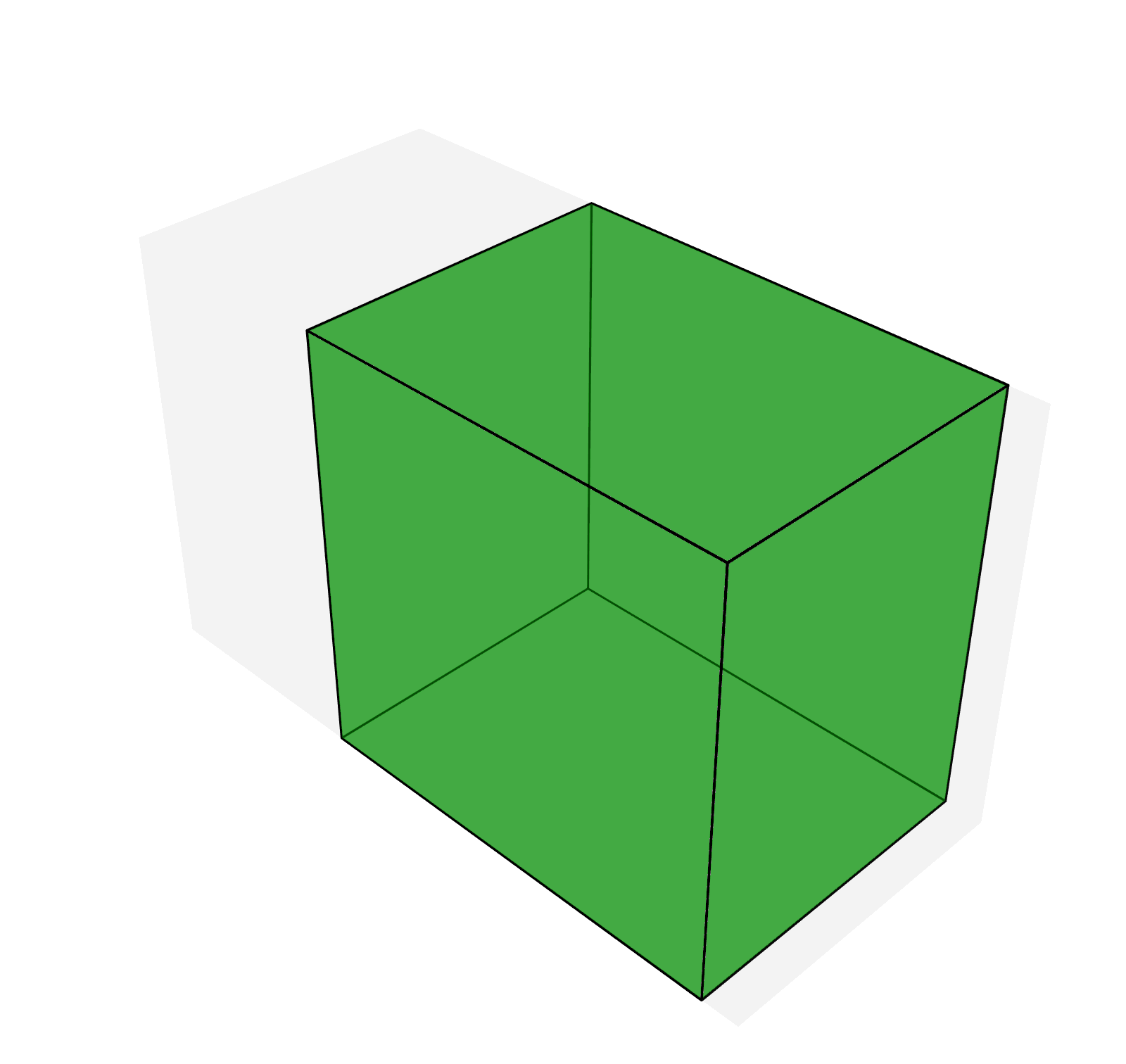} &
        \includegraphics[width=0.124\linewidth]{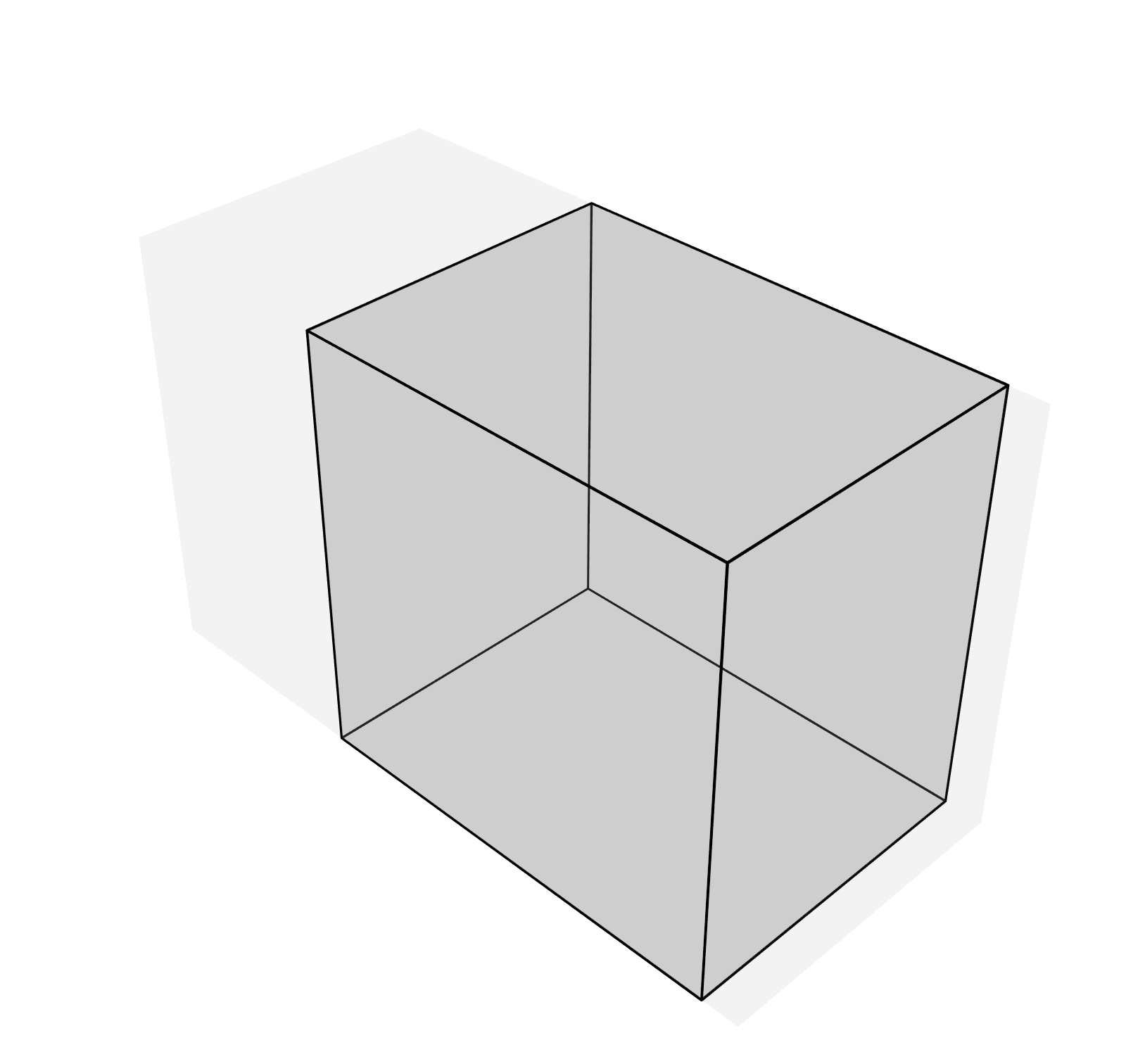} &
        \includegraphics[width=0.124\linewidth]{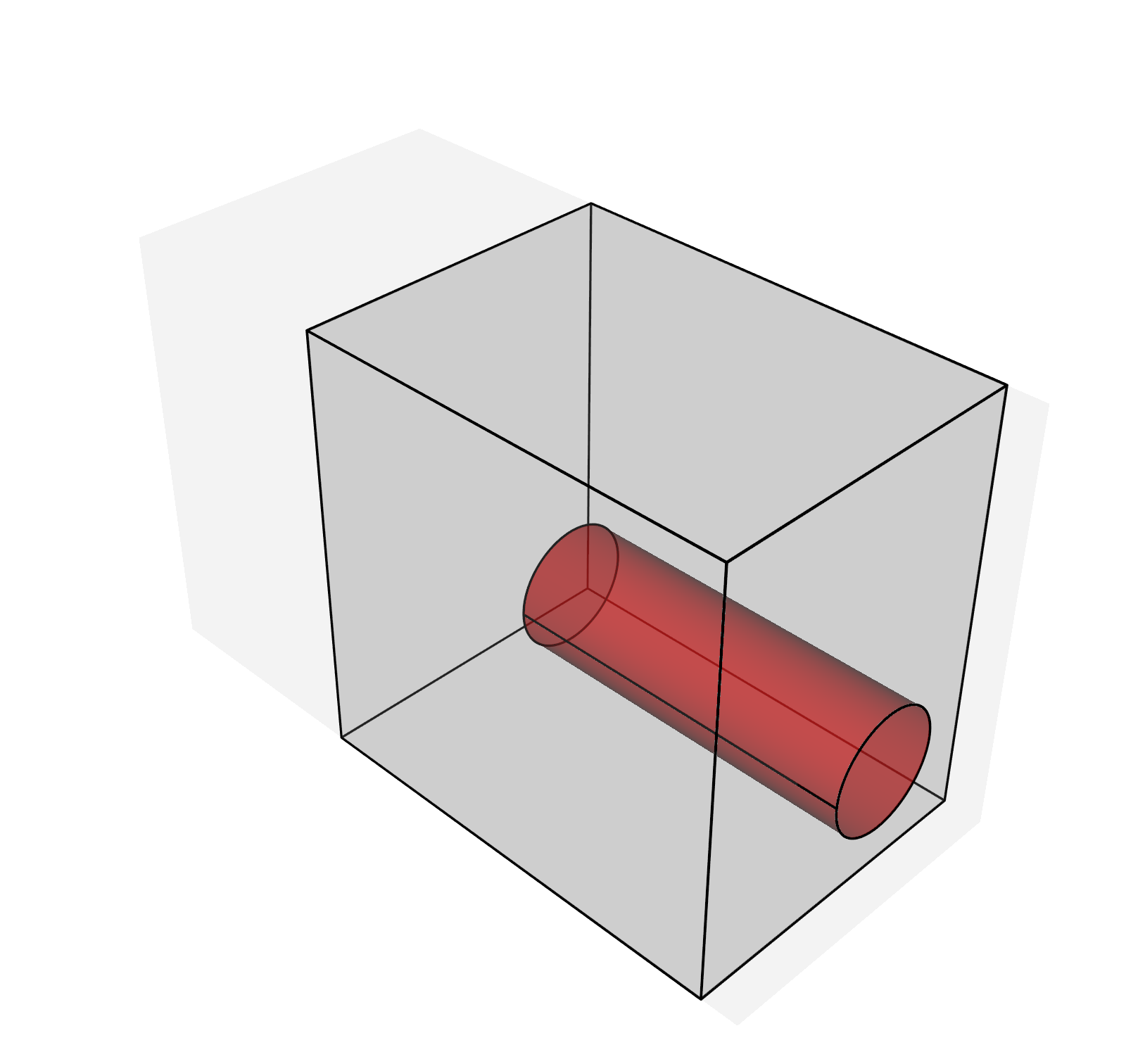} &
        \includegraphics[width=0.124\linewidth]{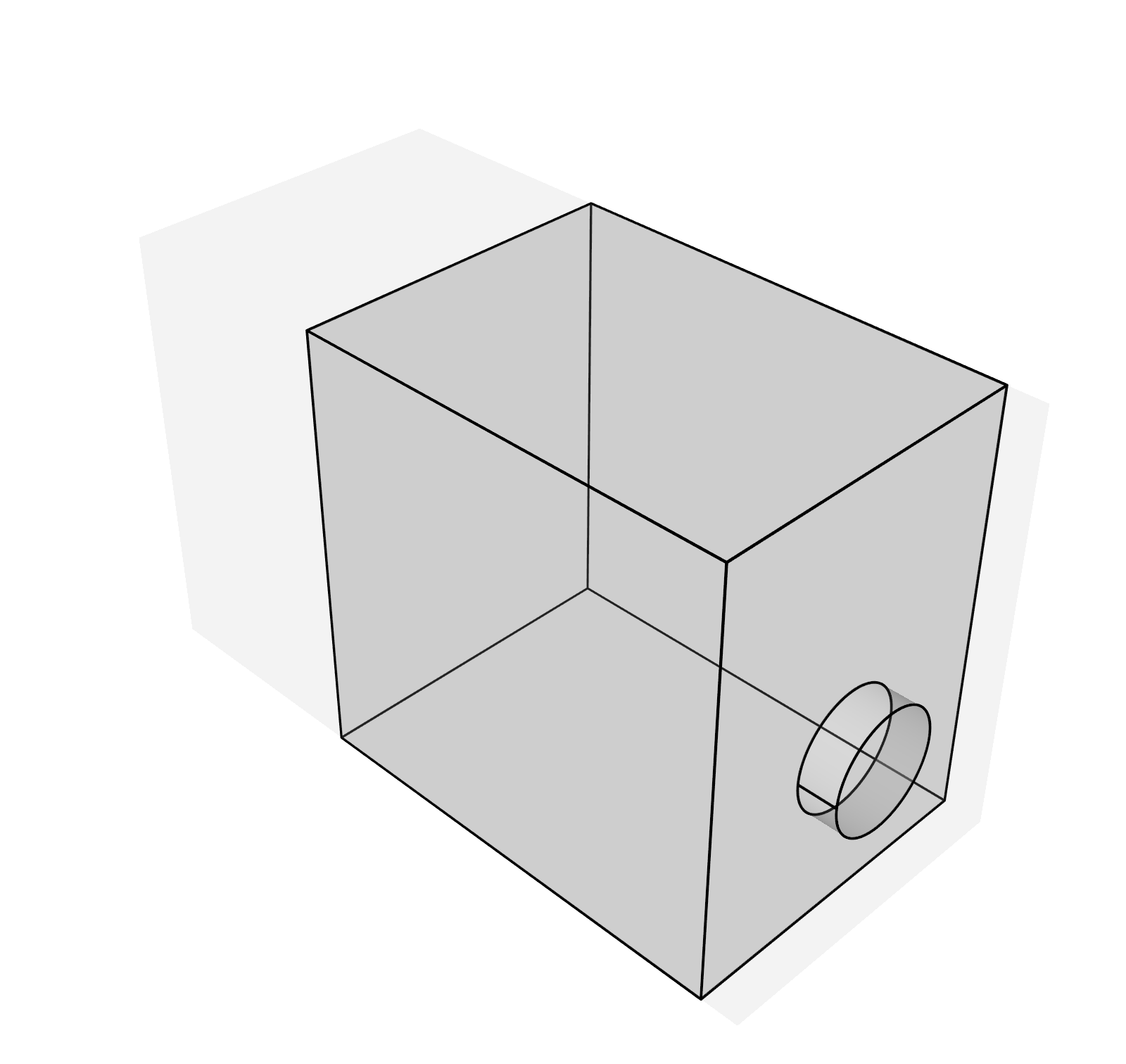} &
        \includegraphics[width=0.124\linewidth]{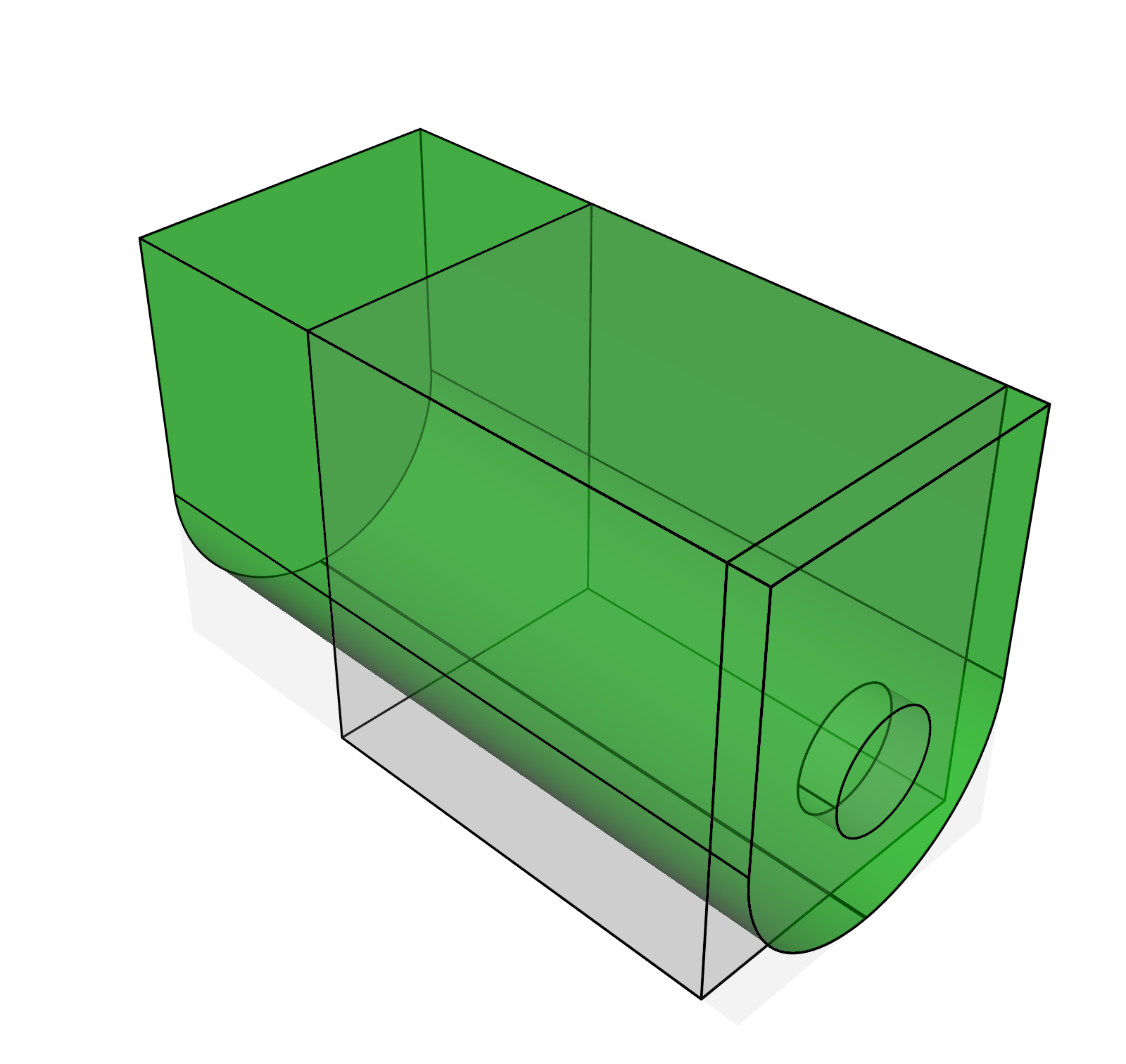} &
        \includegraphics[width=0.124\linewidth]{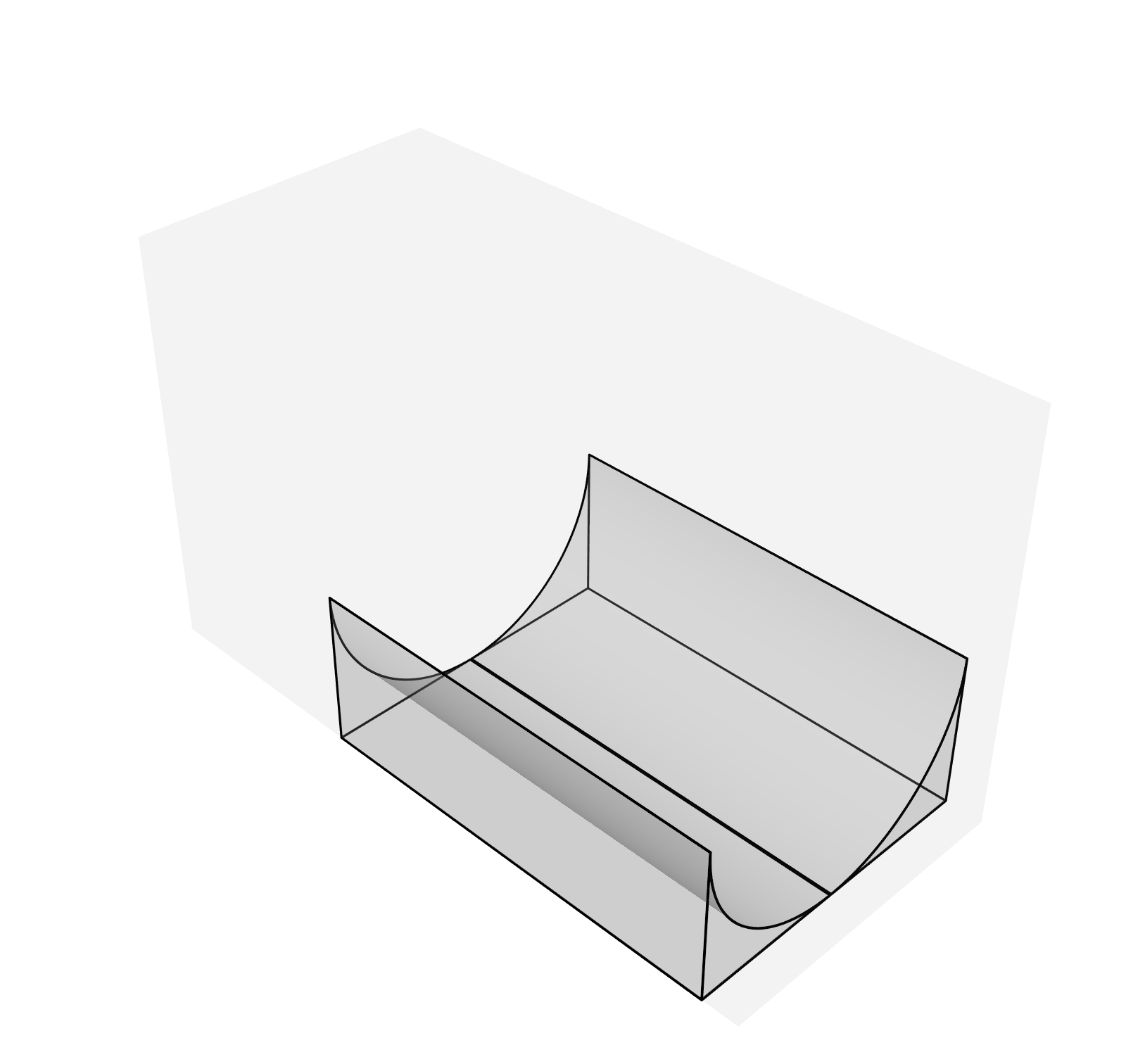} &
        \includegraphics[width=0.124\linewidth]{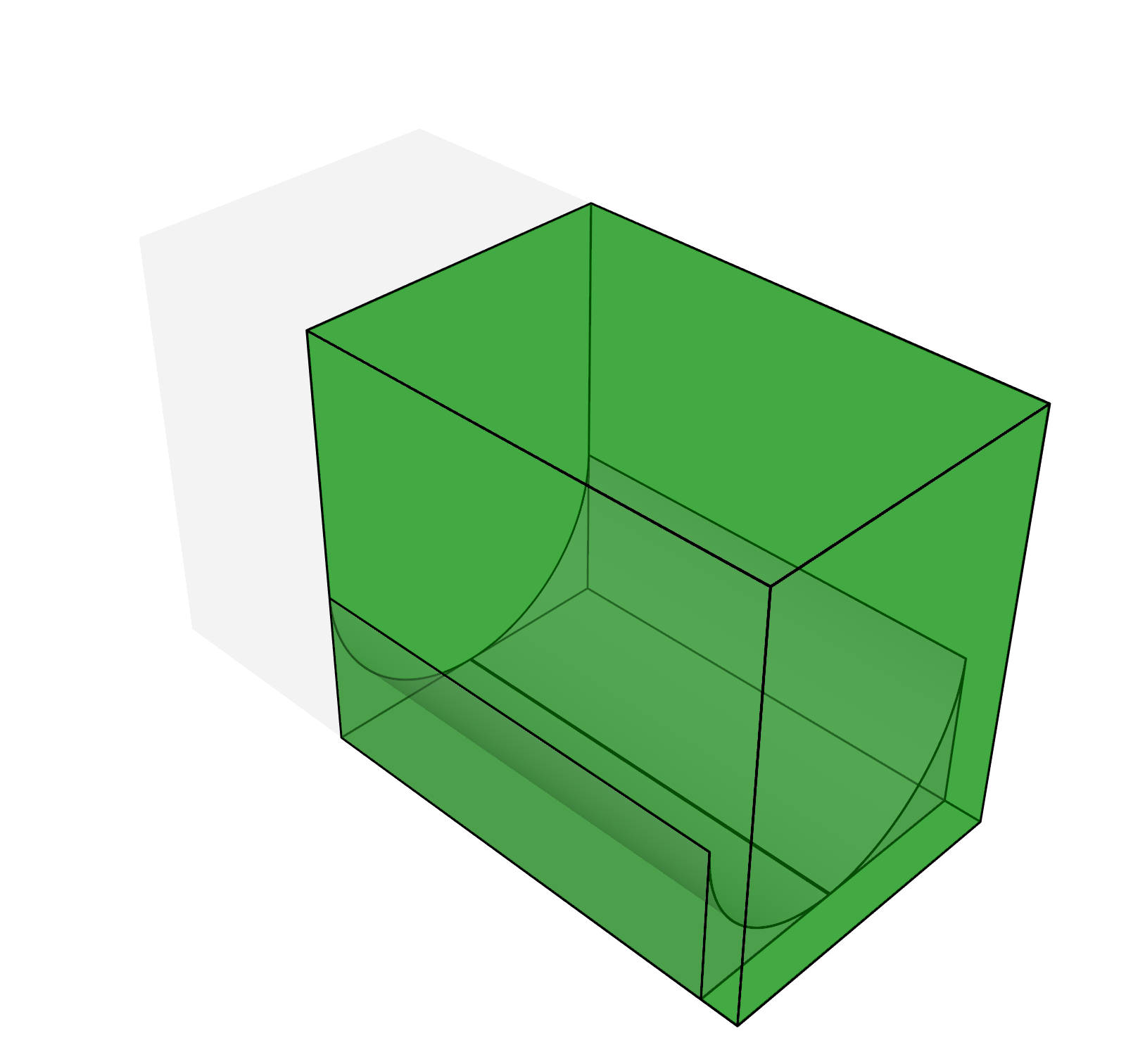} &
        \includegraphics[width=0.124\linewidth]{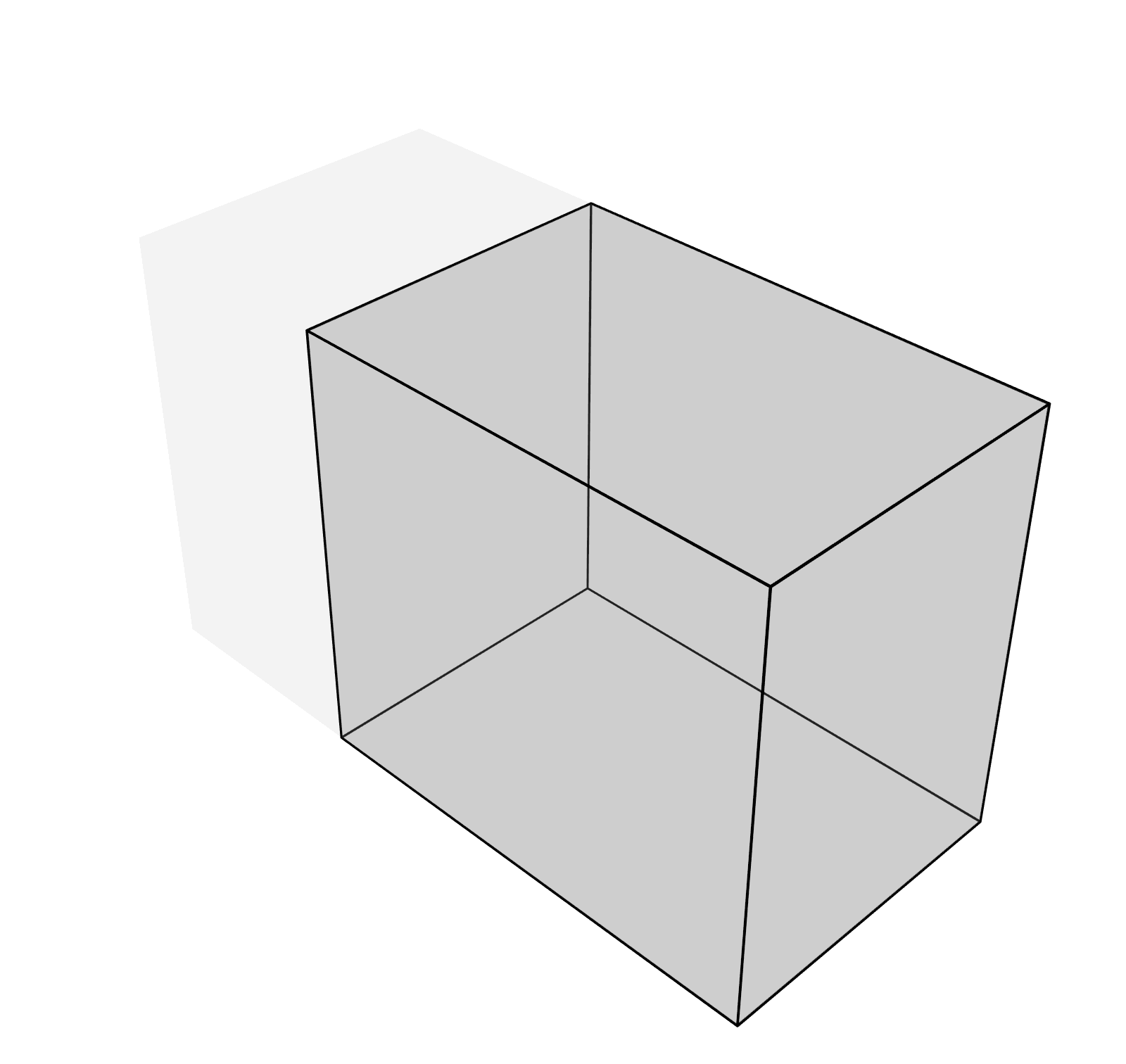}
        \\
        \multicolumn{1}{l}{Ours Heur} & & & & &
        \\
        \includegraphics[width=0.124\linewidth]{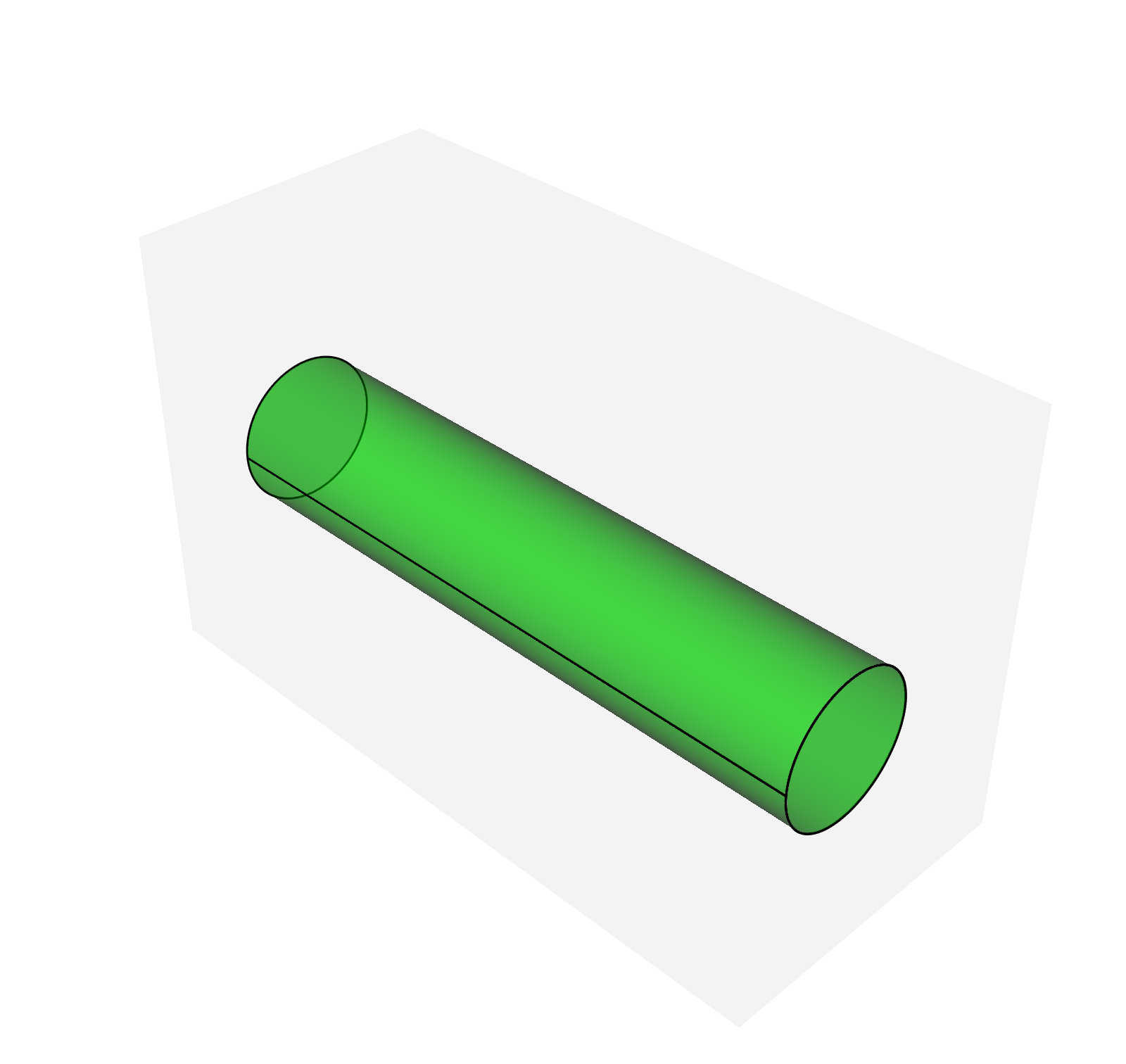} &
        \includegraphics[width=0.124\linewidth]{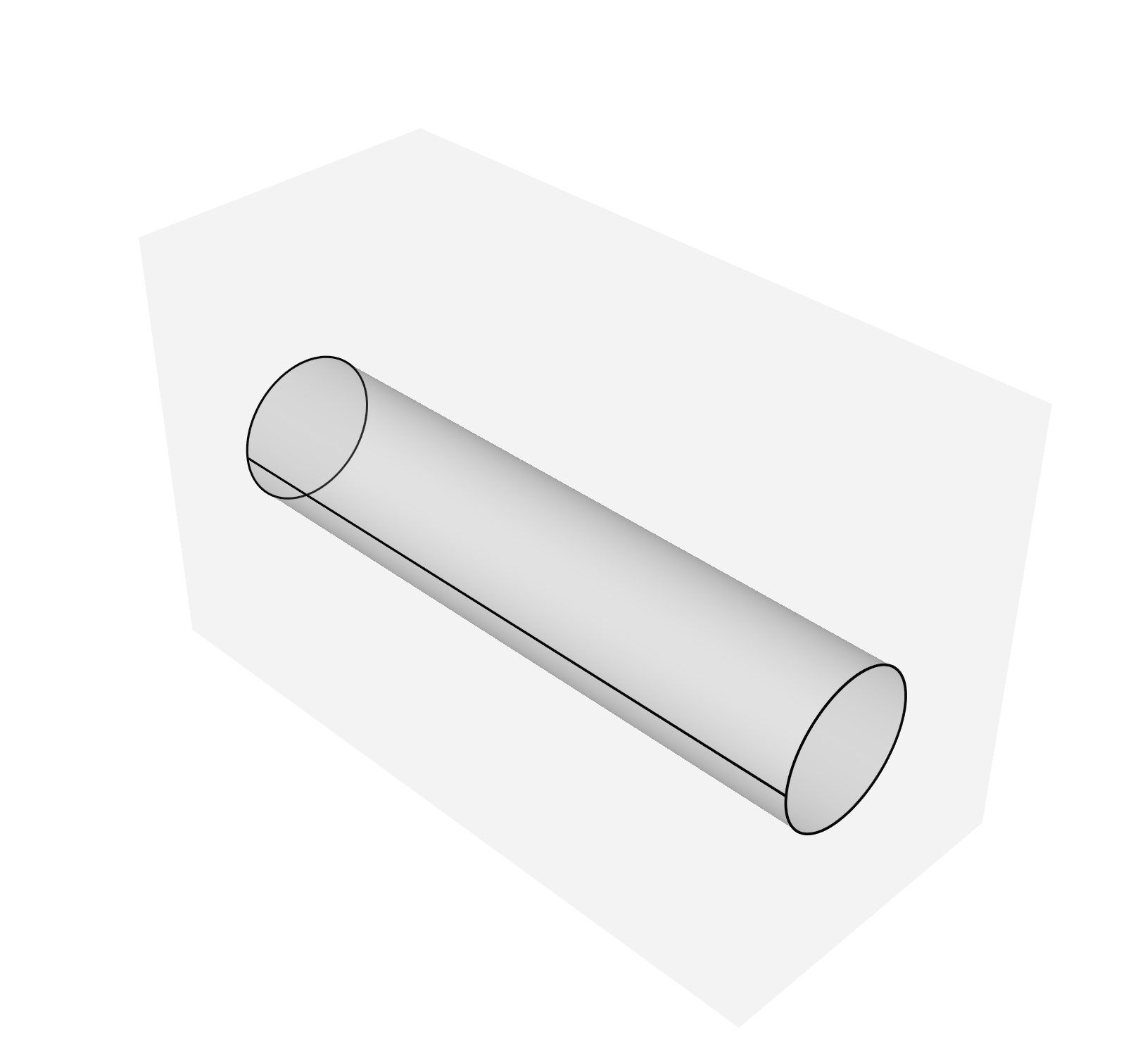} &
        \includegraphics[width=0.124\linewidth]{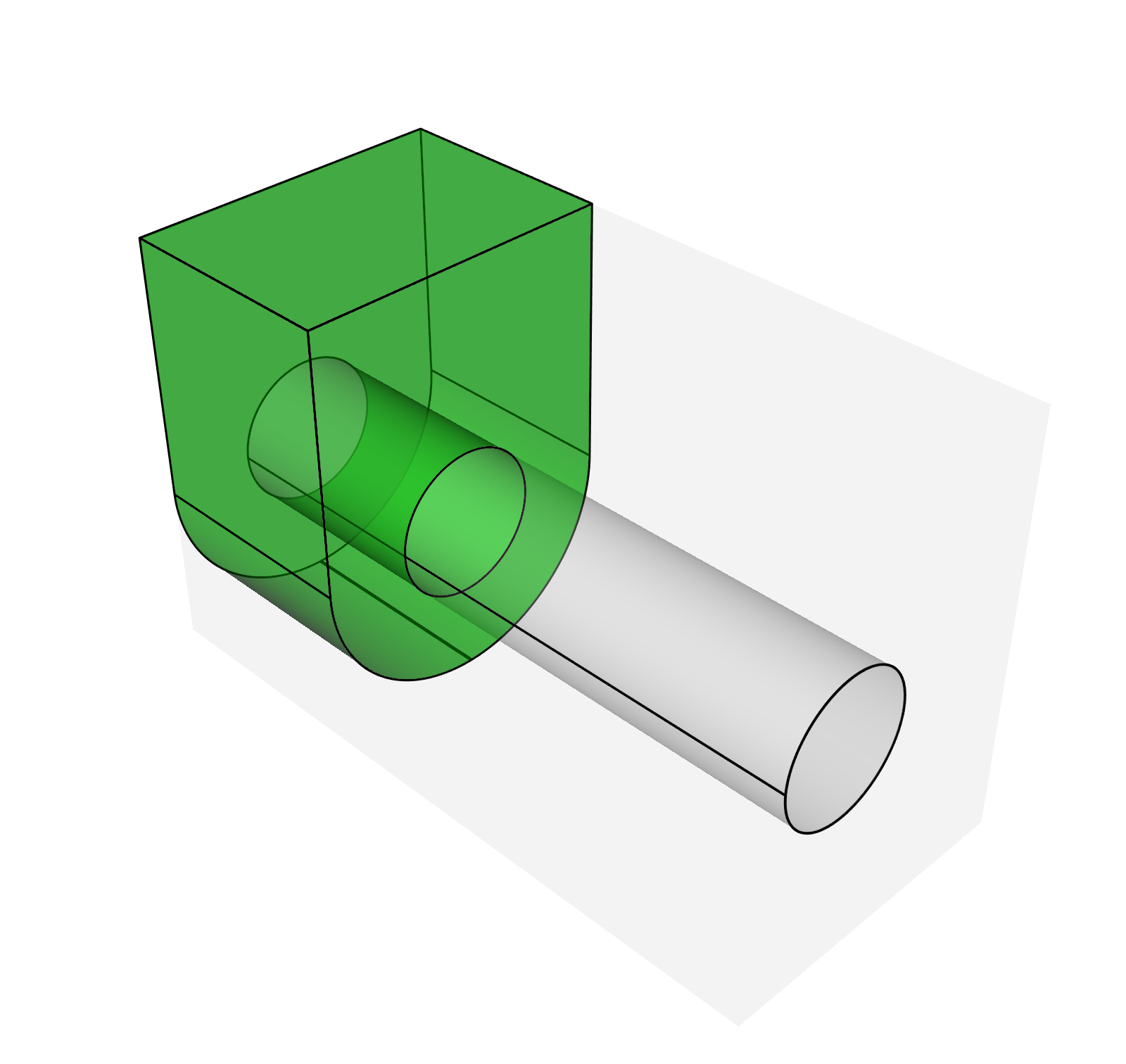} &
        \includegraphics[width=0.124\linewidth]{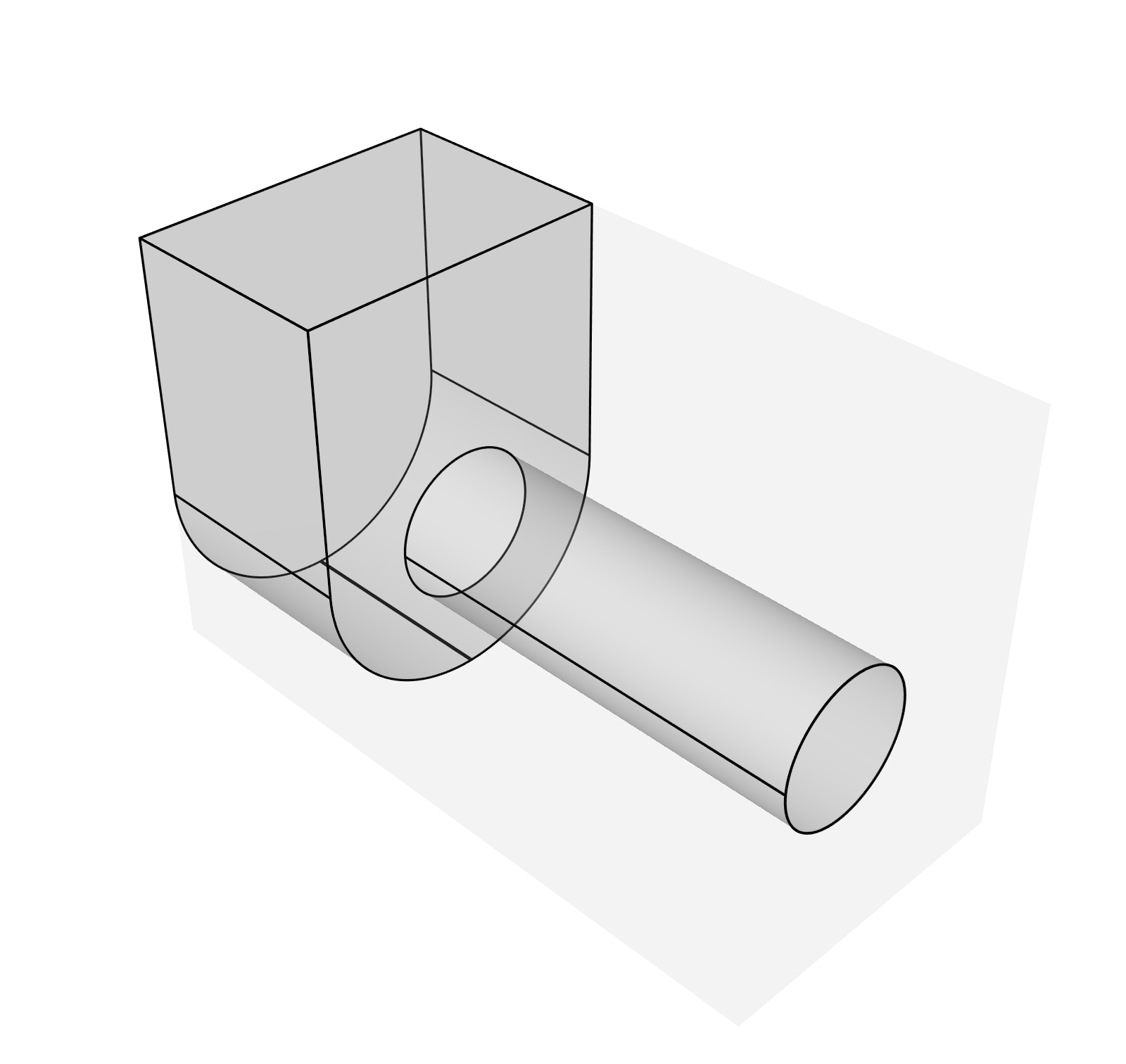} &
        \includegraphics[width=0.124\linewidth]{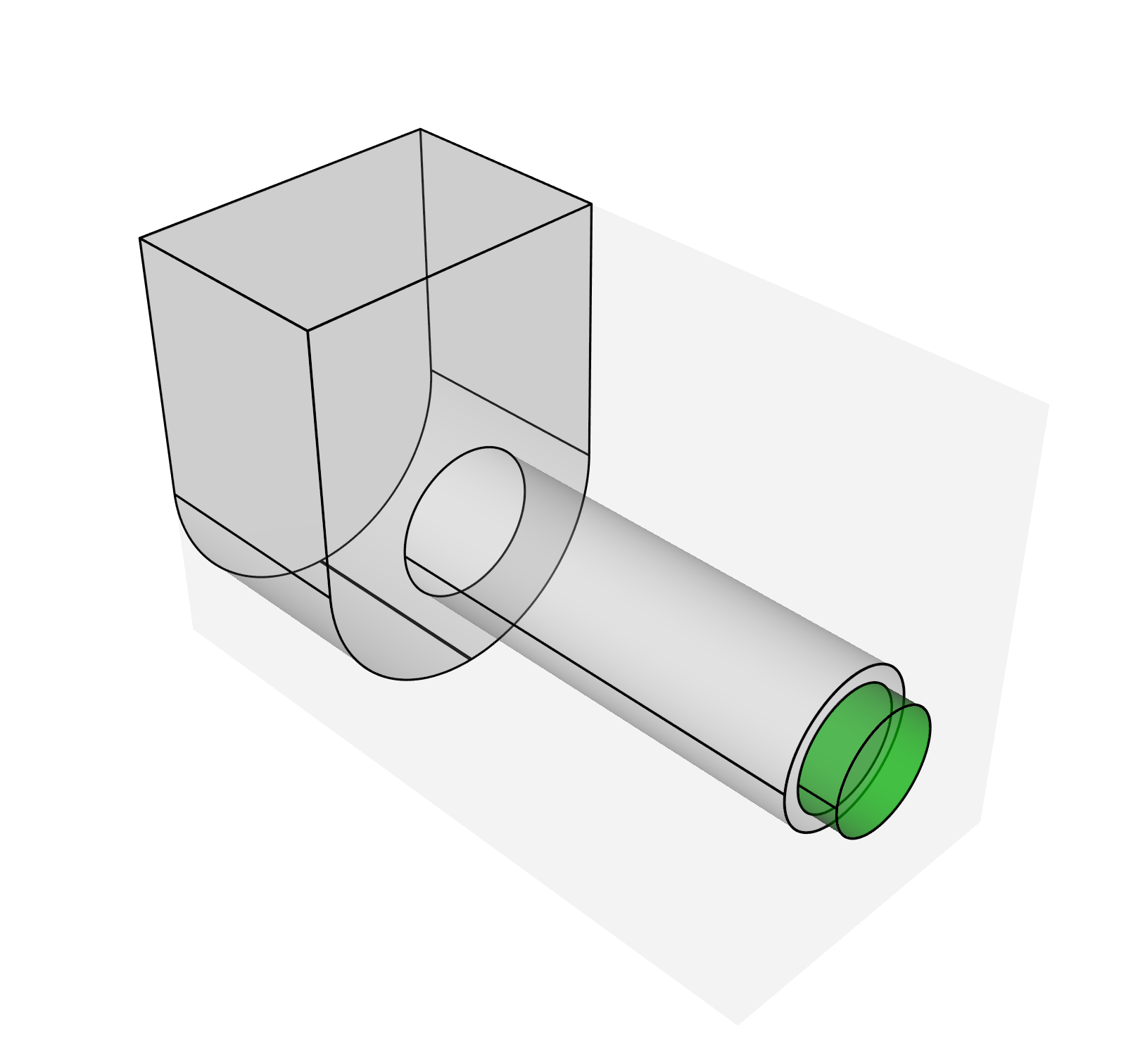} &
        \includegraphics[width=0.124\linewidth]{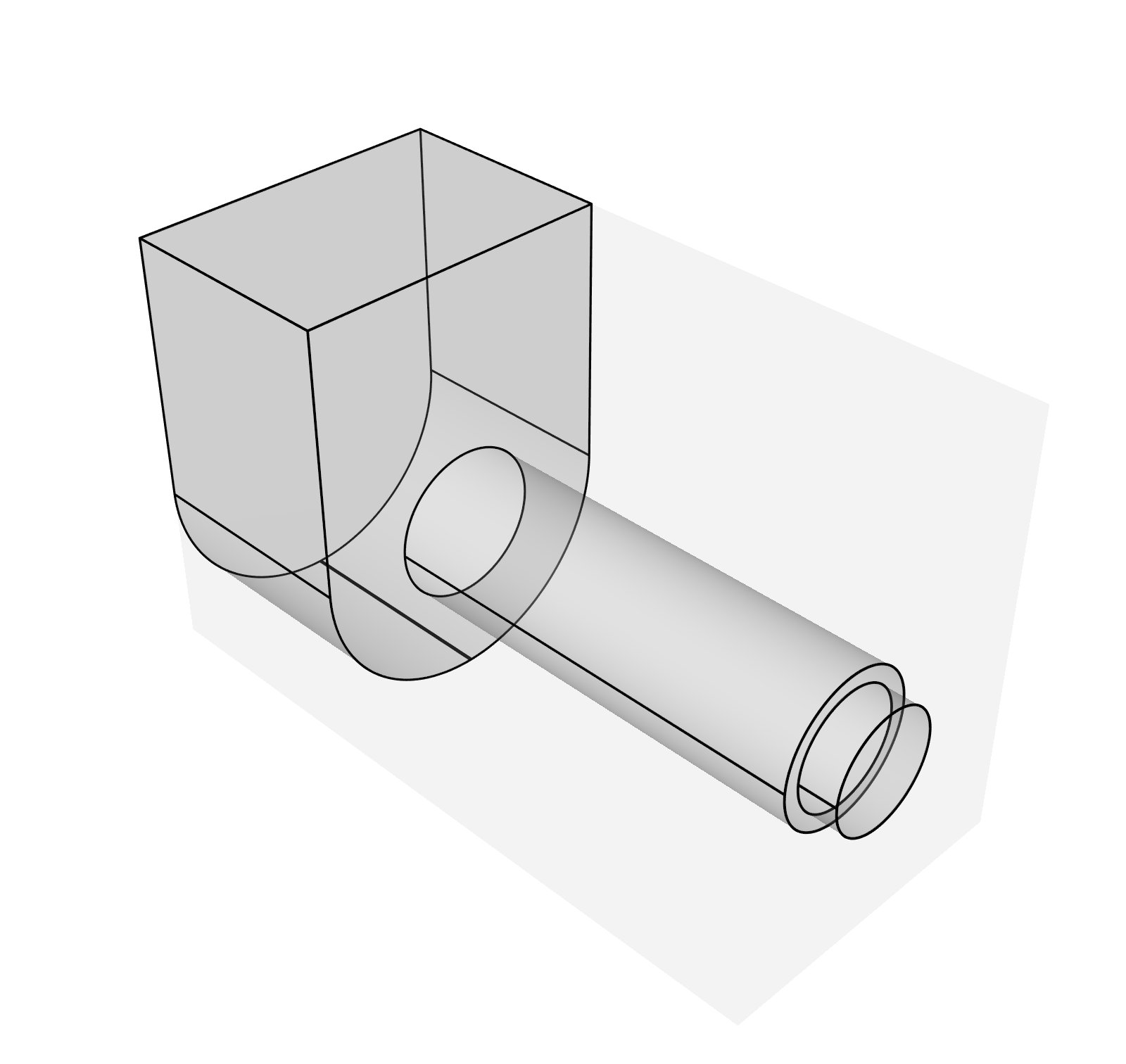} &
        \includegraphics[width=0.124\linewidth]{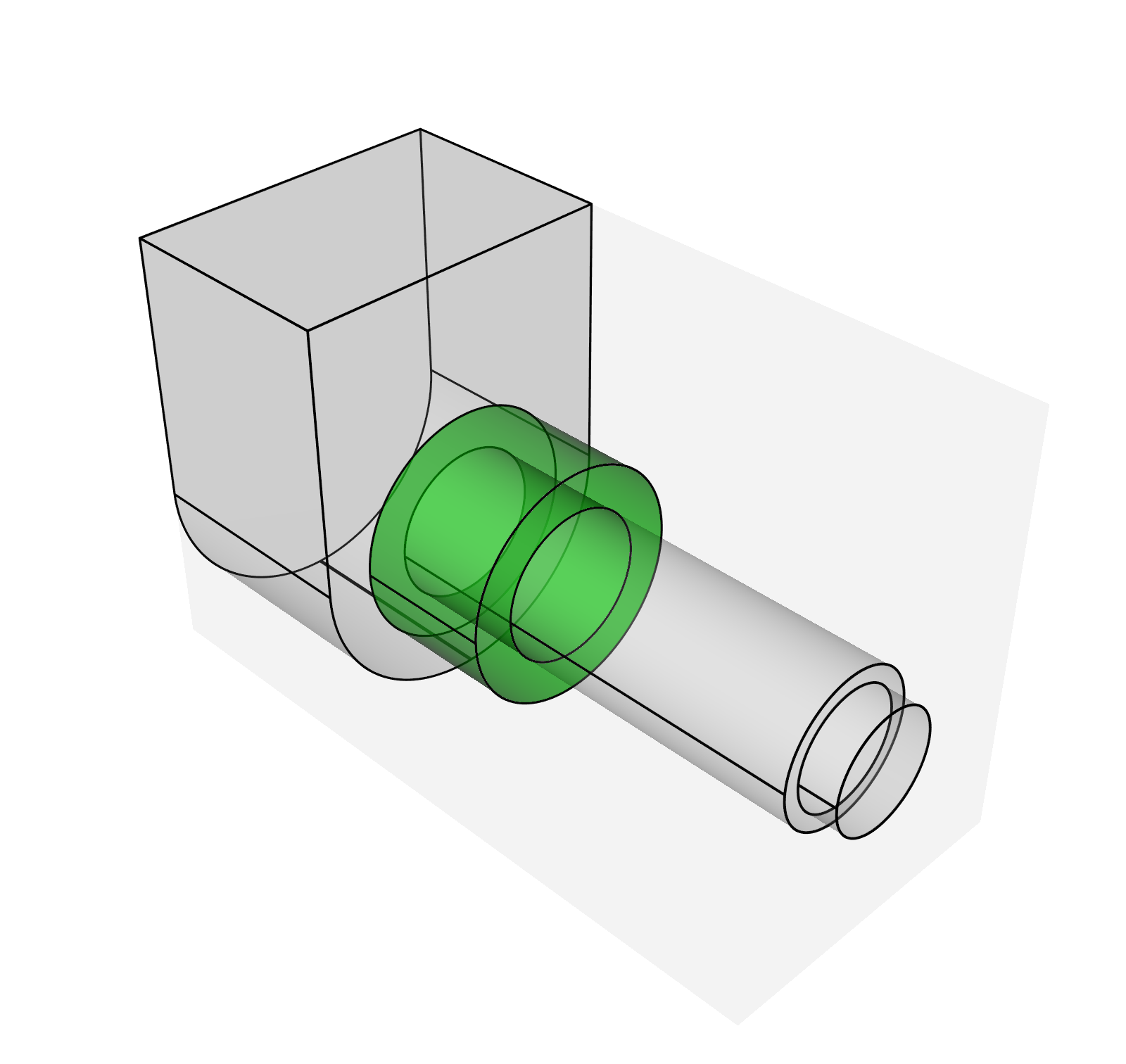} &
        \includegraphics[width=0.124\linewidth]{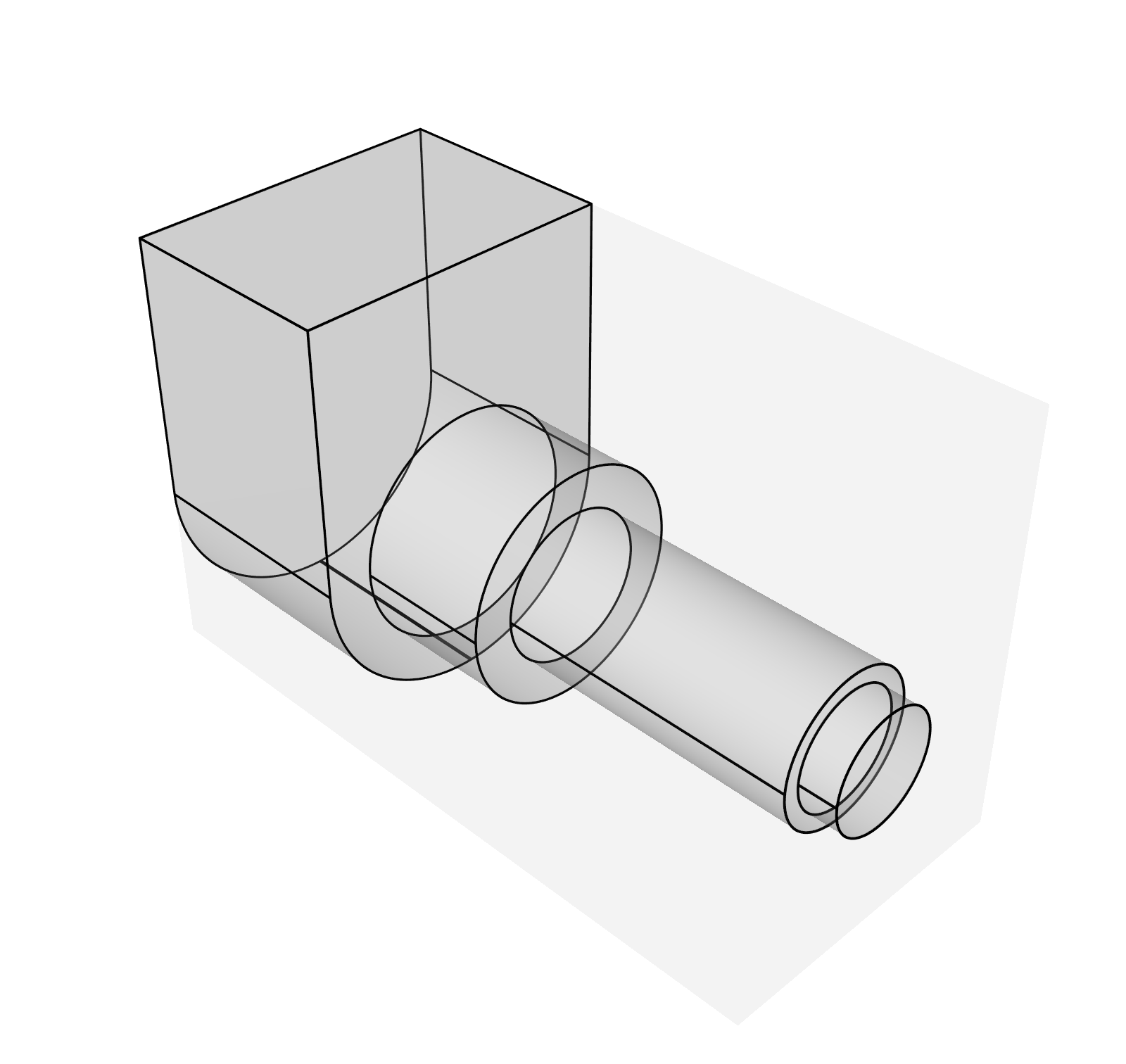}
        \\
        \multicolumn{1}{l}{Ours Net} & & & & &
        \\
        \includegraphics[width=0.124\linewidth]{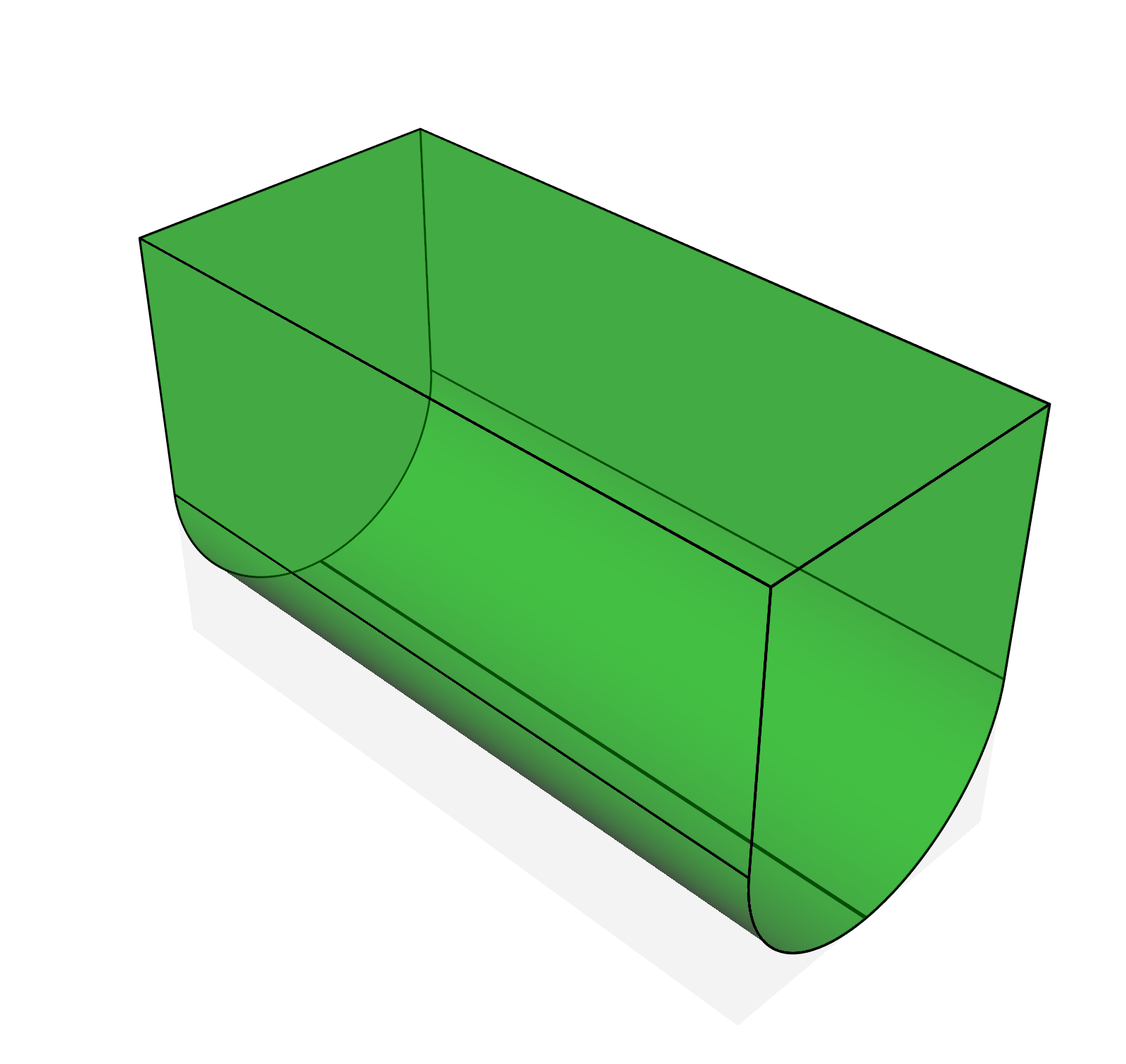} &
        \includegraphics[width=0.124\linewidth]{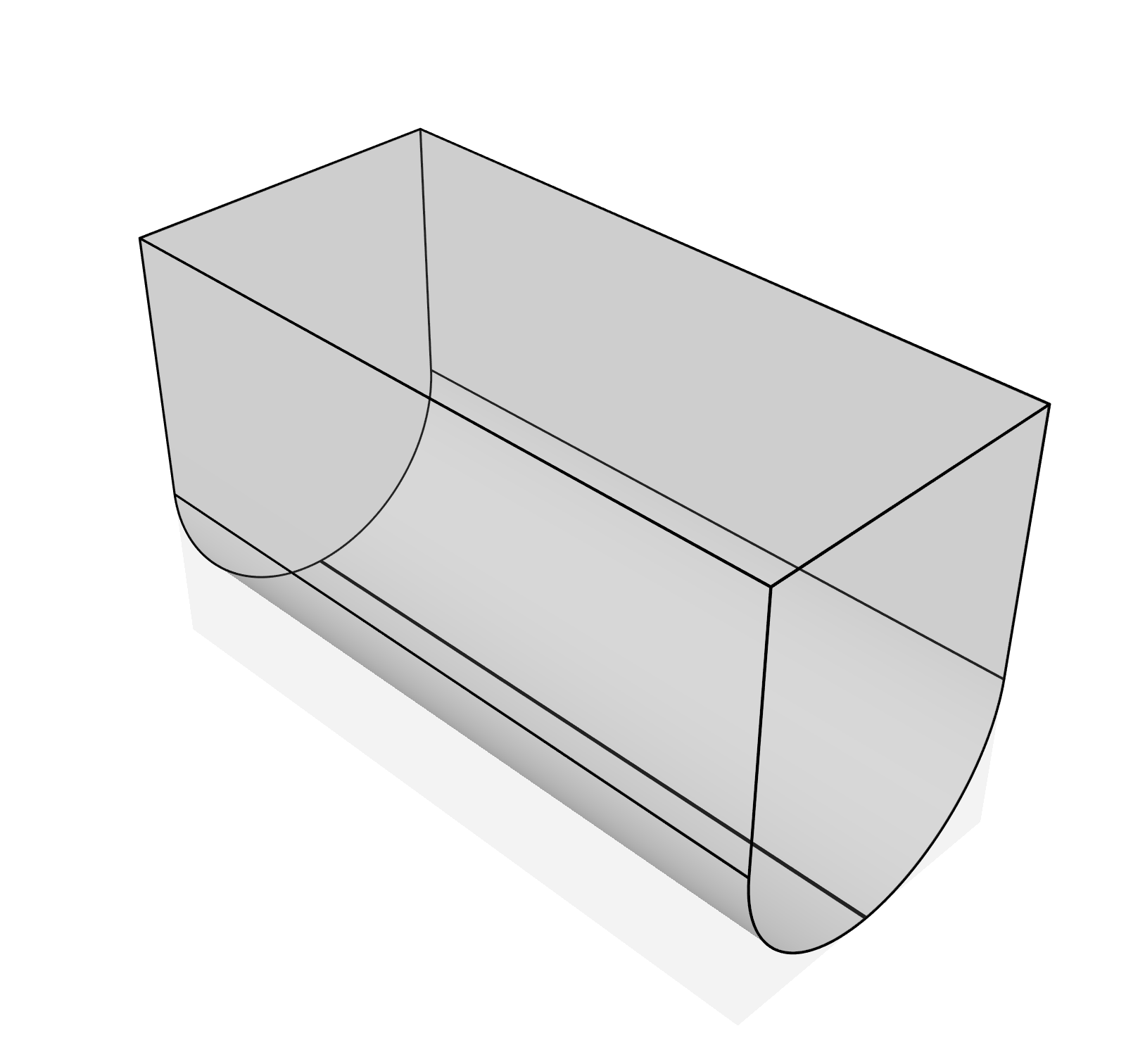} &
        \includegraphics[width=0.124\linewidth]{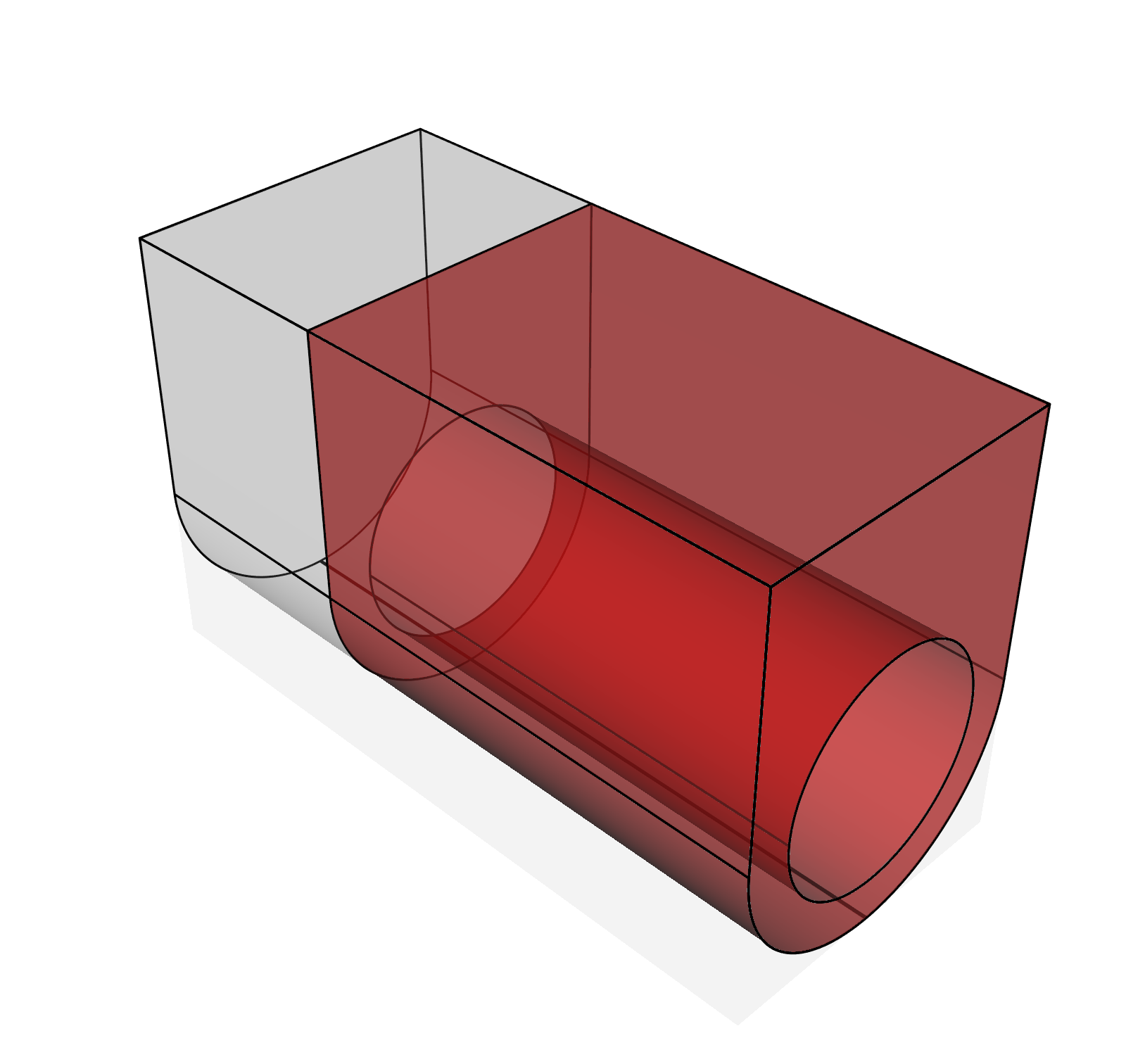} &
        \includegraphics[width=0.124\linewidth]{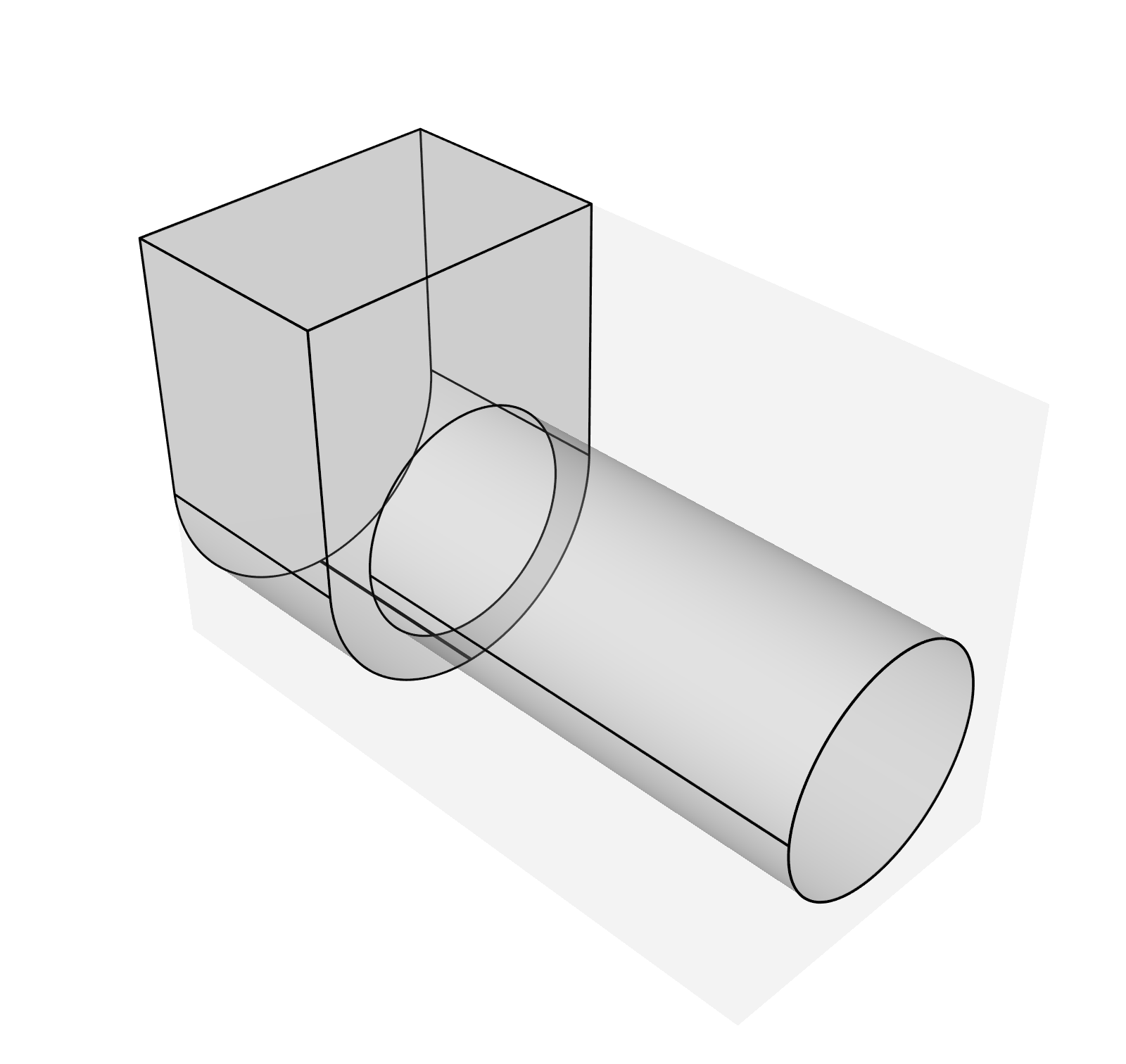} &
        \includegraphics[width=0.124\linewidth]{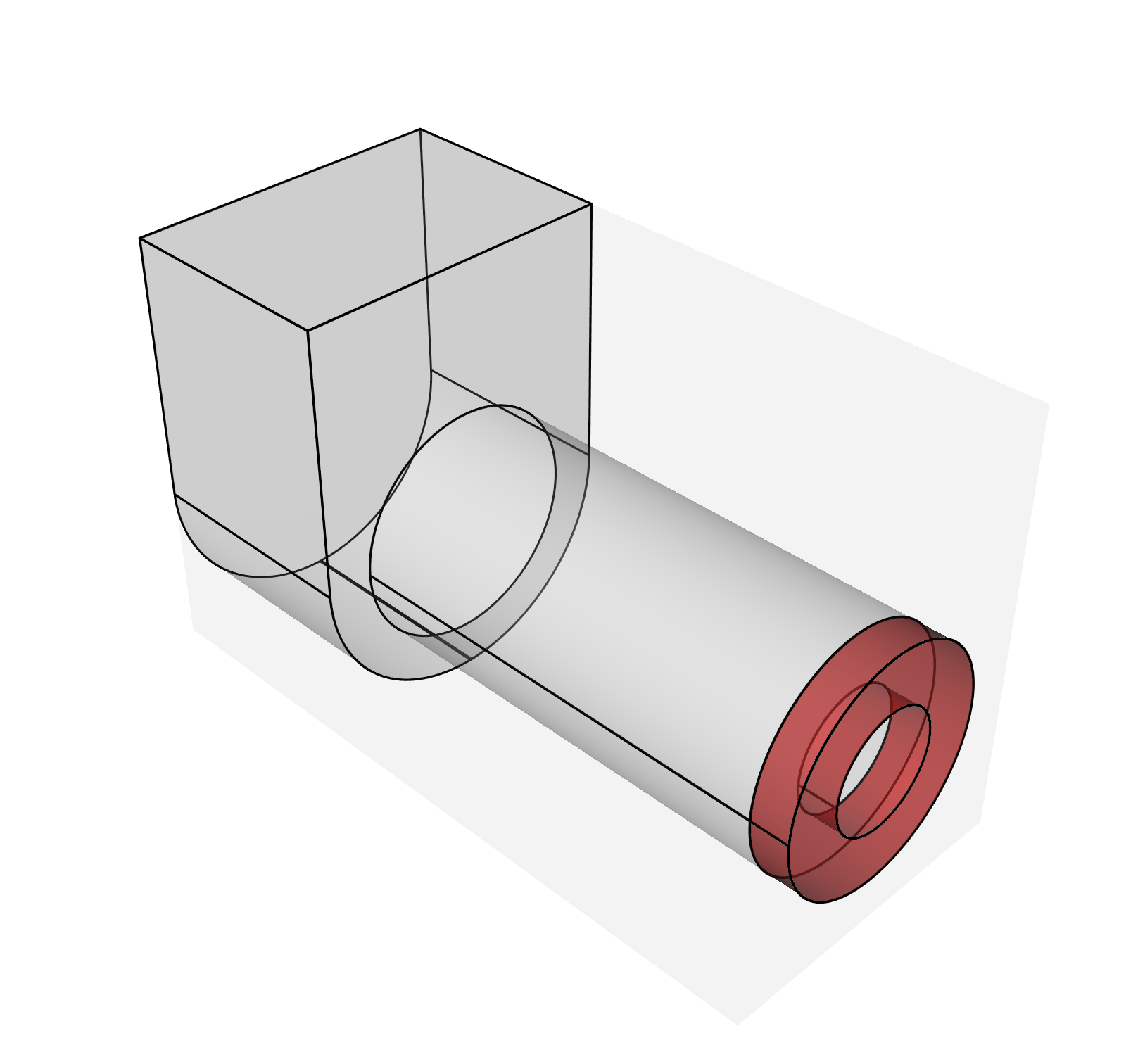} &
        \includegraphics[width=0.124\linewidth]{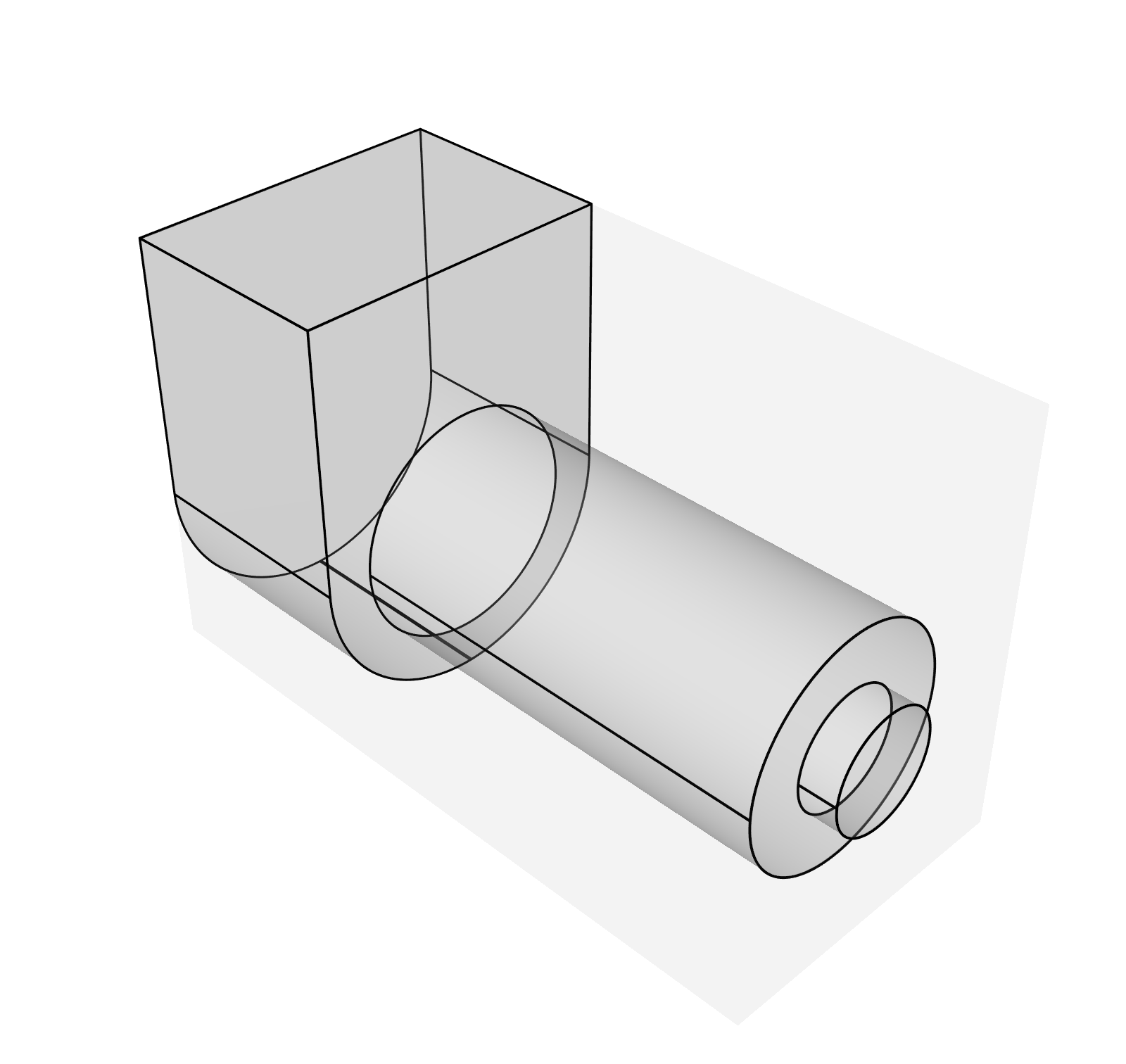} &
        \includegraphics[width=0.124\linewidth]{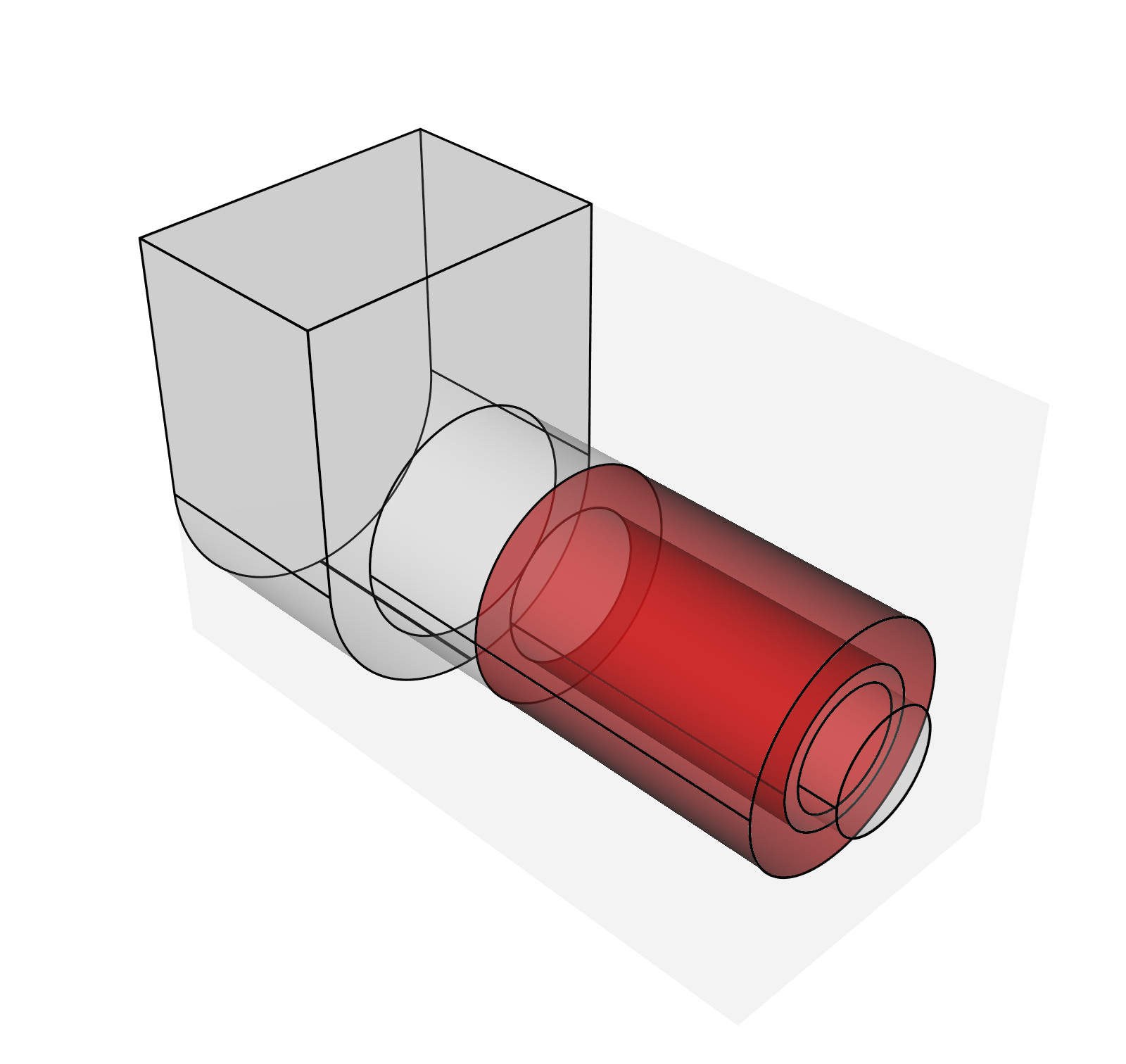} &
        \includegraphics[width=0.124\linewidth]{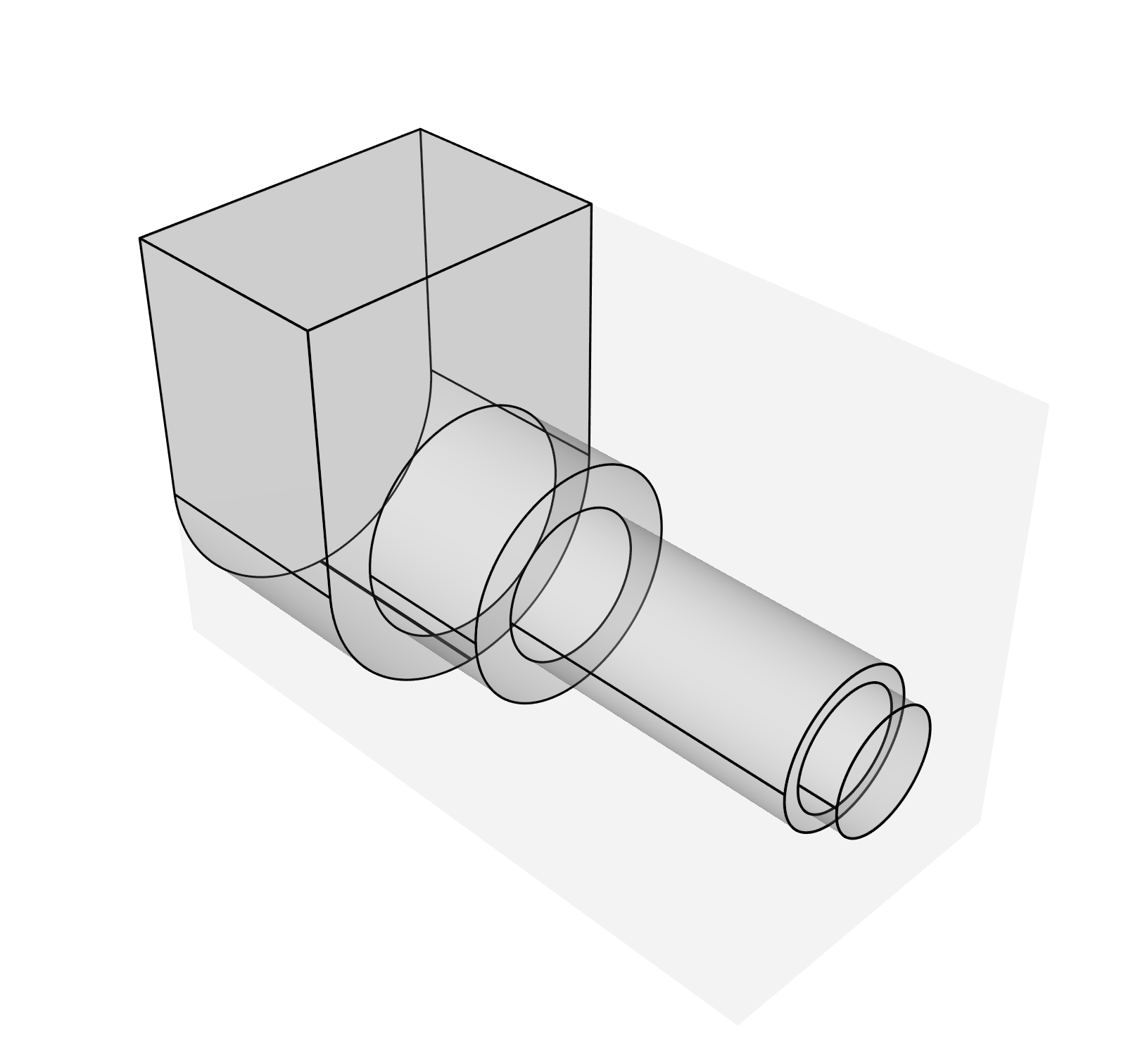}
        \\
    \end{tabular}
    \caption{
    Qualitative comparison of the output of our model's inferred programs (Ours Net) vs. those of Random and Ours Heur. Green: addition, Red: subtraction, Grey: current. (Case 2)
    }
    \label{figure:qualitative_comparision2}
\end{figure*}

Please see Figure~\ref{figure:qualitative_comparision1} and \ref{figure:qualitative_comparision2} for additional examples of the qualitative comparison of the output of our model's inferred programs (Network) vs. those of Random and Heuristics. 

\subsubsection{Ours vs. InverseCSG}
\begin{figure*}[h!]
    \centering
    \setlength{\tabcolsep}{1pt}
    \begin{tabular}{cccccccc}
        \multicolumn{2}{c}{Target} & & & &
        \\
        \multicolumn{2}{c}{\includegraphics[width=0.25\linewidth]{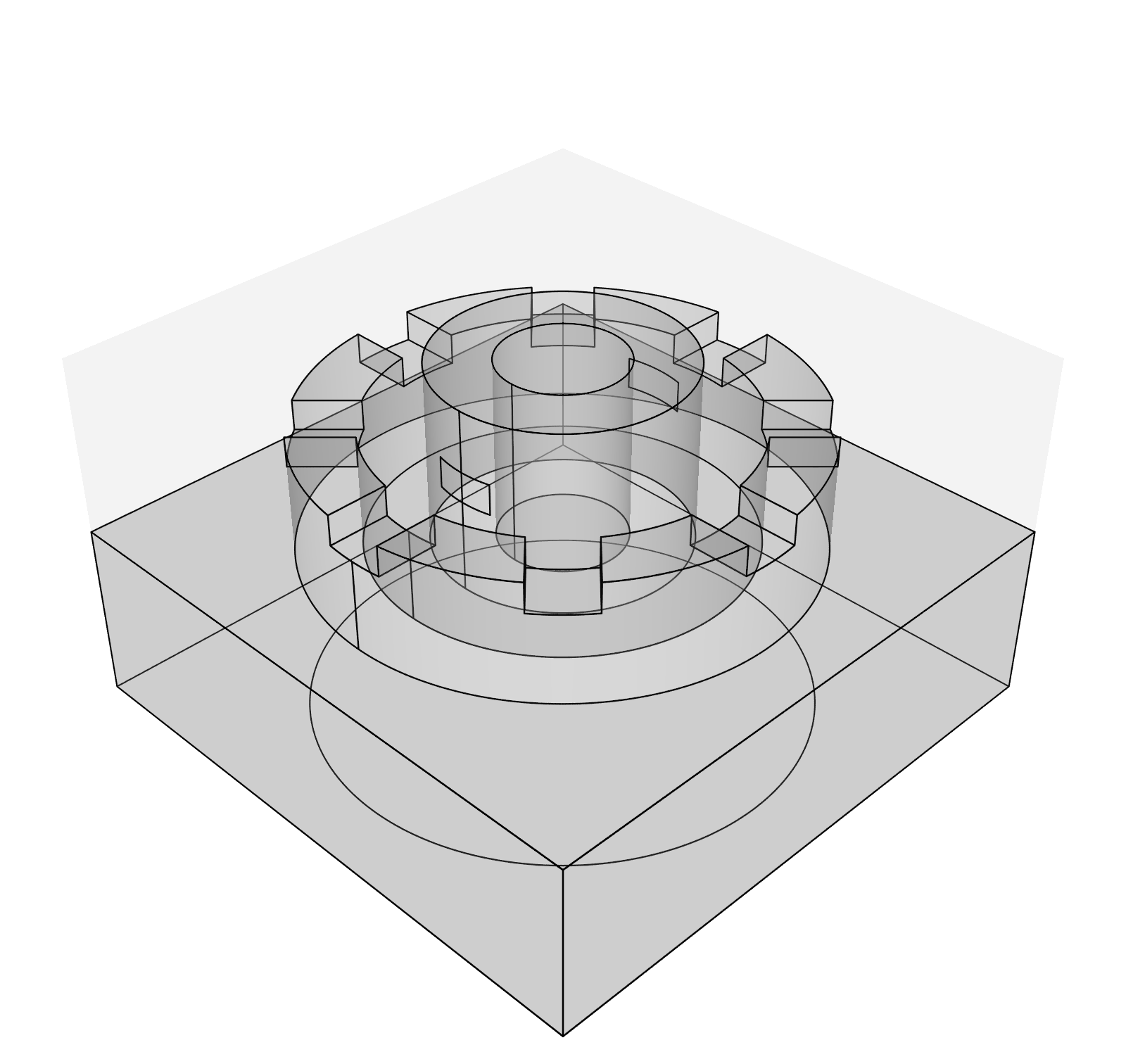}} & & & &
        \\
        \multicolumn{1}{l}{InverseCSG} & & & & &
        \\
        \includegraphics[width=0.124\linewidth]{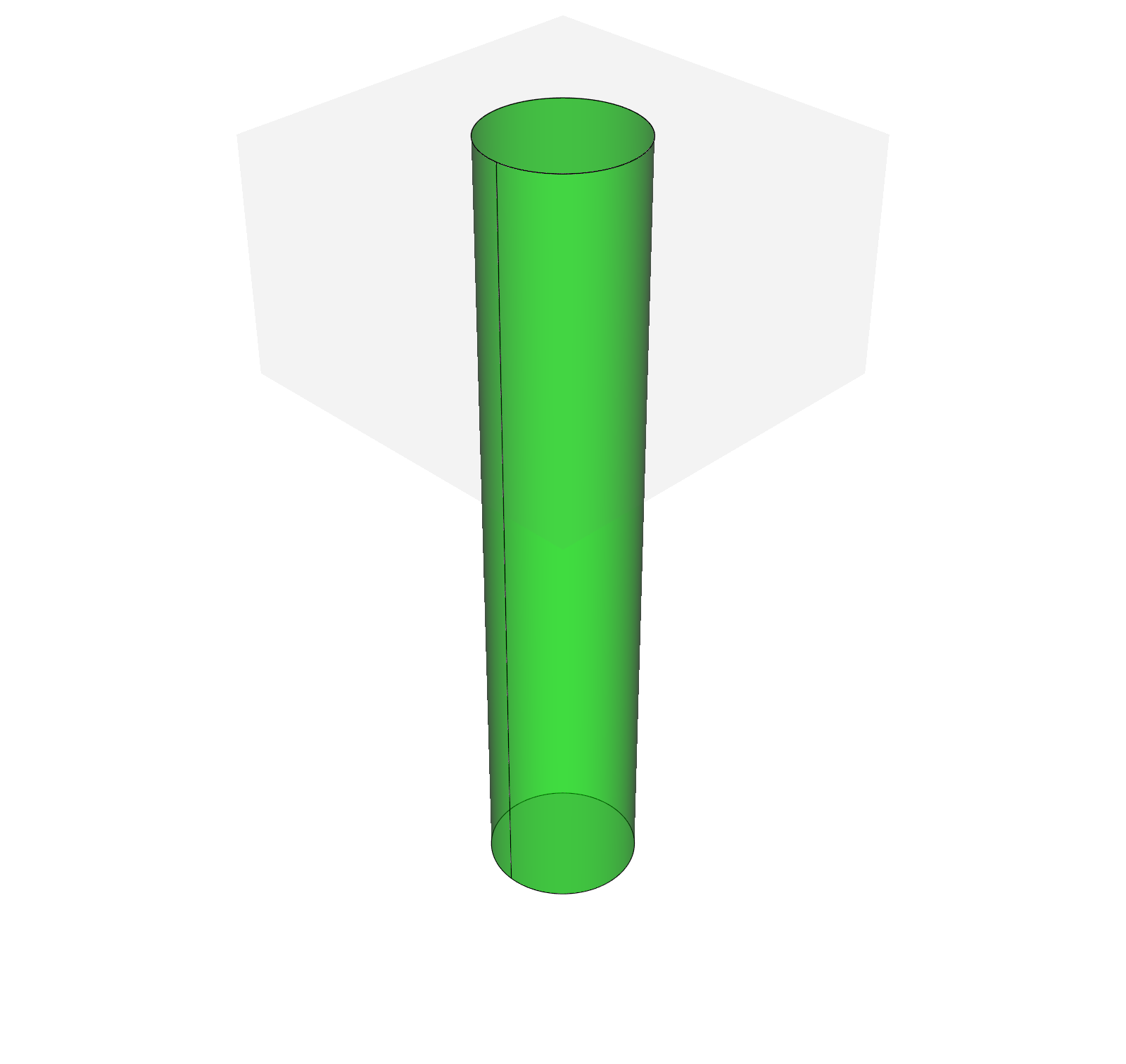} &
        \includegraphics[width=0.124\linewidth]{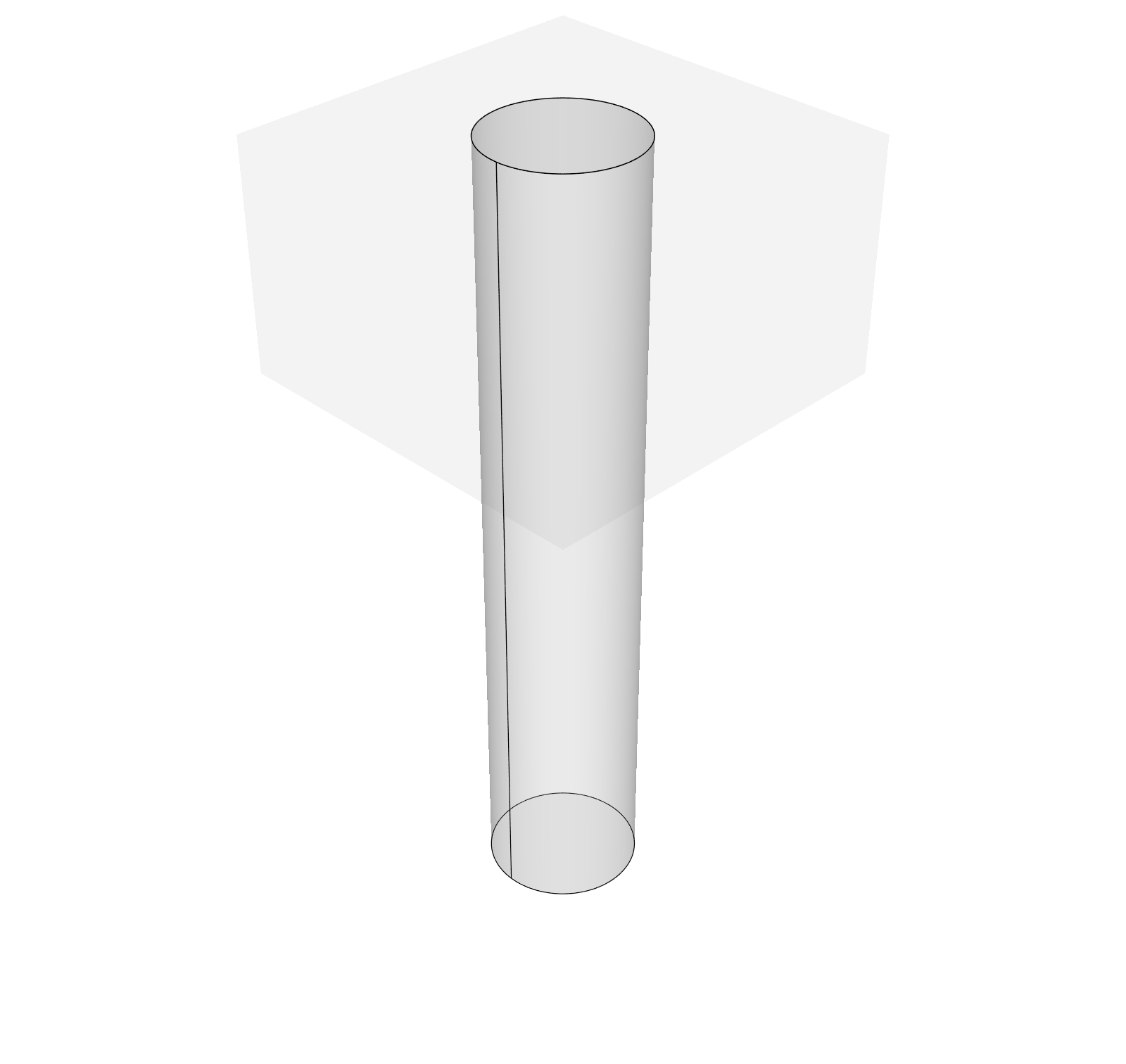} &
        \includegraphics[width=0.124\linewidth]{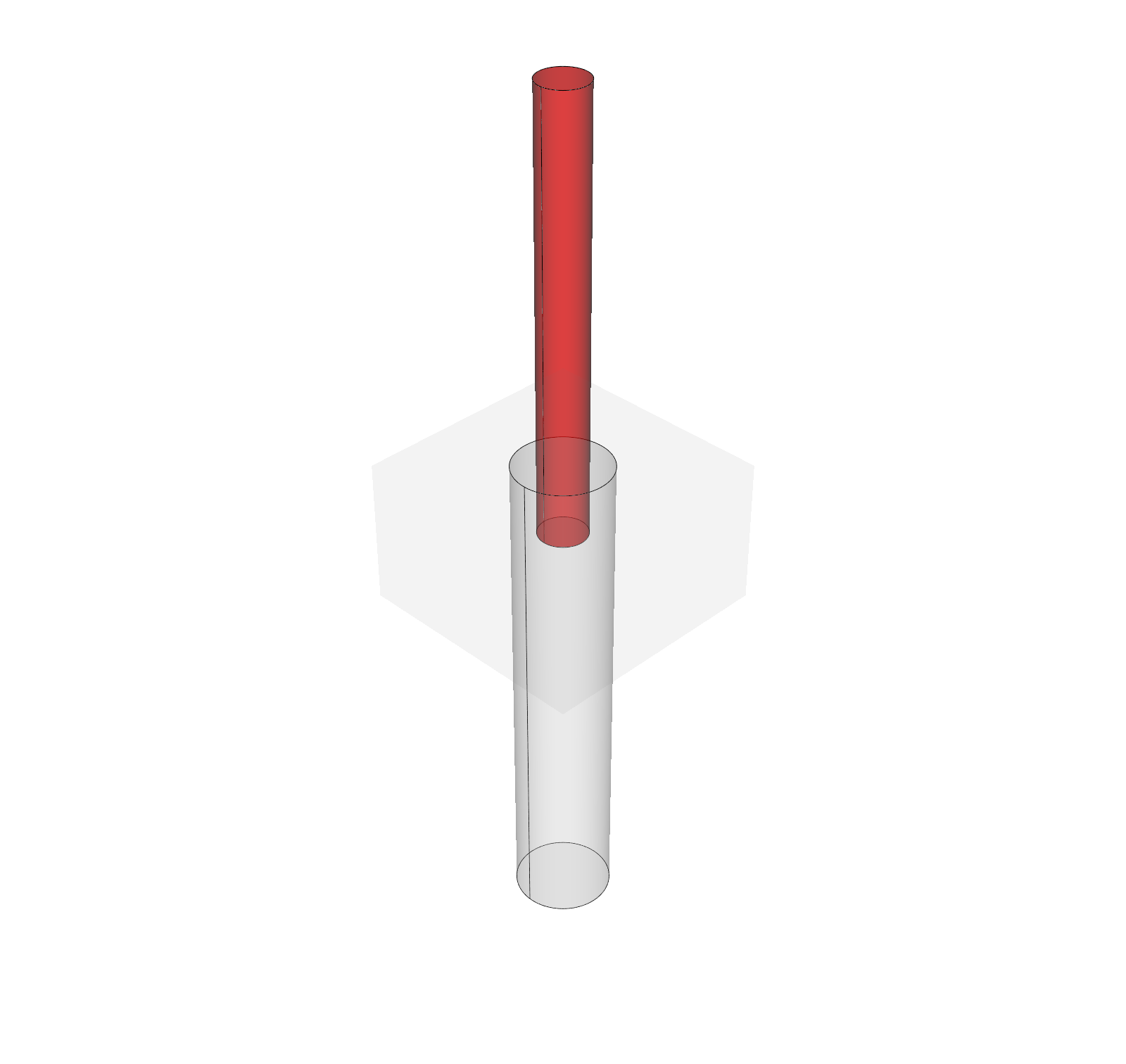} &
        \includegraphics[width=0.124\linewidth]{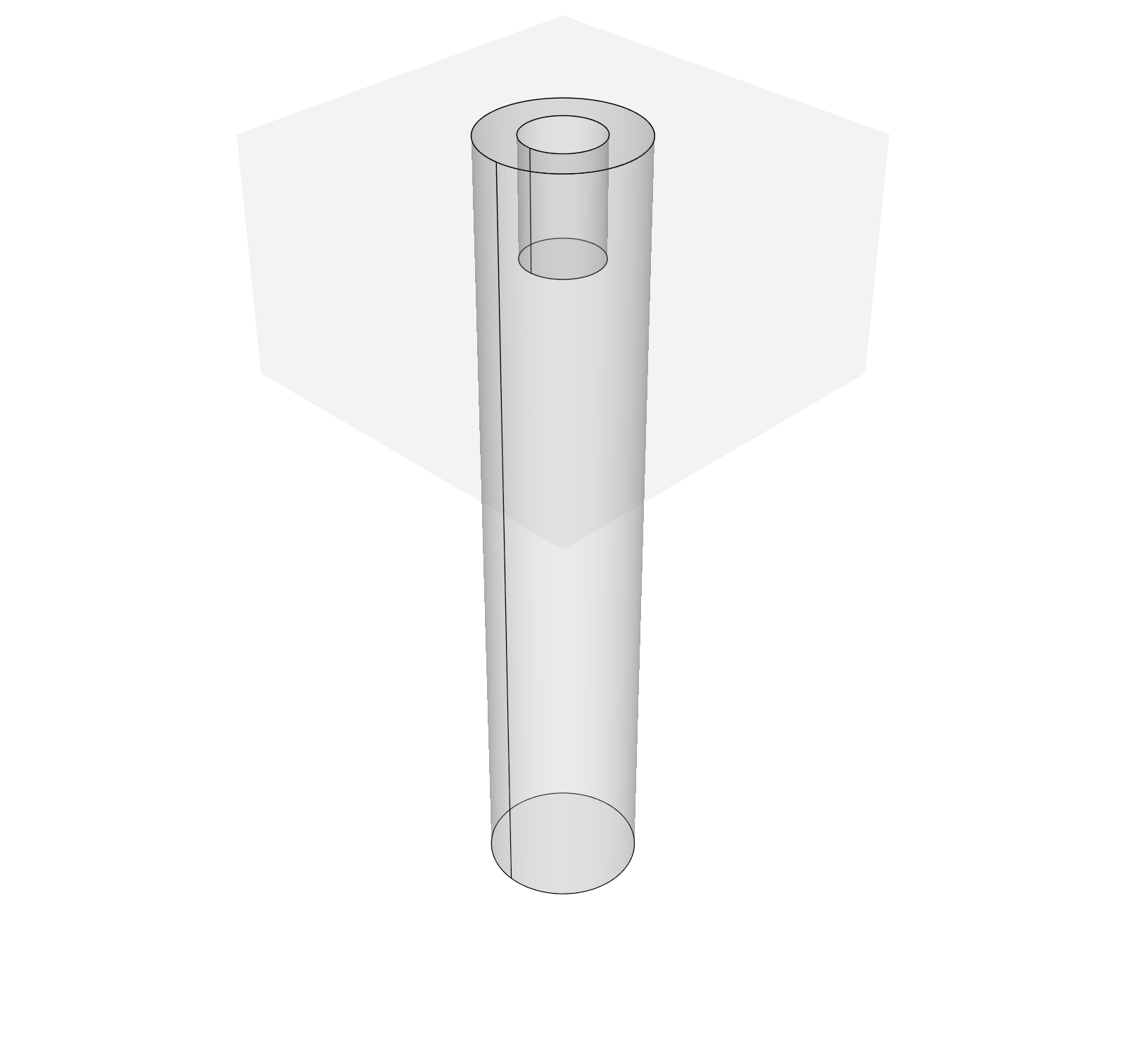} &
        \includegraphics[width=0.124\linewidth]{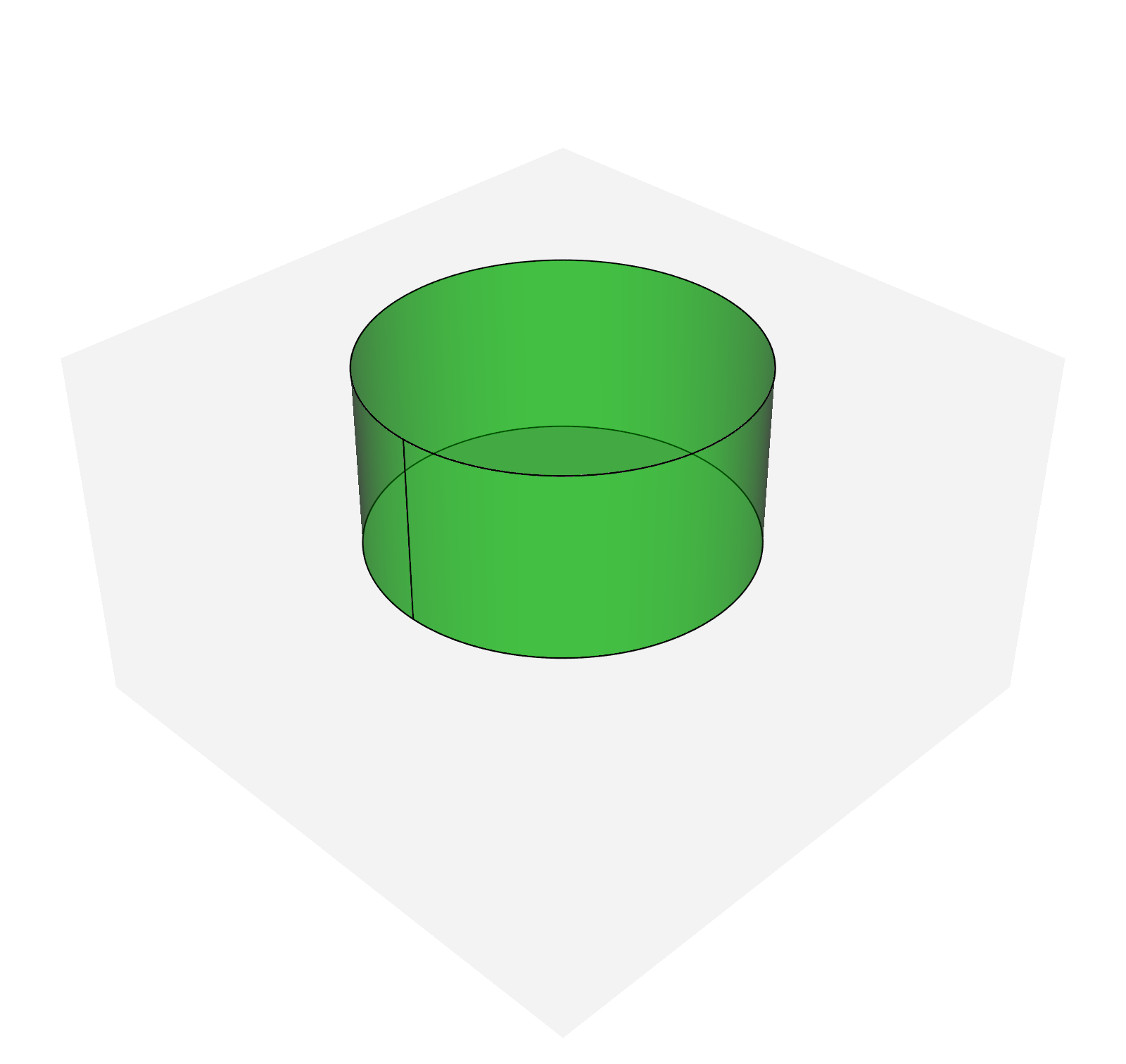} &
        \includegraphics[width=0.124\linewidth]{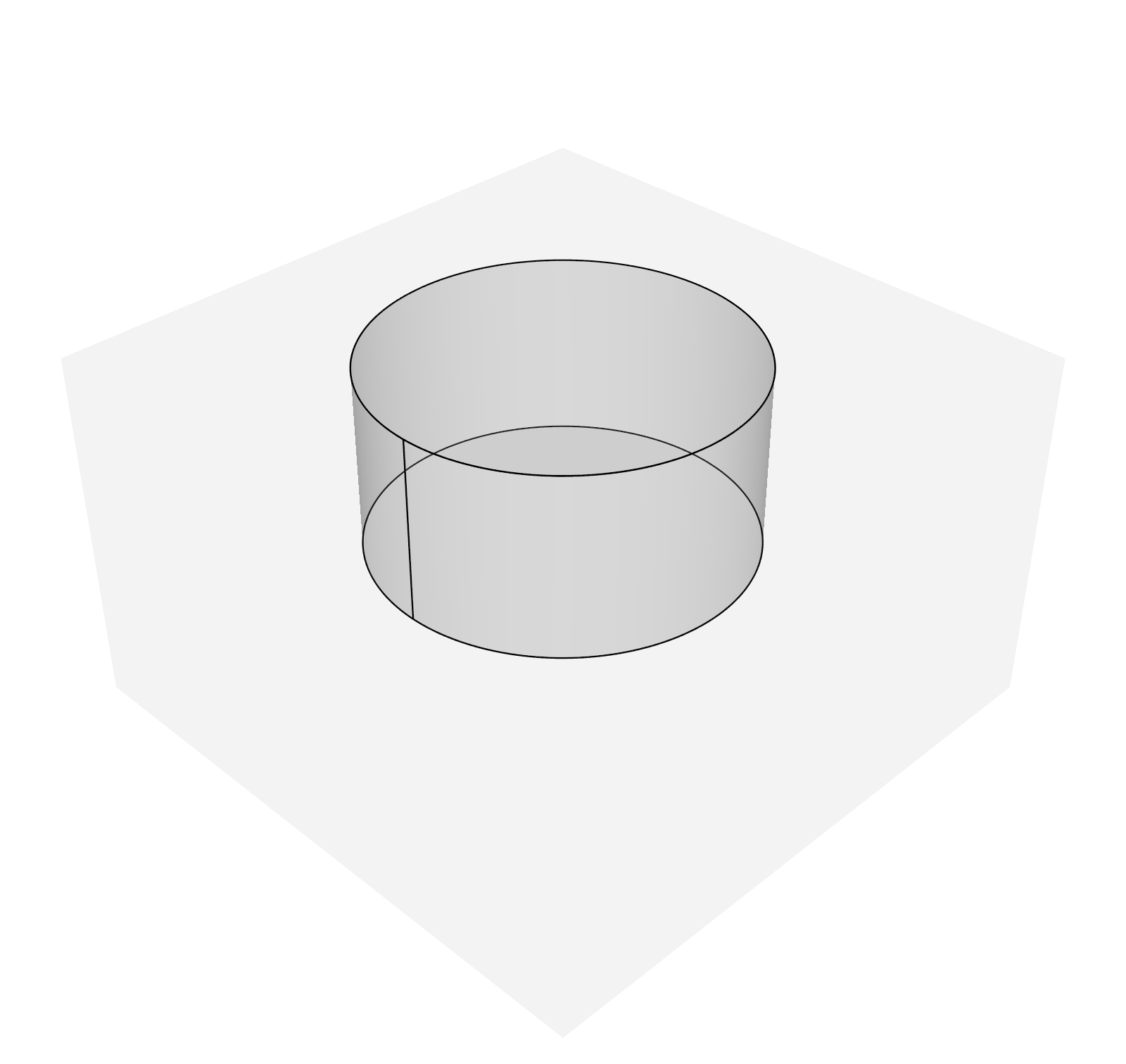} &
        \includegraphics[width=0.124\linewidth]{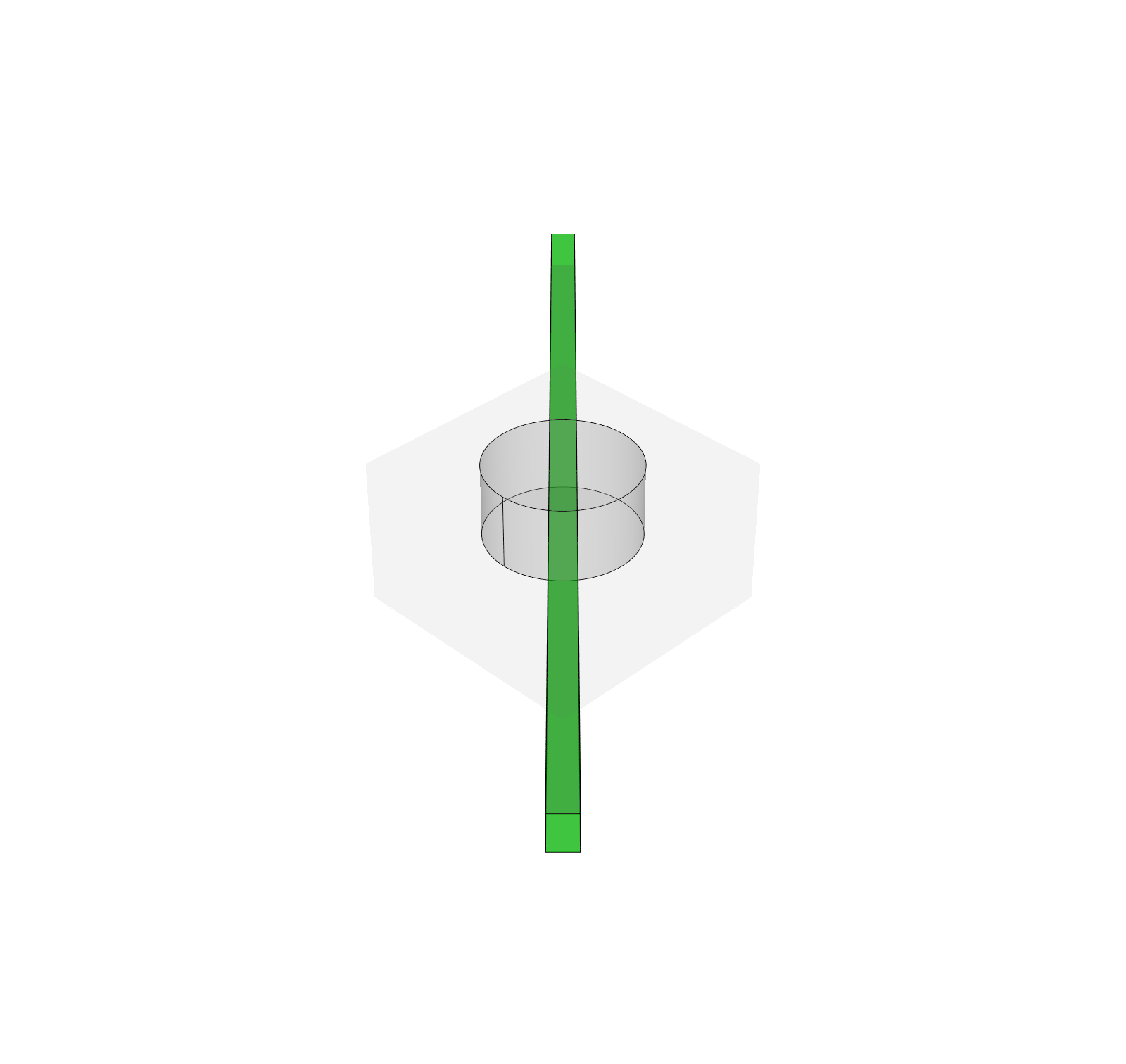} &
        \includegraphics[width=0.124\linewidth]{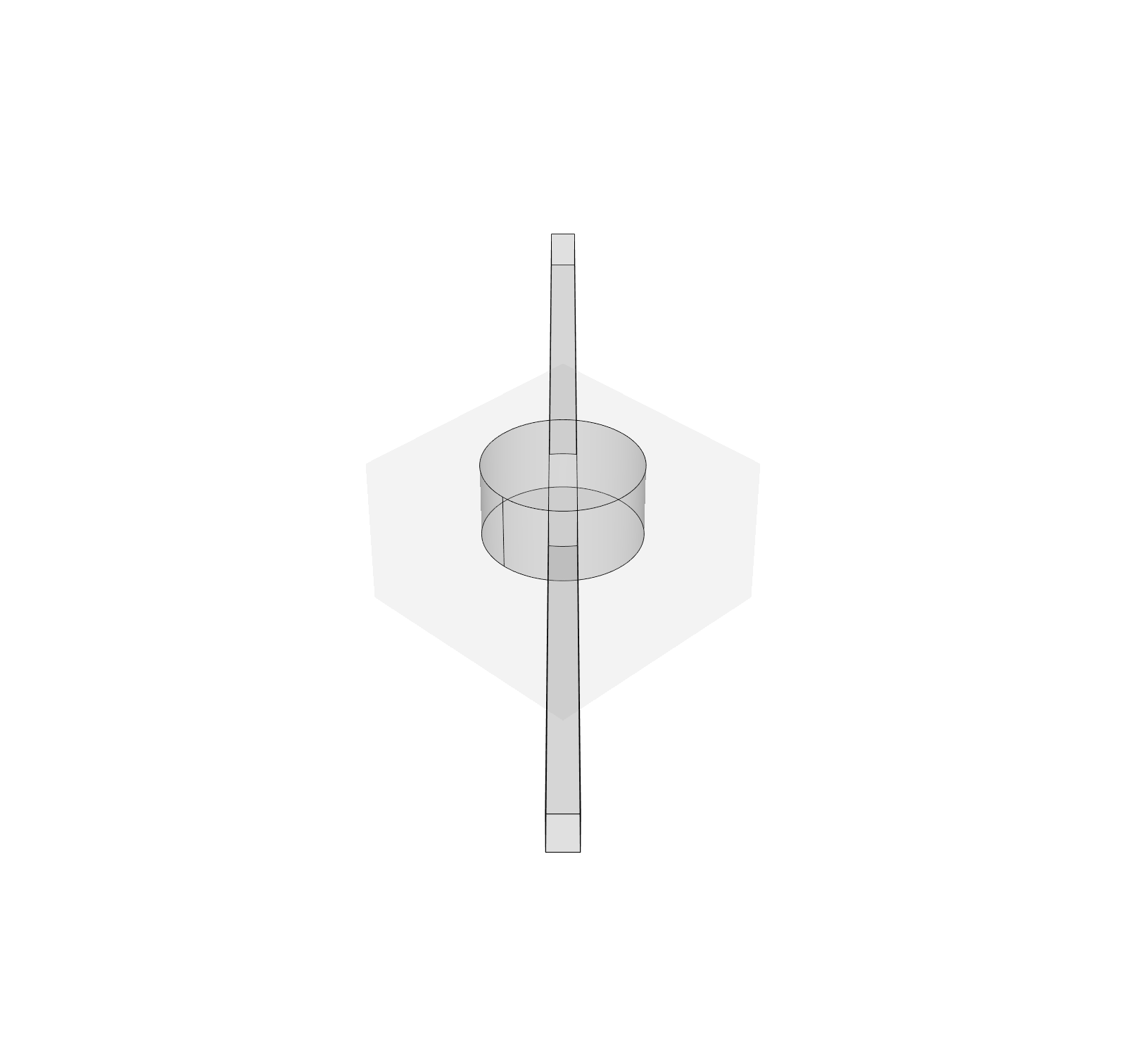}
        \\
        \includegraphics[width=0.124\linewidth]{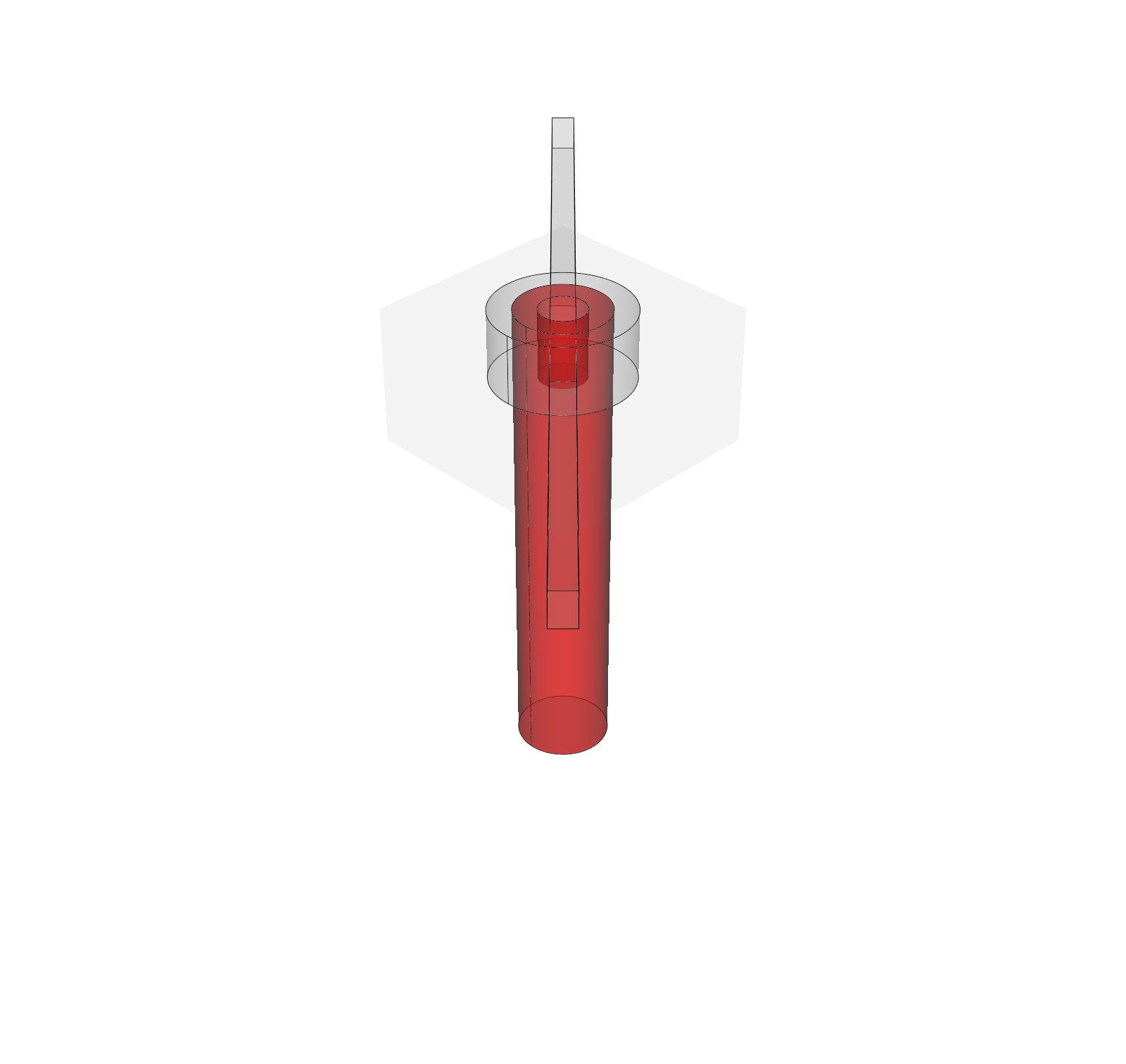} &
        \includegraphics[width=0.124\linewidth]{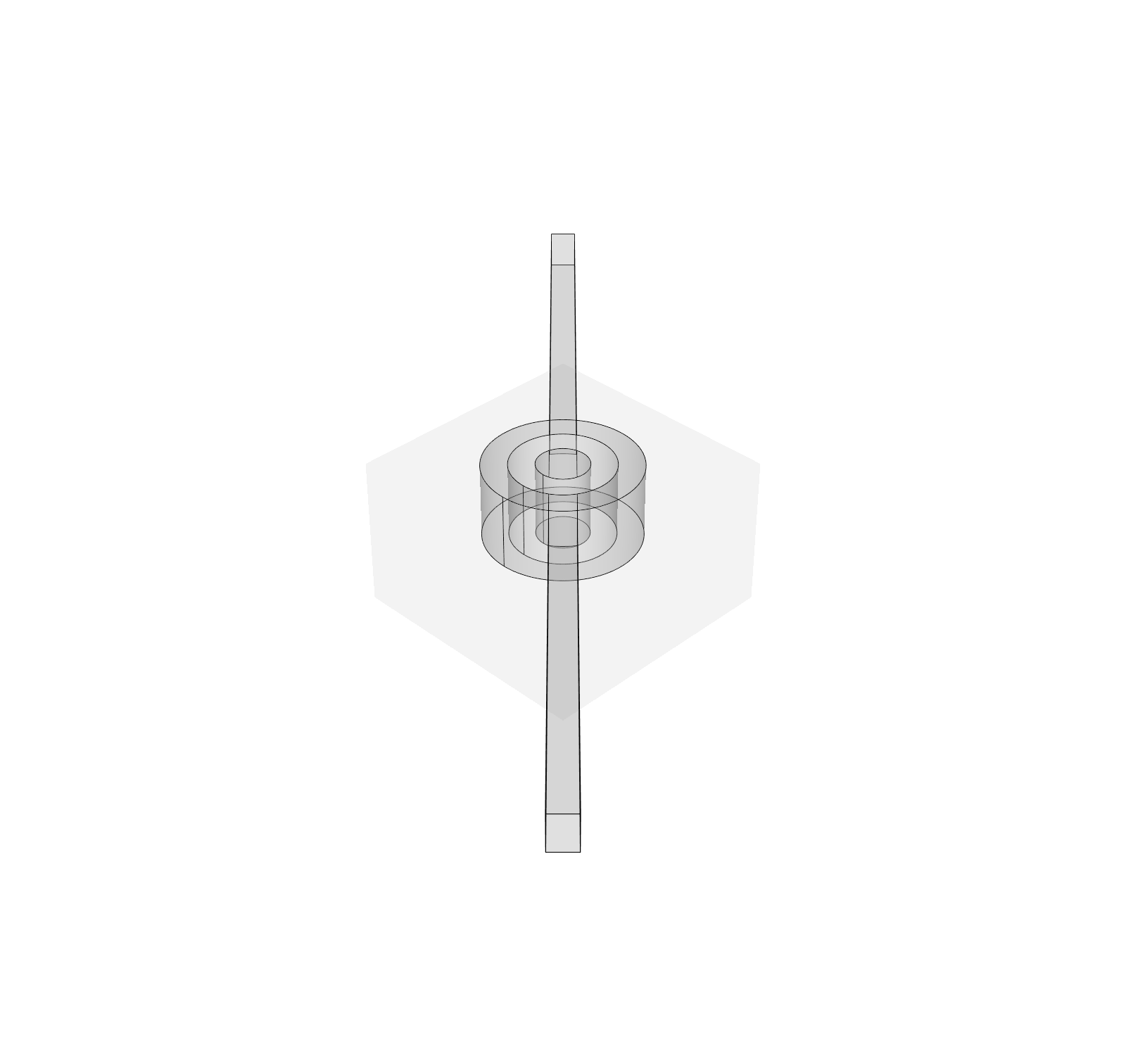} &
        \includegraphics[width=0.124\linewidth]{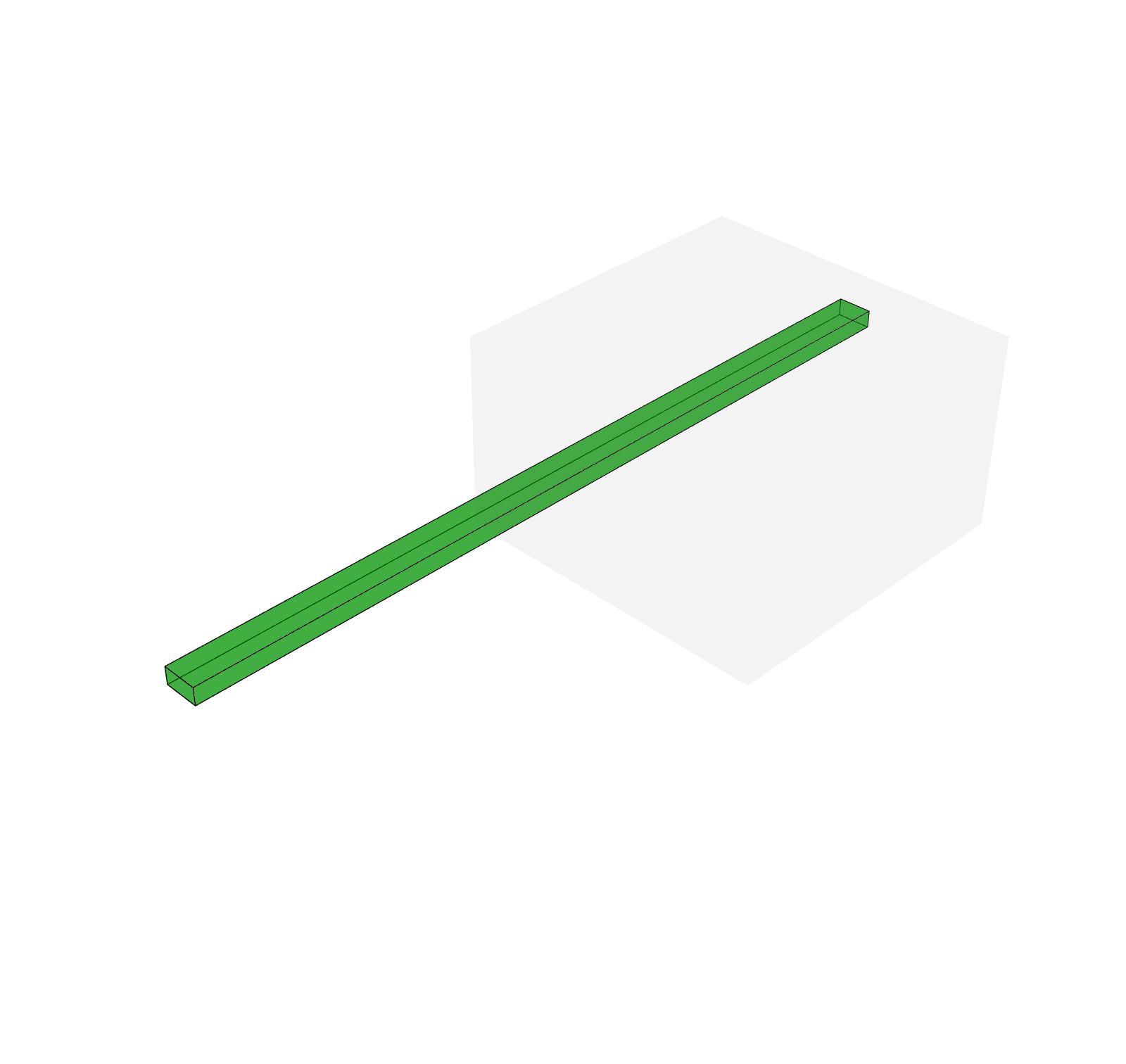} &
        \includegraphics[width=0.124\linewidth]{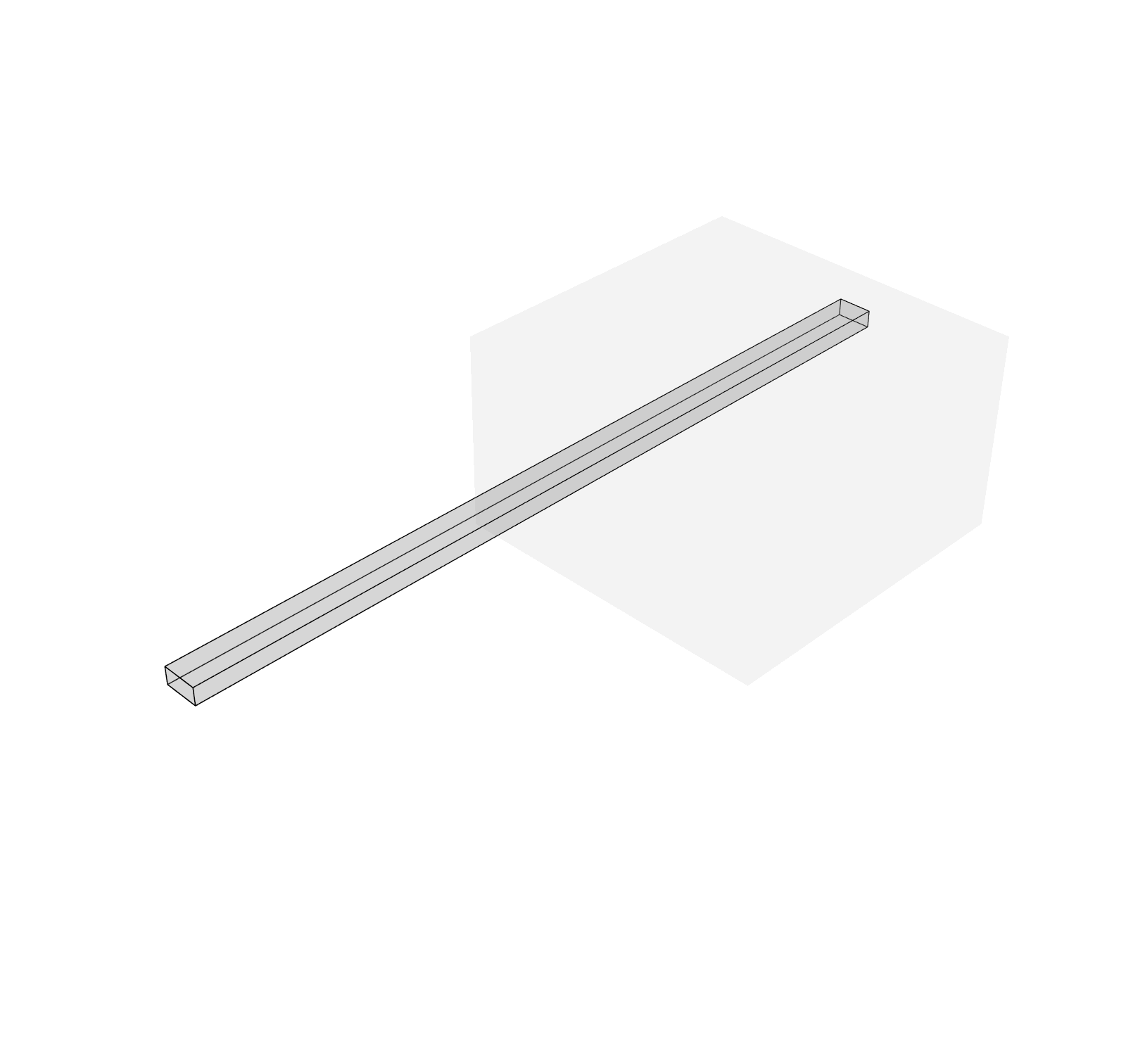} &
        \includegraphics[width=0.124\linewidth]{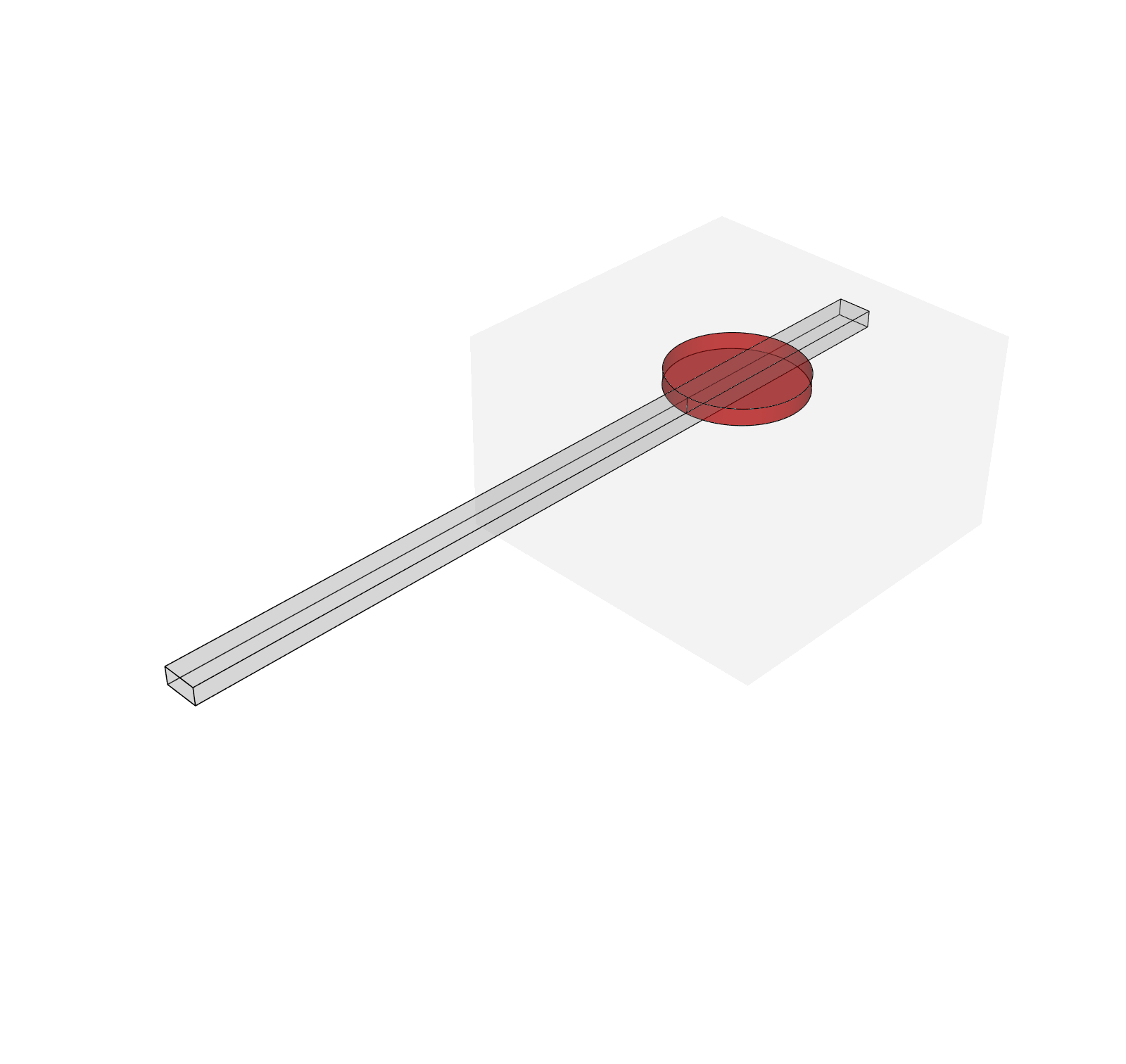} &
        \includegraphics[width=0.124\linewidth]{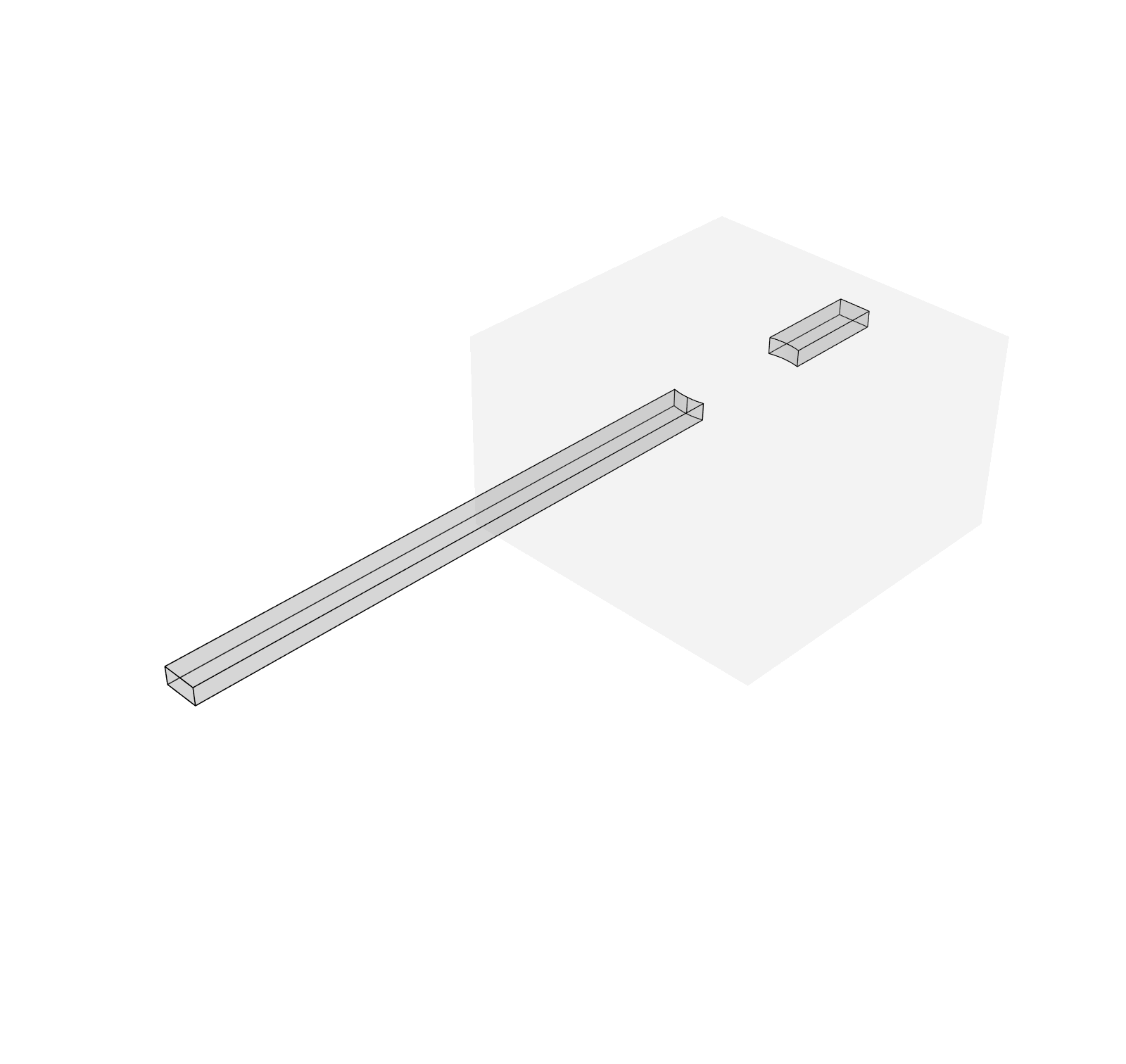} &
        \includegraphics[width=0.124\linewidth]{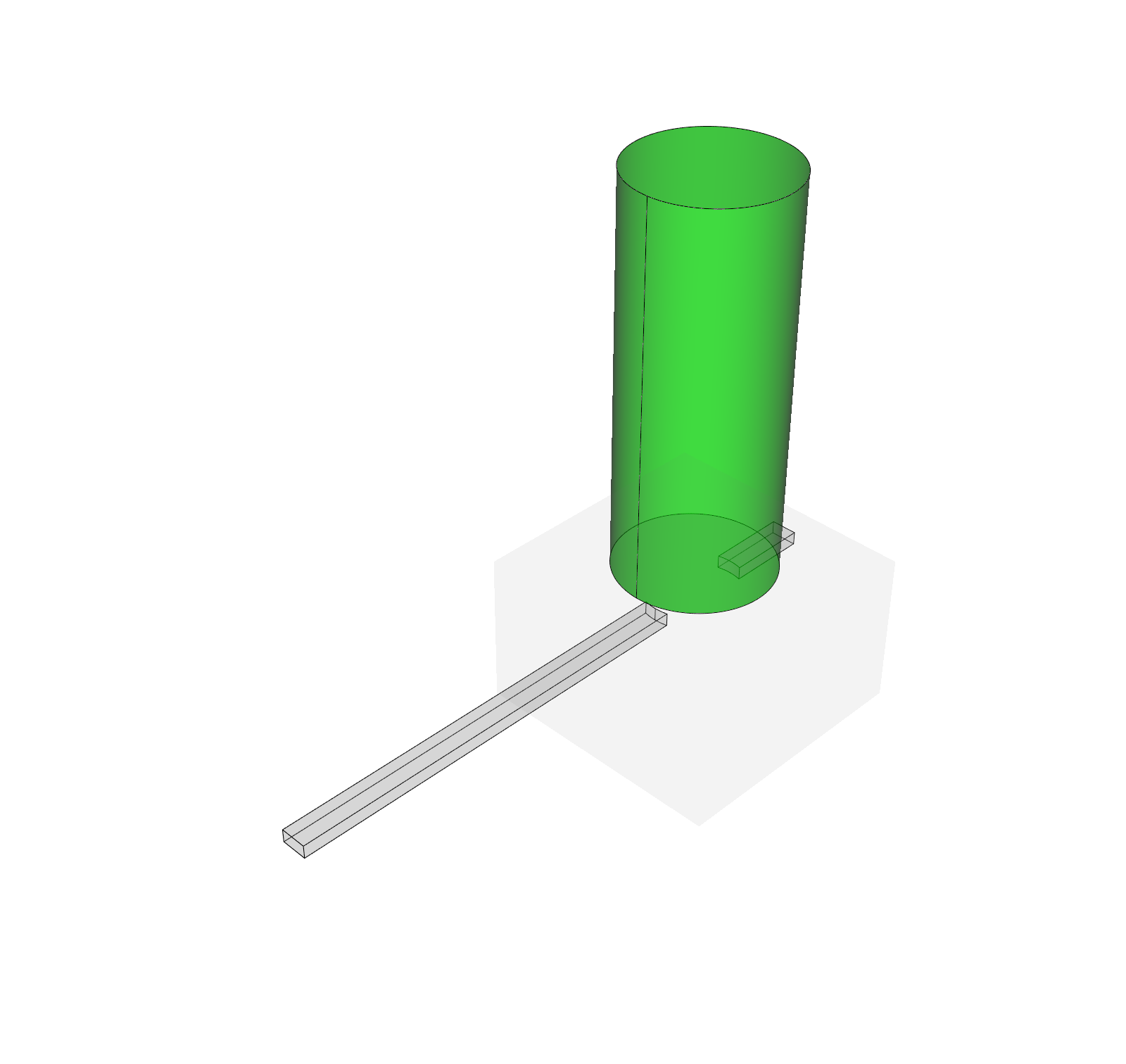} &
        \includegraphics[width=0.124\linewidth]{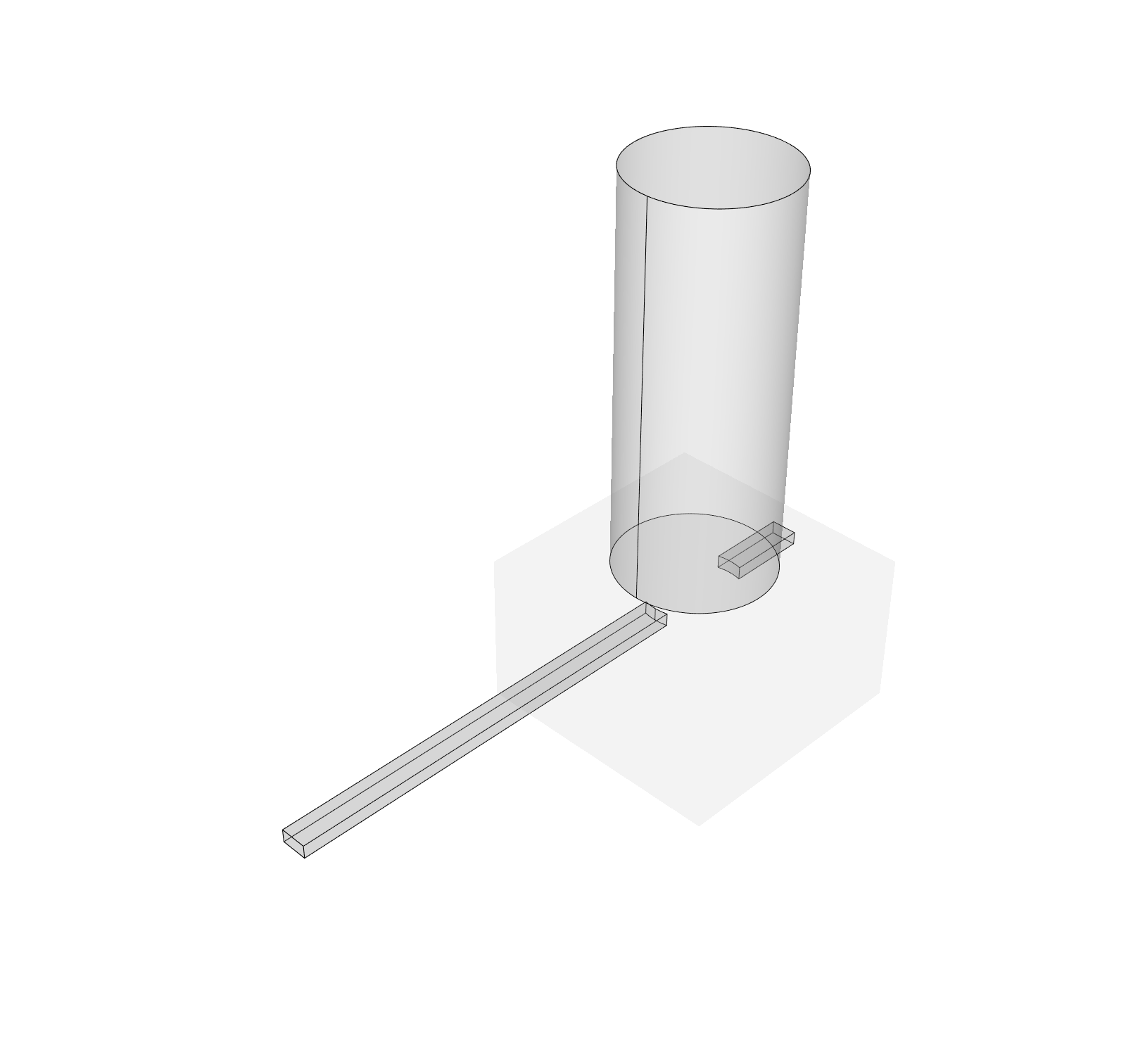}
        \\
        \includegraphics[width=0.124\linewidth]{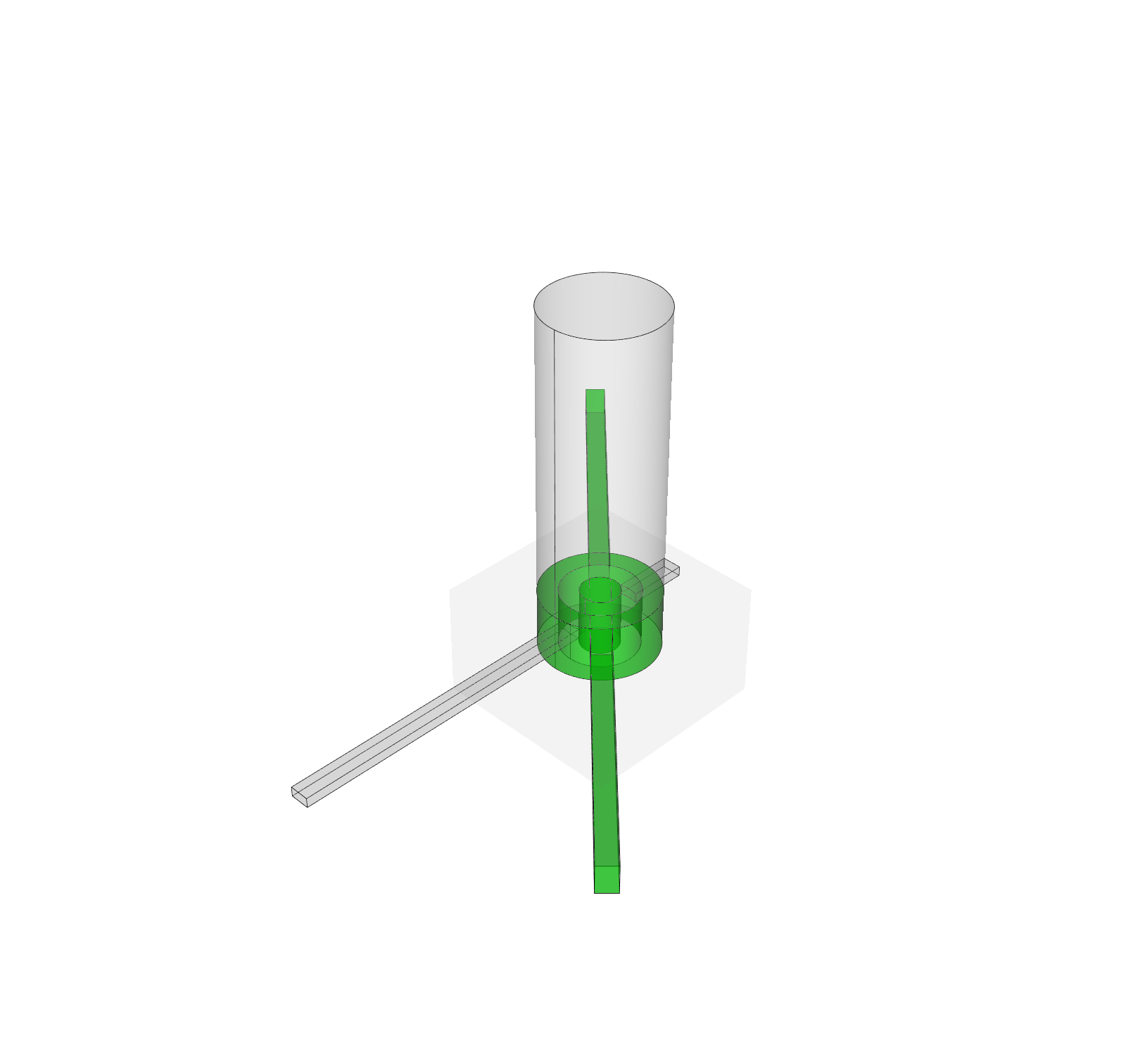} &
        \includegraphics[width=0.124\linewidth]{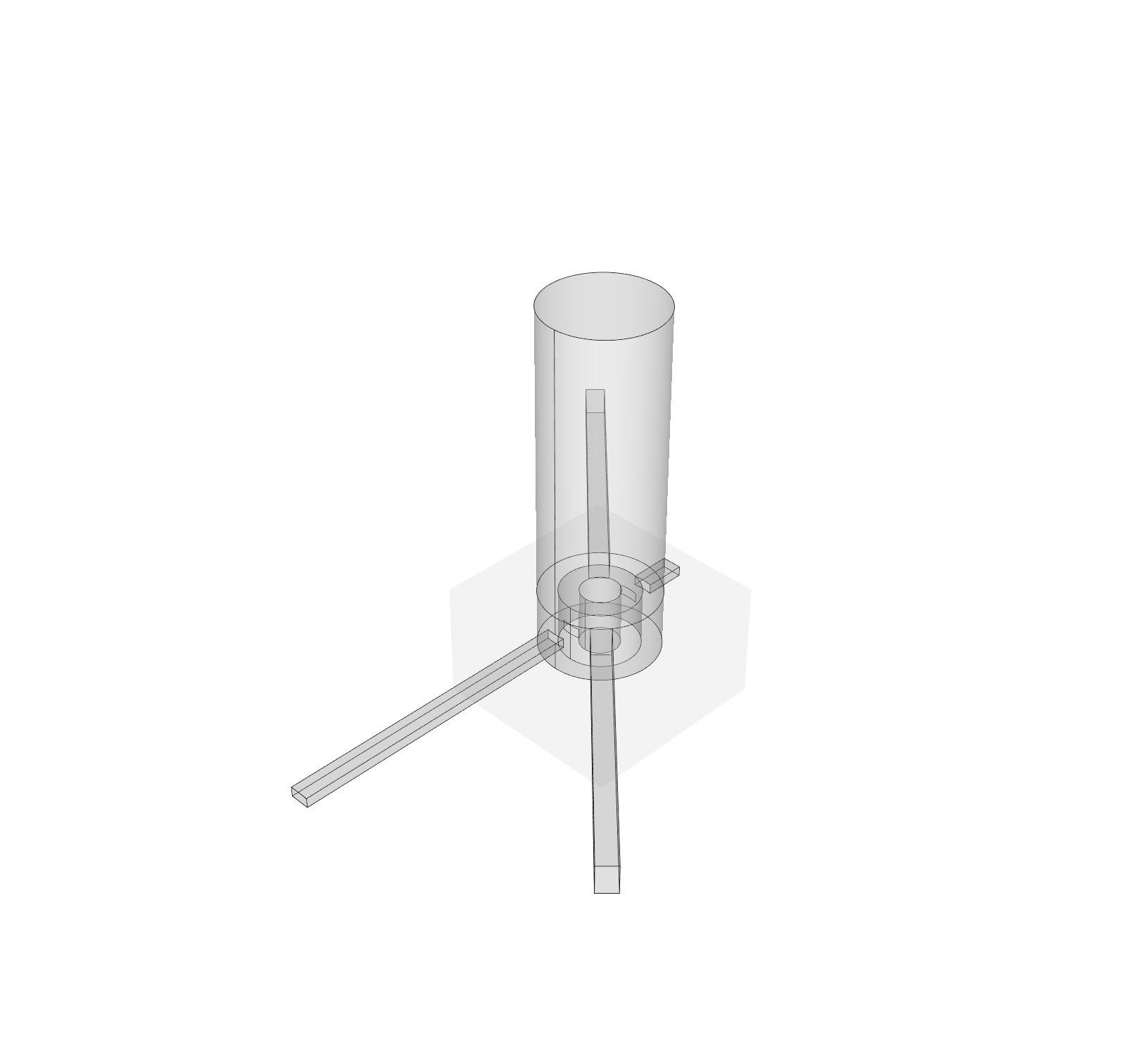} &
        \includegraphics[width=0.124\linewidth]{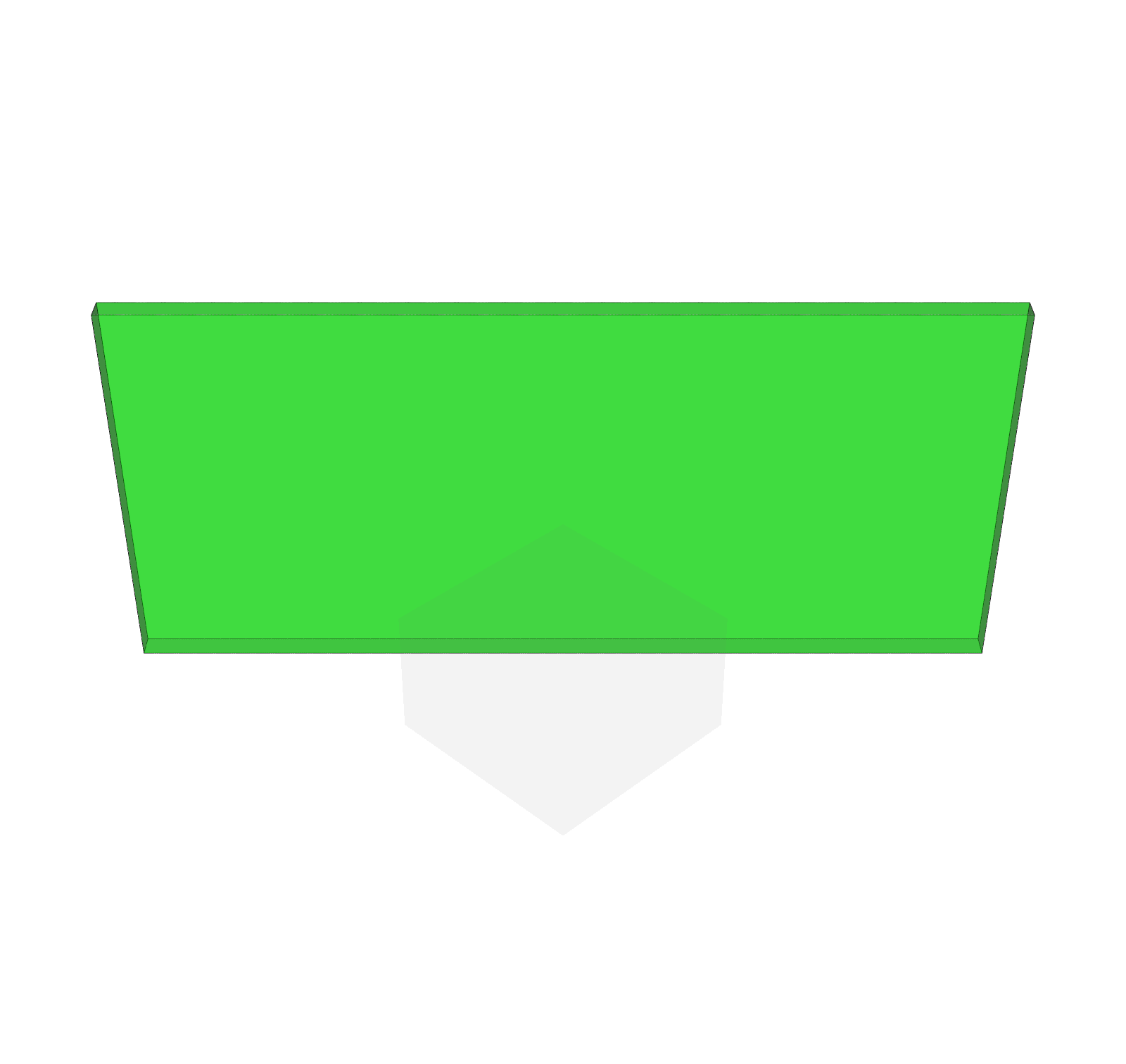} &
        \includegraphics[width=0.124\linewidth]{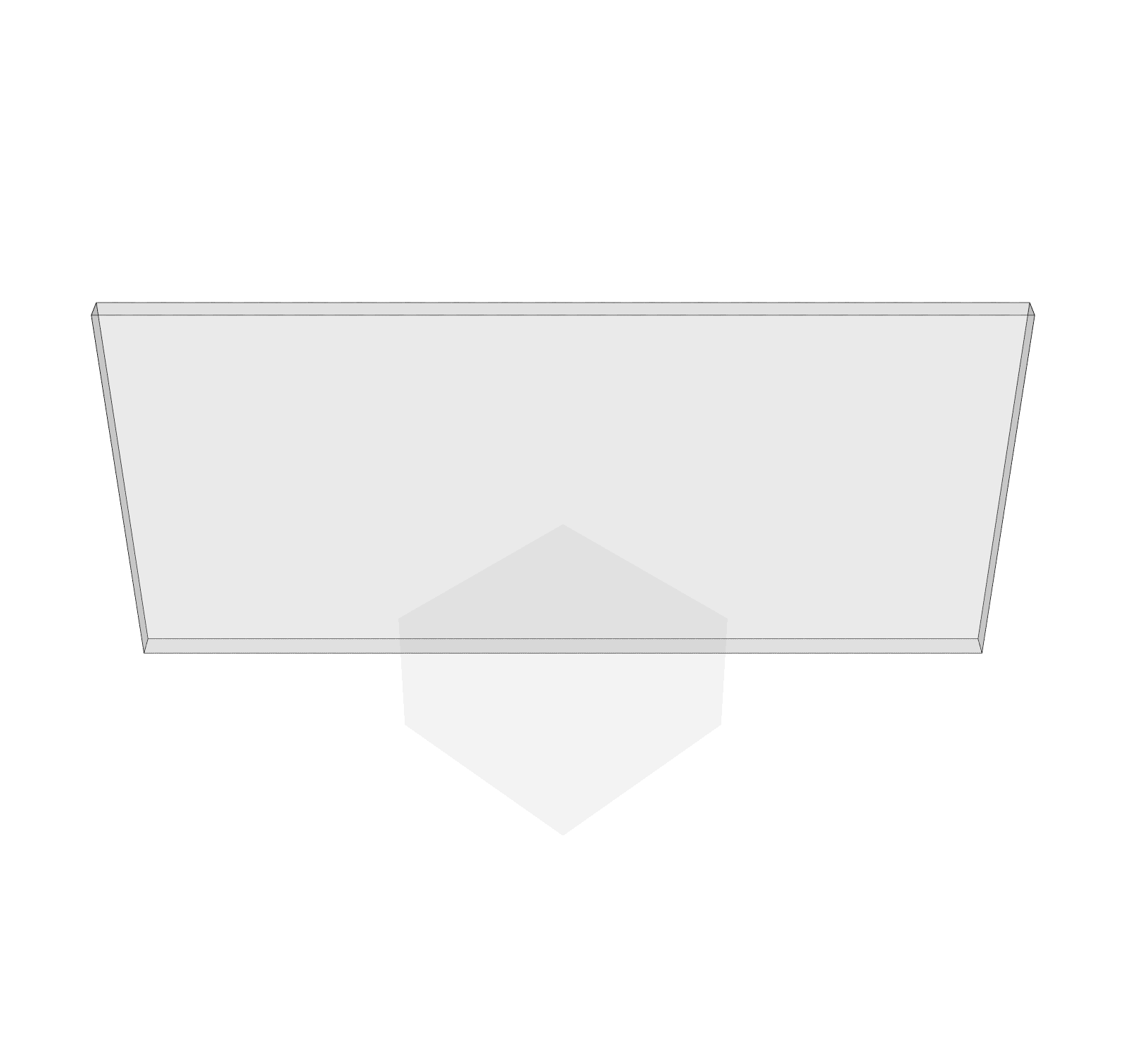} &
        \includegraphics[width=0.124\linewidth]{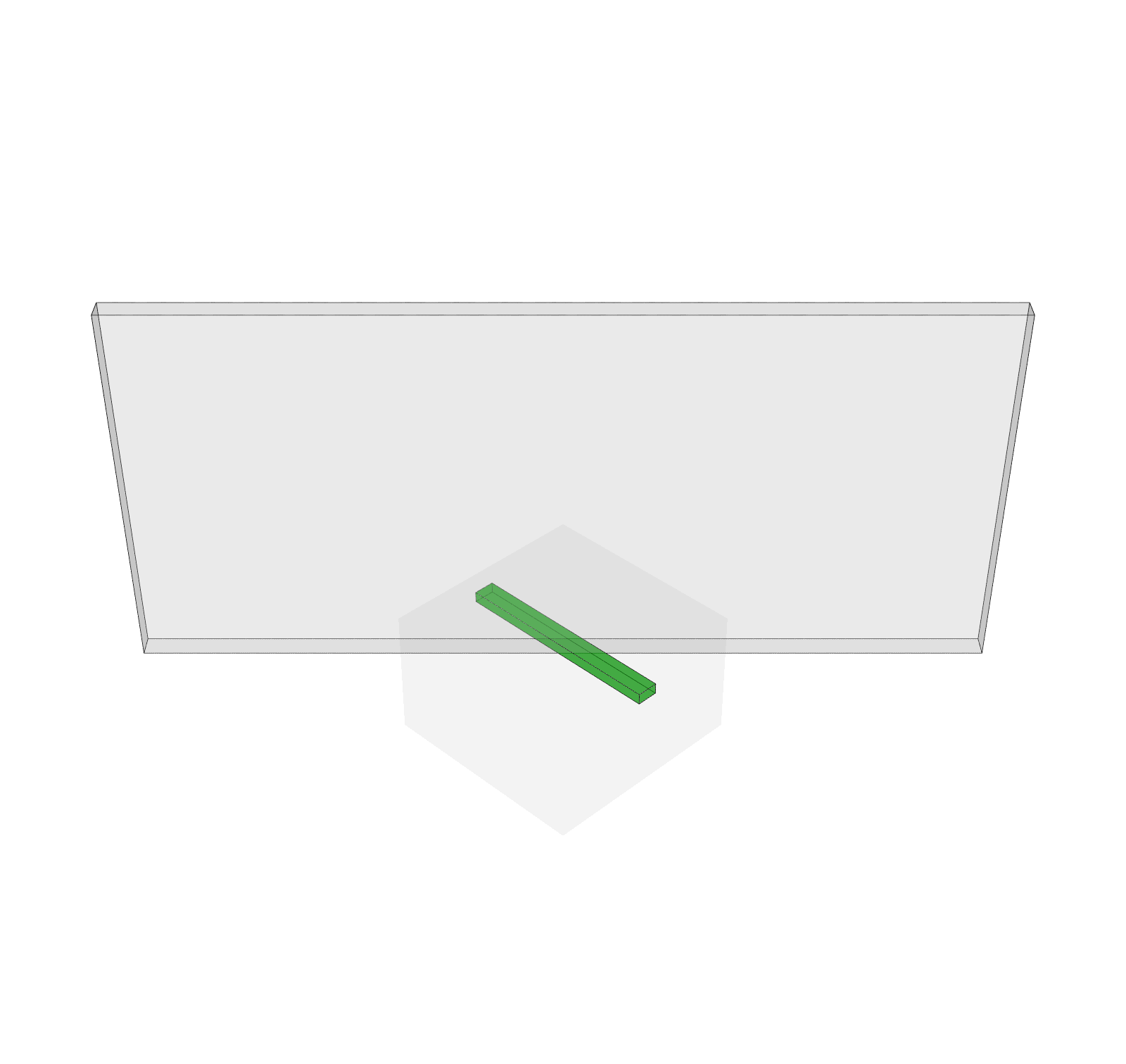} &
        \includegraphics[width=0.124\linewidth]{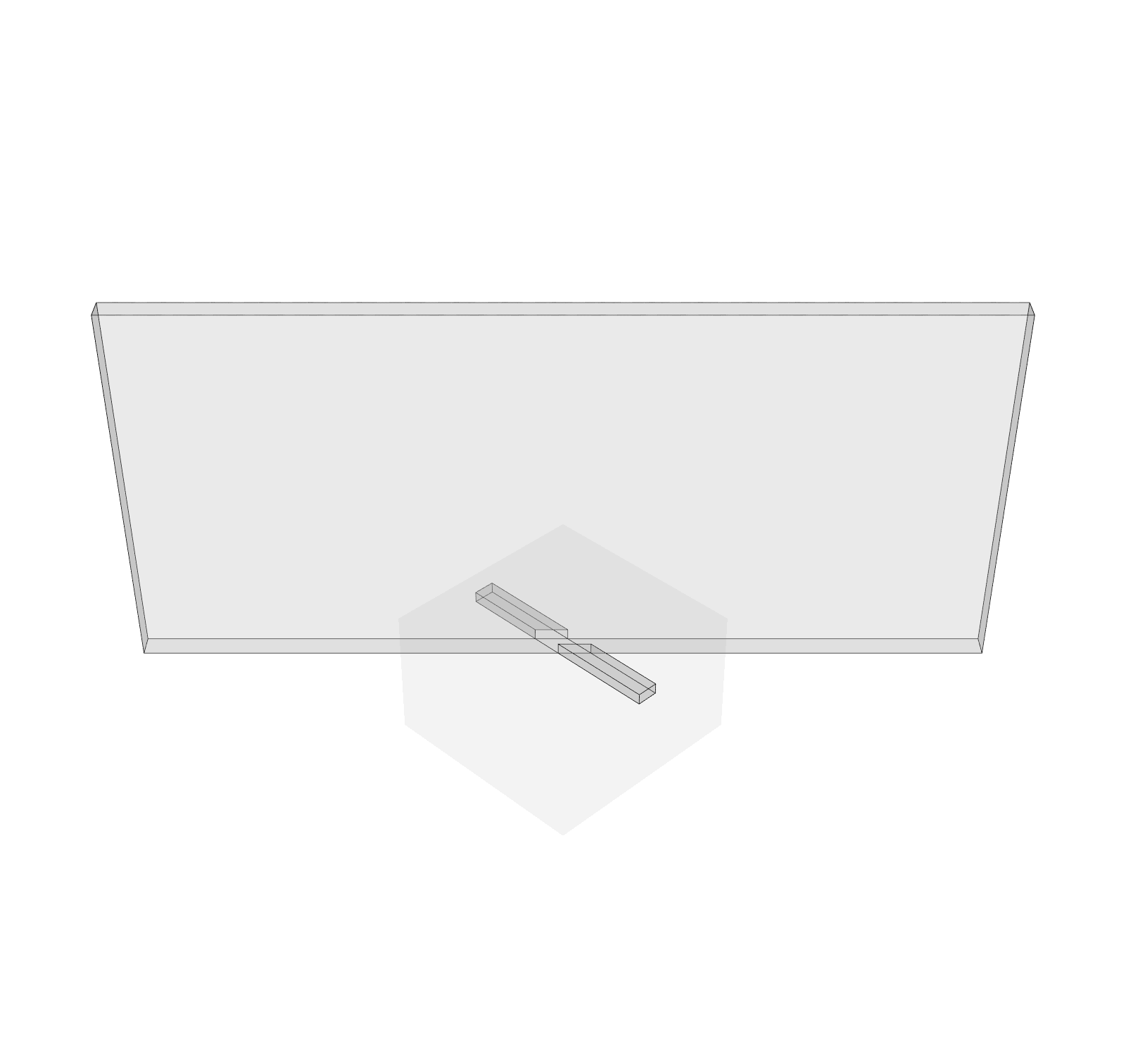} &
        \includegraphics[width=0.124\linewidth]{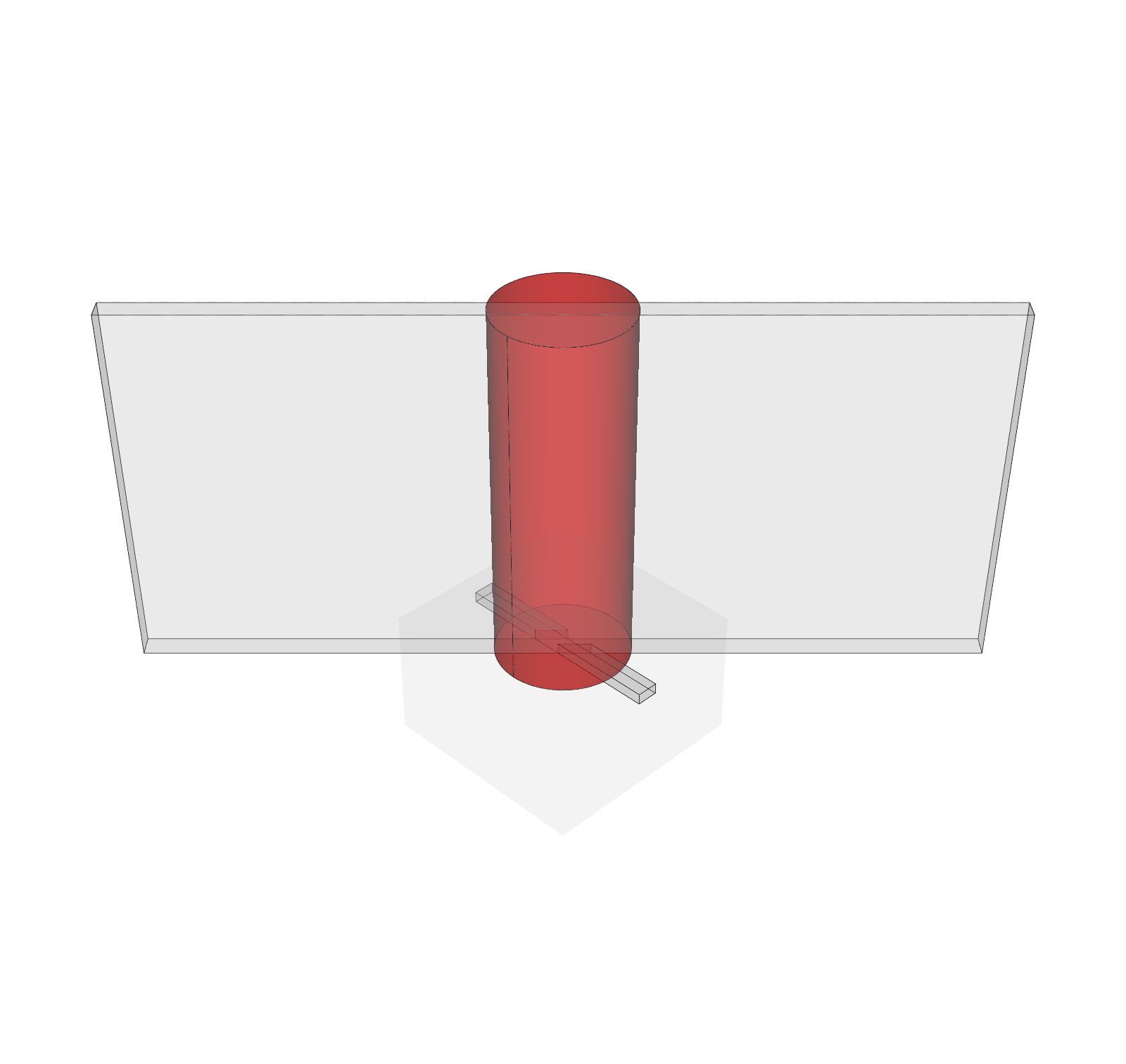} &
        \includegraphics[width=0.124\linewidth]{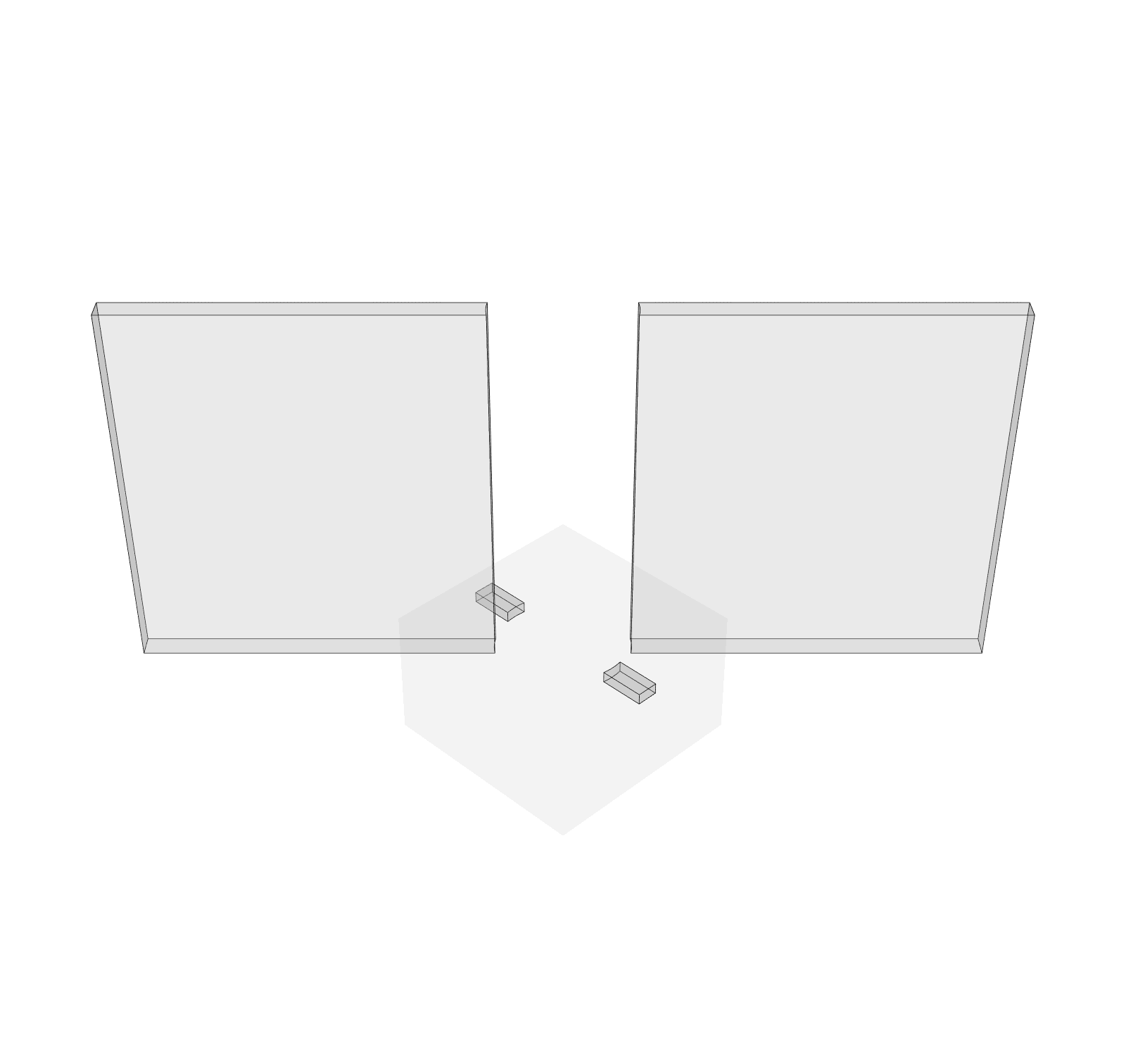}
        \\
        \includegraphics[width=0.124\linewidth]{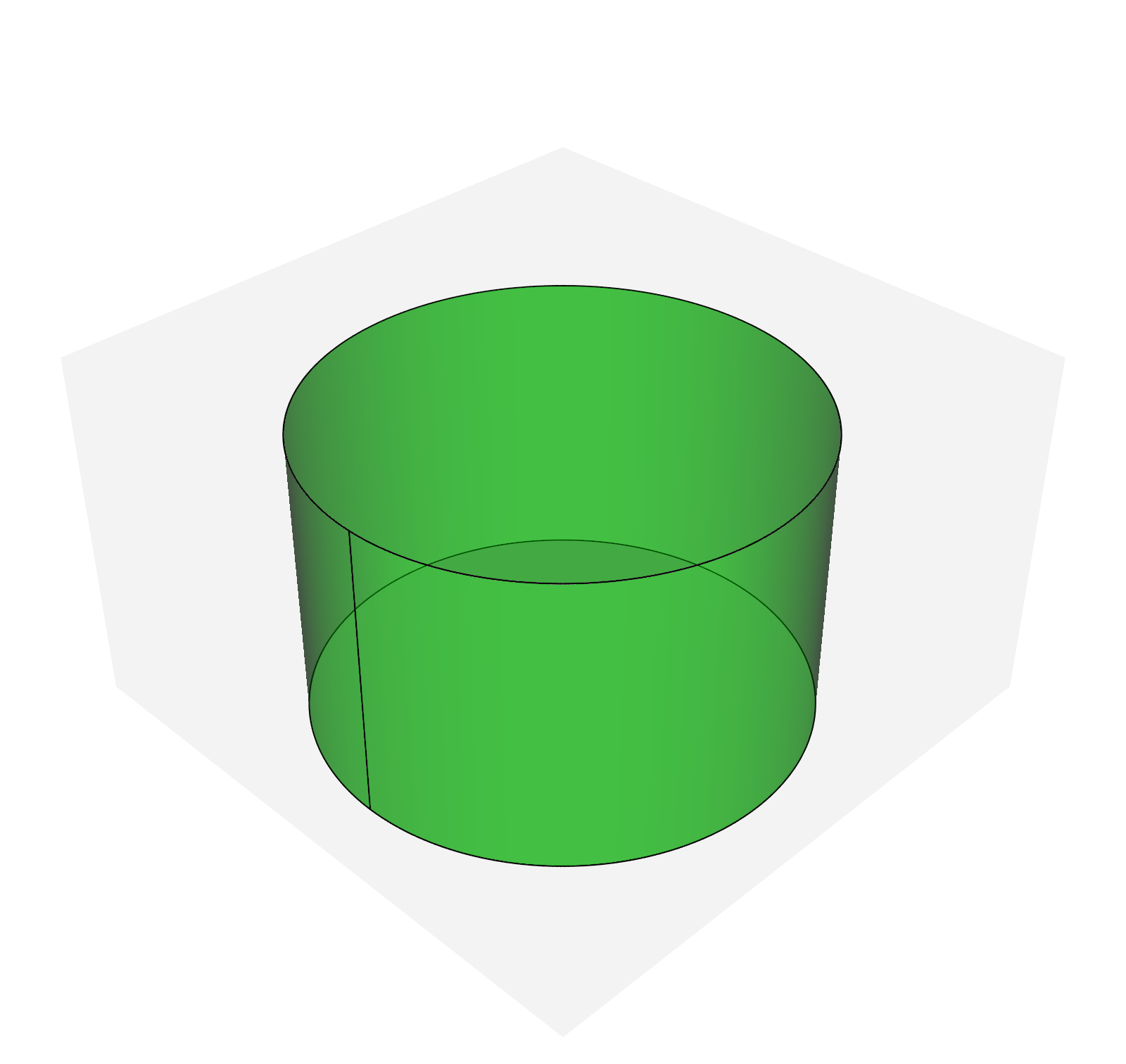} &
        \includegraphics[width=0.124\linewidth]{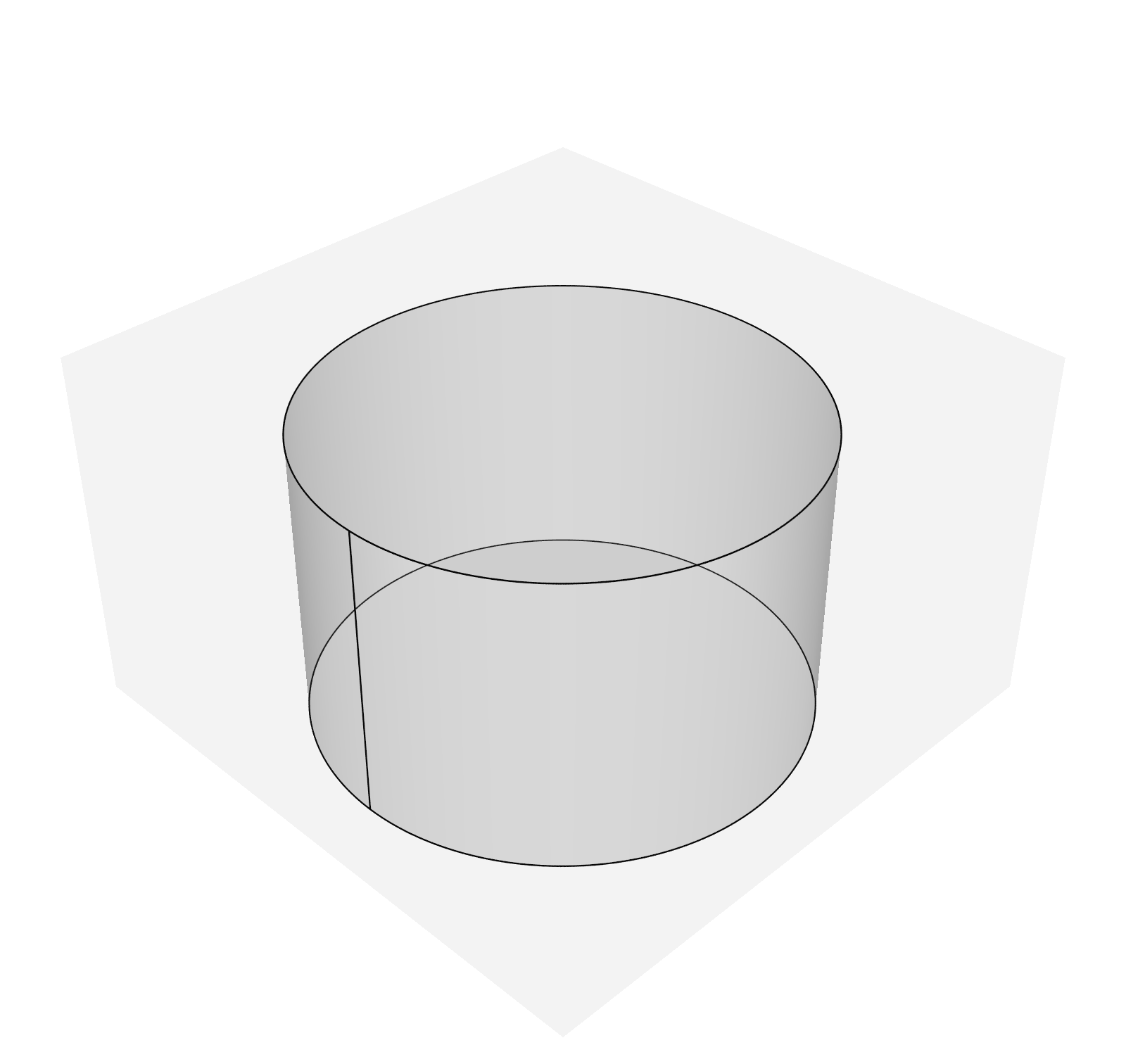} &
        \includegraphics[width=0.124\linewidth]{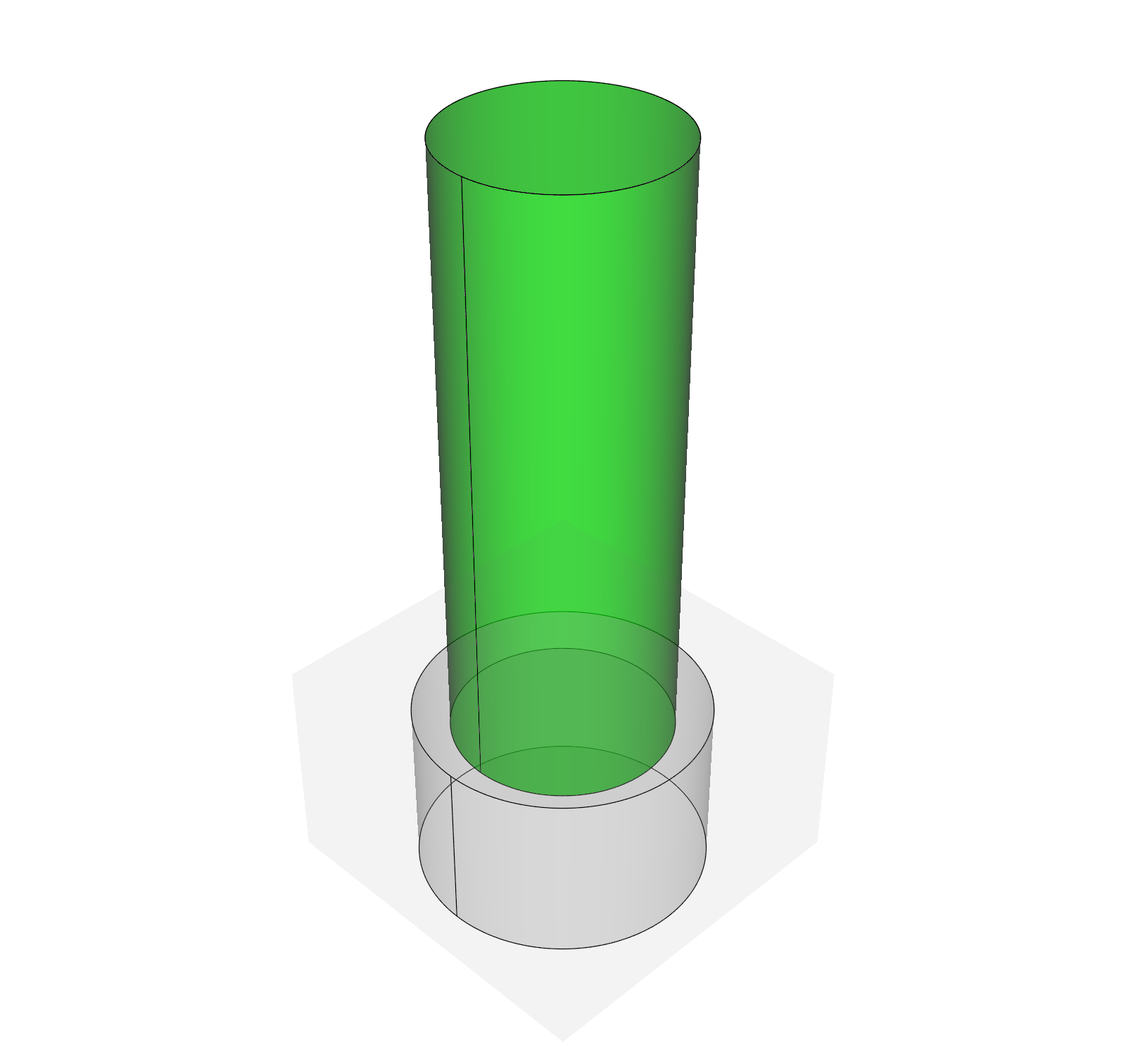} &
        \includegraphics[width=0.124\linewidth]{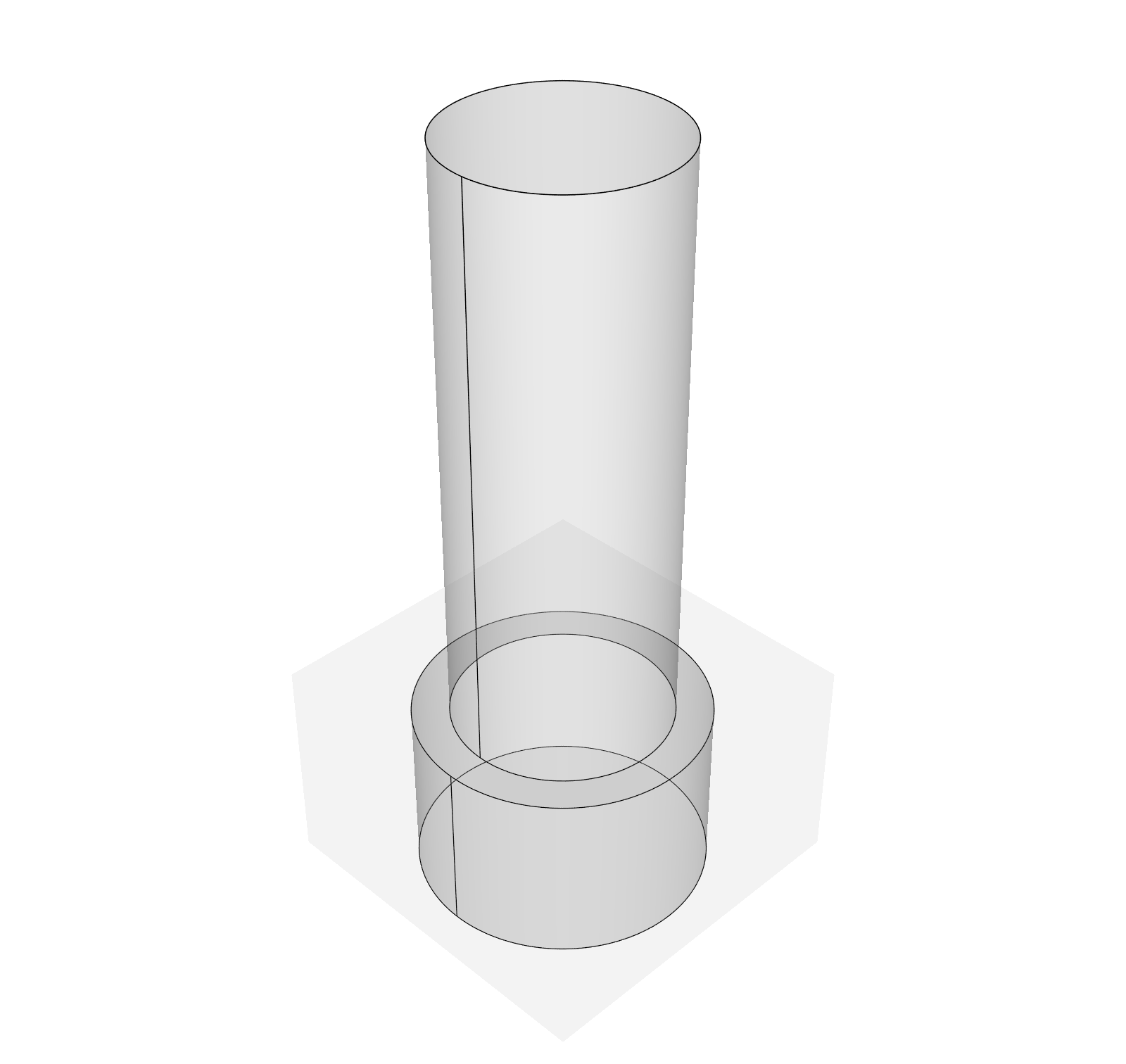} &
        \includegraphics[width=0.124\linewidth]{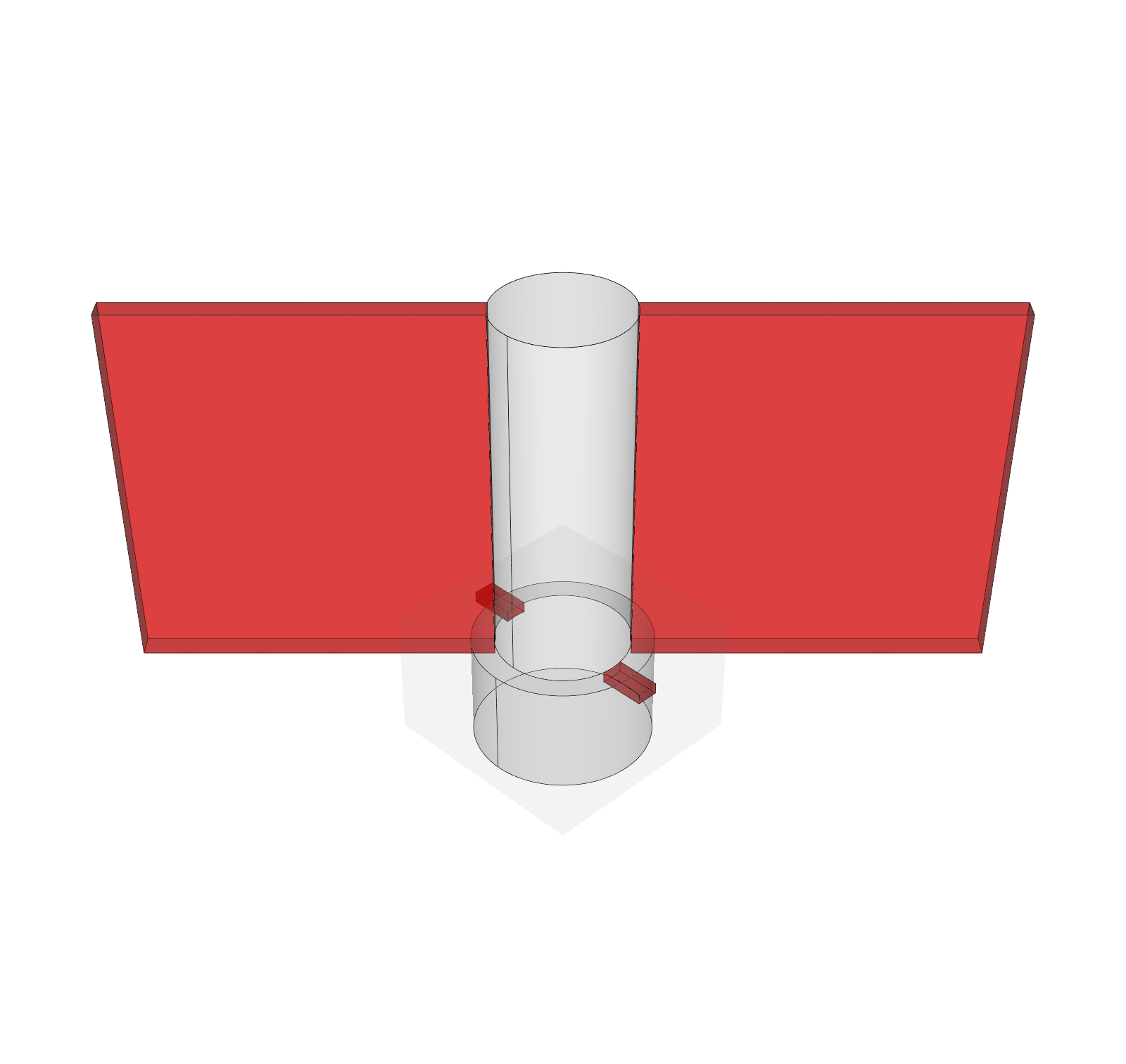} &
        \includegraphics[width=0.124\linewidth]{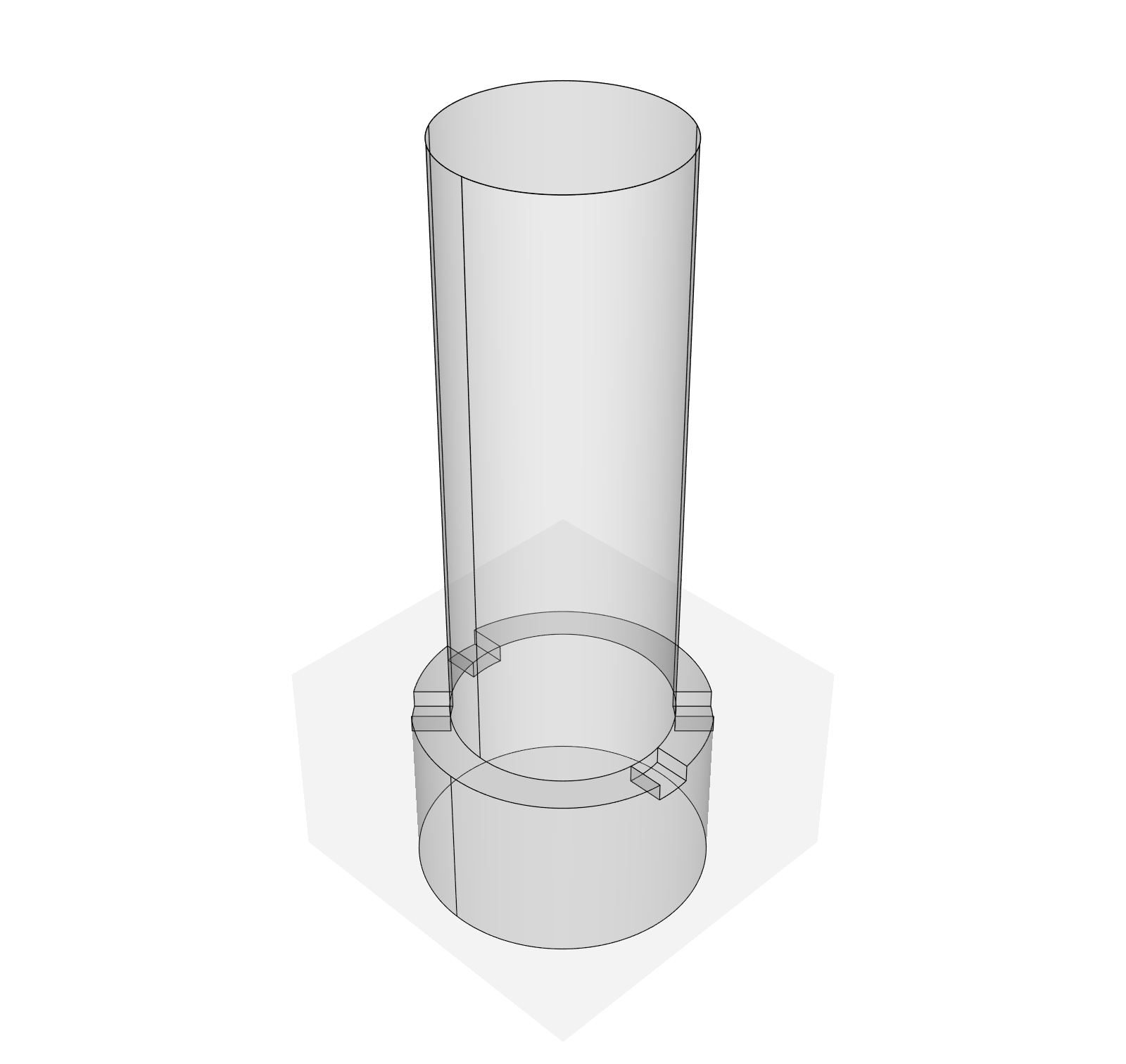} &
        \includegraphics[width=0.124\linewidth]{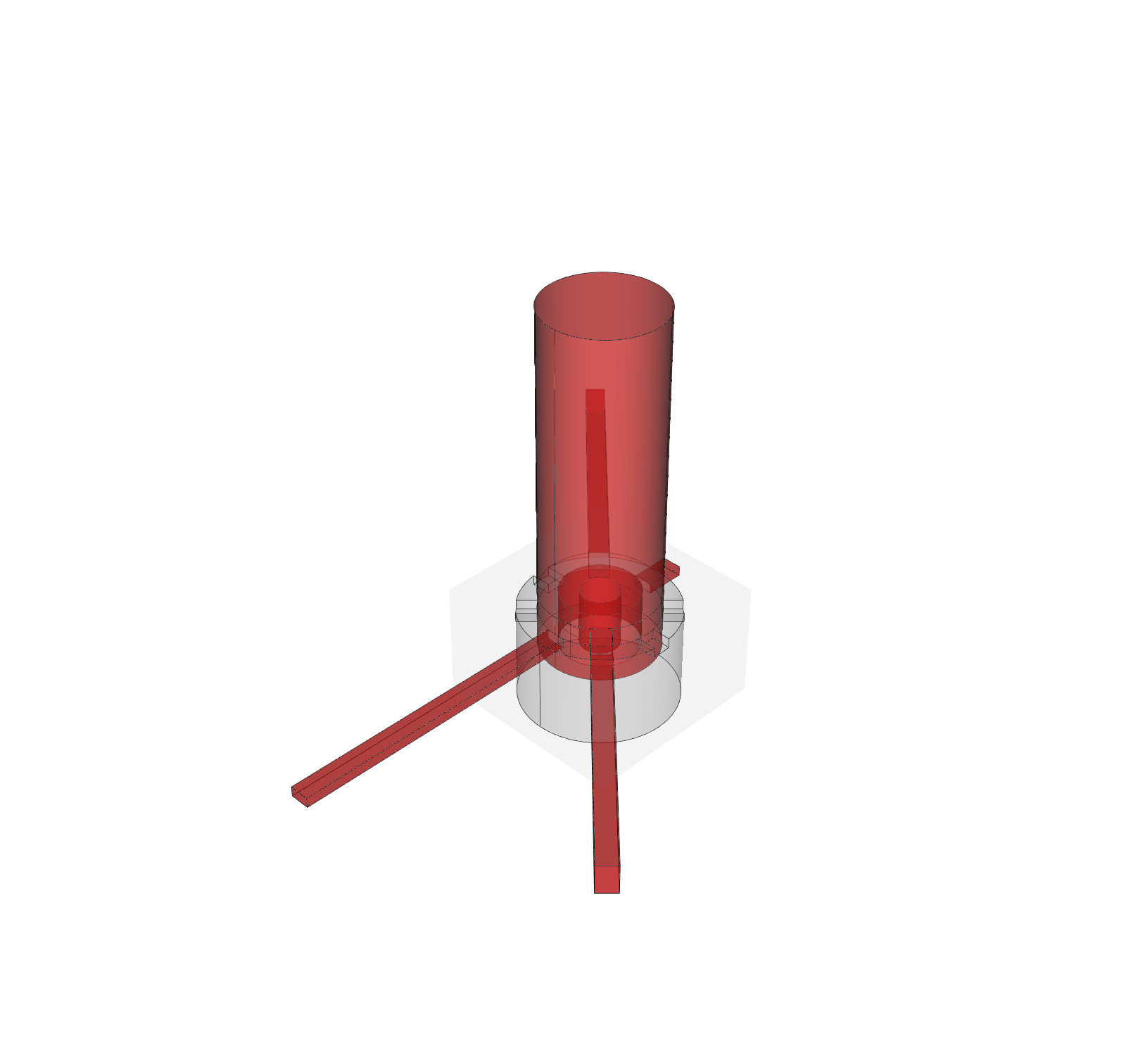} &
        \includegraphics[width=0.124\linewidth]{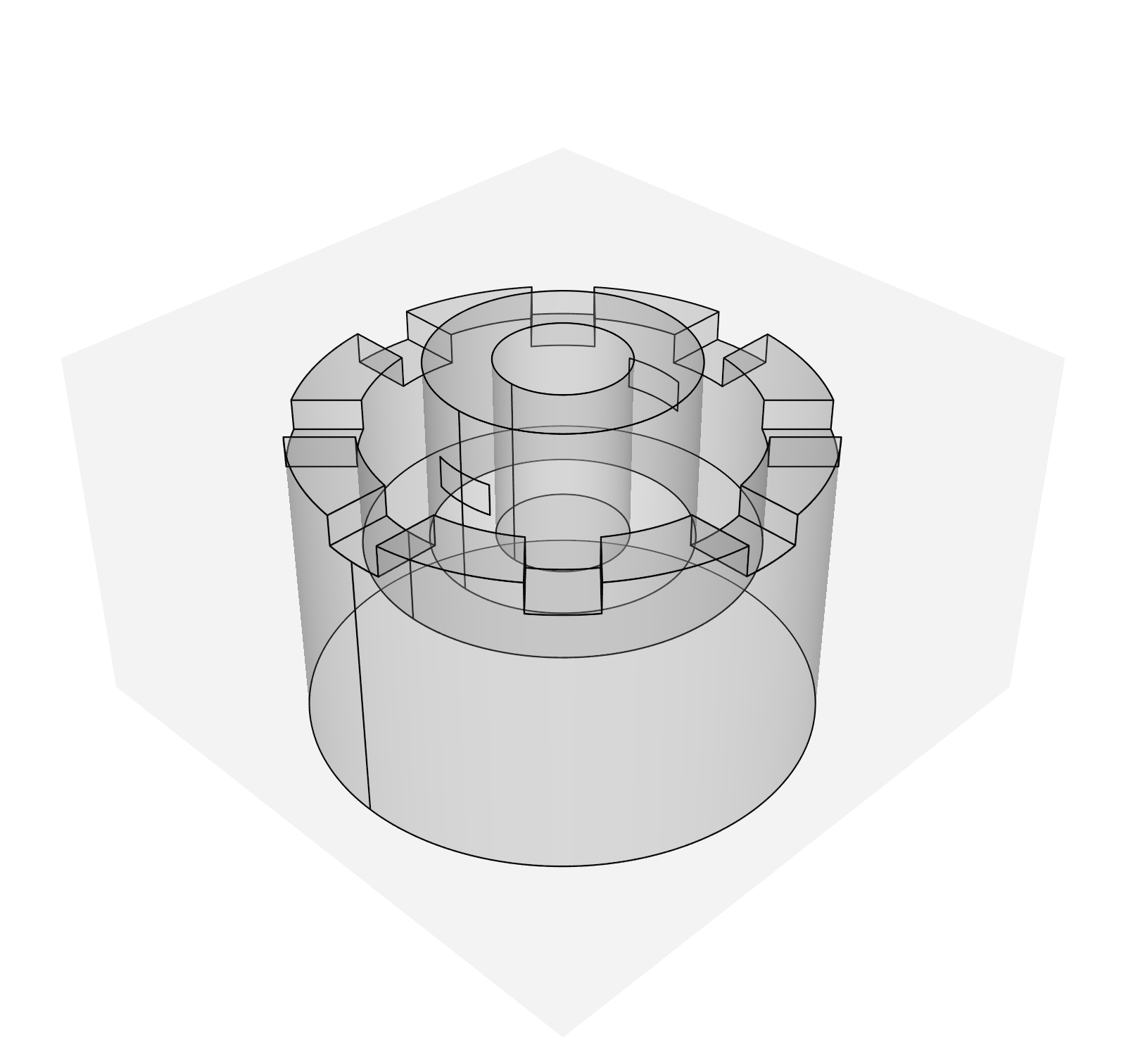}
        \\
        \includegraphics[width=0.124\linewidth]{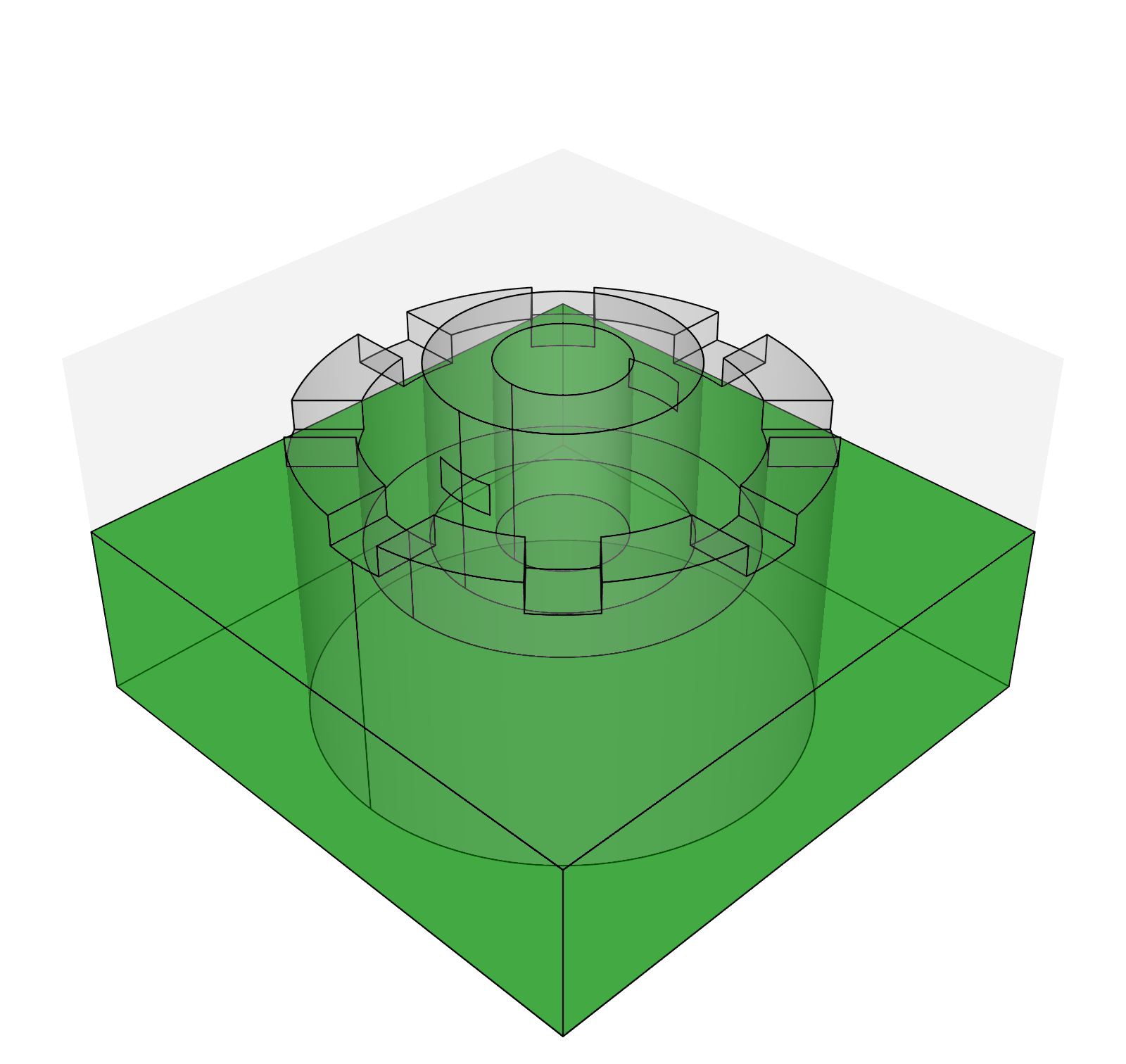} &
        \includegraphics[width=0.124\linewidth]{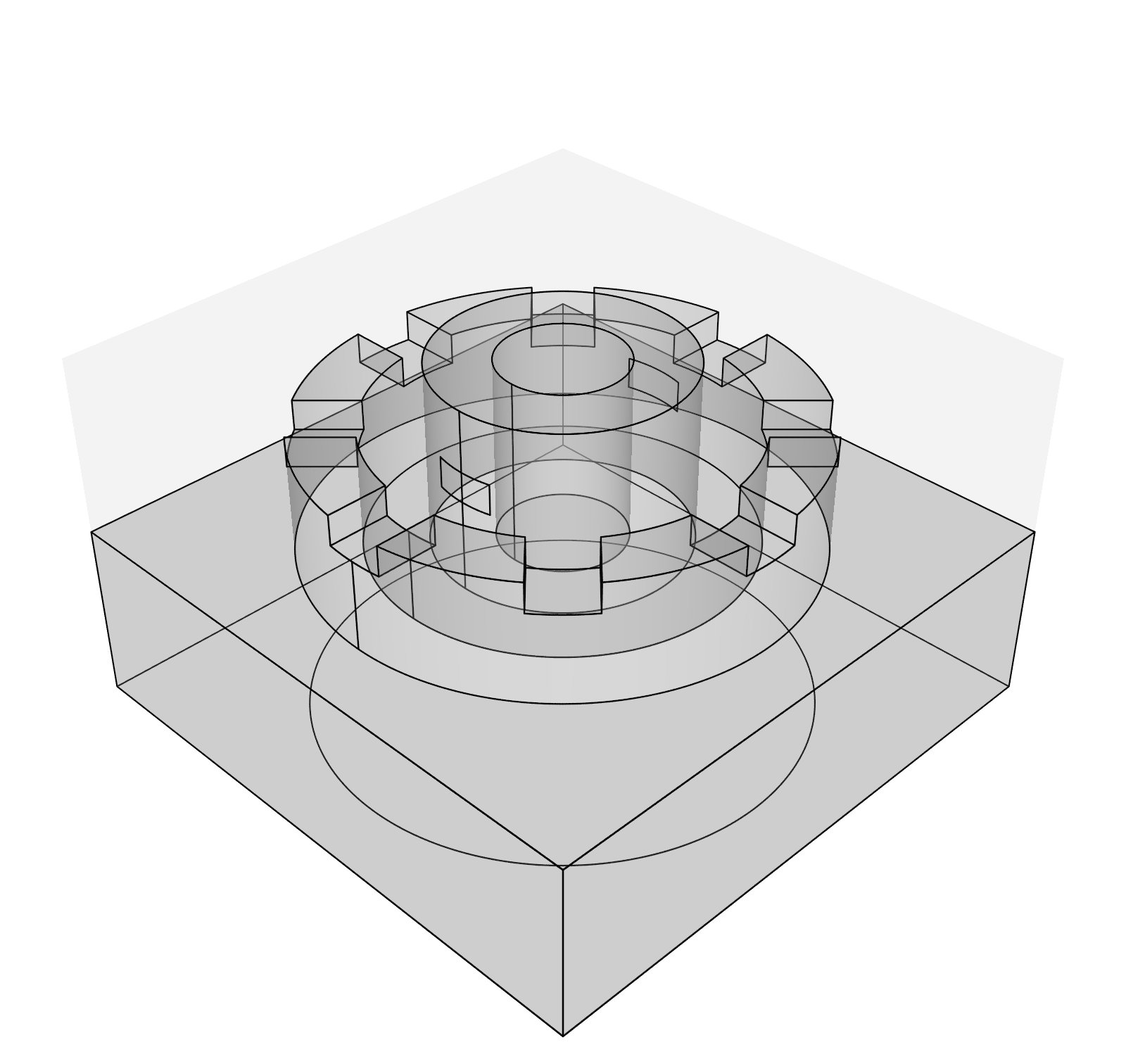} &
        &
        &
        &
        &
        &
        \\

        \multicolumn{1}{l}{Ours Net} & & & & &
        \\
        \includegraphics[width=0.124\linewidth]{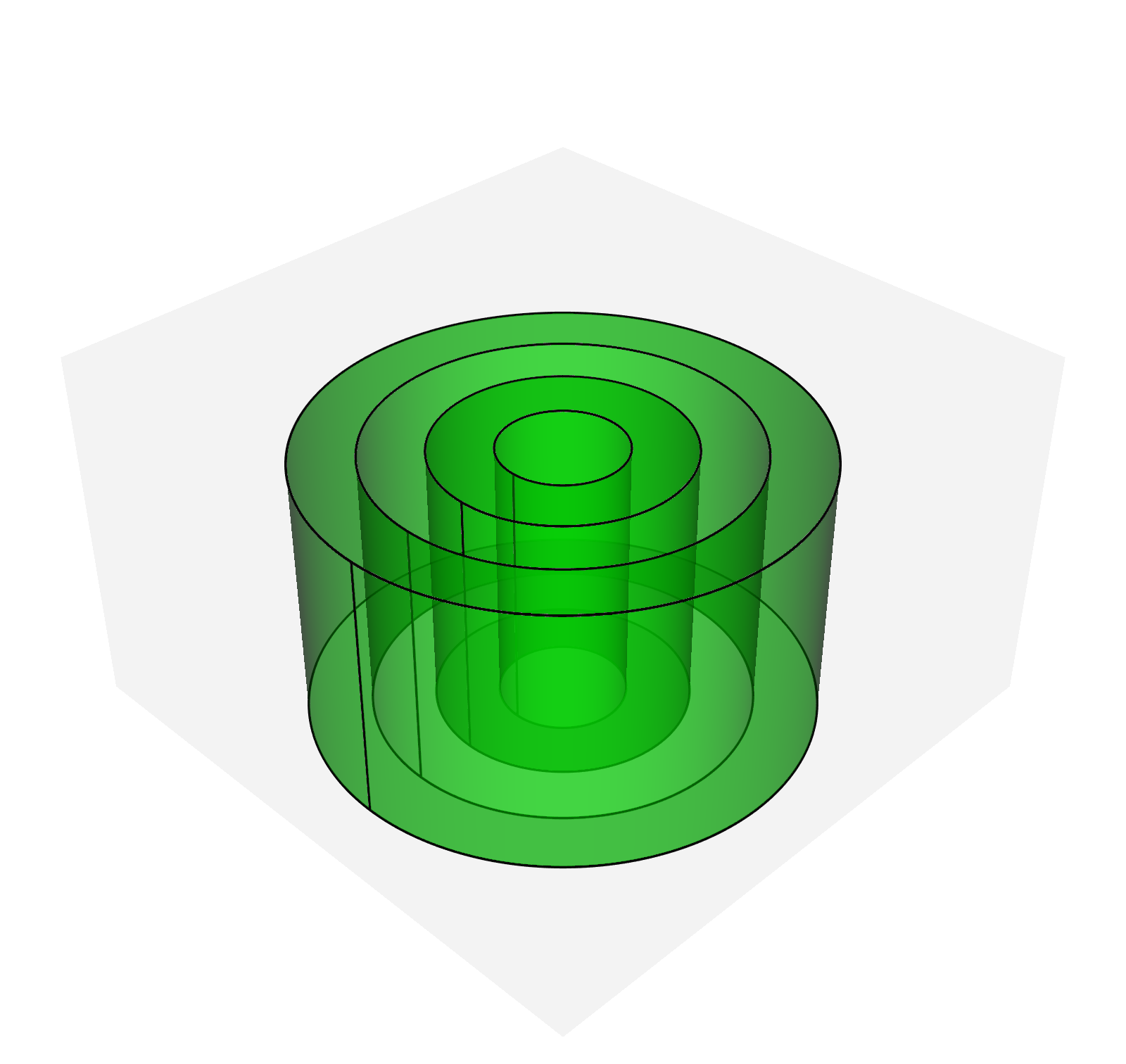} &
        \includegraphics[width=0.124\linewidth]{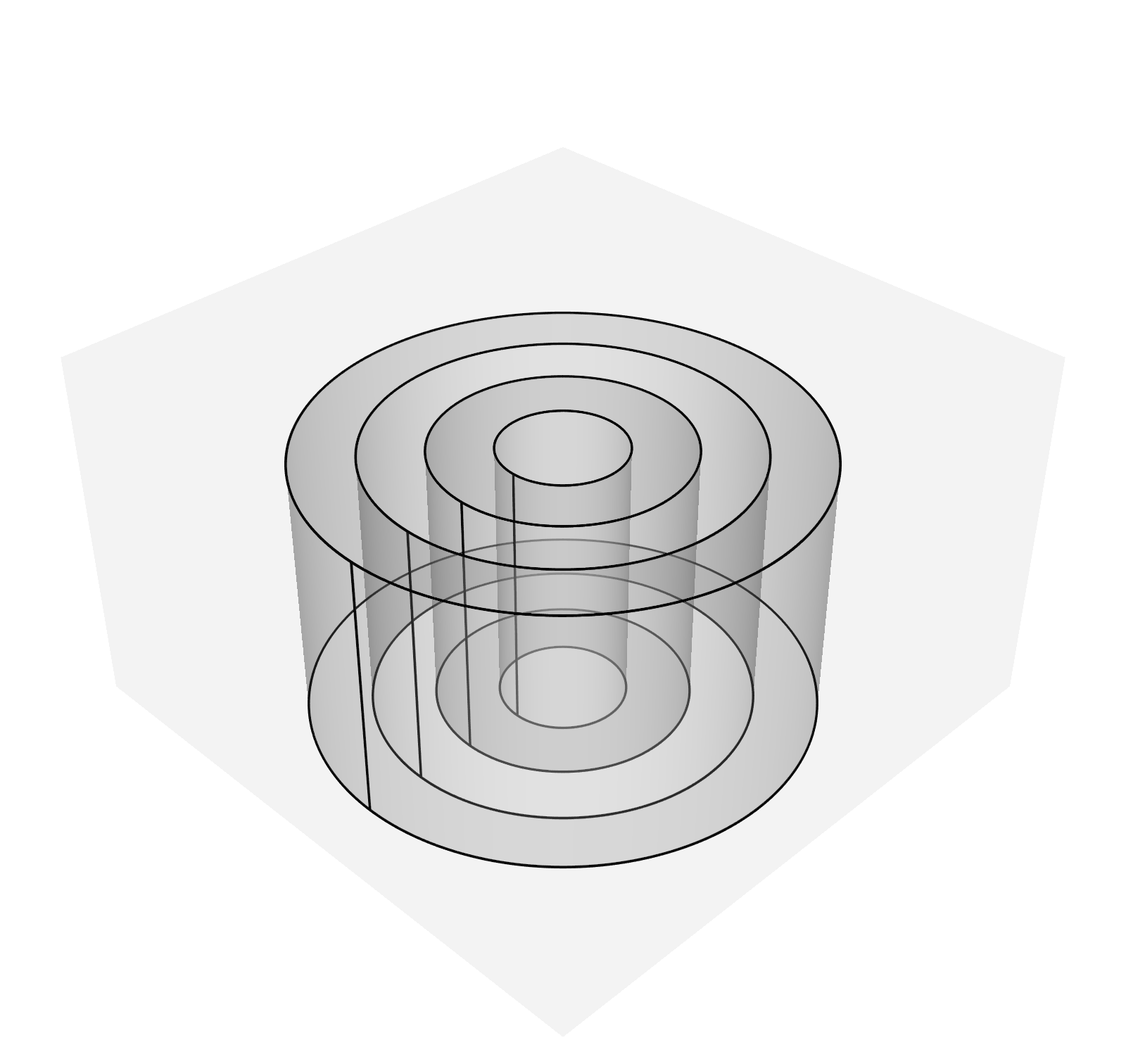} &
        \includegraphics[width=0.124\linewidth]{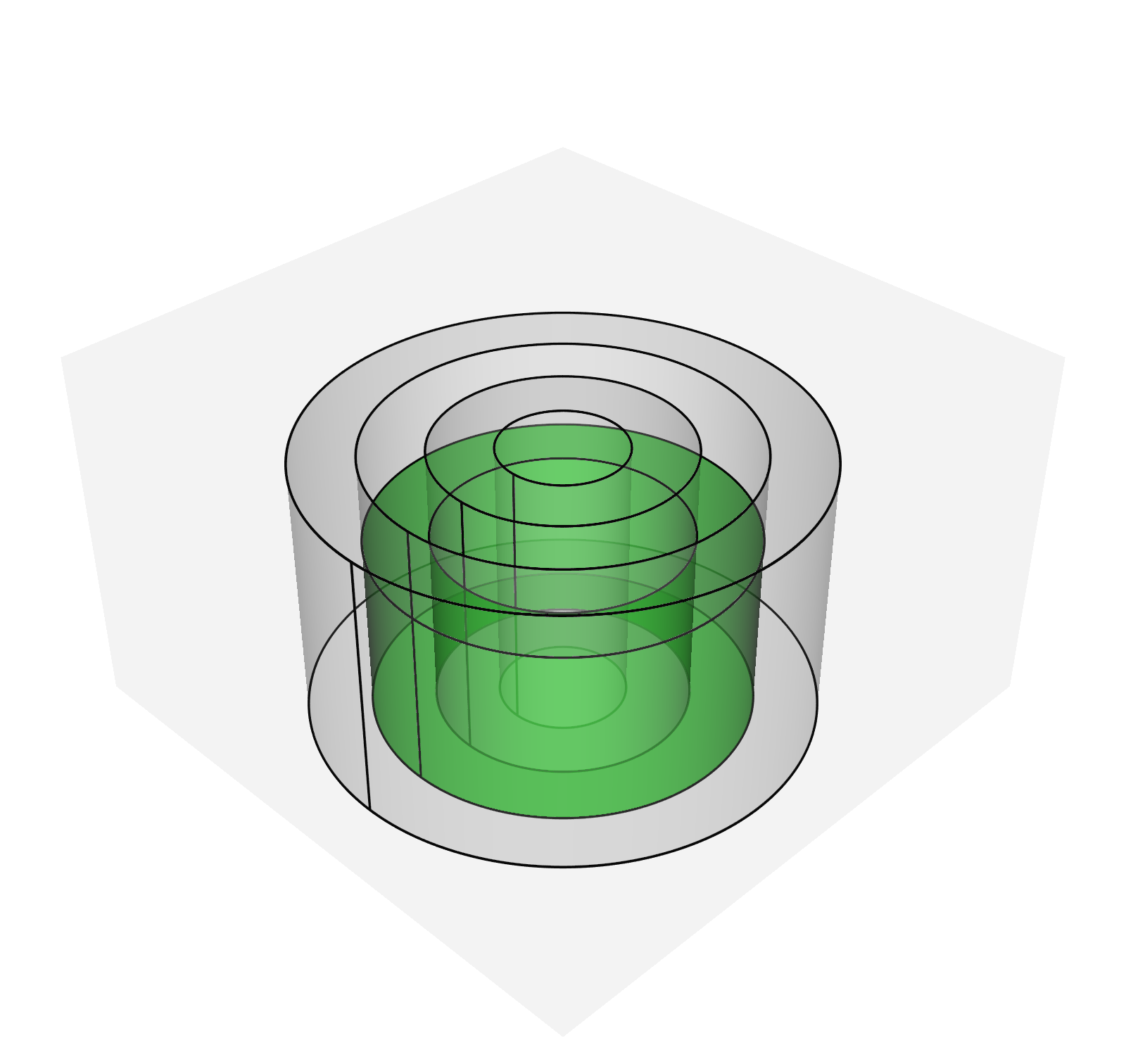} &
        \includegraphics[width=0.124\linewidth]{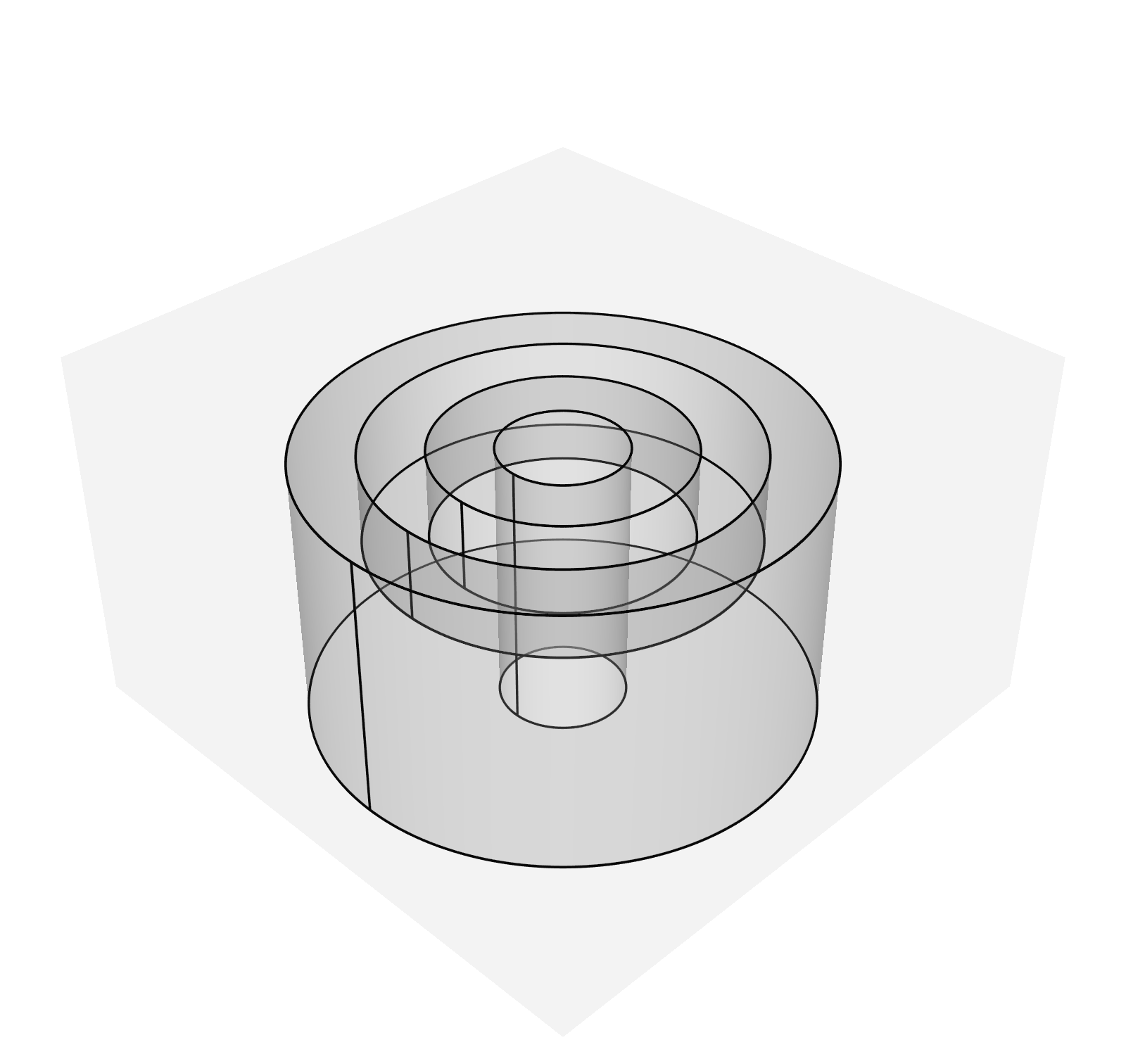} &
        \includegraphics[width=0.124\linewidth]{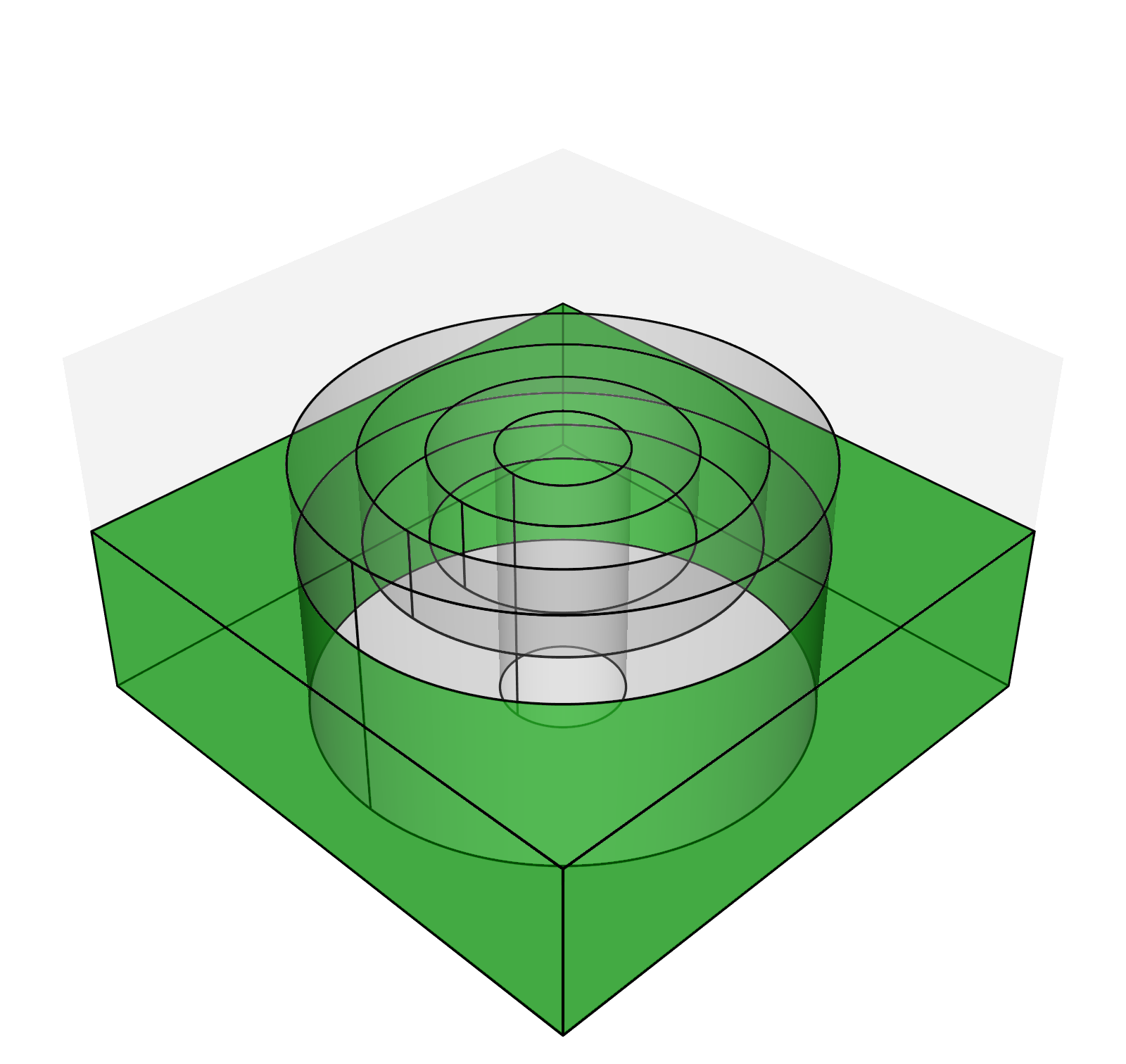} &
        \includegraphics[width=0.124\linewidth]{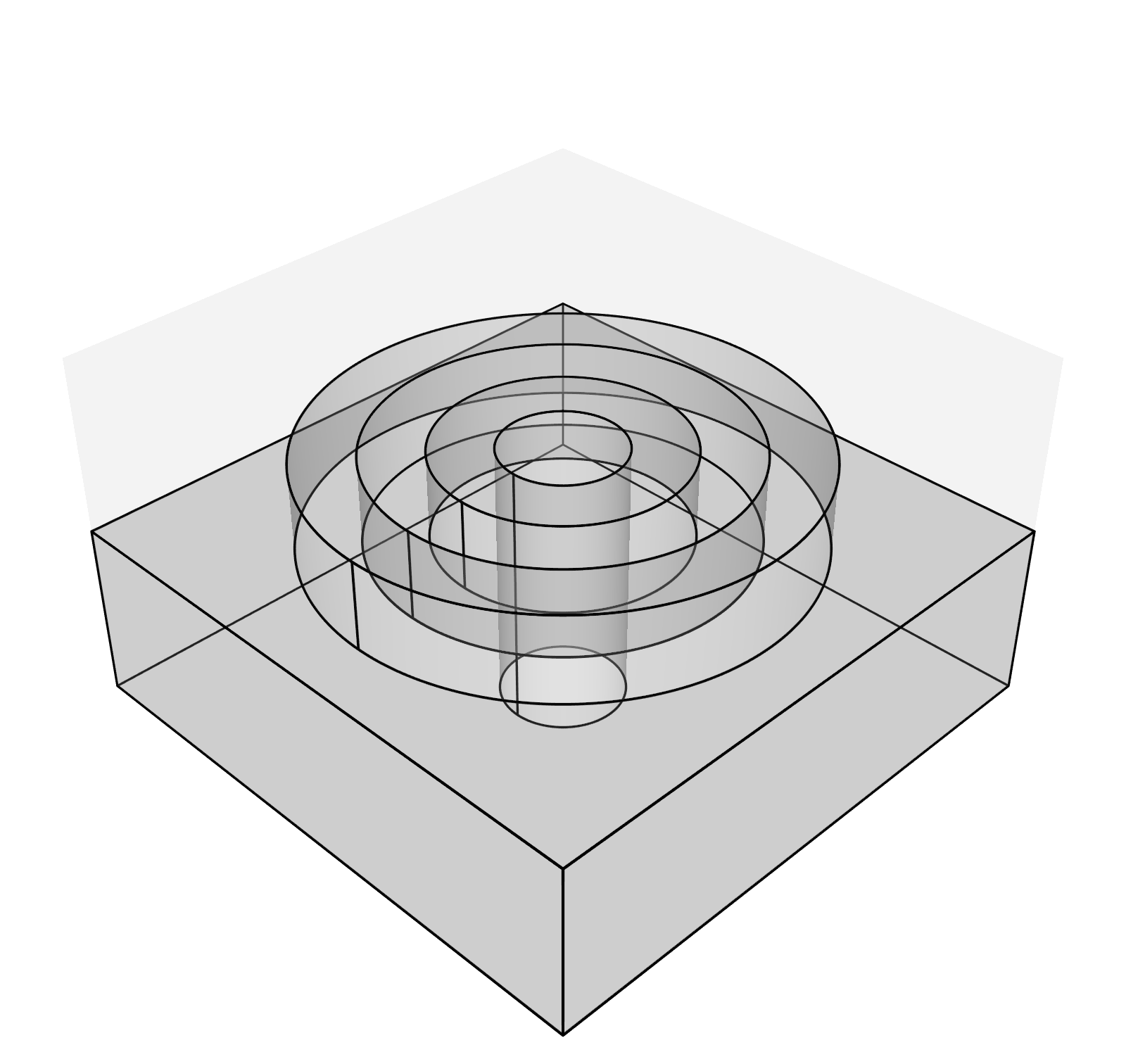} &
        \includegraphics[width=0.124\linewidth]{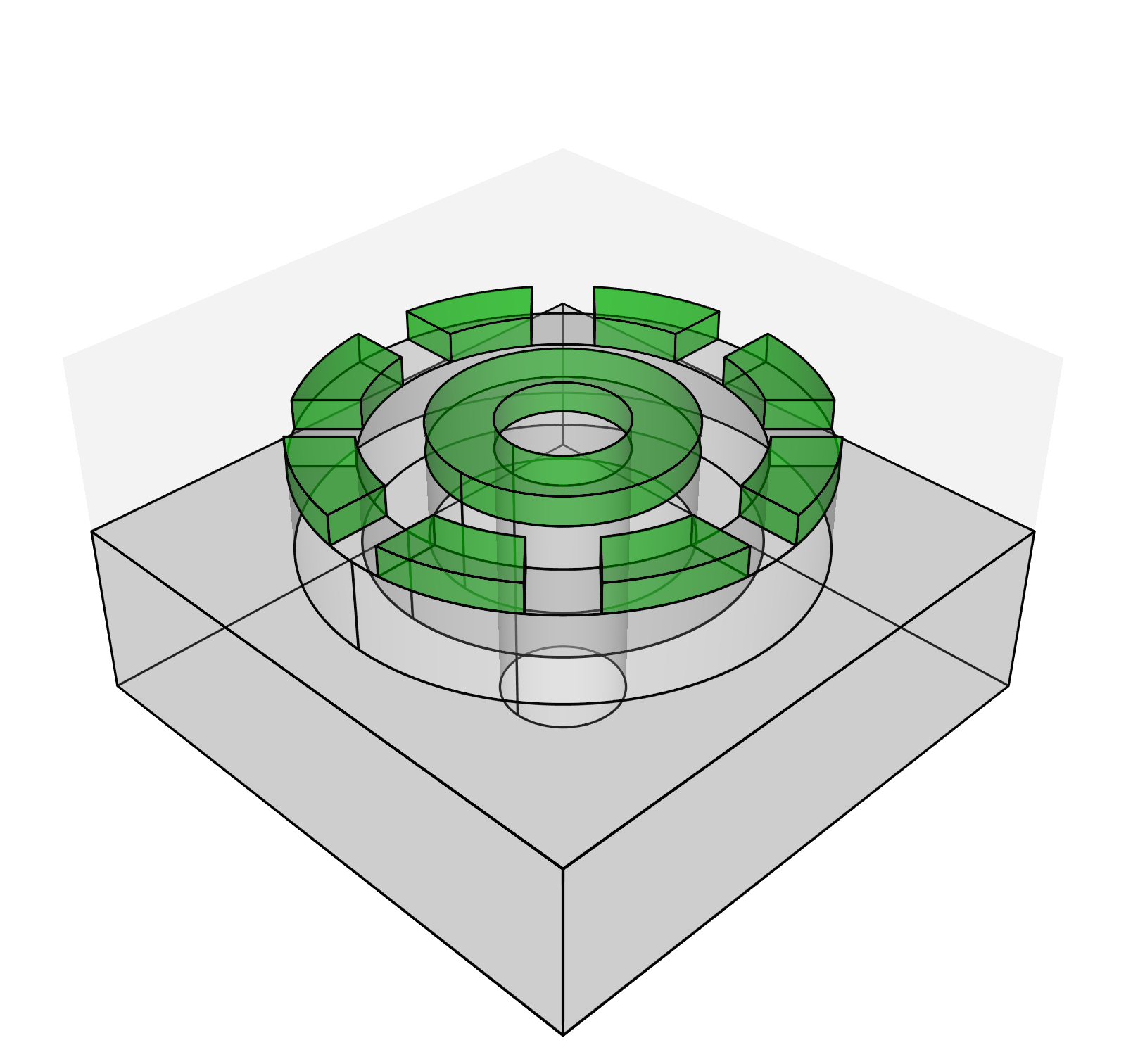} &
        \includegraphics[width=0.124\linewidth]{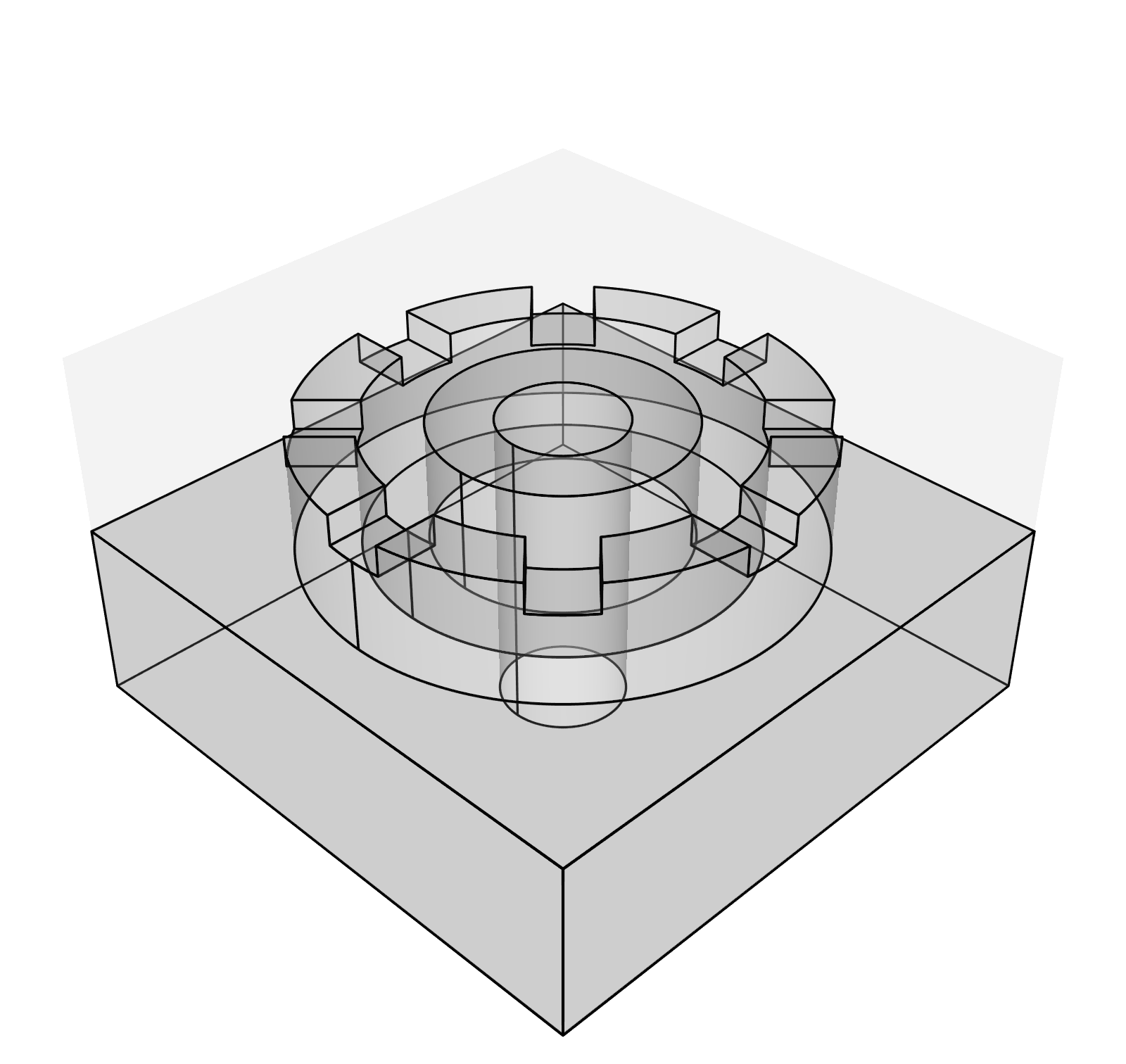}
        \\
        \includegraphics[width=0.124\linewidth]{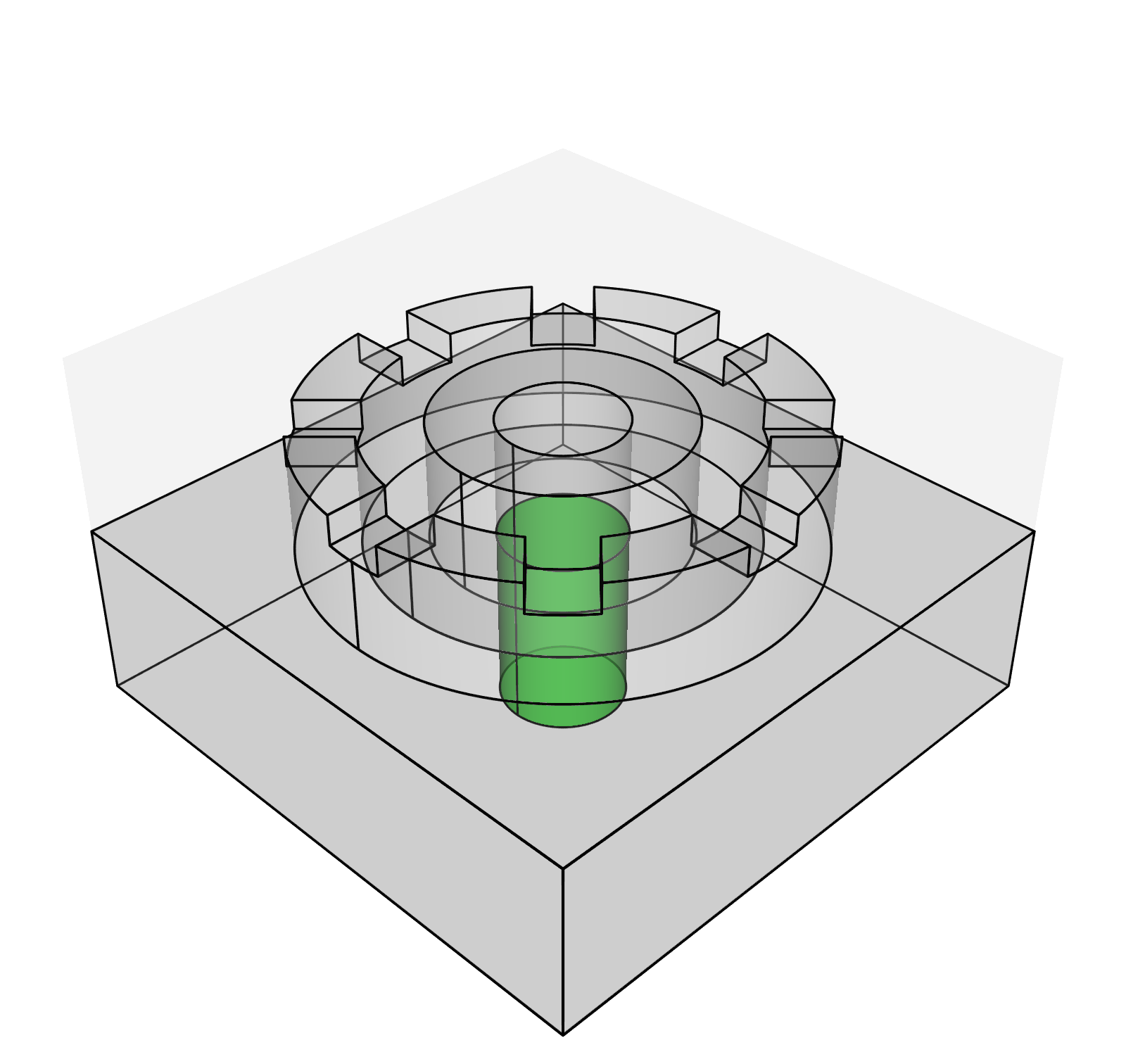} &
        \includegraphics[width=0.124\linewidth]{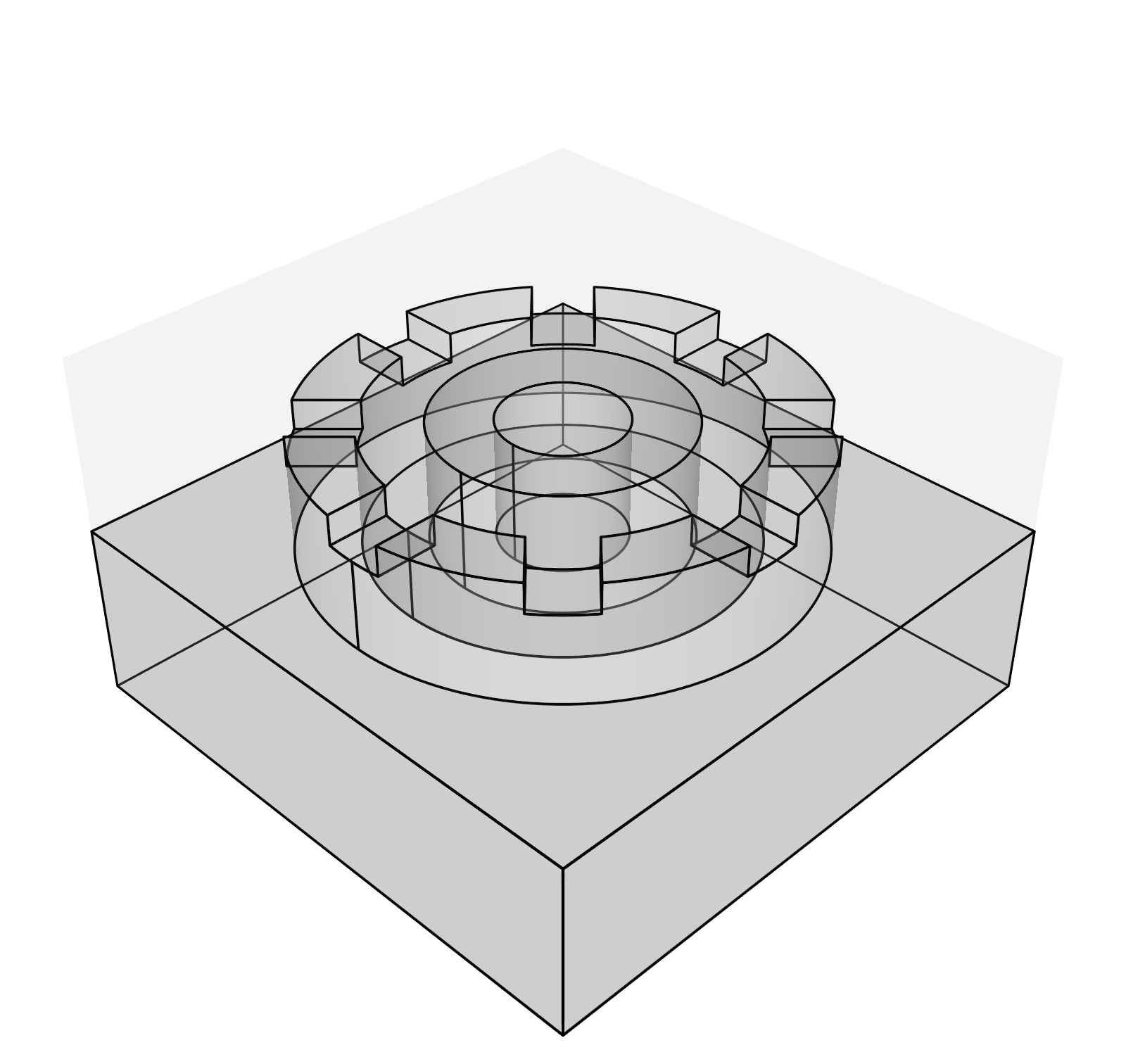} &
        \includegraphics[width=0.124\linewidth]{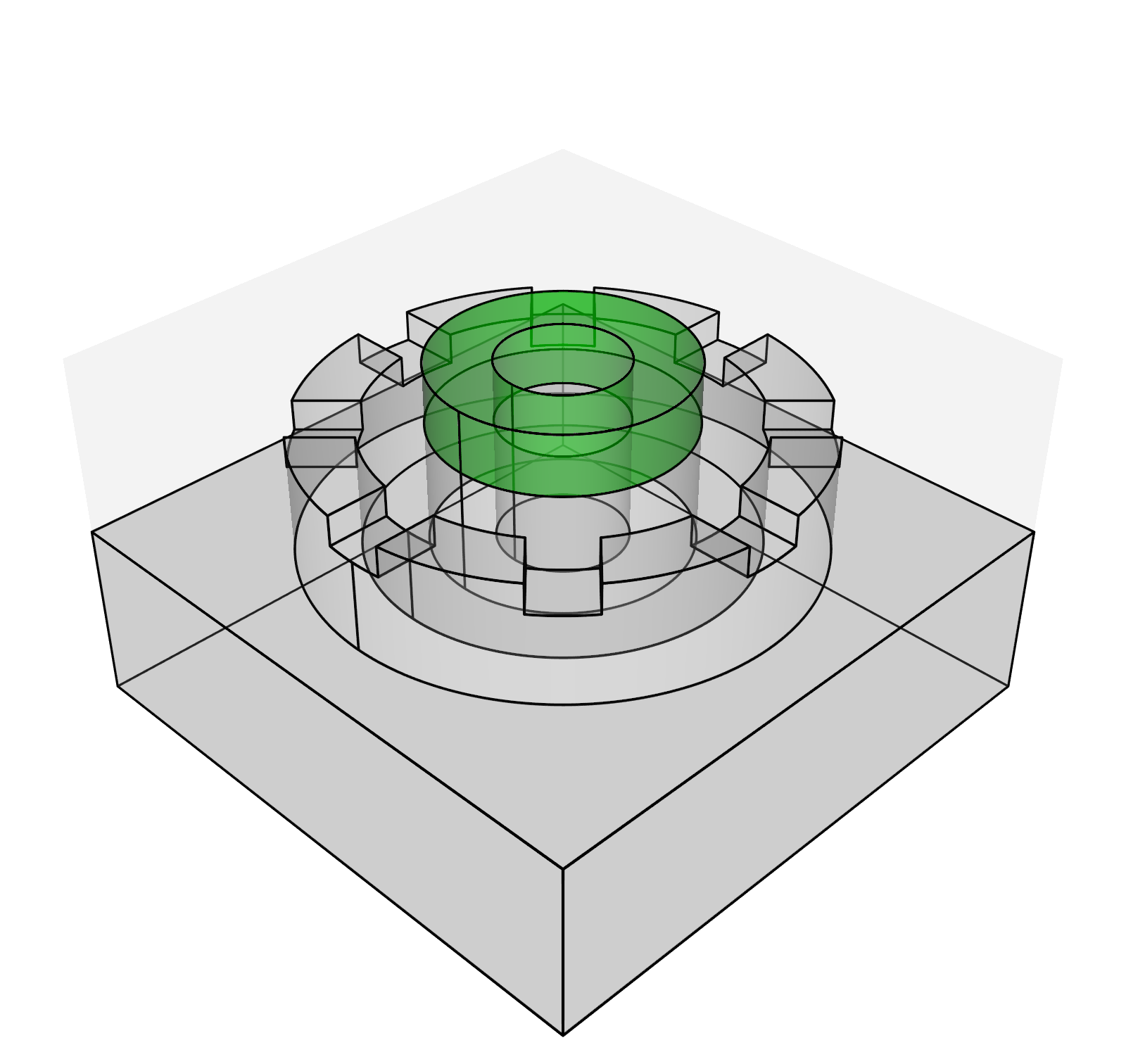} &
        \includegraphics[width=0.124\linewidth]{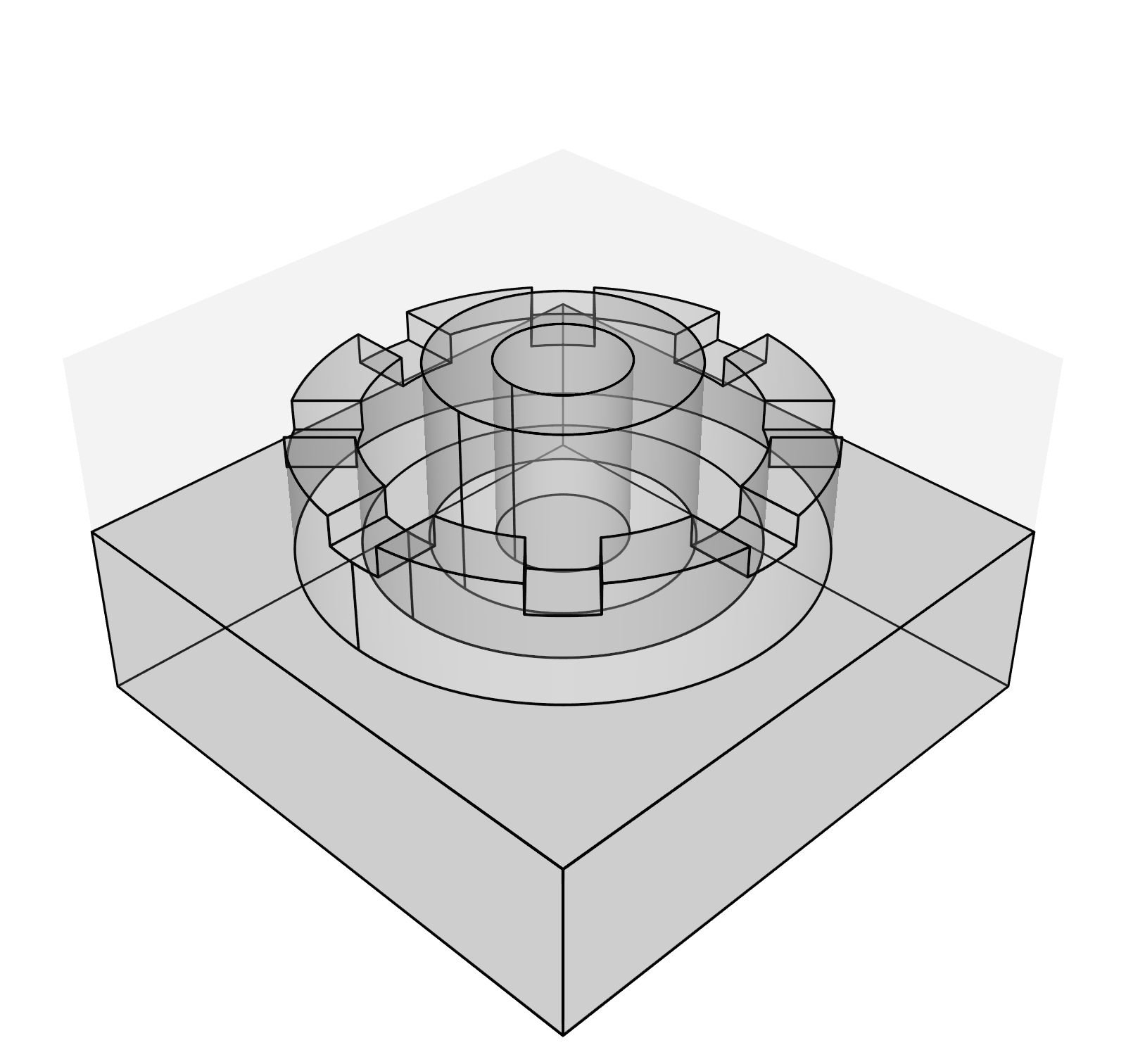} &
        &
        &
        &
        \\

    \end{tabular}
    \caption{
    Qualitative comparison of the output of our model's inferred programs (Ours Net) vs. InverseCSG. Green: addition, Red: subtraction, Blue: intersection, Grey: current. (Case 1)
    }
    \label{figure:qualitative_comparision_inv1}
\end{figure*}

\begin{figure*}[h!]
    \centering
    \setlength{\tabcolsep}{1pt}
    \begin{tabular}{cccccccc}
        \multicolumn{2}{c}{Target} & & & &
        \\
        \multicolumn{2}{c}{\includegraphics[width=0.25\linewidth]{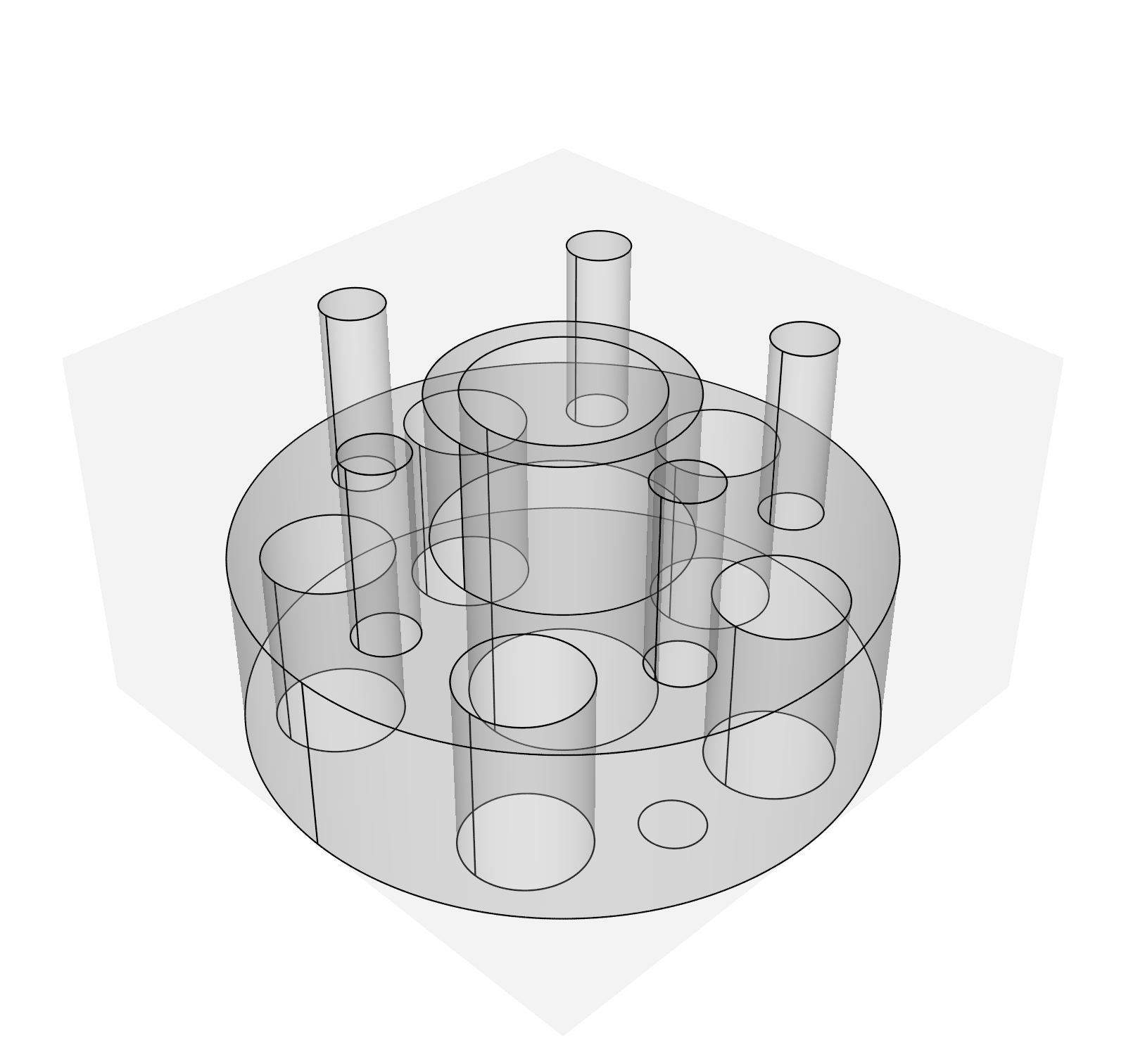}} & & & &
        \\
        \multicolumn{1}{l}{InverseCSG} & & & & &
        \\
        \includegraphics[width=0.124\linewidth]{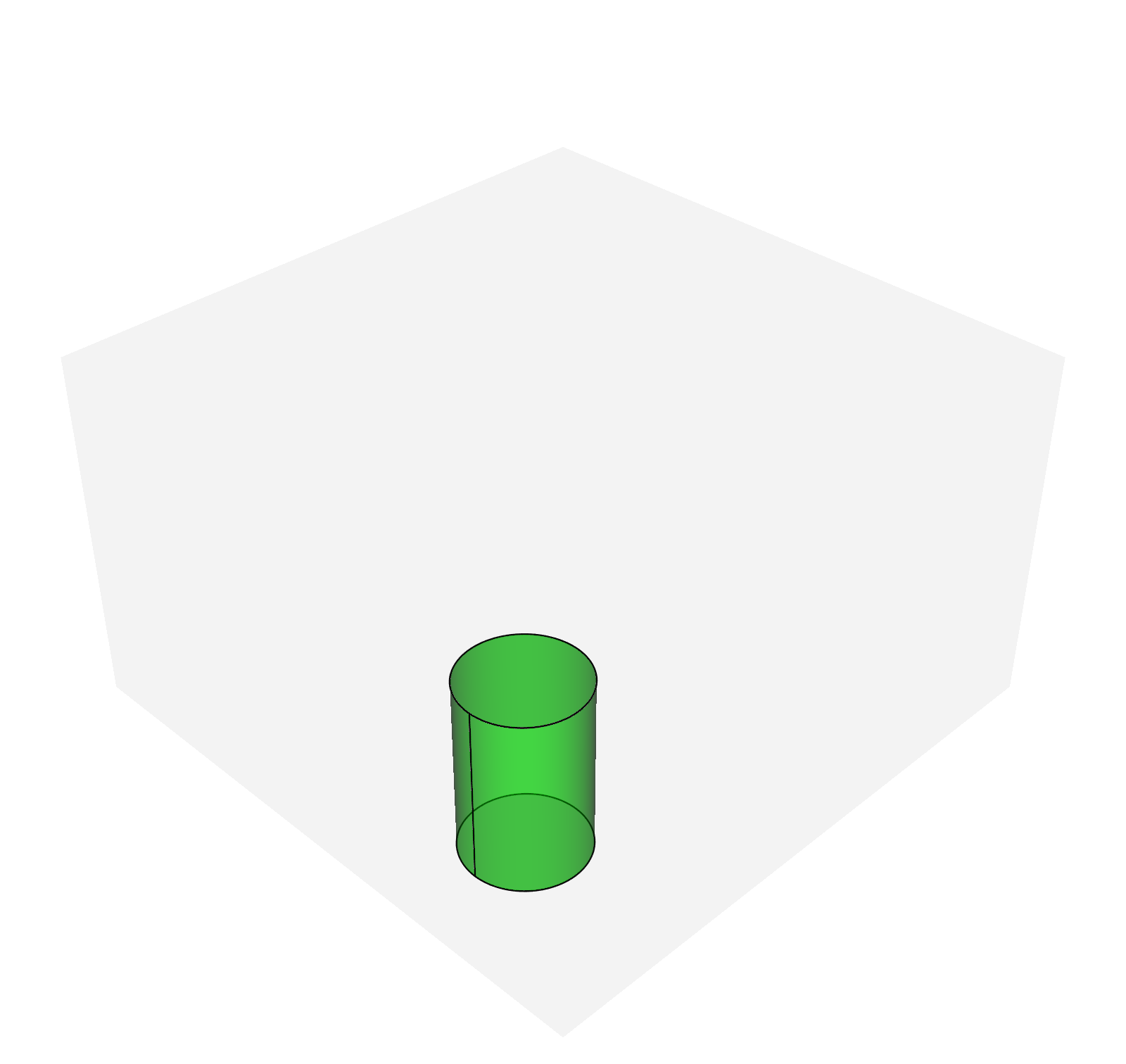} &
        \includegraphics[width=0.124\linewidth]{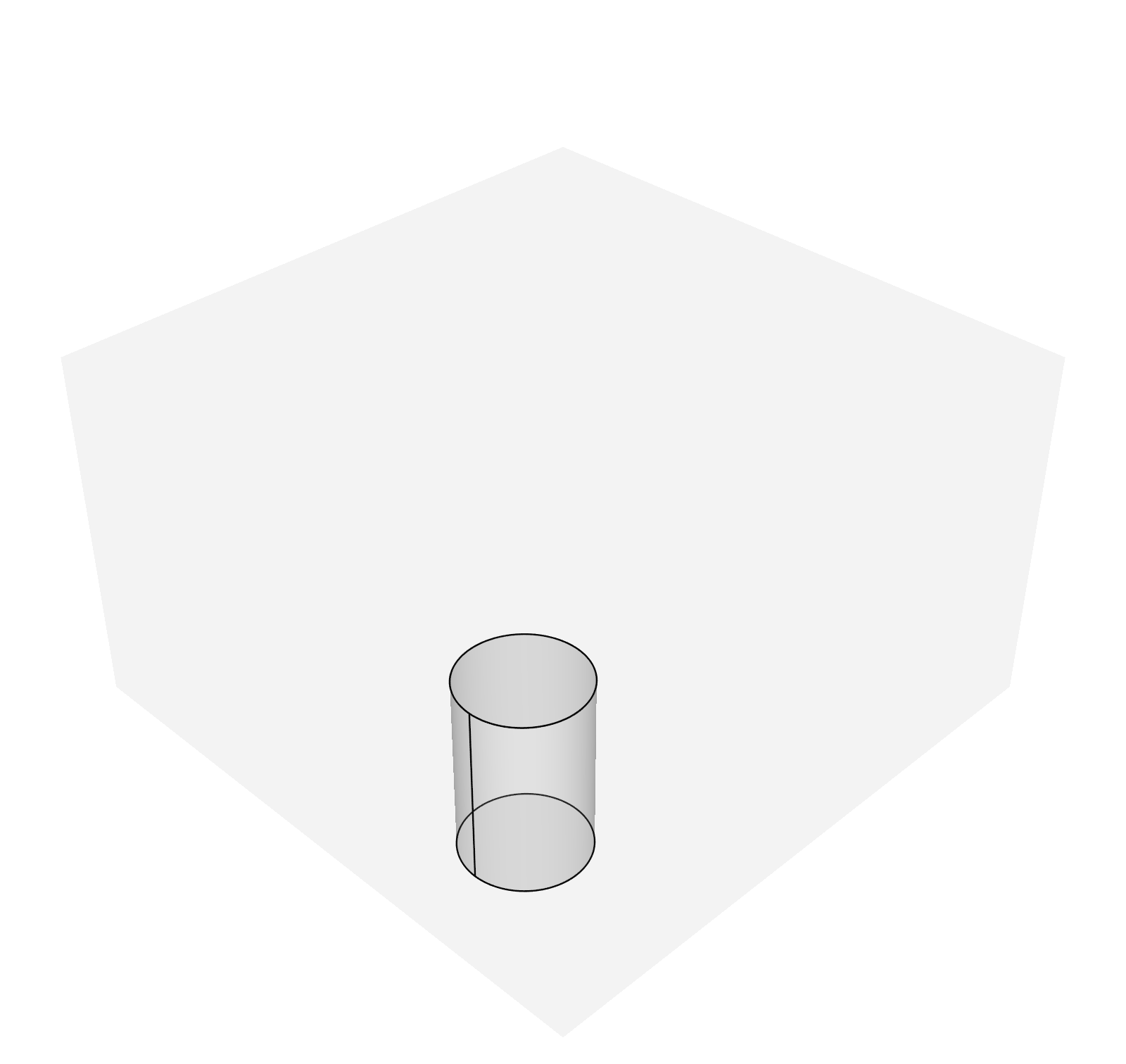} &
        \includegraphics[width=0.124\linewidth]{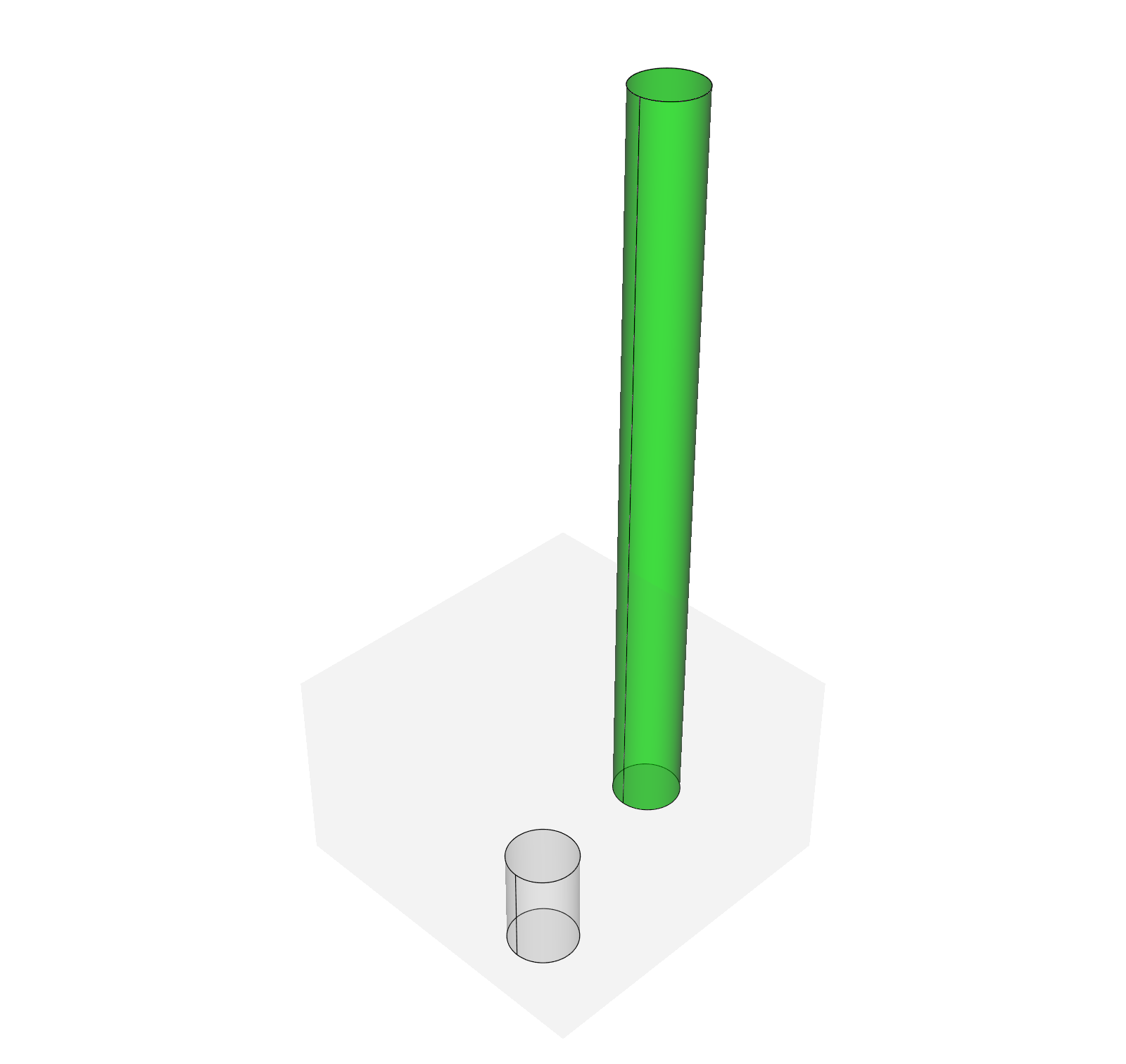} &
        \includegraphics[width=0.124\linewidth]{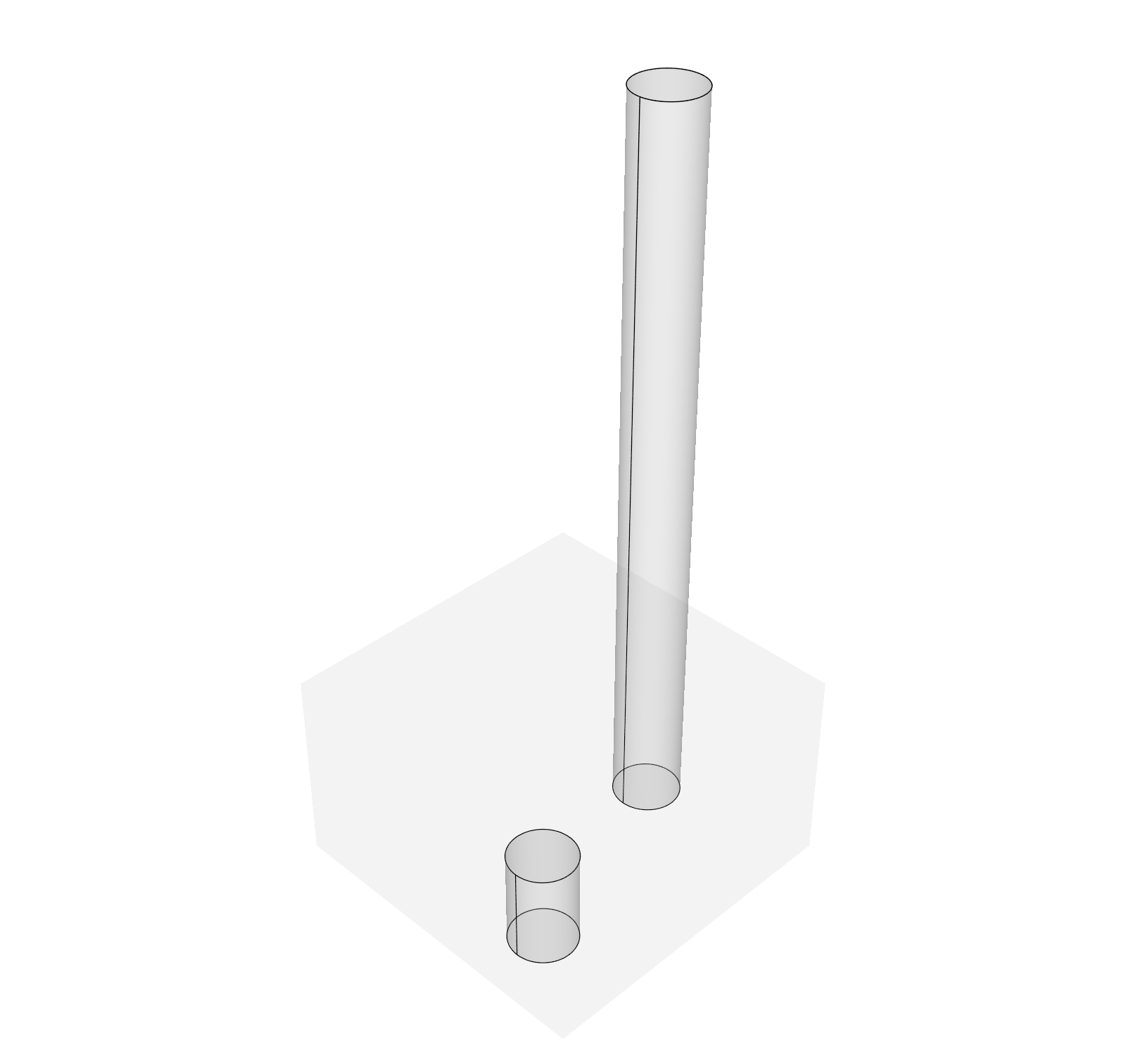} &
        \includegraphics[width=0.124\linewidth]{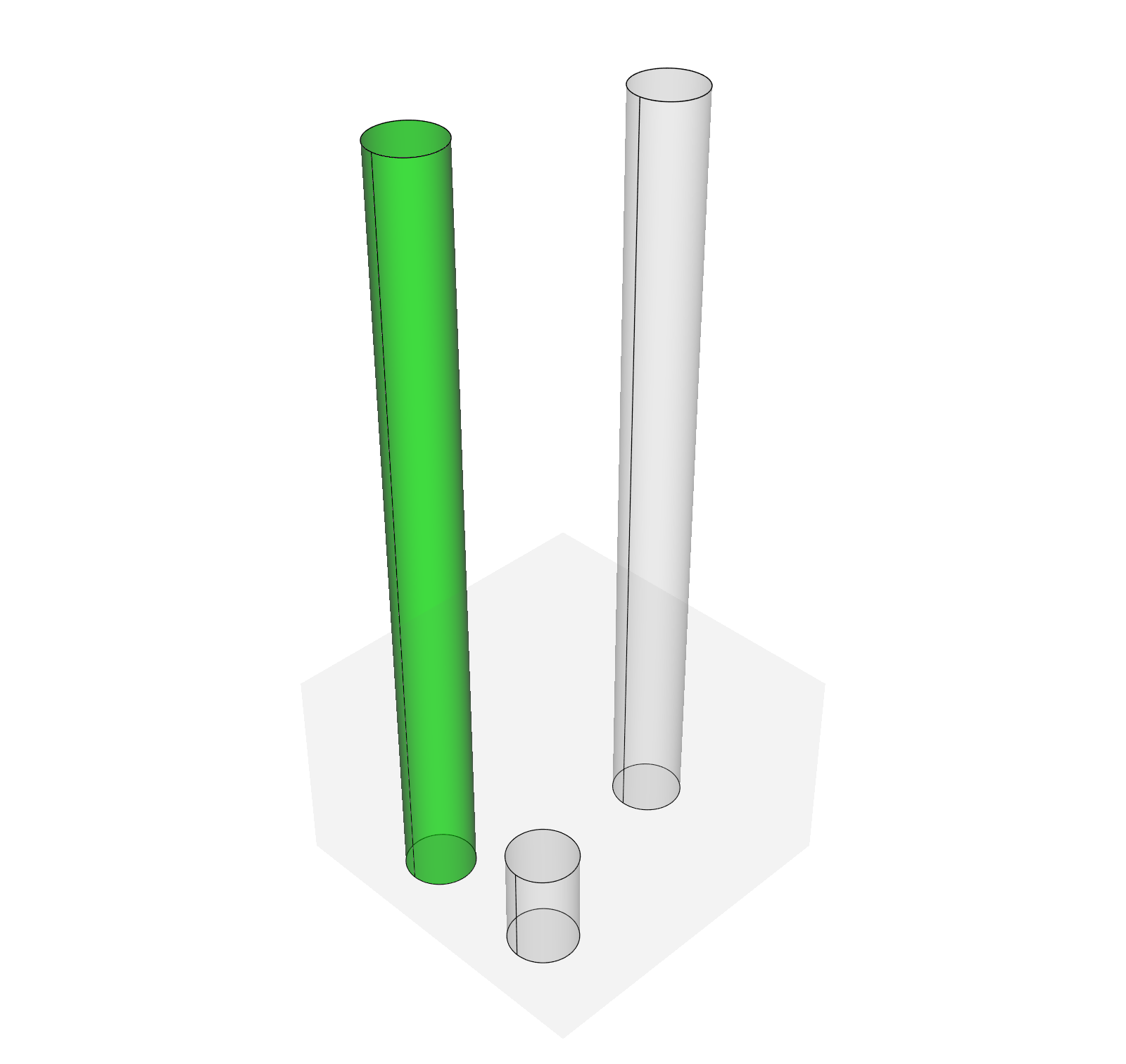} &
        \includegraphics[width=0.124\linewidth]{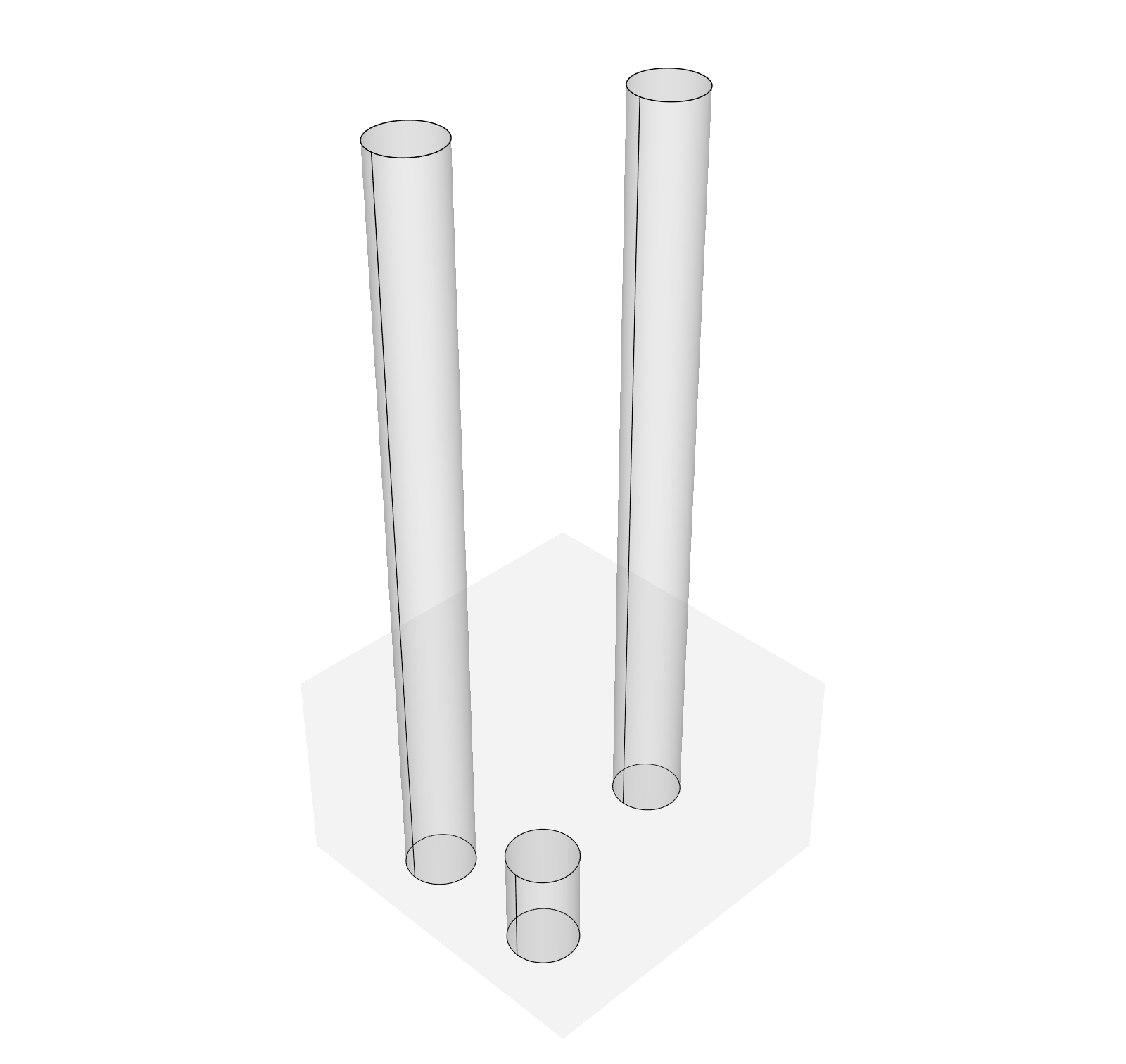} &
        \includegraphics[width=0.124\linewidth]{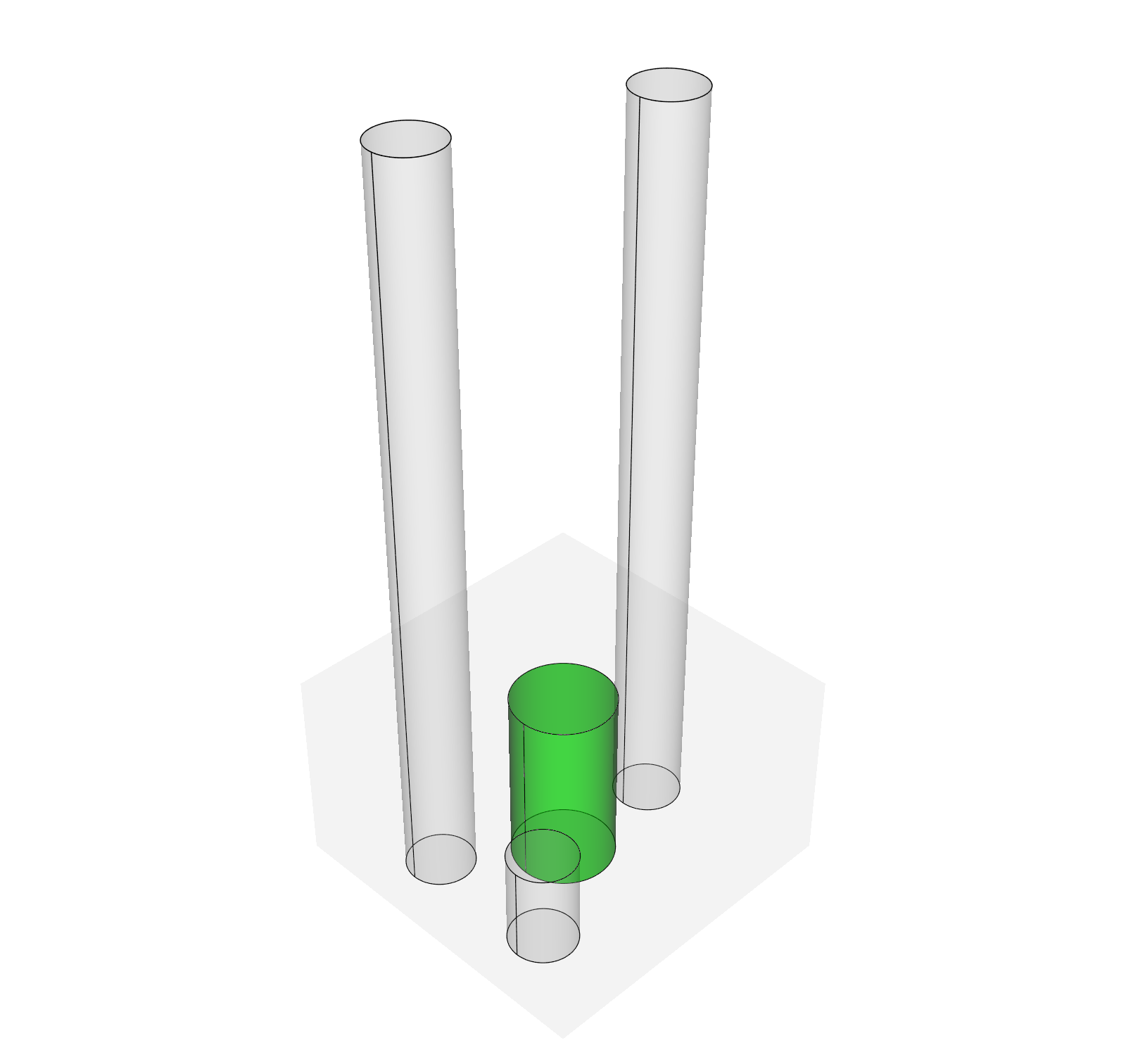} &
        \includegraphics[width=0.124\linewidth]{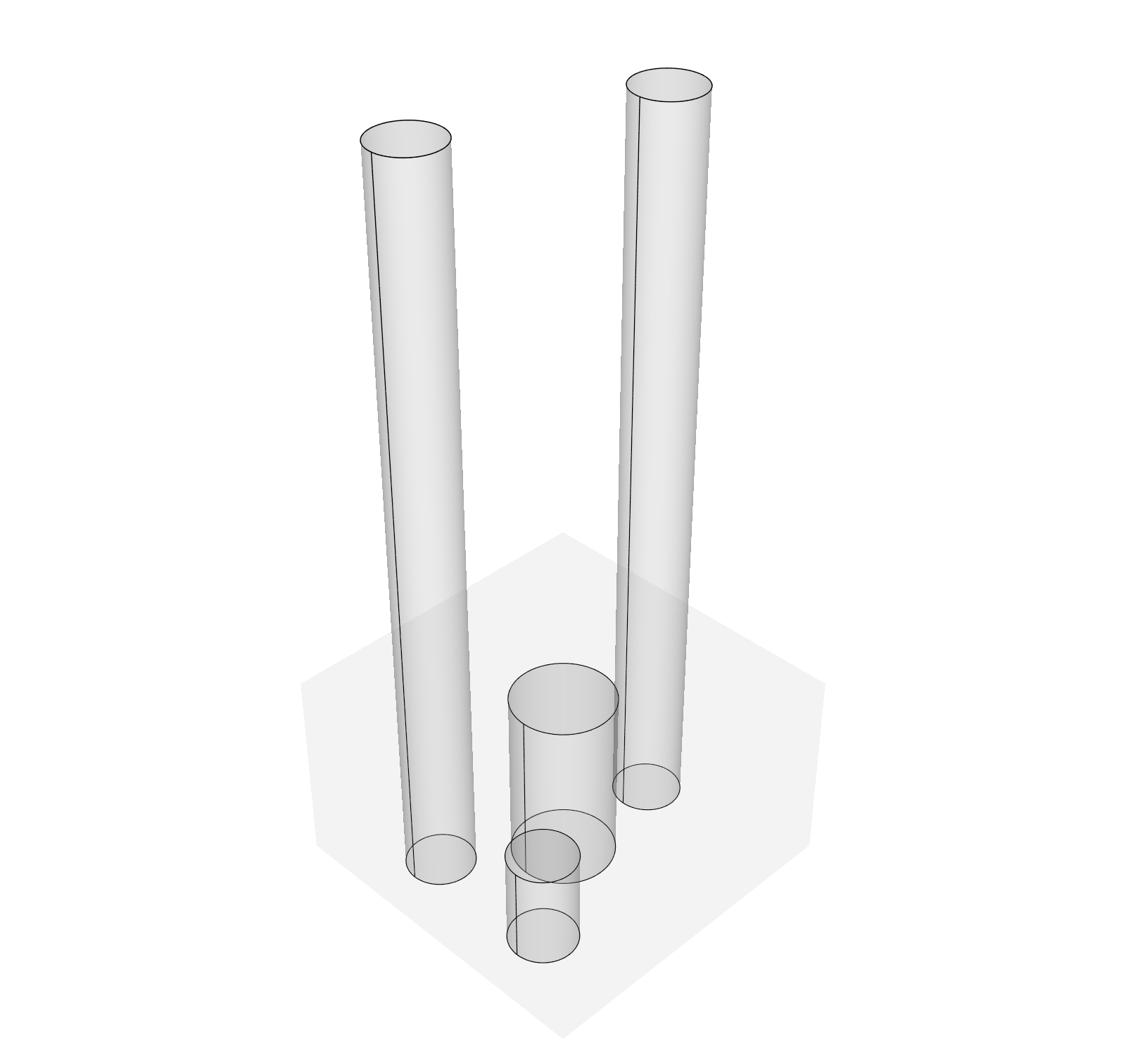}
        \\
        \includegraphics[width=0.124\linewidth]{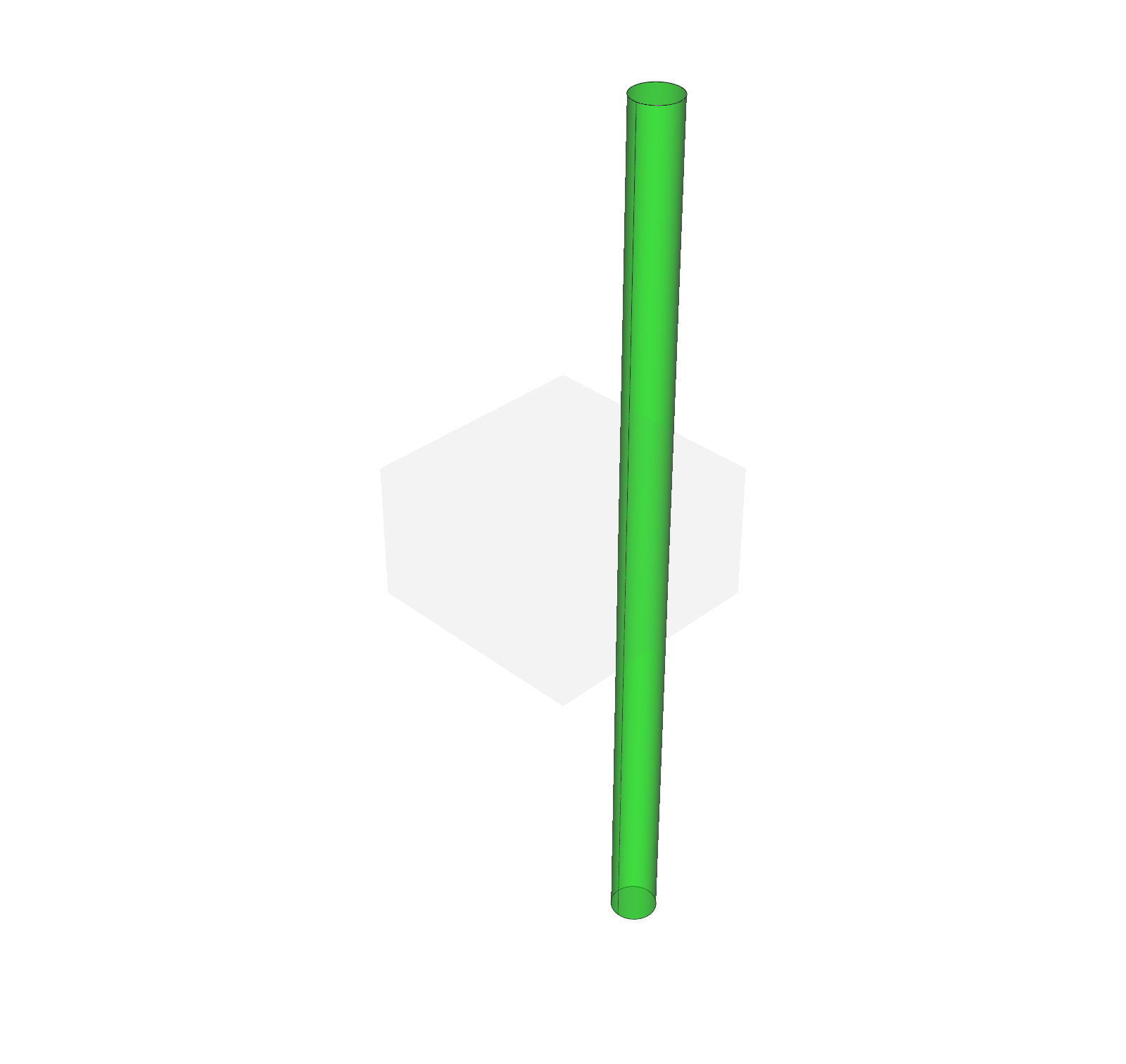} &
        \includegraphics[width=0.124\linewidth]{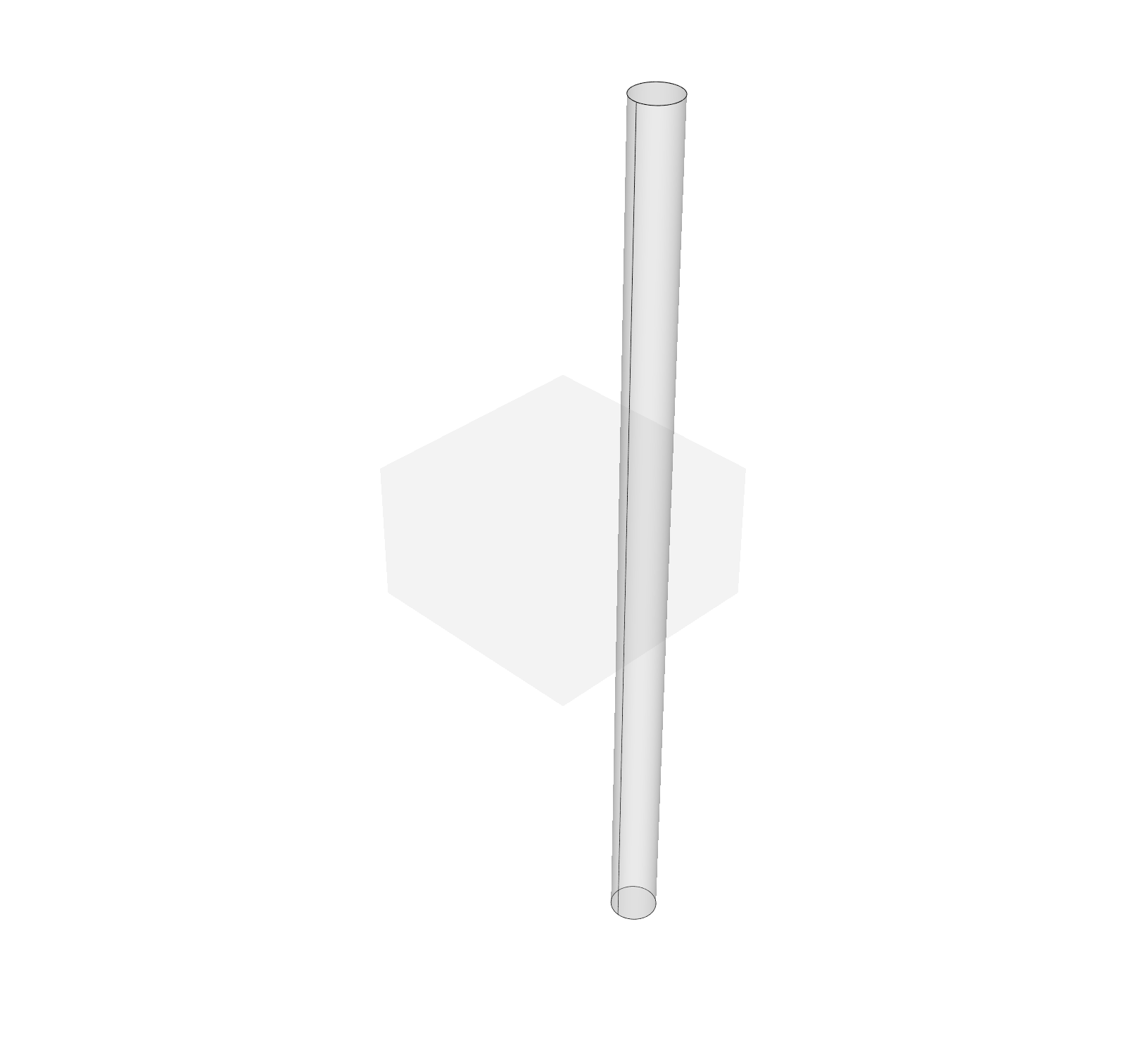} &
        \includegraphics[width=0.124\linewidth]{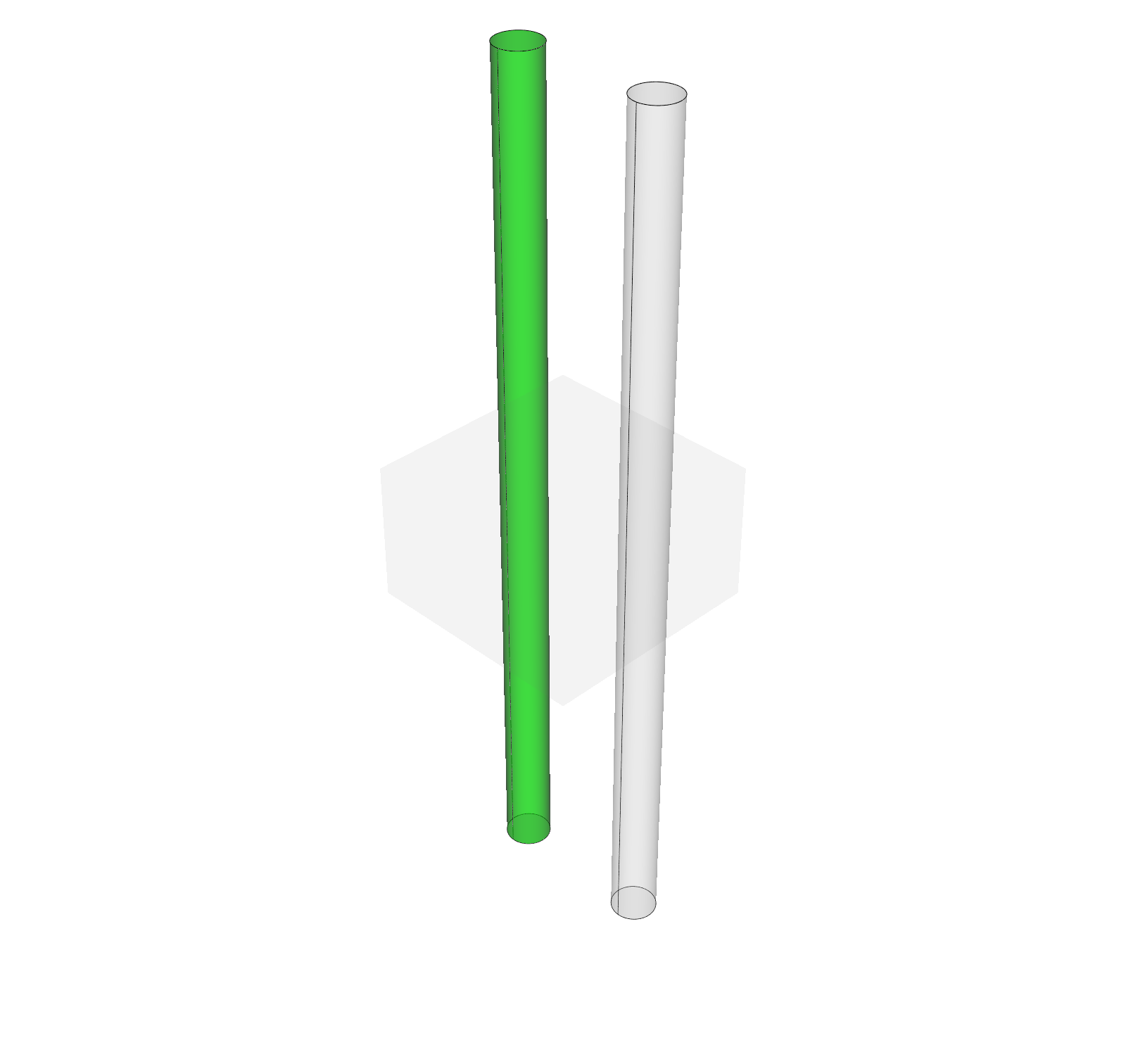} &
        \includegraphics[width=0.124\linewidth]{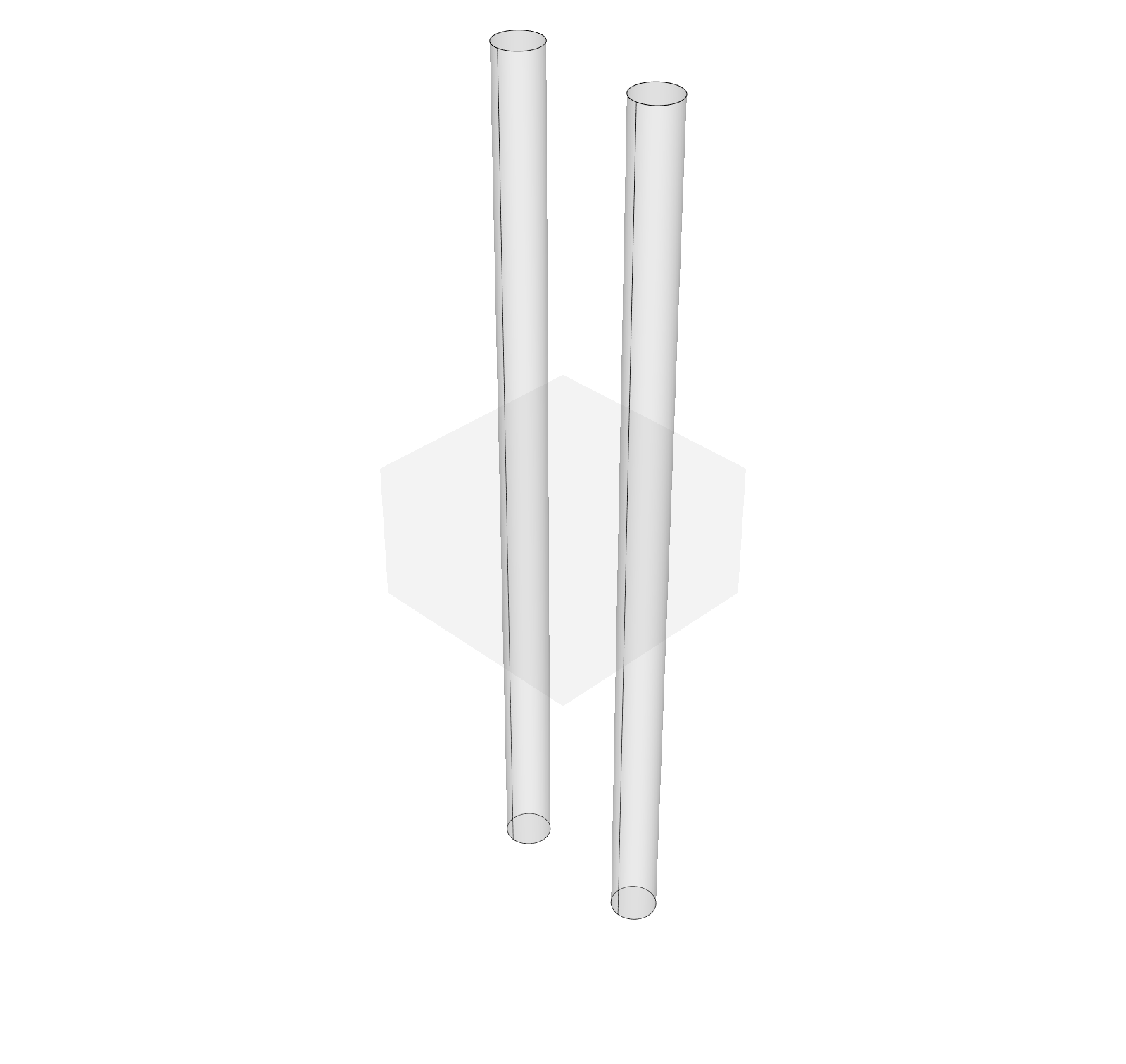} &
        \includegraphics[width=0.124\linewidth]{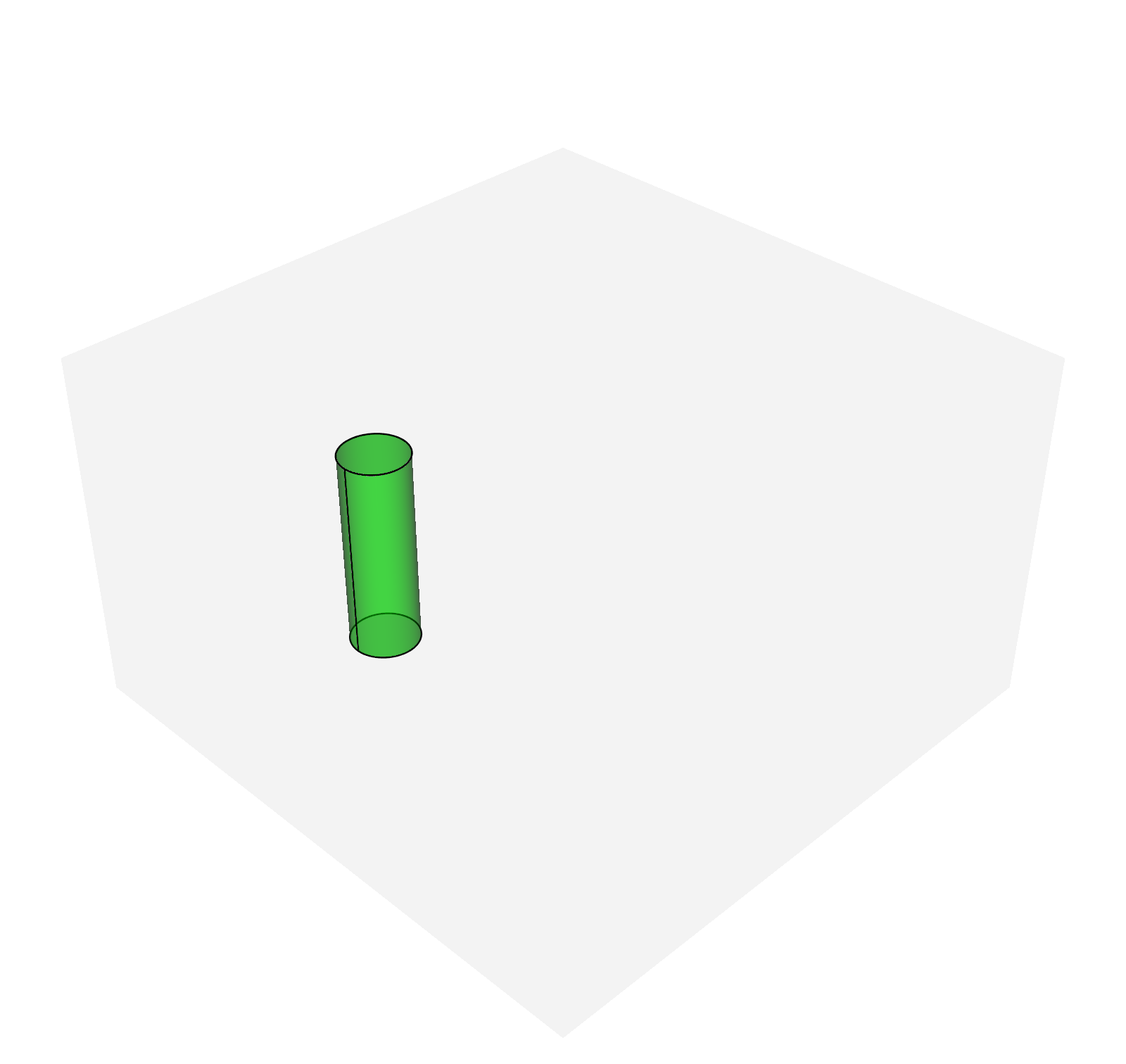} &
        \includegraphics[width=0.124\linewidth]{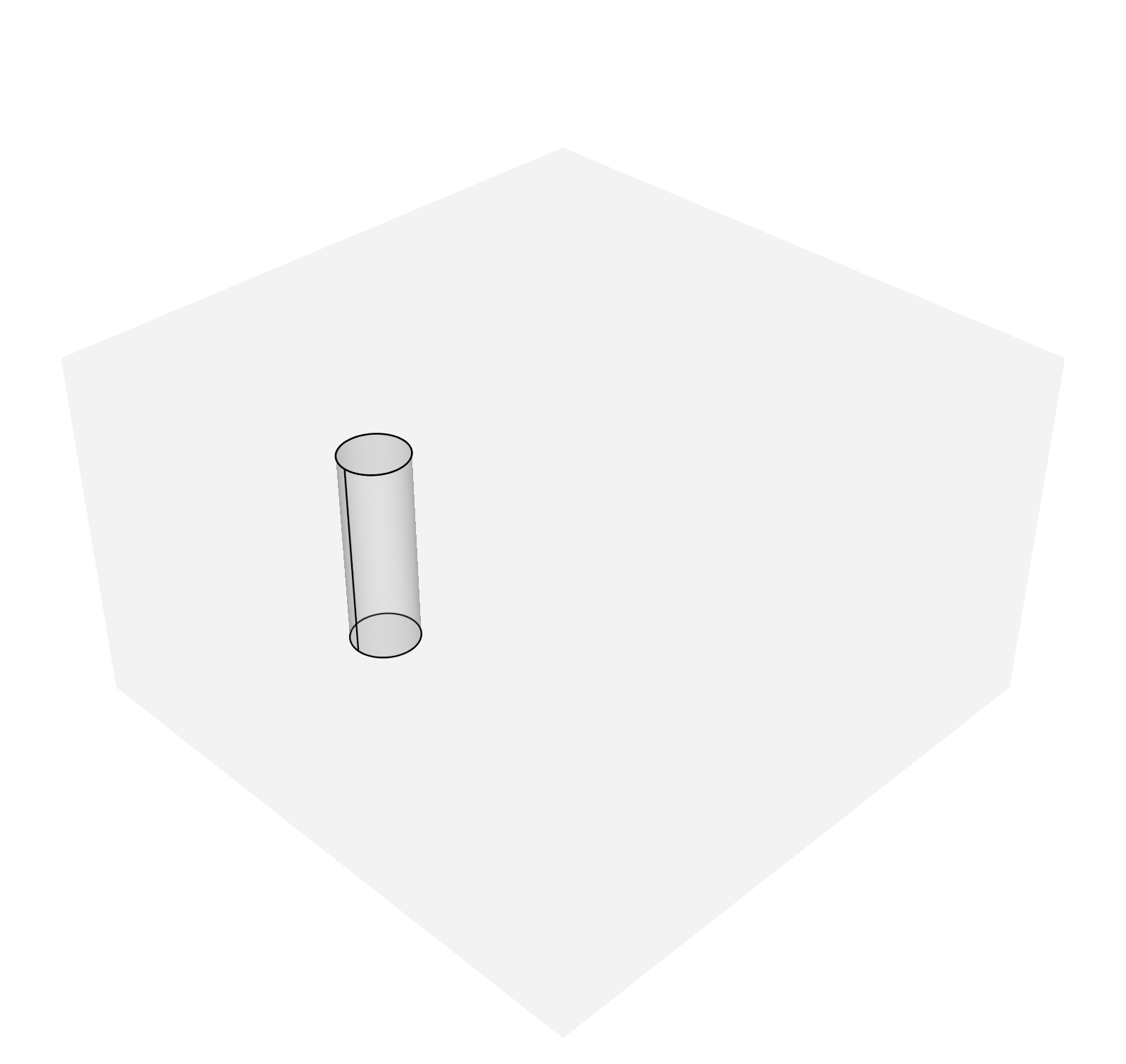} &
        \includegraphics[width=0.124\linewidth]{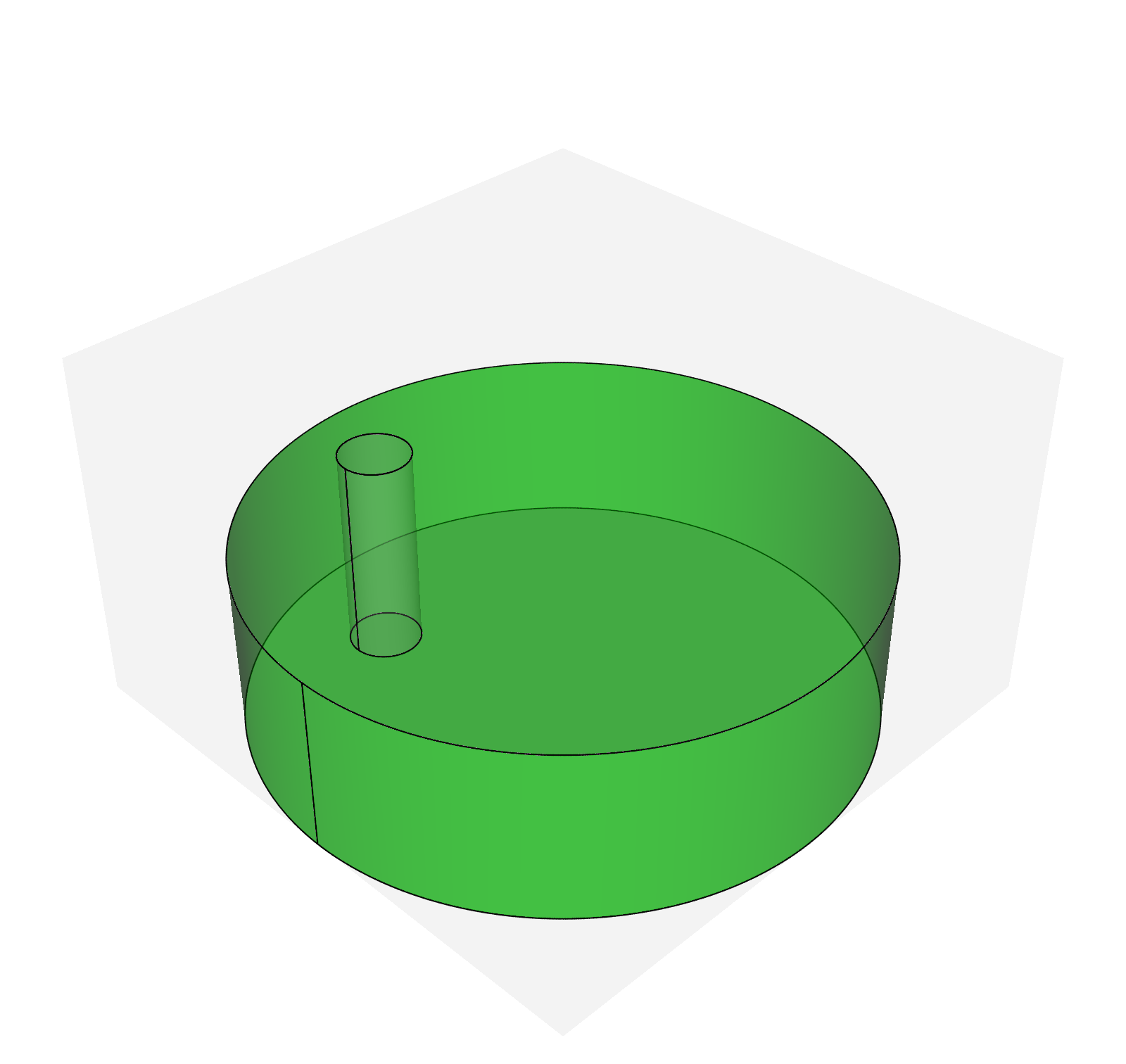} &
        \includegraphics[width=0.124\linewidth]{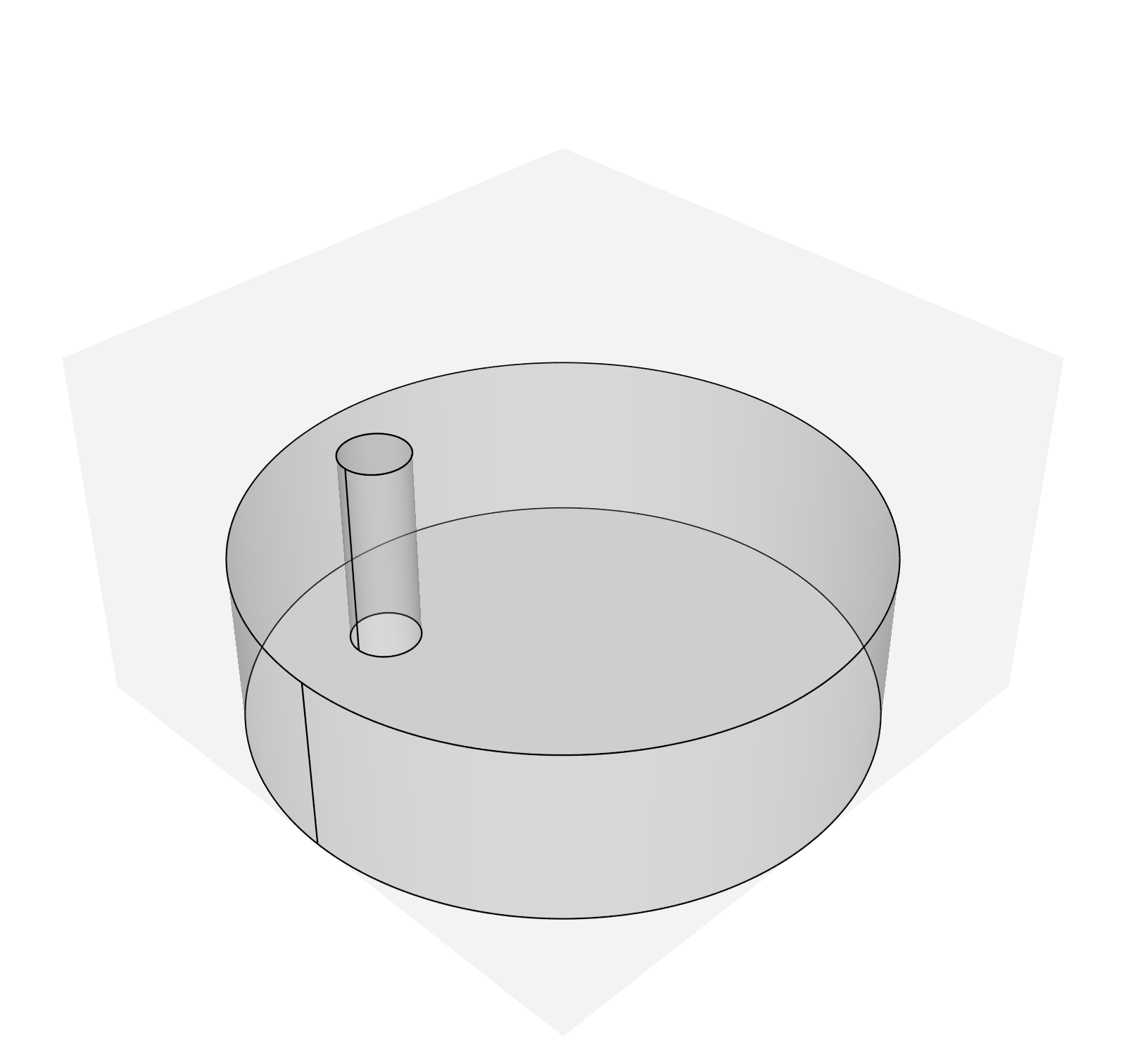}
        \\
        \includegraphics[width=0.124\linewidth]{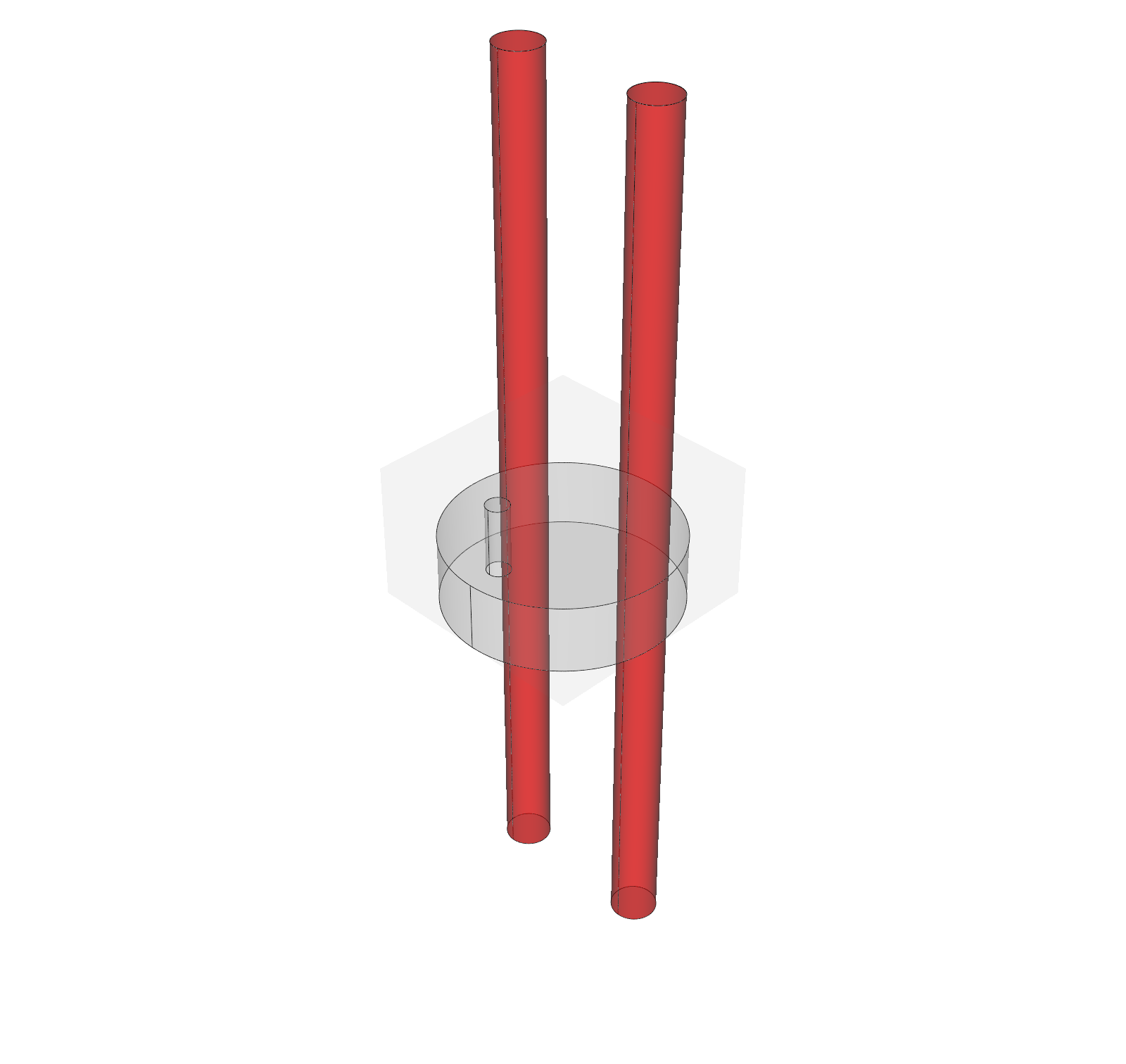} &
        \includegraphics[width=0.124\linewidth]{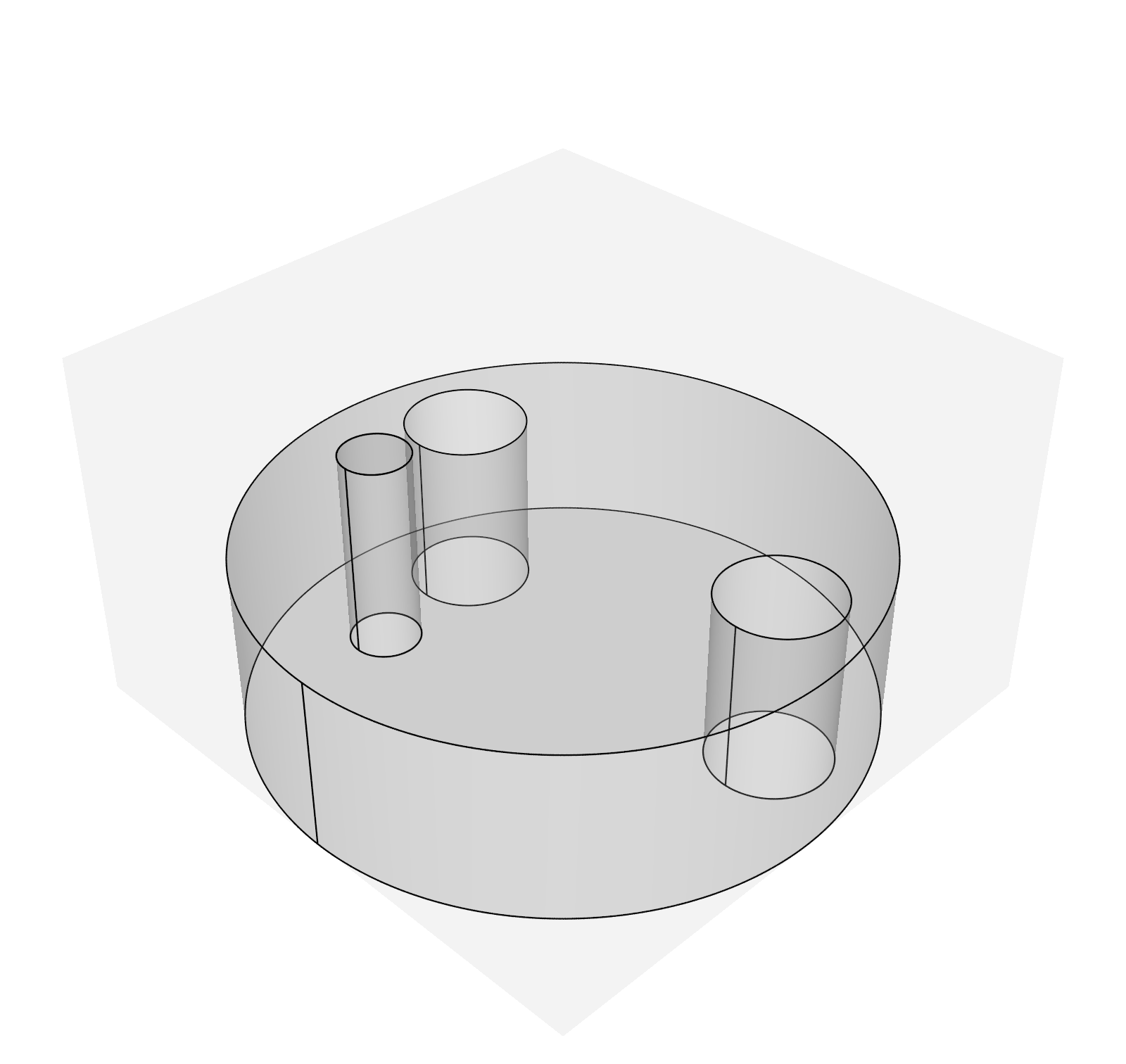} &
        \includegraphics[width=0.124\linewidth]{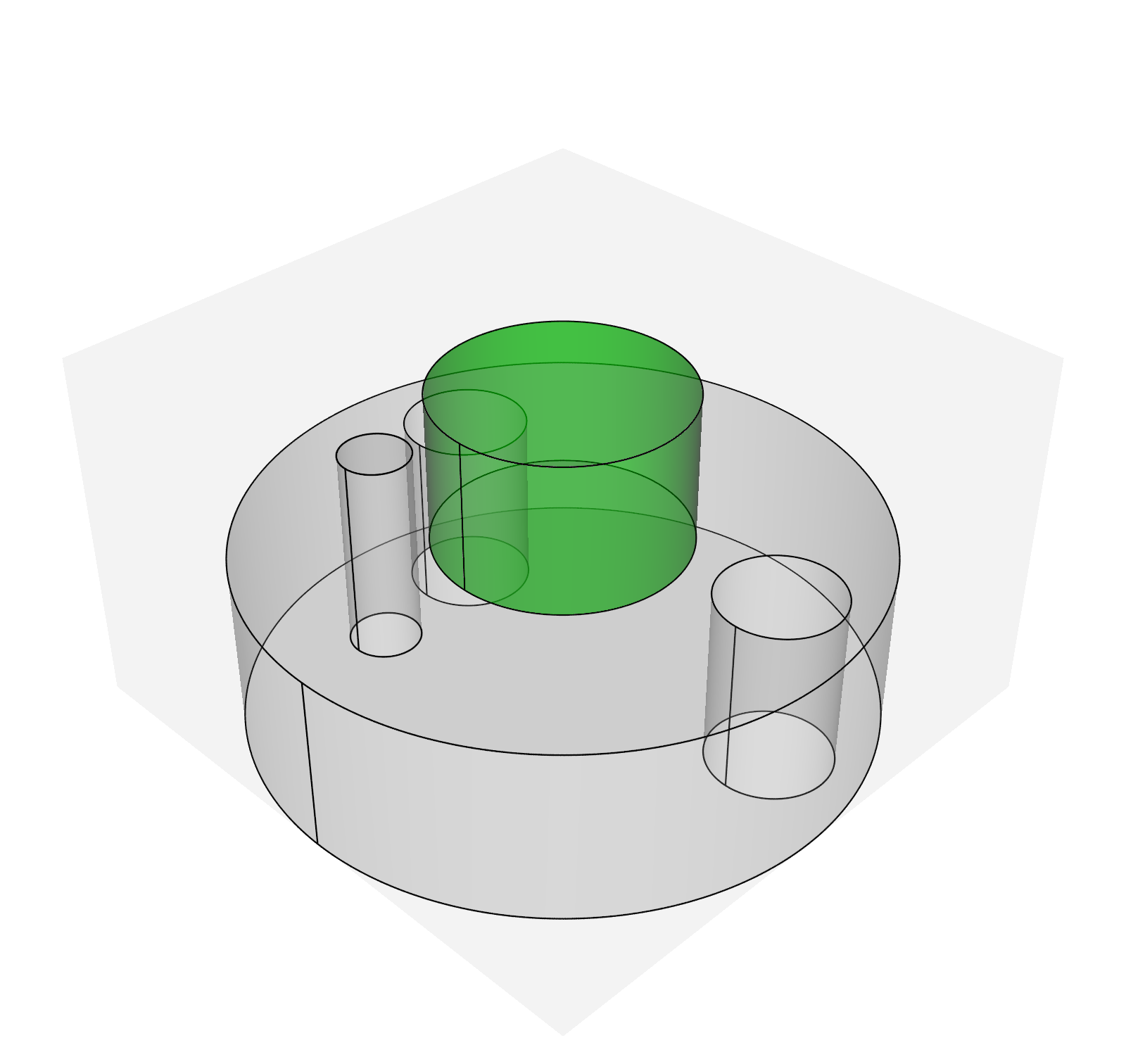} &
        \includegraphics[width=0.124\linewidth]{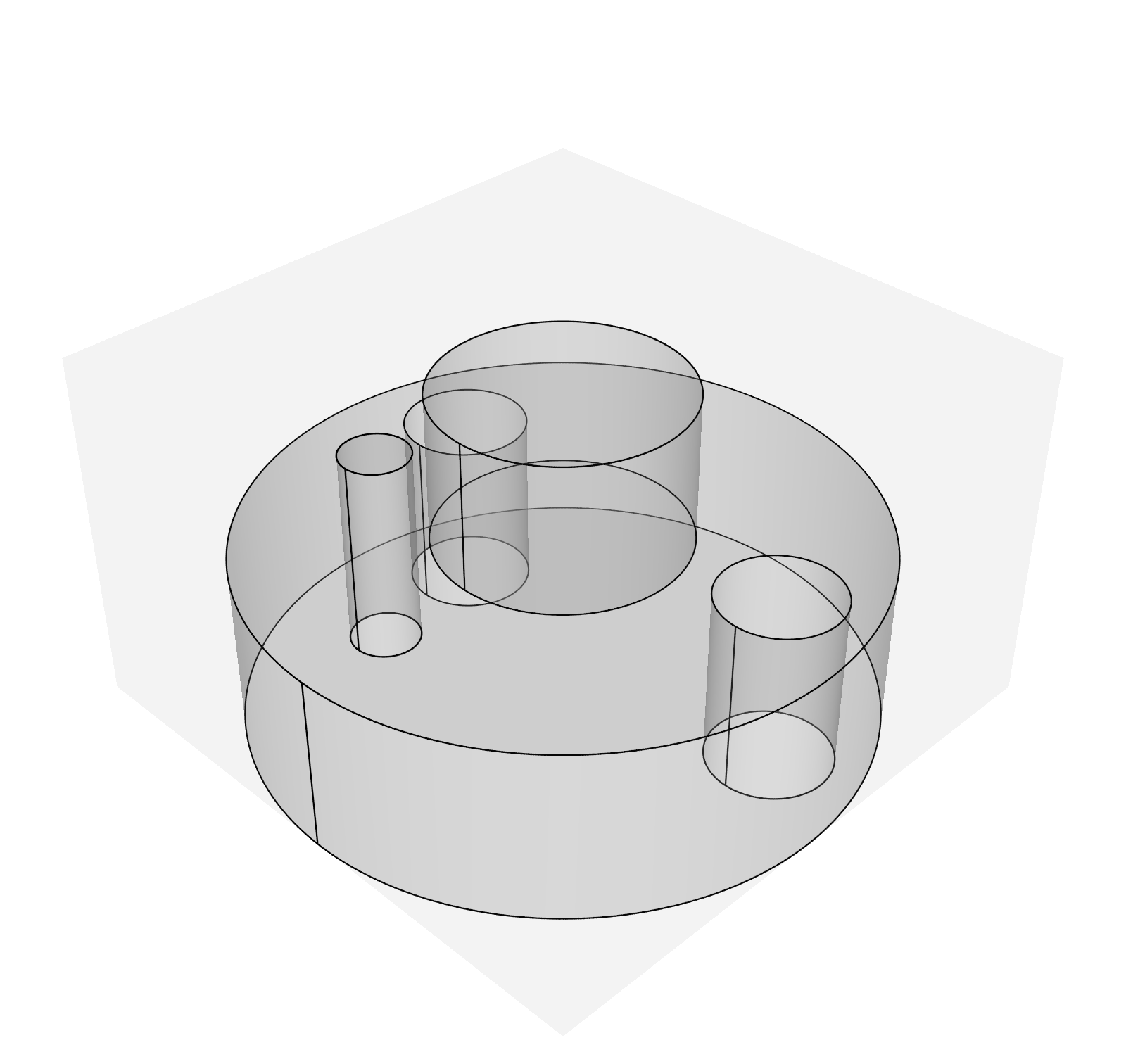} &
        \includegraphics[width=0.124\linewidth]{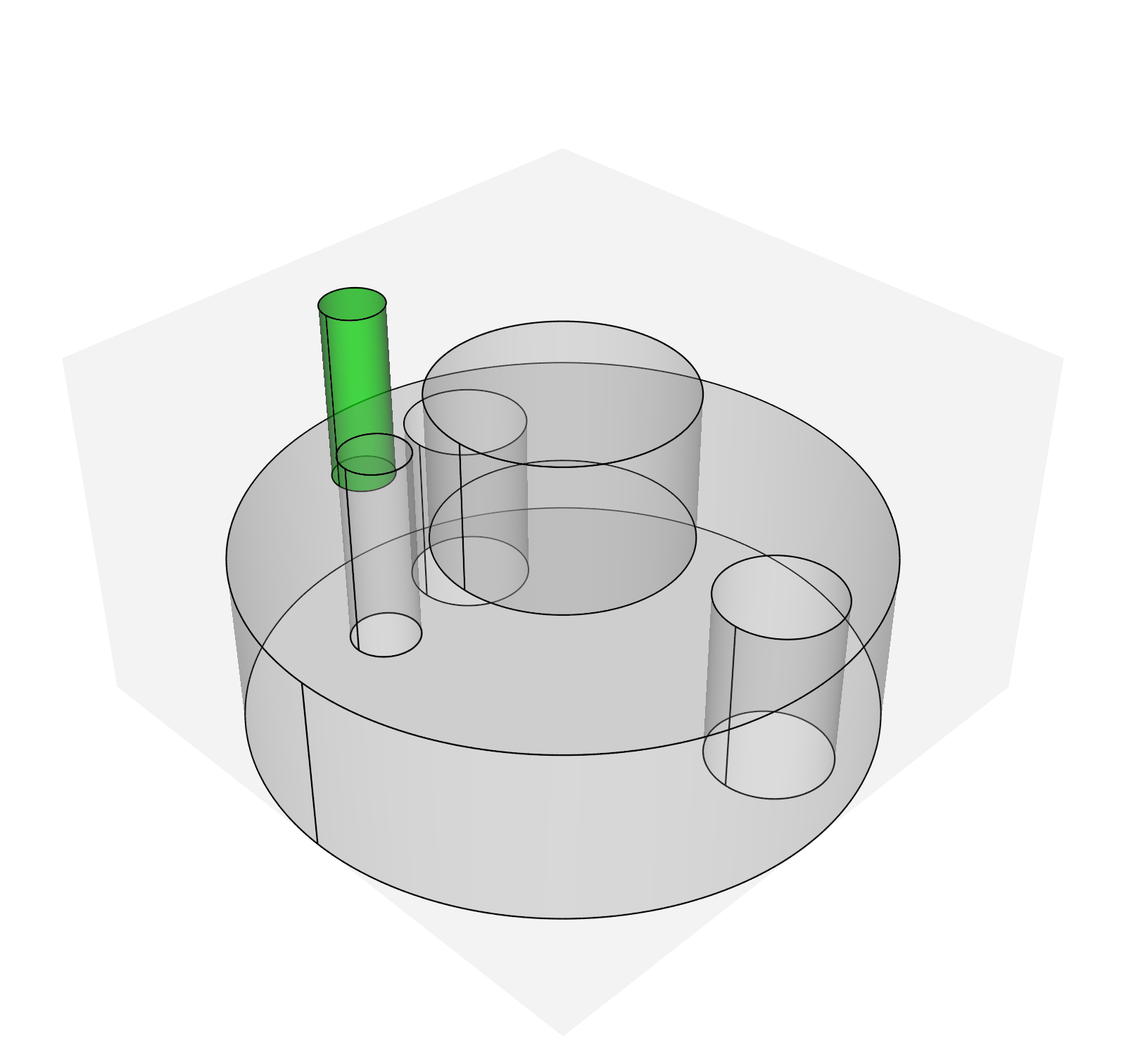} &
        \includegraphics[width=0.124\linewidth]{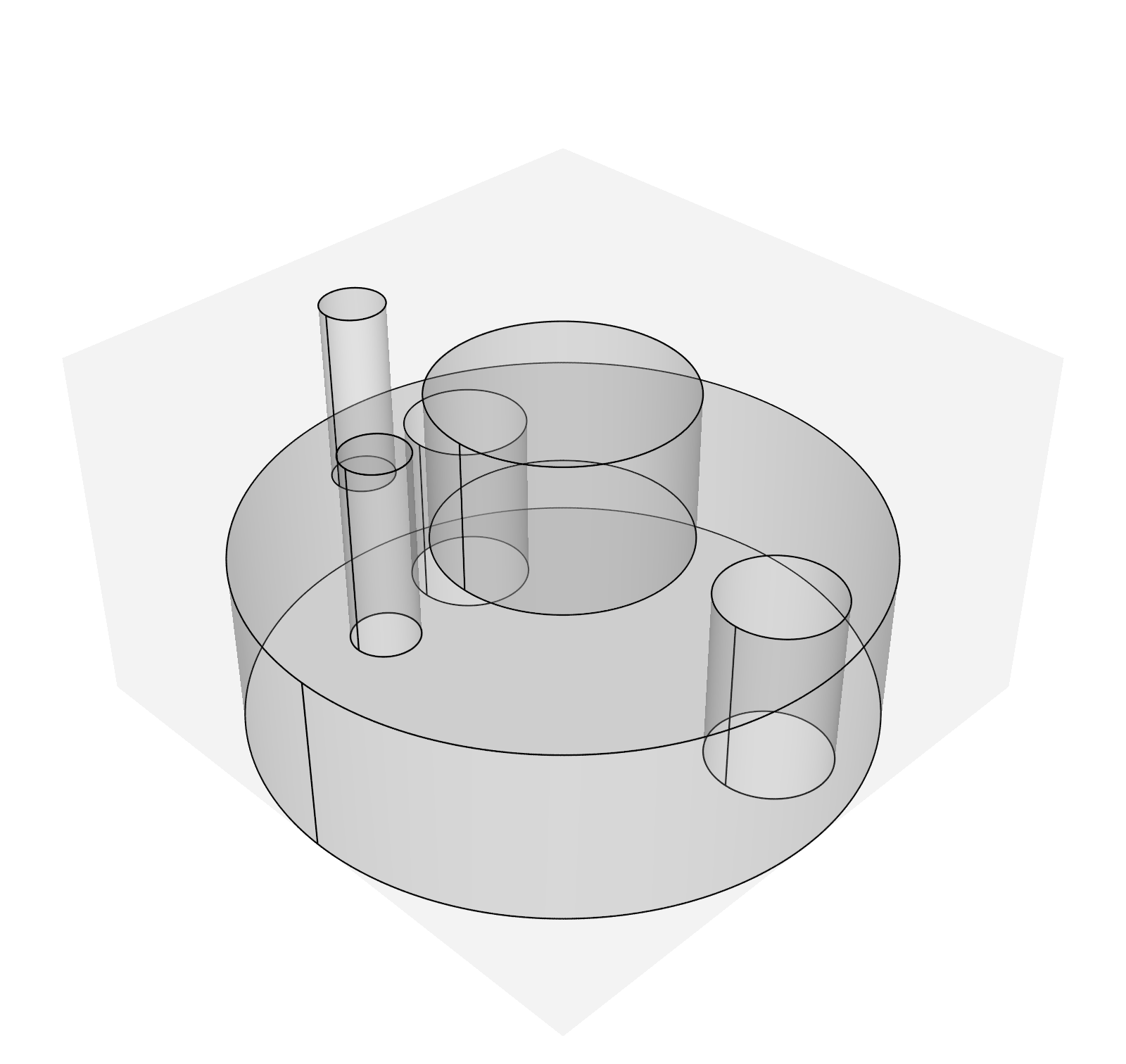} &
        \includegraphics[width=0.124\linewidth]{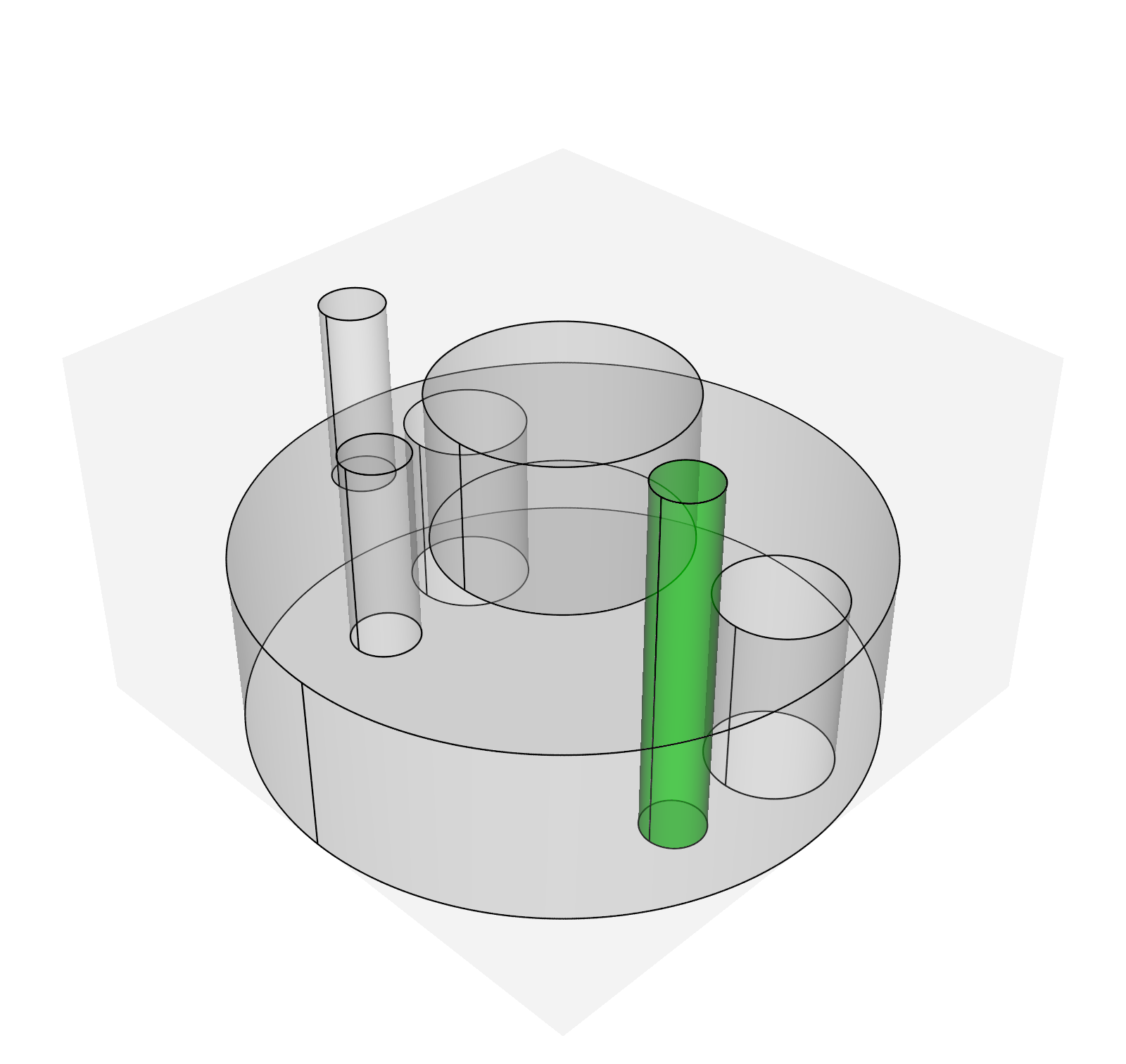} &
        \includegraphics[width=0.124\linewidth]{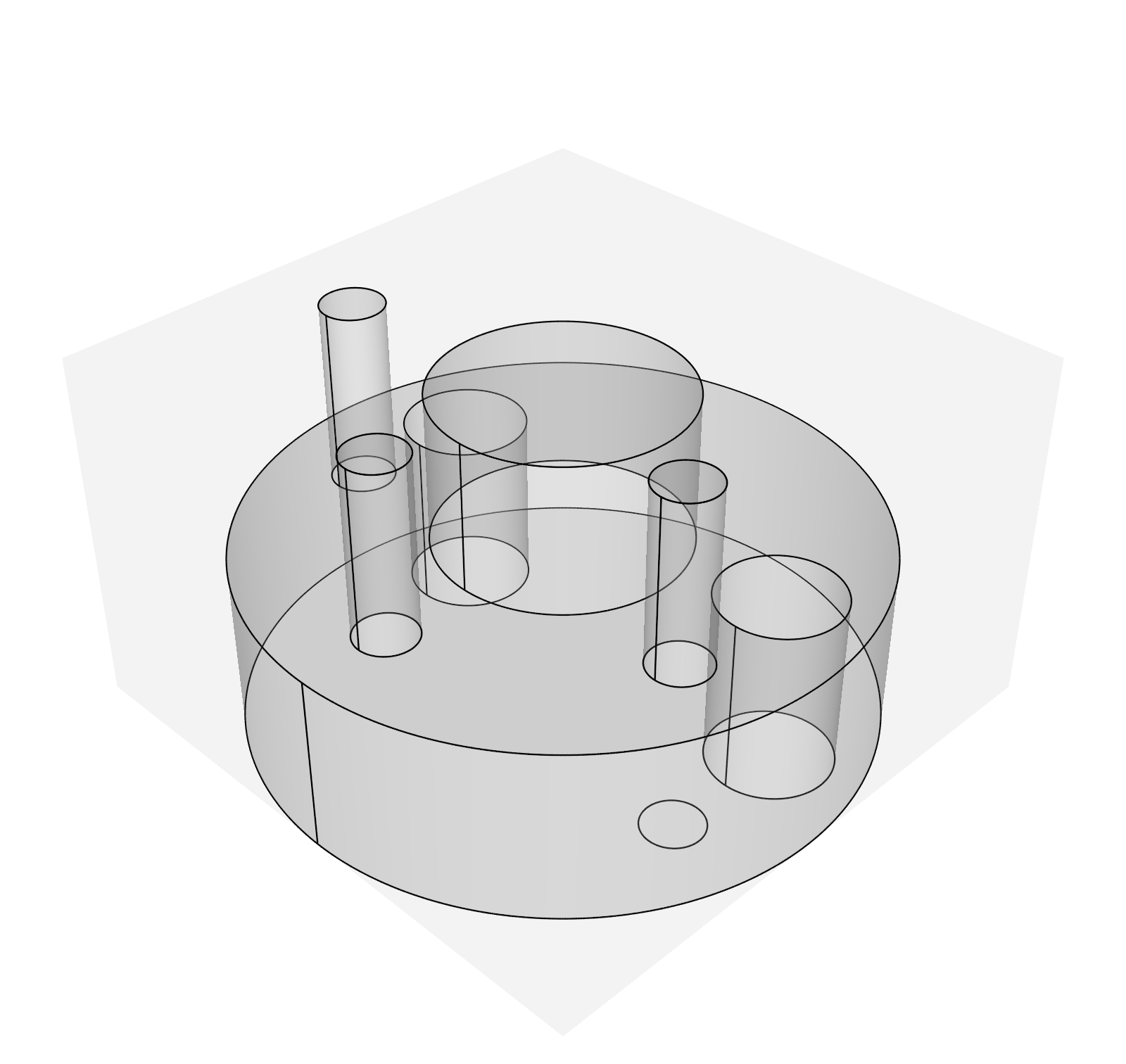}
        \\
        \includegraphics[width=0.124\linewidth]{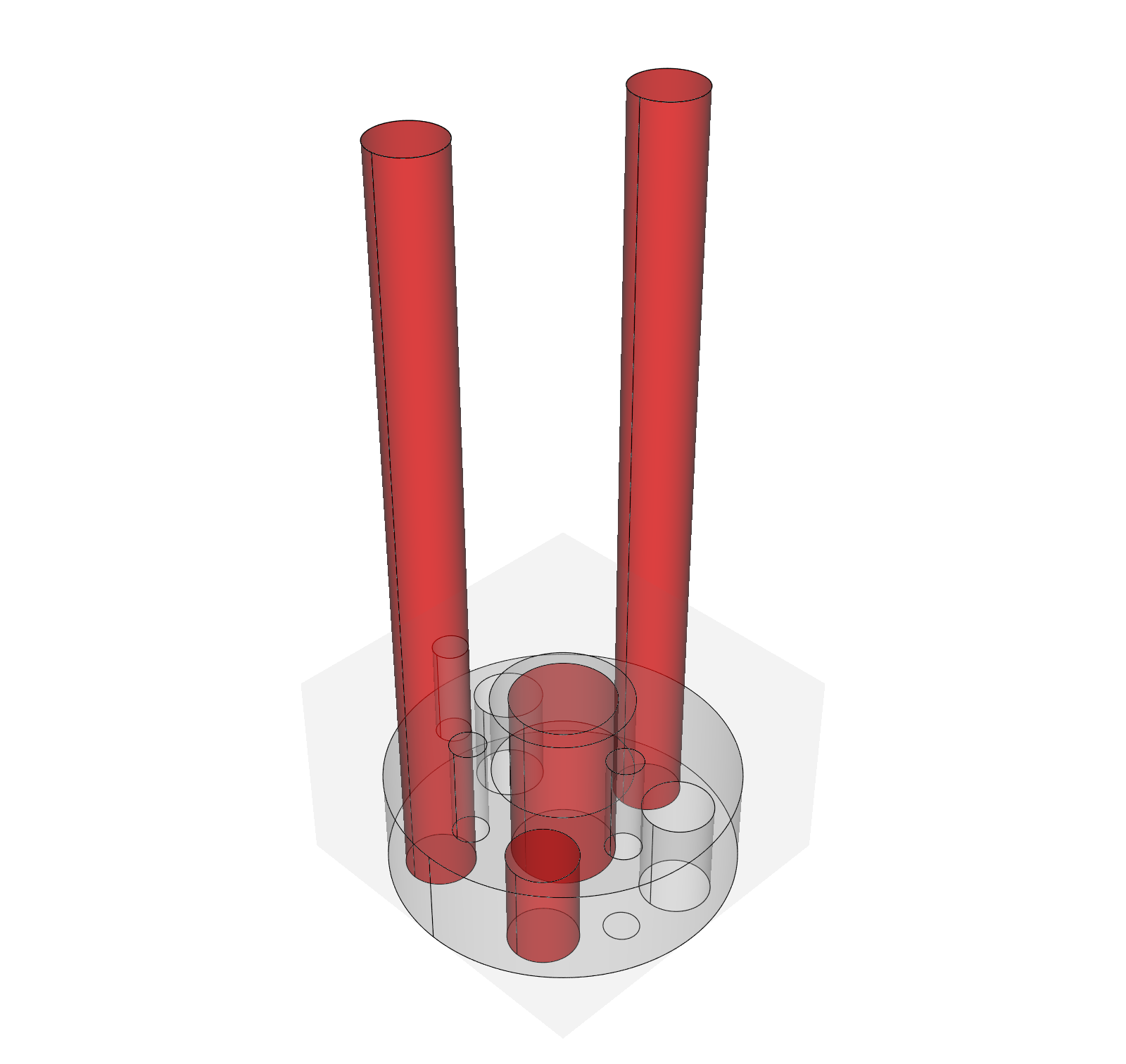} &
        \includegraphics[width=0.124\linewidth]{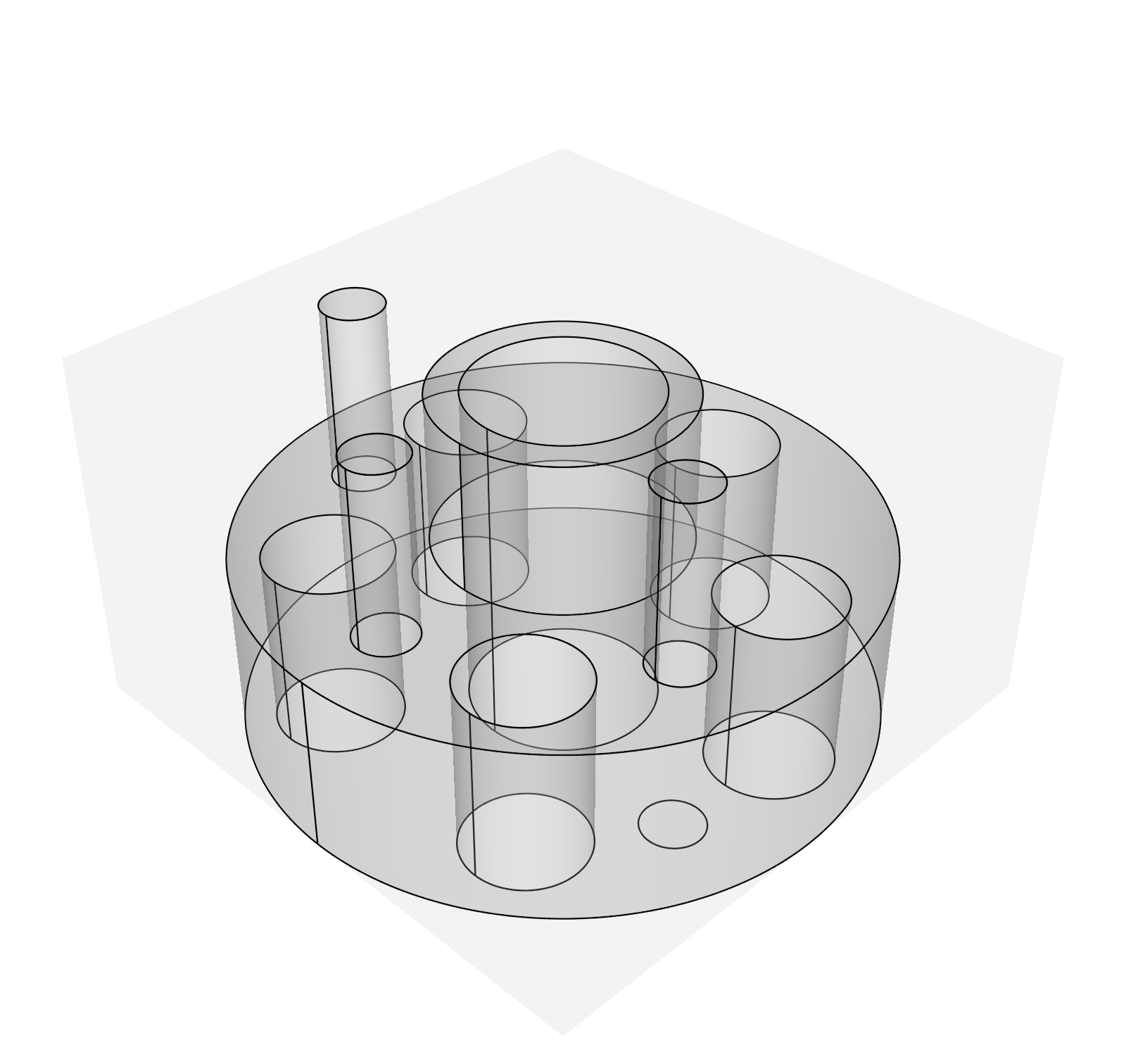} &
        \includegraphics[width=0.124\linewidth]{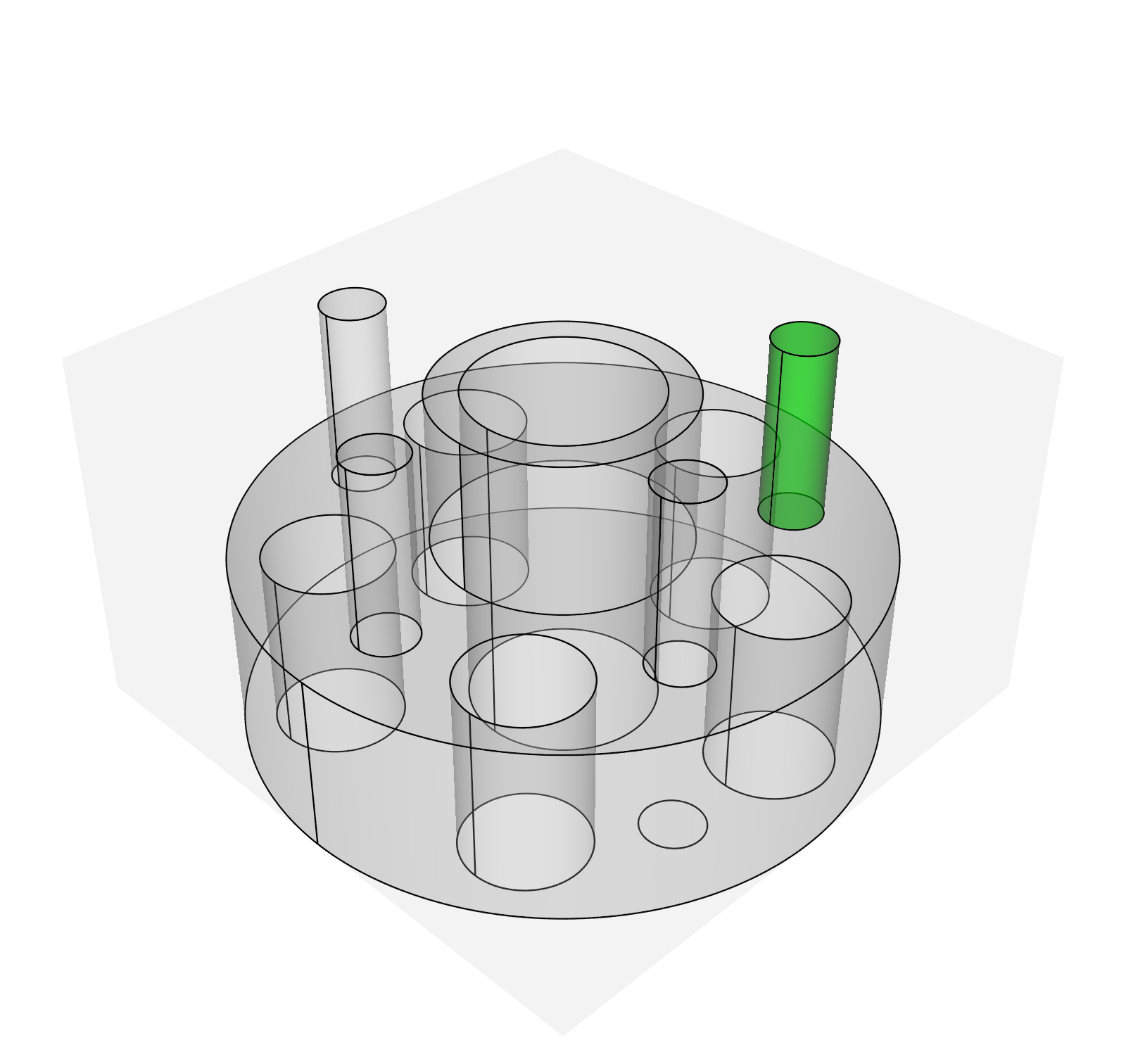} &
        \includegraphics[width=0.124\linewidth]{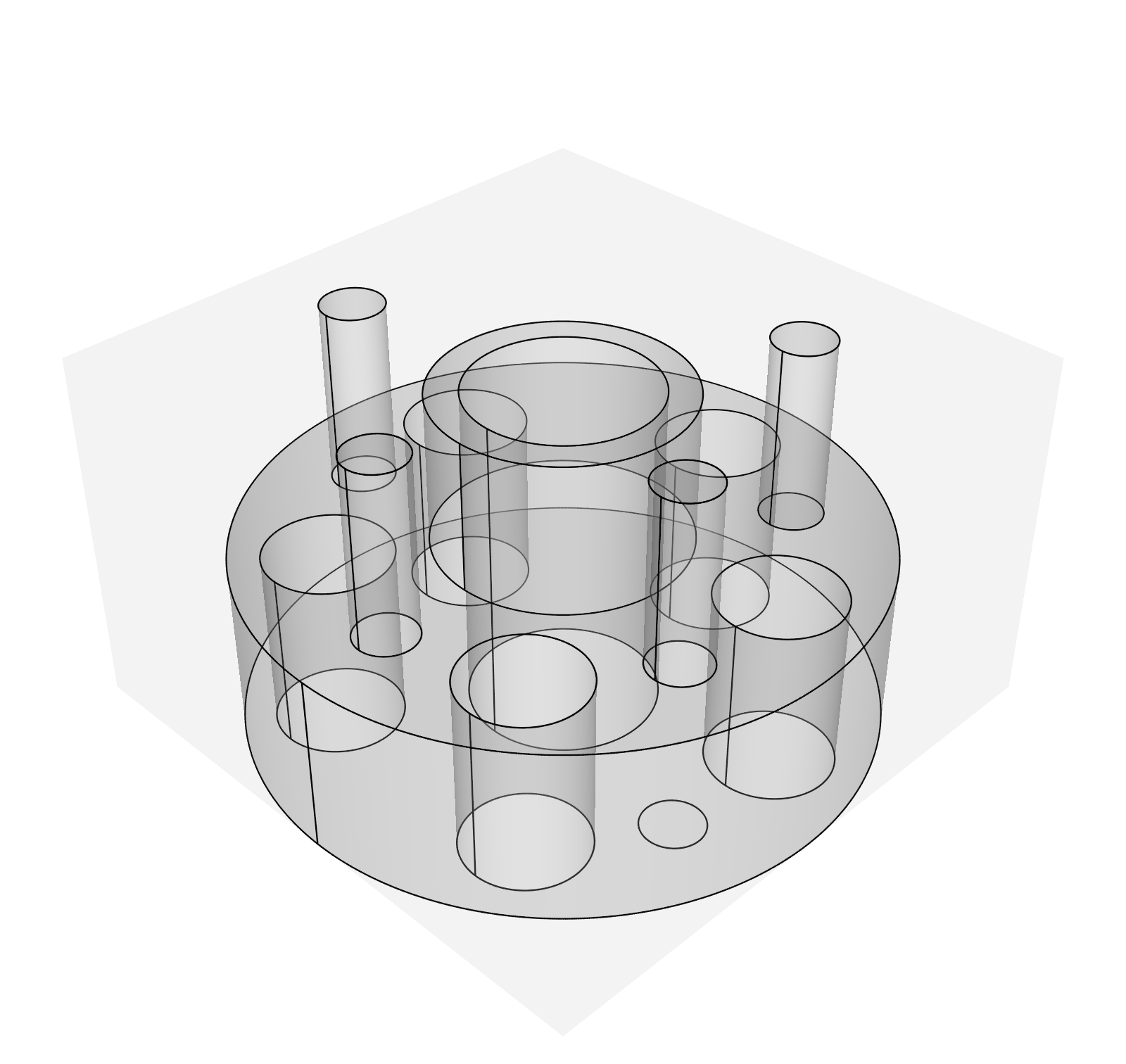} &
        \includegraphics[width=0.124\linewidth]{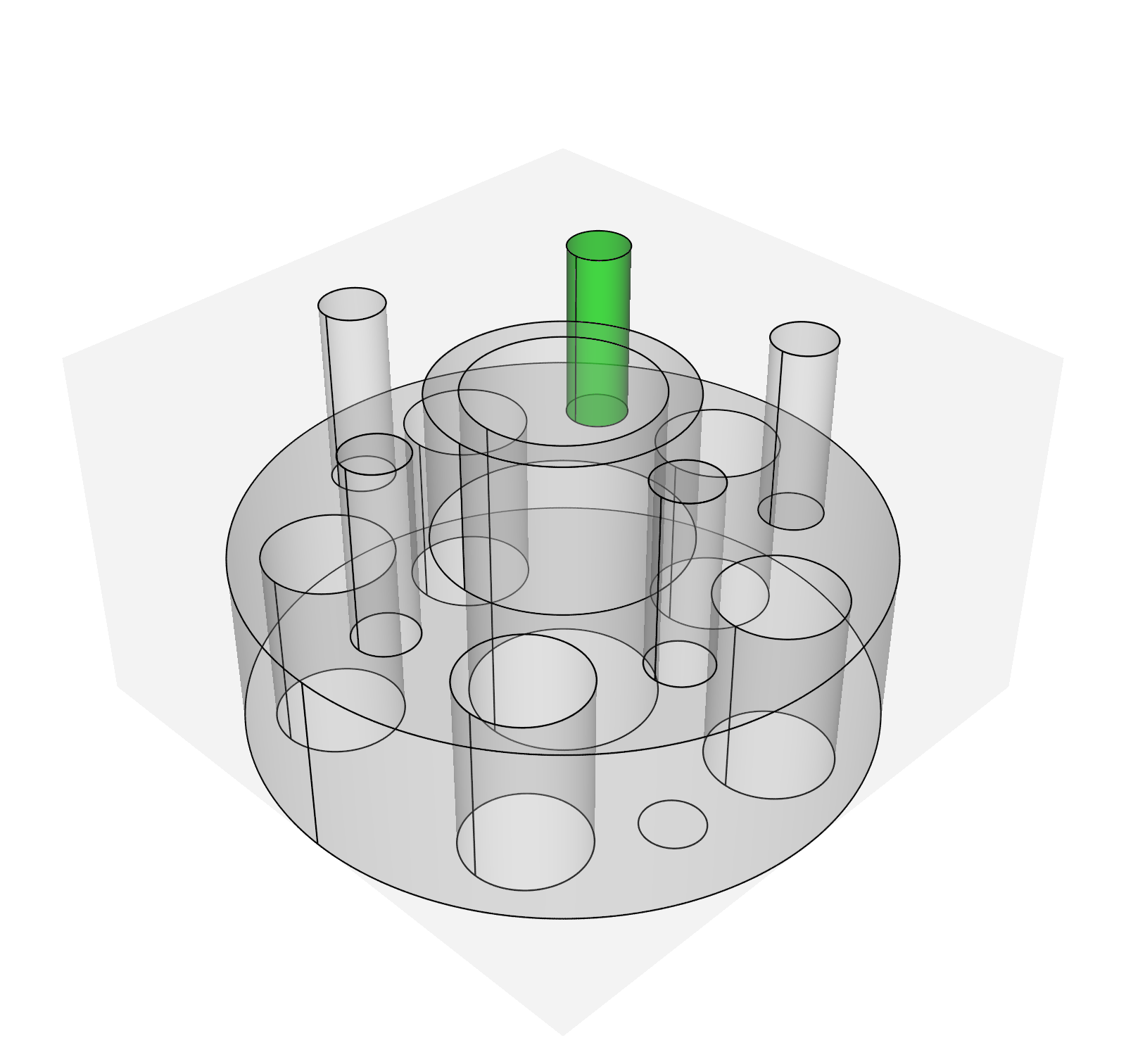} &
        \includegraphics[width=0.124\linewidth]{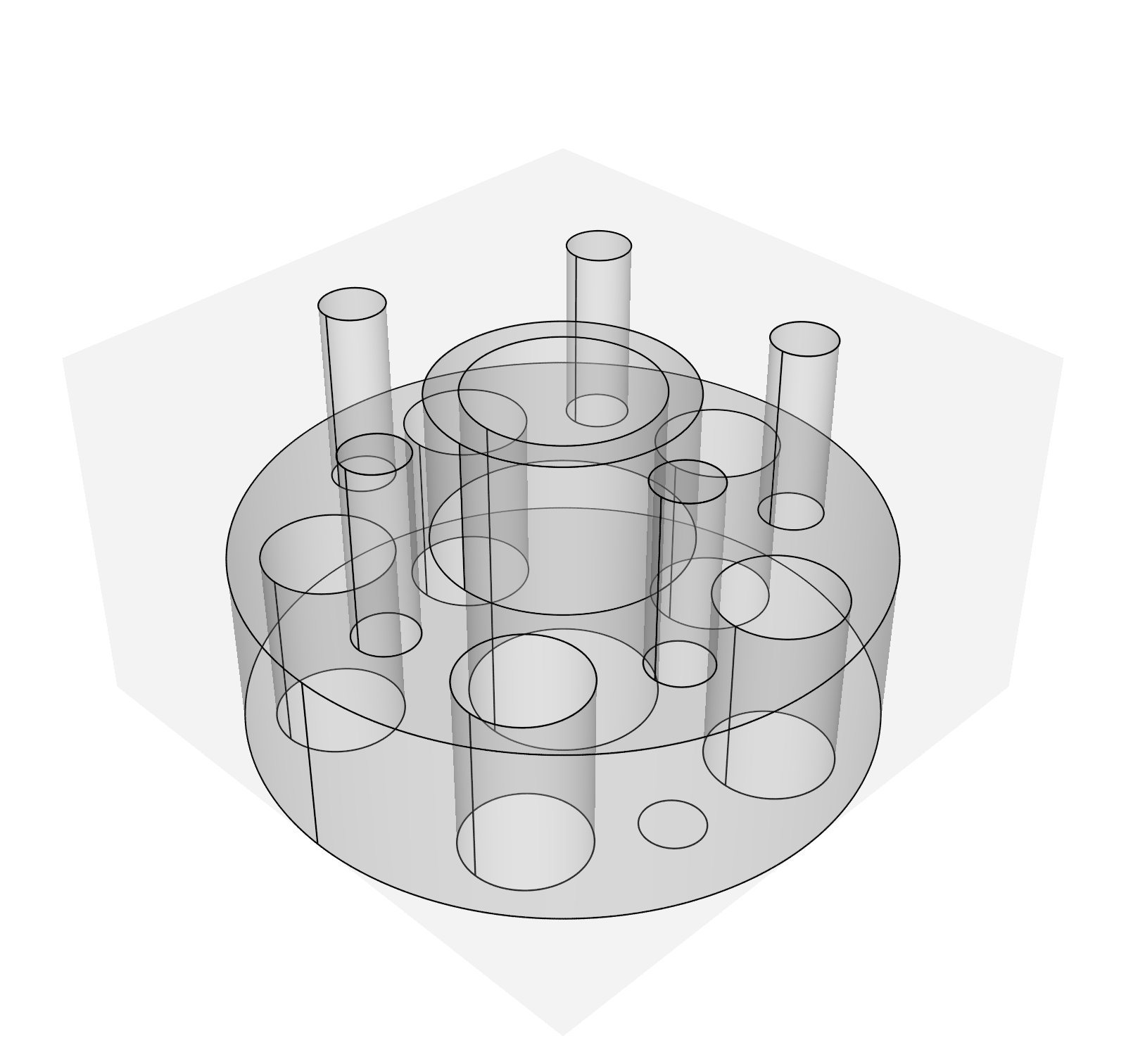}&
        &
        \\
        \multicolumn{1}{l}{Ours Net} & & & & &
        \\
        \includegraphics[width=0.124\linewidth]{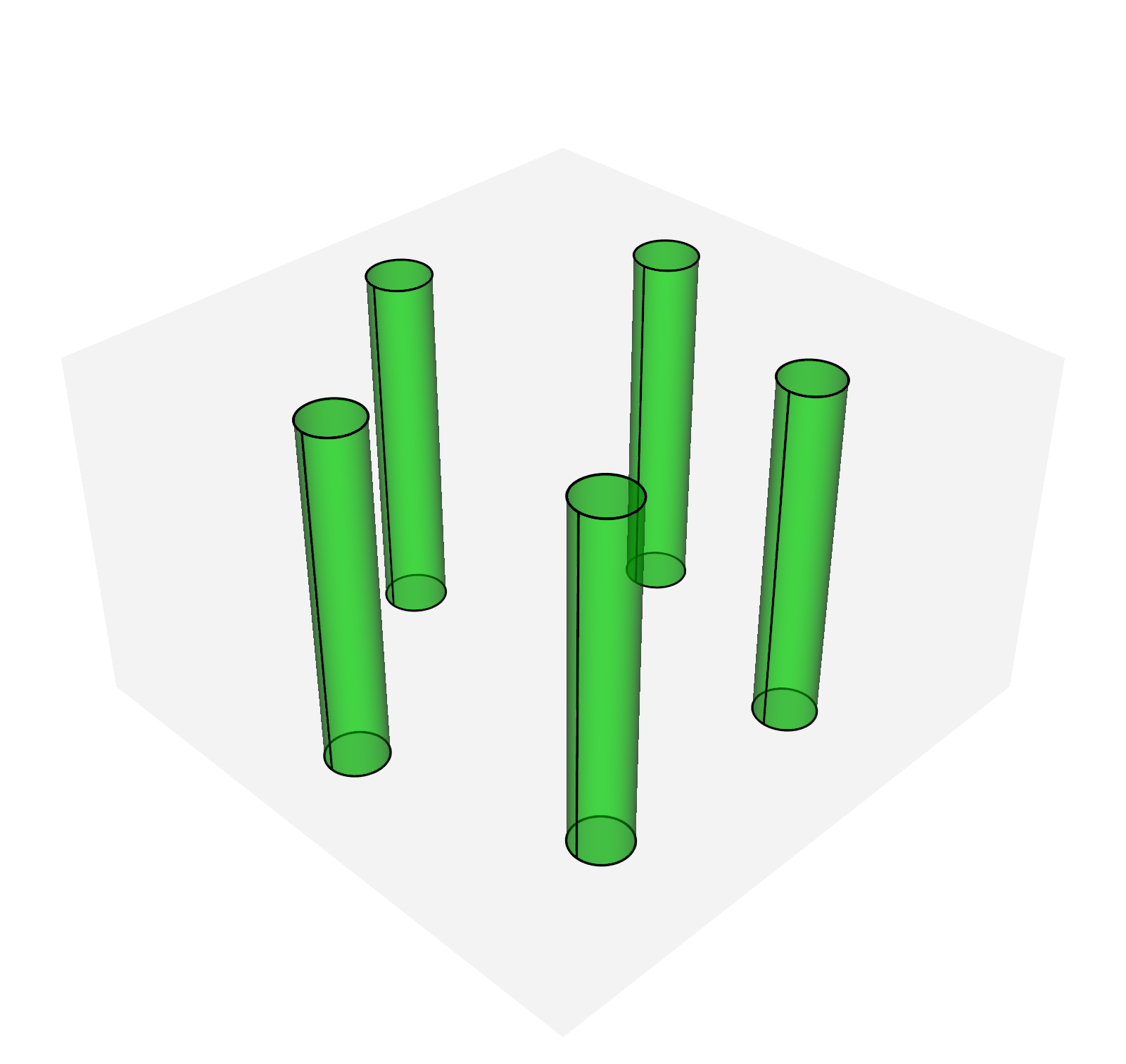} &
        \includegraphics[width=0.124\linewidth]{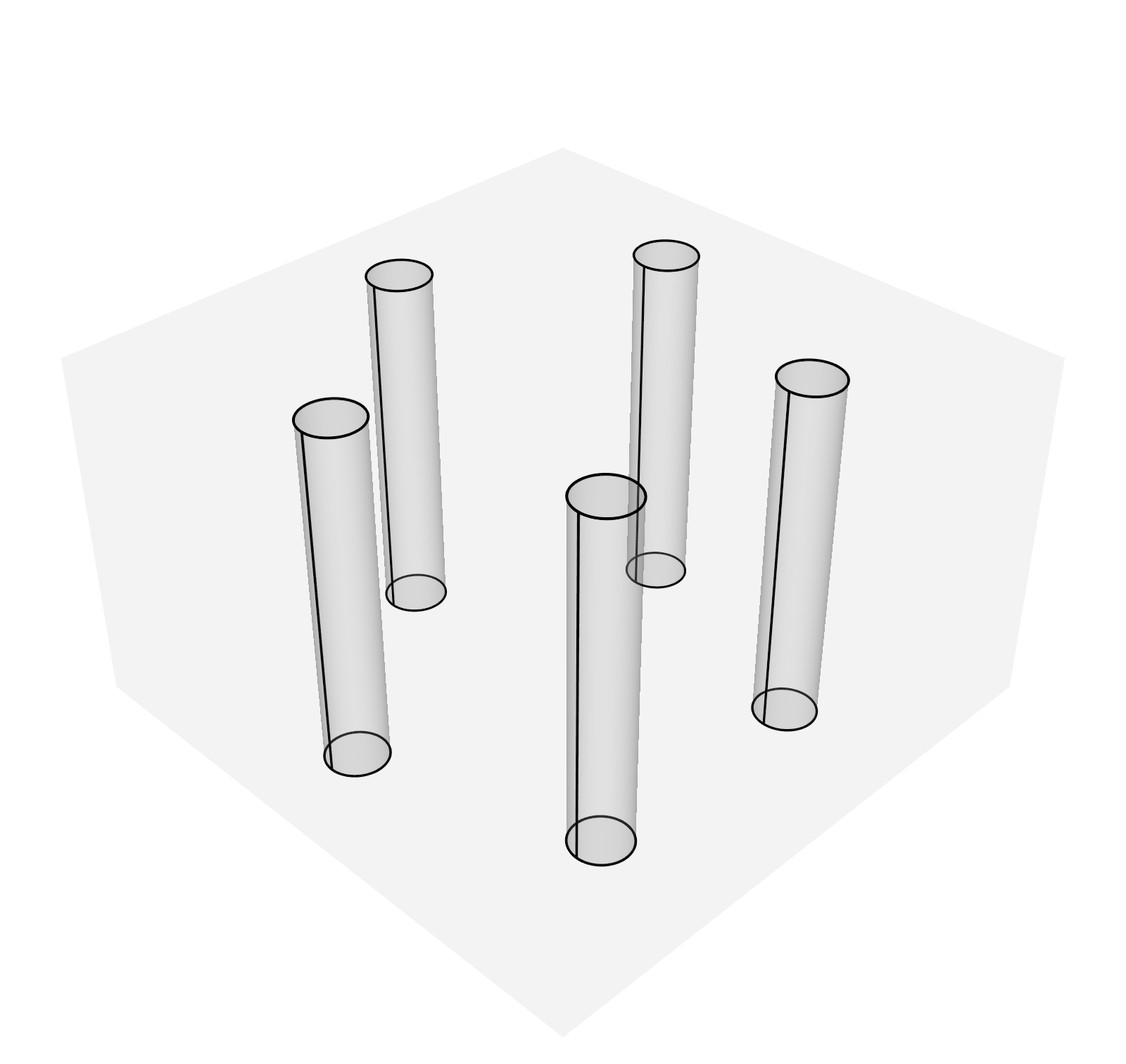} &
        \includegraphics[width=0.124\linewidth]{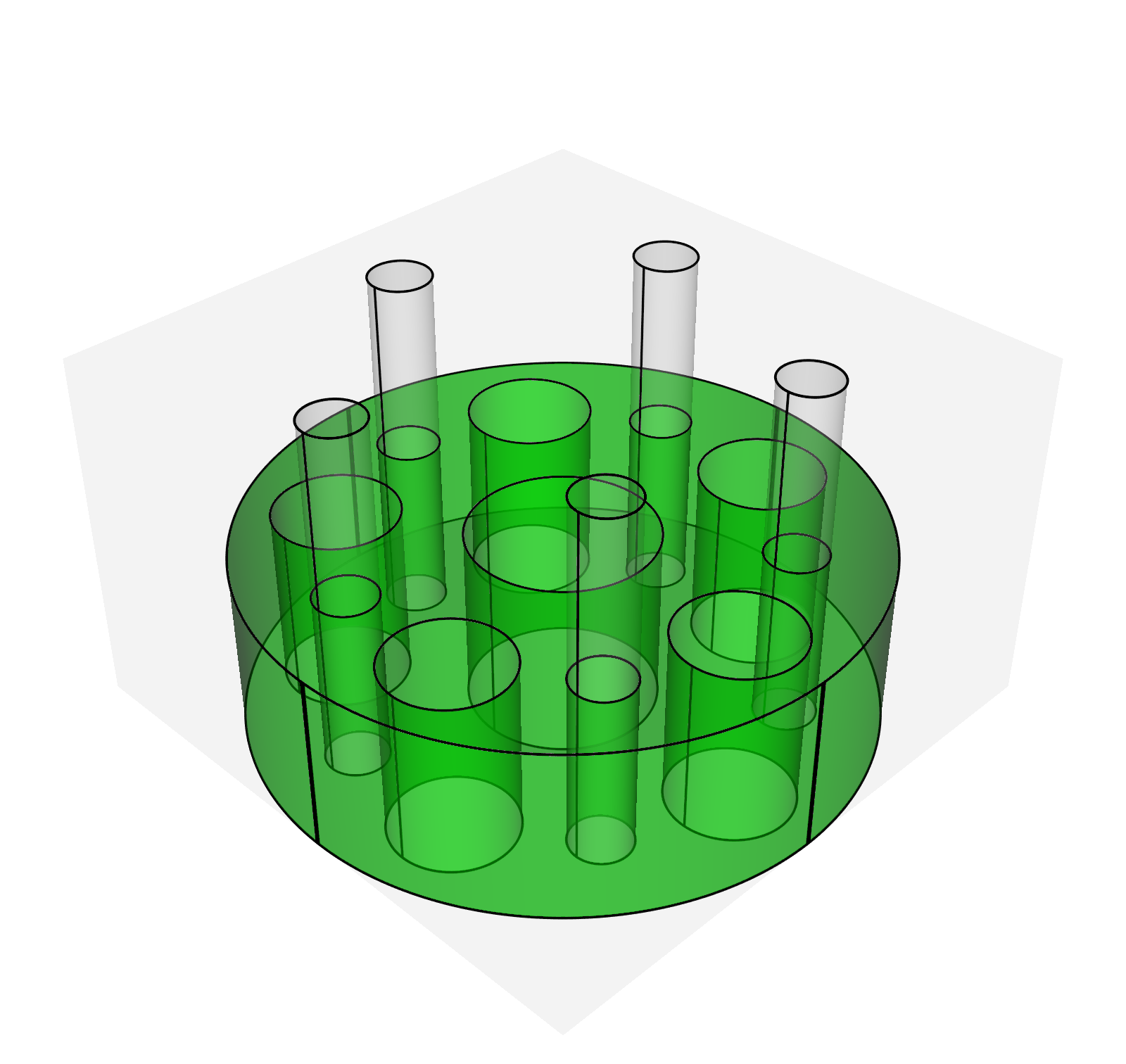} &
        \includegraphics[width=0.124\linewidth]{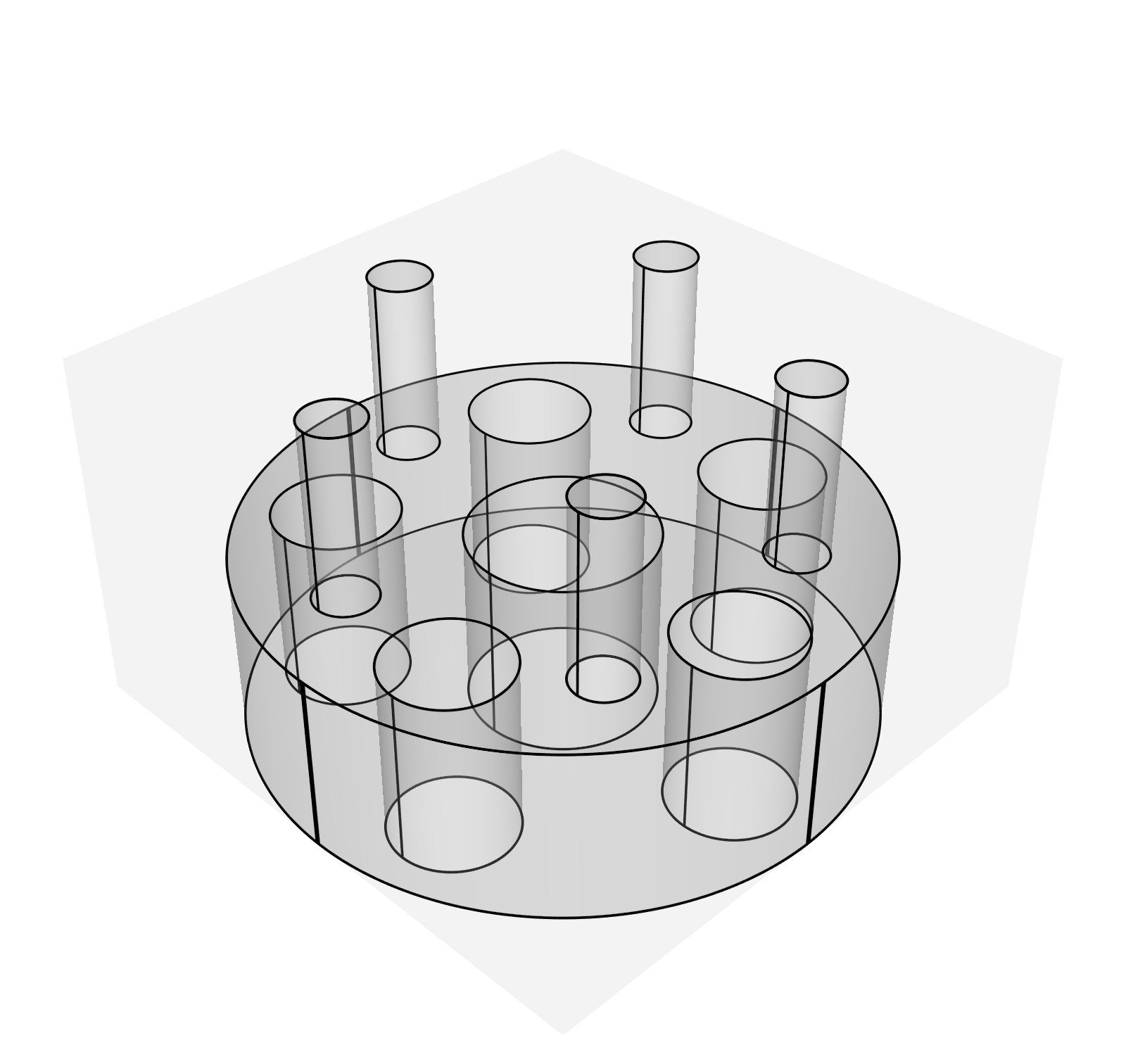} &
        \includegraphics[width=0.124\linewidth]{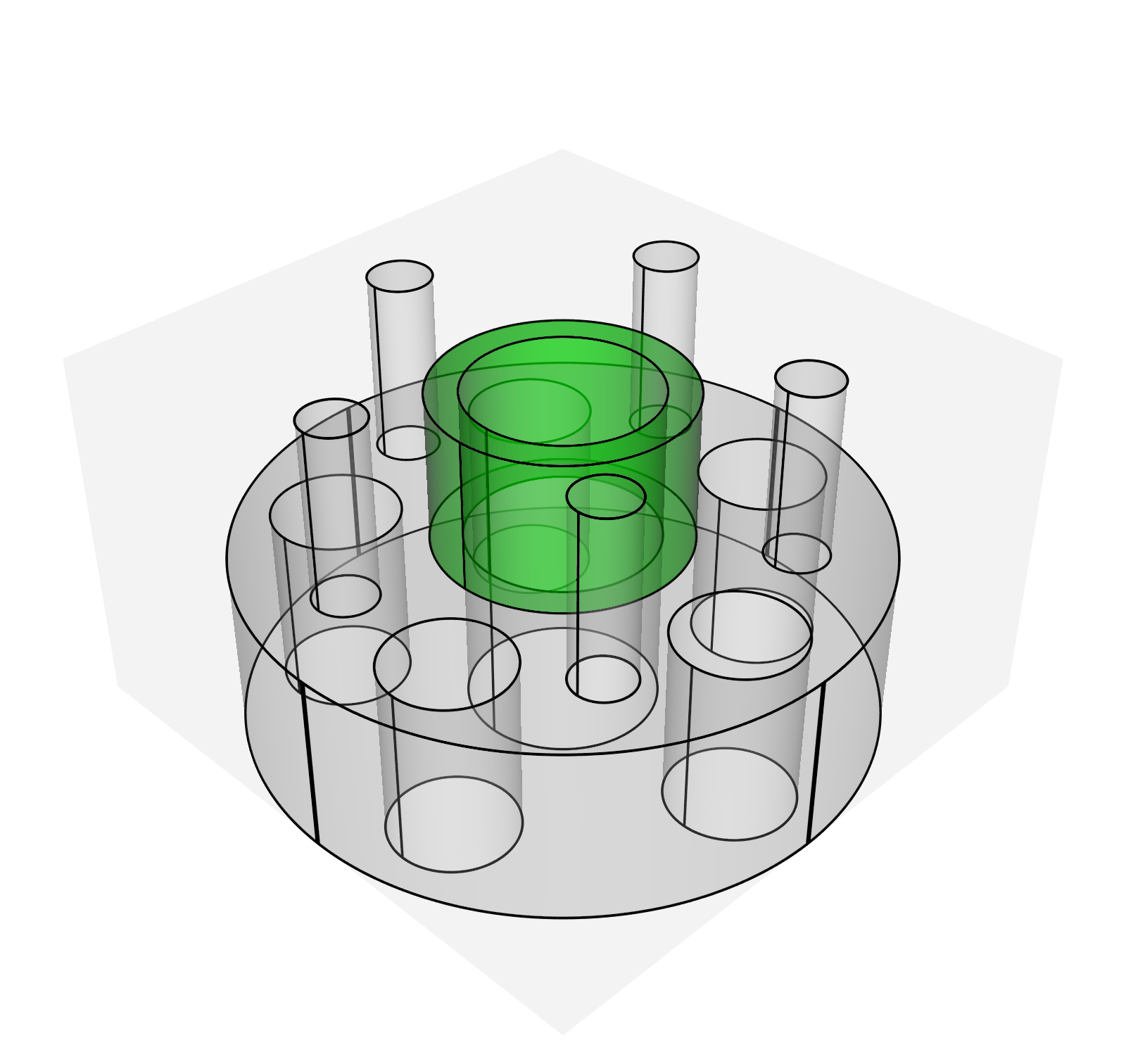} &
        \includegraphics[width=0.124\linewidth]{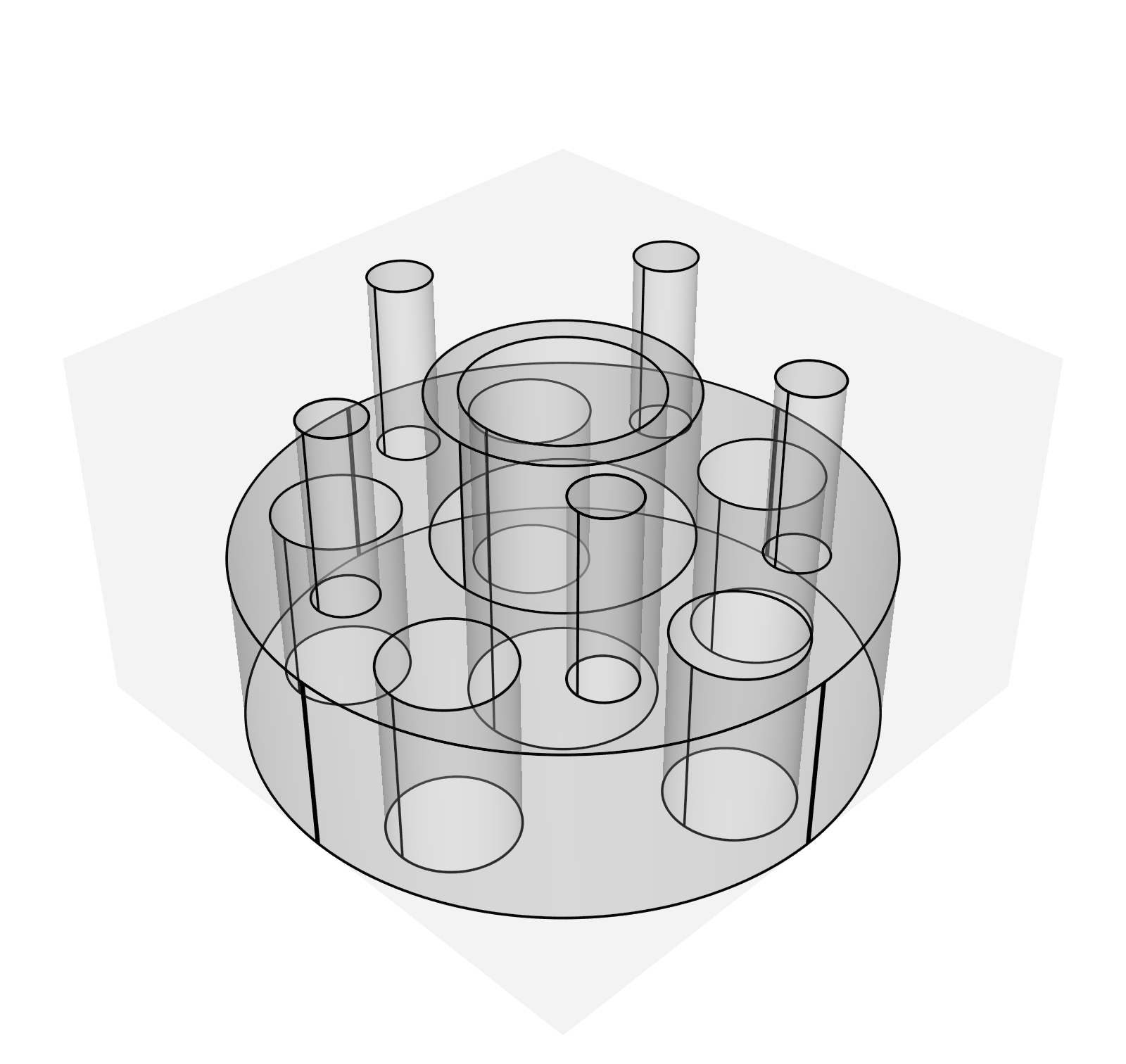}&
        &
        \\

    \end{tabular}
    \caption{
    Qualitative comparison of the output of our model's inferred programs (Ours Net) vs. InverseCSG. Green: addition, Red: subtraction, Blue: intersection, Grey: current. (Case 3)
    }
    \label{figure:qualitative_comparision_inv3}
\end{figure*}

\begin{figure*}[!h]
    \centering
    \setlength{\tabcolsep}{1pt}
    \begin{tabular}{ccc|ccc}
        Target & InverseCSG & Ours Heur & Target & InverseCSG & Ours Heur
        \\
        \includegraphics[width=0.15\linewidth]{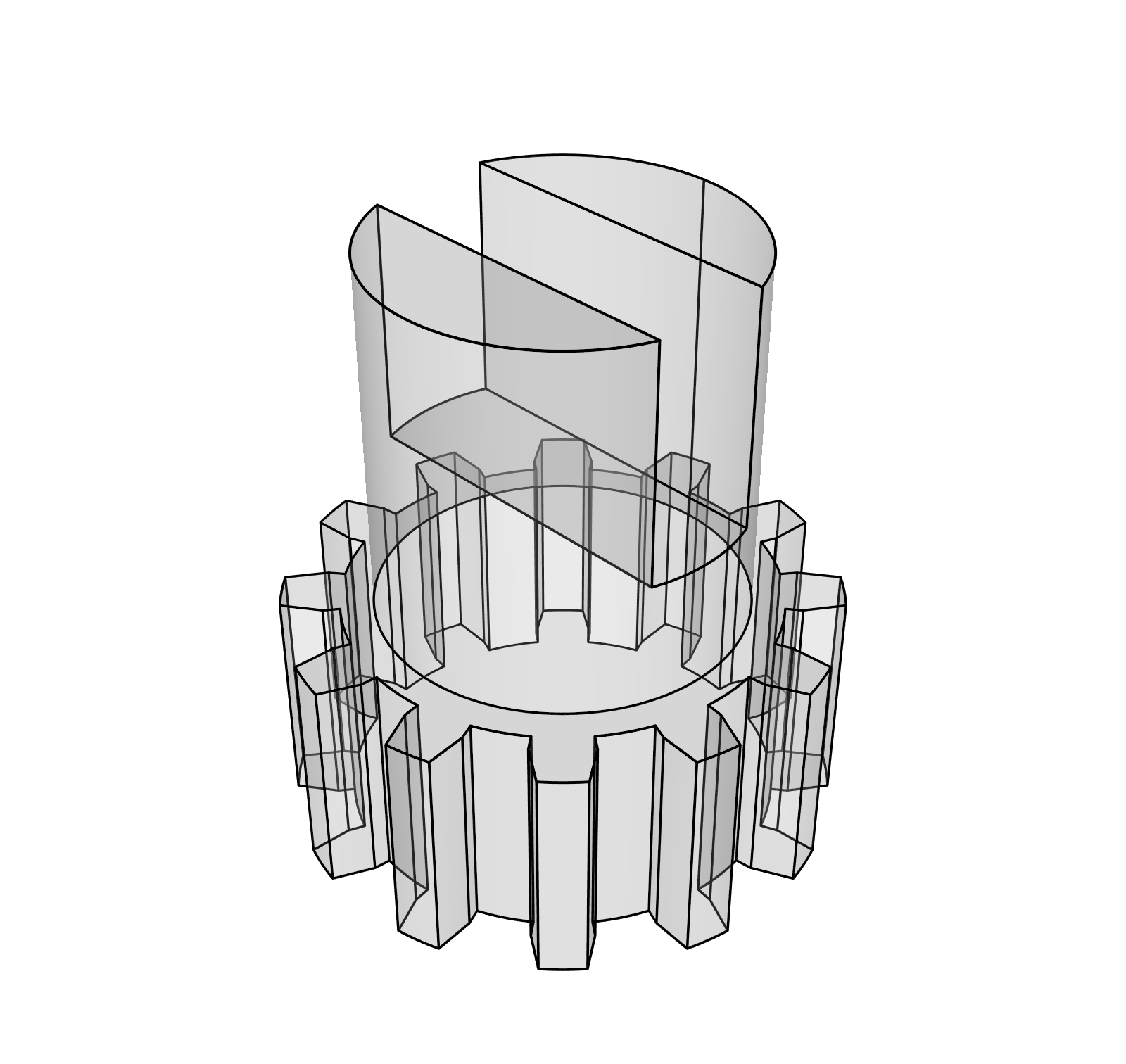} &
        \includegraphics[width=0.15\linewidth]{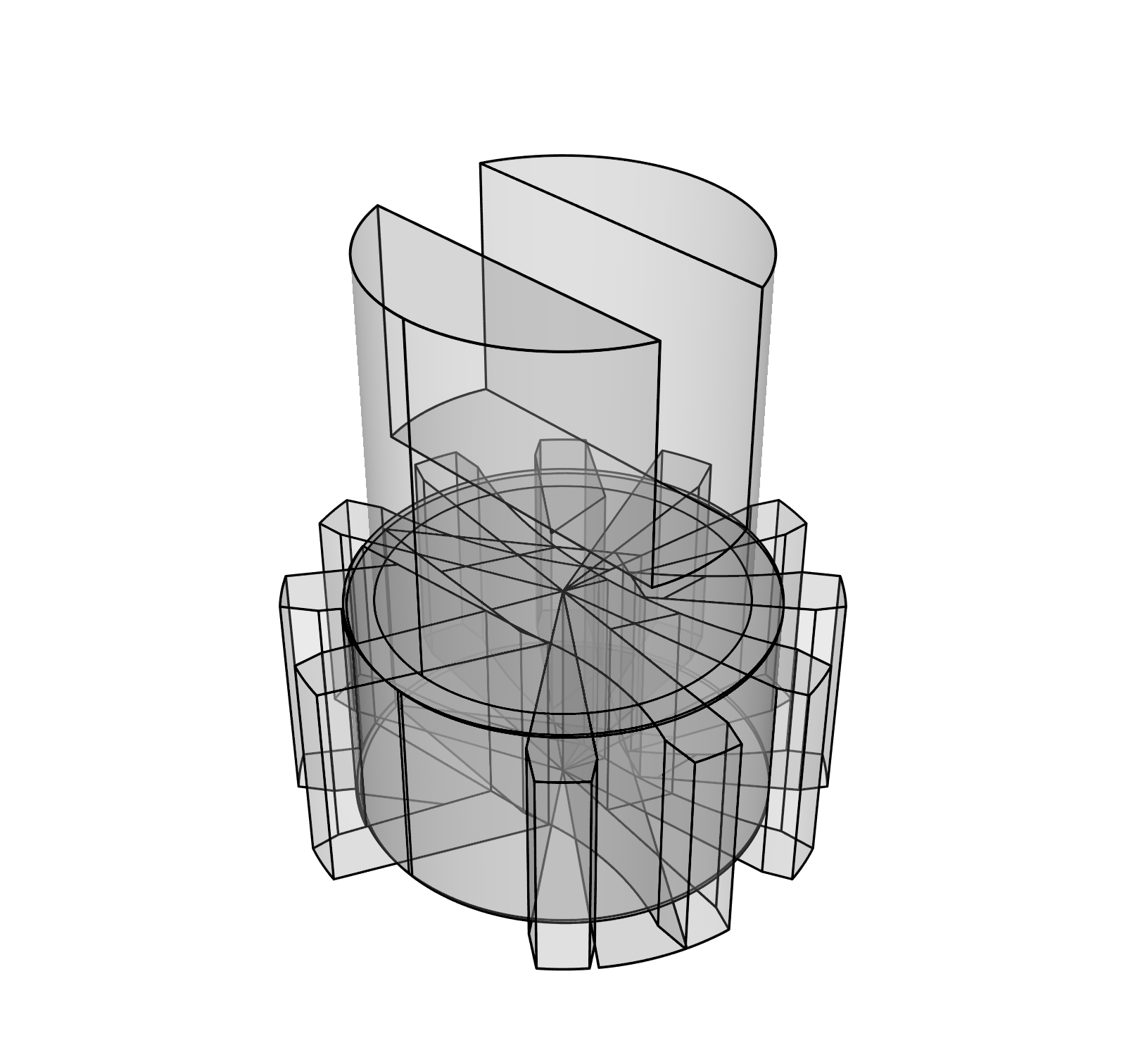} &
        \includegraphics[width=0.15\linewidth]{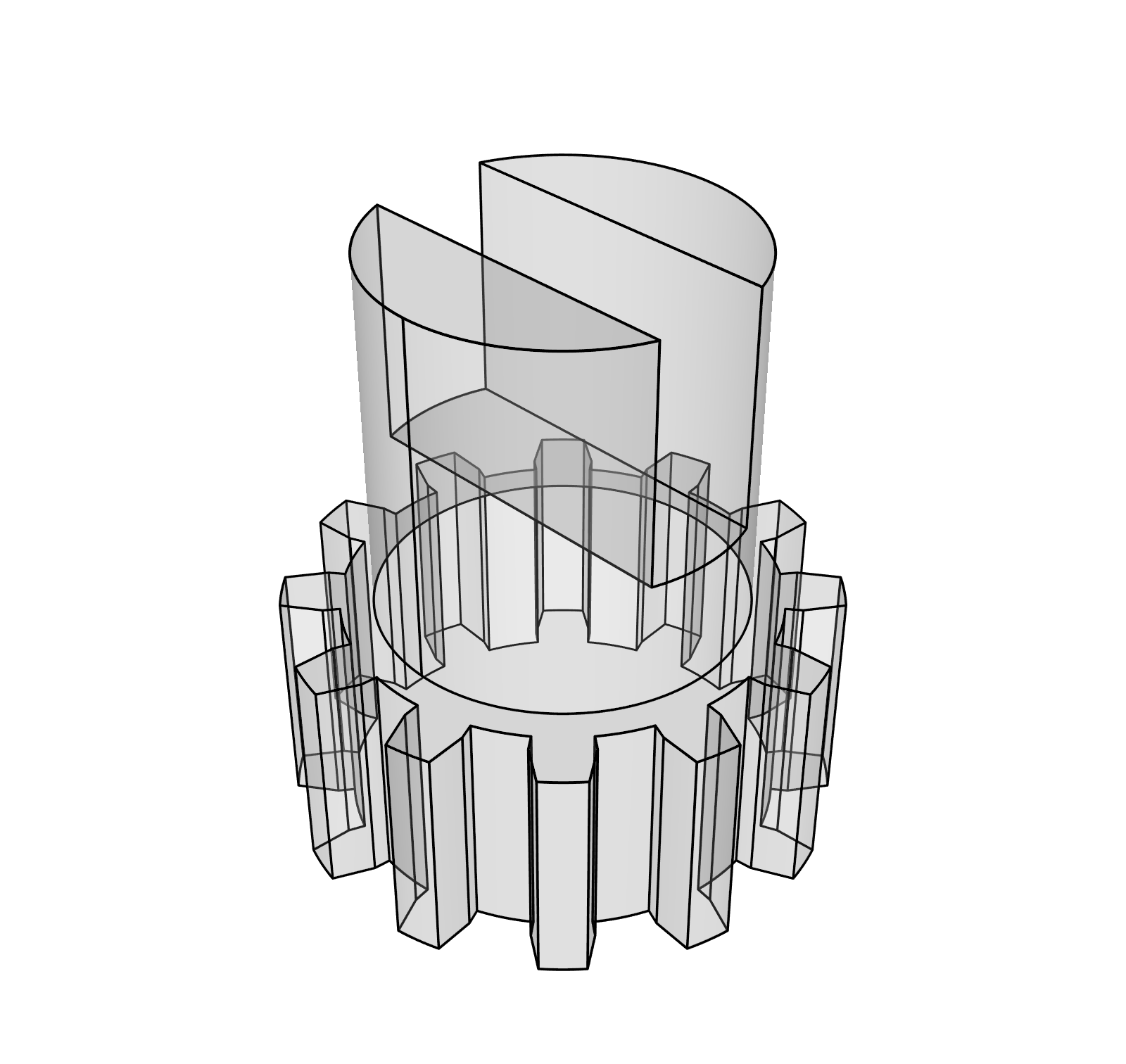} &
        \includegraphics[width=0.15\linewidth]{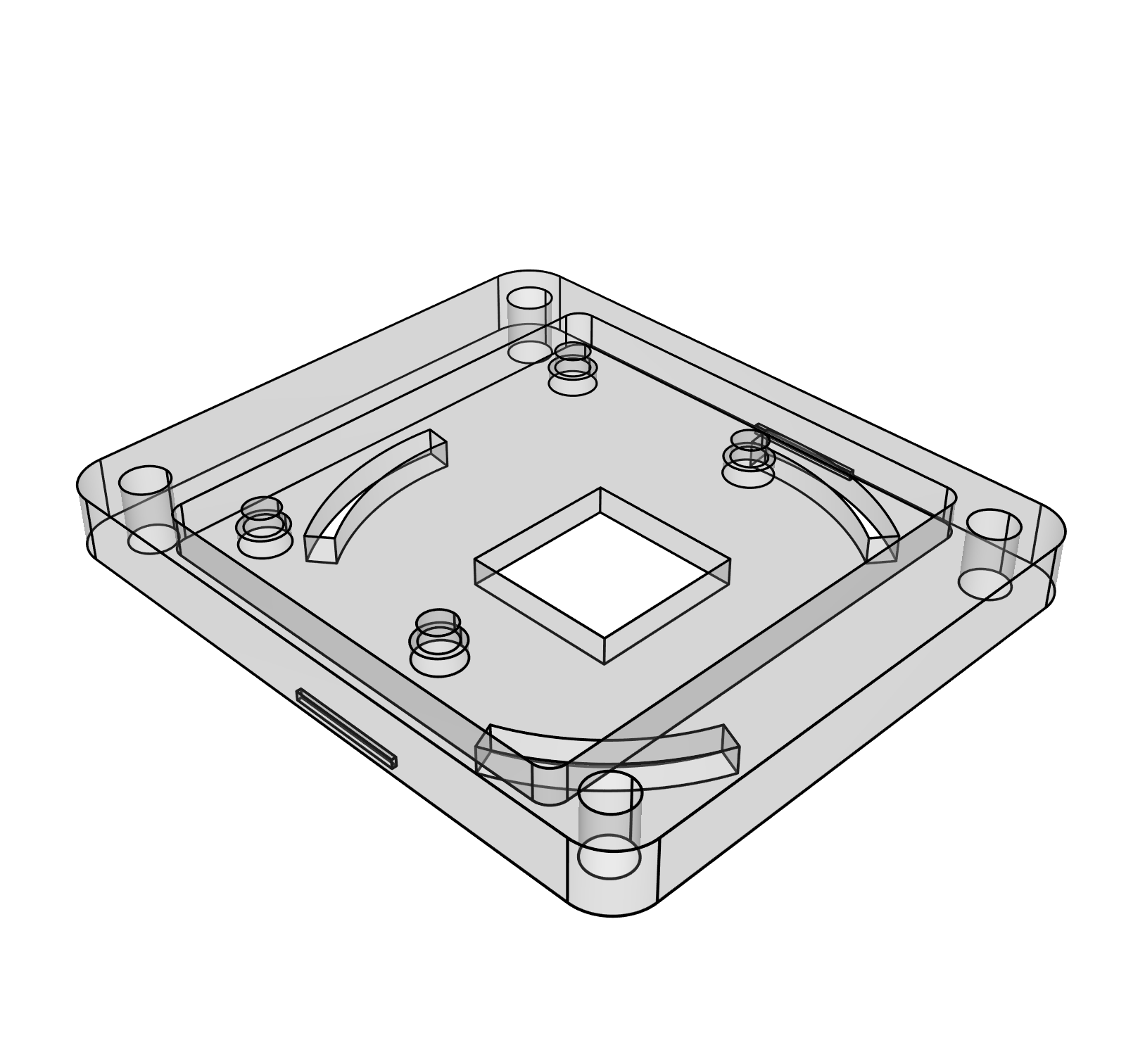} &
        \includegraphics[width=0.15\linewidth]{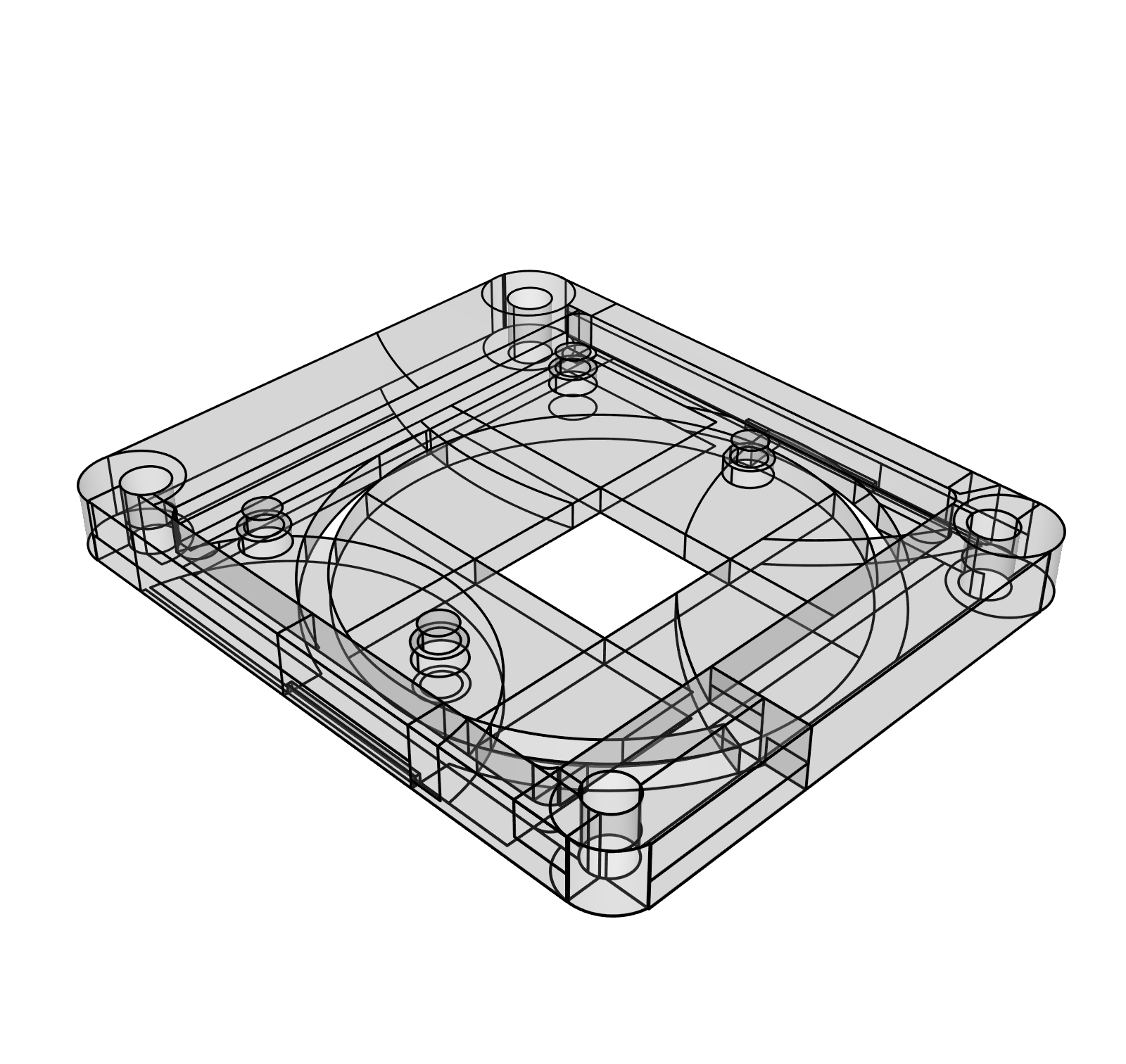} &
        \includegraphics[width=0.15\linewidth]{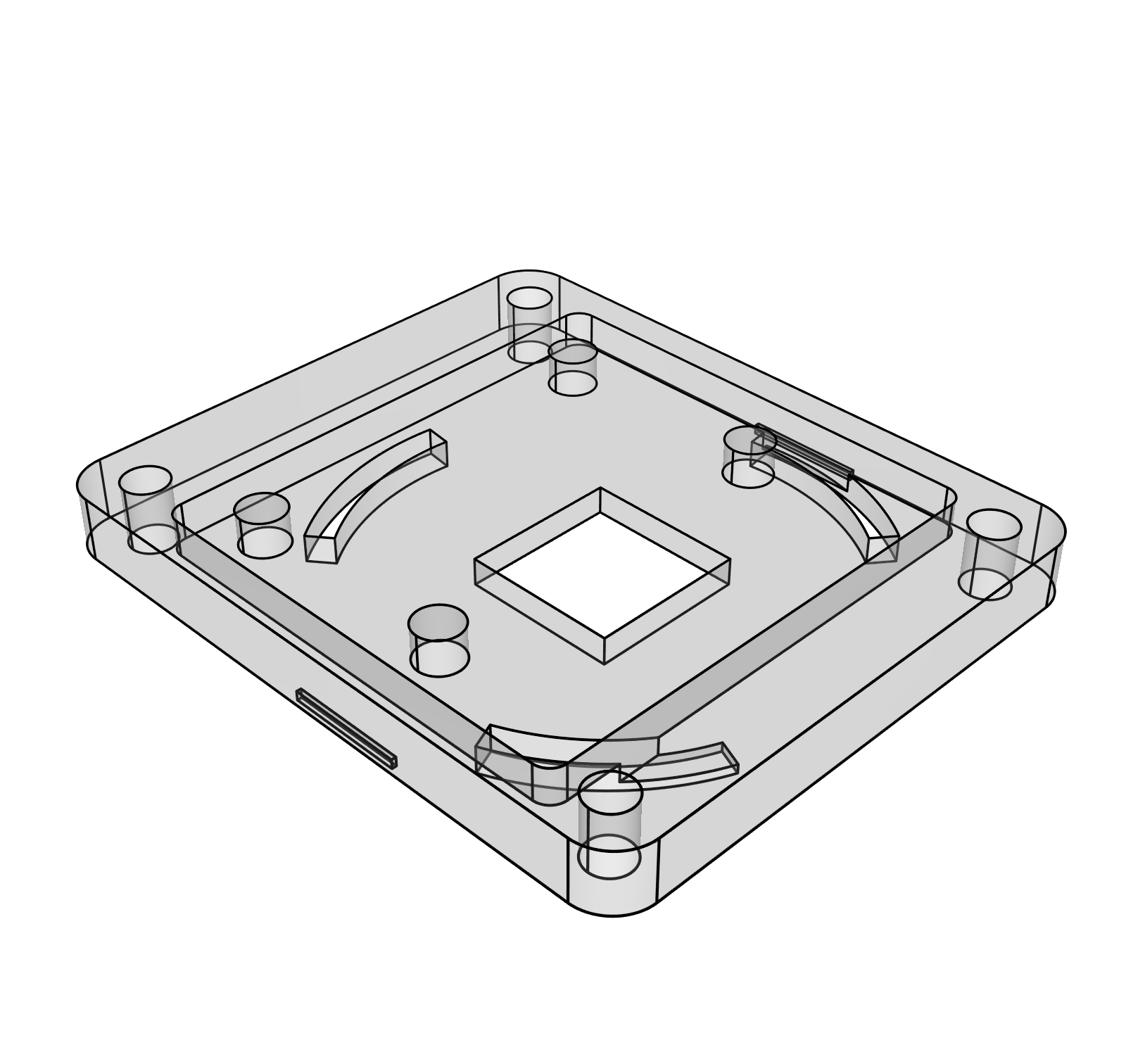}
        \\
        \includegraphics[width=0.15\linewidth]{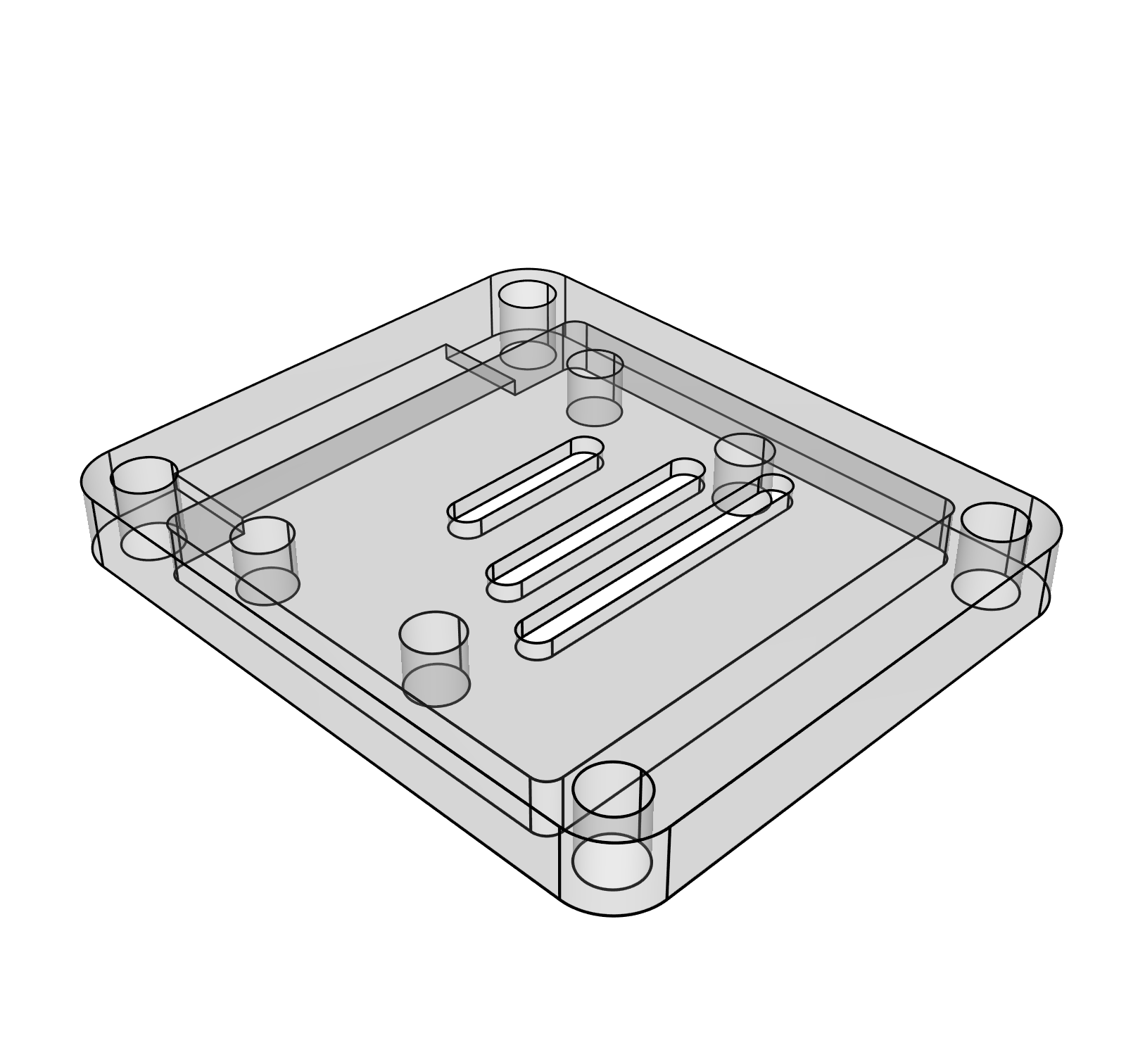} &
        \includegraphics[width=0.15\linewidth]{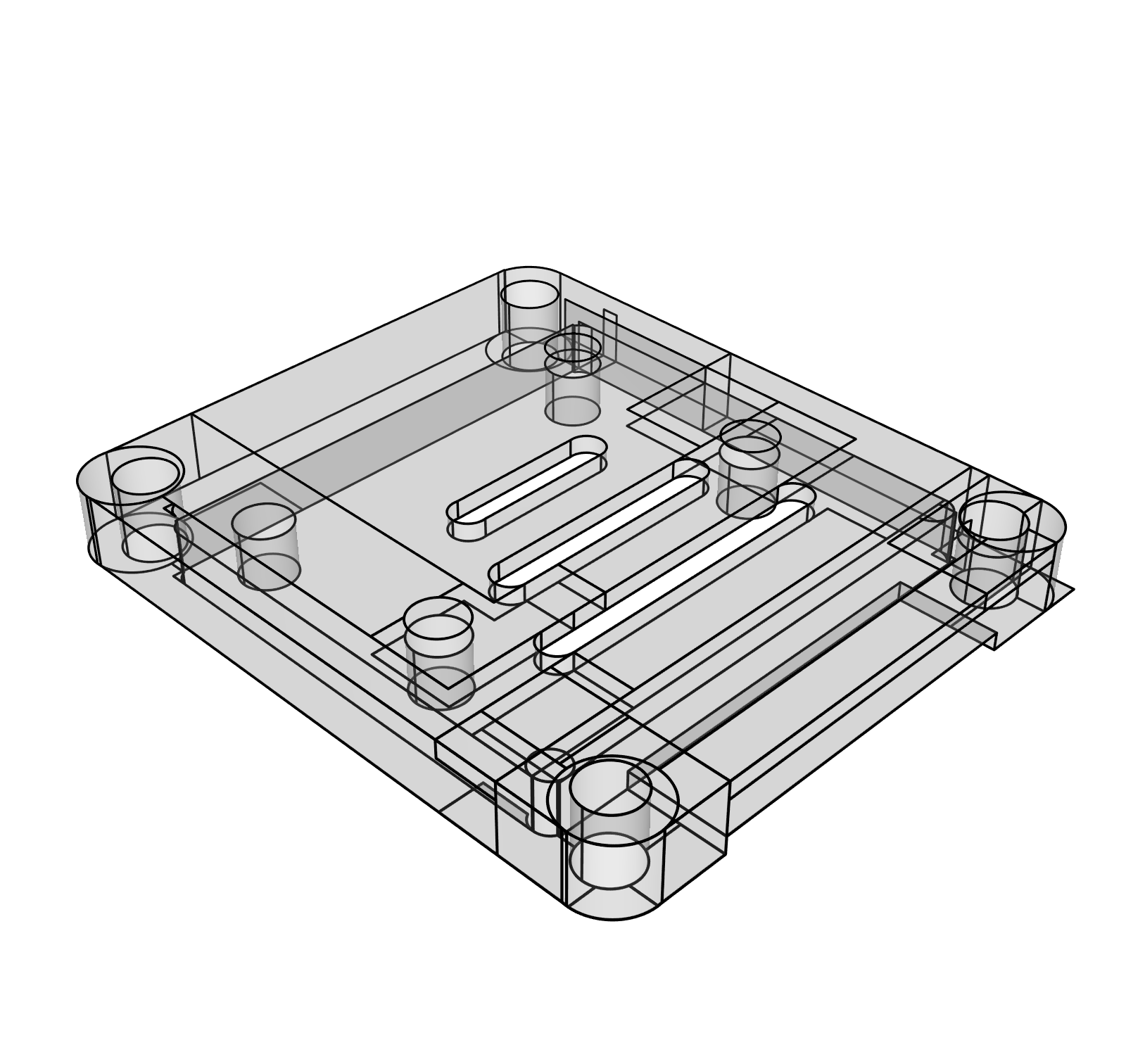} &
        \includegraphics[width=0.15\linewidth]{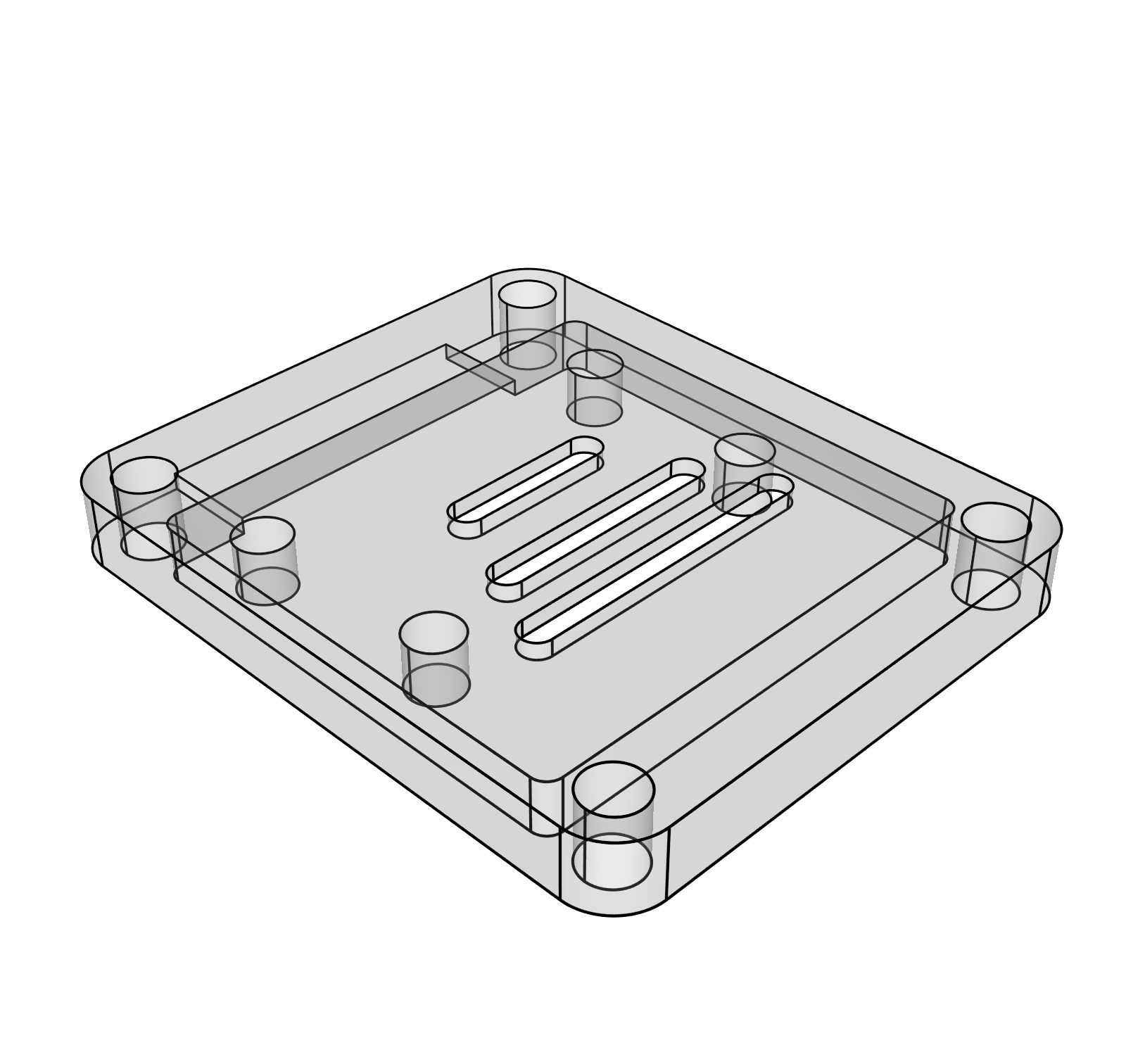} &
        \includegraphics[width=0.15\linewidth]{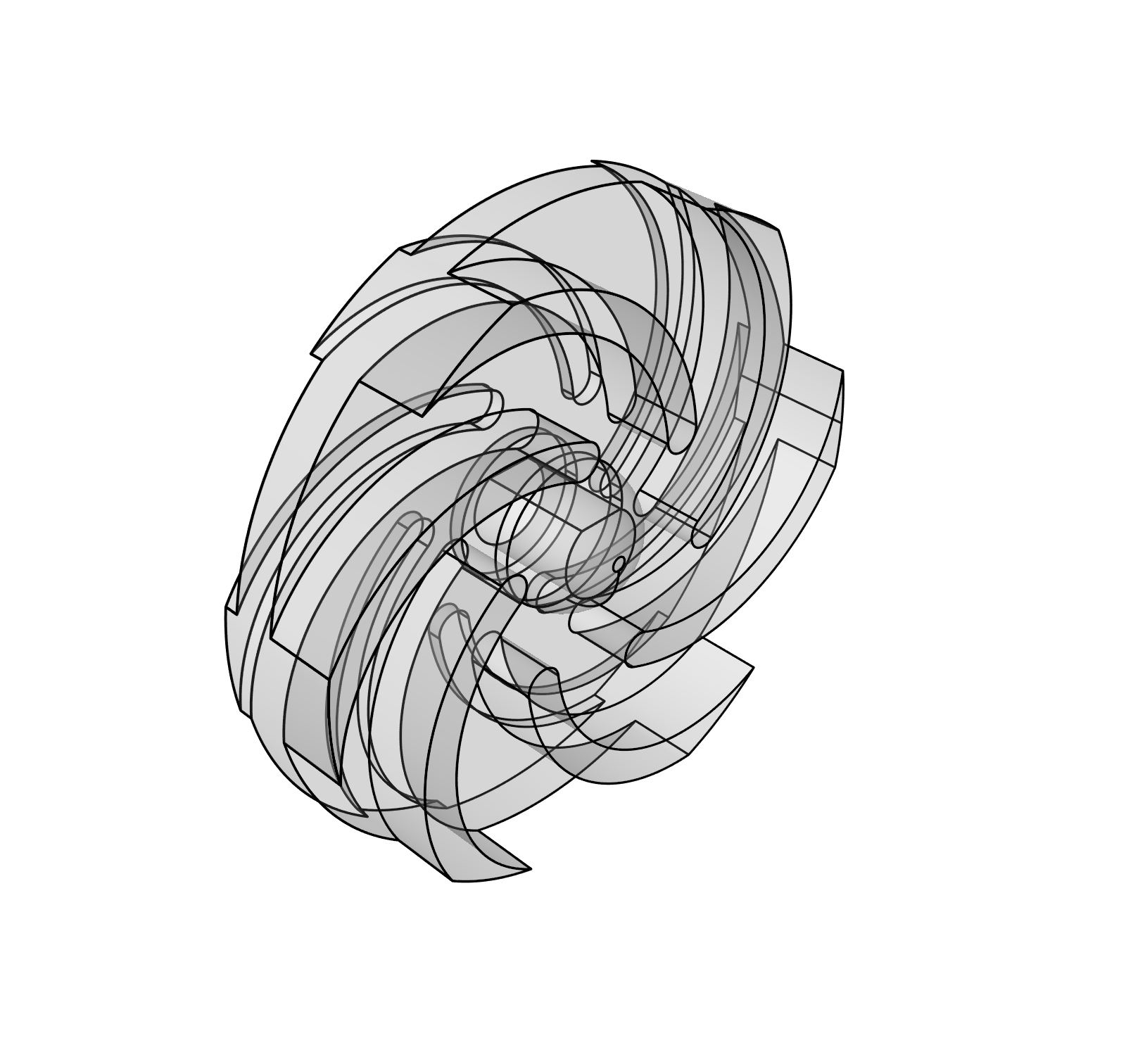} &
        \includegraphics[width=0.15\linewidth]{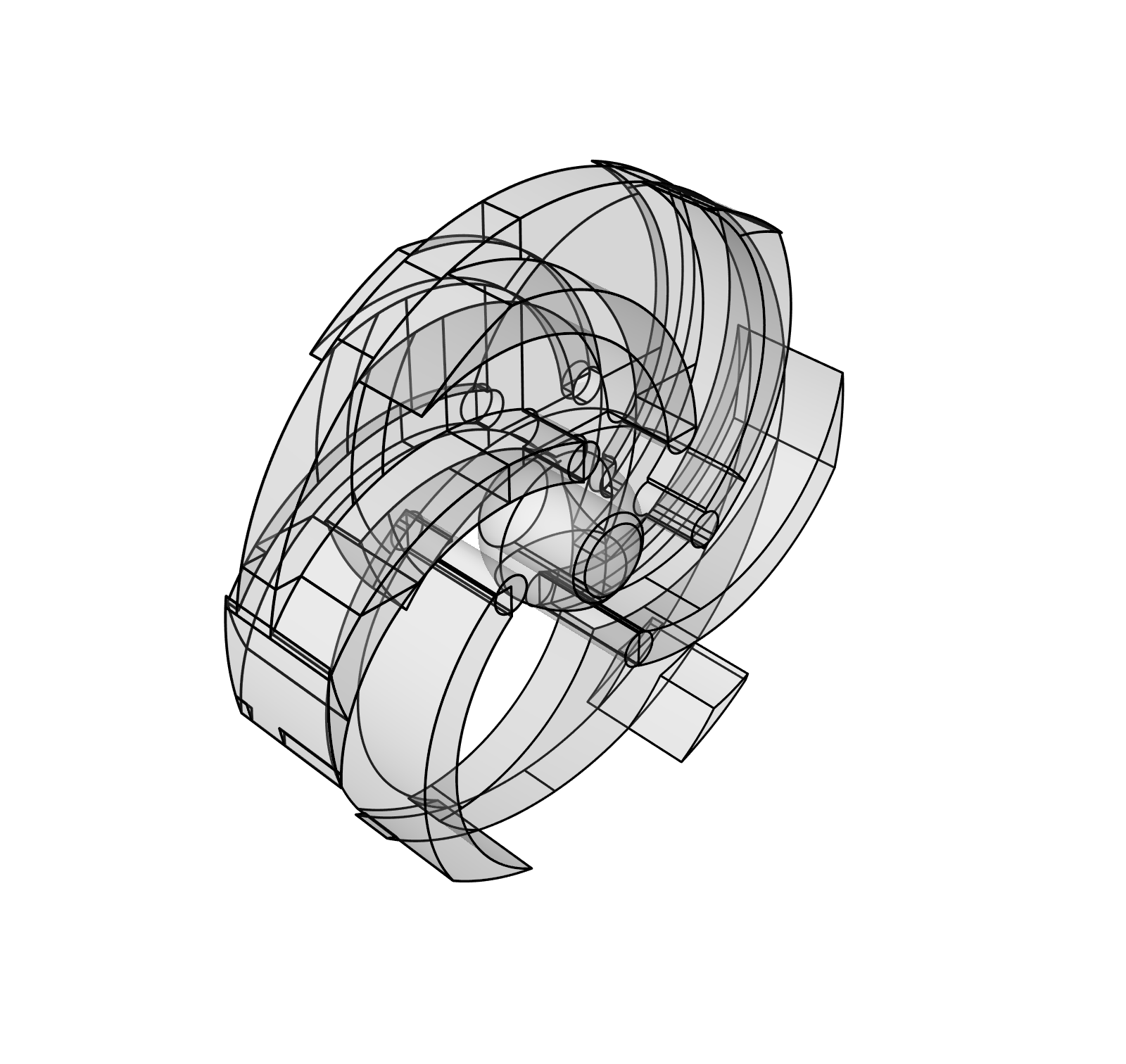} &
        \includegraphics[width=0.15\linewidth]{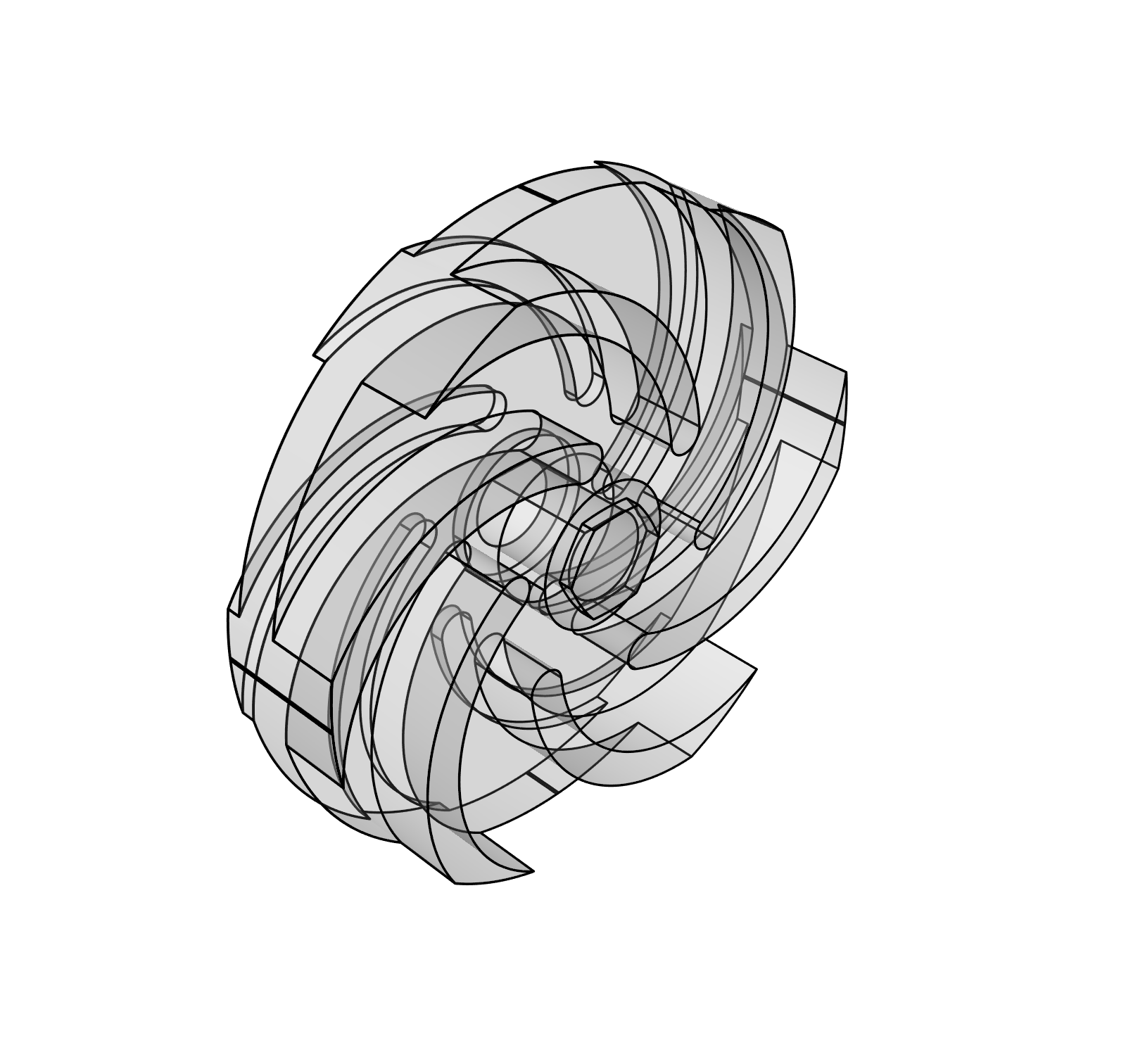}
        \\
        \includegraphics[width=0.15\linewidth]{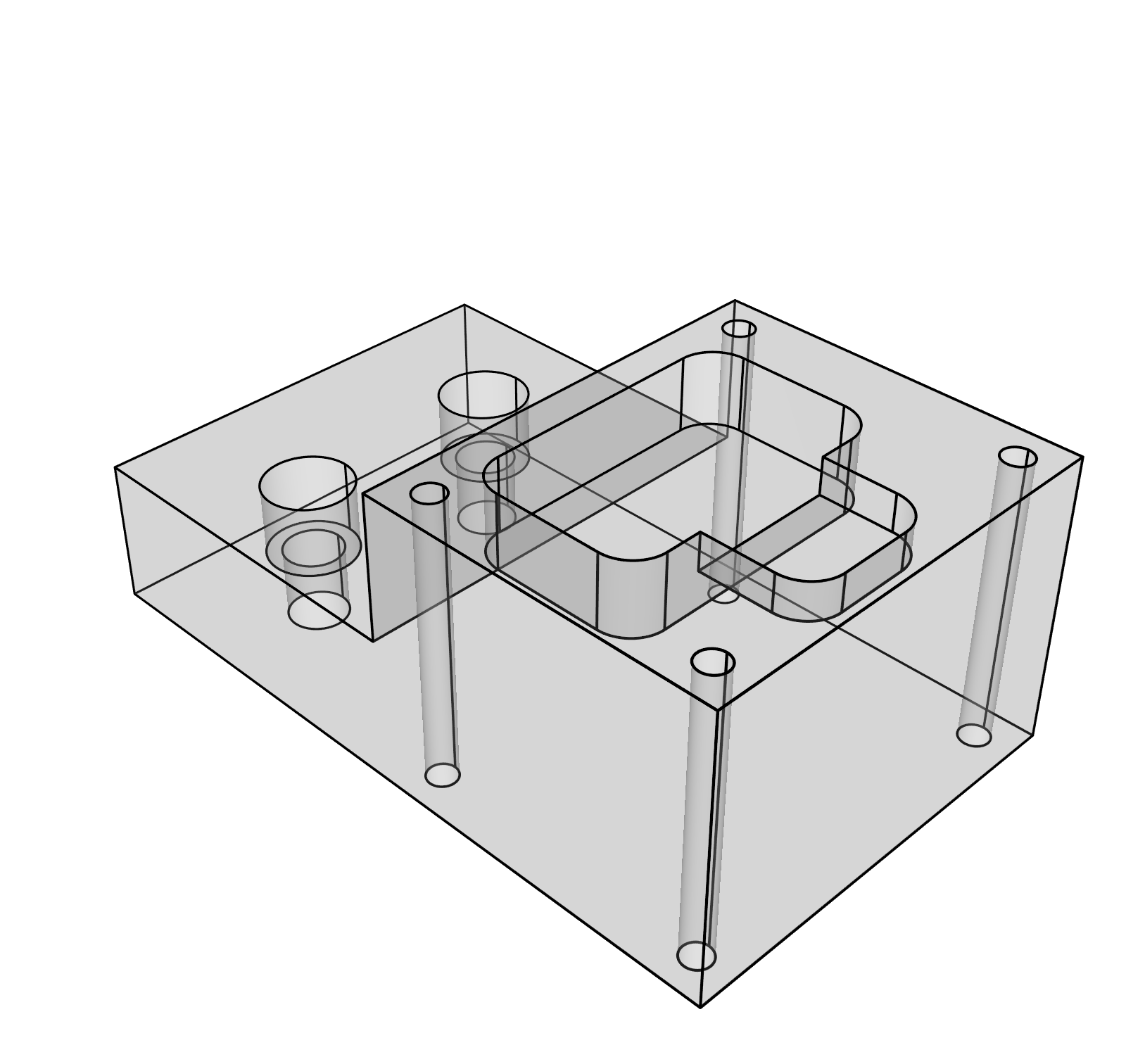} &
        \includegraphics[width=0.15\linewidth]{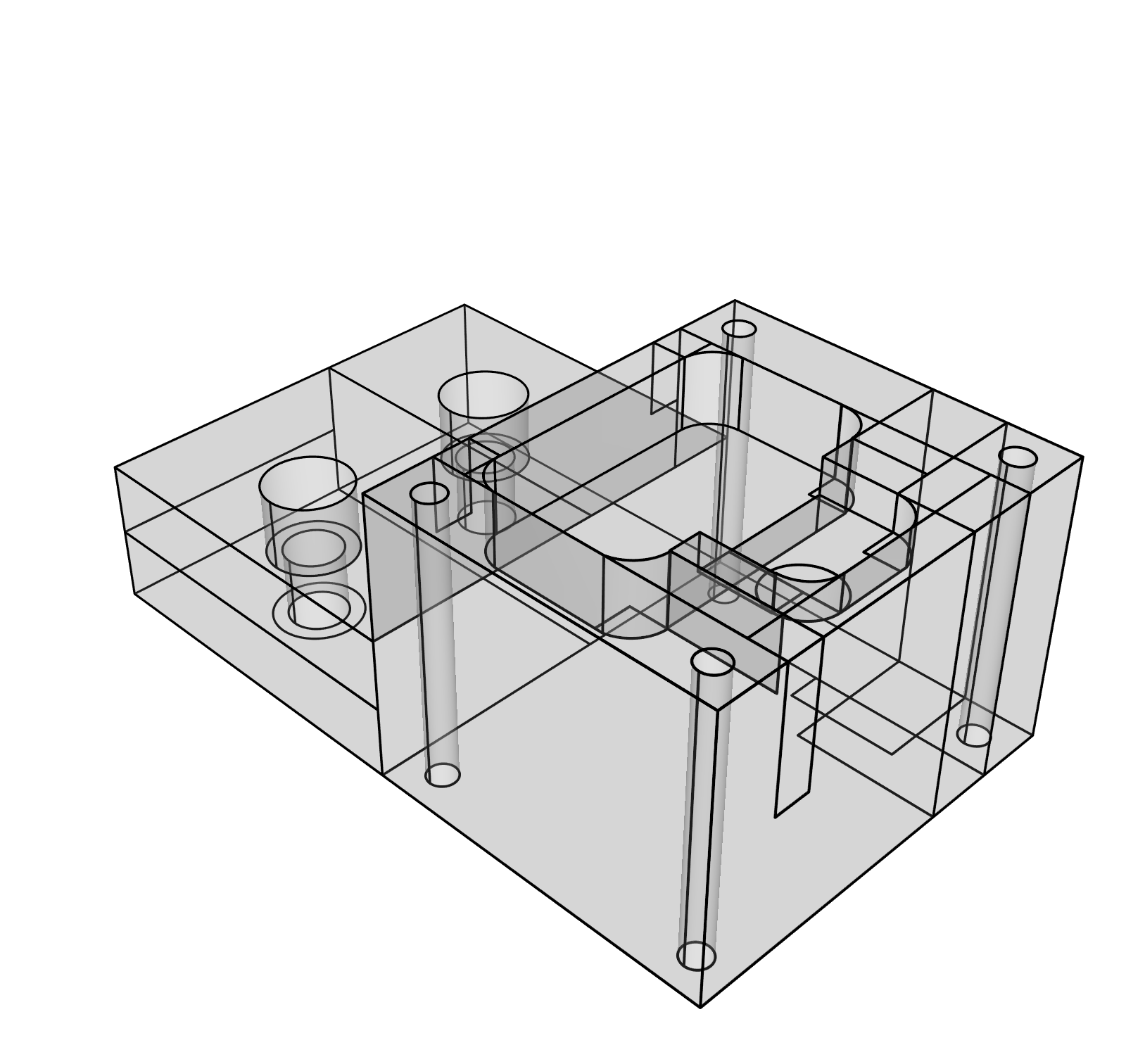} &
        \includegraphics[width=0.15\linewidth]{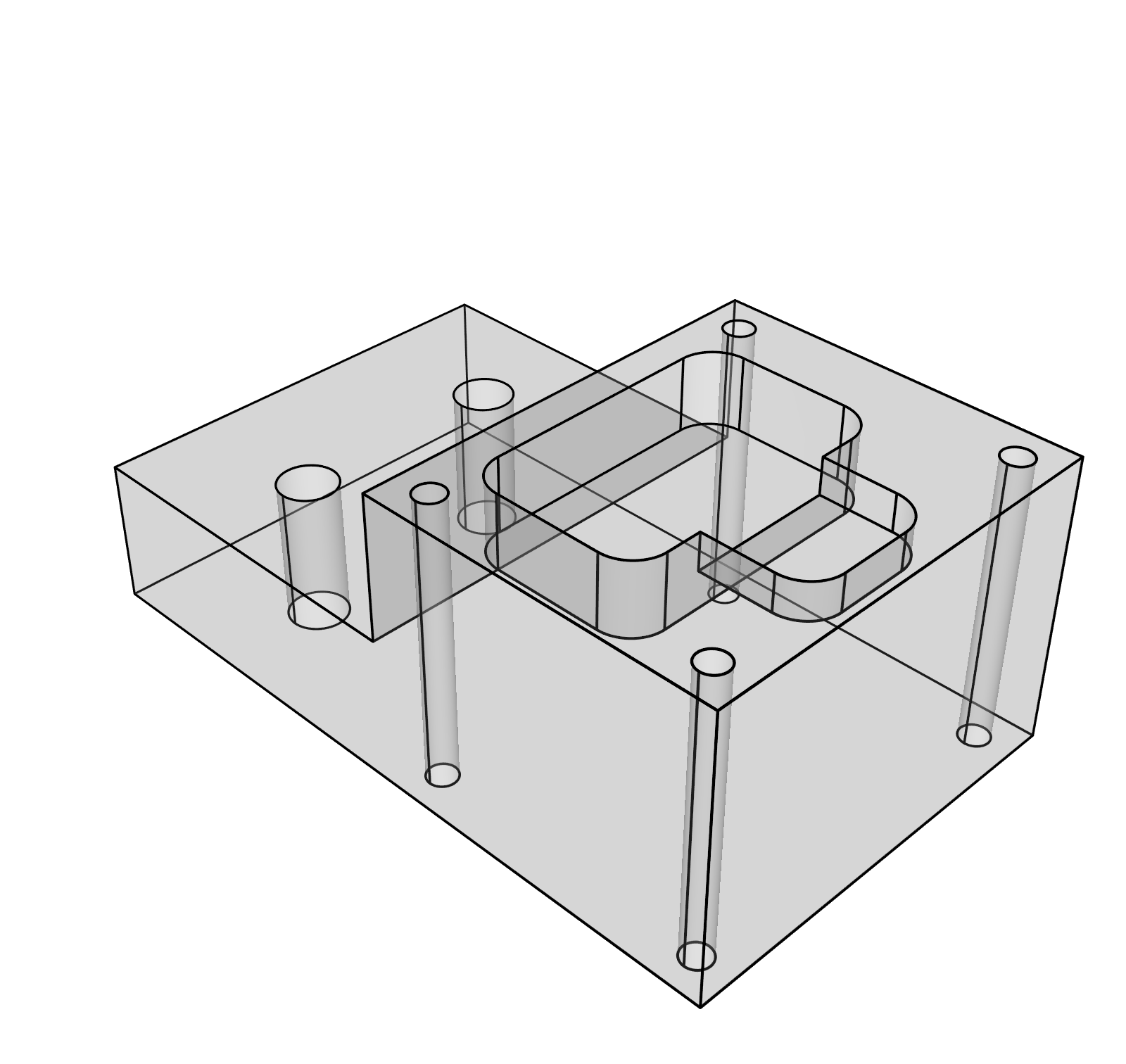}&
        \includegraphics[width=0.15\linewidth]{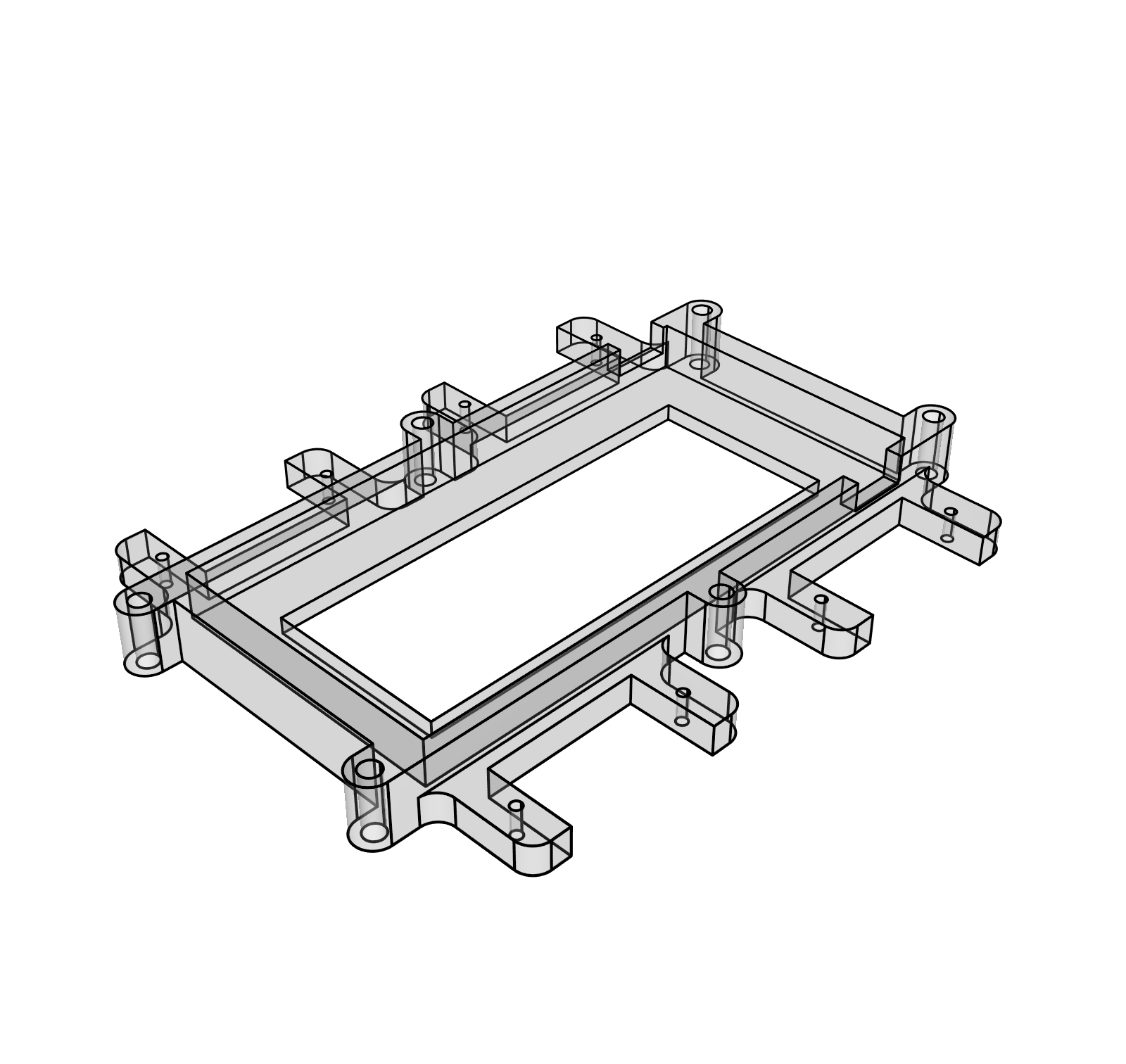} &
        \includegraphics[width=0.15\linewidth]{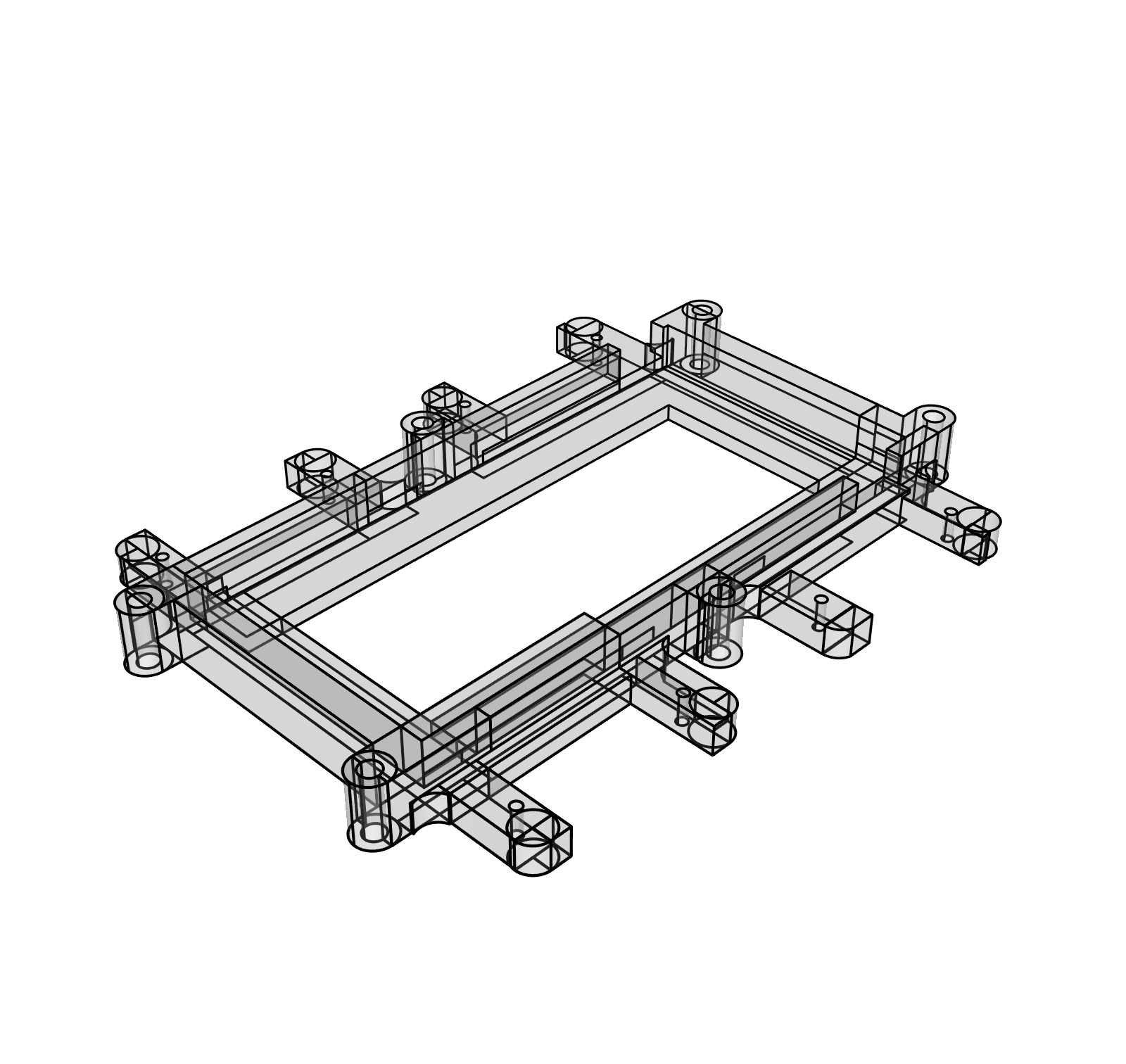} &
        \includegraphics[width=0.15\linewidth]{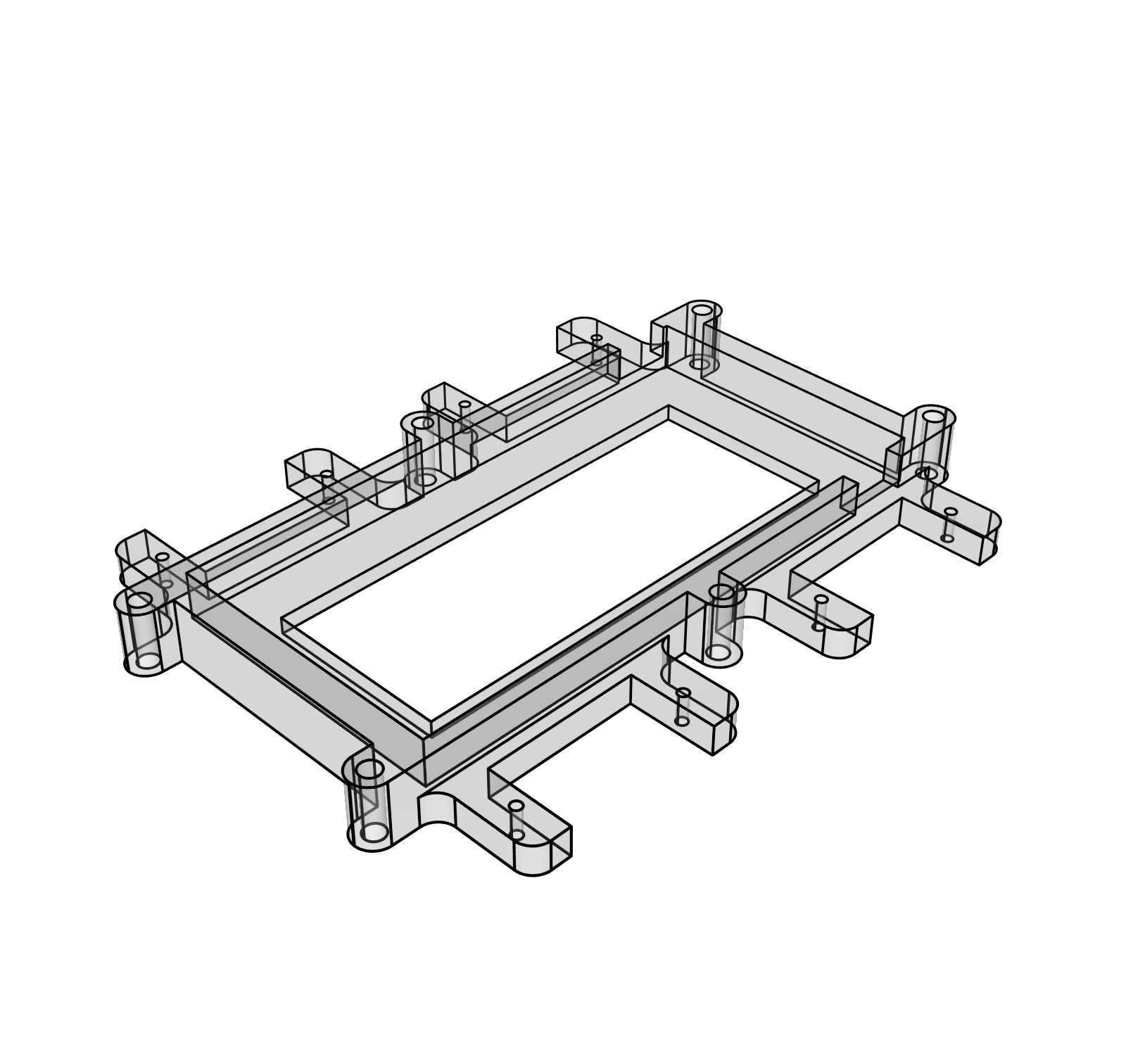}
        \\
        \includegraphics[width=0.15\linewidth]{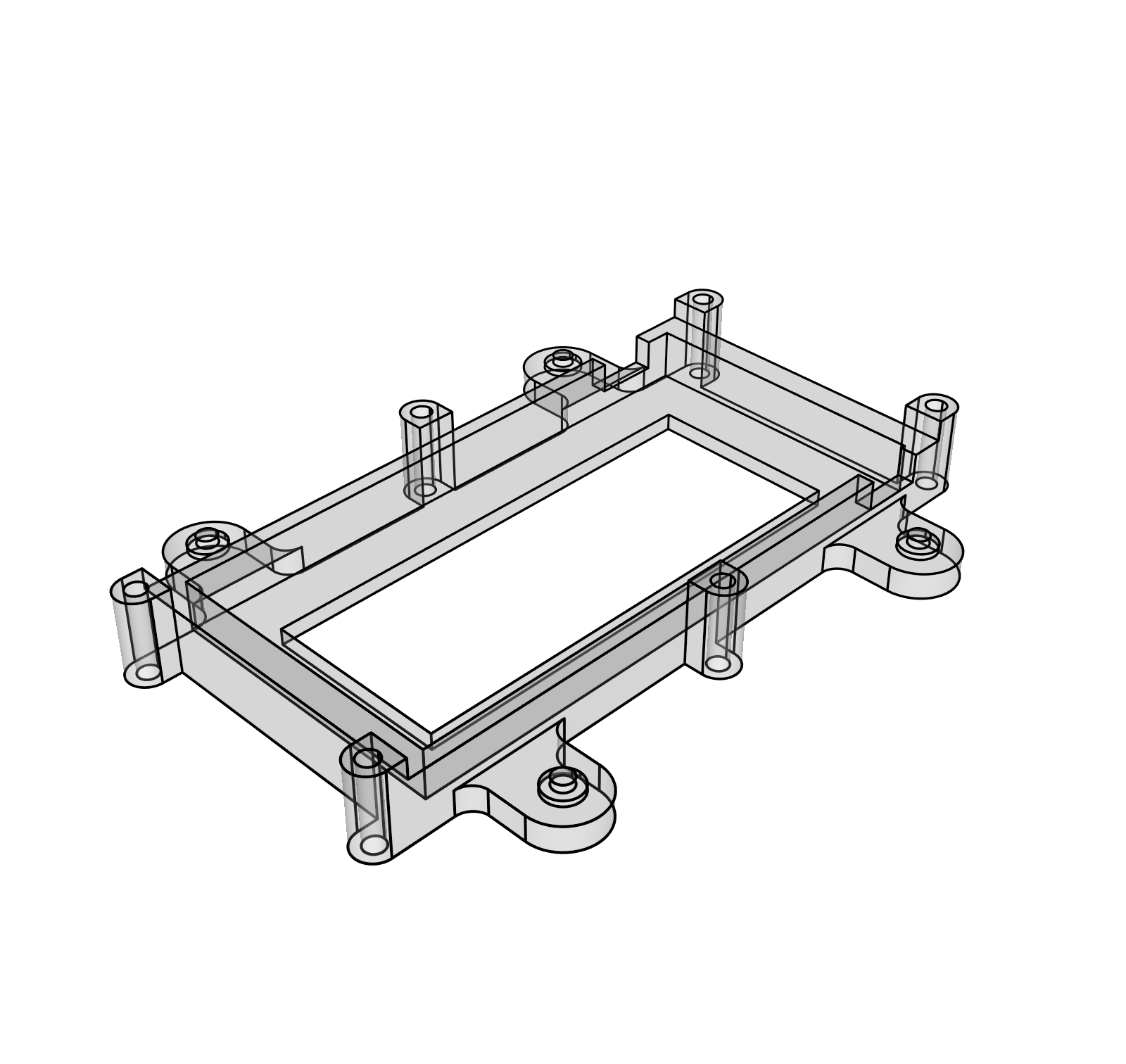} &
        \includegraphics[width=0.15\linewidth]{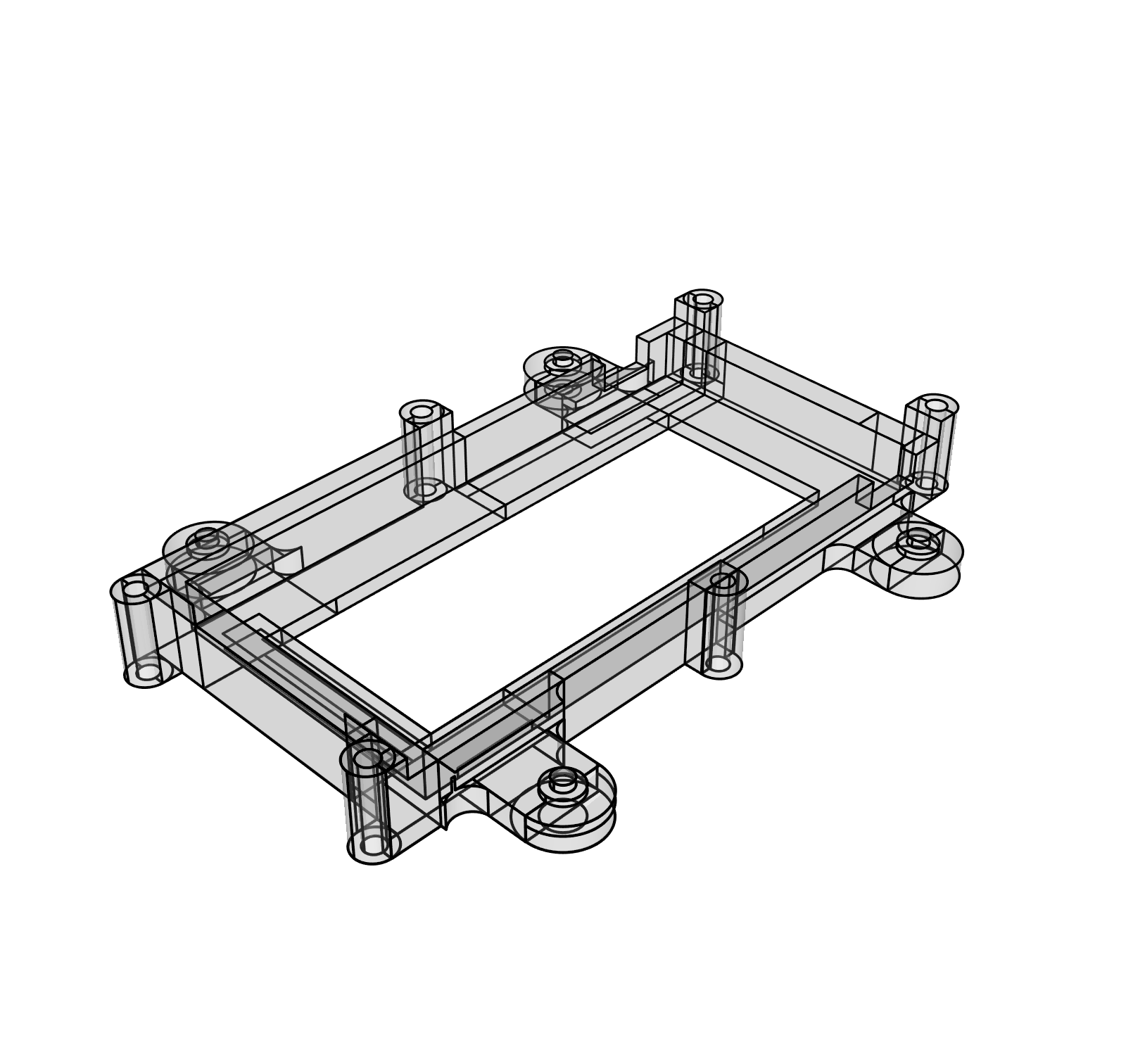} &
        \includegraphics[width=0.15\linewidth]{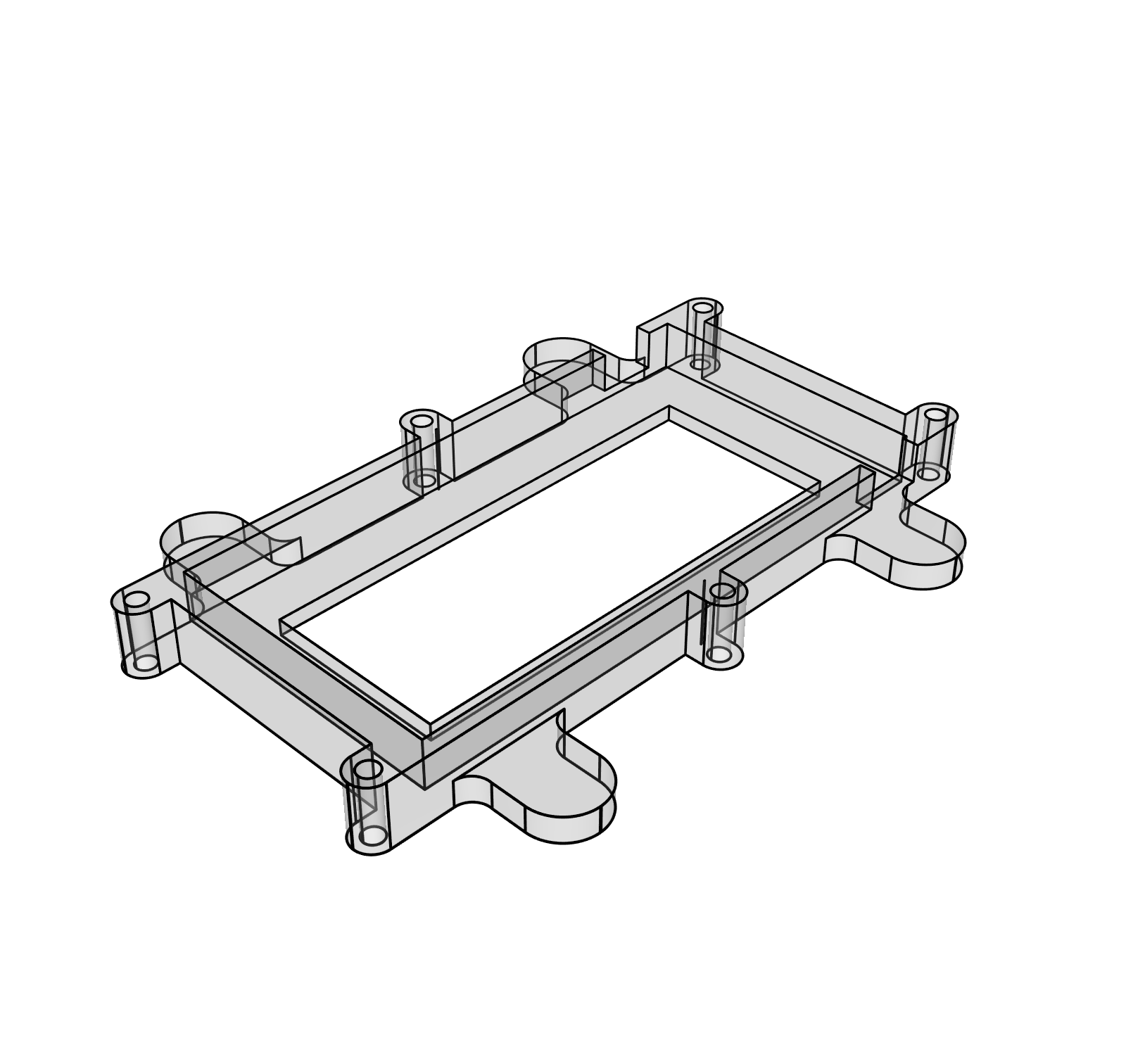} &
        \includegraphics[width=0.15\linewidth]{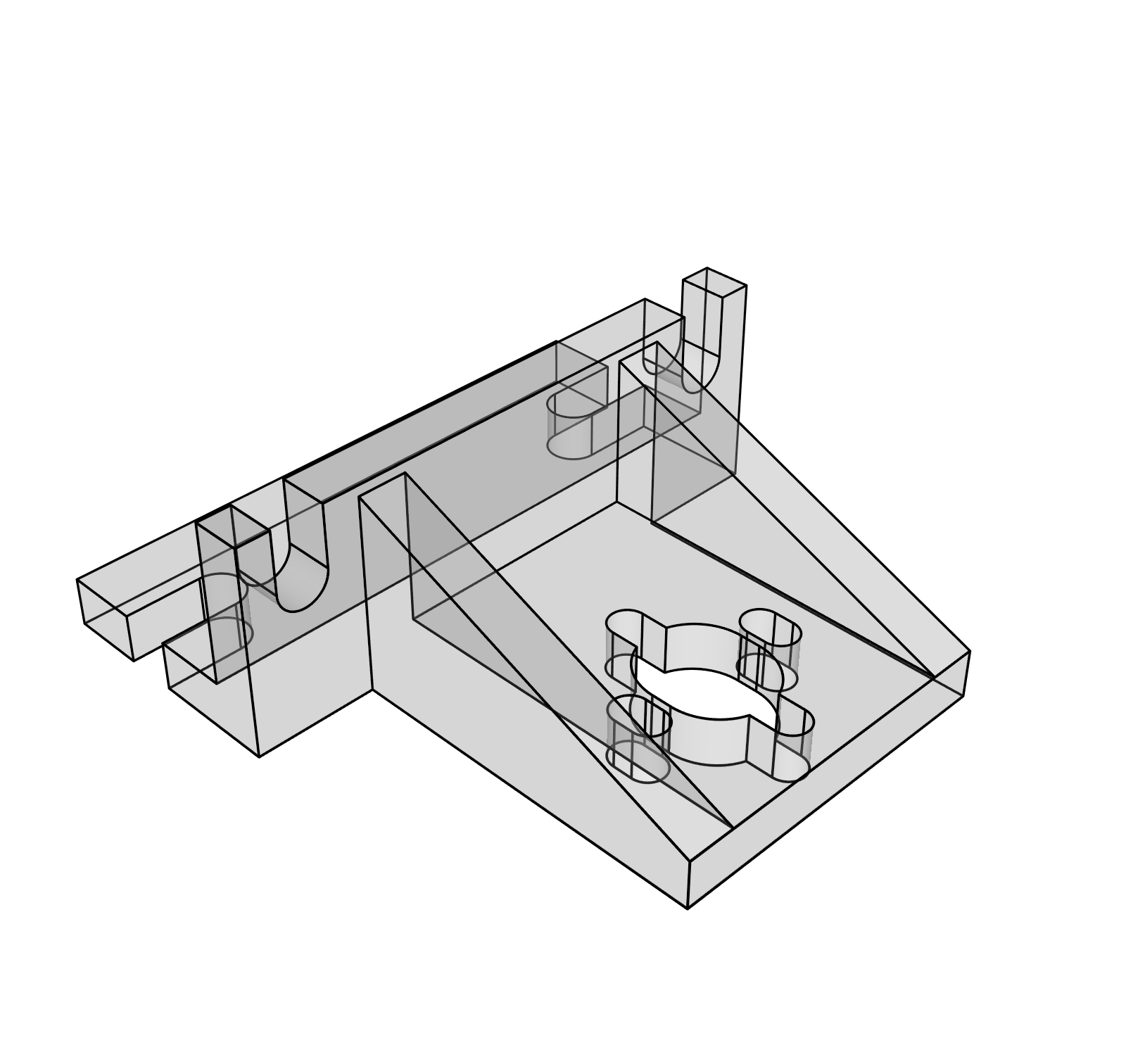} &
        \includegraphics[width=0.15\linewidth]{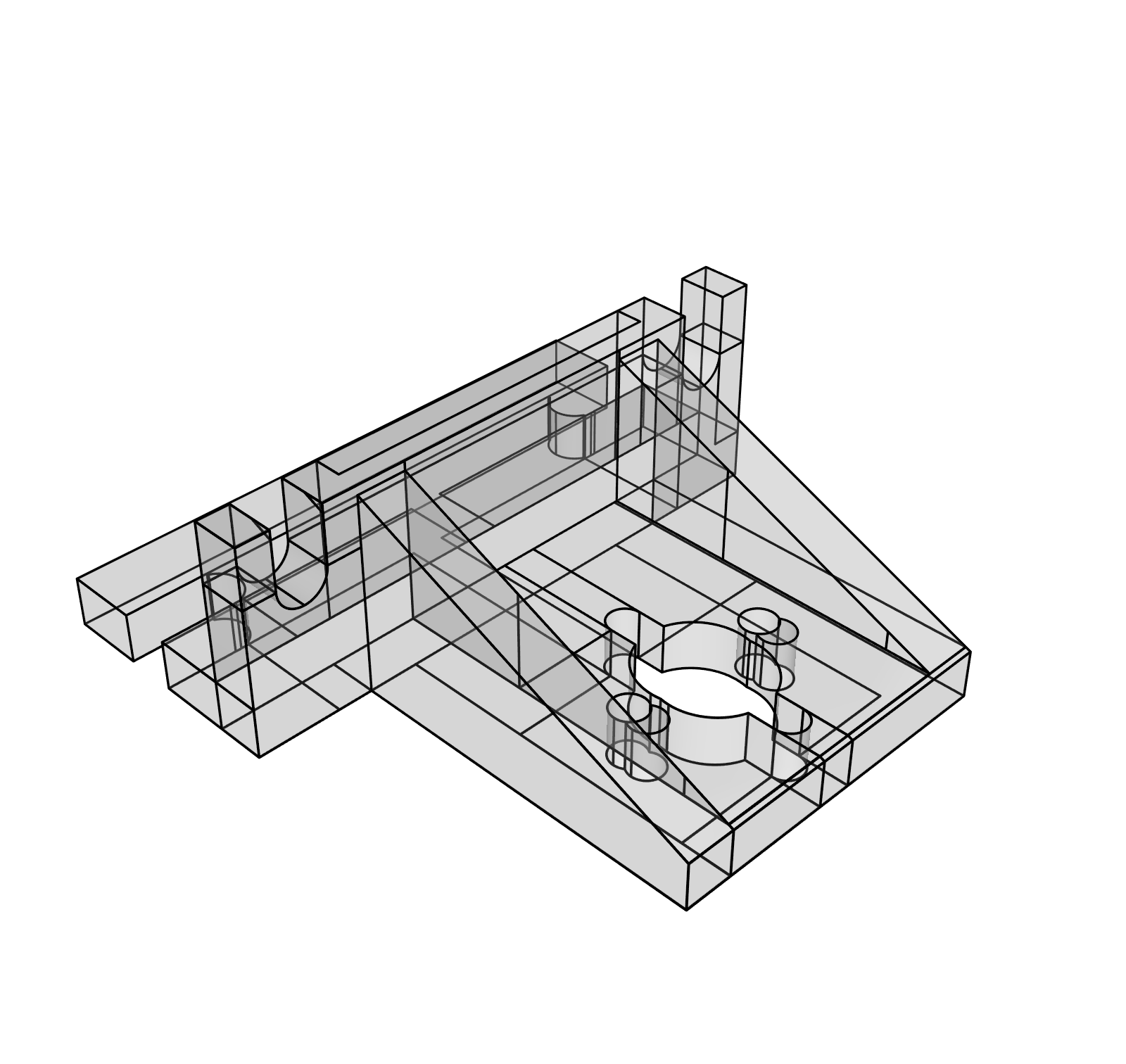} &
        \includegraphics[width=0.15\linewidth]{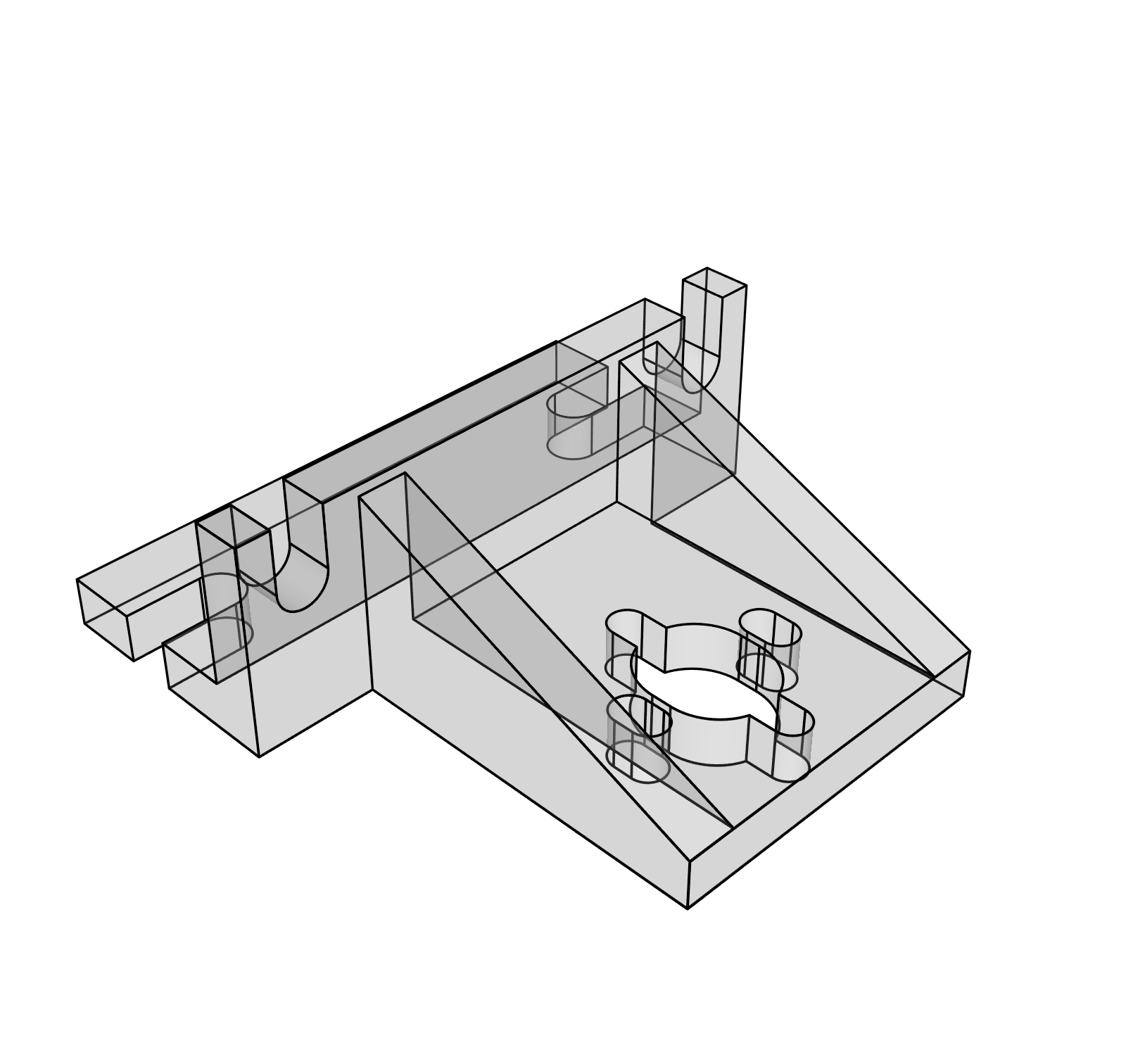}
        \\
        \includegraphics[width=0.15\linewidth]{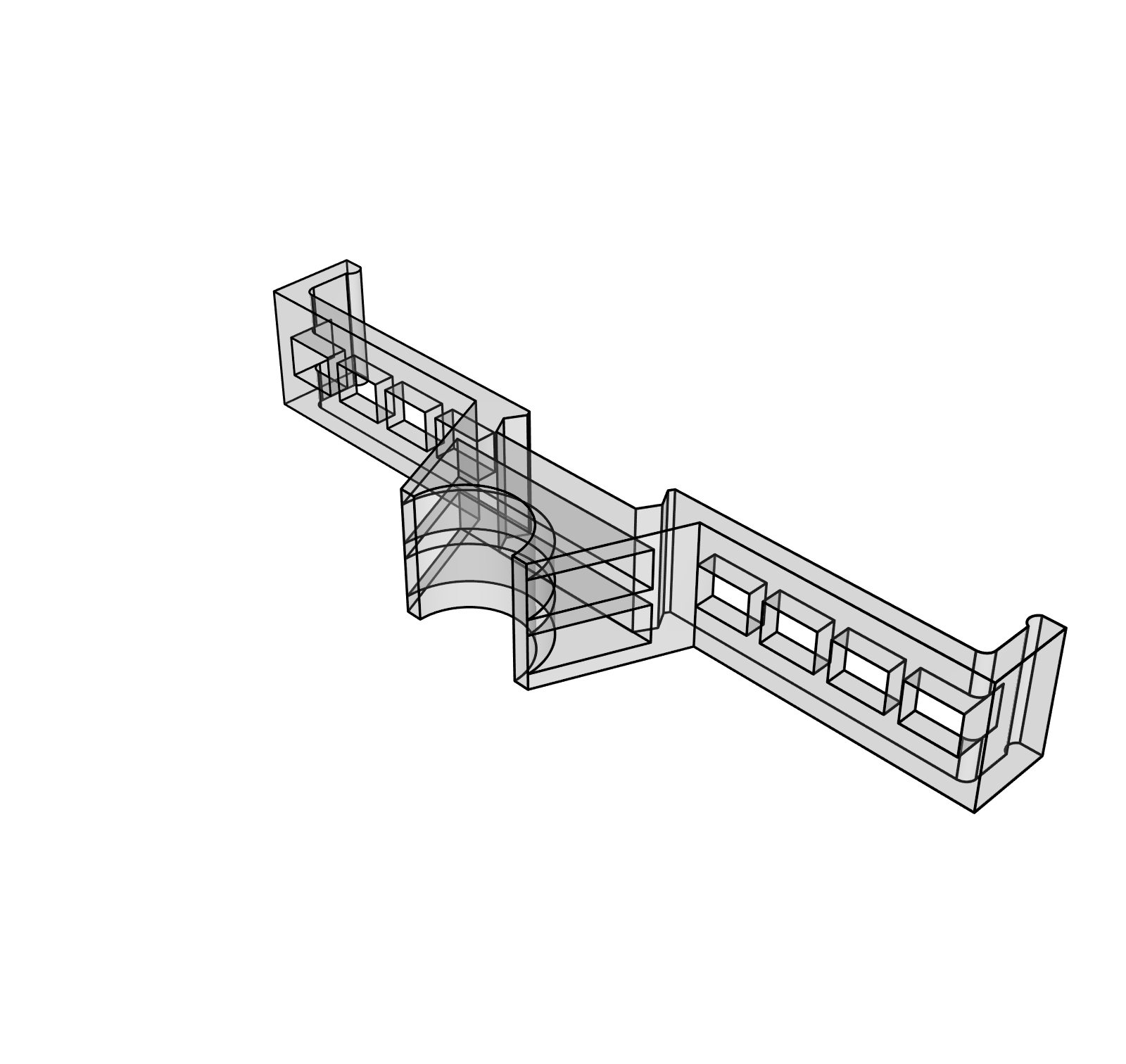} &
        \includegraphics[width=0.15\linewidth]{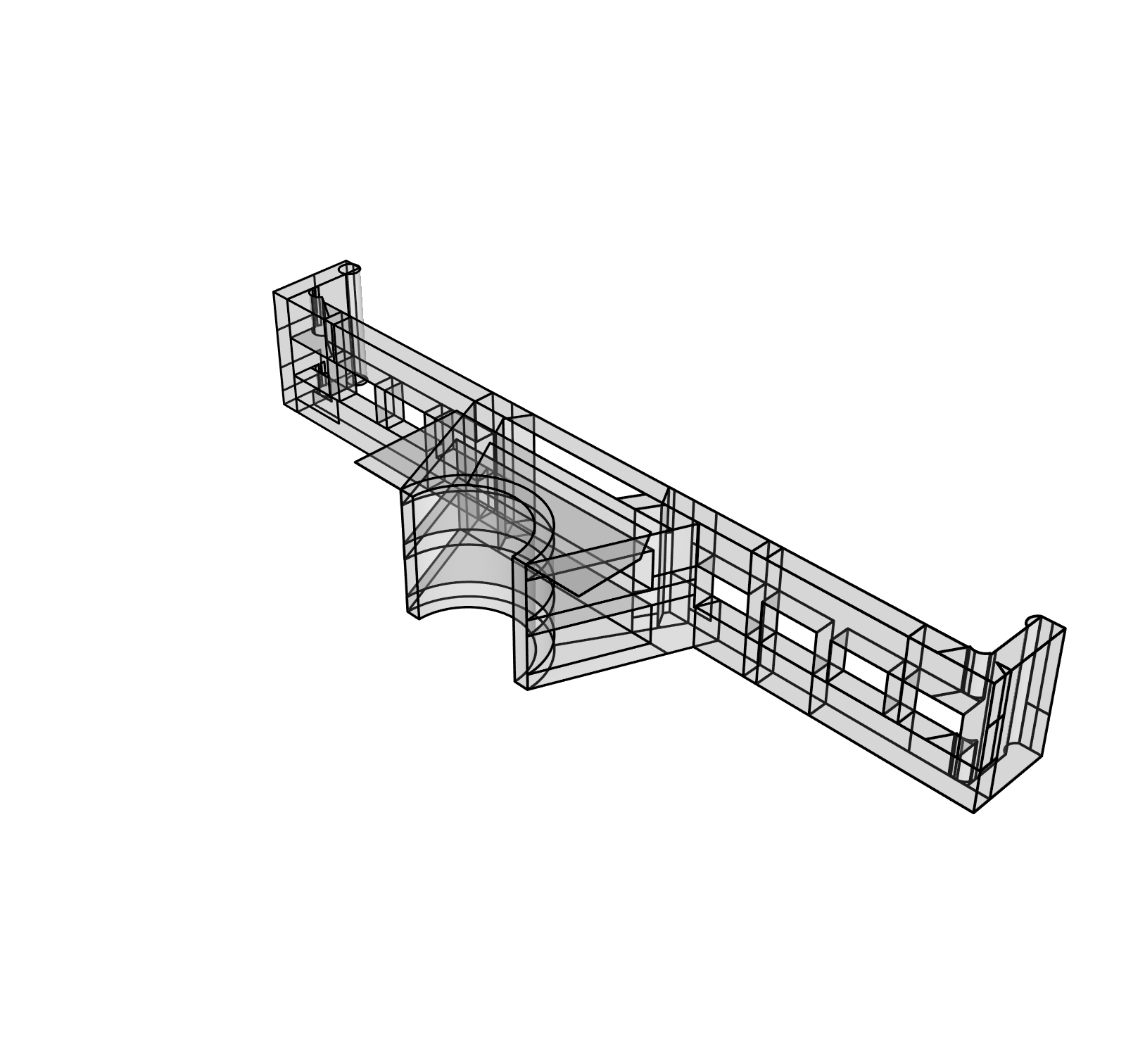} &
        \includegraphics[width=0.15\linewidth]{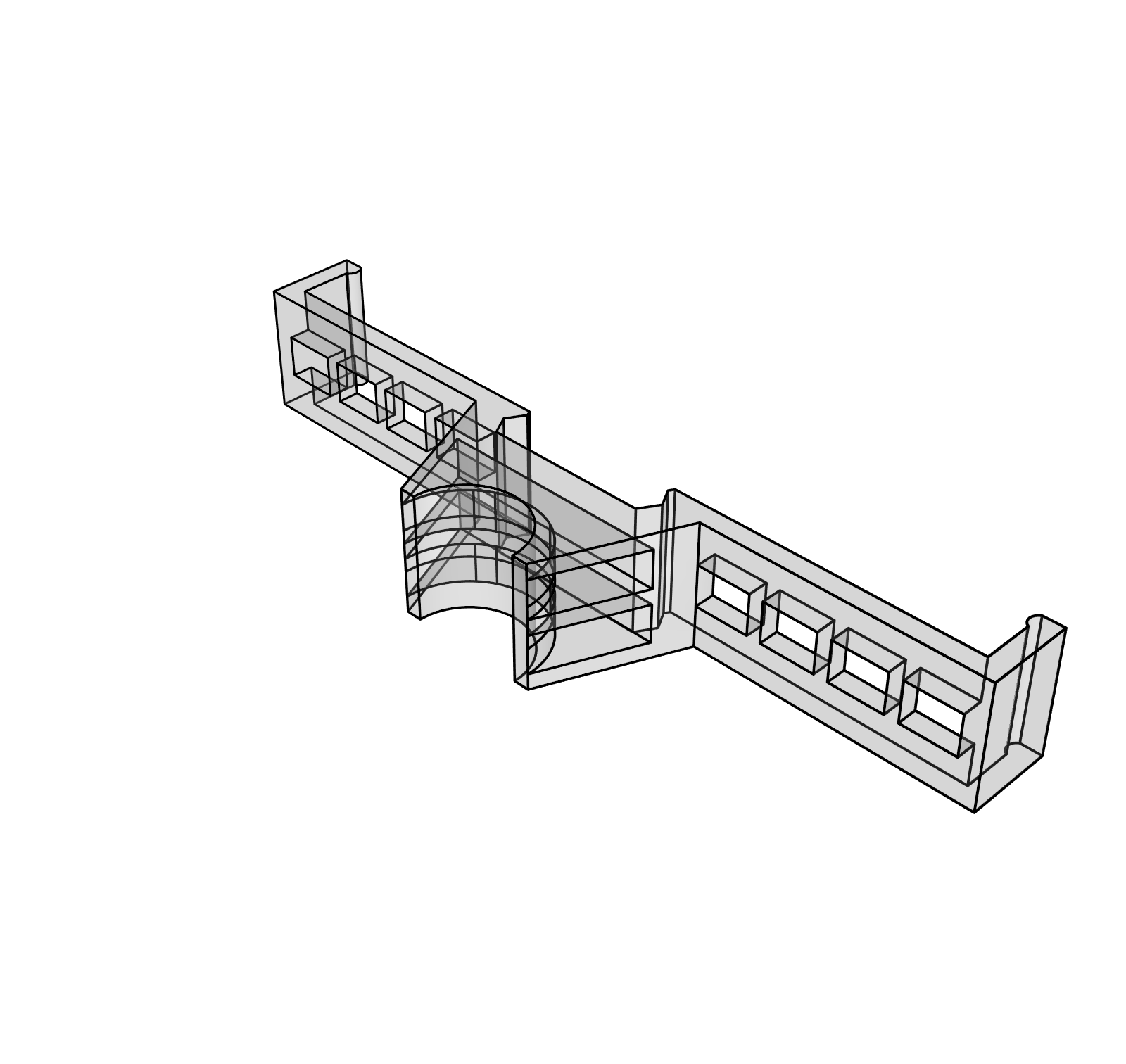} &
        \includegraphics[width=0.15\linewidth]{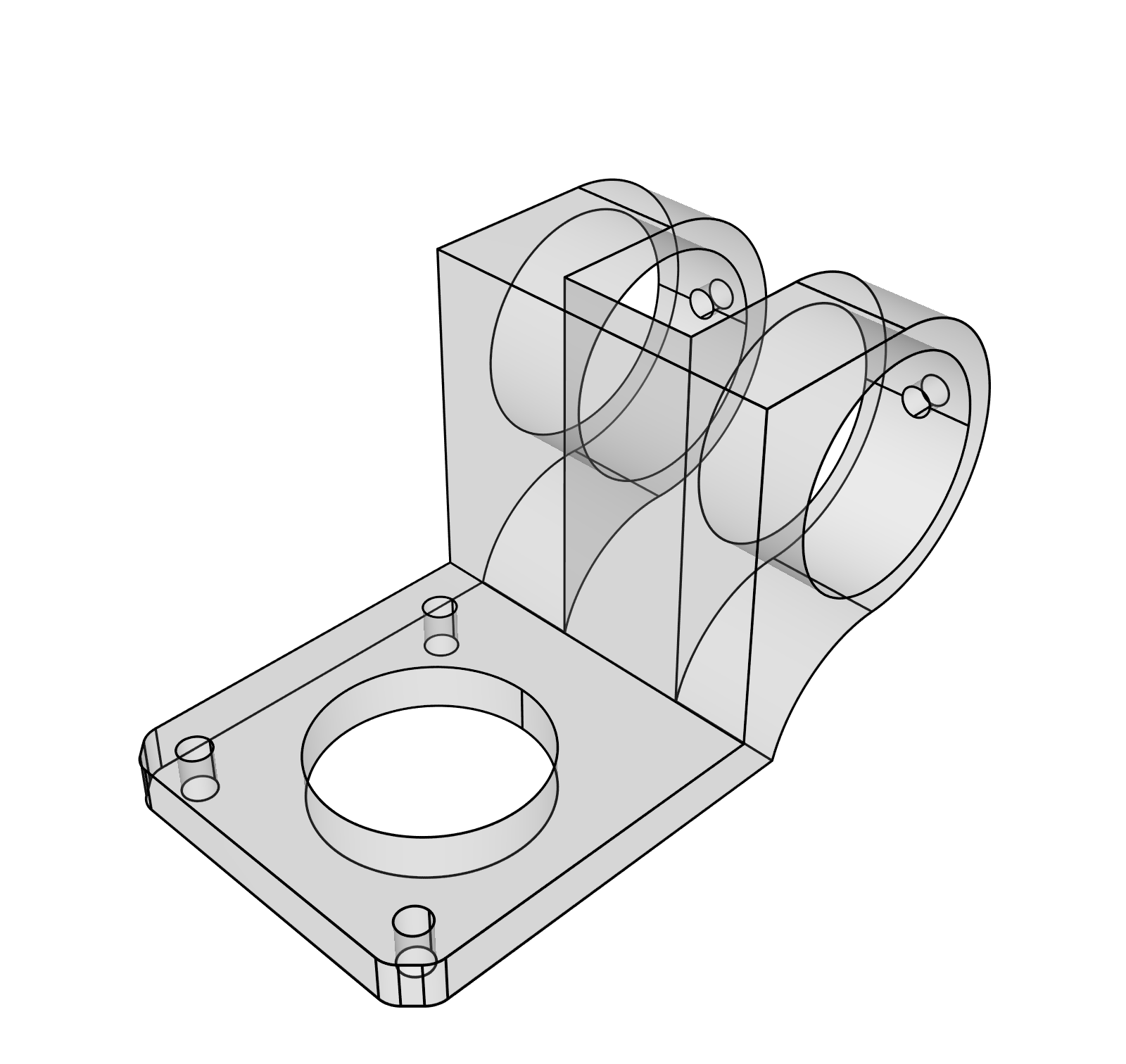} &
        \includegraphics[width=0.15\linewidth]{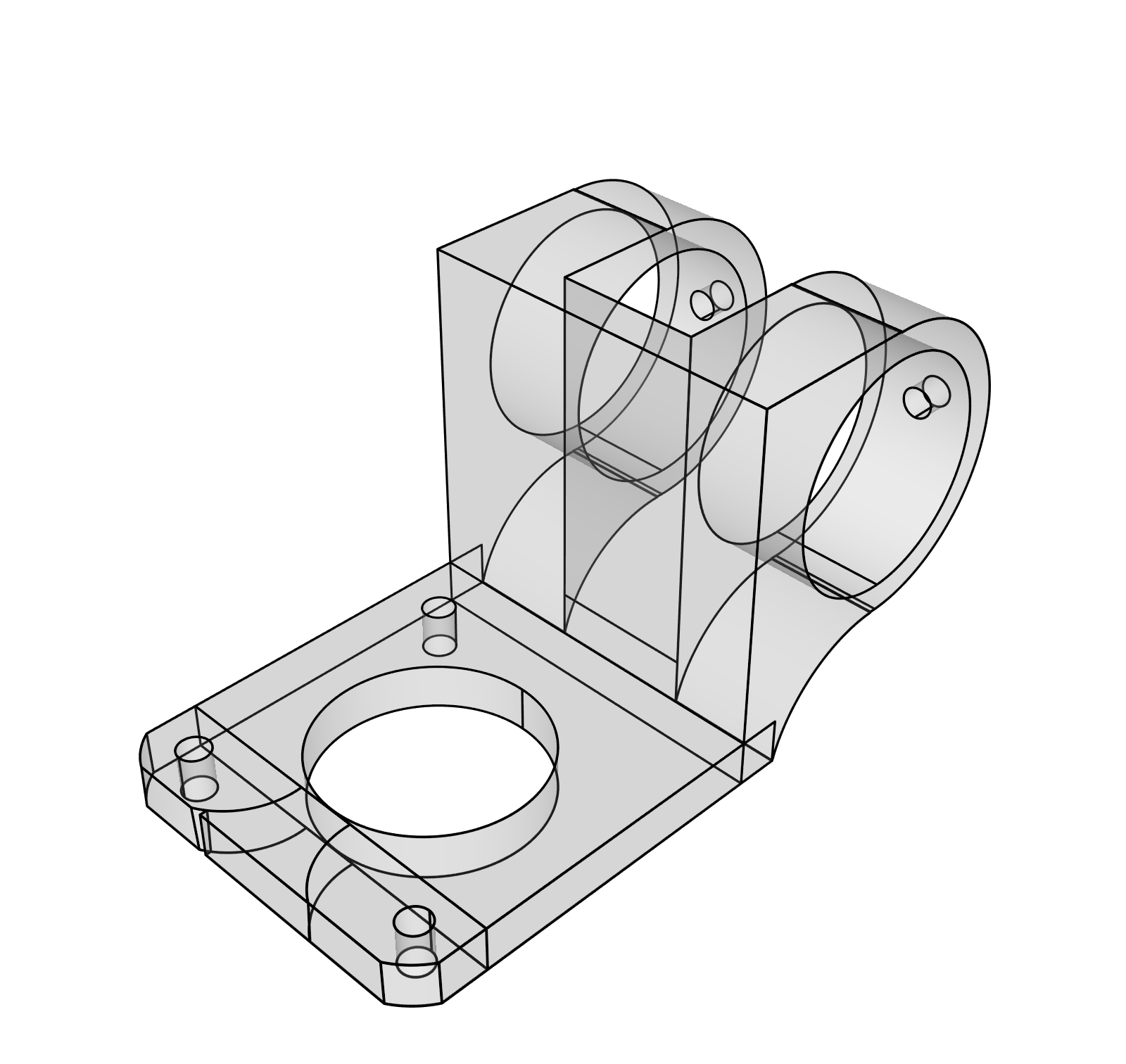} &
        \includegraphics[width=0.15\linewidth]{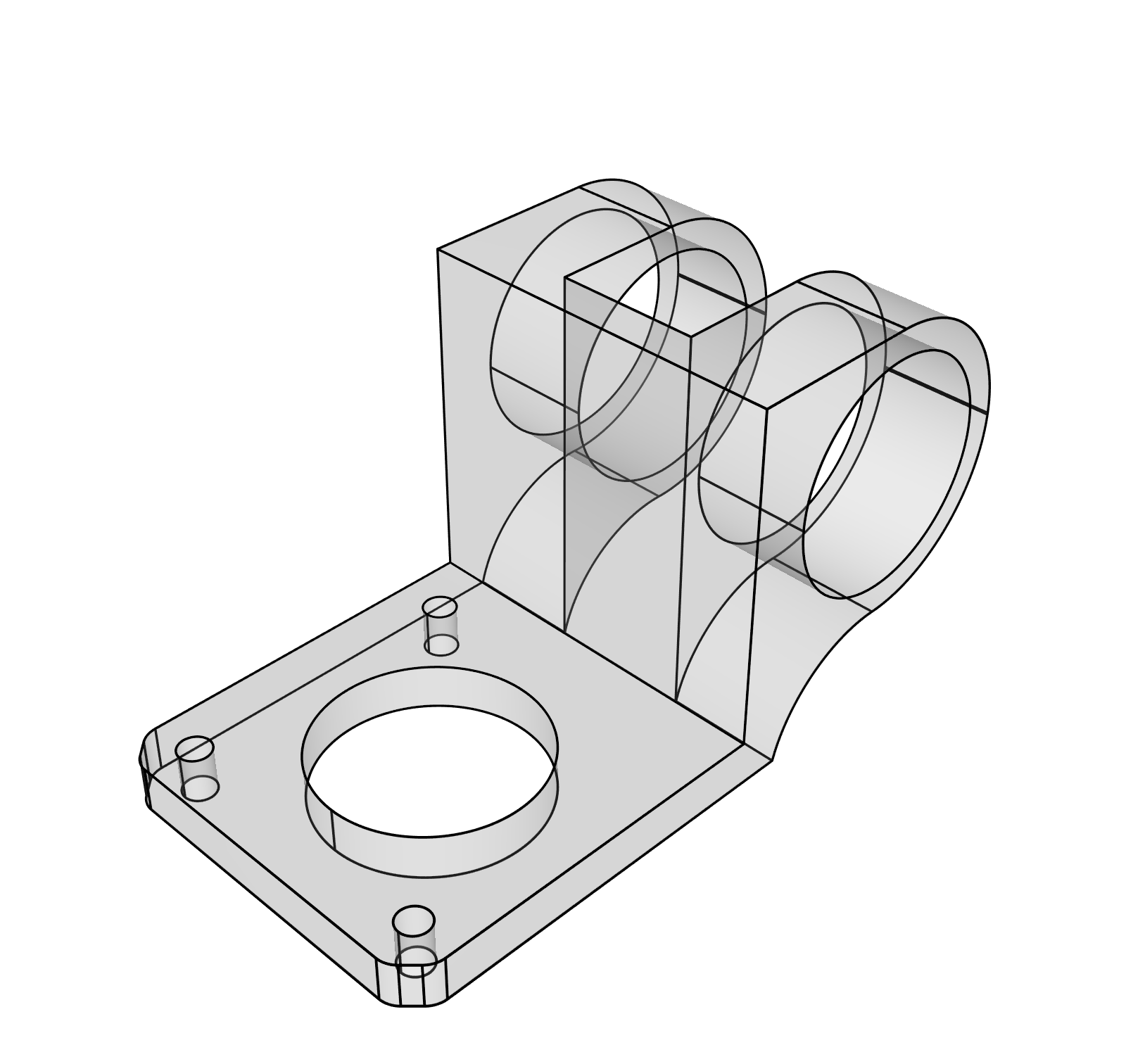}
        \\
        \includegraphics[width=0.15\linewidth]{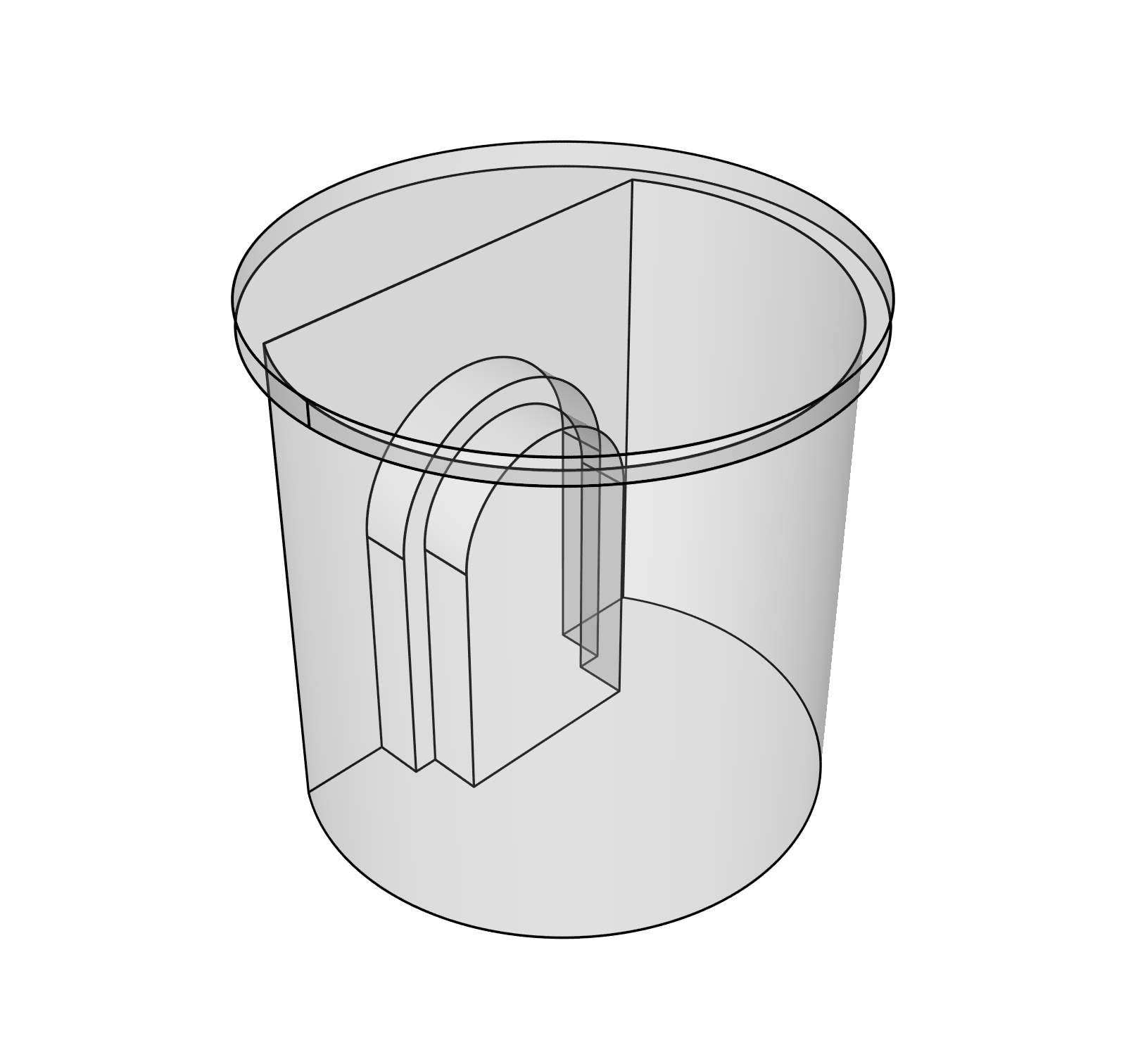} &
        \includegraphics[width=0.15\linewidth]{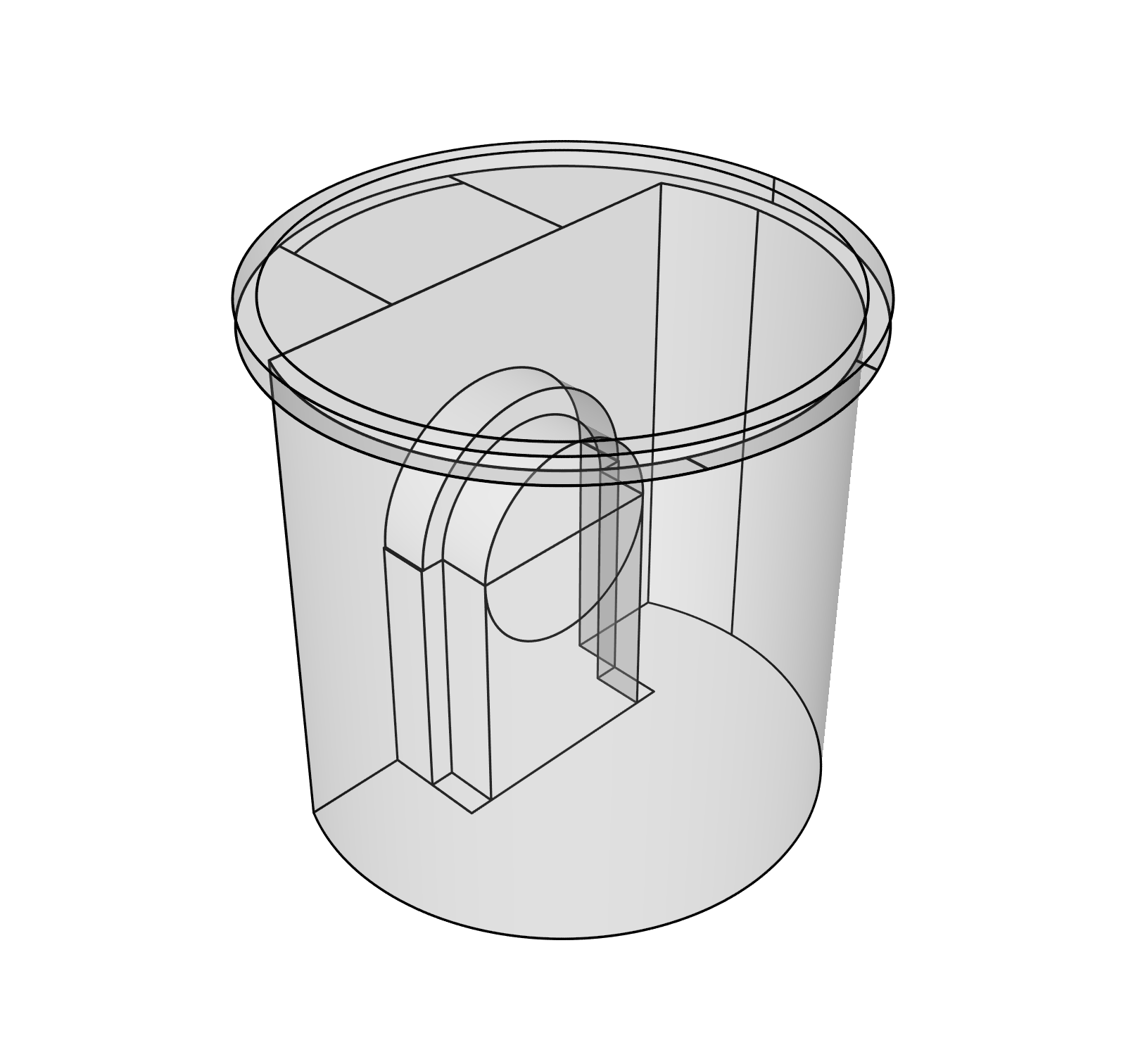} &
        \includegraphics[width=0.15\linewidth]{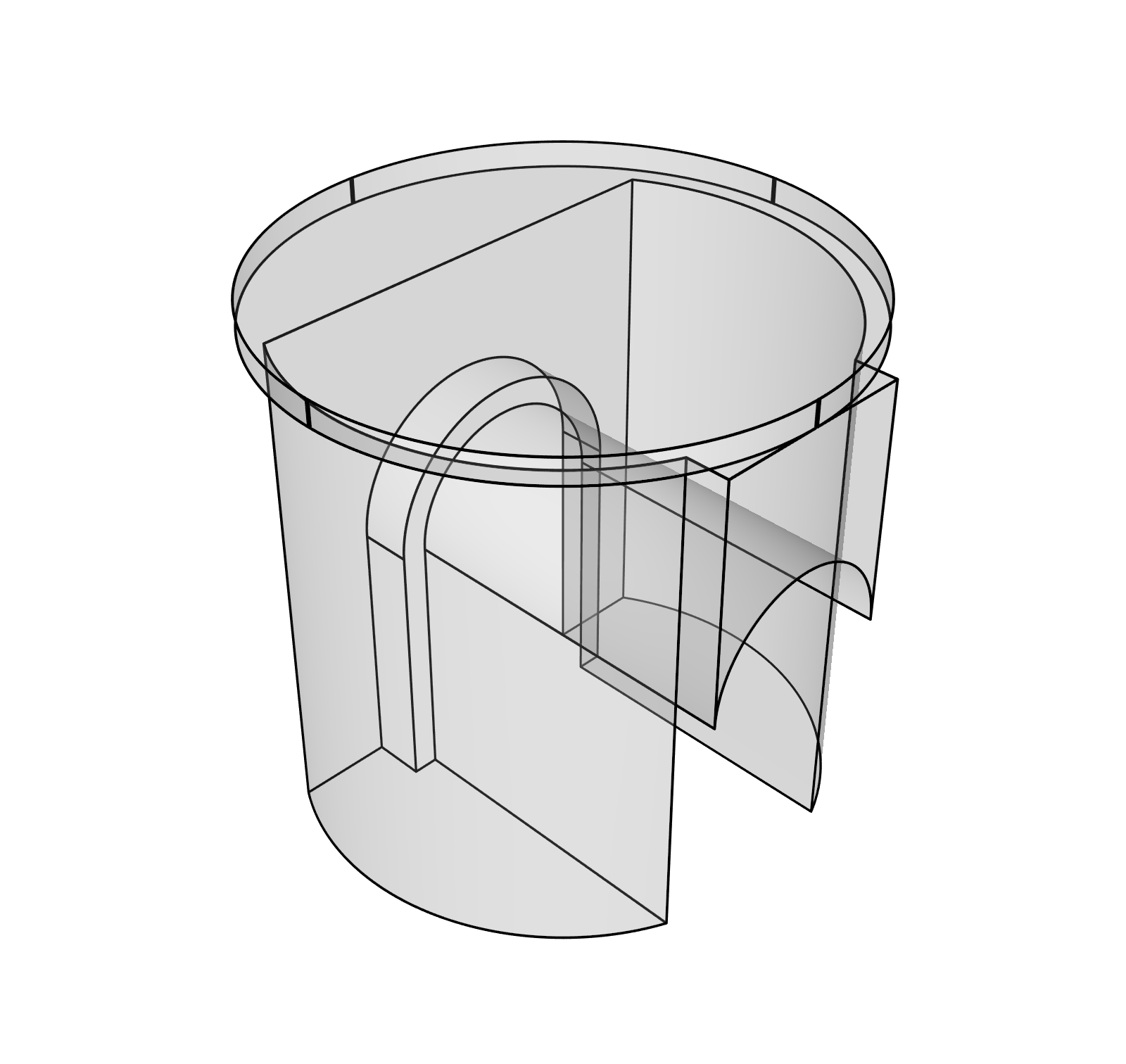} &
        \includegraphics[width=0.15\linewidth]{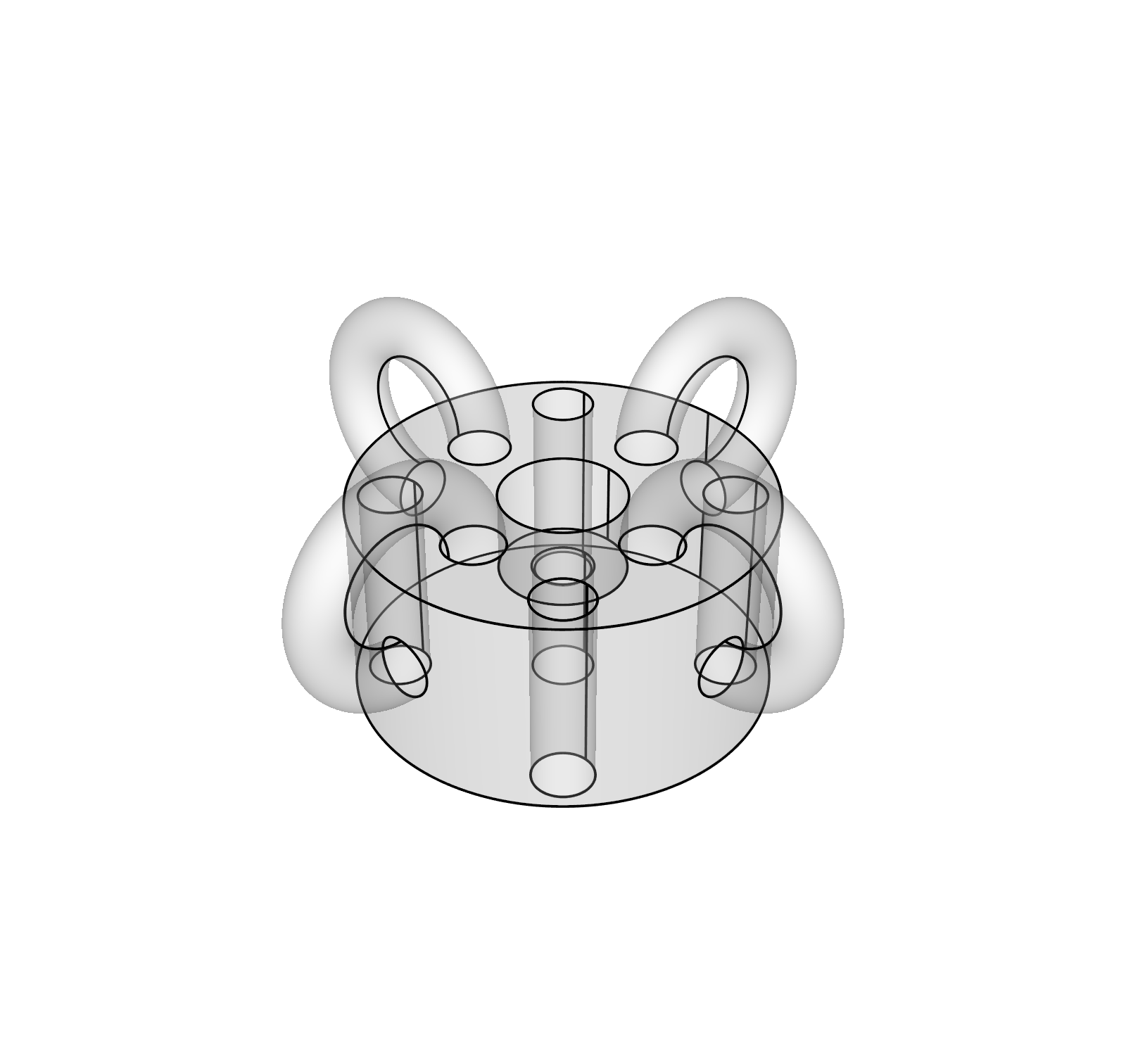} &
        \includegraphics[width=0.15\linewidth]{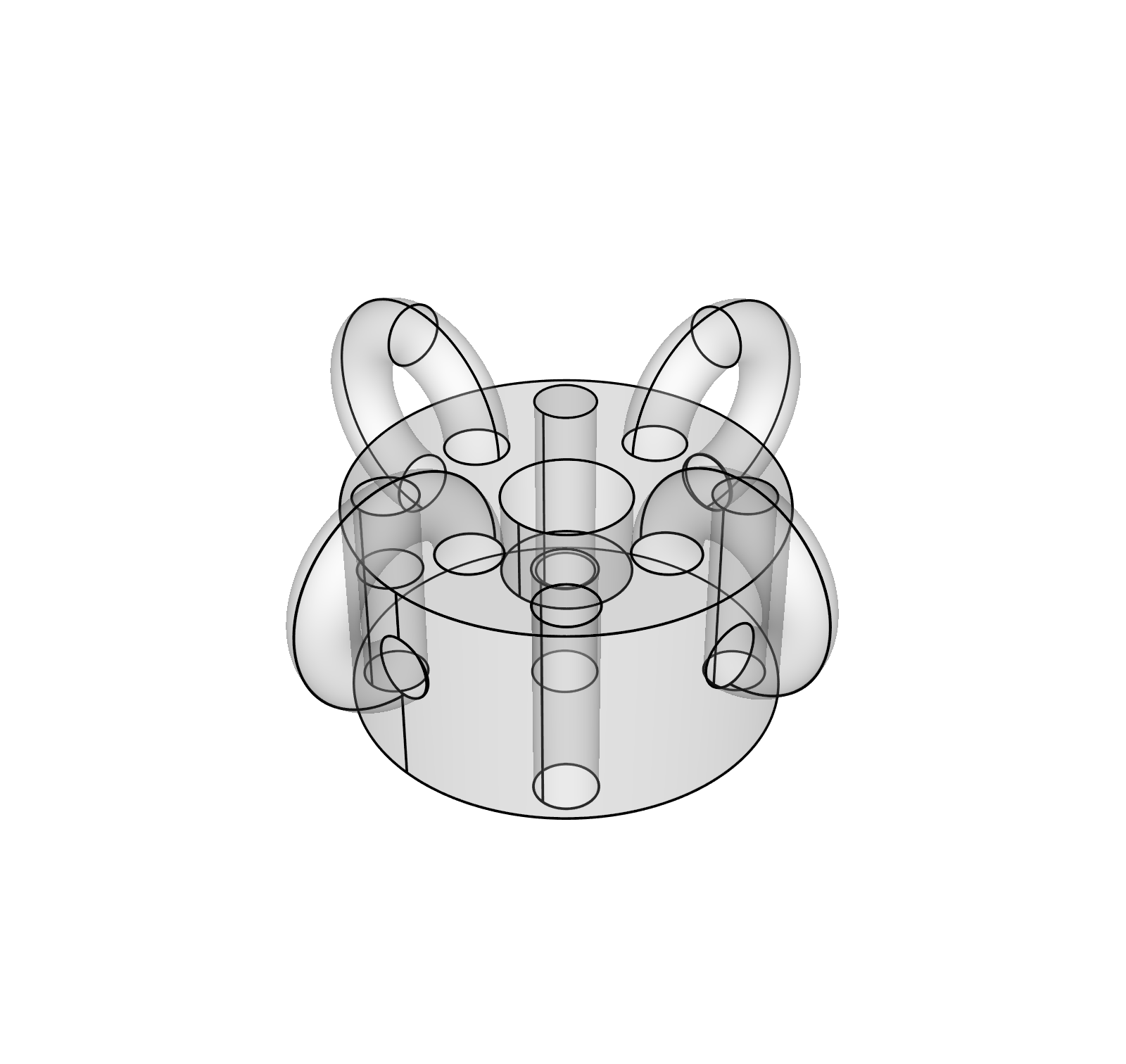} &
        \includegraphics[width=0.15\linewidth]{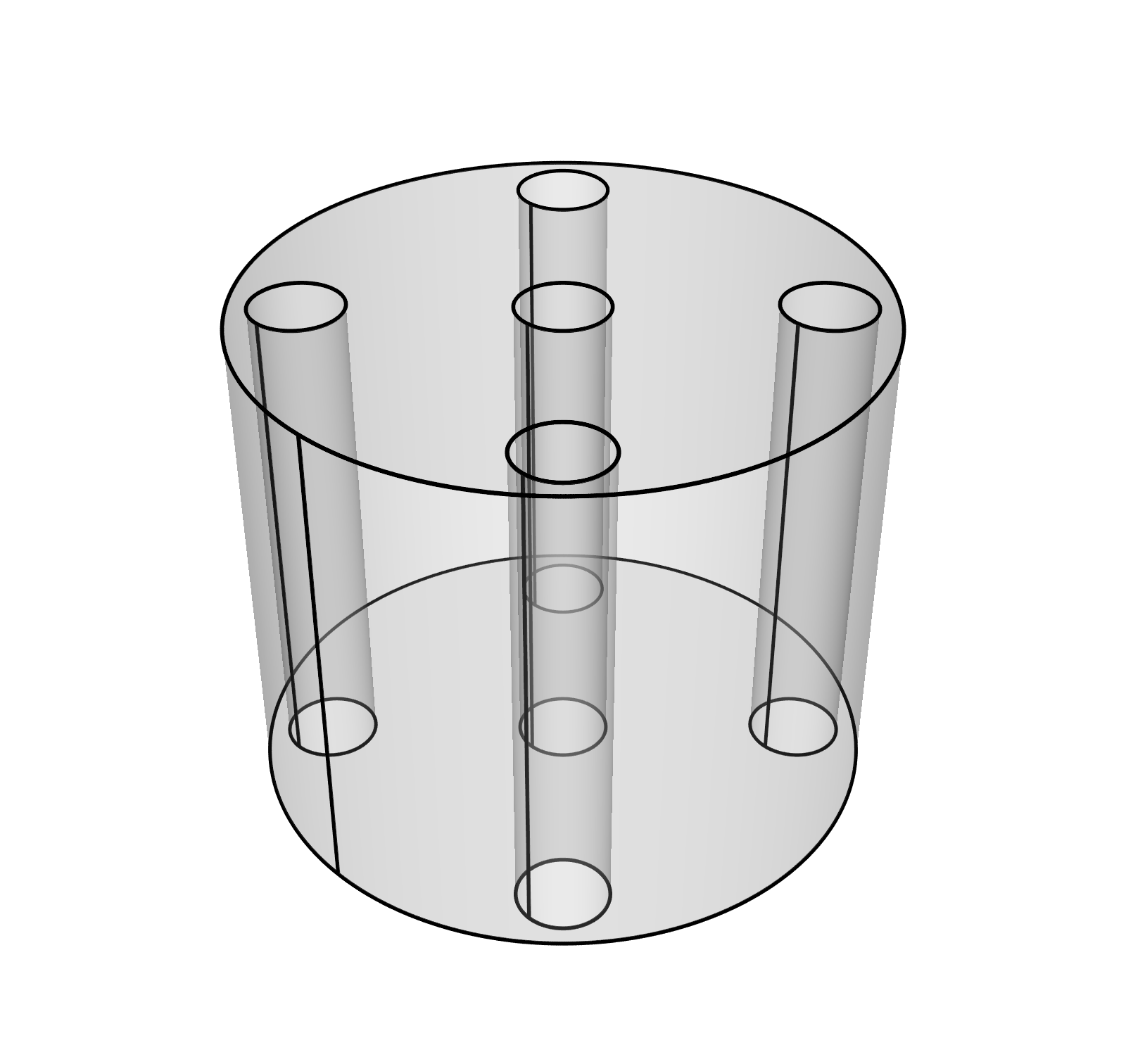}
        \\
        \includegraphics[width=0.15\linewidth]{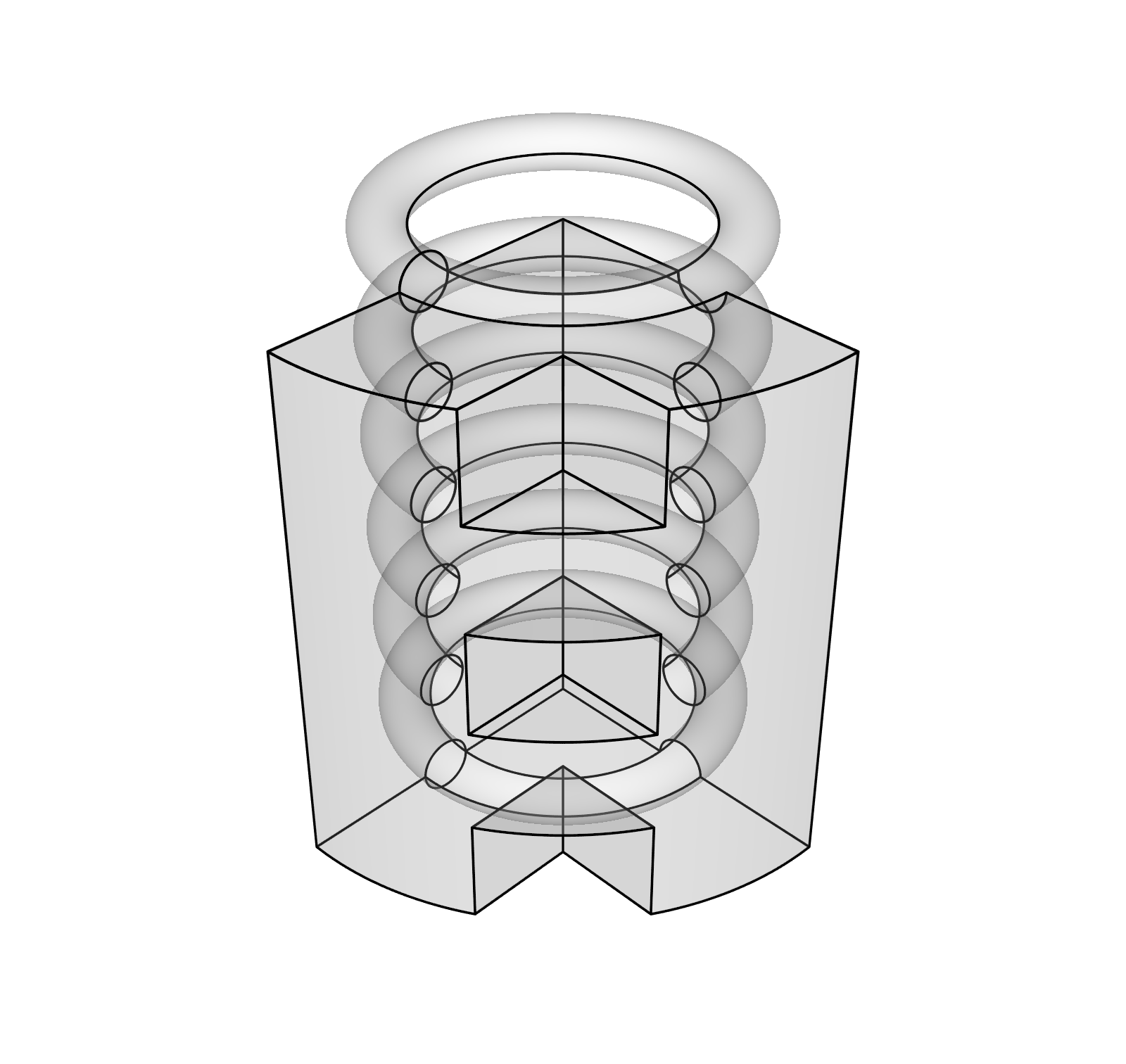} &
        \includegraphics[width=0.15\linewidth]{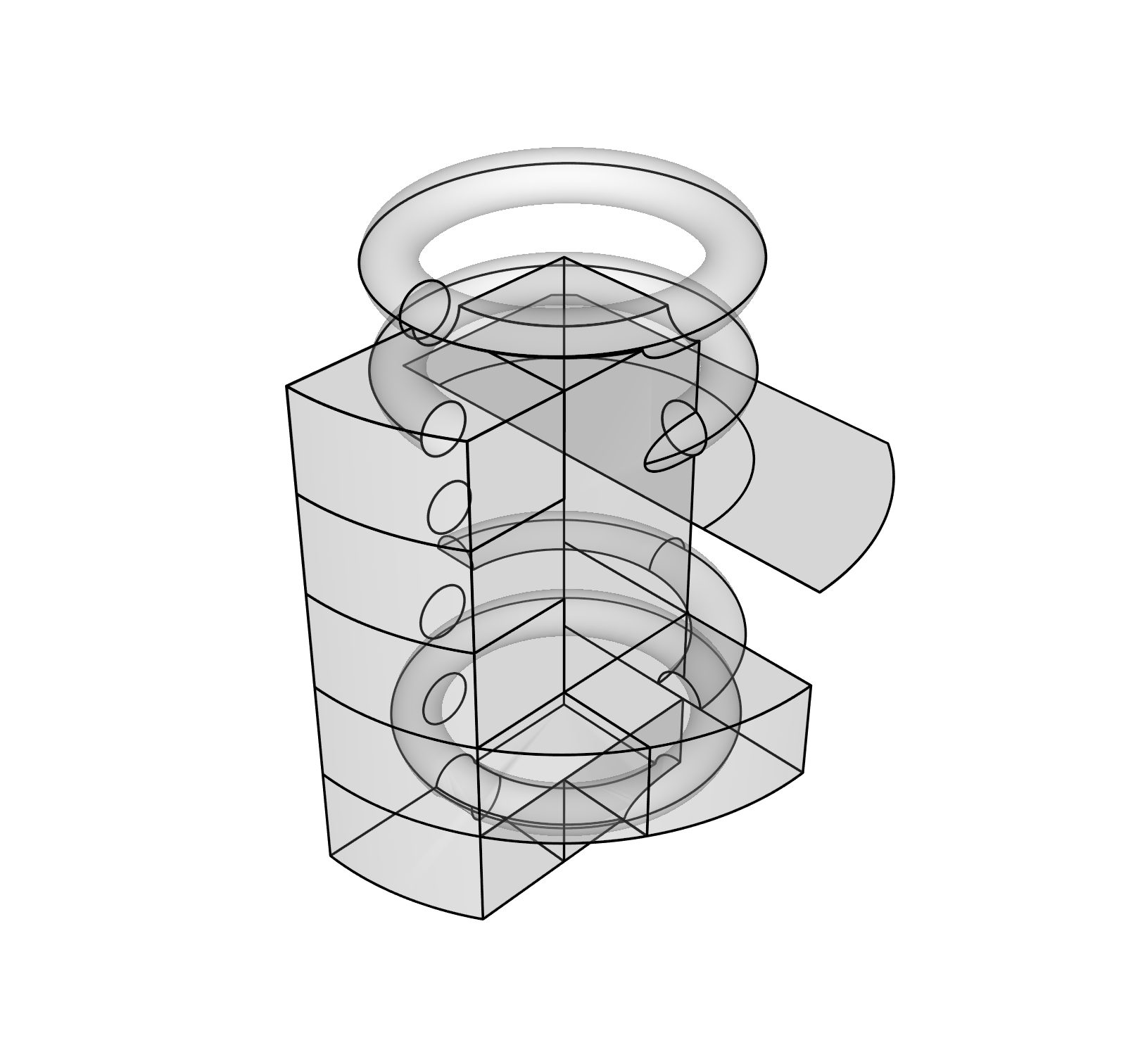} &
        \includegraphics[width=0.15\linewidth]{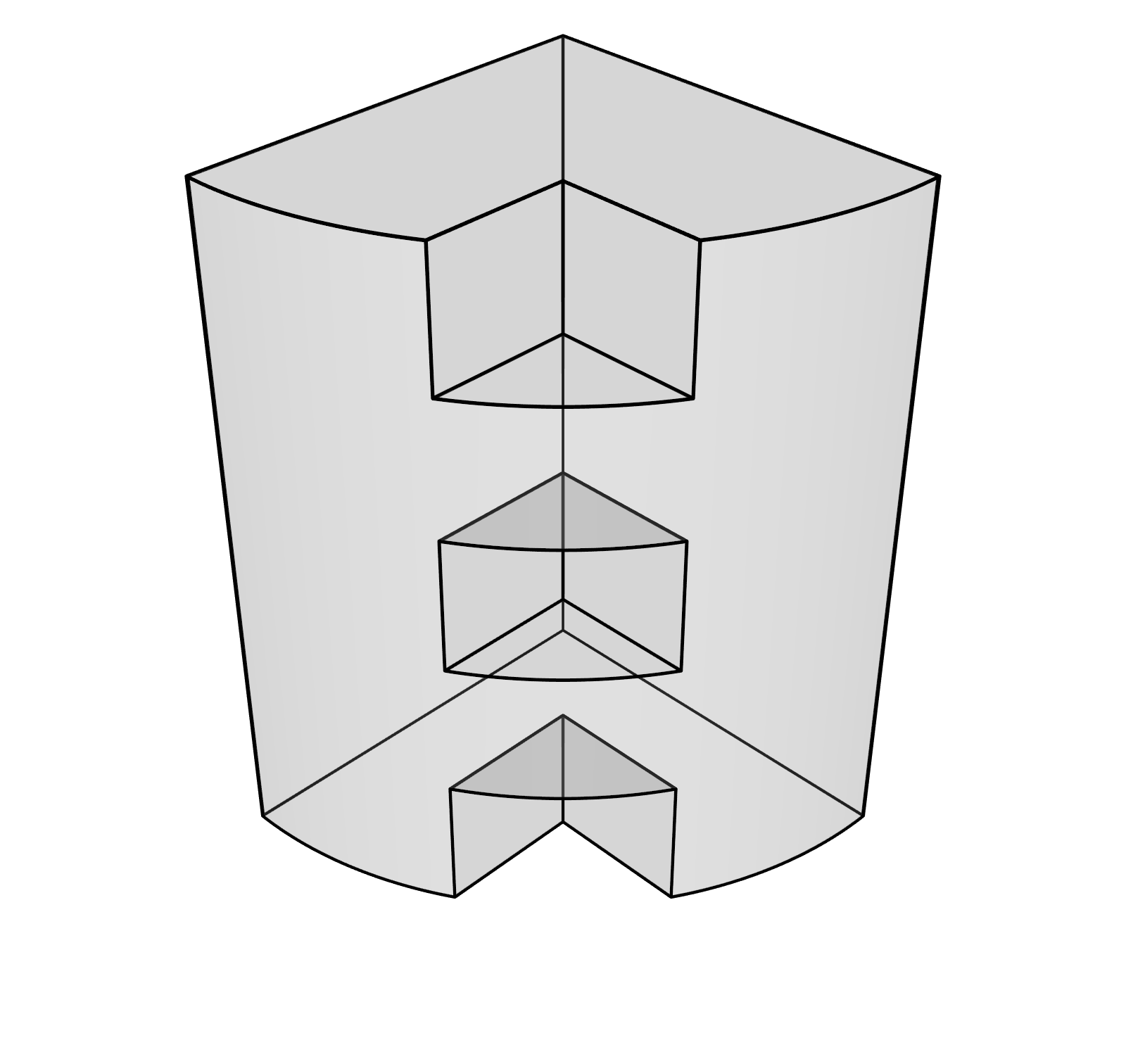} &
        \includegraphics[width=0.15\linewidth]{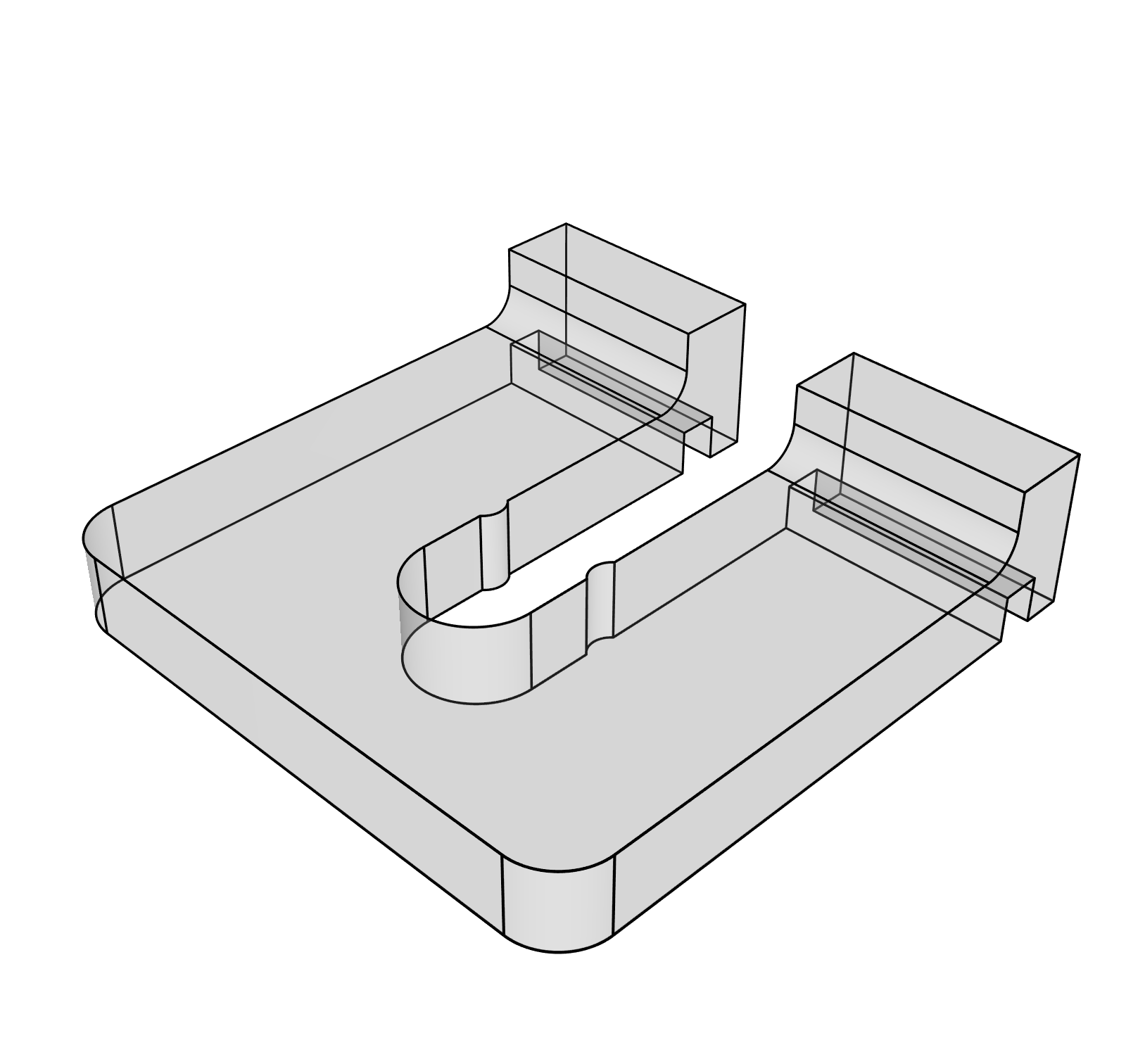} &
        \includegraphics[width=0.15\linewidth]{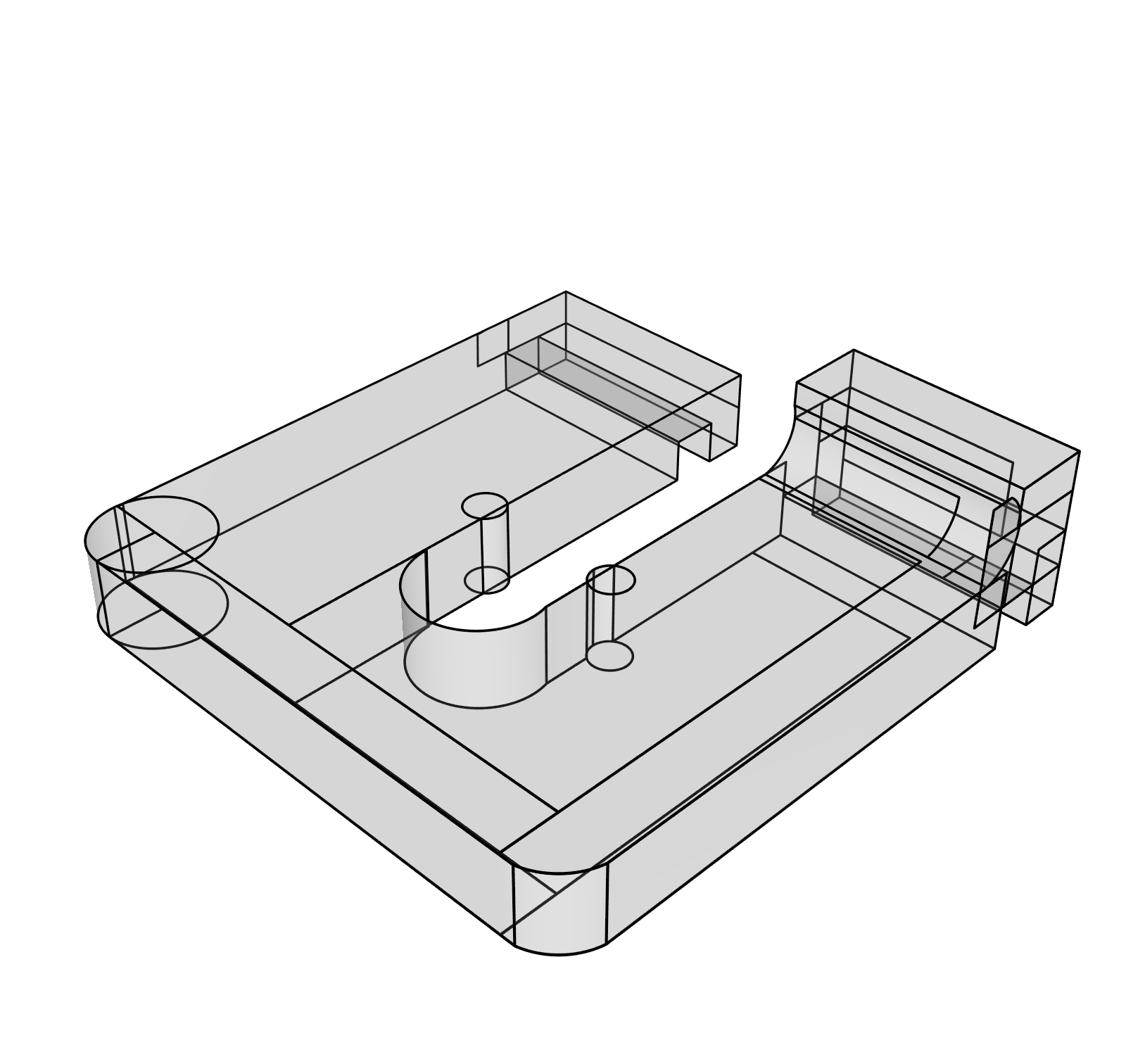} &
        \includegraphics[width=0.15\linewidth]{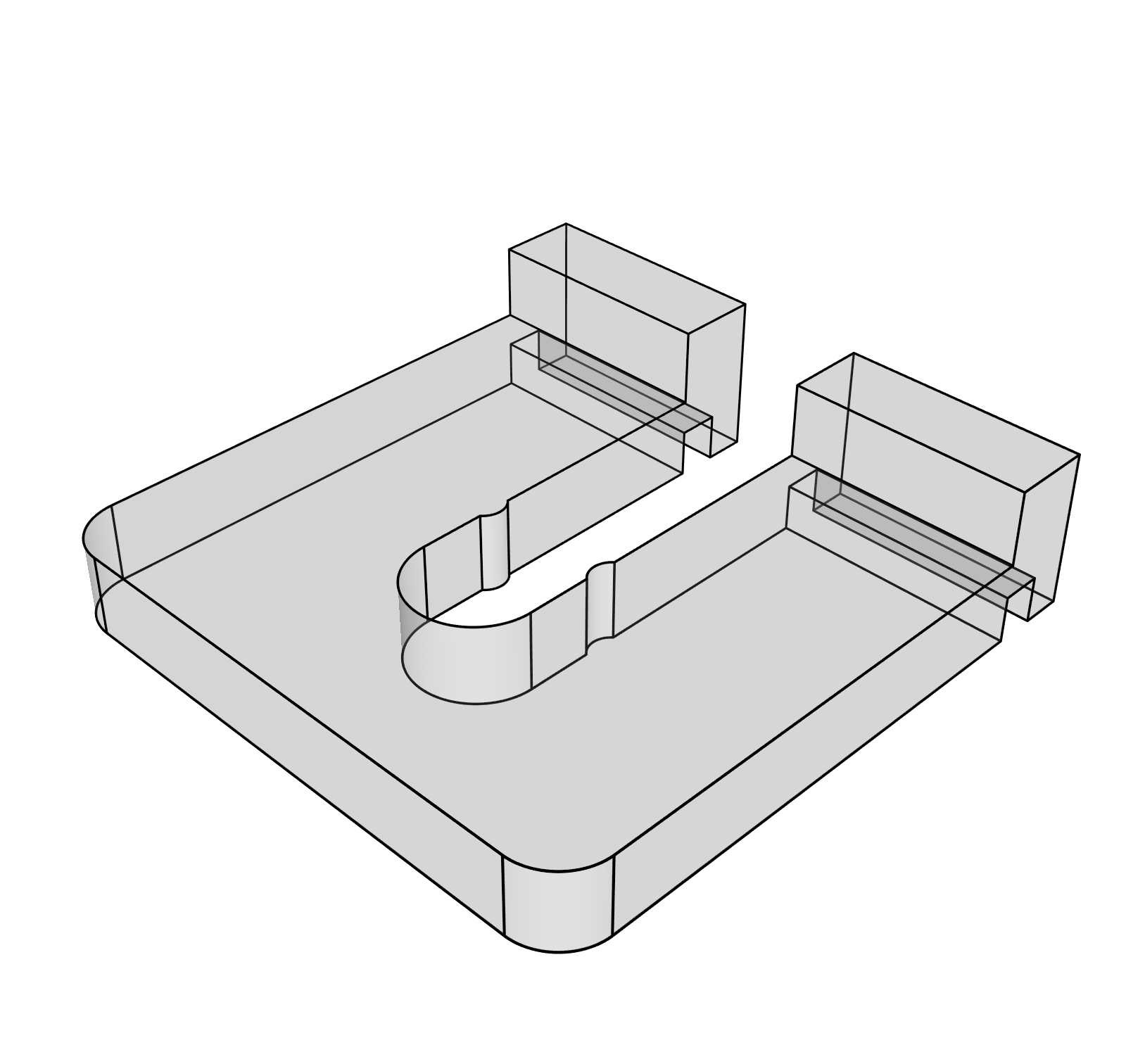} 
        \\
        \includegraphics[width=0.15\linewidth]{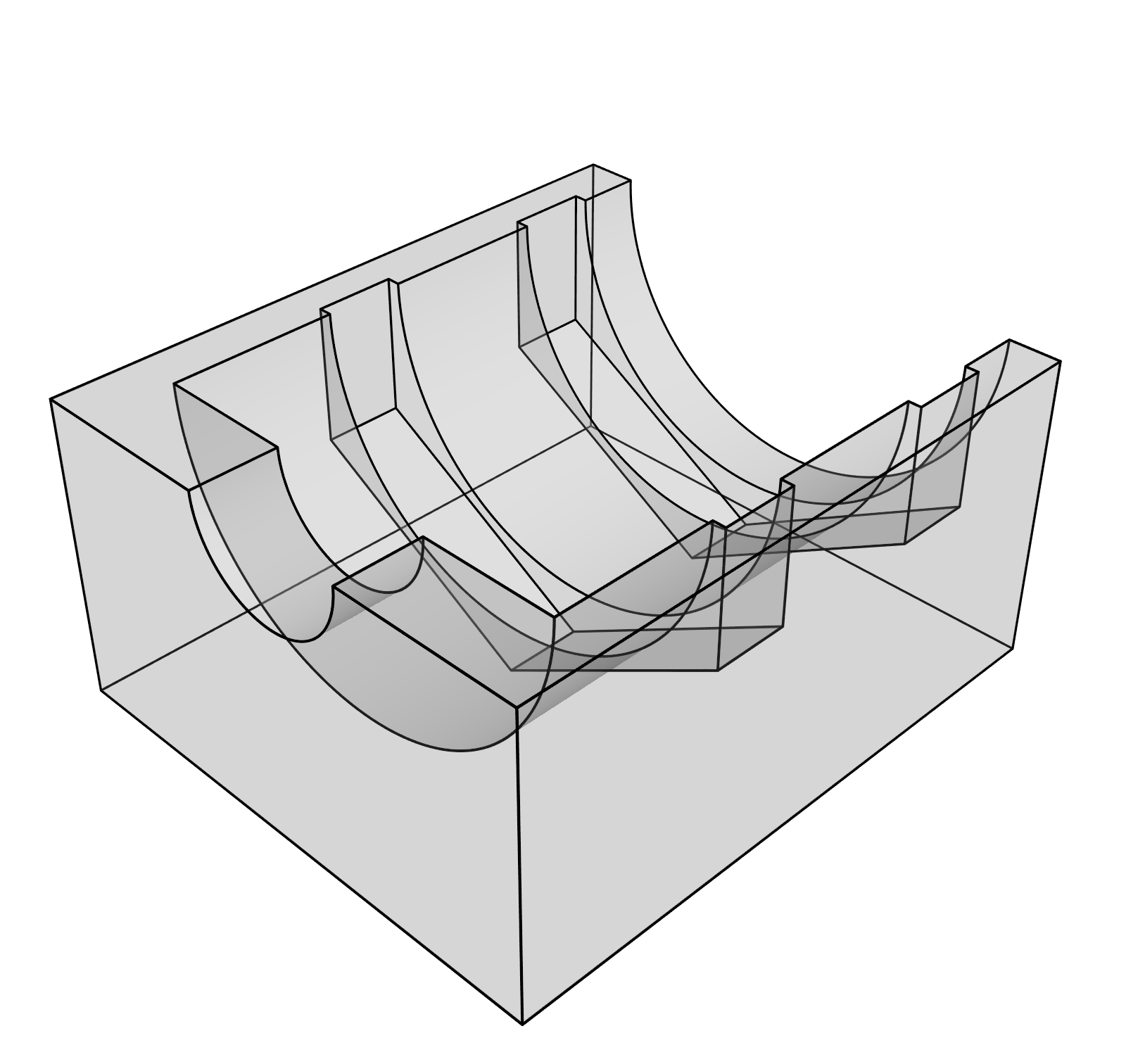} &
        \includegraphics[width=0.15\linewidth]{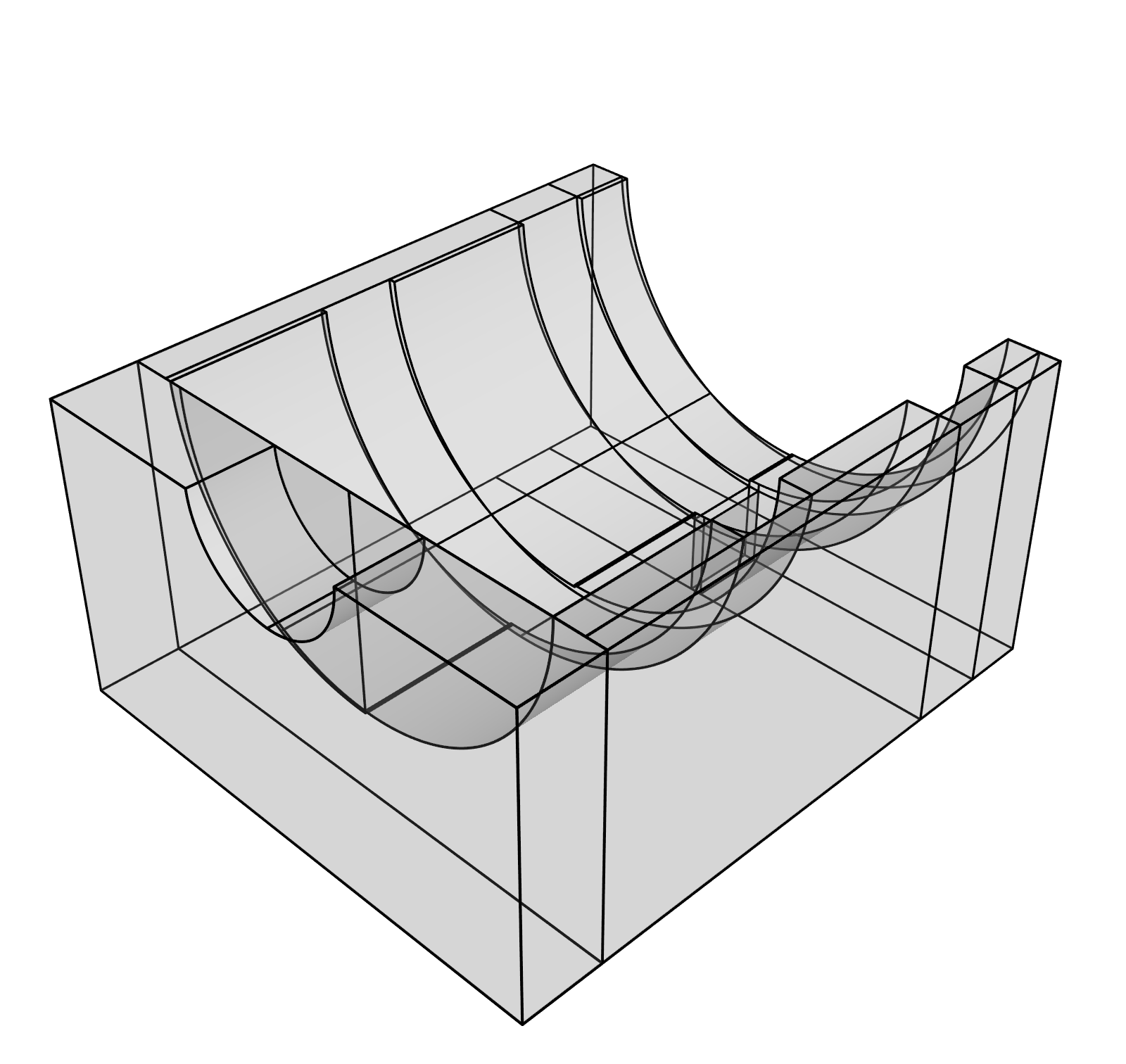} &
        \includegraphics[width=0.15\linewidth]{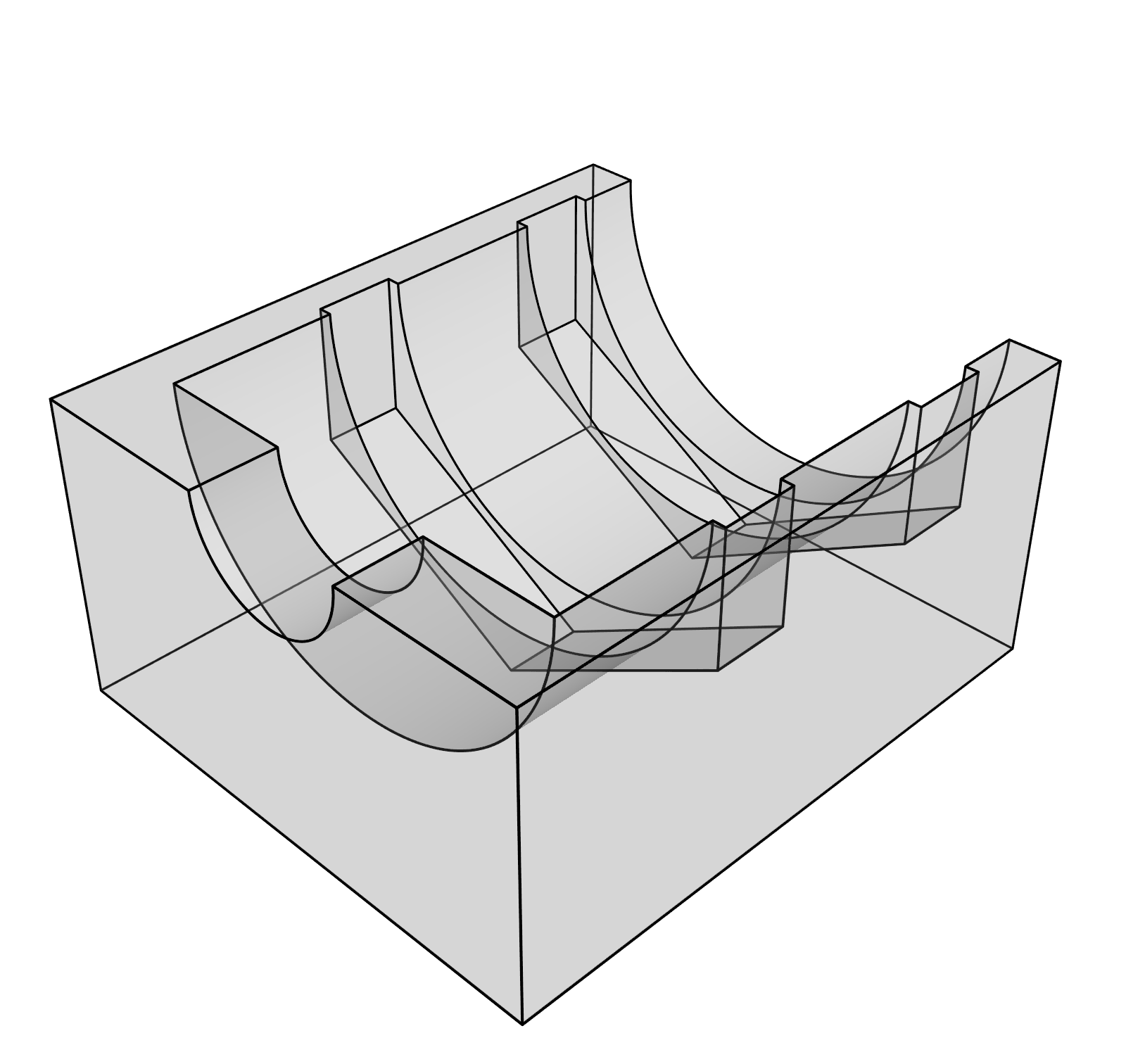} & & &
    \end{tabular}
    \caption{
    Reconstruction result comparison of InverseCSG vs Ours.
    }
    \label{figure:reconstruction_comparison}
\end{figure*}

Please see Figure~\ref{figure:qualitative_comparision_inv1}, \ref{figure:qualitative_comparision_inv2} and \ref{figure:qualitative_comparision_inv3} for examples of the qualitative comparison of the output of our model's inferred programs vs. those of InverseCSG. See Figure \ref{figure:reconstruction_comparison} for examples of the reconstruction results of our model's reconstruction vs. those of InverseCSG.

Figure~\ref{figure:comparison_histogram}, top compares the reconstruction error of our inferred programs to those of InverseCSG and Figure~\ref{figure:comparison_histogram}, bottom compares the search time of each method. Reconstruction error is calculated using 1-IoU. The search time for InverseCSG is taken from their paper and the InverseCSG error rate is calculated using the reconstructed models released by the InverseCSG authors.

\begin{figure*}[t!]
    \centering
    \setlength{\tabcolsep}{1pt}
    \begin{tabular}{cccccccc}
        \multicolumn{2}{c}{Target} & & & &
        \\
        \multicolumn{2}{c}{\includegraphics[width=0.25\linewidth]{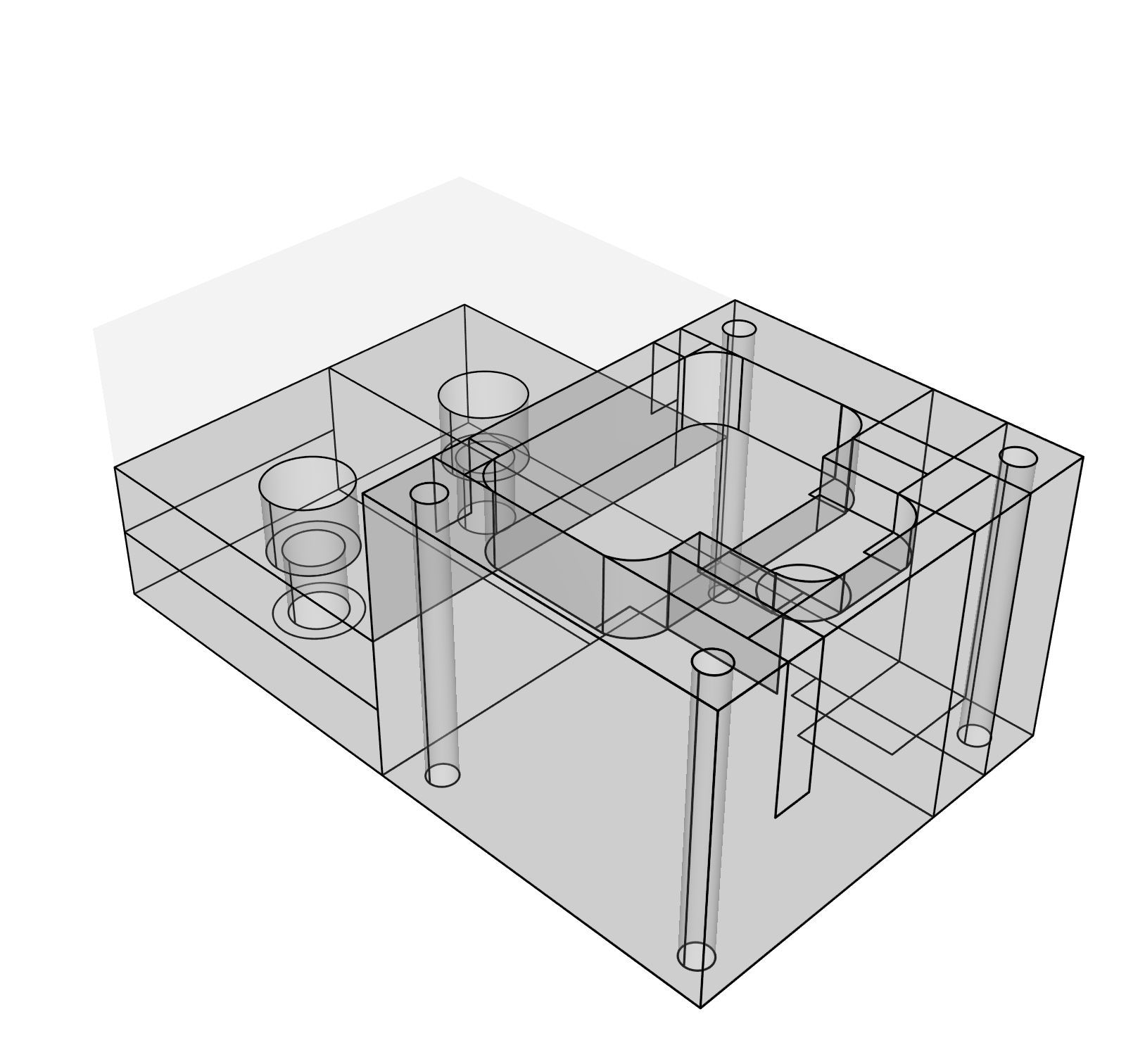}} & & & &
        \\
        \multicolumn{1}{l}{InverseCSG} & & & & &
        \\
        \includegraphics[width=0.124\linewidth]{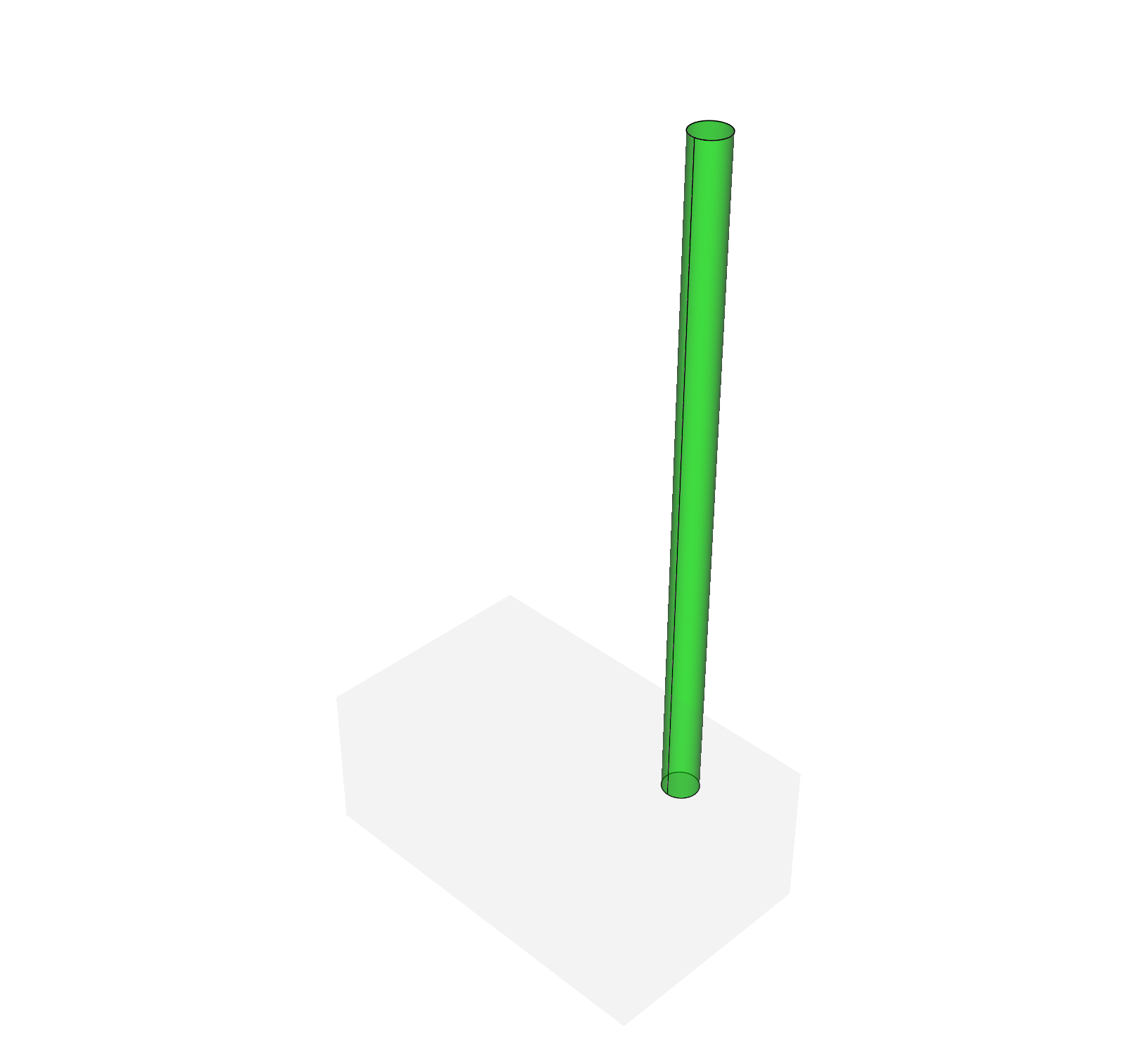} &
        \includegraphics[width=0.124\linewidth]{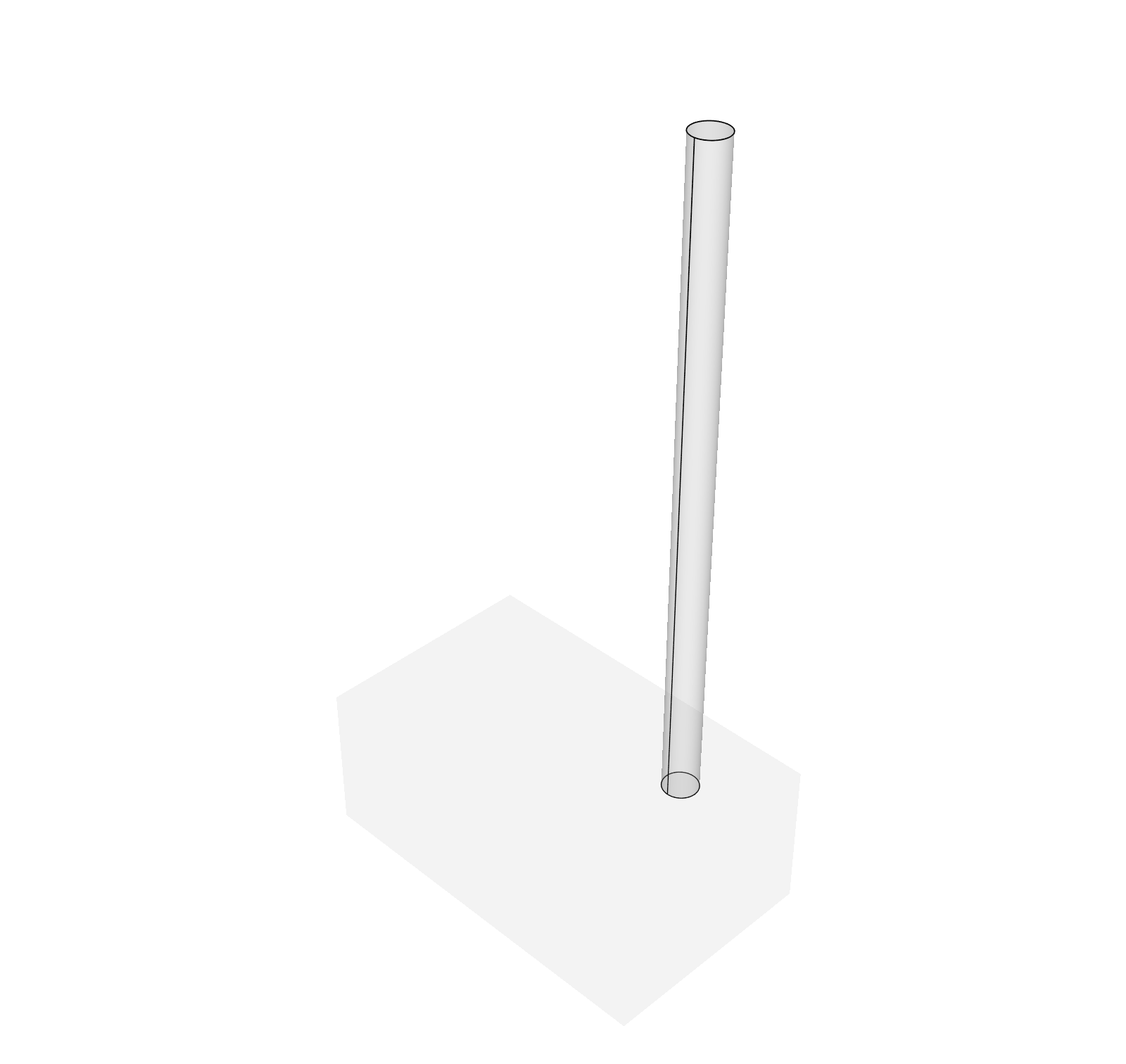} &
        \includegraphics[width=0.124\linewidth]{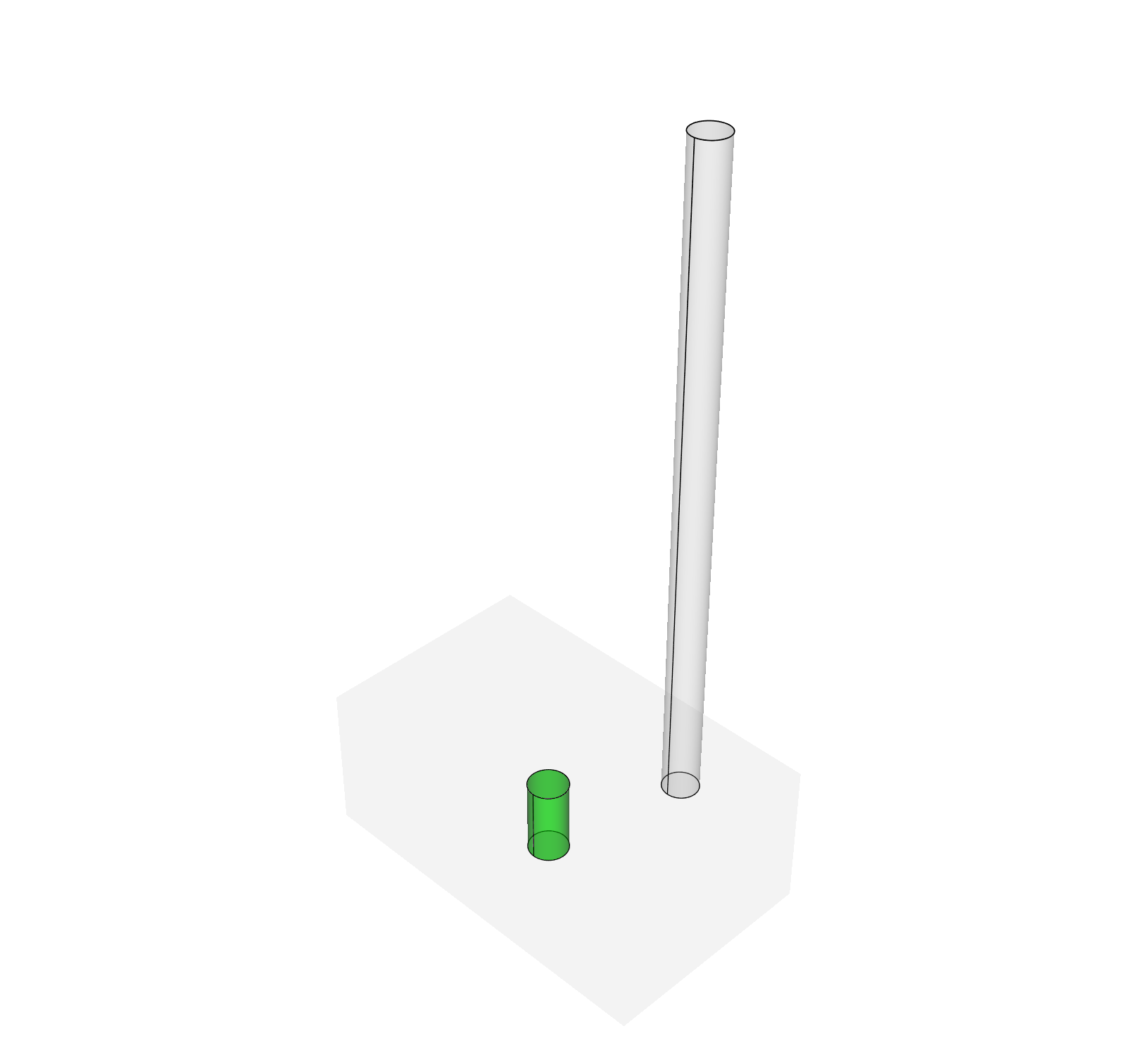} &
        \includegraphics[width=0.124\linewidth]{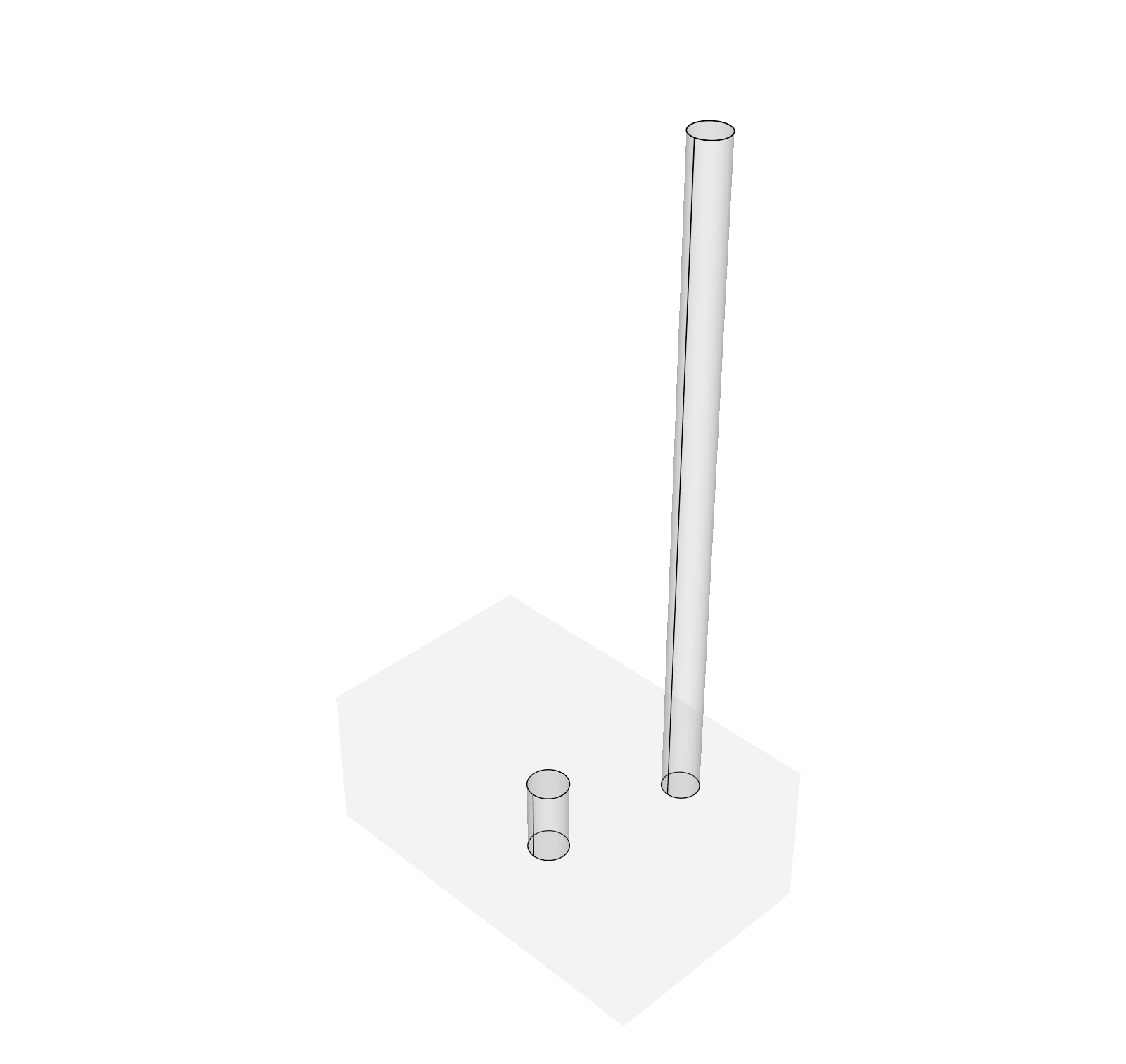} &
        \includegraphics[width=0.124\linewidth]{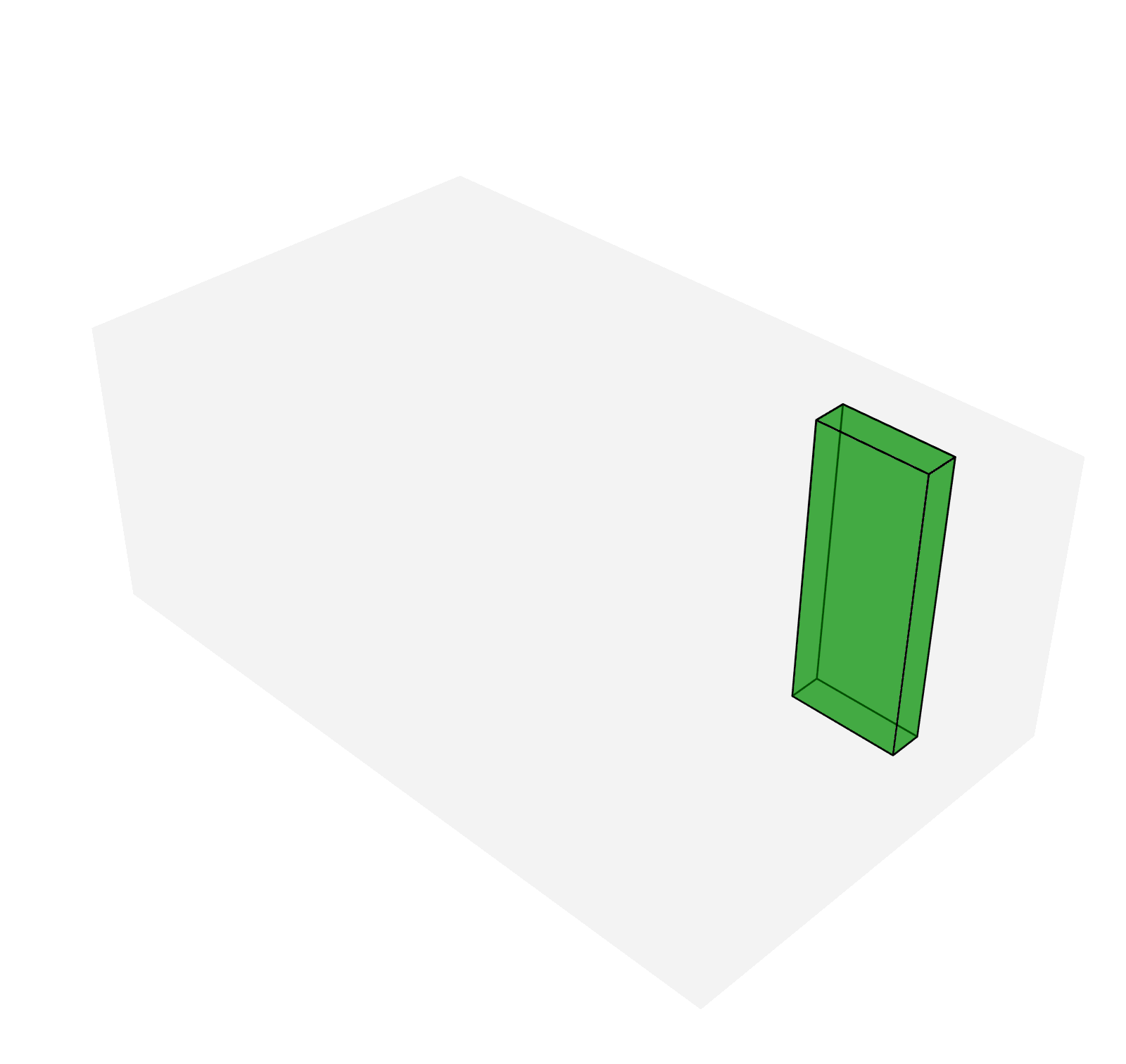} &
        \includegraphics[width=0.124\linewidth]{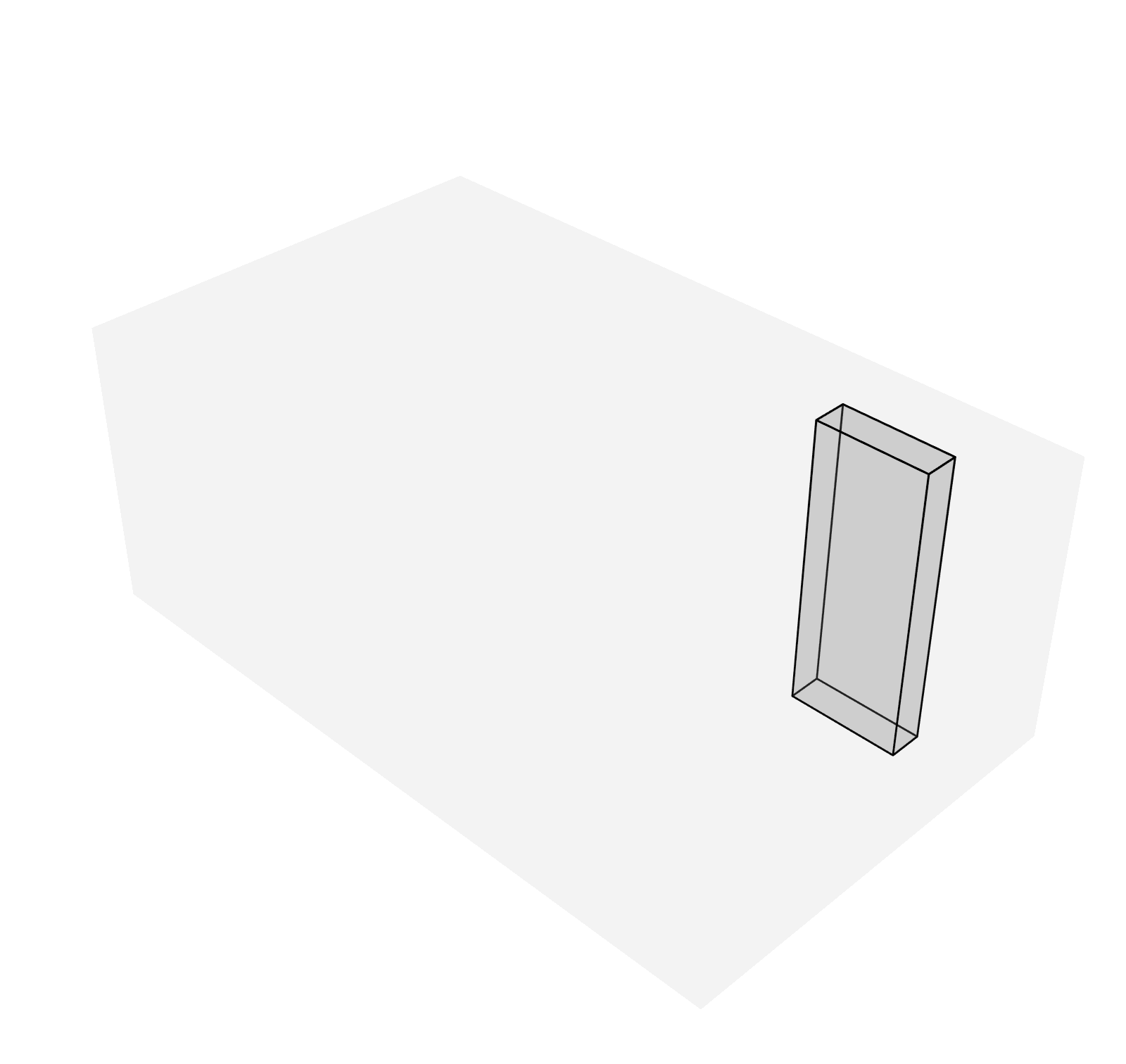} &
        \includegraphics[width=0.124\linewidth]{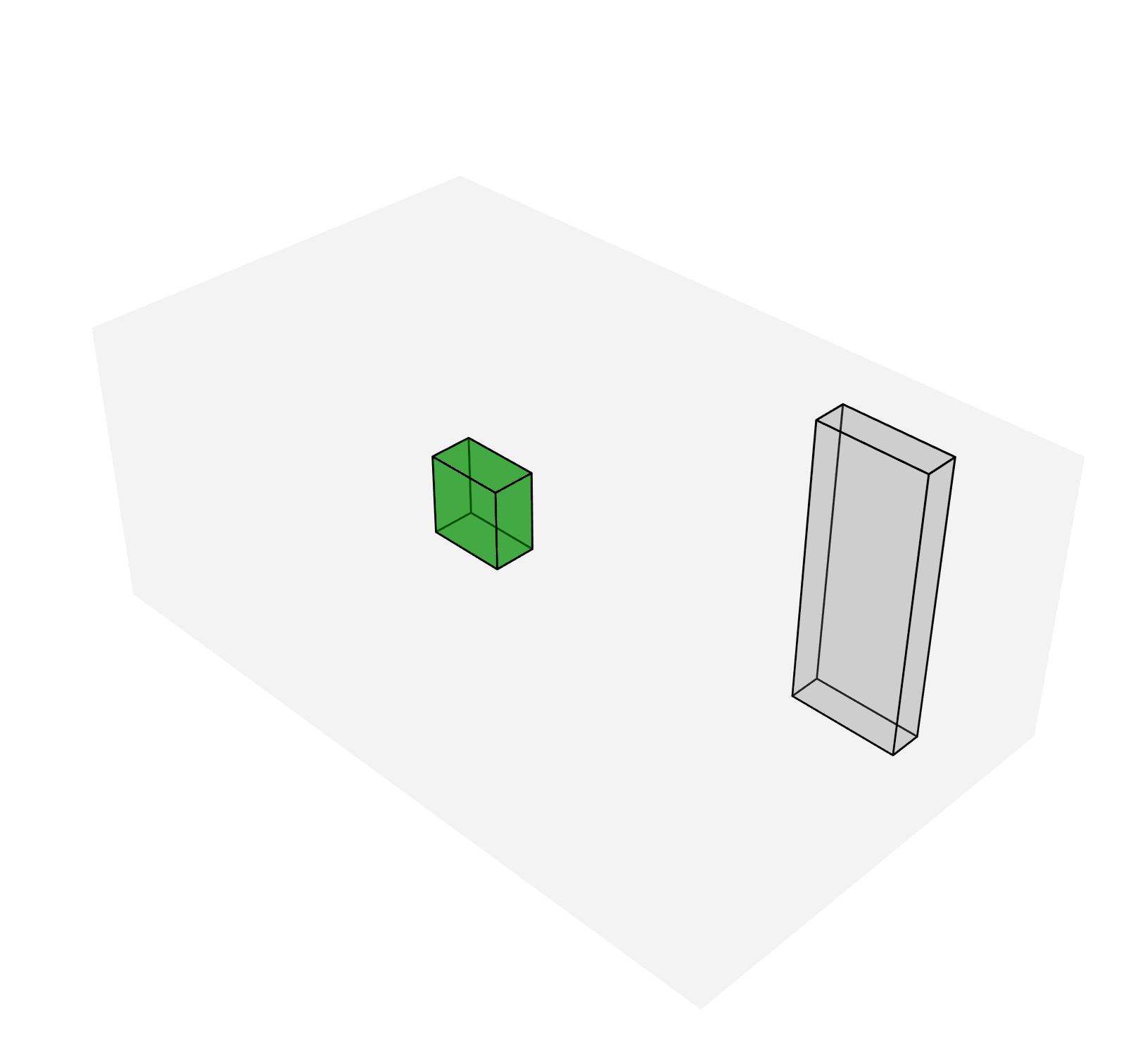} &
        \includegraphics[width=0.124\linewidth]{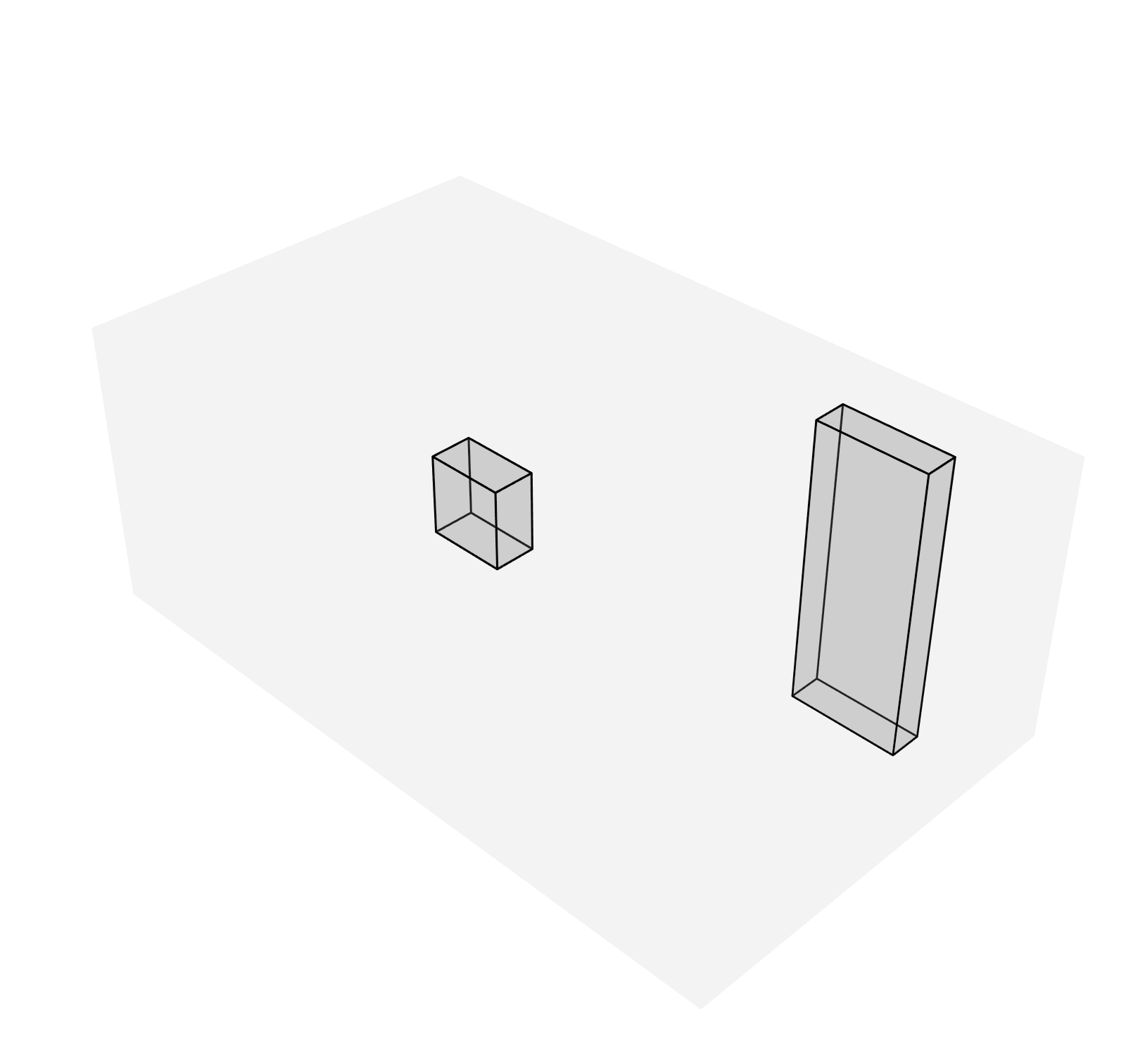}
        \\
        \includegraphics[width=0.124\linewidth]{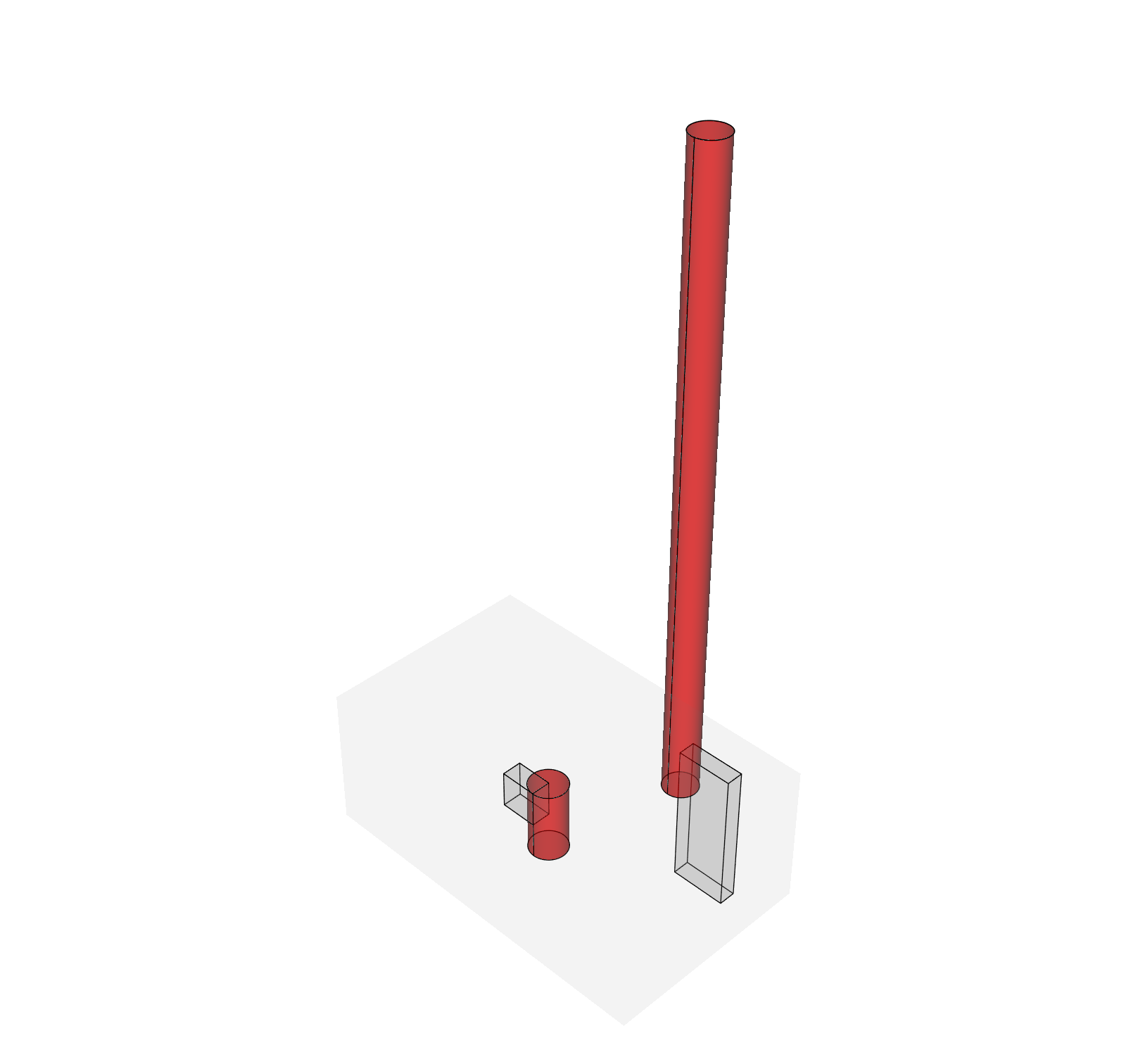} &
        \includegraphics[width=0.124\linewidth]{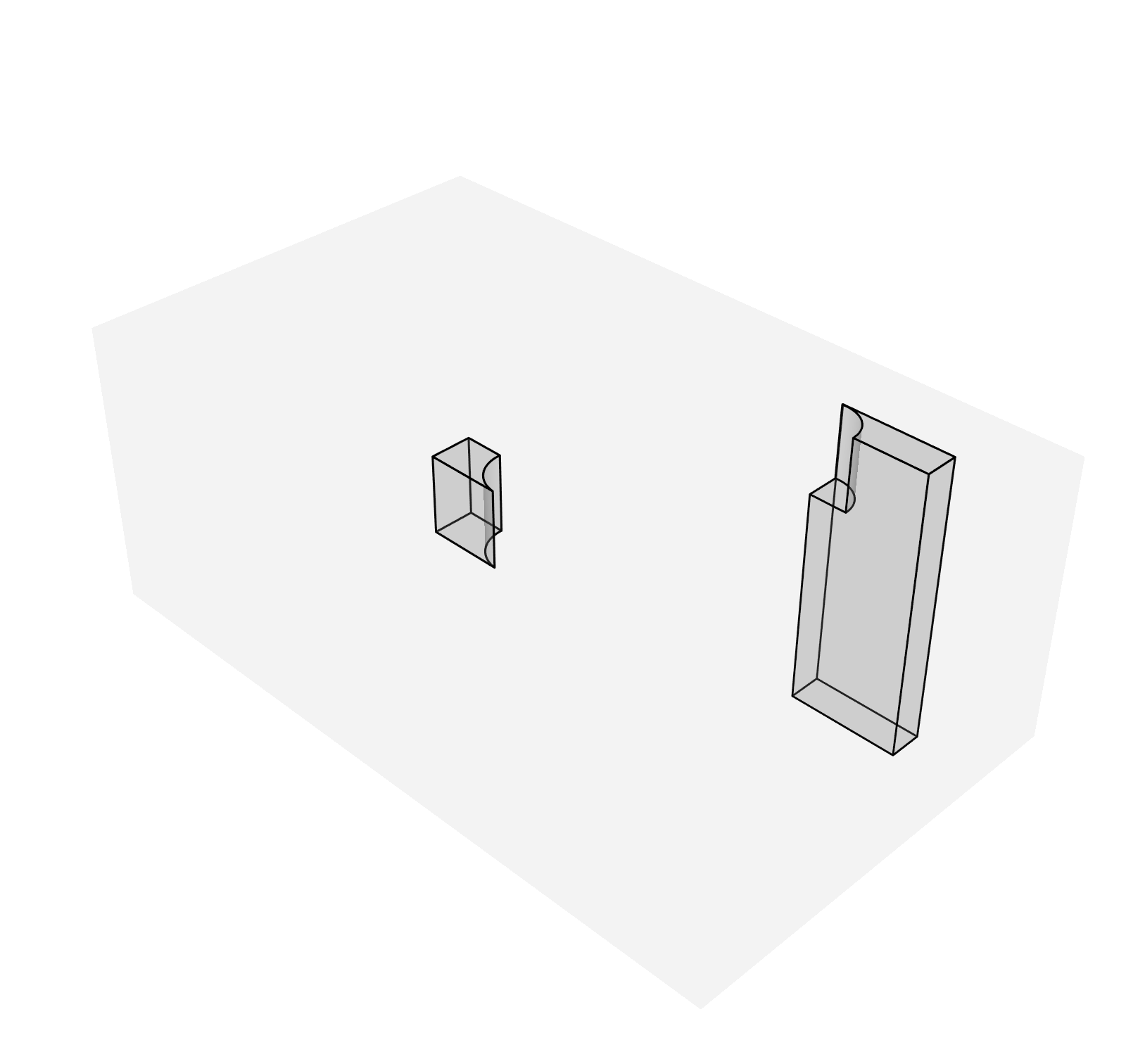} &
        \includegraphics[width=0.124\linewidth]{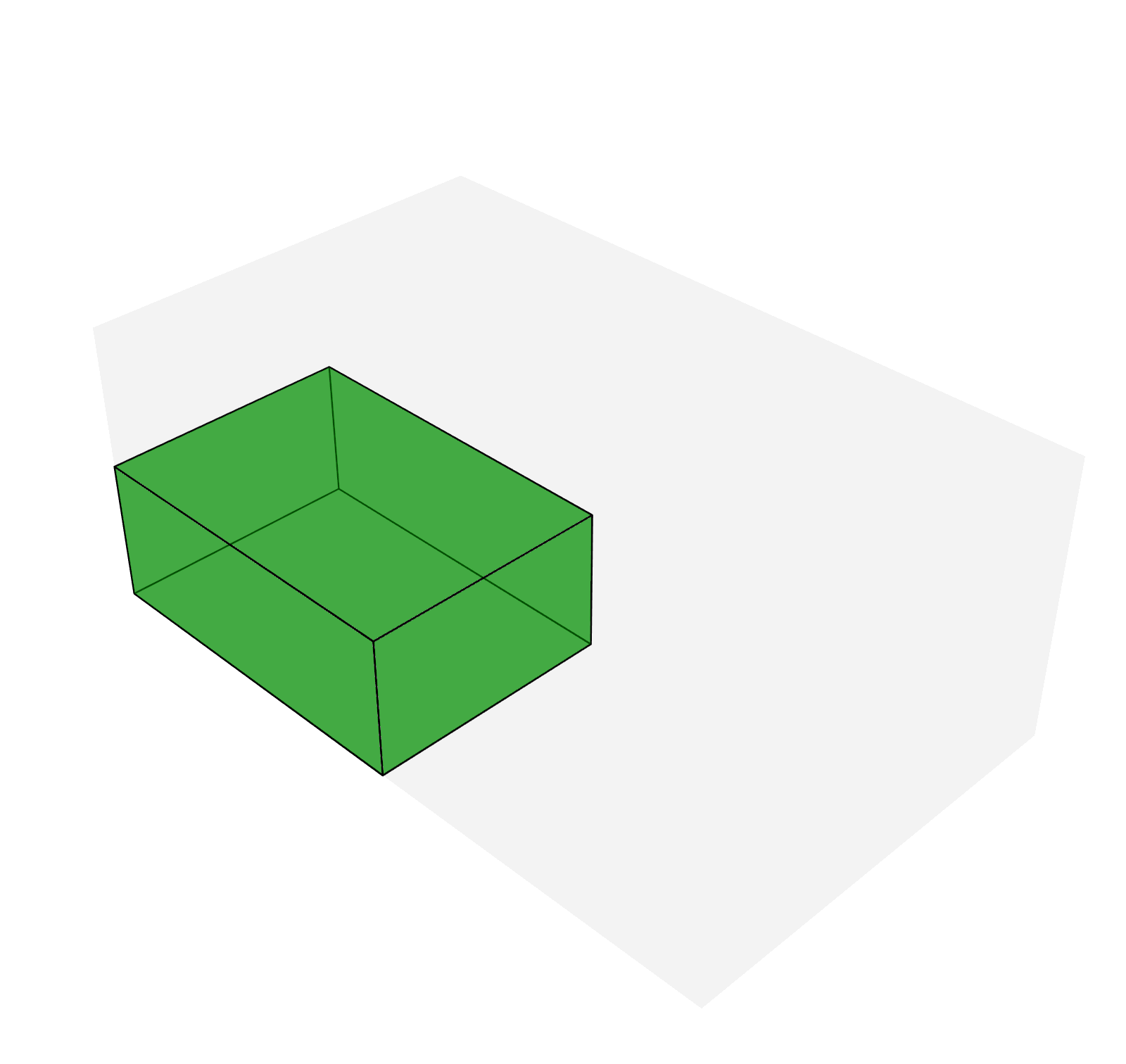} &
        \includegraphics[width=0.124\linewidth]{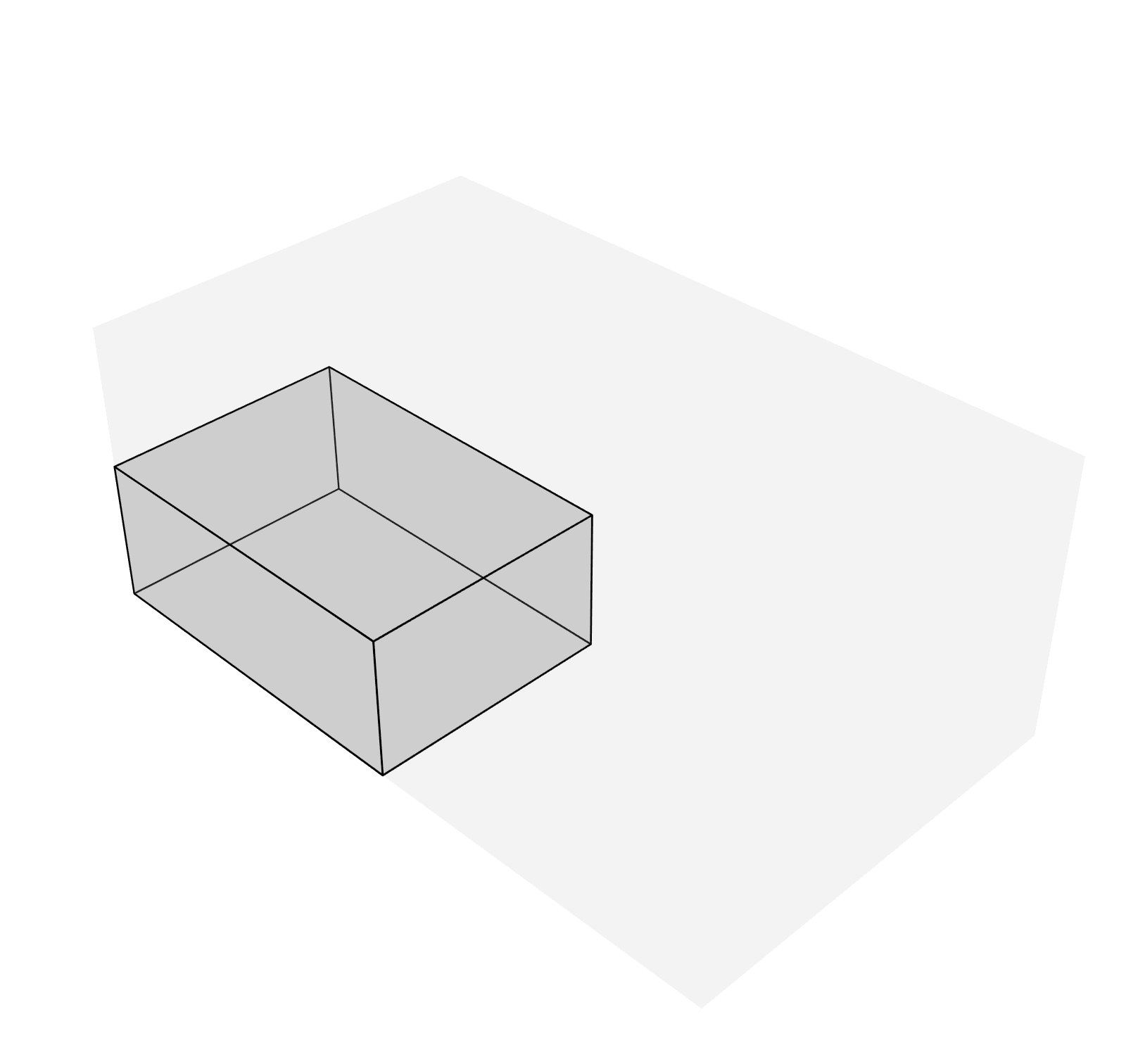} &
        \includegraphics[width=0.124\linewidth]{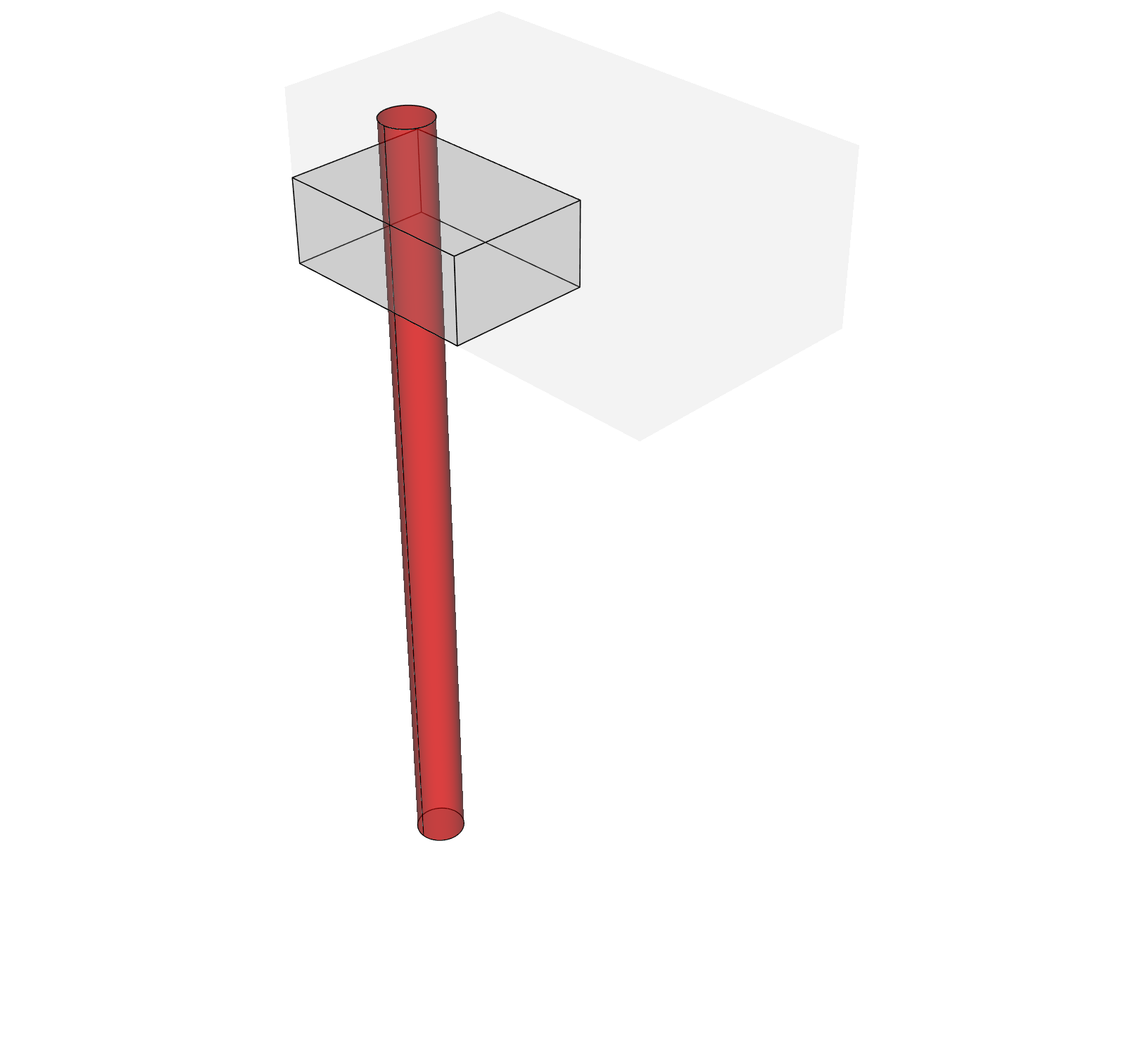} &
        \includegraphics[width=0.124\linewidth]{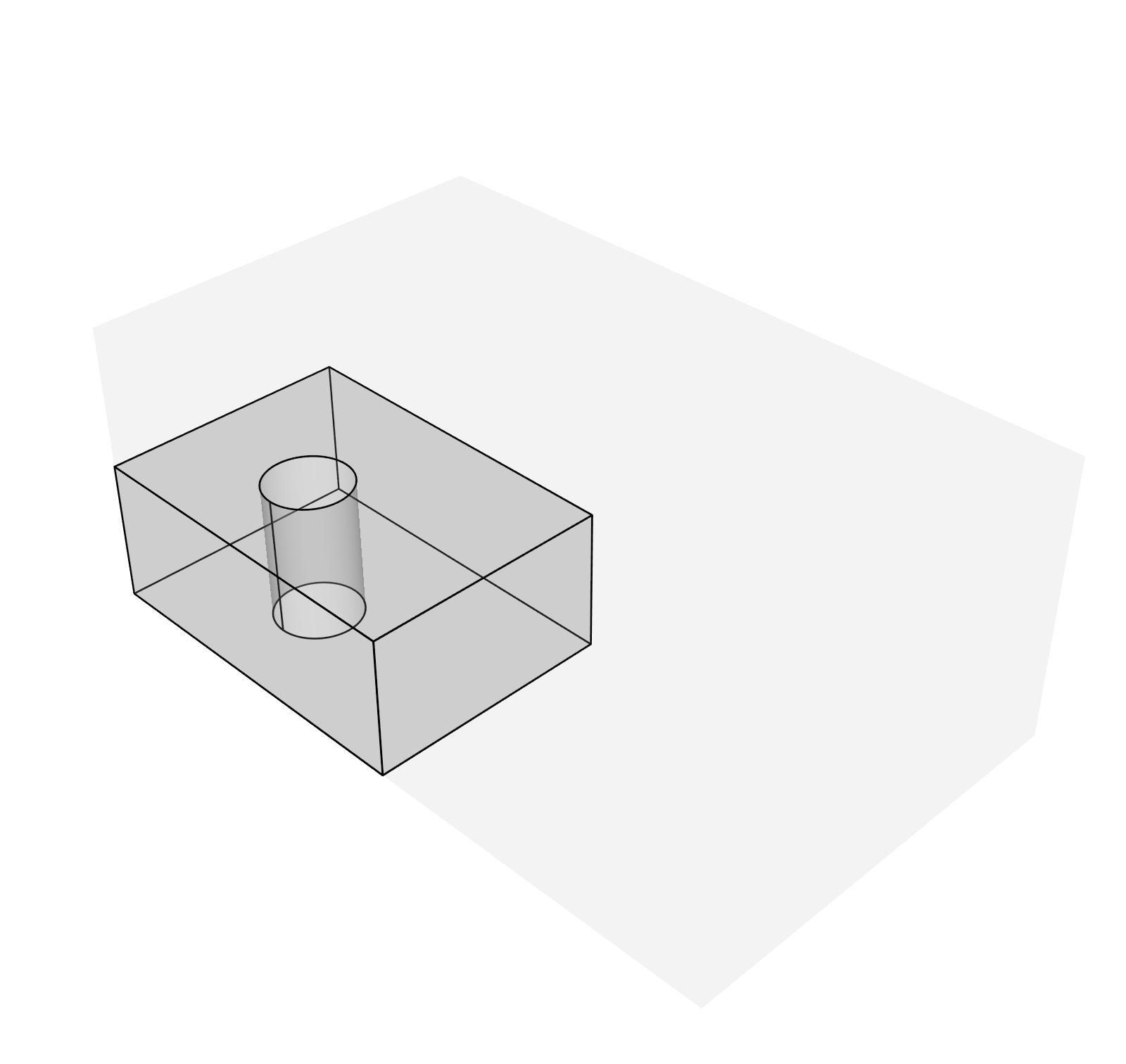} &
        \includegraphics[width=0.124\linewidth]{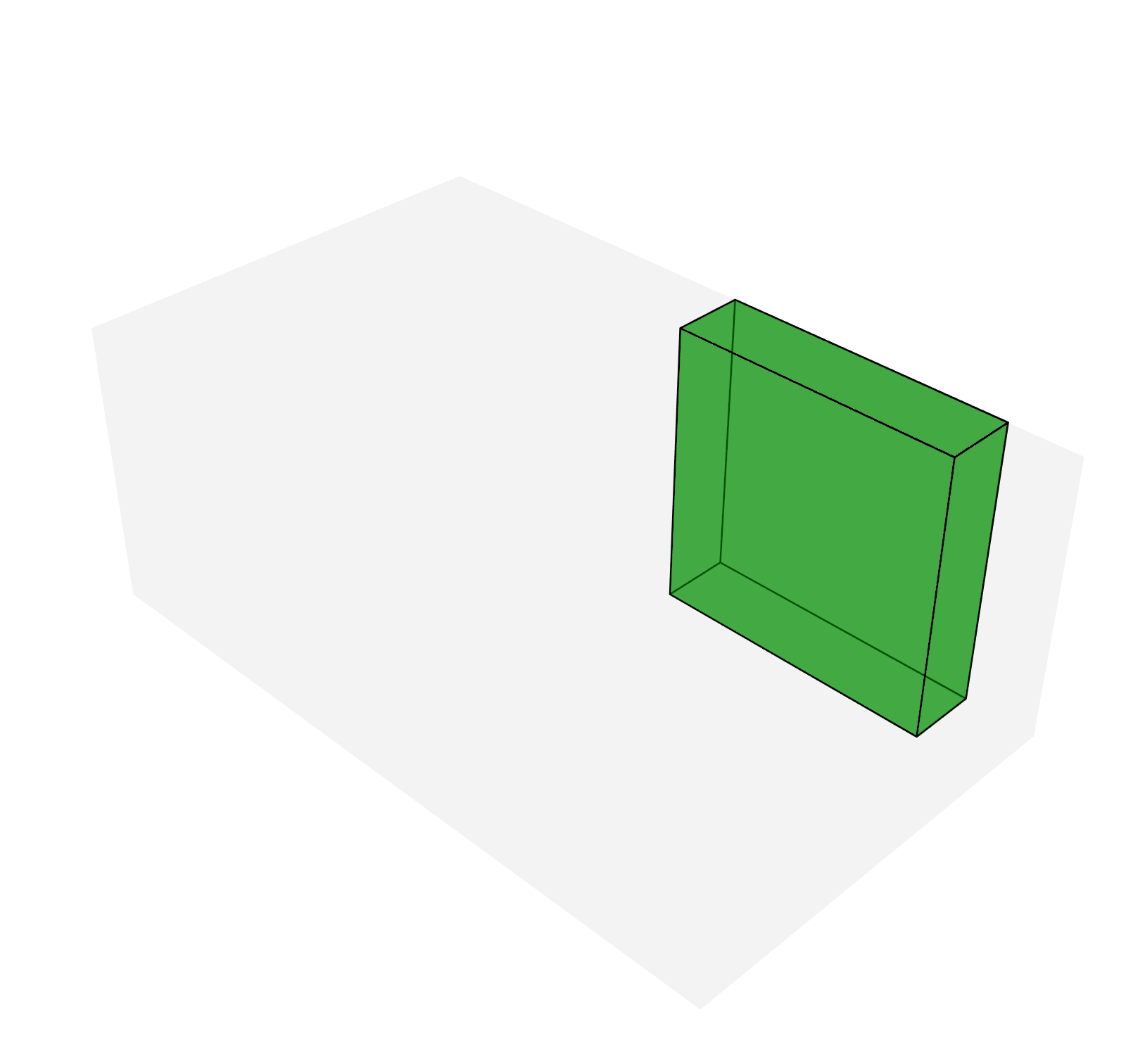} &
        \includegraphics[width=0.124\linewidth]{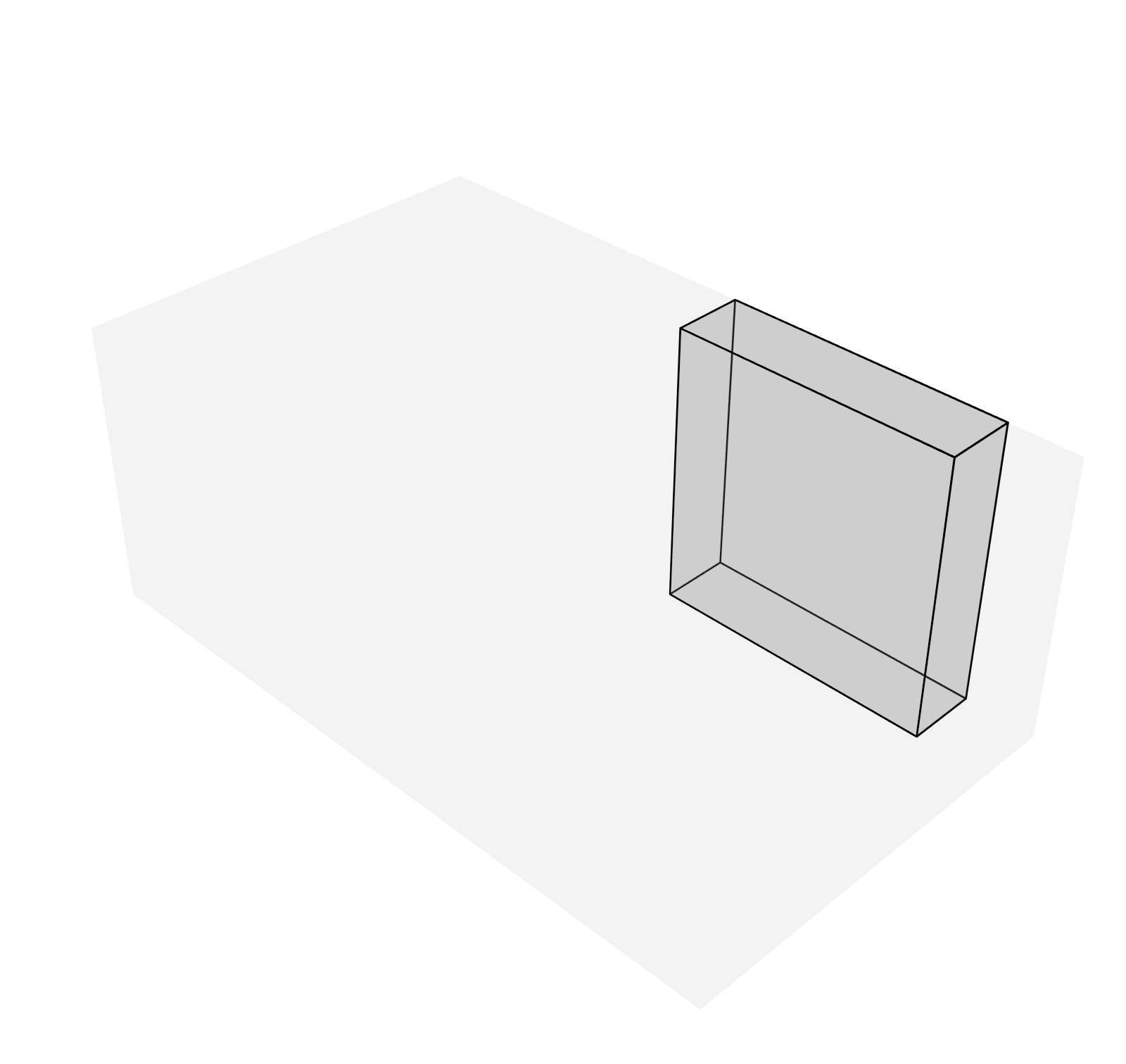}
        \\
        \includegraphics[width=0.124\linewidth]{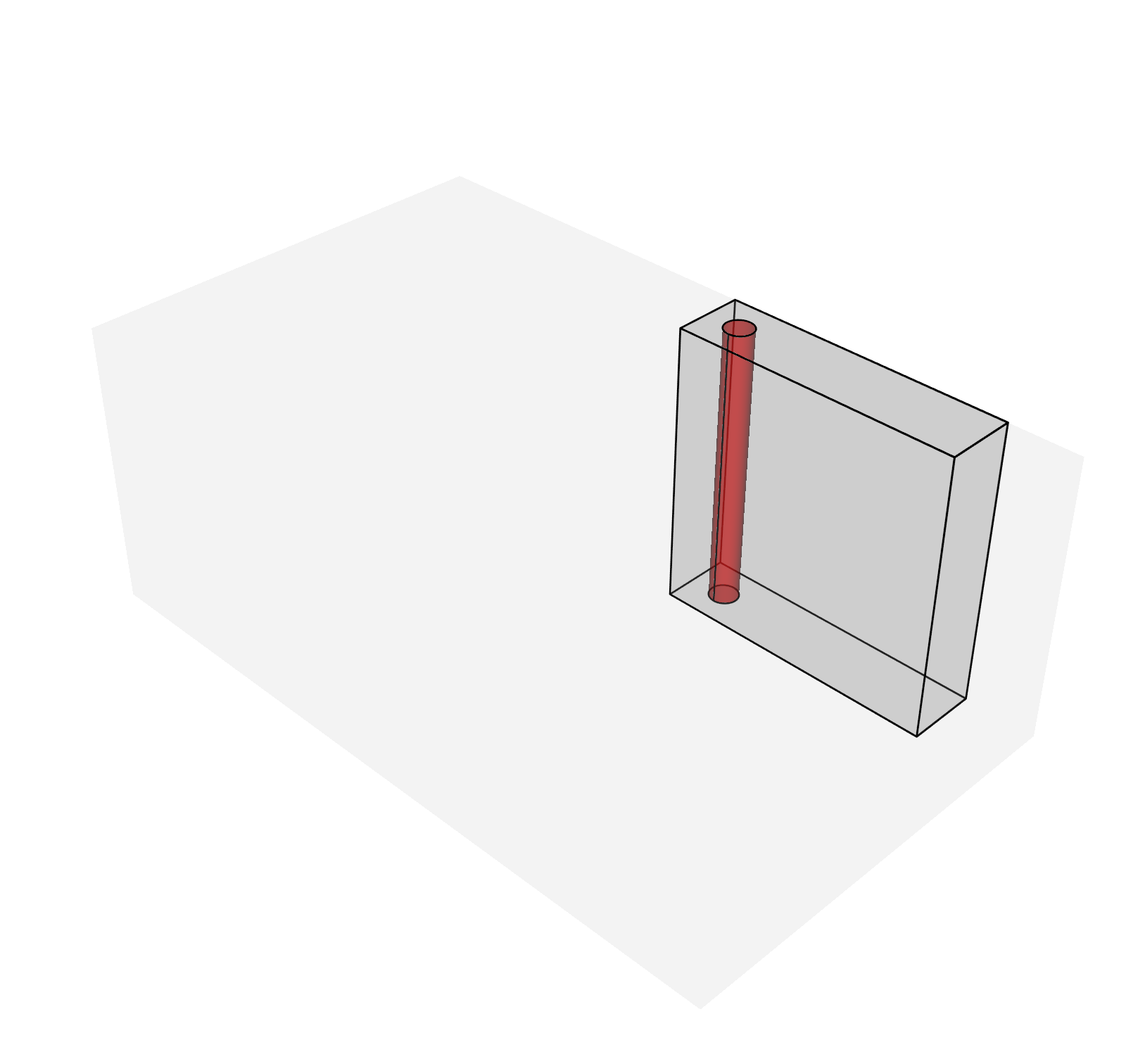} &
        \includegraphics[width=0.124\linewidth]{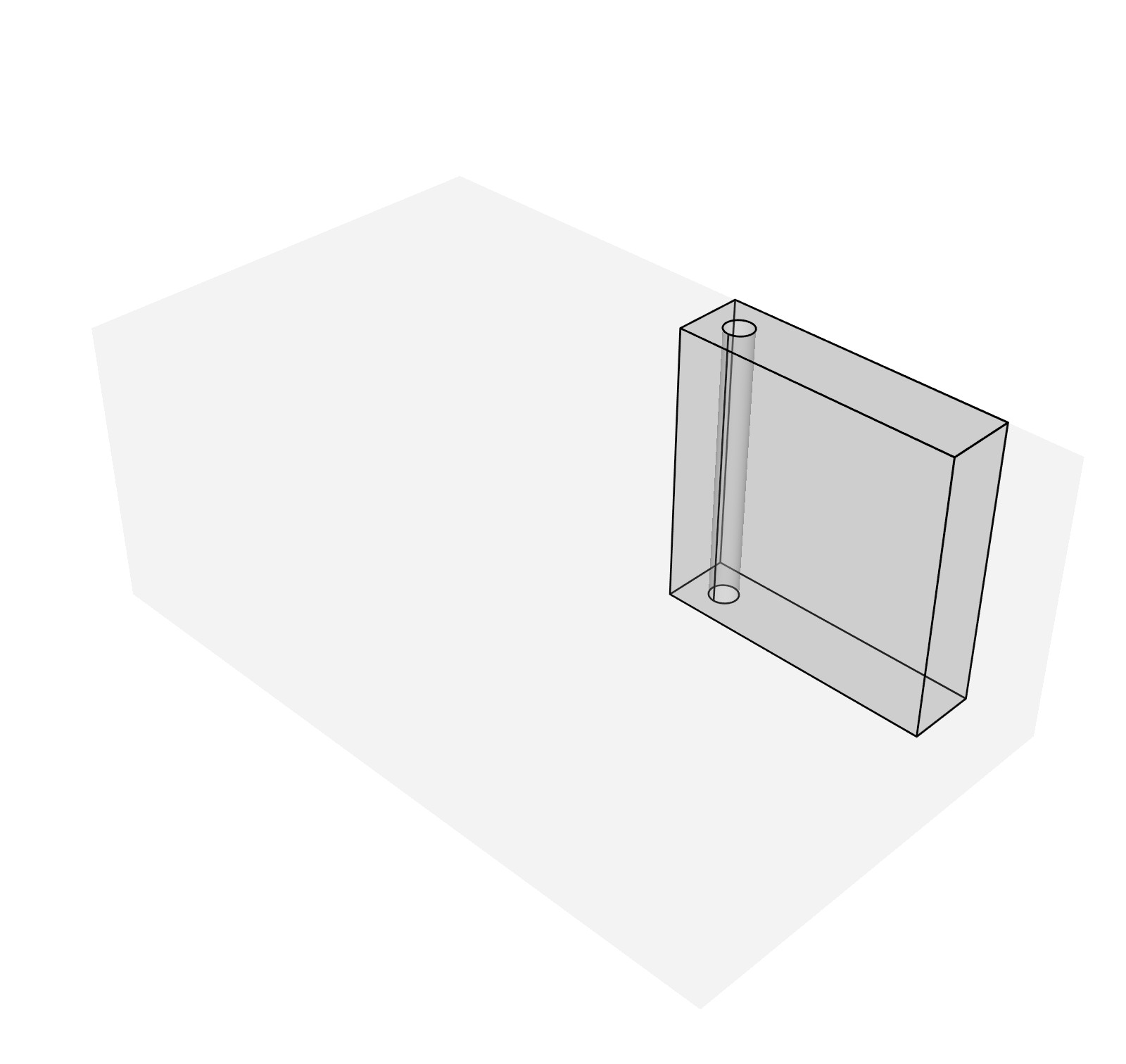} &
        \includegraphics[width=0.124\linewidth]{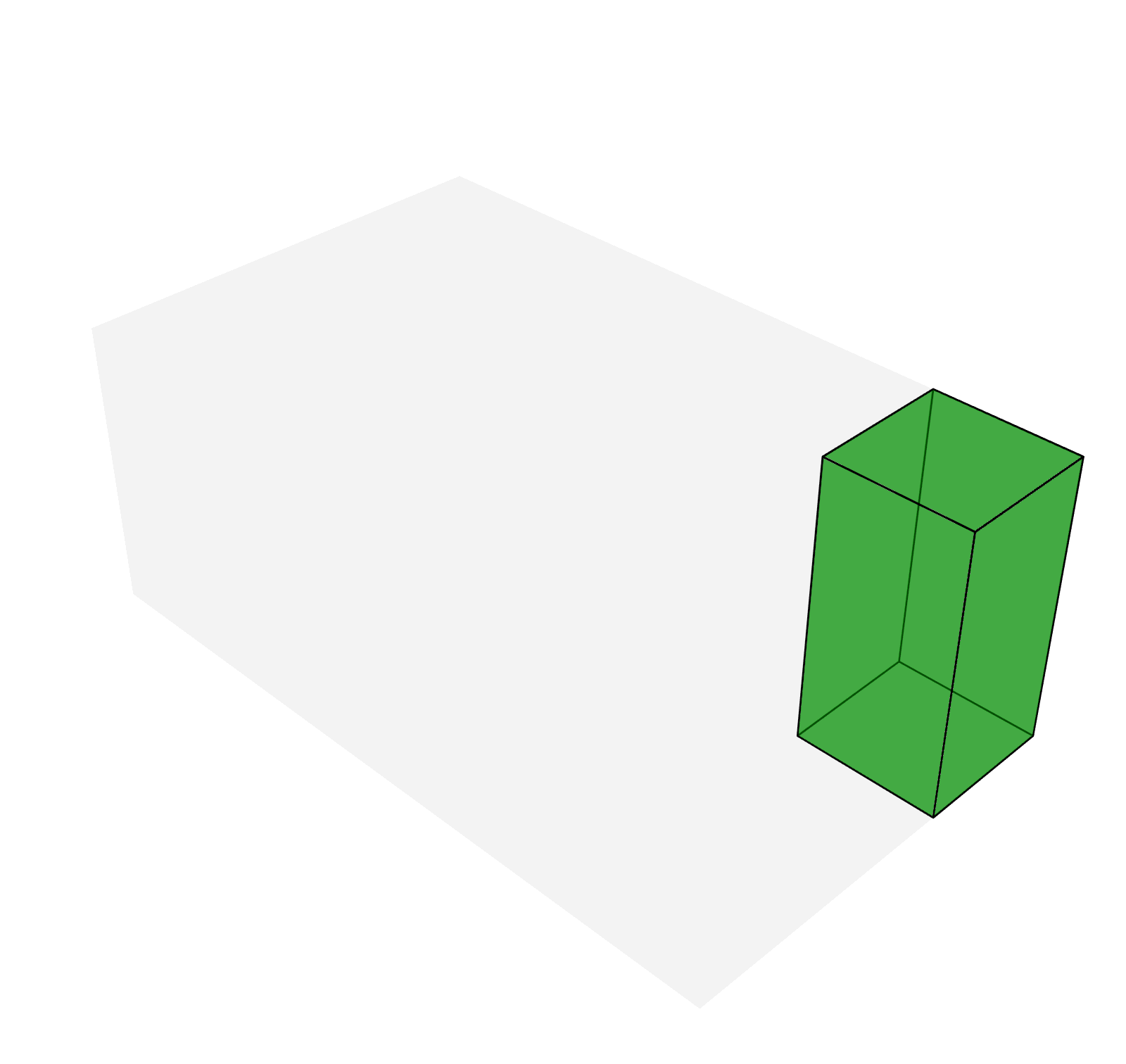} &
        \includegraphics[width=0.124\linewidth]{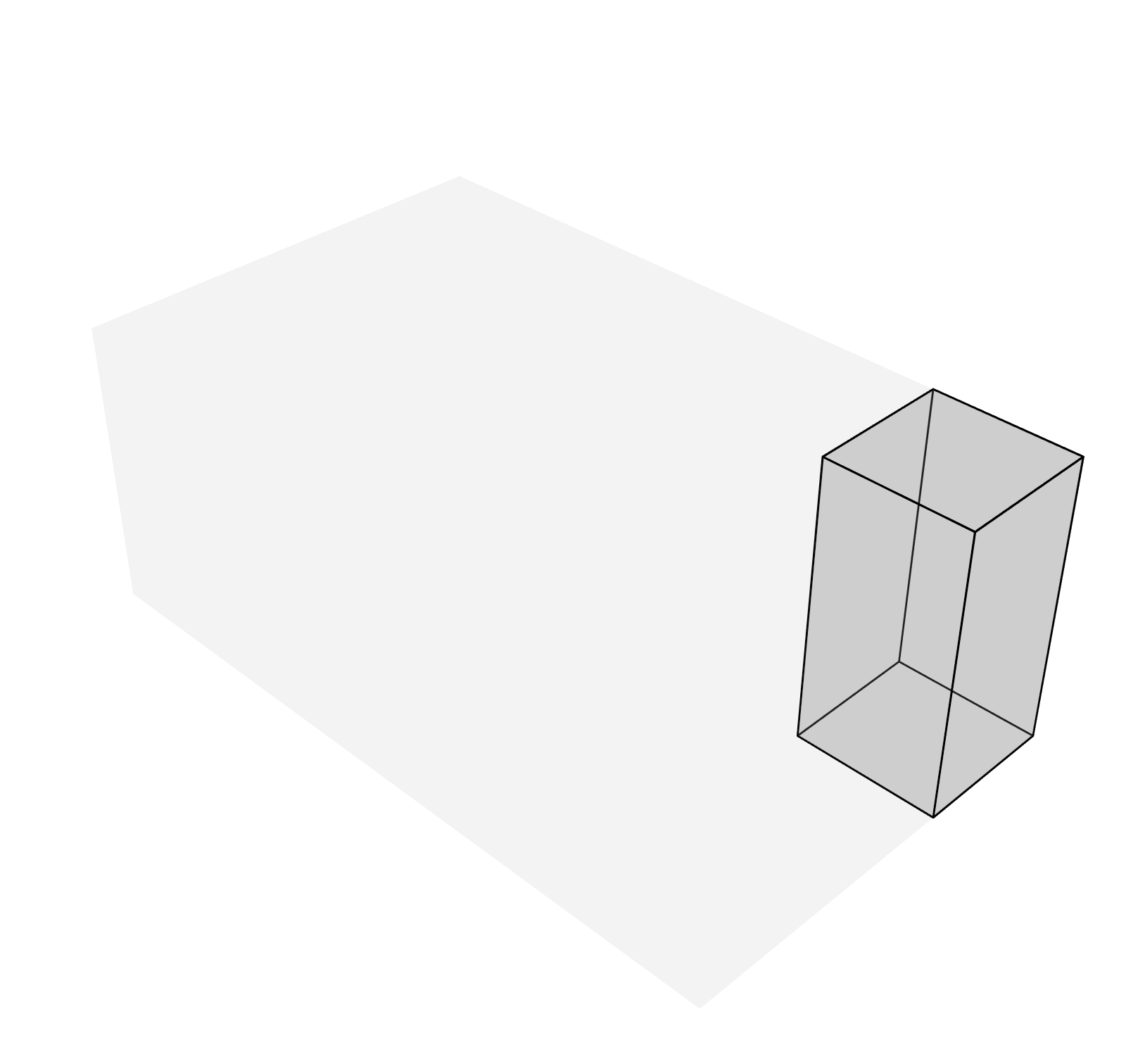} &
        \includegraphics[width=0.124\linewidth]{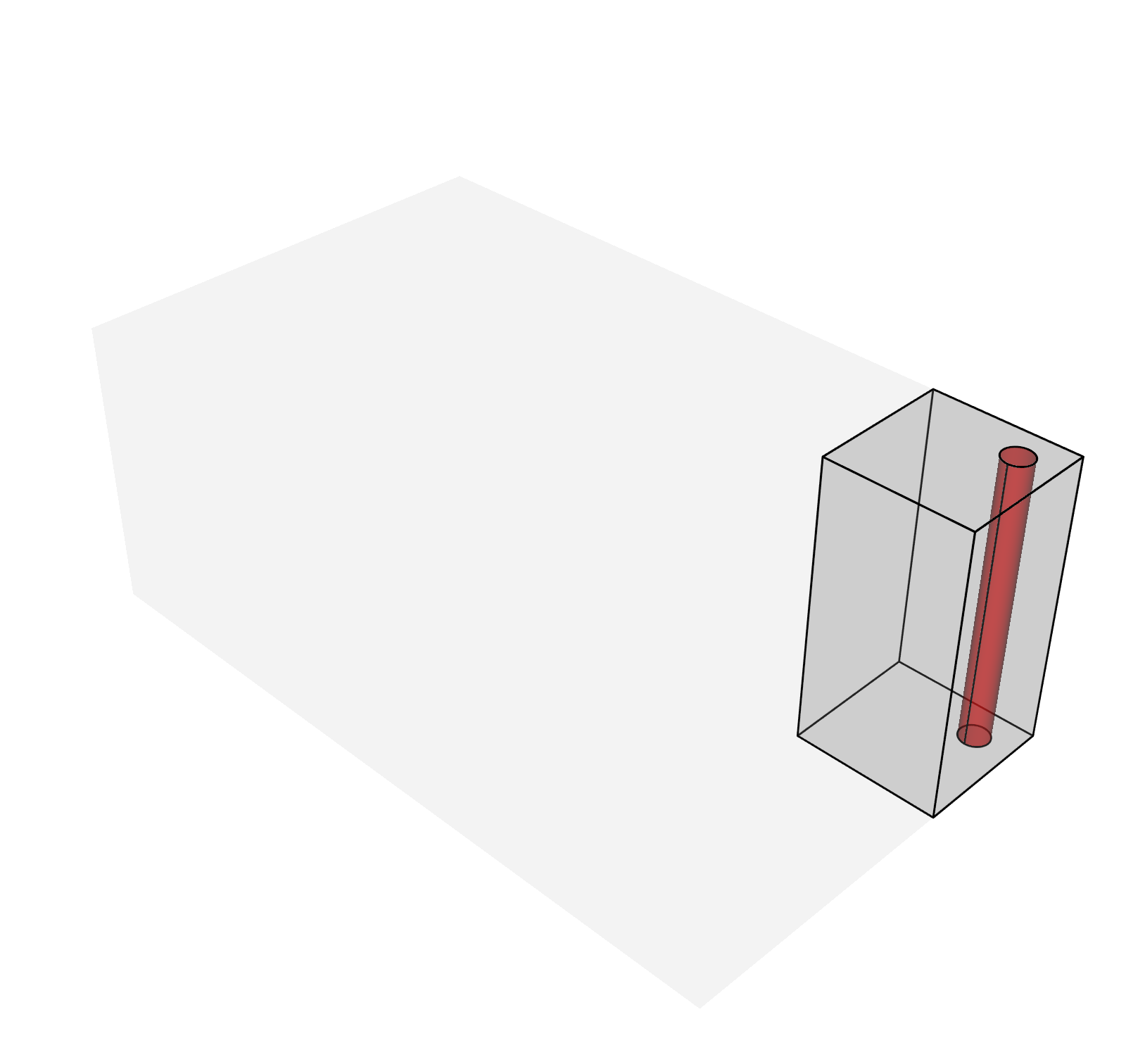} &
        \includegraphics[width=0.124\linewidth]{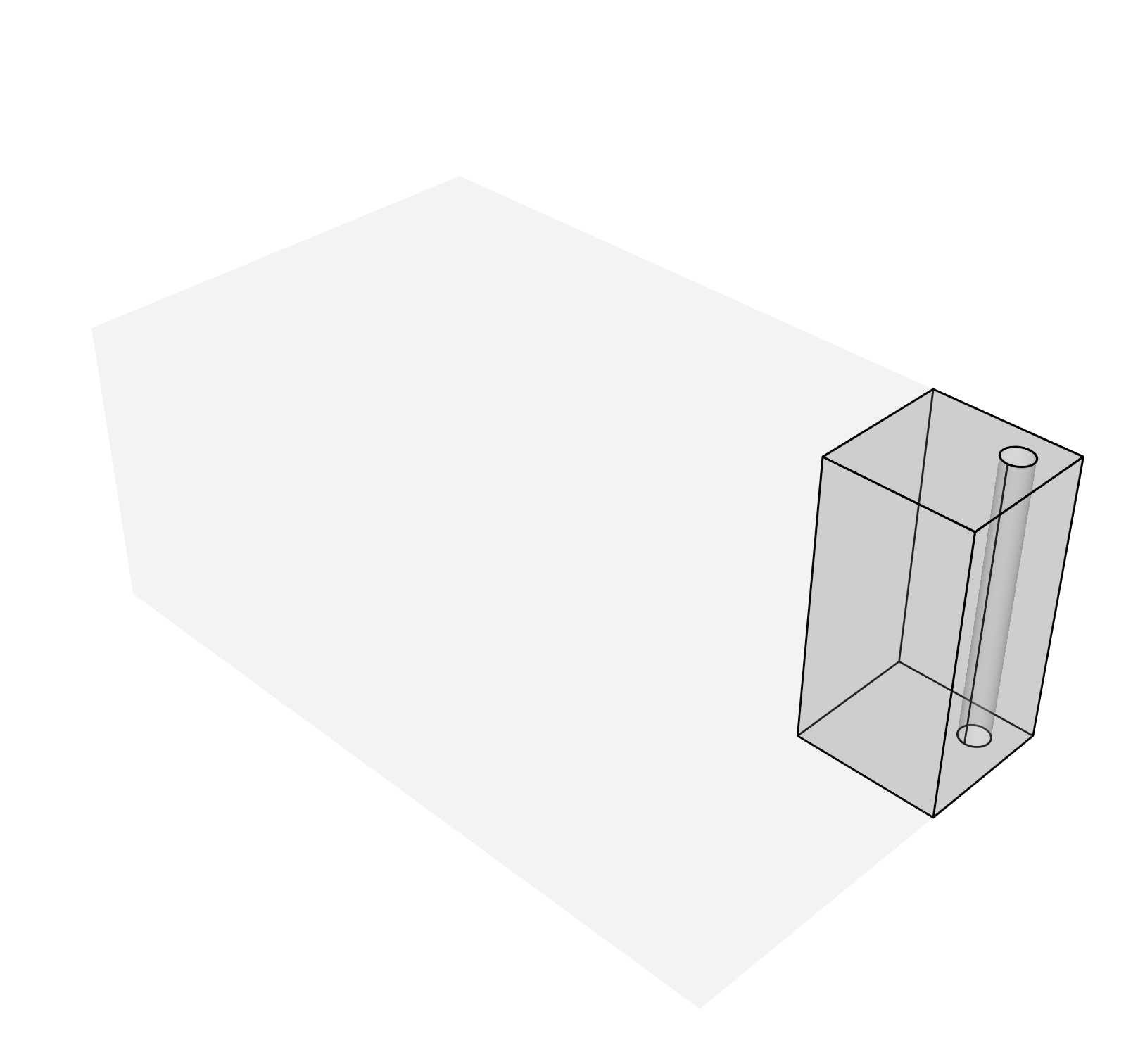} &
        \includegraphics[width=0.124\linewidth]{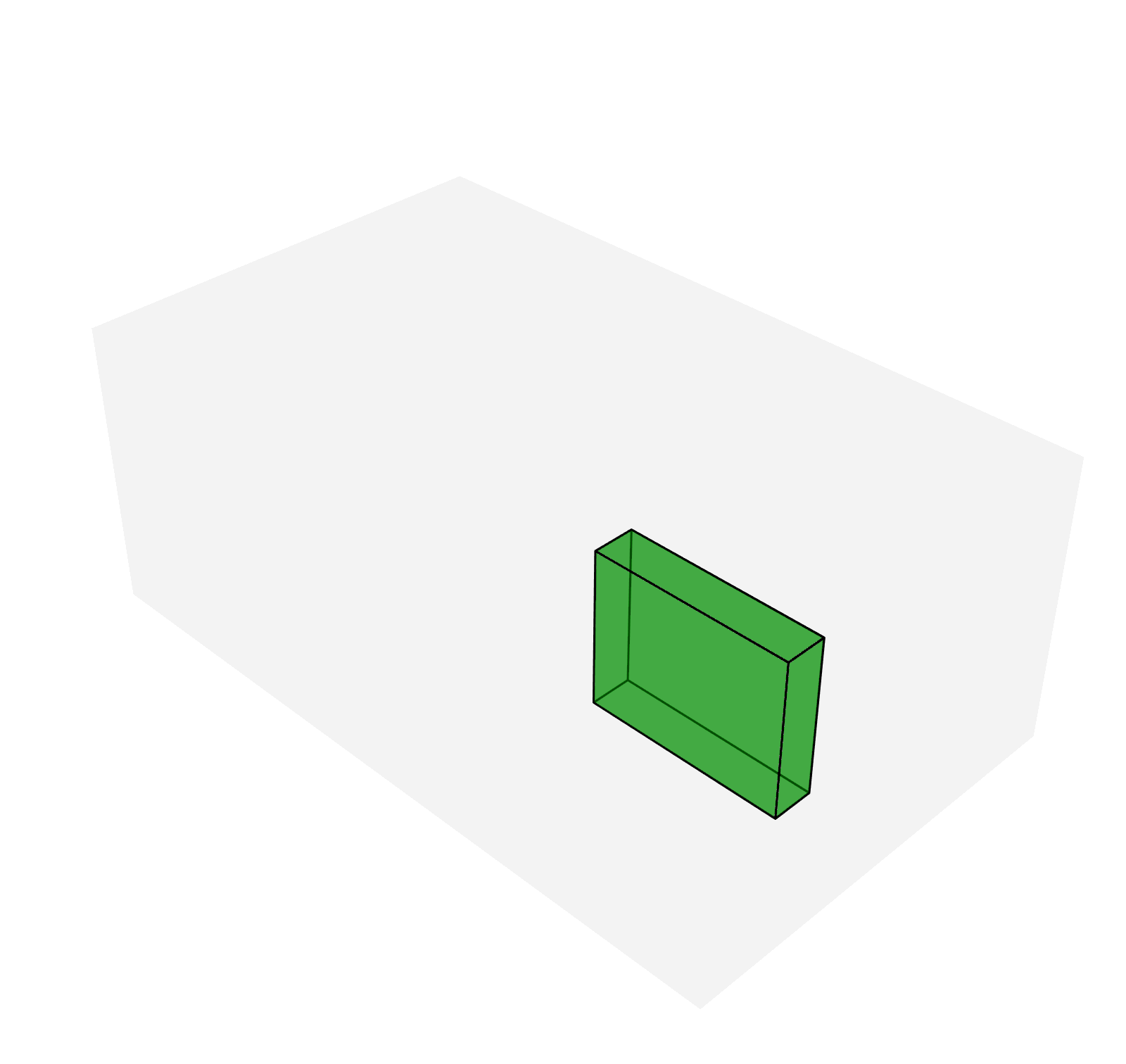} &
        \includegraphics[width=0.124\linewidth]{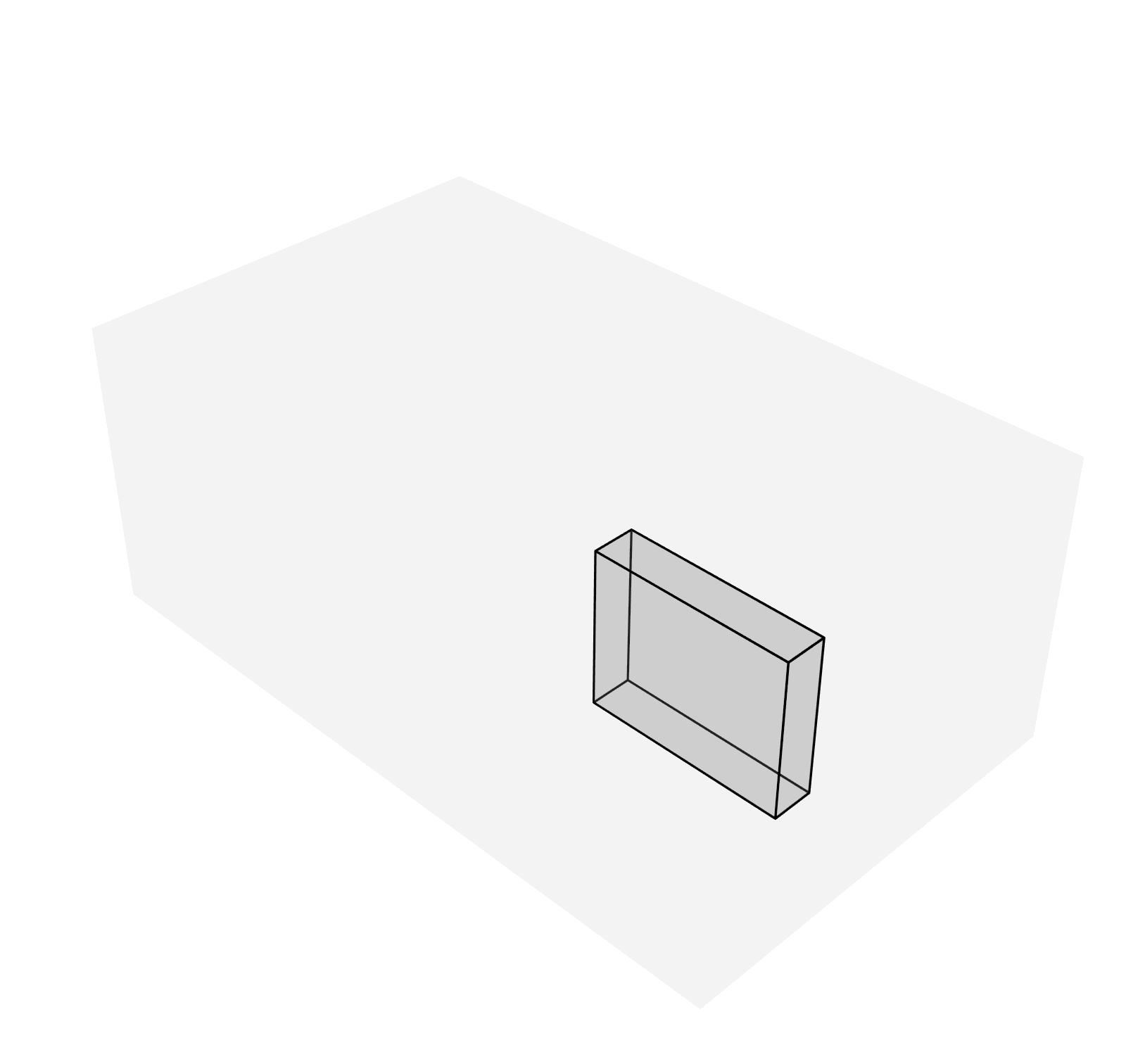}
        \\
        \includegraphics[width=0.124\linewidth]{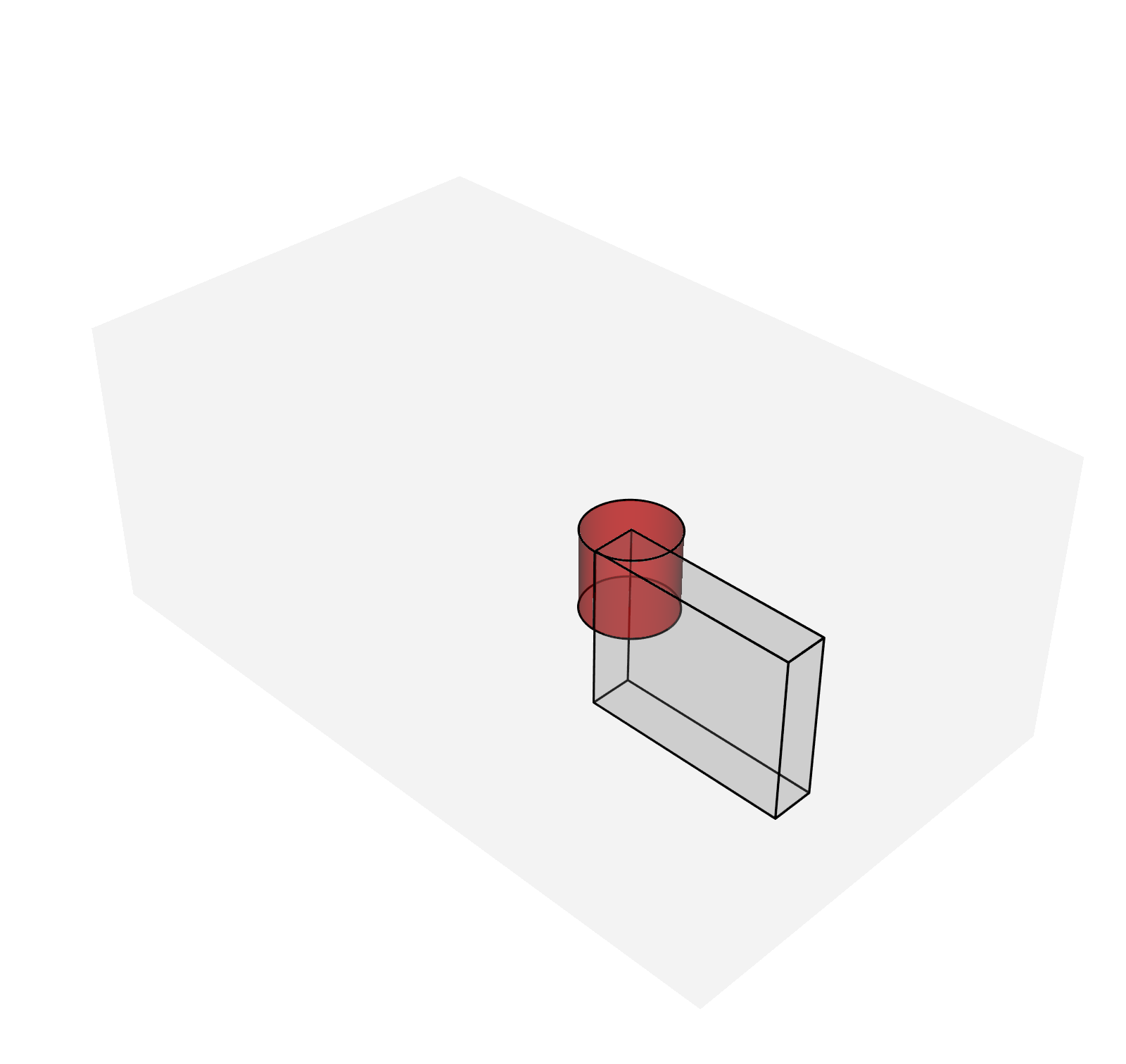} &
        \includegraphics[width=0.124\linewidth]{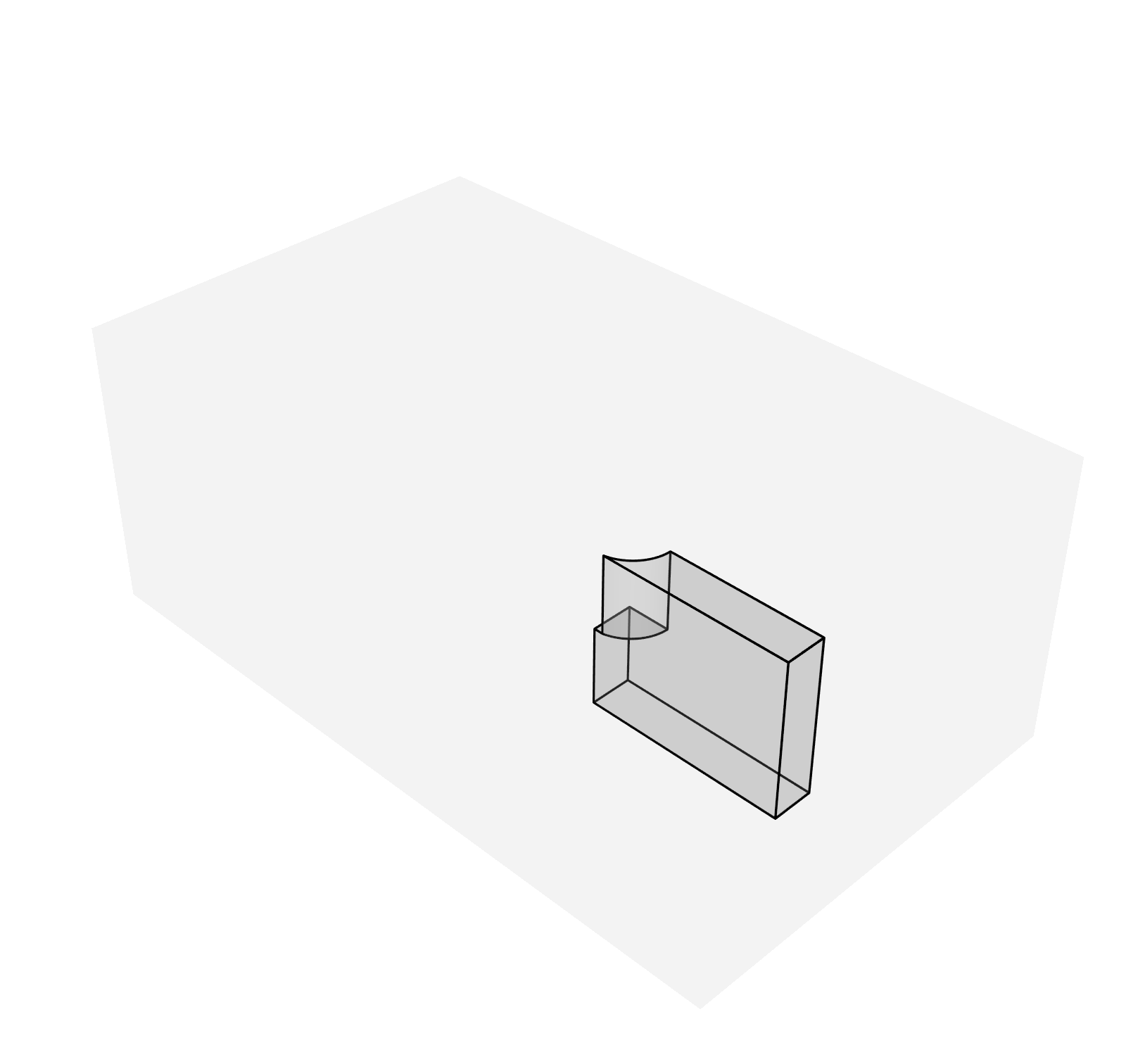} &
        \includegraphics[width=0.124\linewidth]{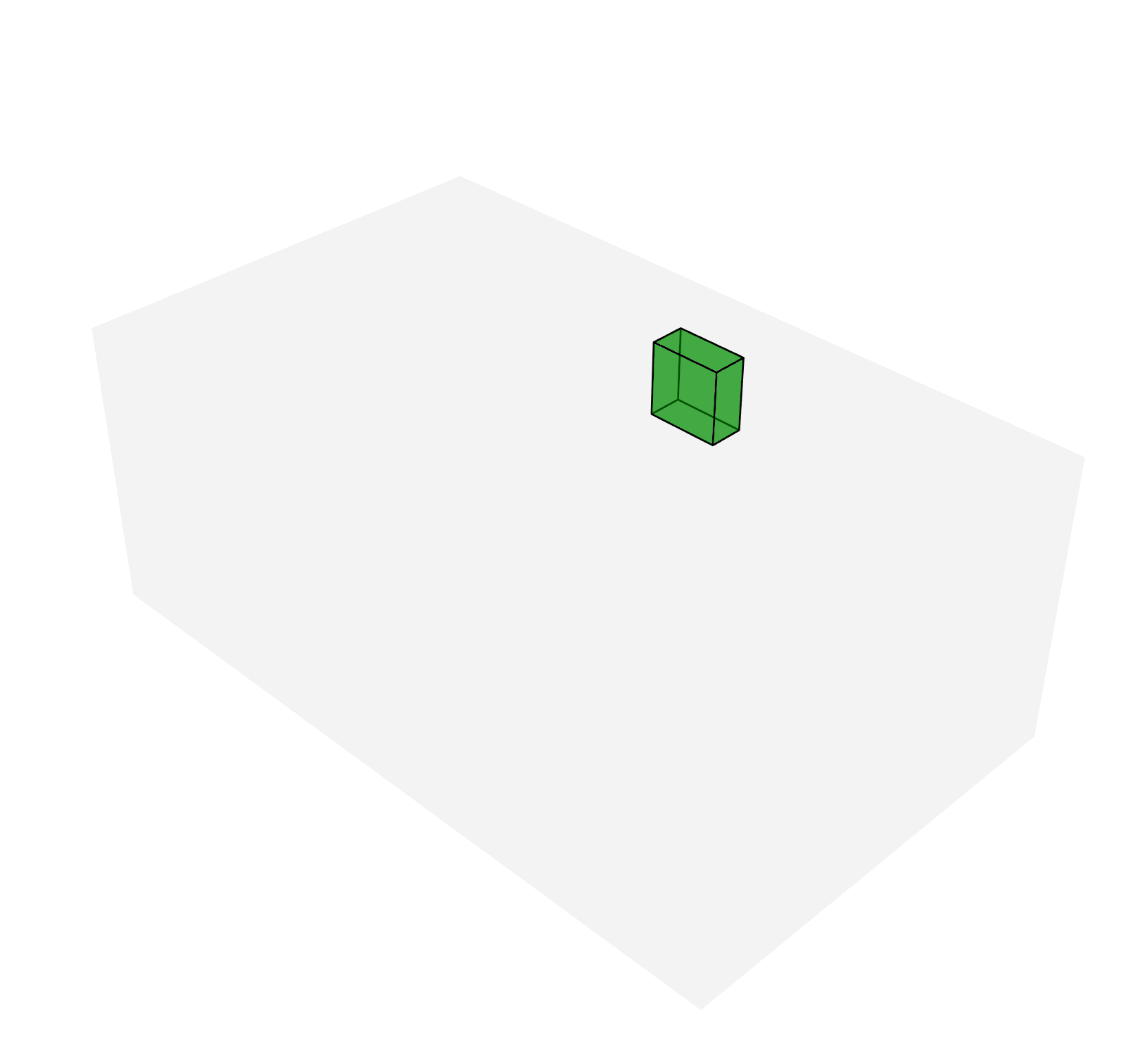} &
        \includegraphics[width=0.124\linewidth]{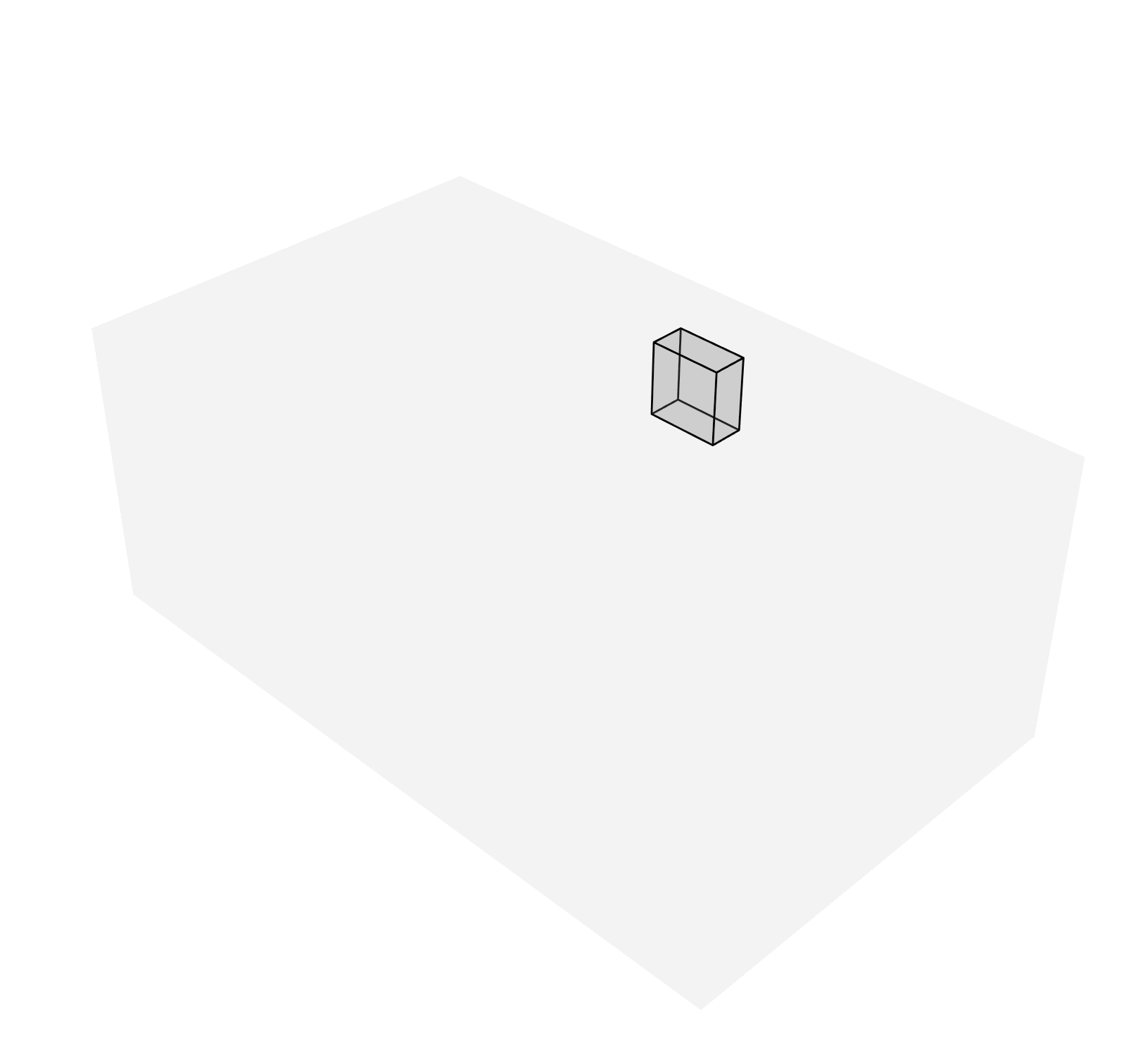} &
        \includegraphics[width=0.124\linewidth]{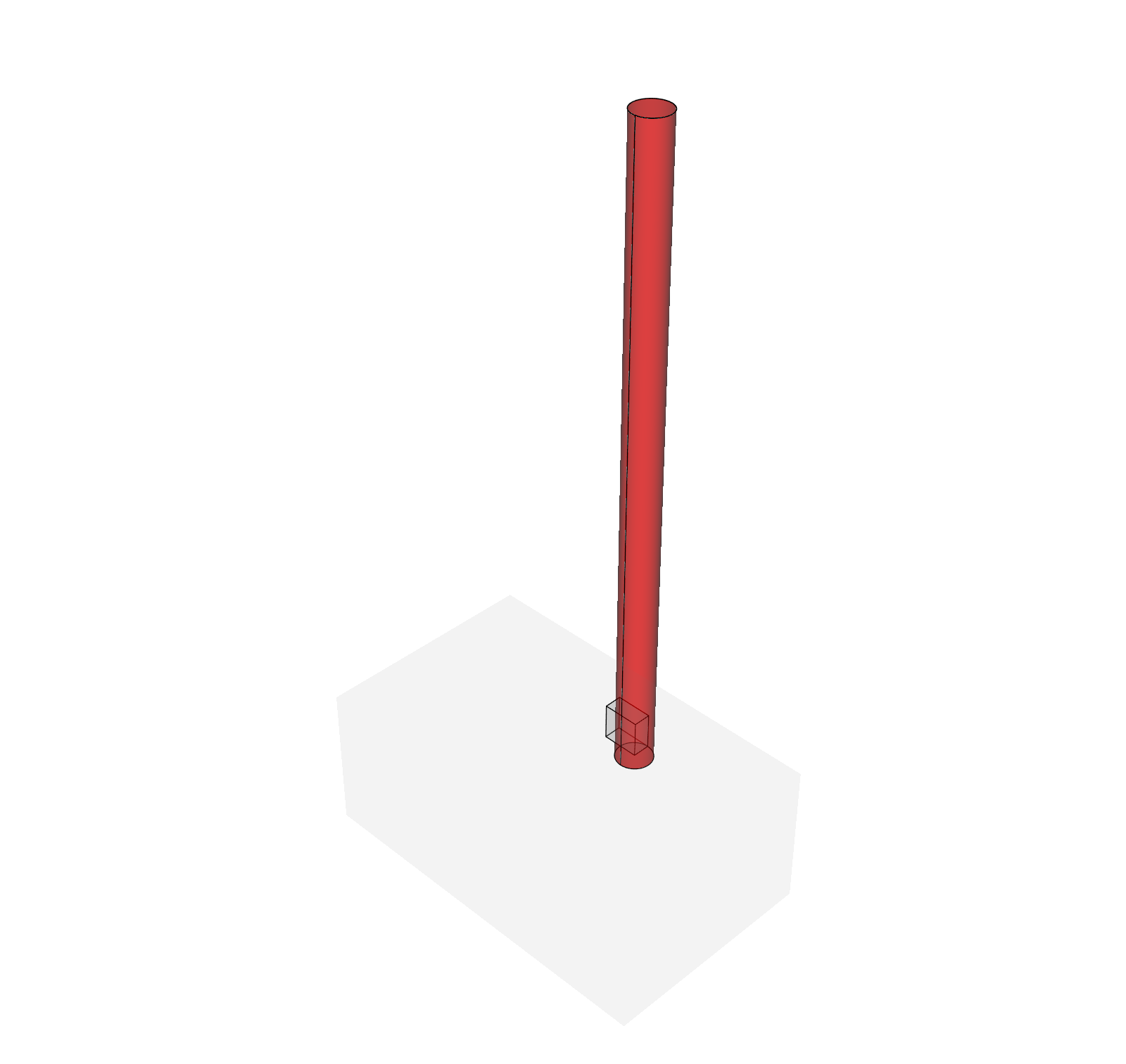} &
        \includegraphics[width=0.124\linewidth]{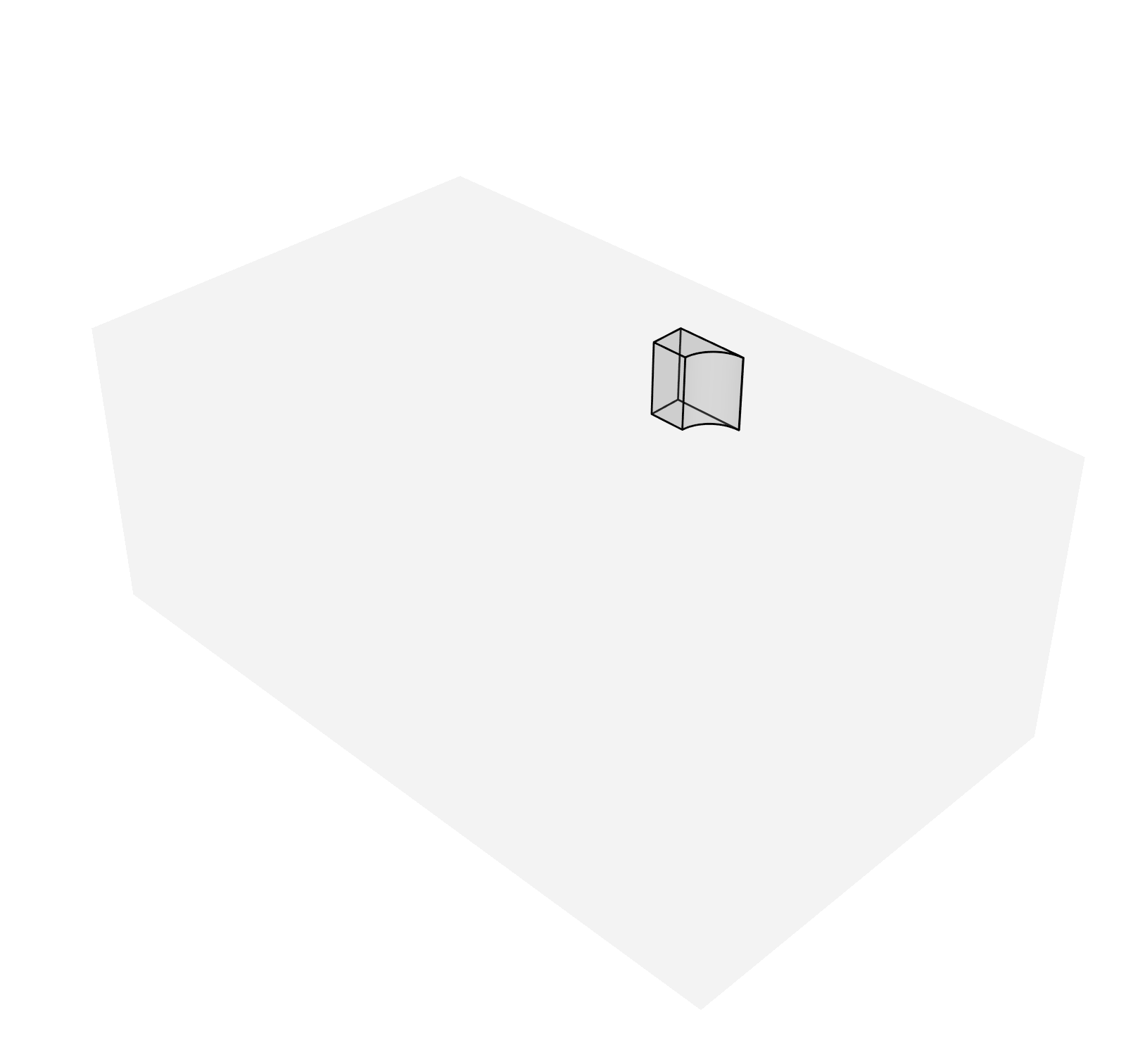} &
        \includegraphics[width=0.124\linewidth]{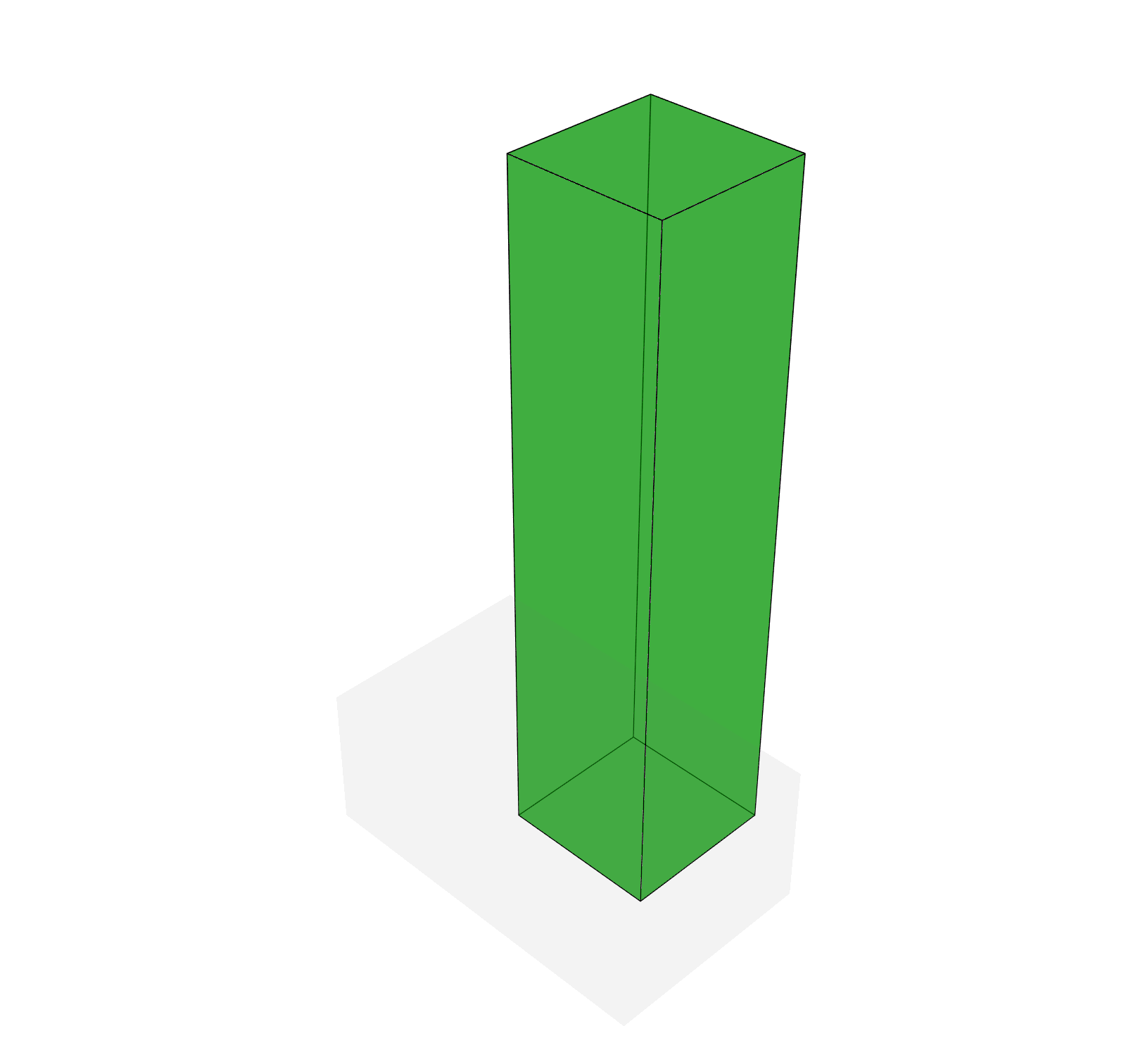} &
        \includegraphics[width=0.124\linewidth]{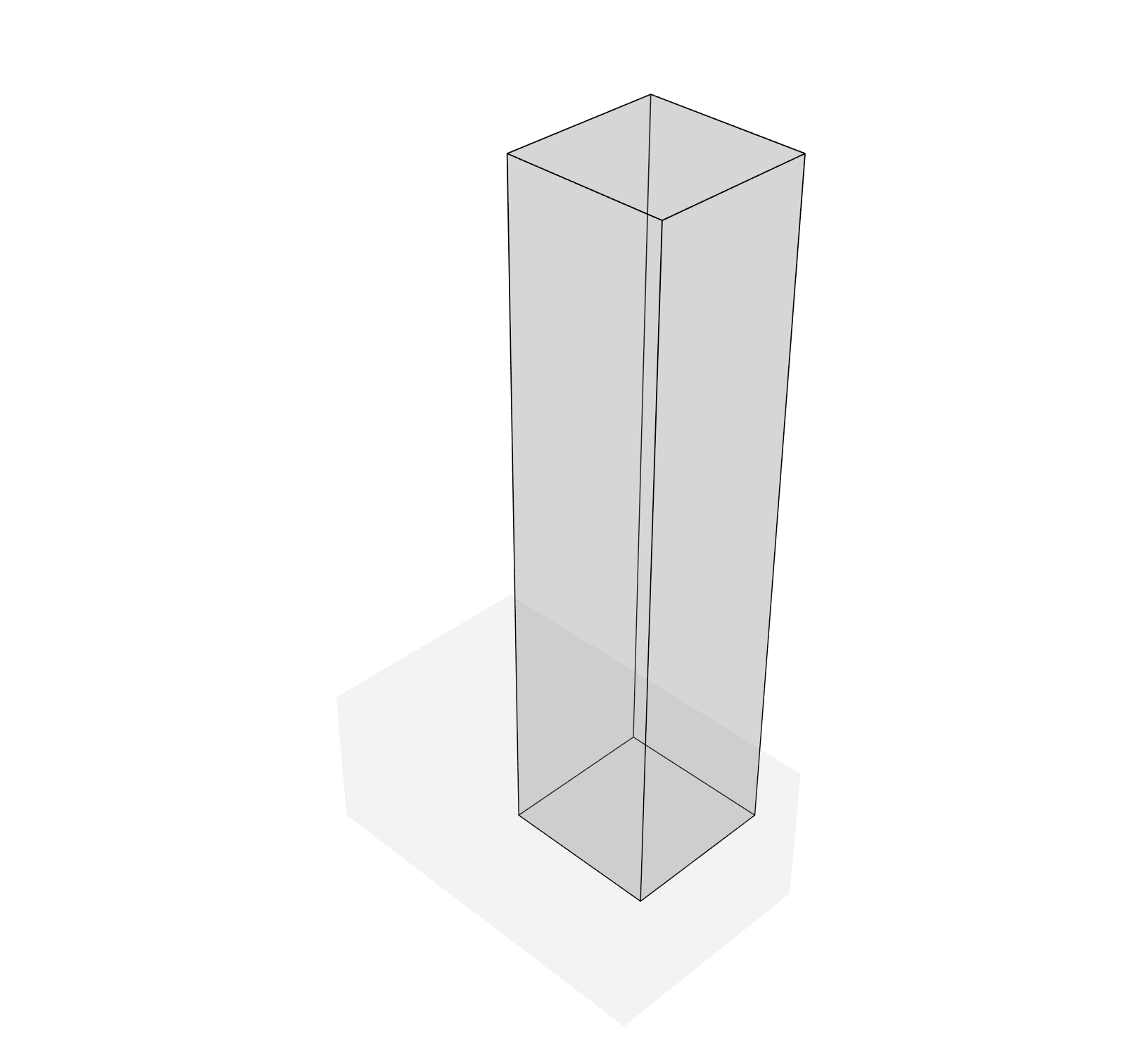}
        \\
        \includegraphics[width=0.124\linewidth]{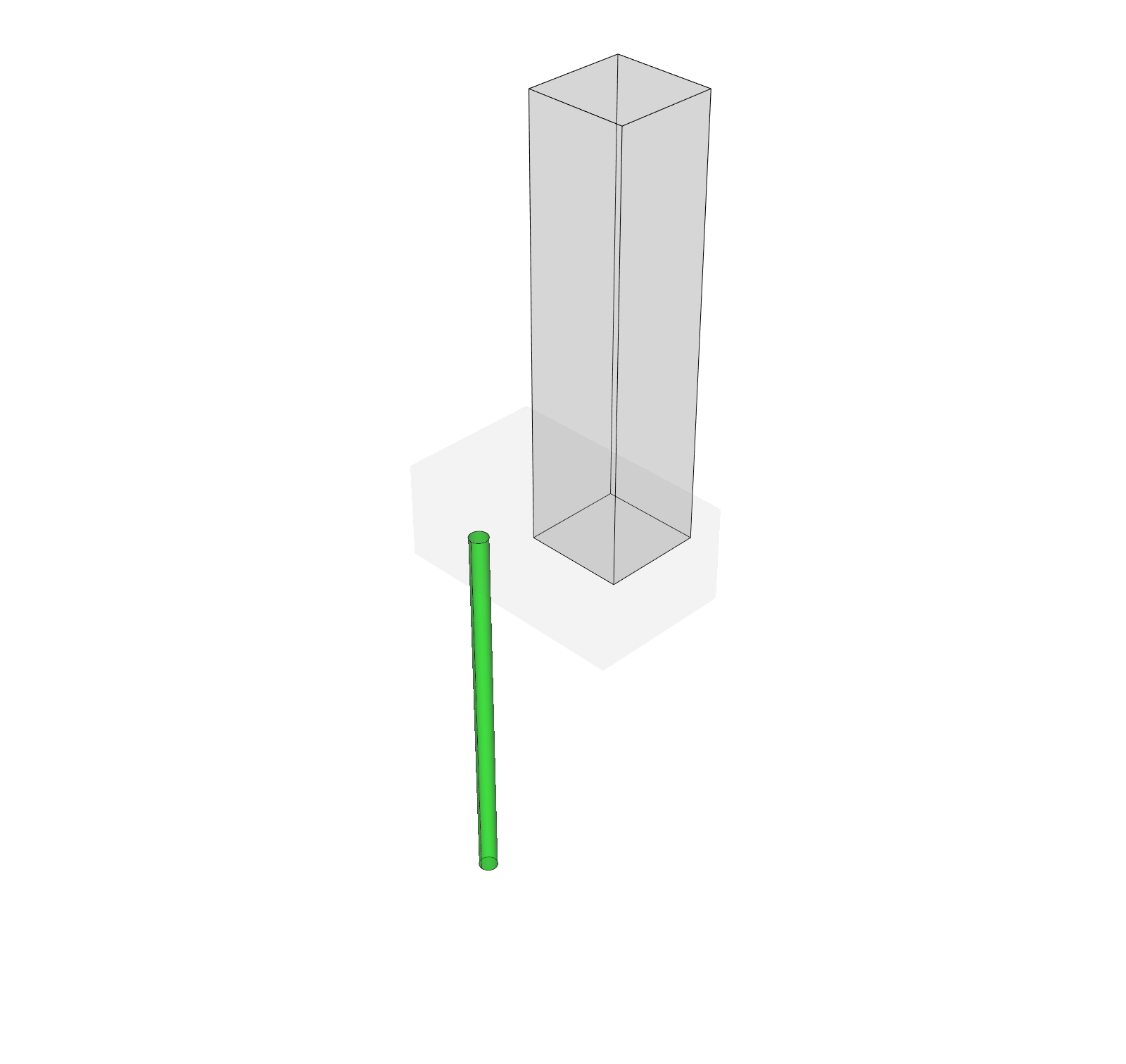} &
        \includegraphics[width=0.124\linewidth]{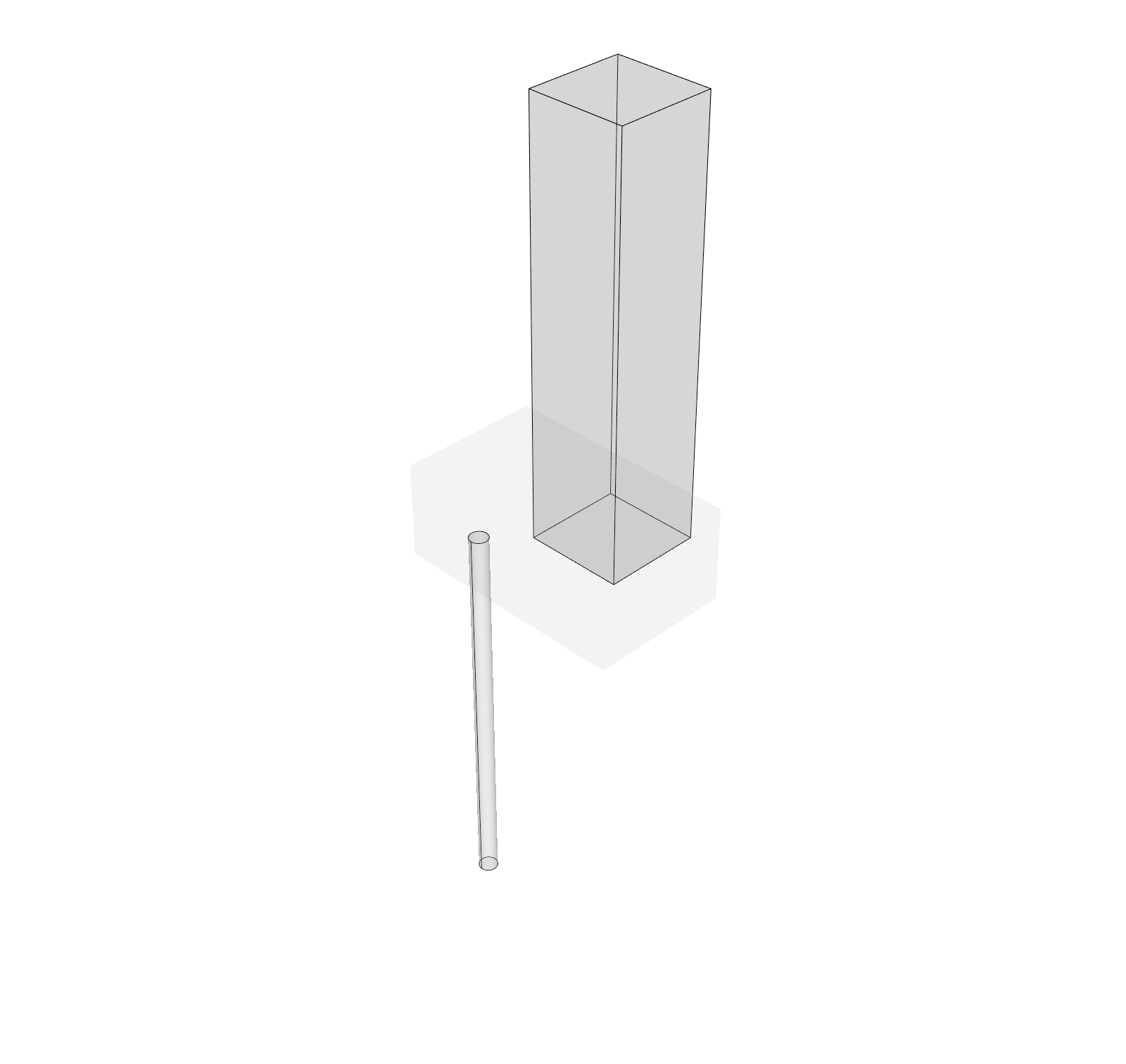} &
        \includegraphics[width=0.124\linewidth]{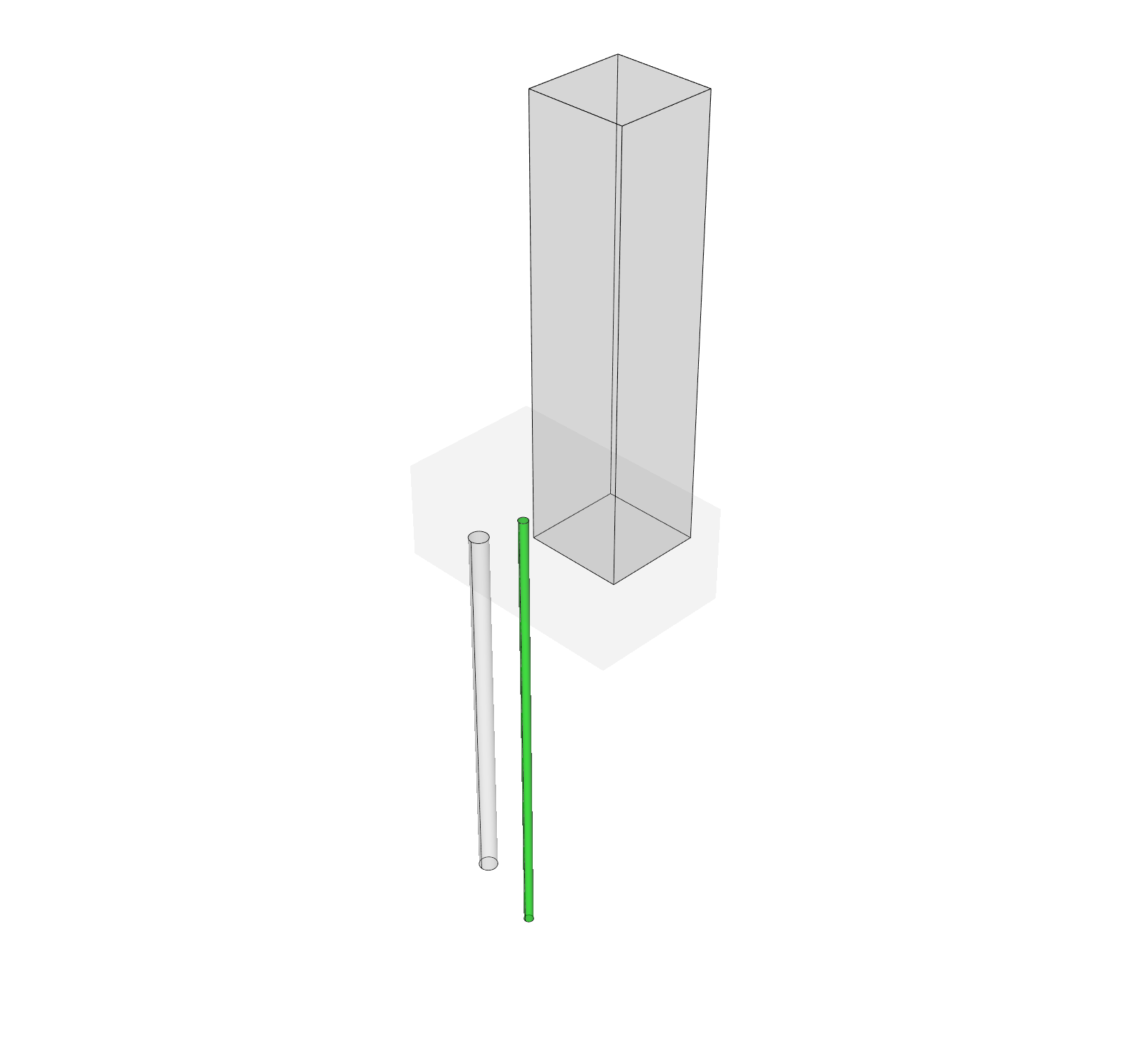} &
        \includegraphics[width=0.124\linewidth]{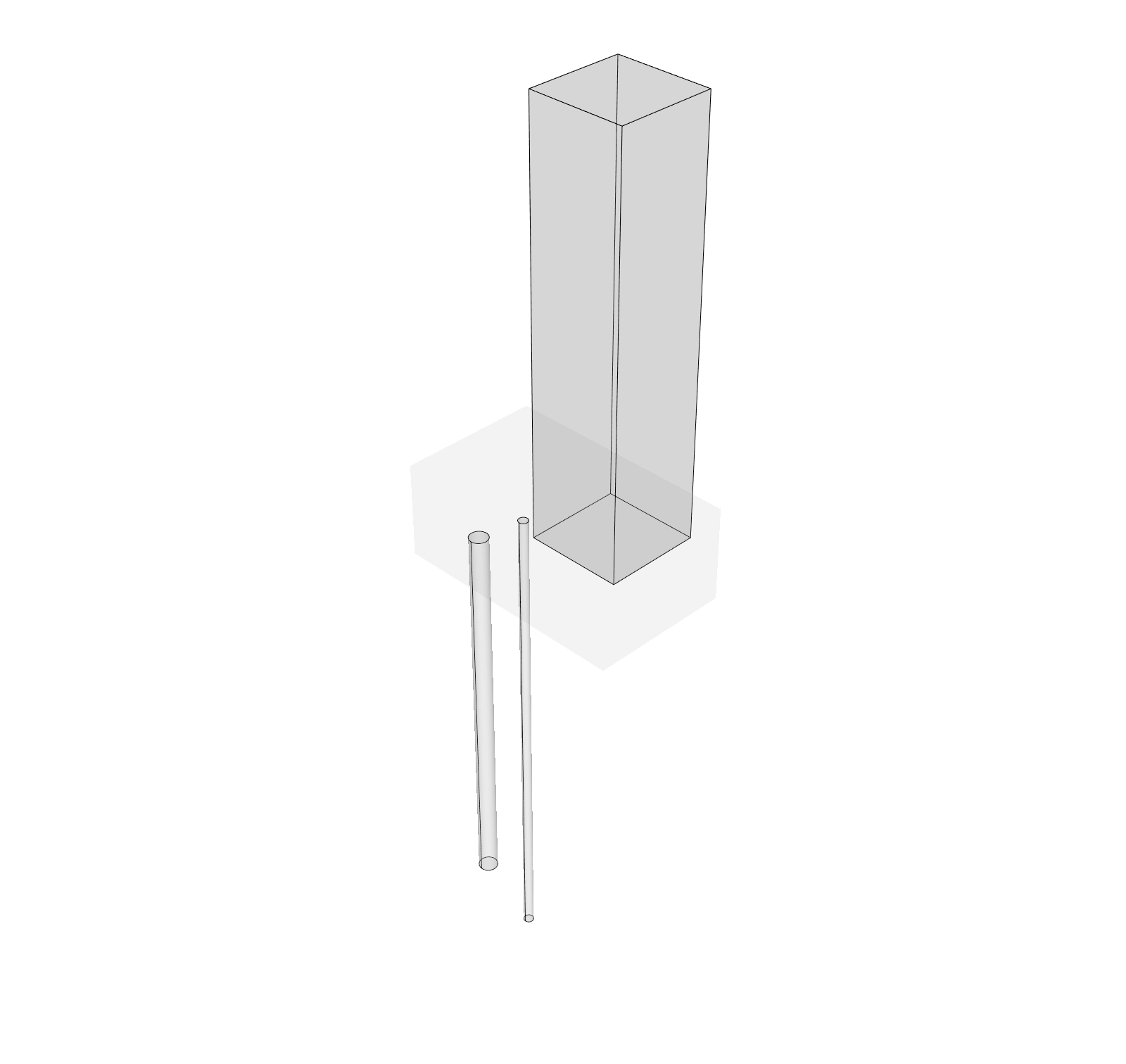} &
        \includegraphics[width=0.124\linewidth]{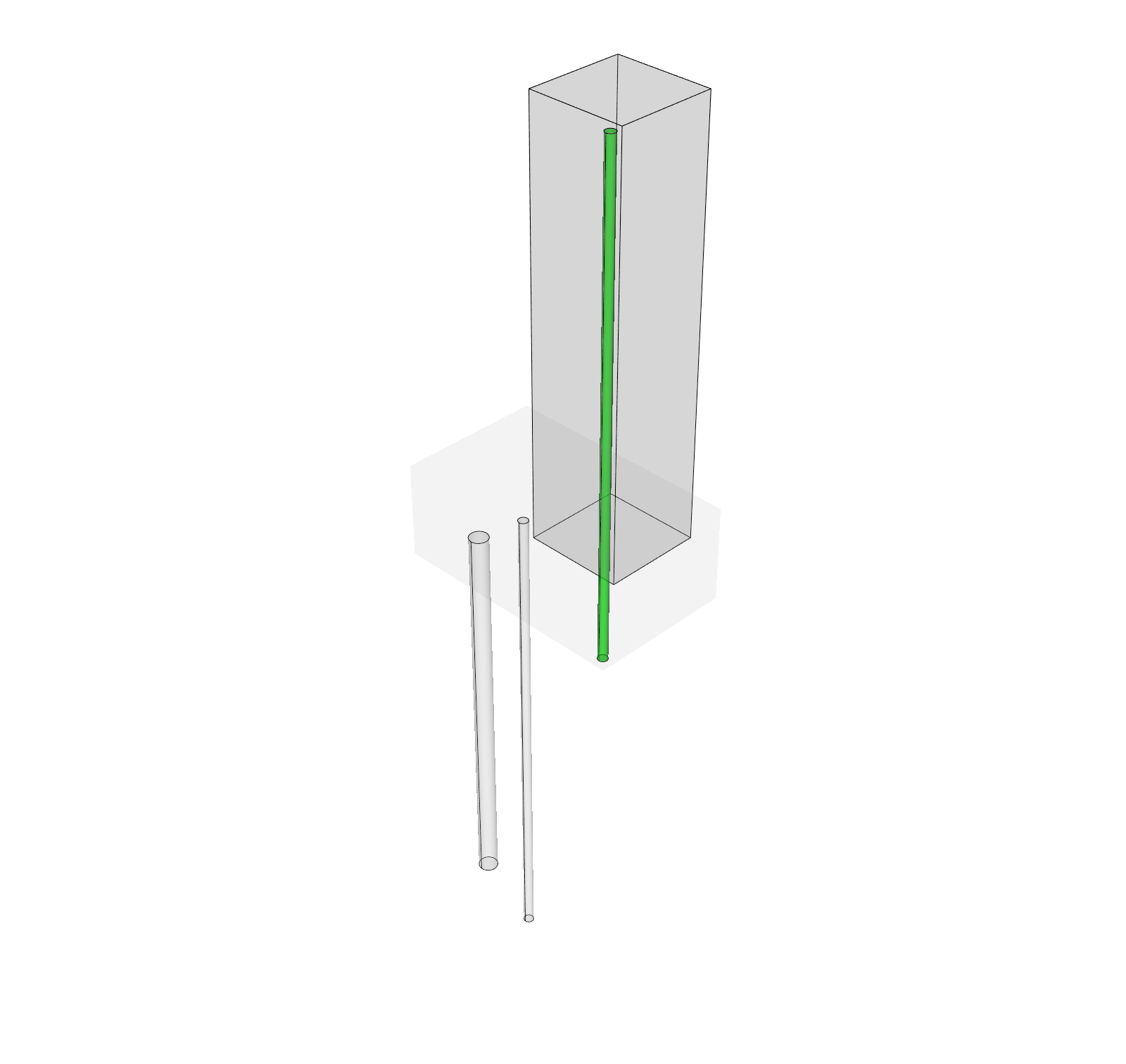} &
        \includegraphics[width=0.124\linewidth]{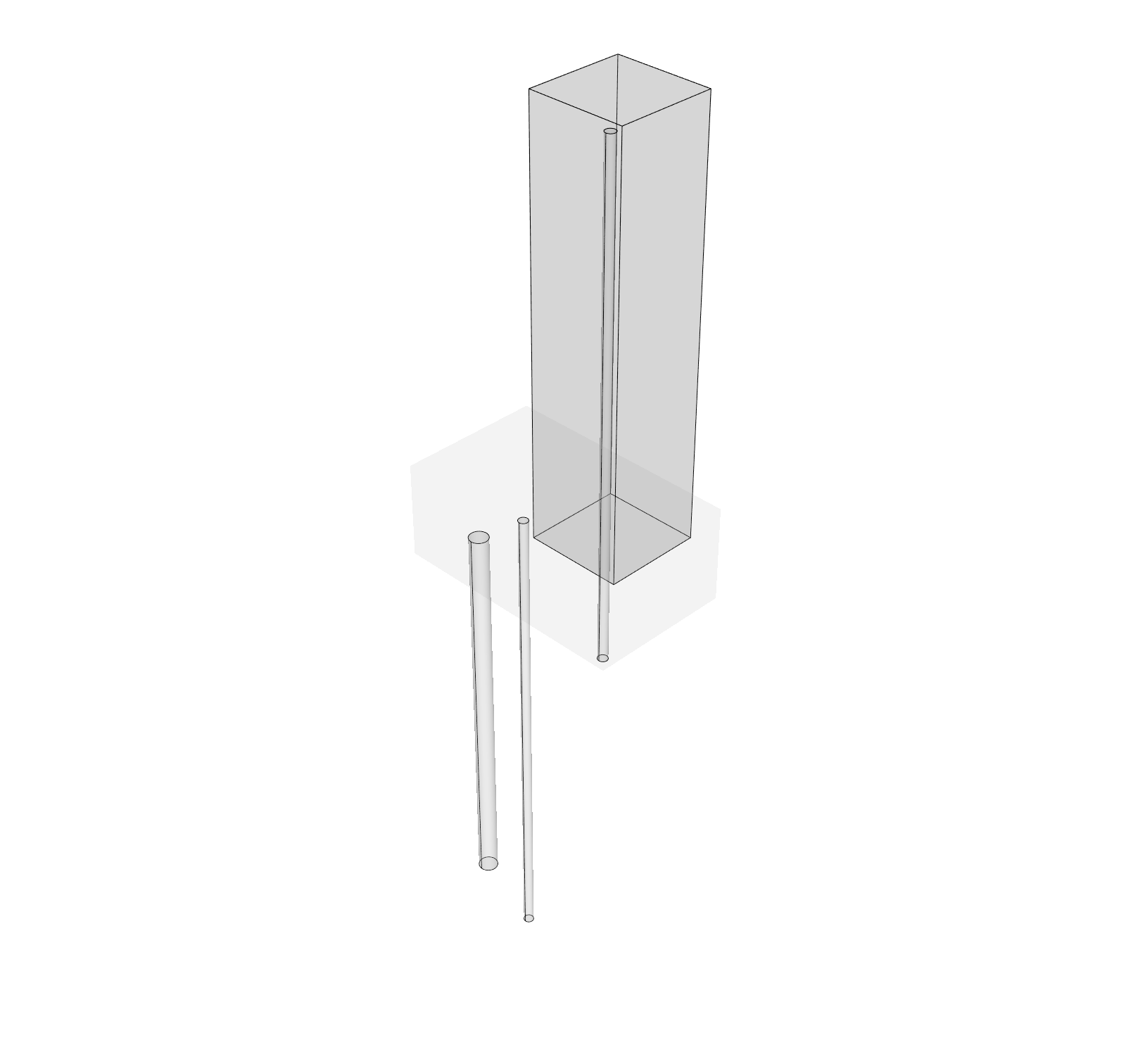} &
        \includegraphics[width=0.124\linewidth]{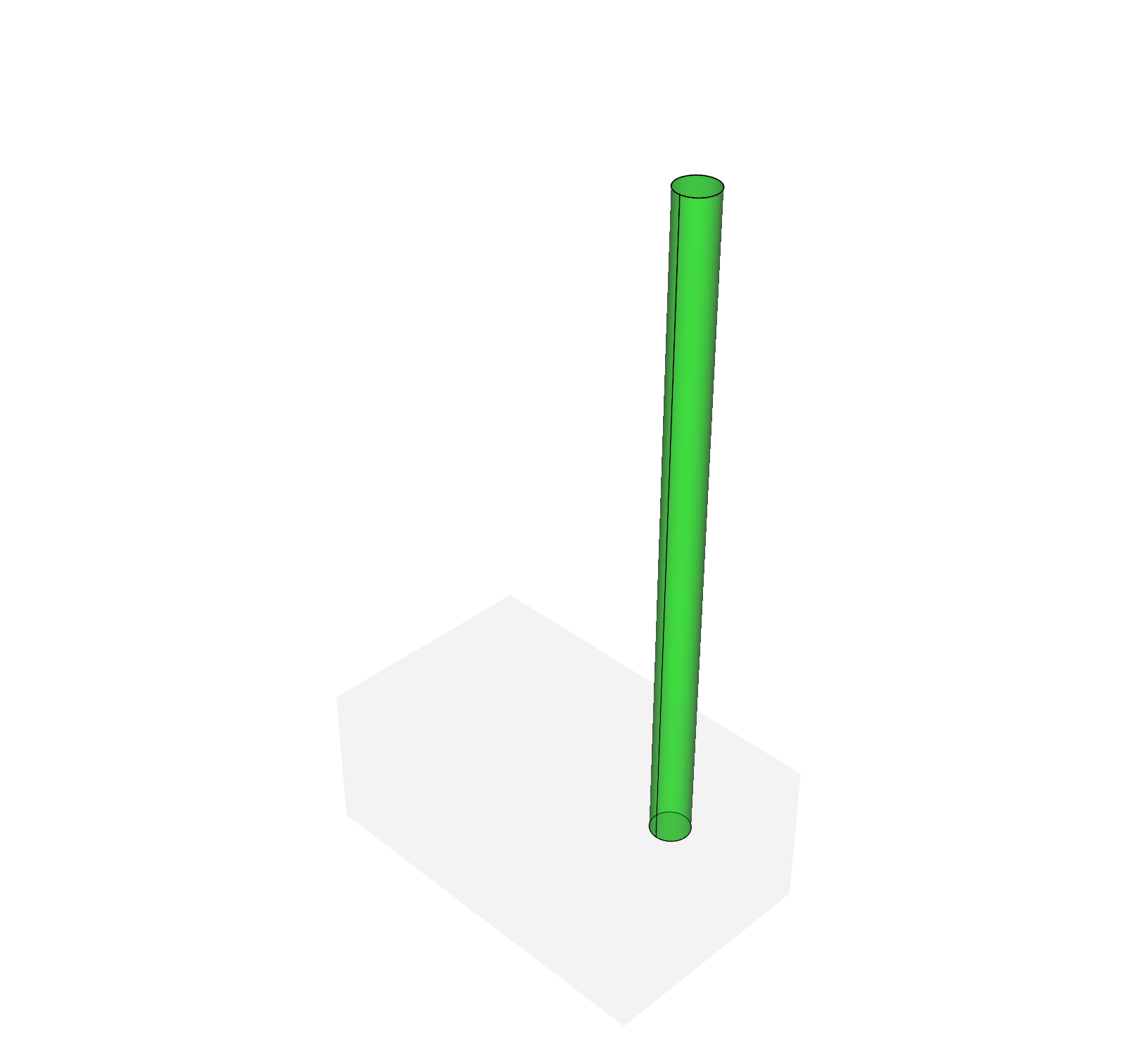} &
        \includegraphics[width=0.124\linewidth]{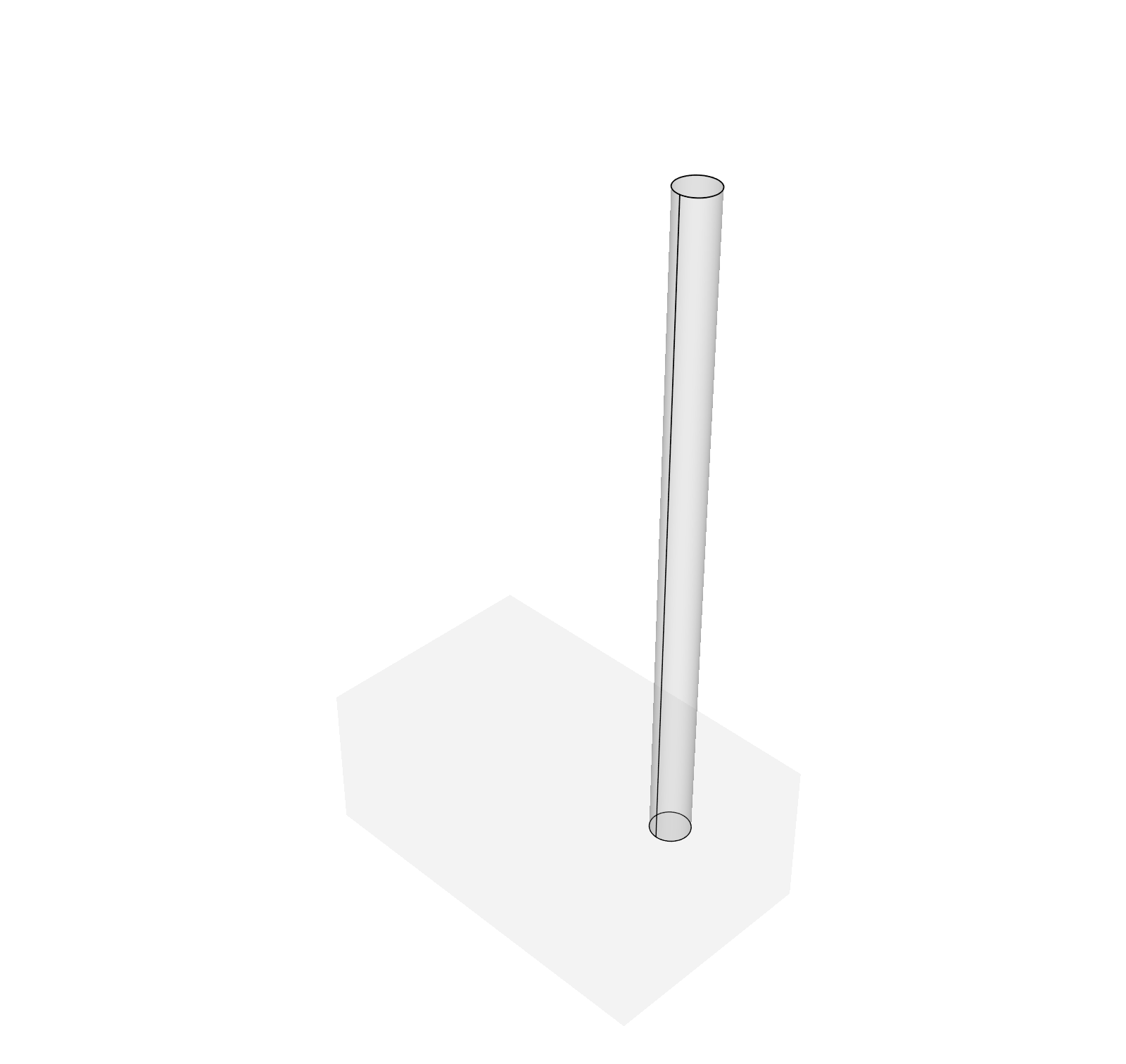}
        \\
        \includegraphics[width=0.124\linewidth]{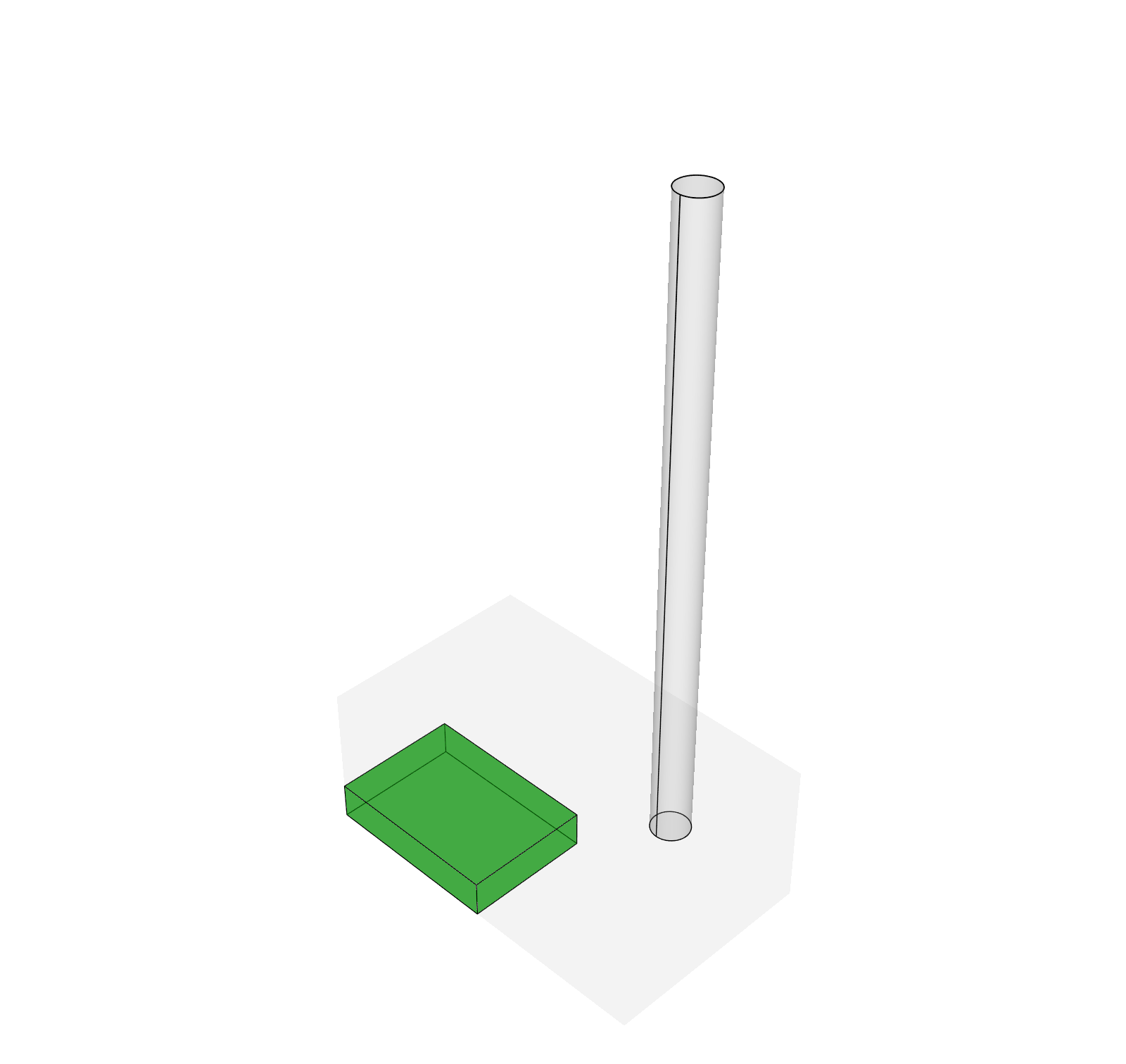} &
        \includegraphics[width=0.124\linewidth]{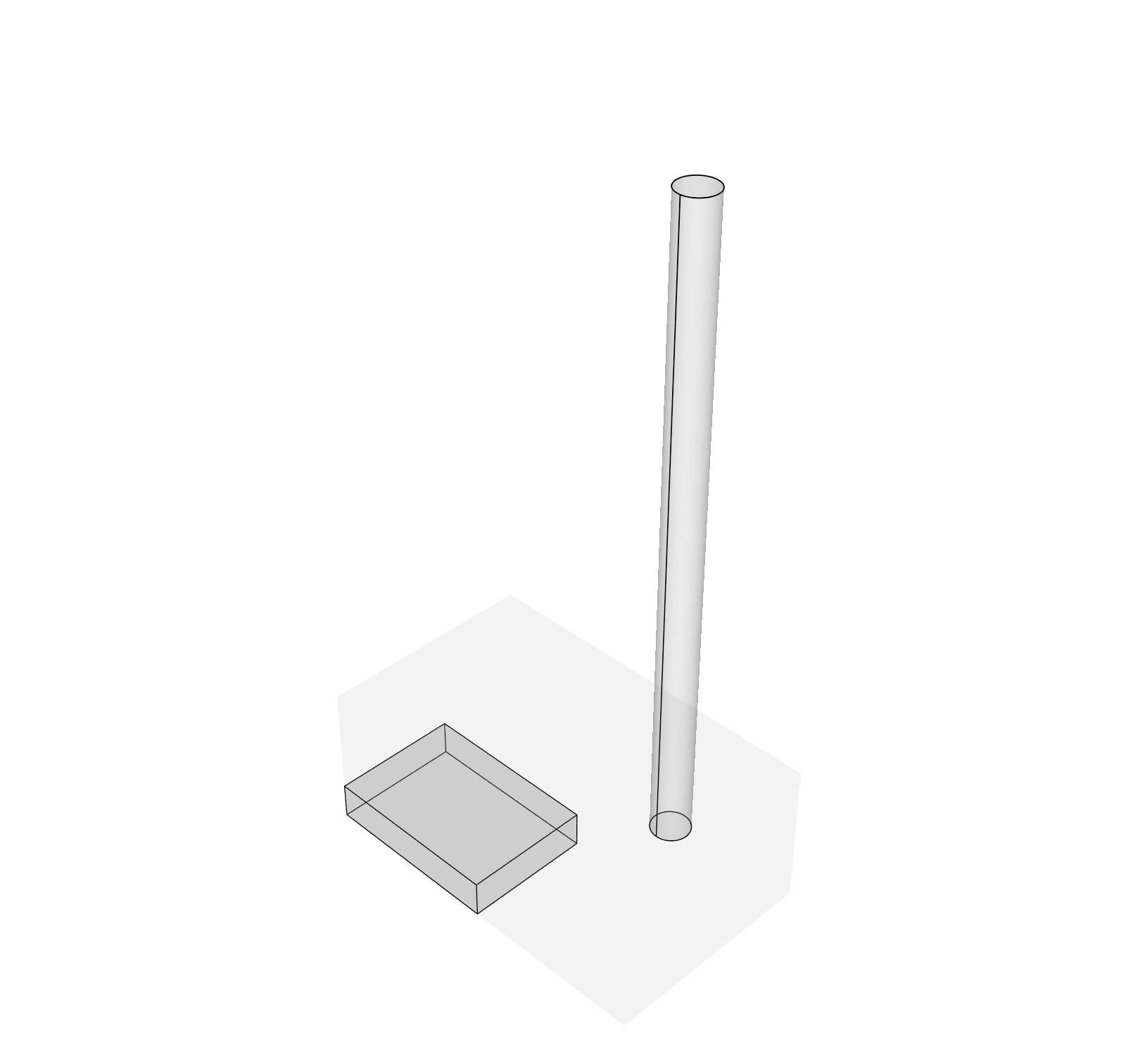} &
        \includegraphics[width=0.124\linewidth]{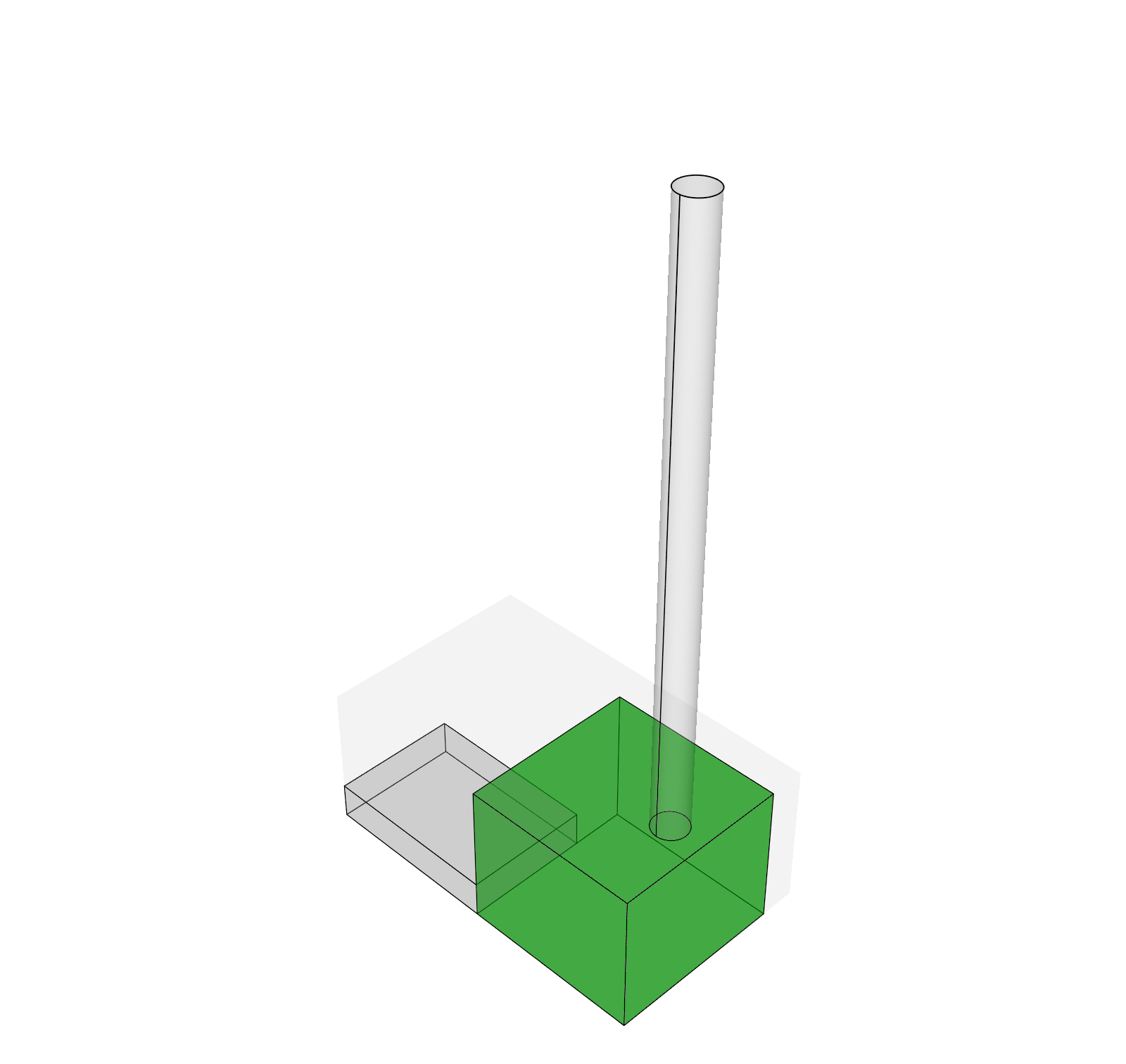} &
        \includegraphics[width=0.124\linewidth]{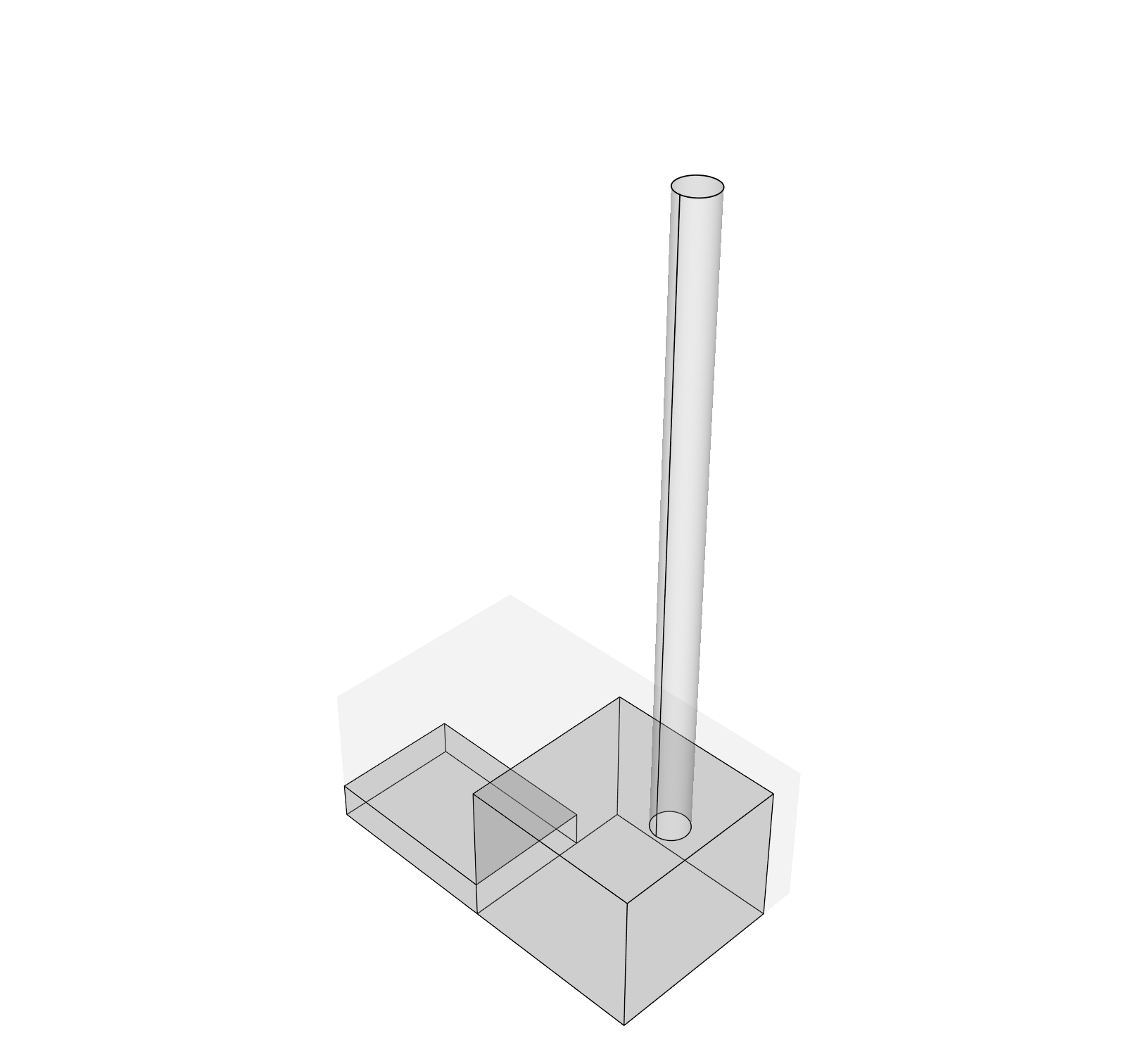} &
        \includegraphics[width=0.124\linewidth]{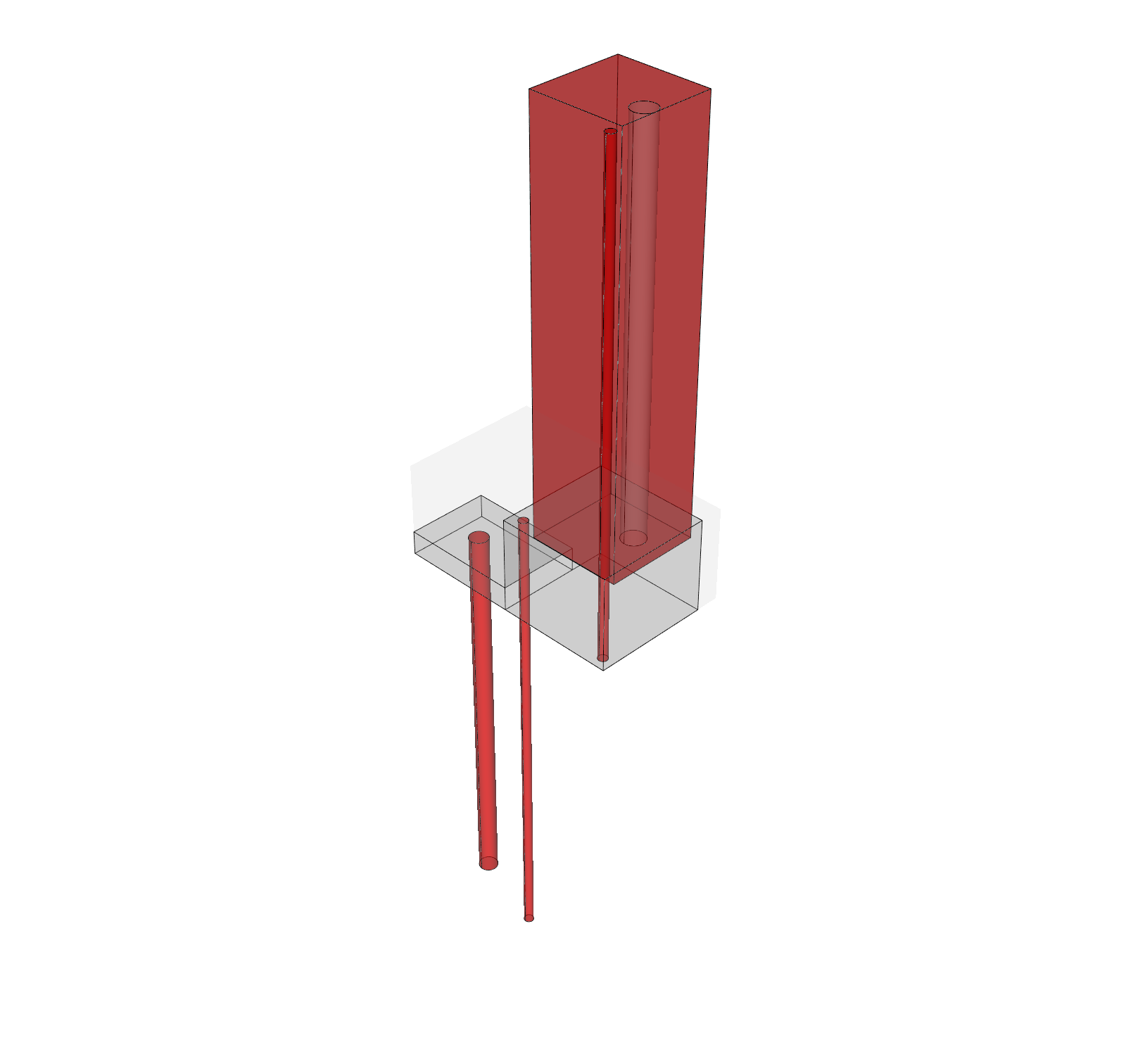} &
        \includegraphics[width=0.124\linewidth]{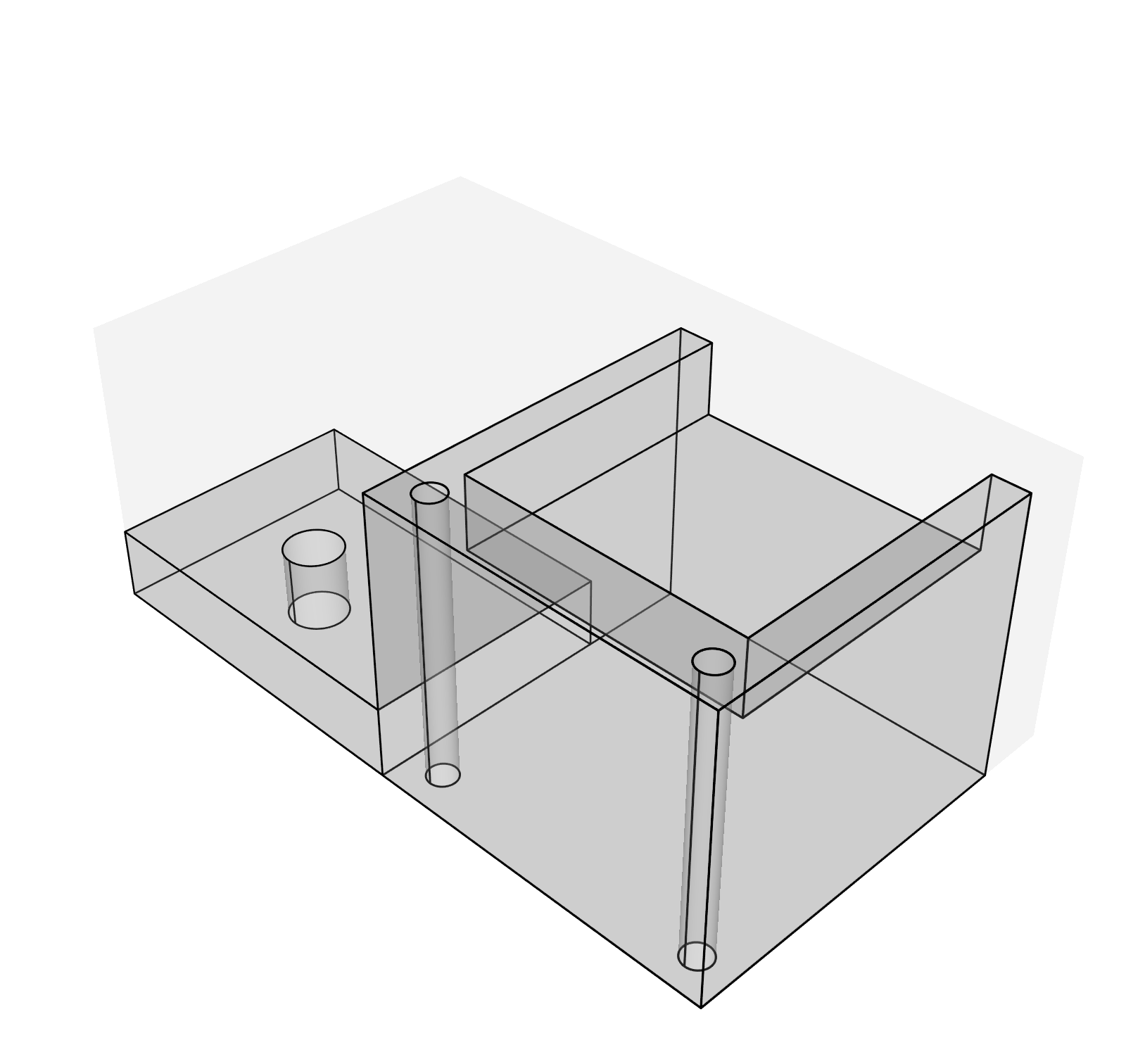} &
        \includegraphics[width=0.124\linewidth]{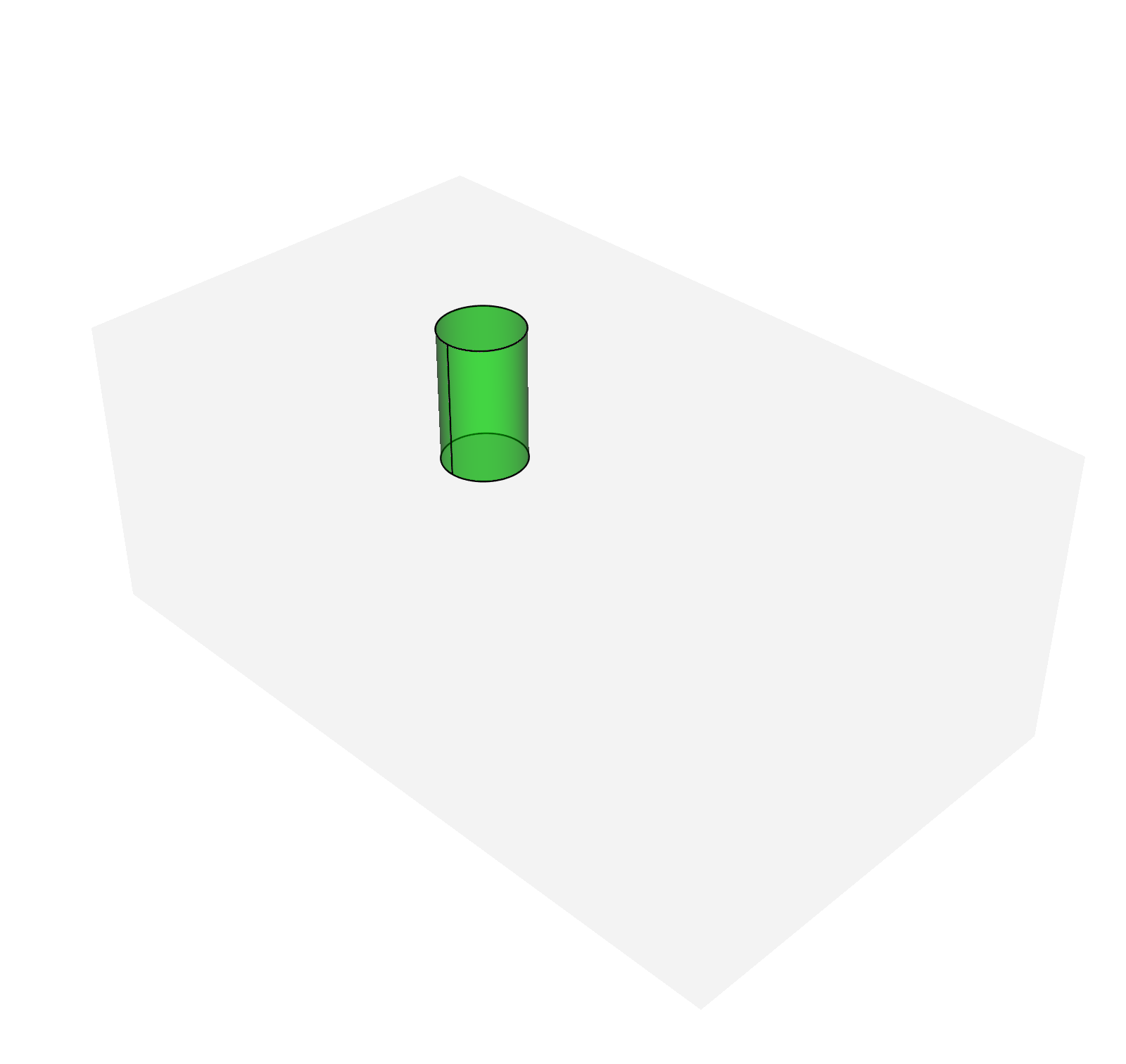} &
        \includegraphics[width=0.124\linewidth]{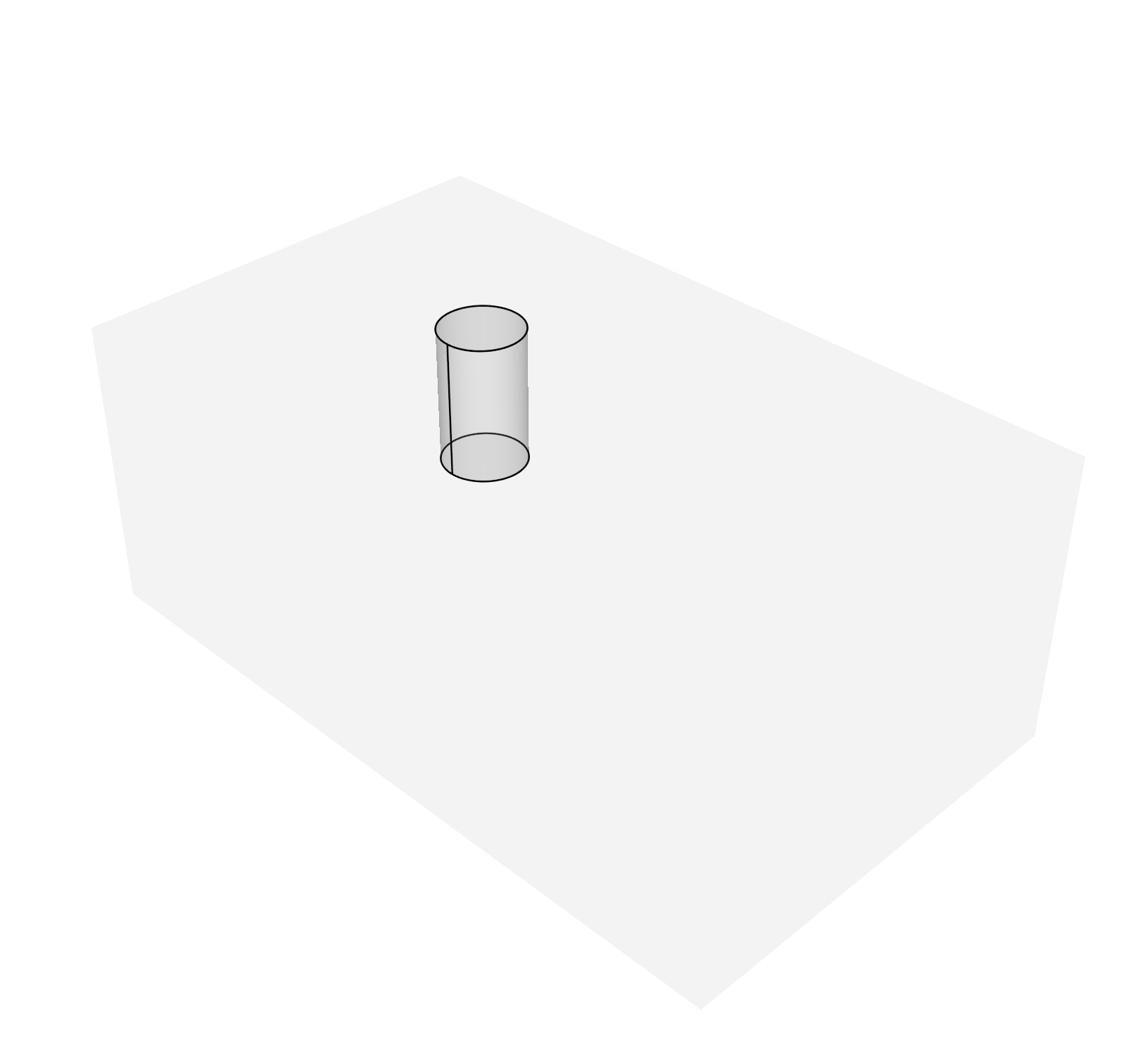}
        \\
        \includegraphics[width=0.124\linewidth]{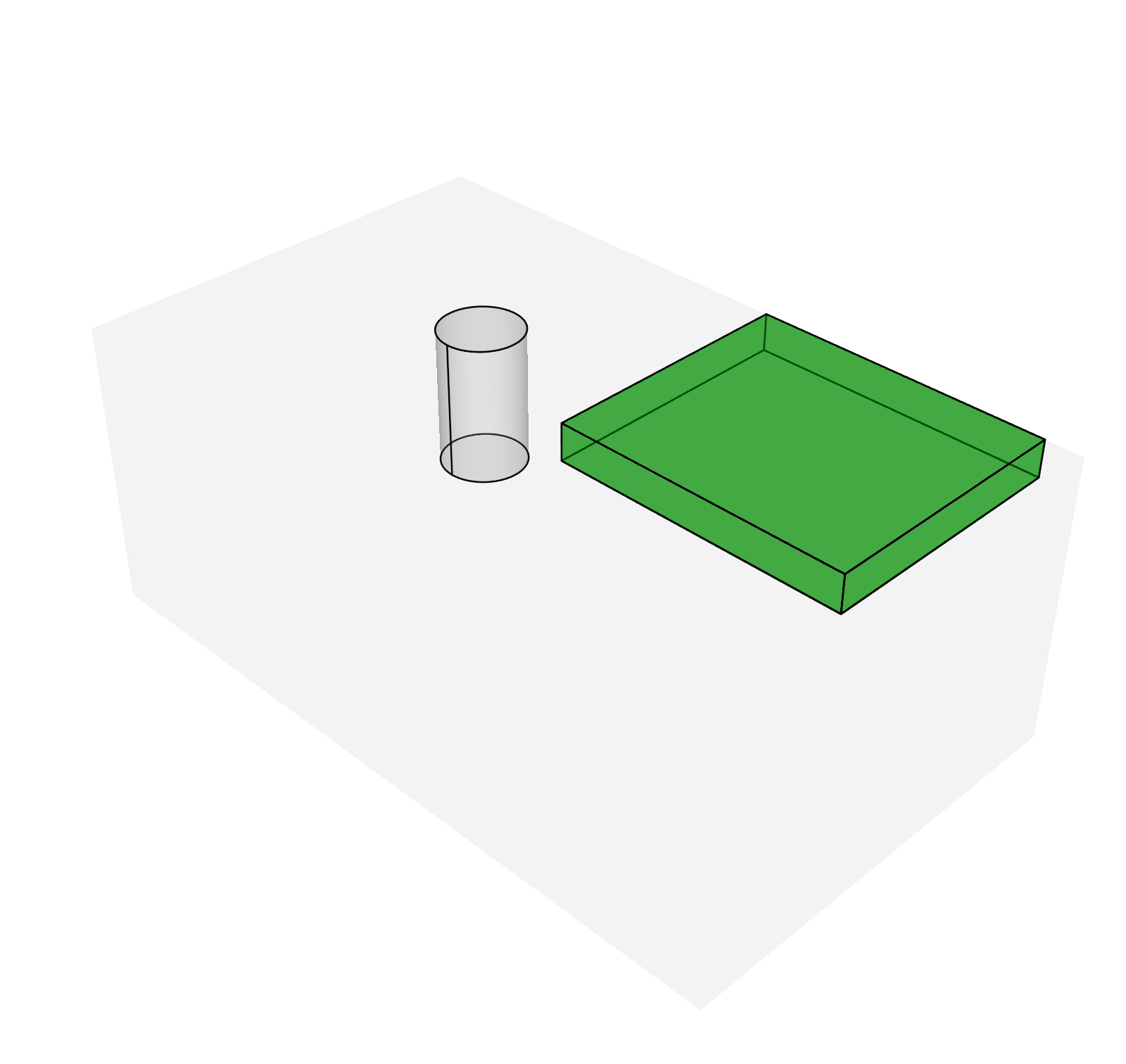} &
        \includegraphics[width=0.124\linewidth]{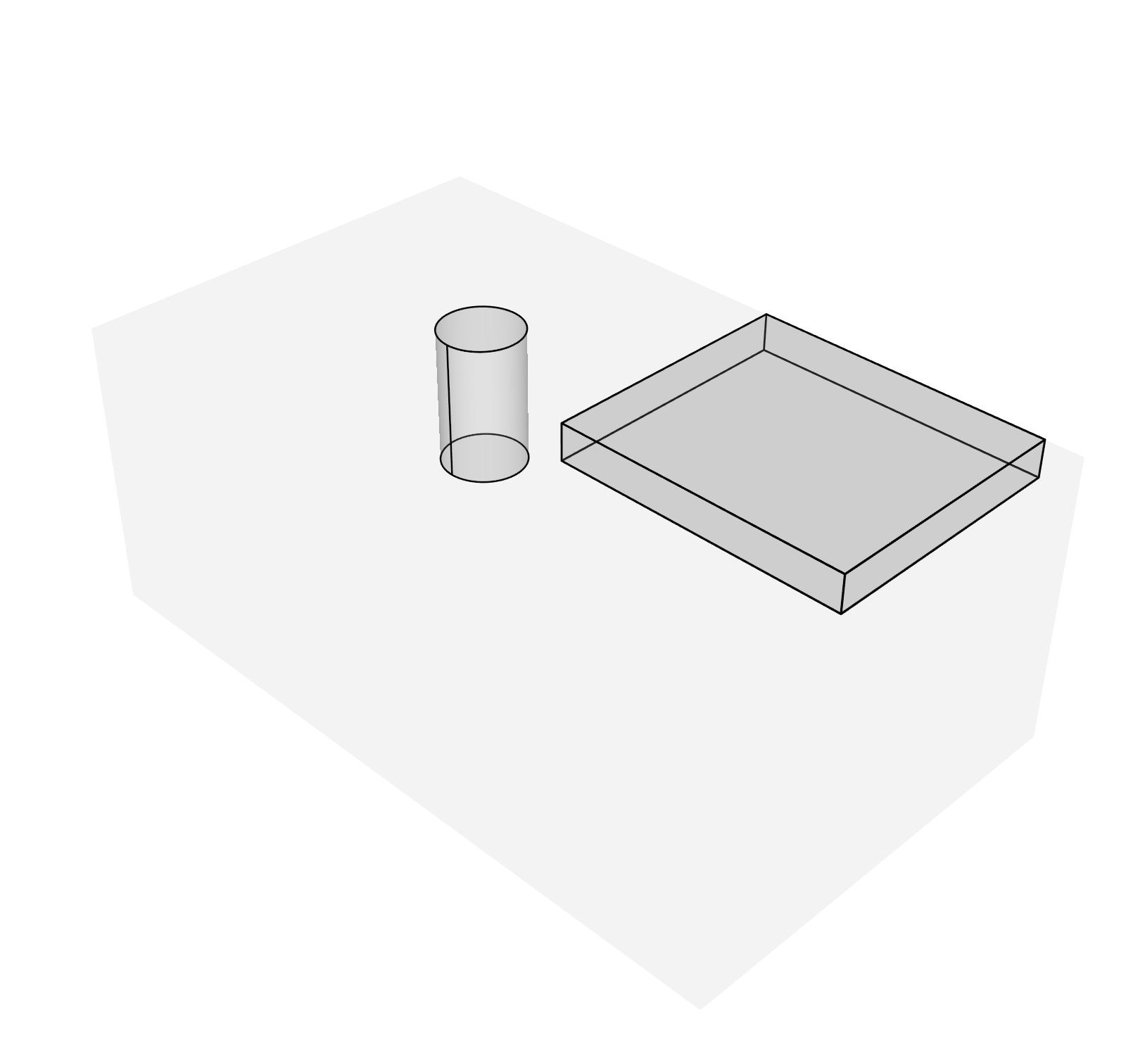} &
        \includegraphics[width=0.124\linewidth]{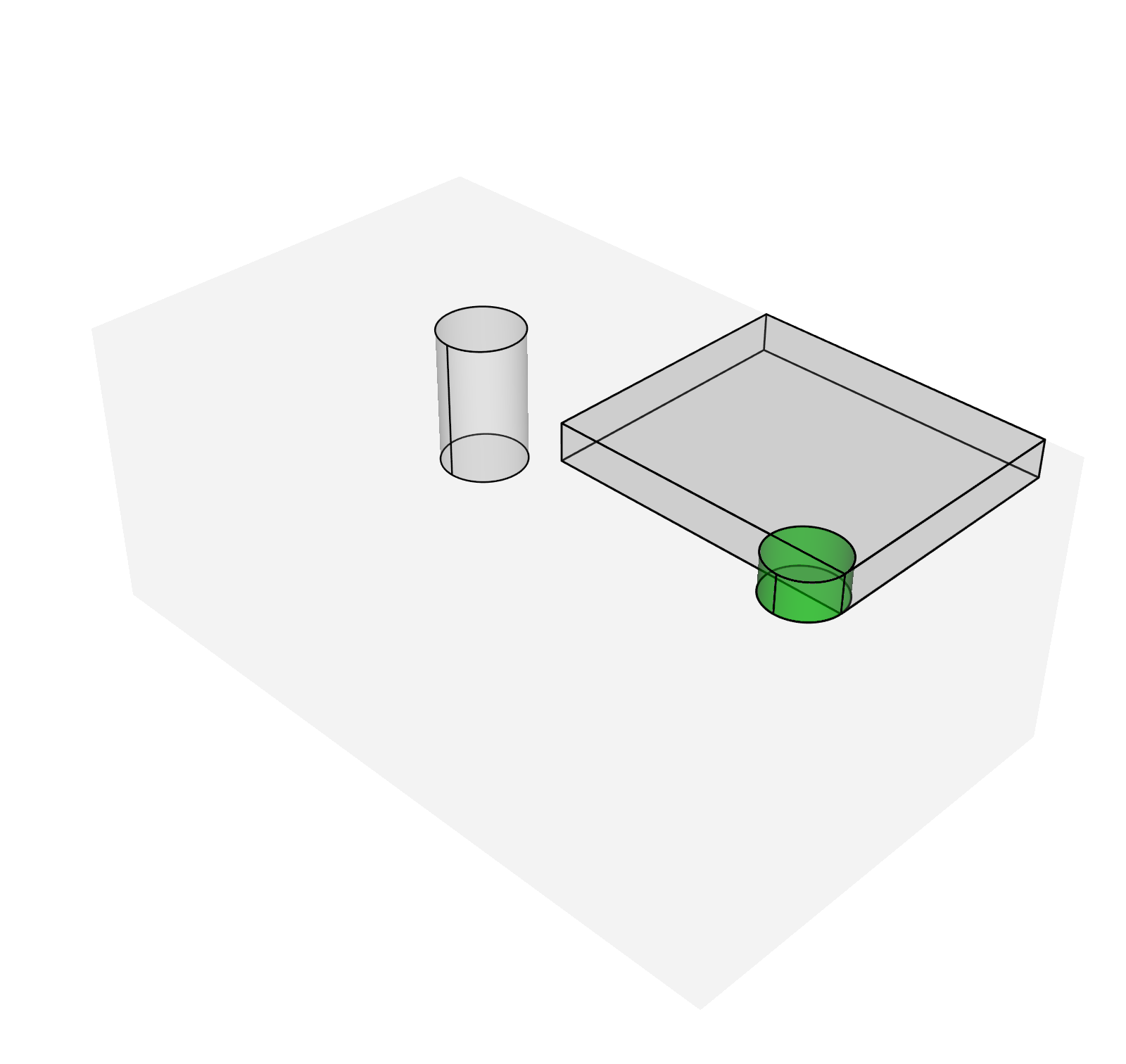} &
        \includegraphics[width=0.124\linewidth]{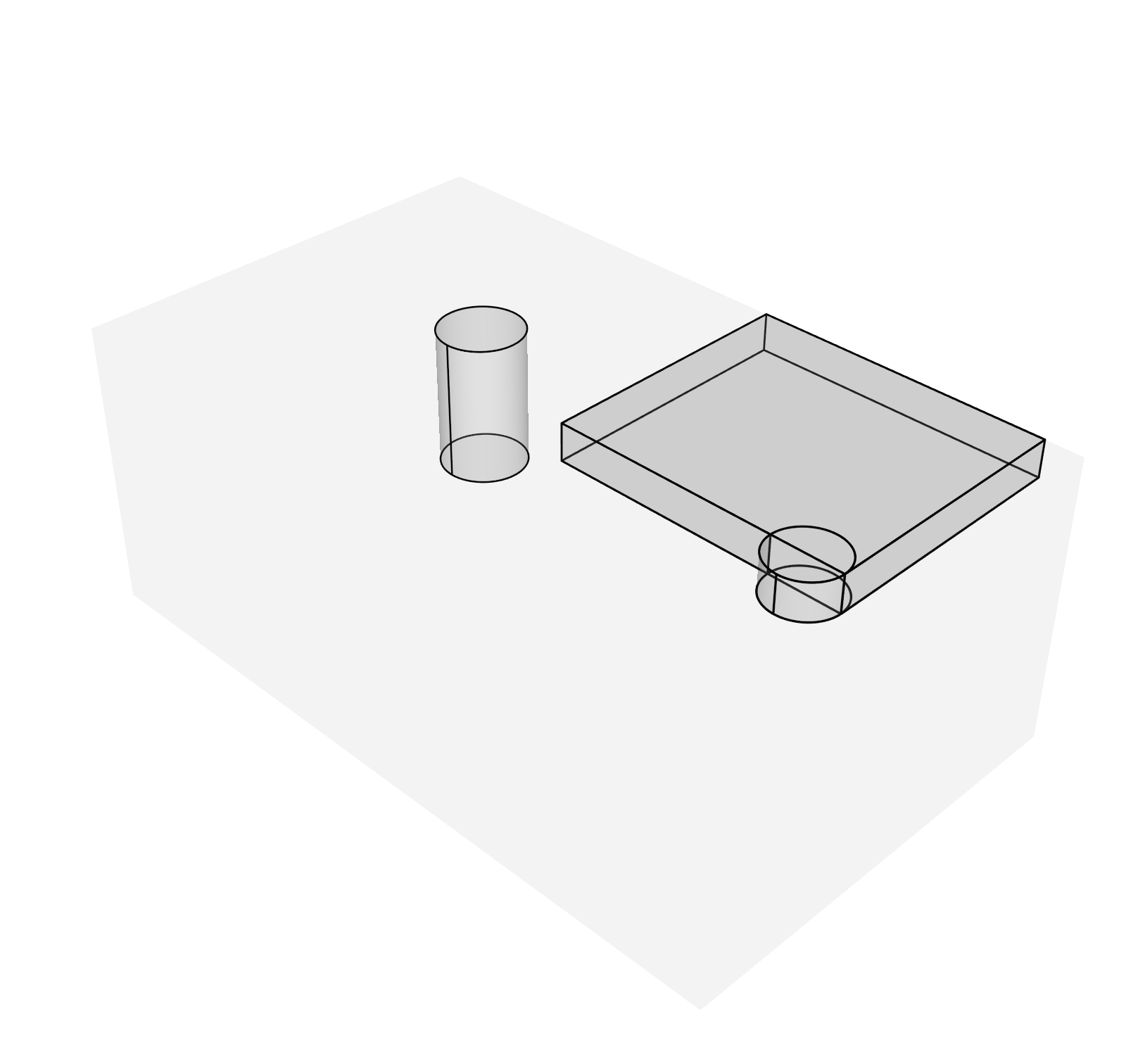} &
        \includegraphics[width=0.124\linewidth]{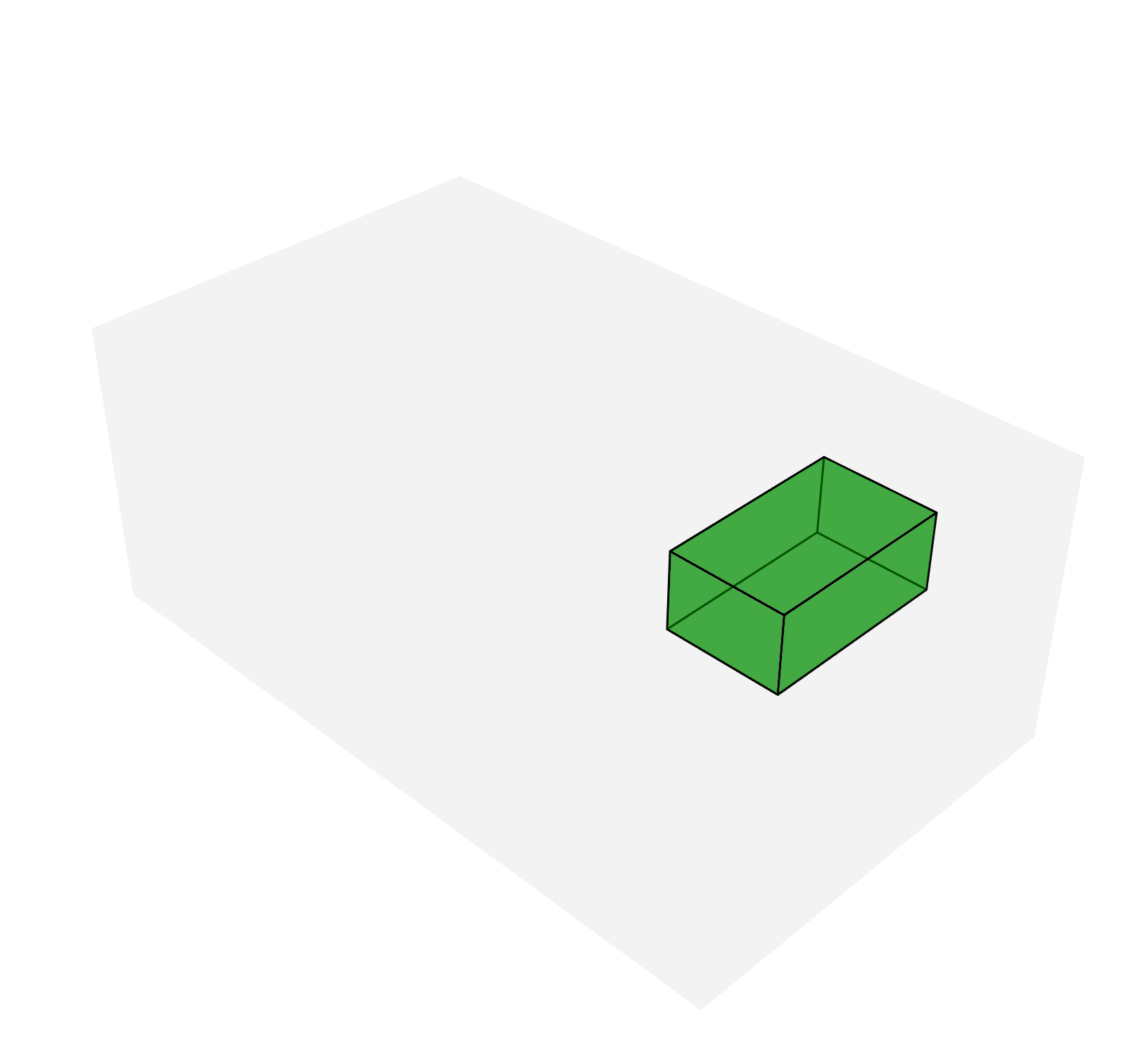} &
        \includegraphics[width=0.124\linewidth]{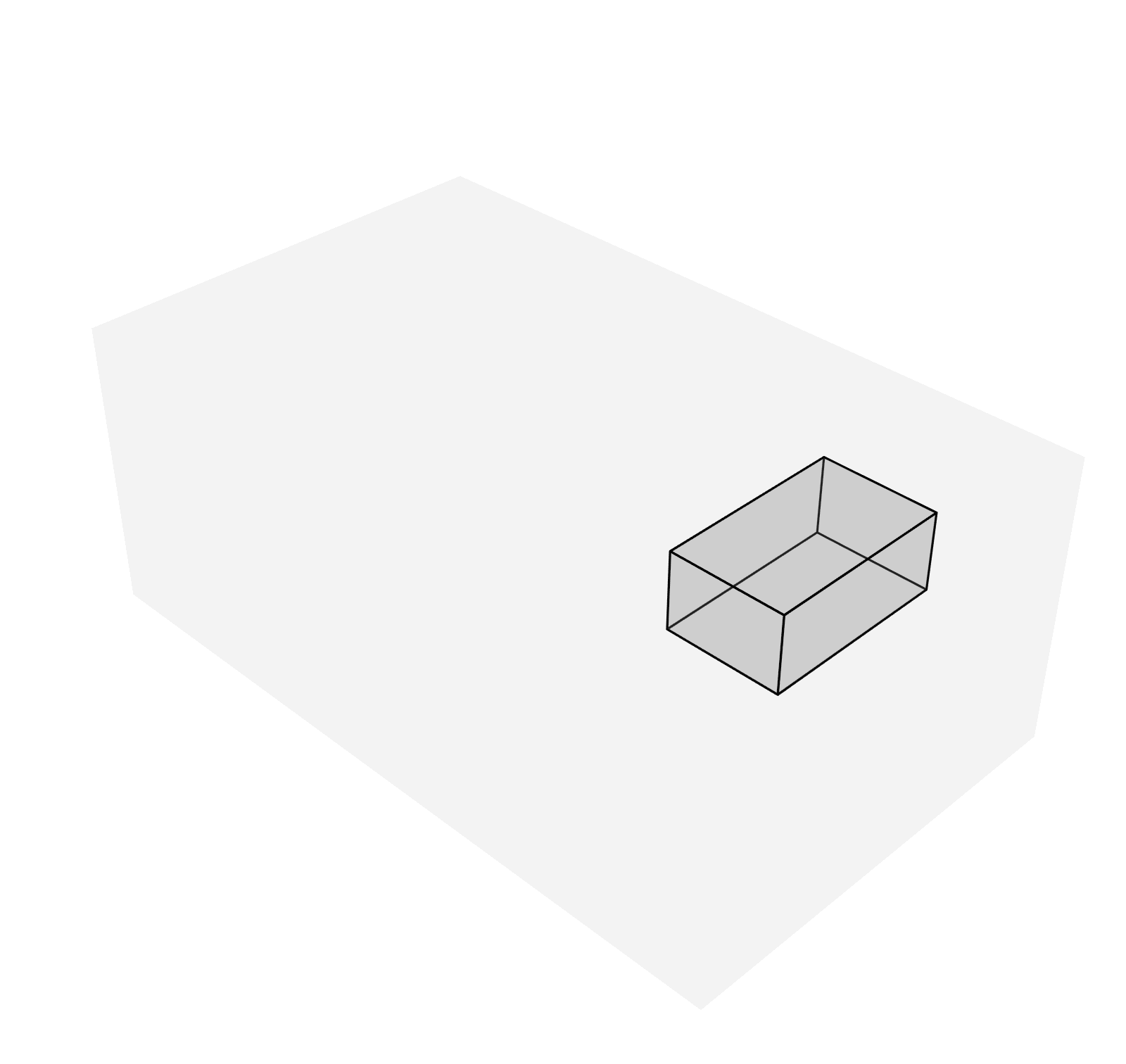} &
        \includegraphics[width=0.124\linewidth]{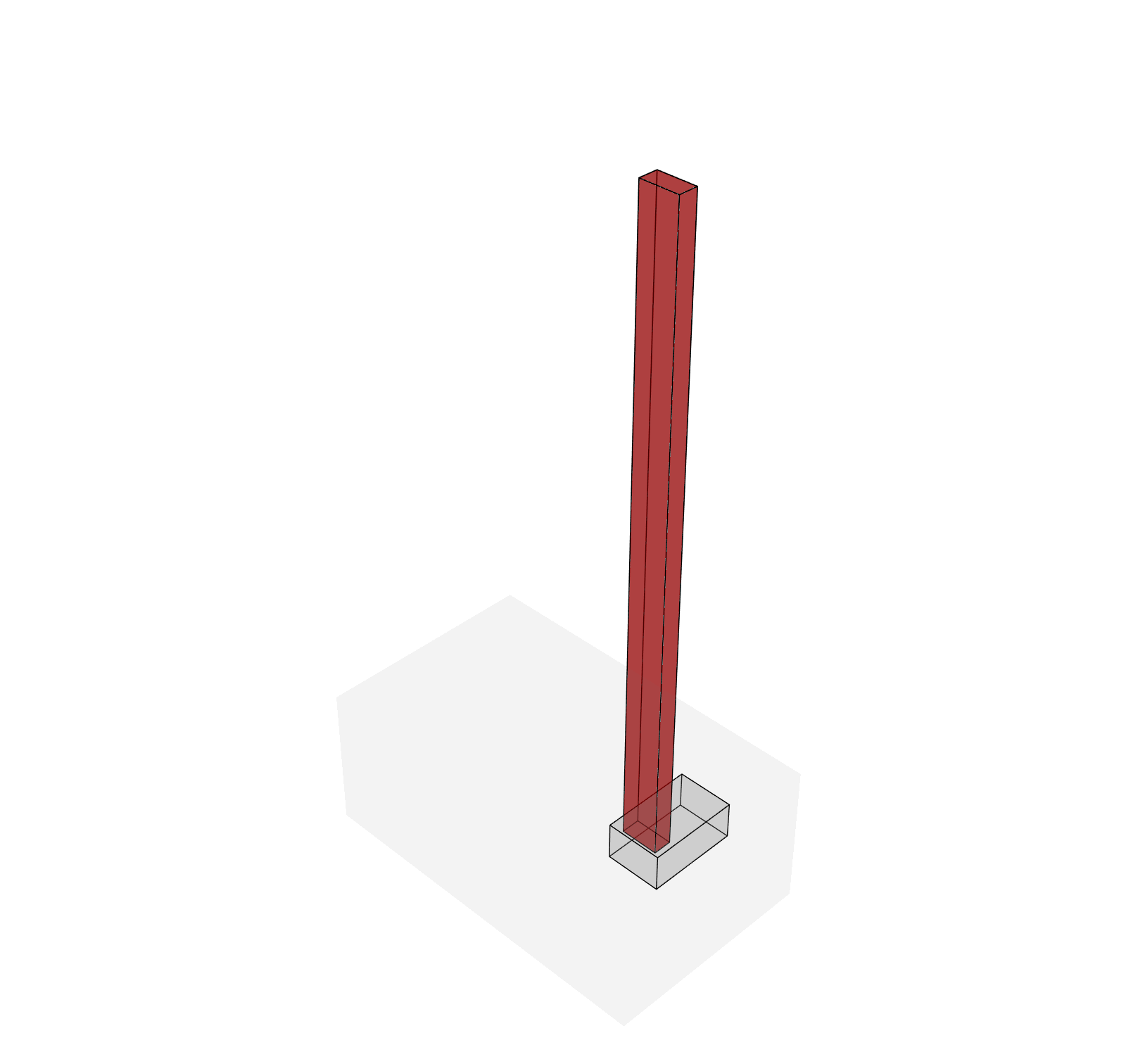} &
        \includegraphics[width=0.124\linewidth]{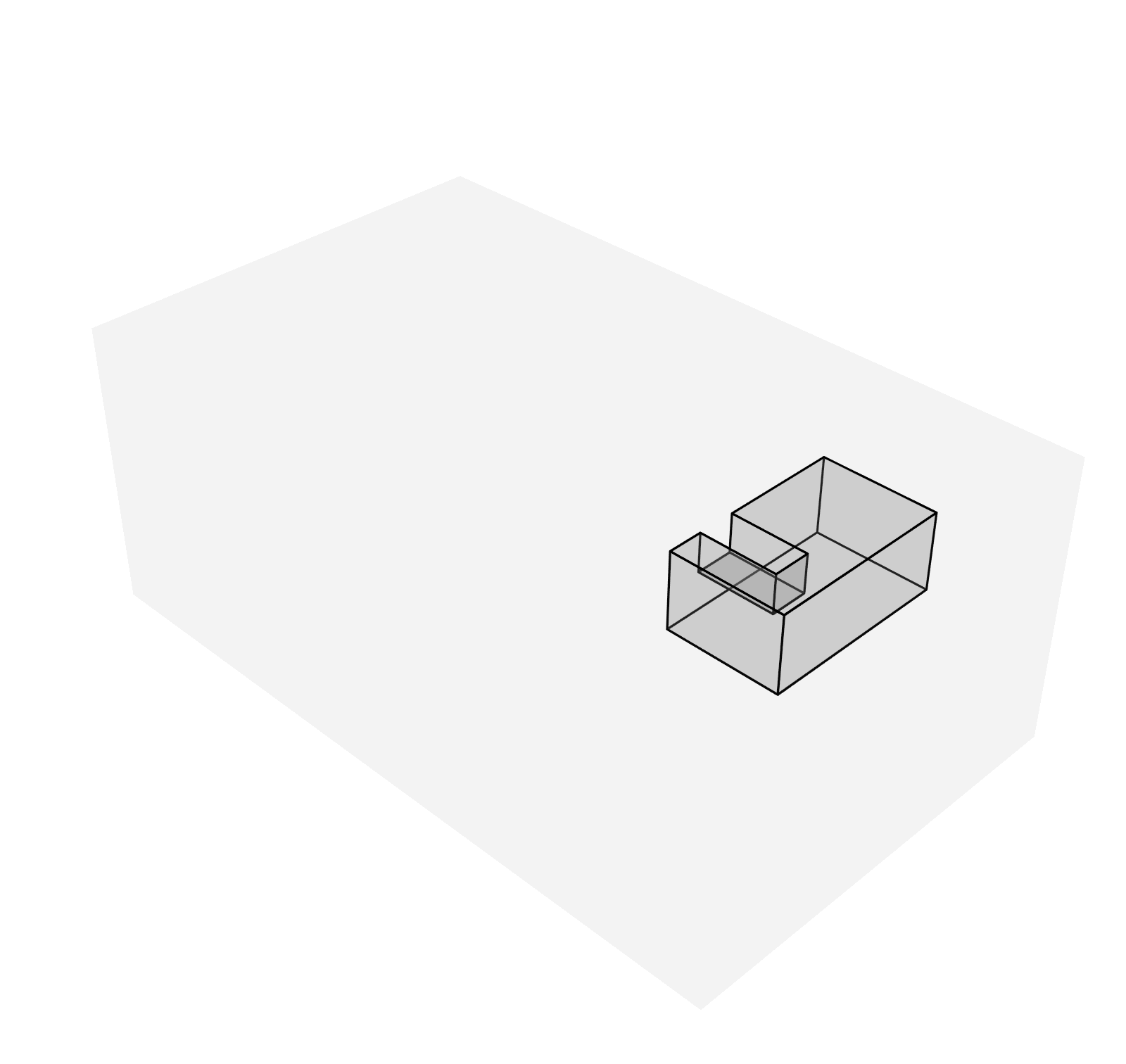}
        \\
    \end{tabular}
    \caption{
    Qualitative comparison of the output of our model's inferred programs (Ours Net) vs. InverseCSG. Green: addition, Red: subtraction, Blue: intersection, Grey: current. (Case 2, Part 1)
    }
    \label{figure:qualitative_comparision_inv2}
\end{figure*}

\begin{figure*}[ht!]
    \centering
    \ContinuedFloat
    \setlength{\tabcolsep}{1pt}
    \begin{tabular}{cccccccc}
        \includegraphics[width=0.124\linewidth]{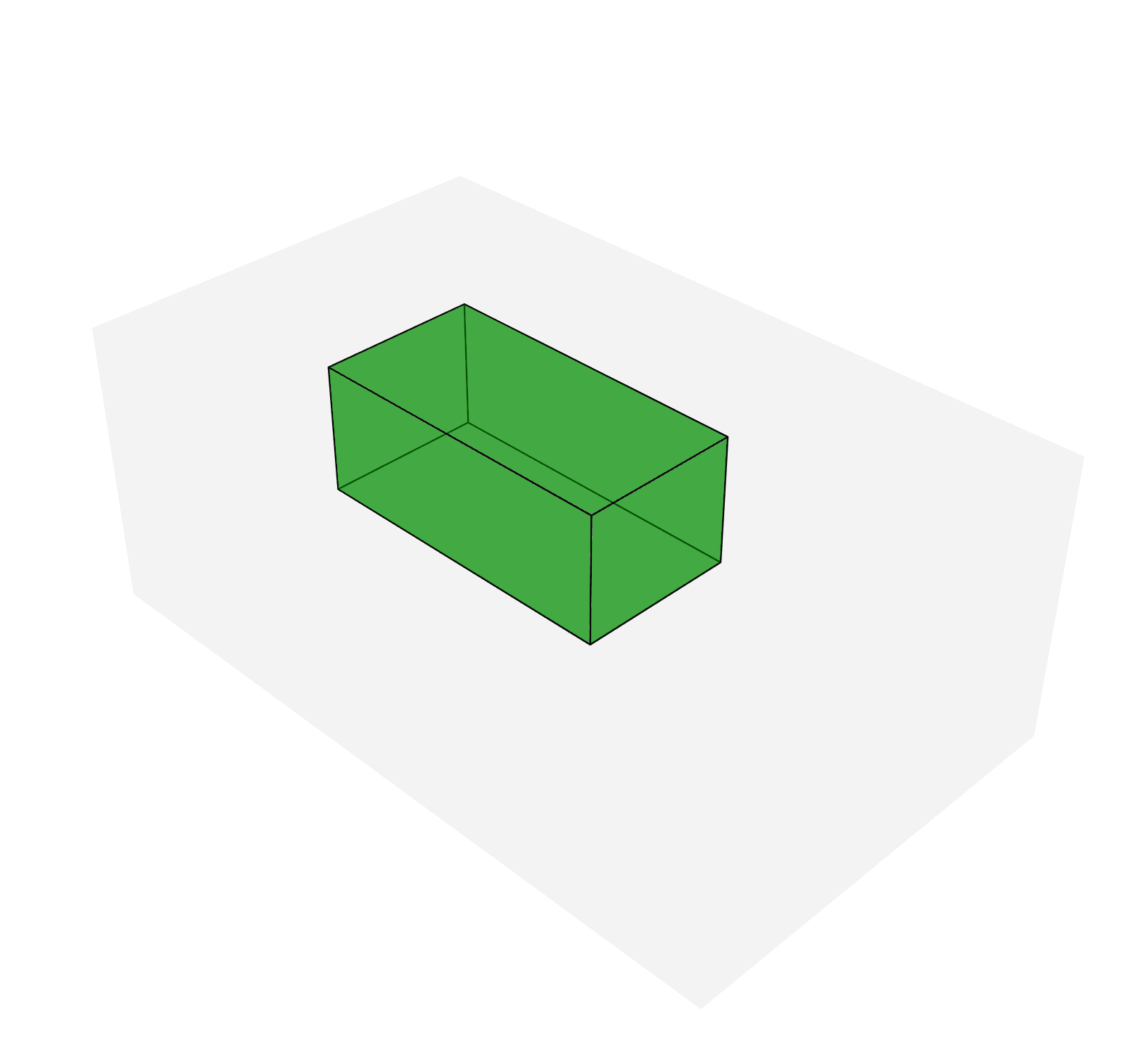} &
        \includegraphics[width=0.124\linewidth]{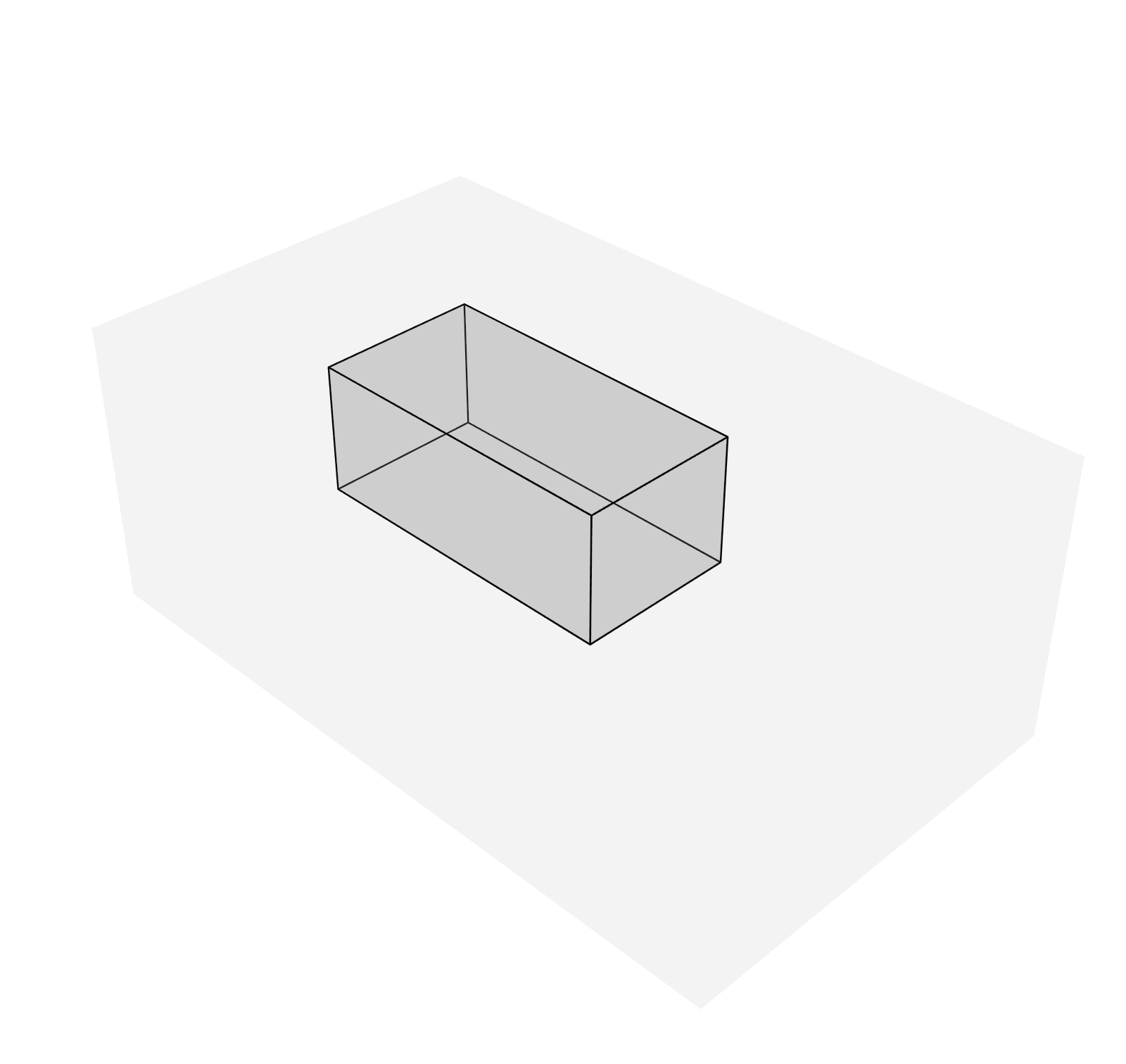} &
        \includegraphics[width=0.124\linewidth]{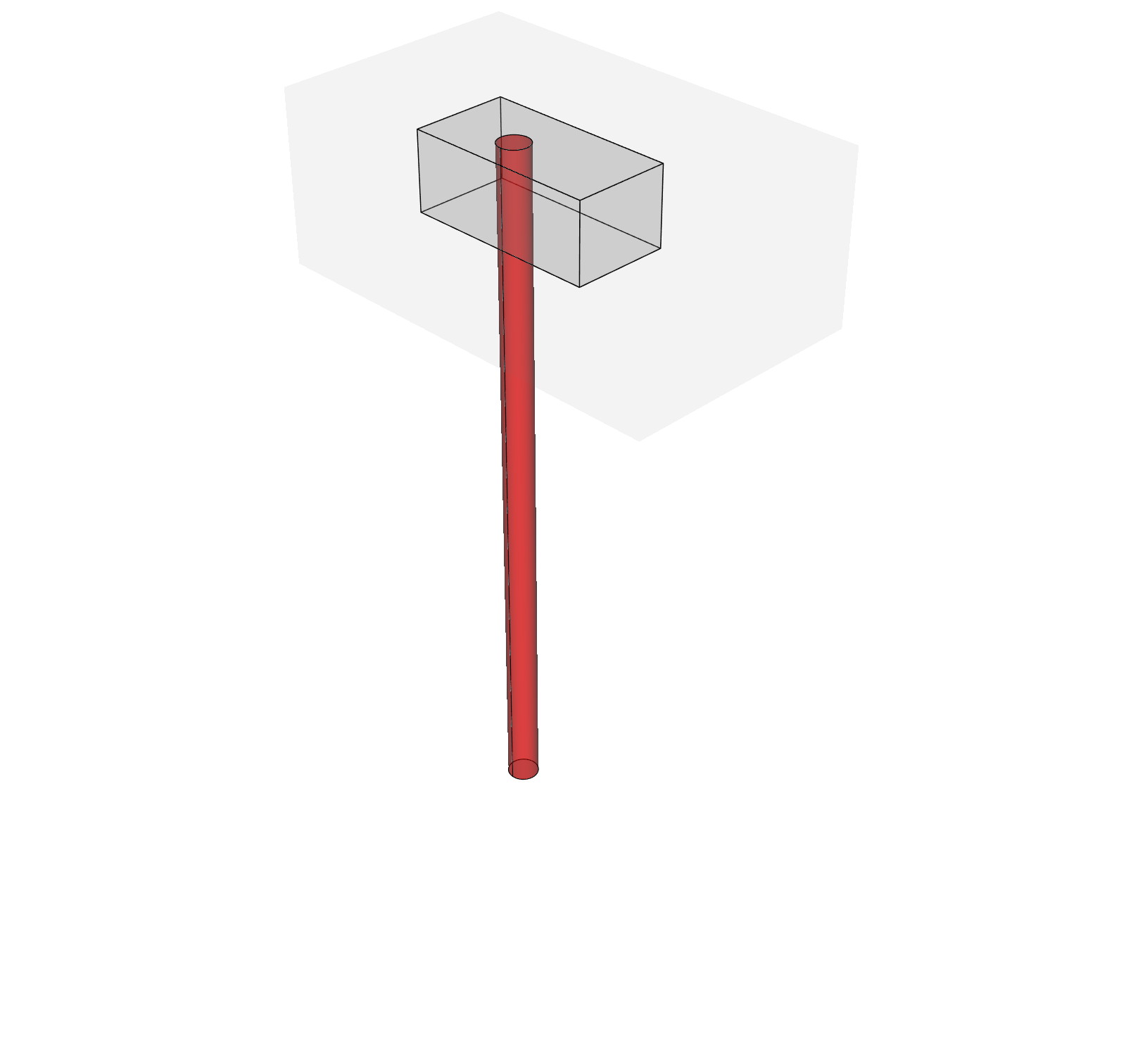} &
        \includegraphics[width=0.124\linewidth]{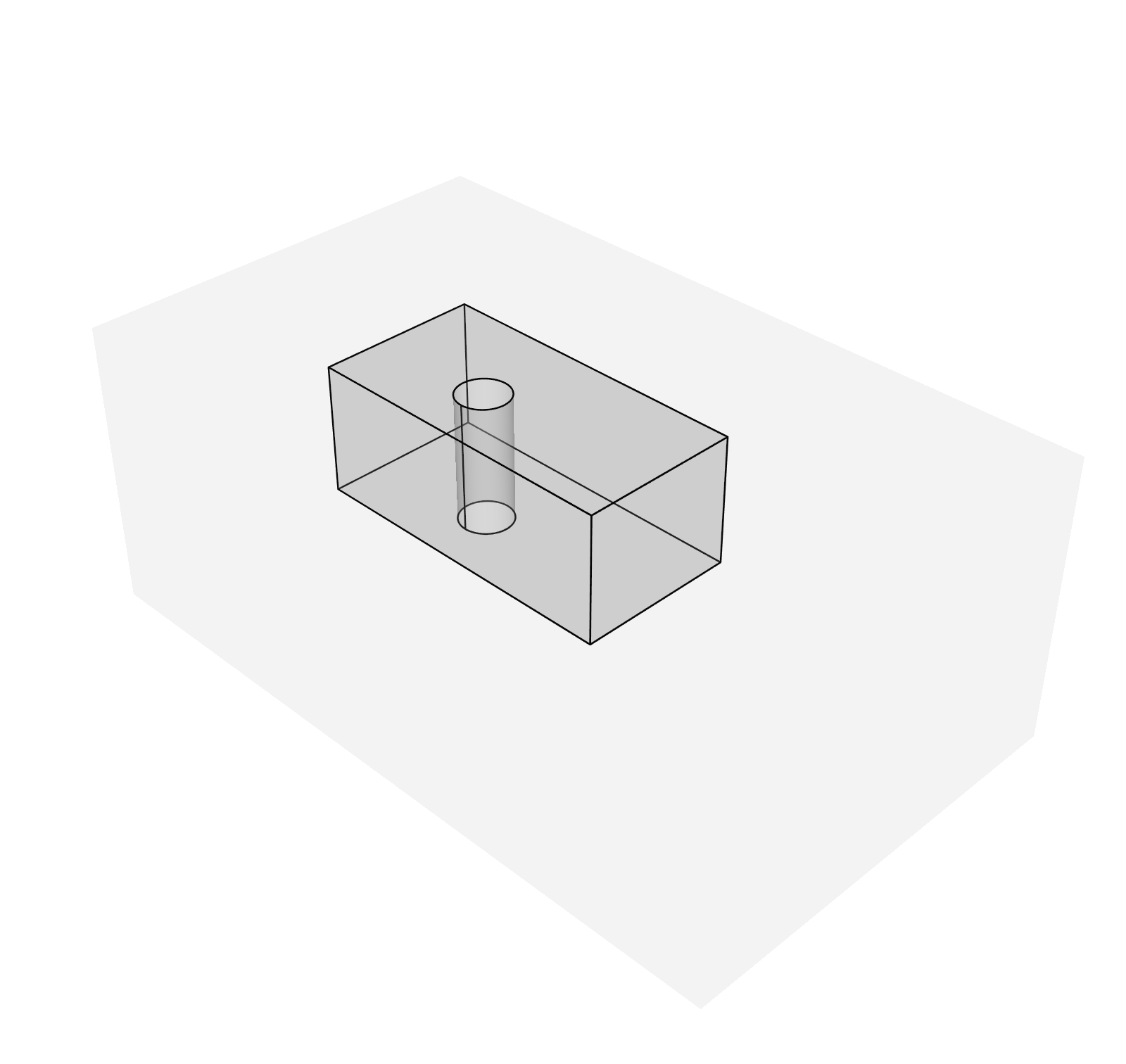} &
        \includegraphics[width=0.124\linewidth]{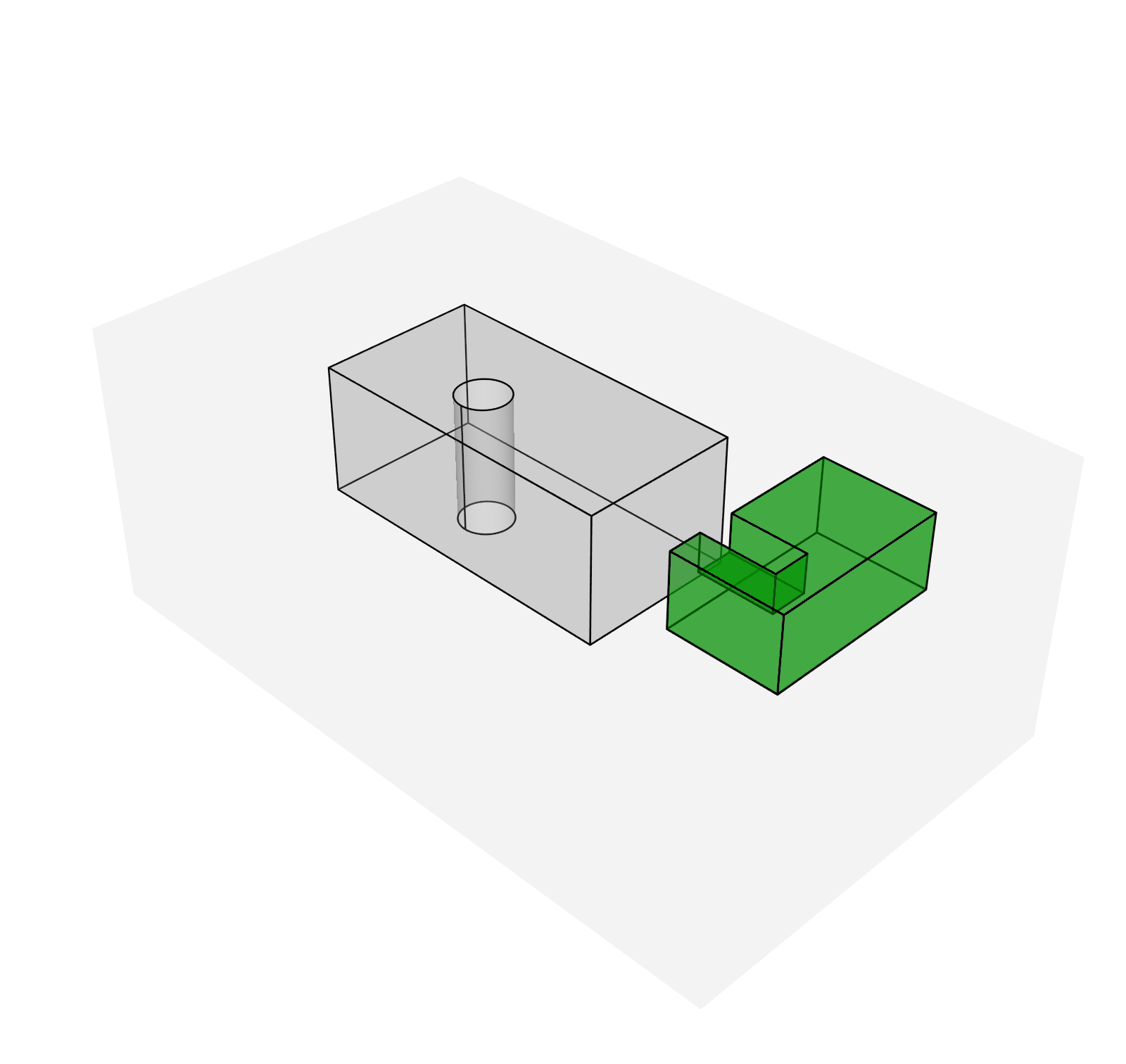} &
        \includegraphics[width=0.124\linewidth]{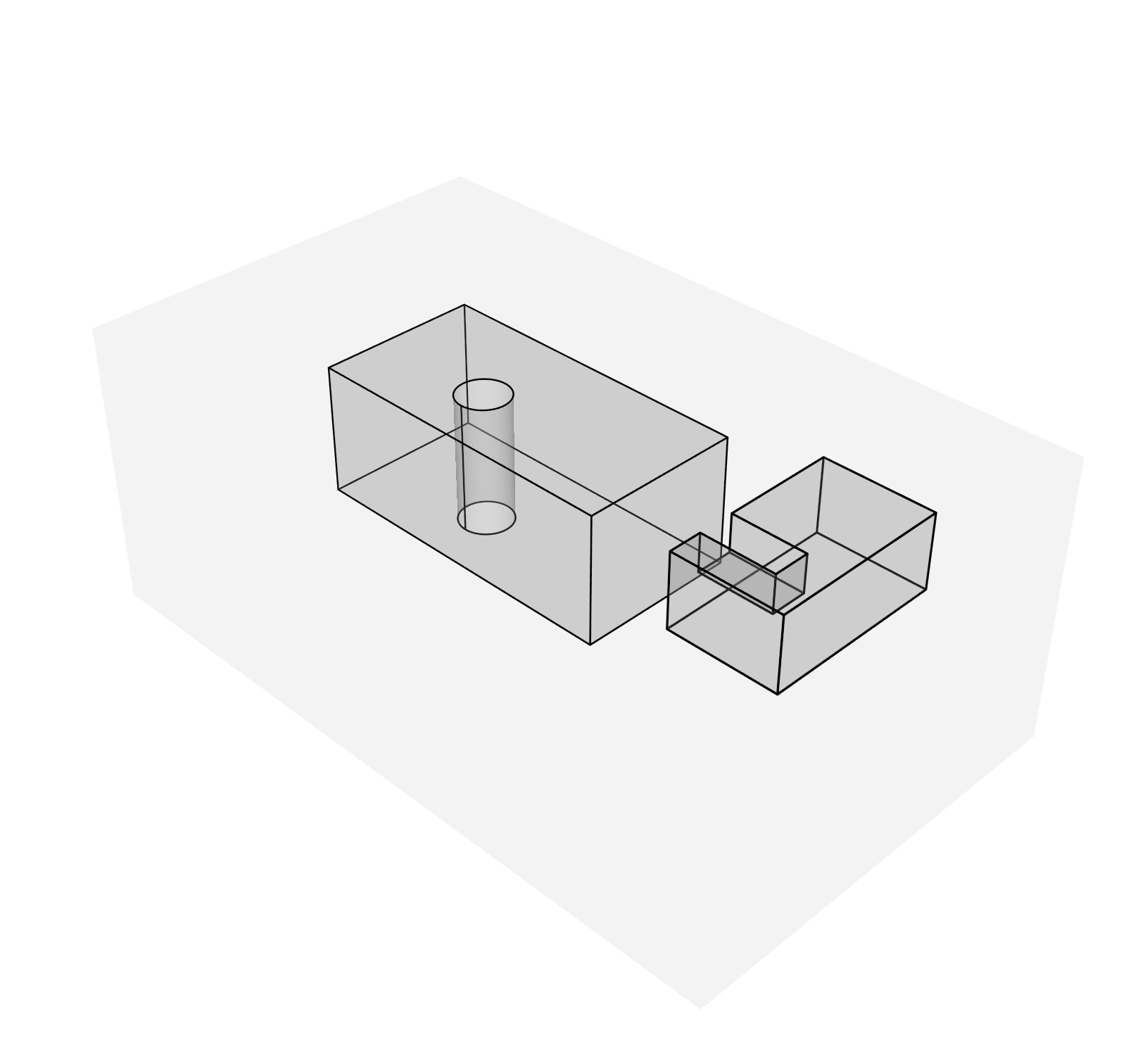} &
        \includegraphics[width=0.124\linewidth]{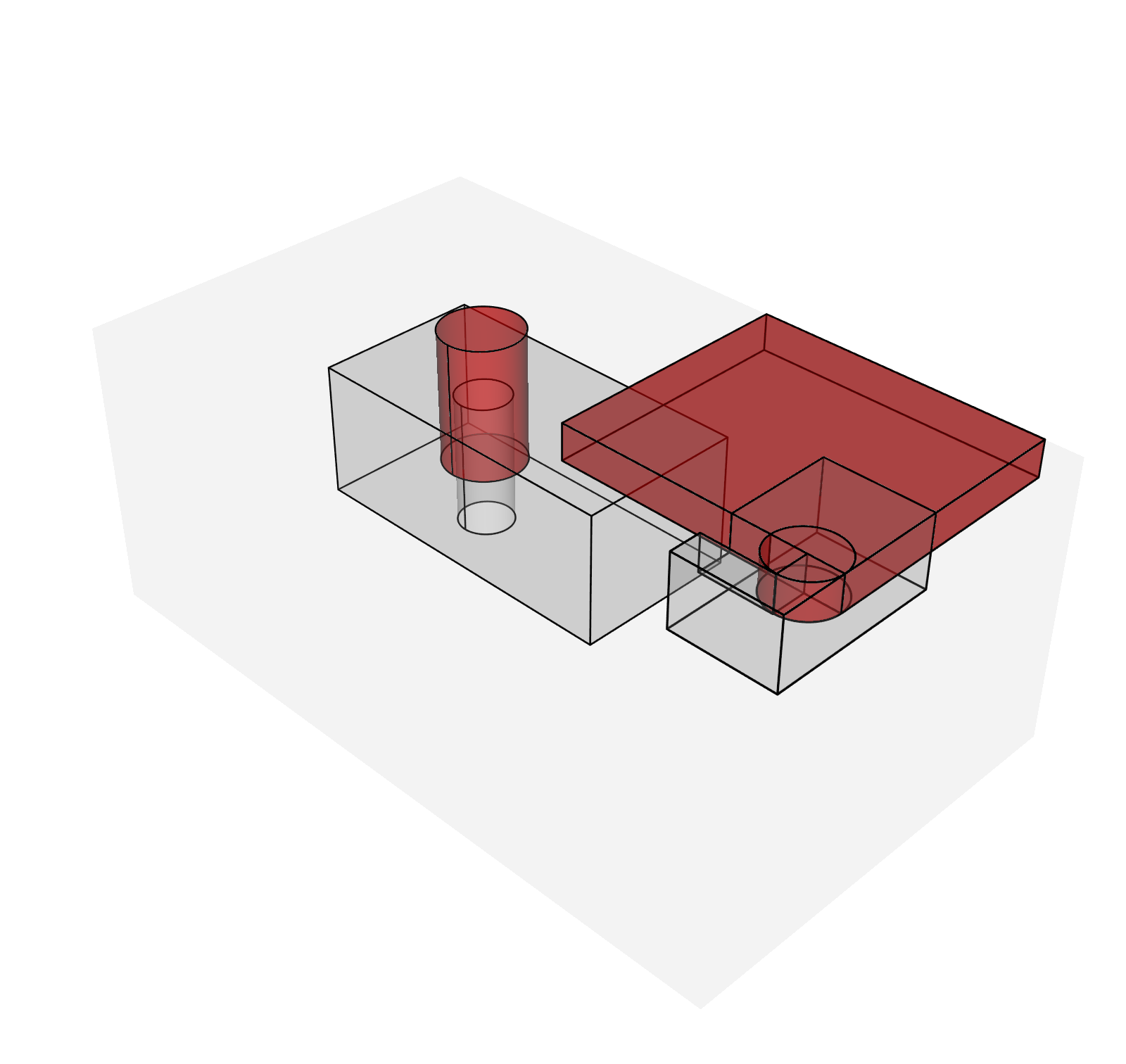} &
        \includegraphics[width=0.124\linewidth]{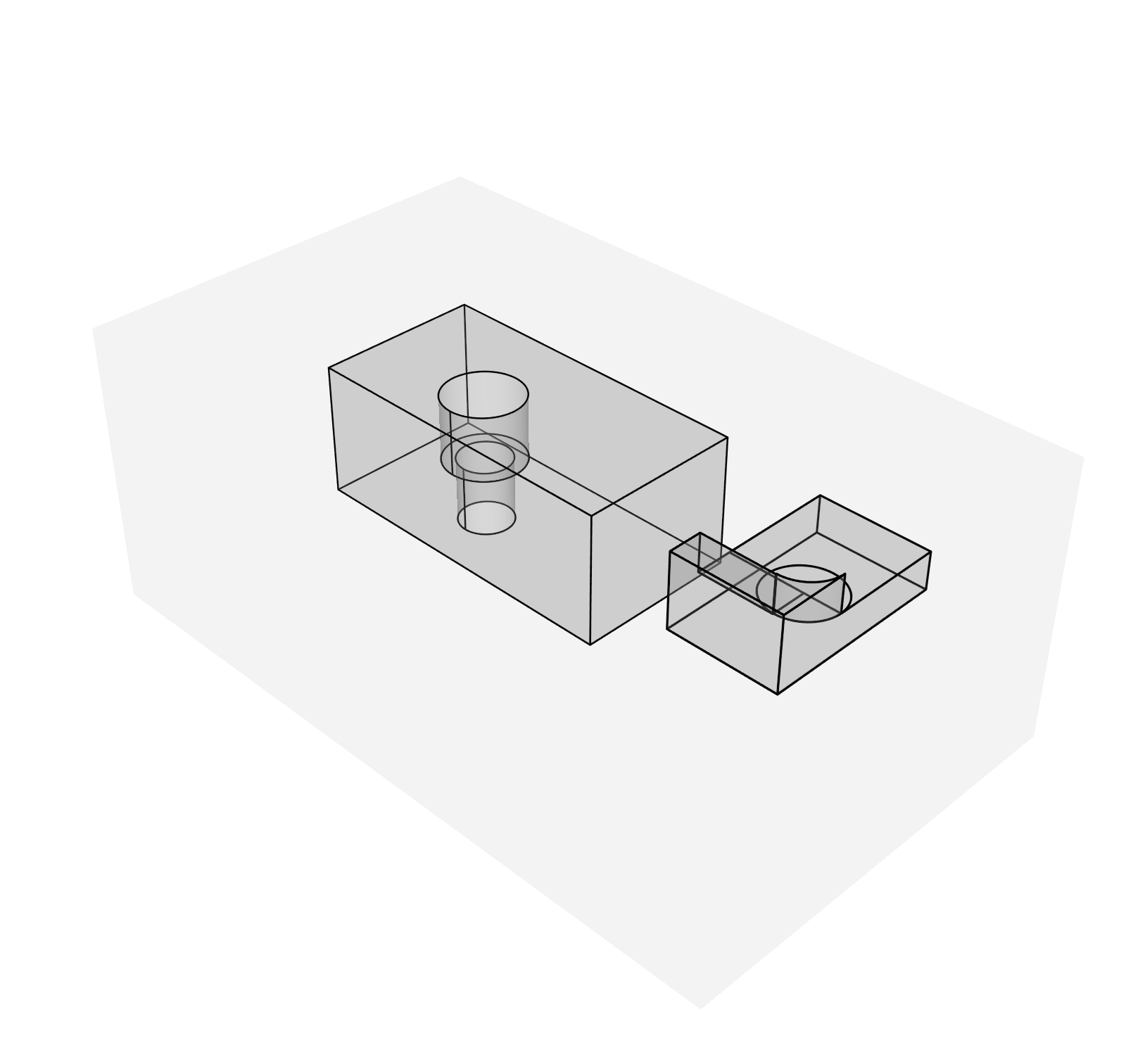}
        \\
        \includegraphics[width=0.124\linewidth]{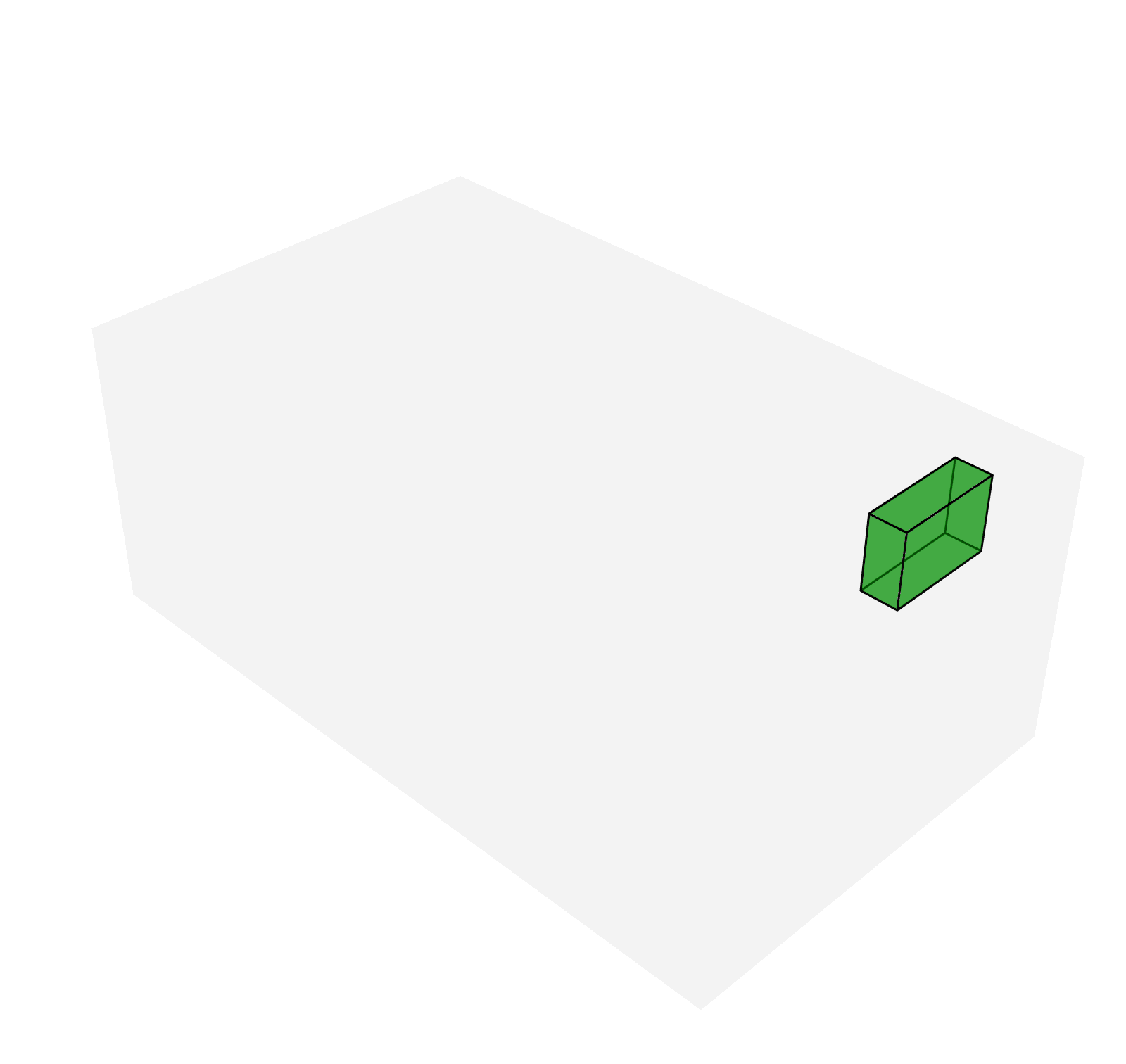} &
        \includegraphics[width=0.124\linewidth]{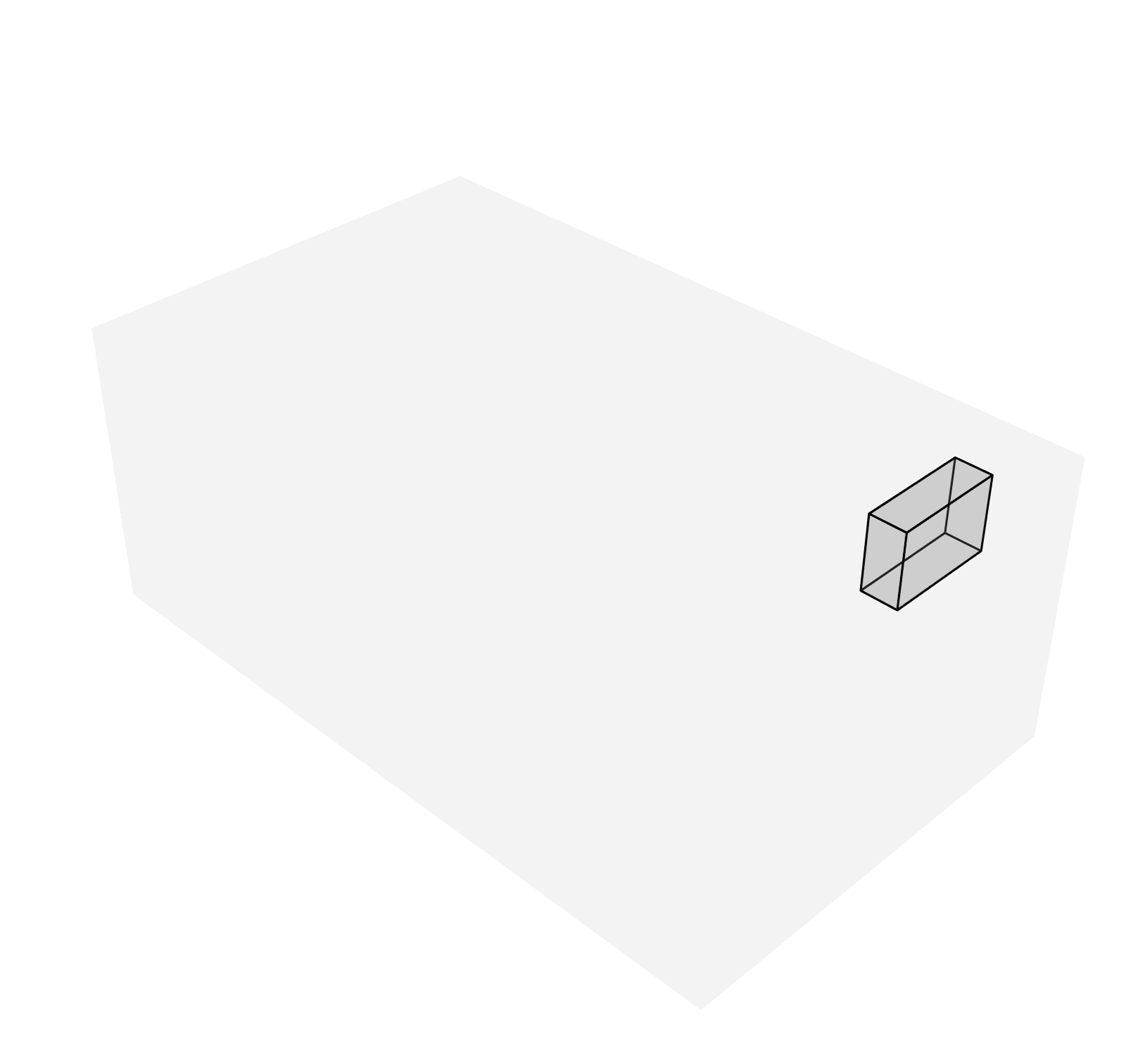} &
        \includegraphics[width=0.124\linewidth]{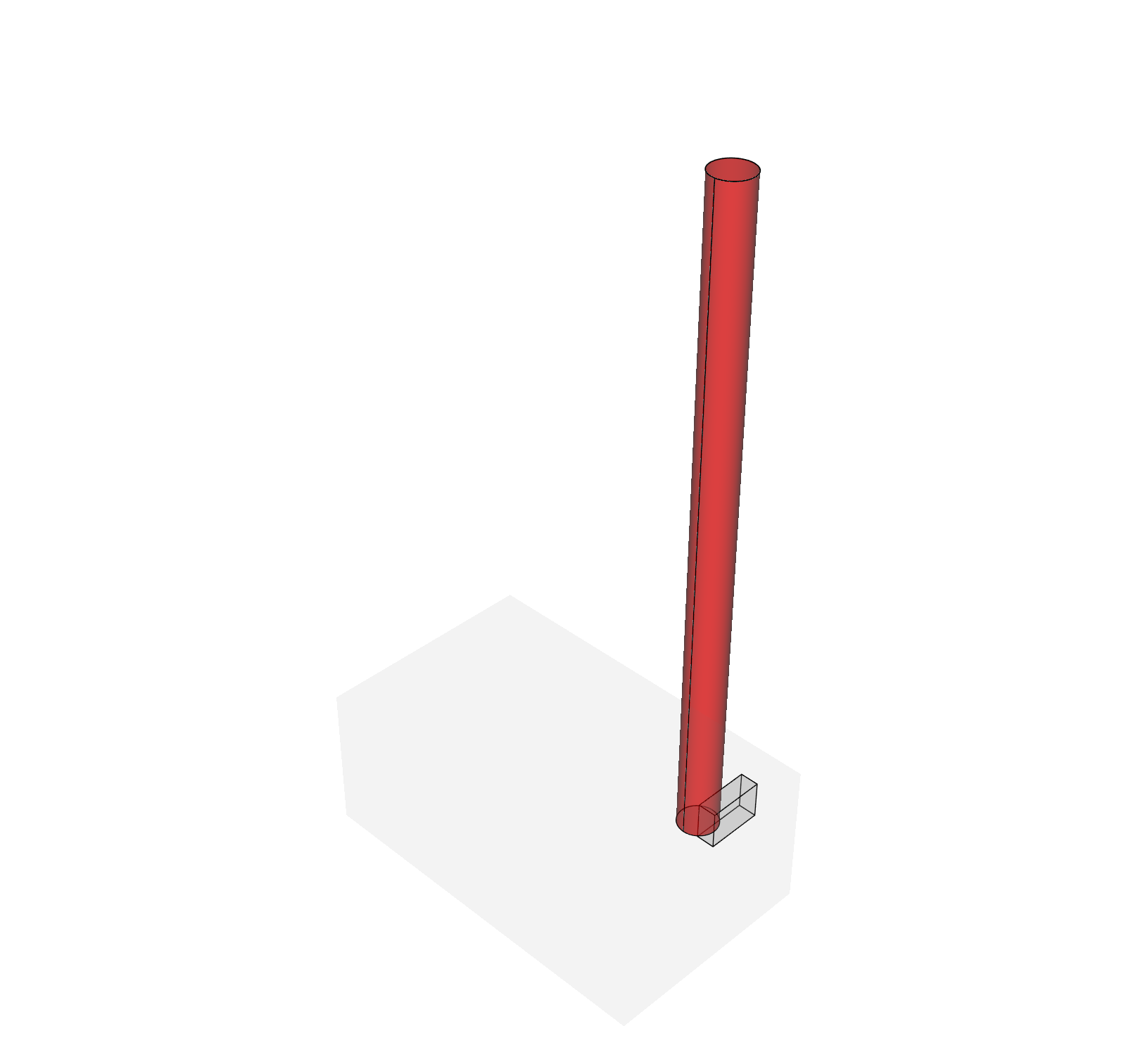} &
        \includegraphics[width=0.124\linewidth]{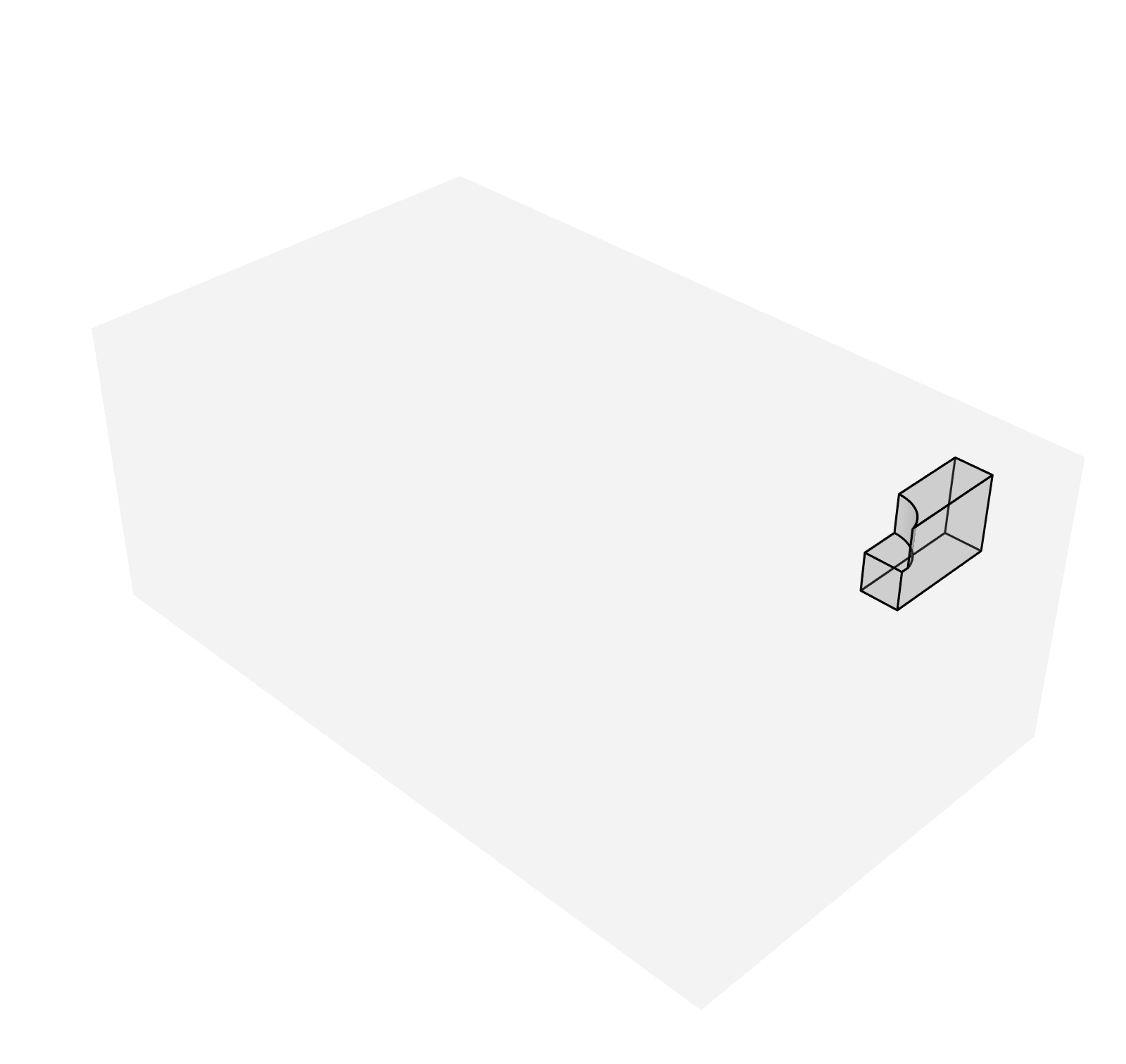} &
        \includegraphics[width=0.124\linewidth]{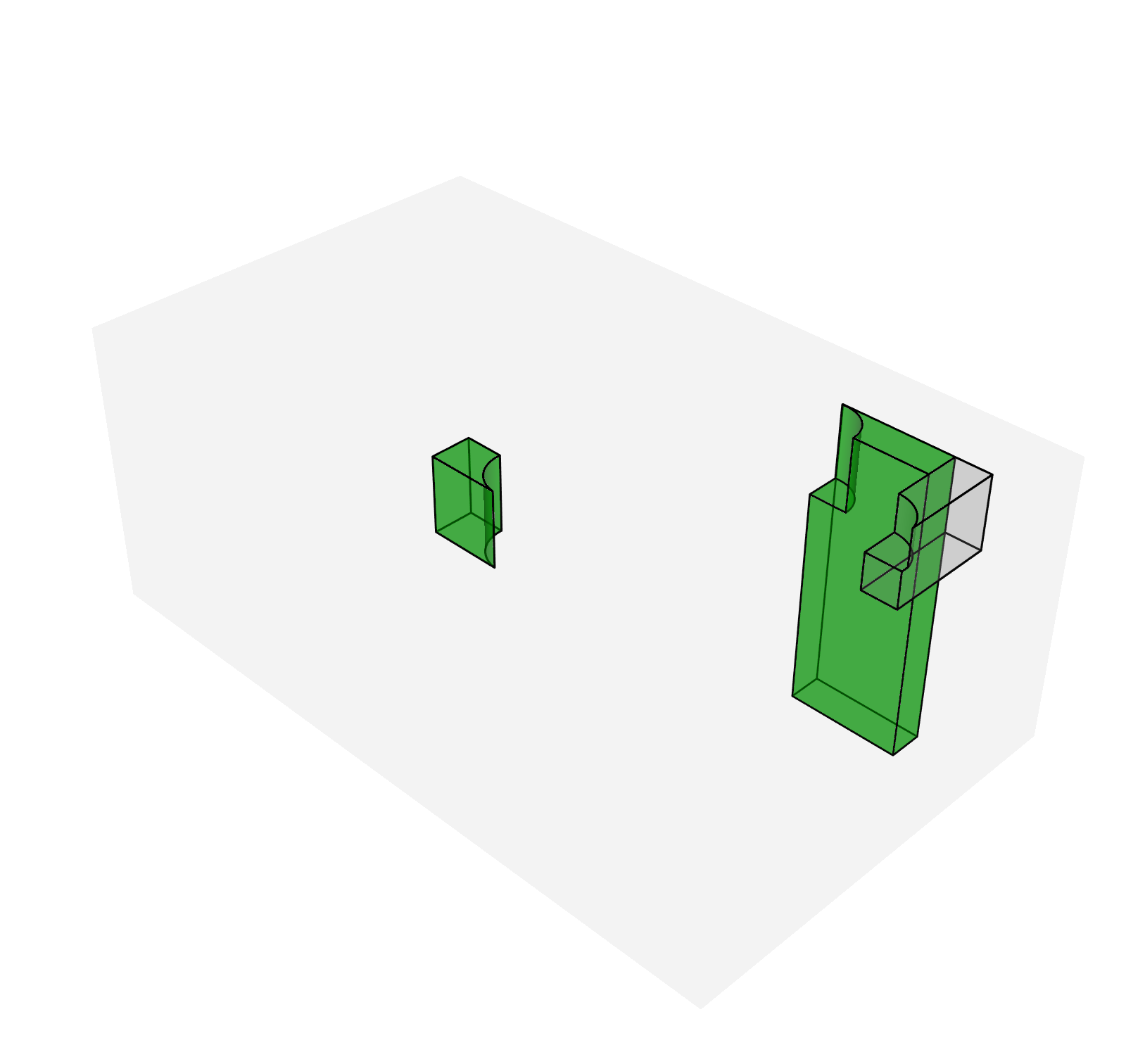} &
        \includegraphics[width=0.124\linewidth]{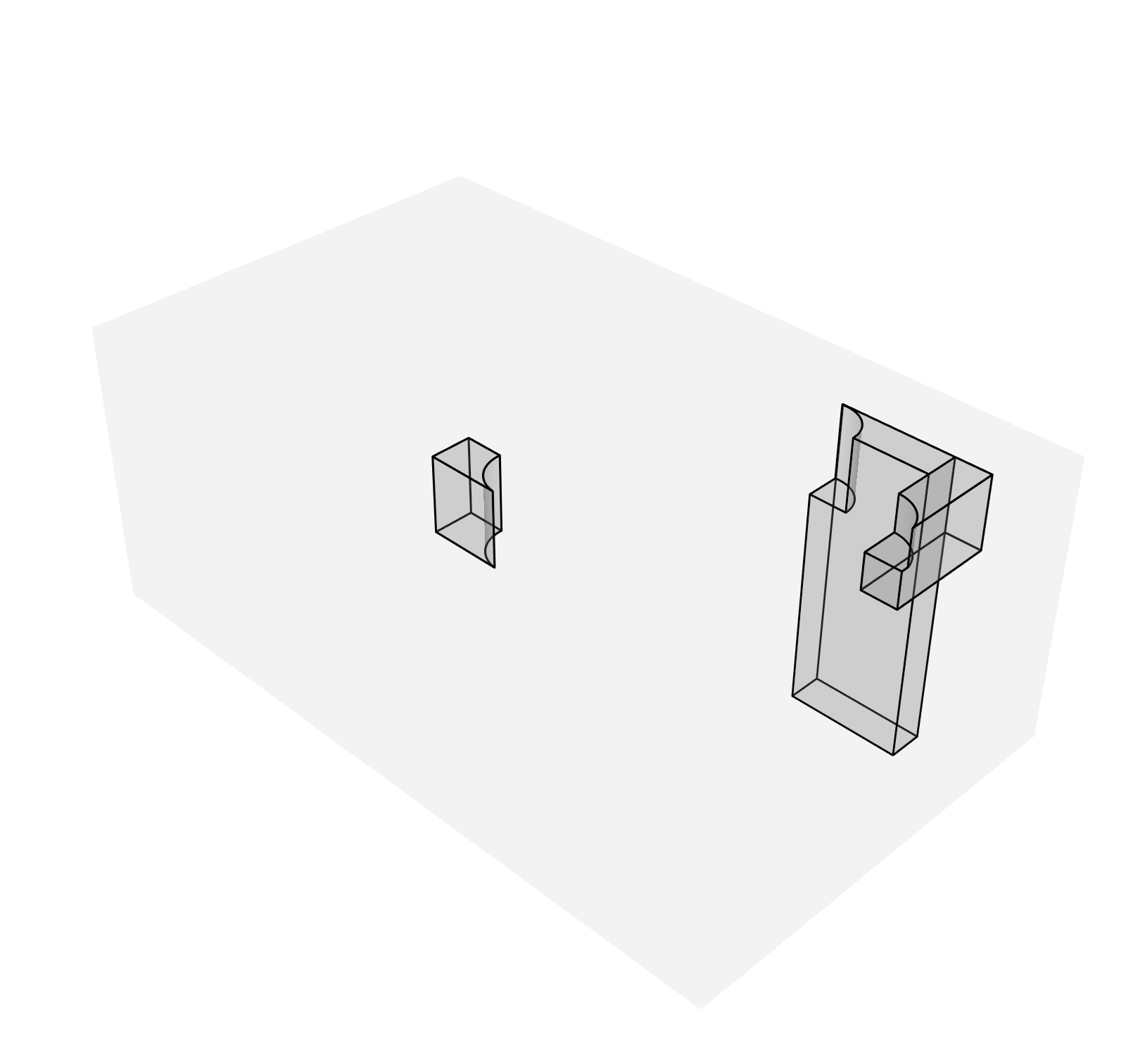} &
        \includegraphics[width=0.124\linewidth]{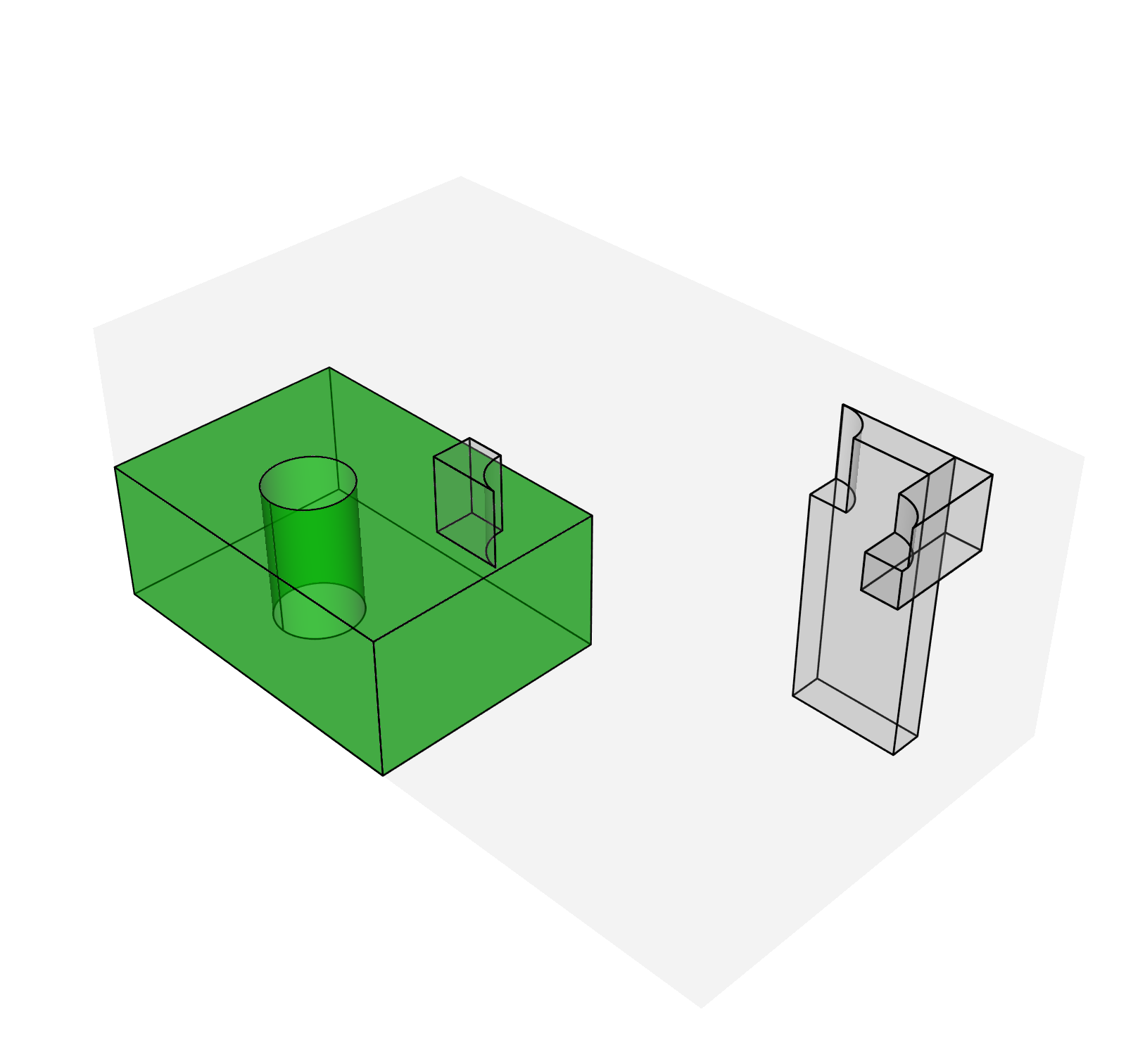} &
        \includegraphics[width=0.124\linewidth]{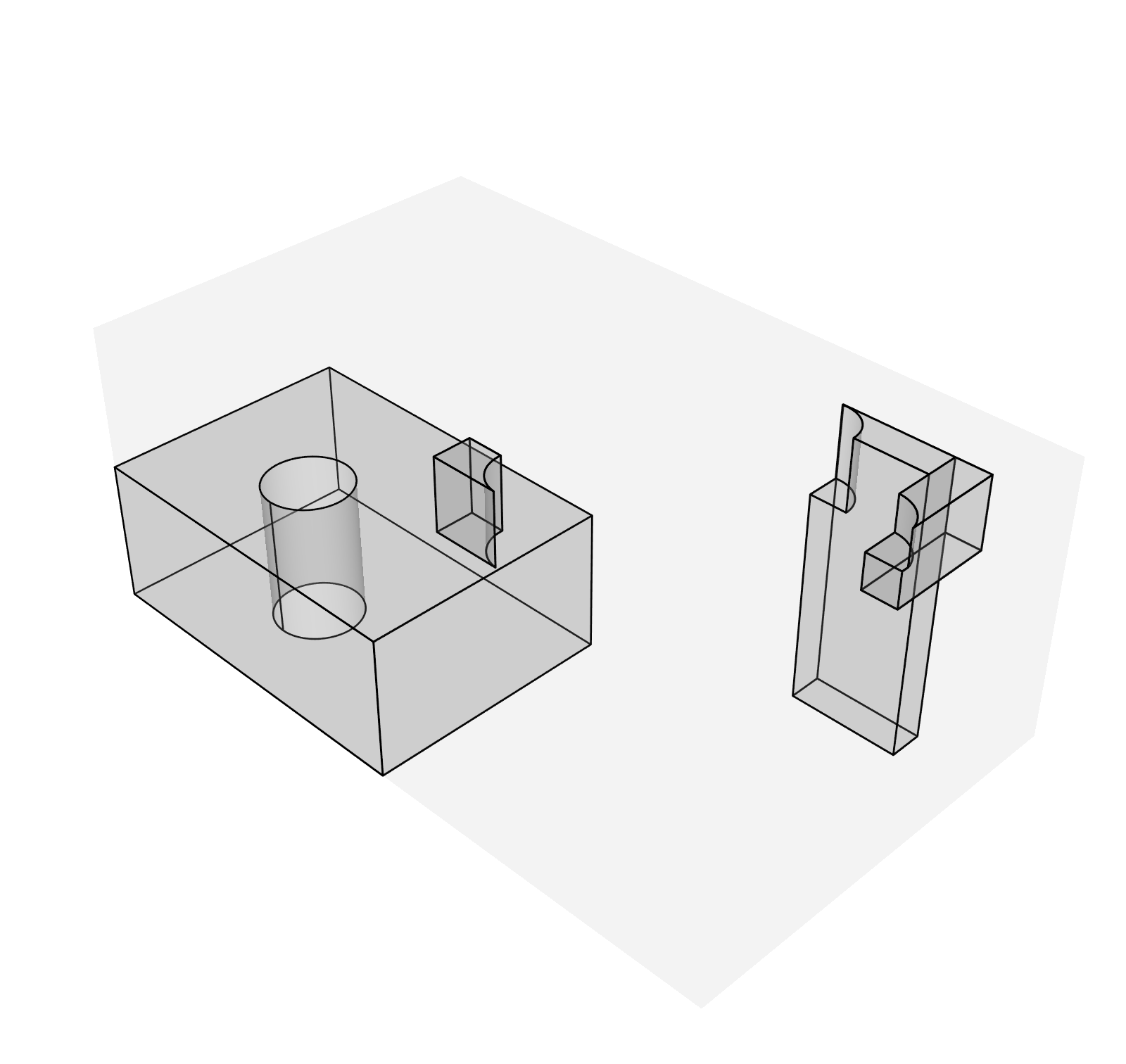}
        \\
        \includegraphics[width=0.124\linewidth]{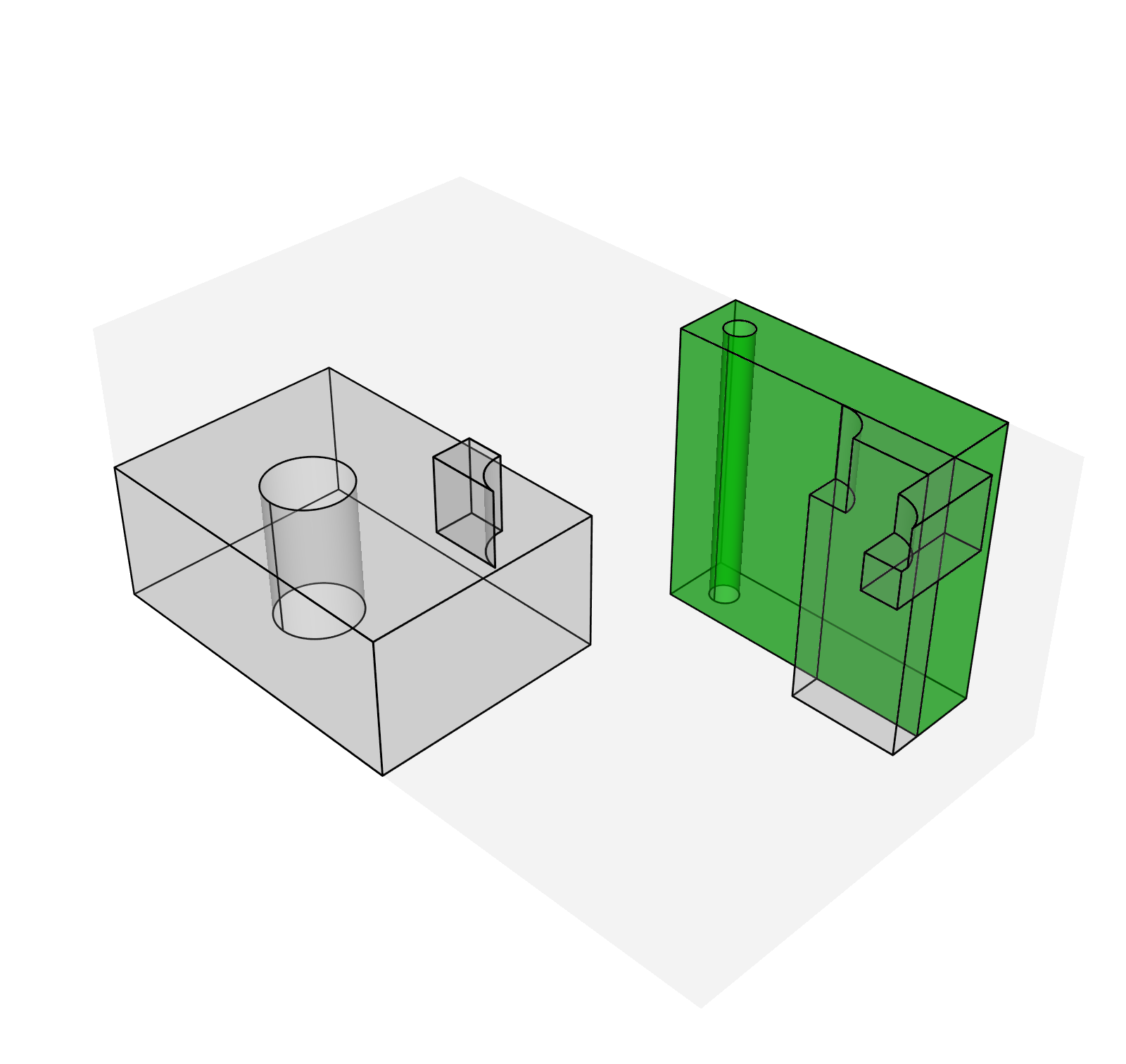} &
        \includegraphics[width=0.124\linewidth]{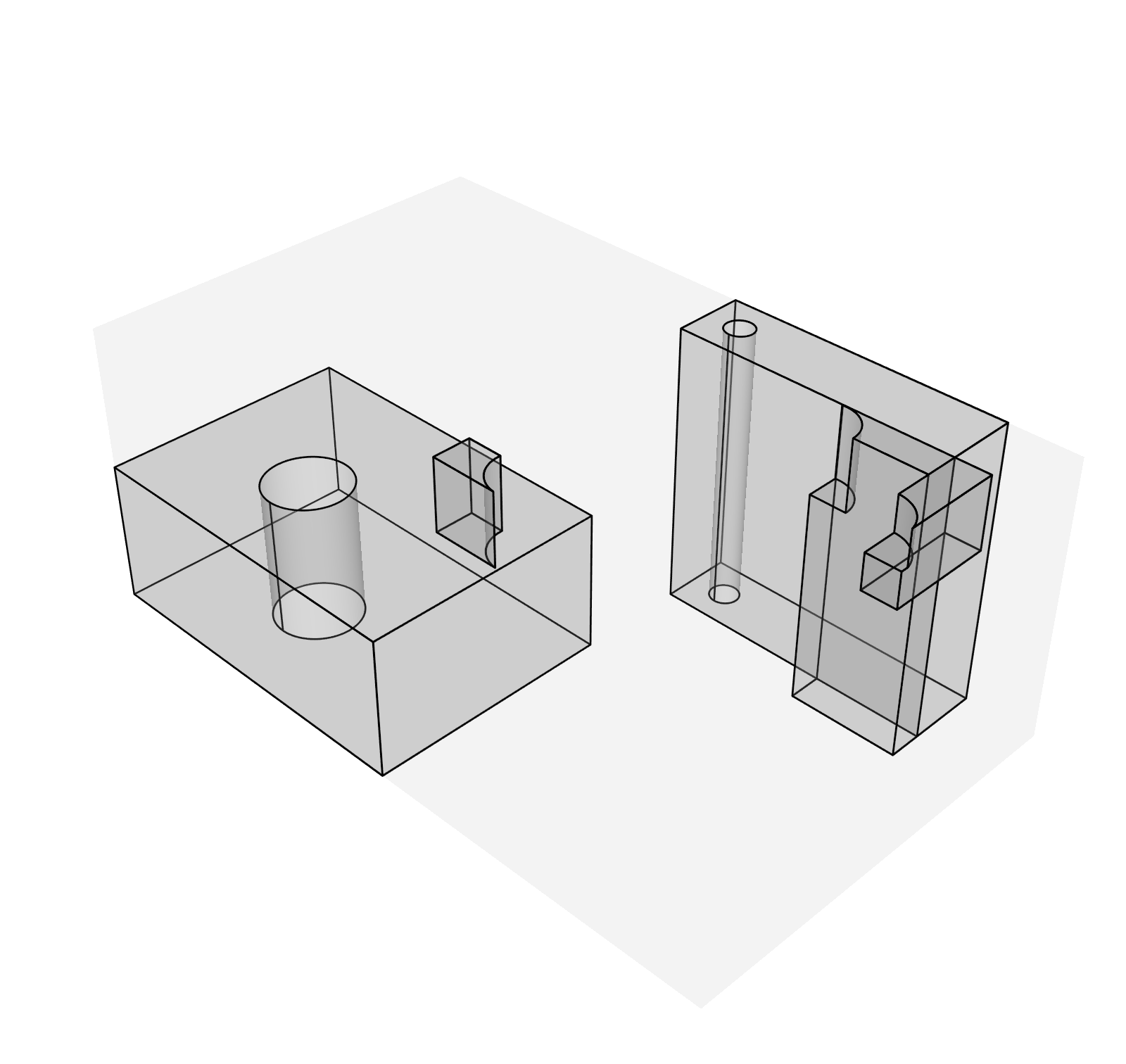} &
        \includegraphics[width=0.124\linewidth]{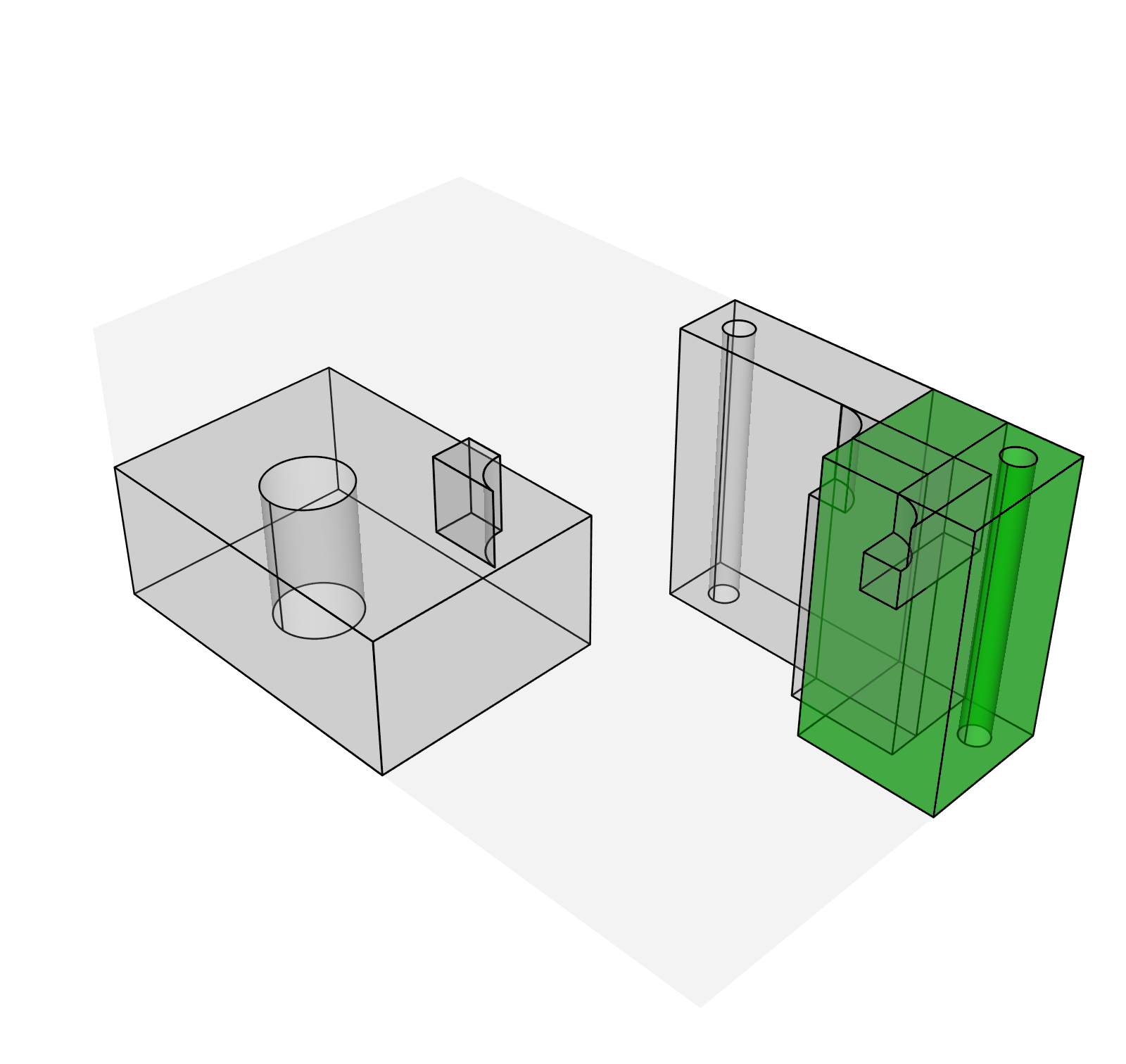} &
        \includegraphics[width=0.124\linewidth]{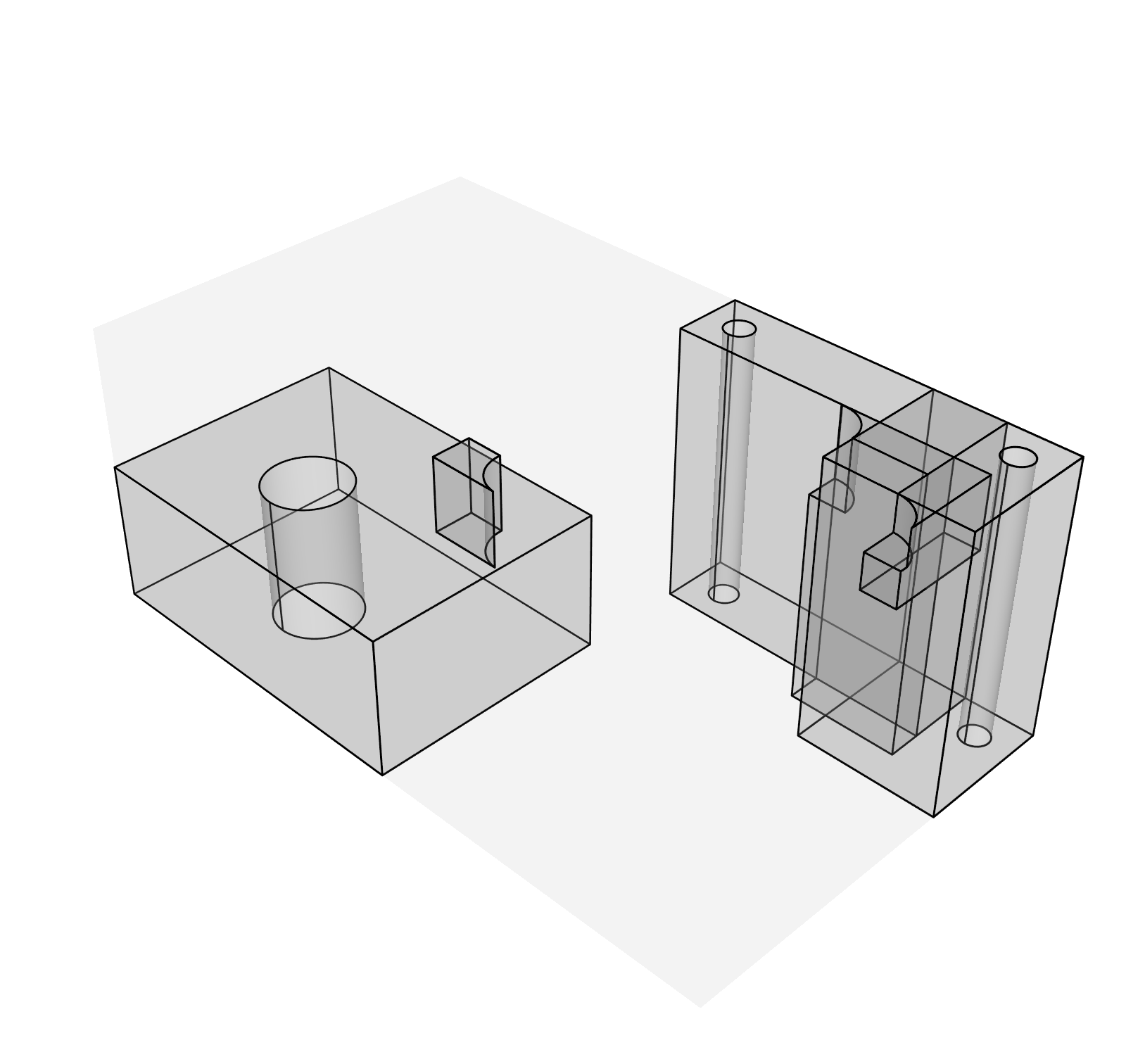} &
        \includegraphics[width=0.124\linewidth]{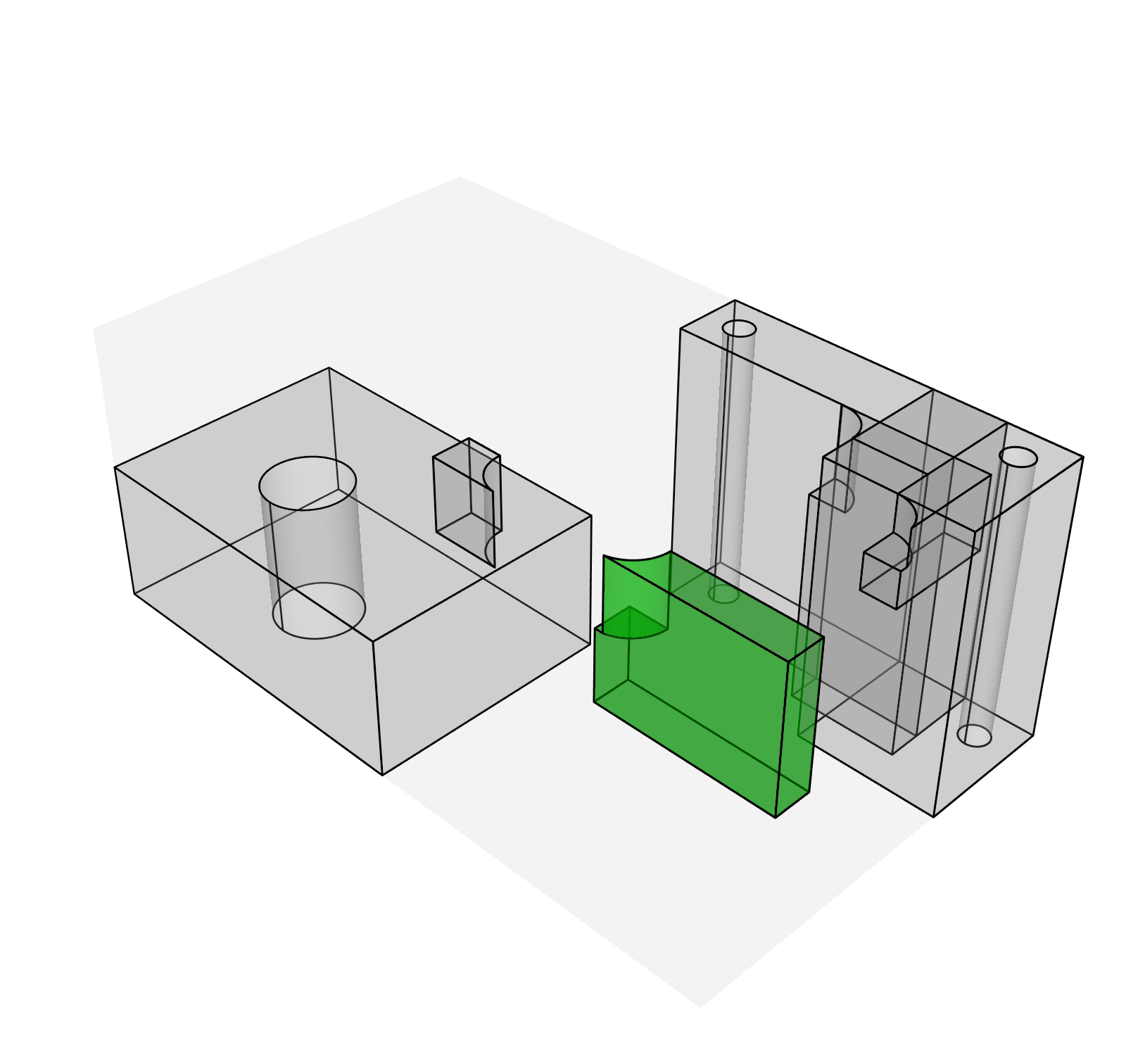} &
        \includegraphics[width=0.124\linewidth]{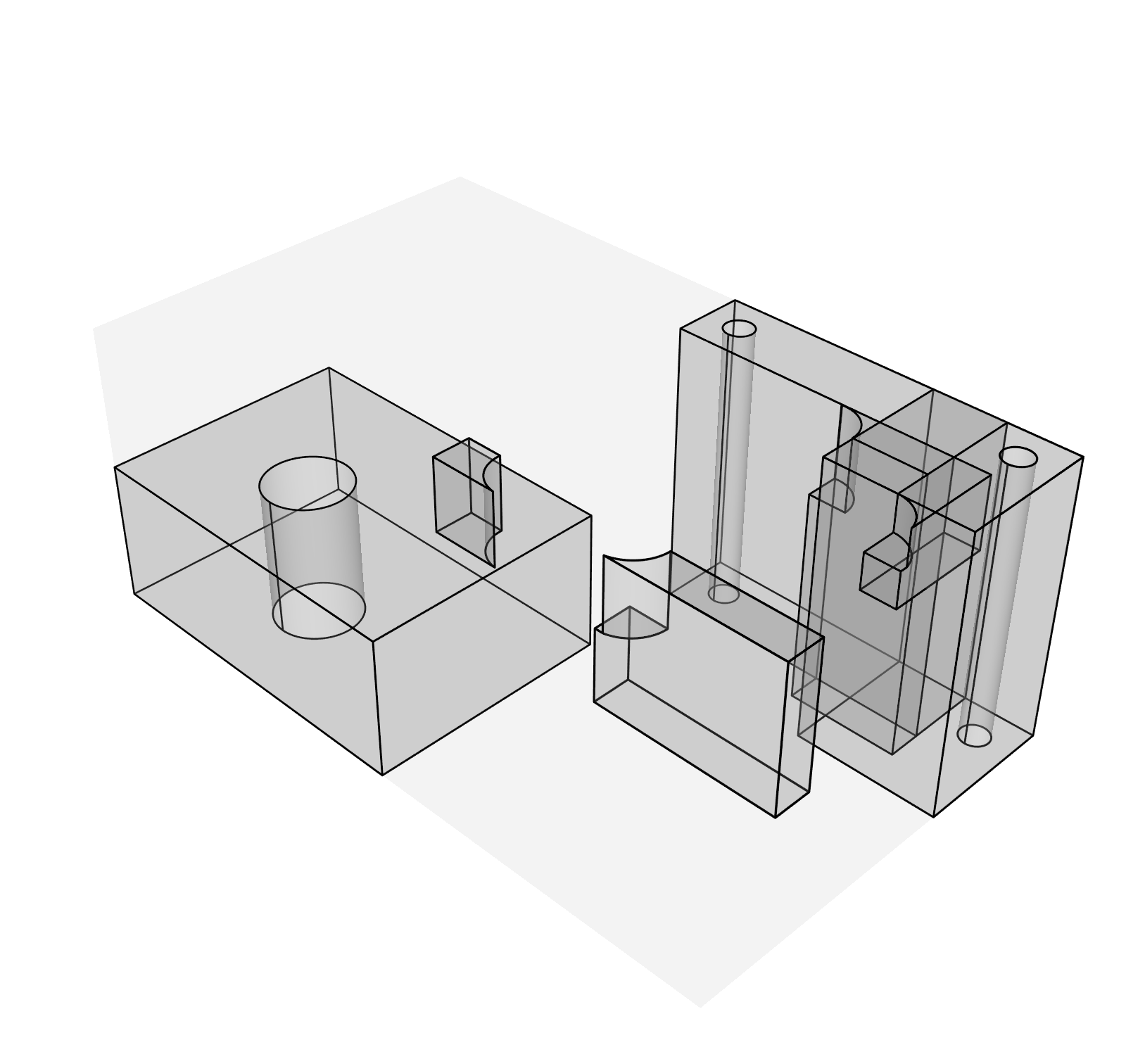} &
        \includegraphics[width=0.124\linewidth]{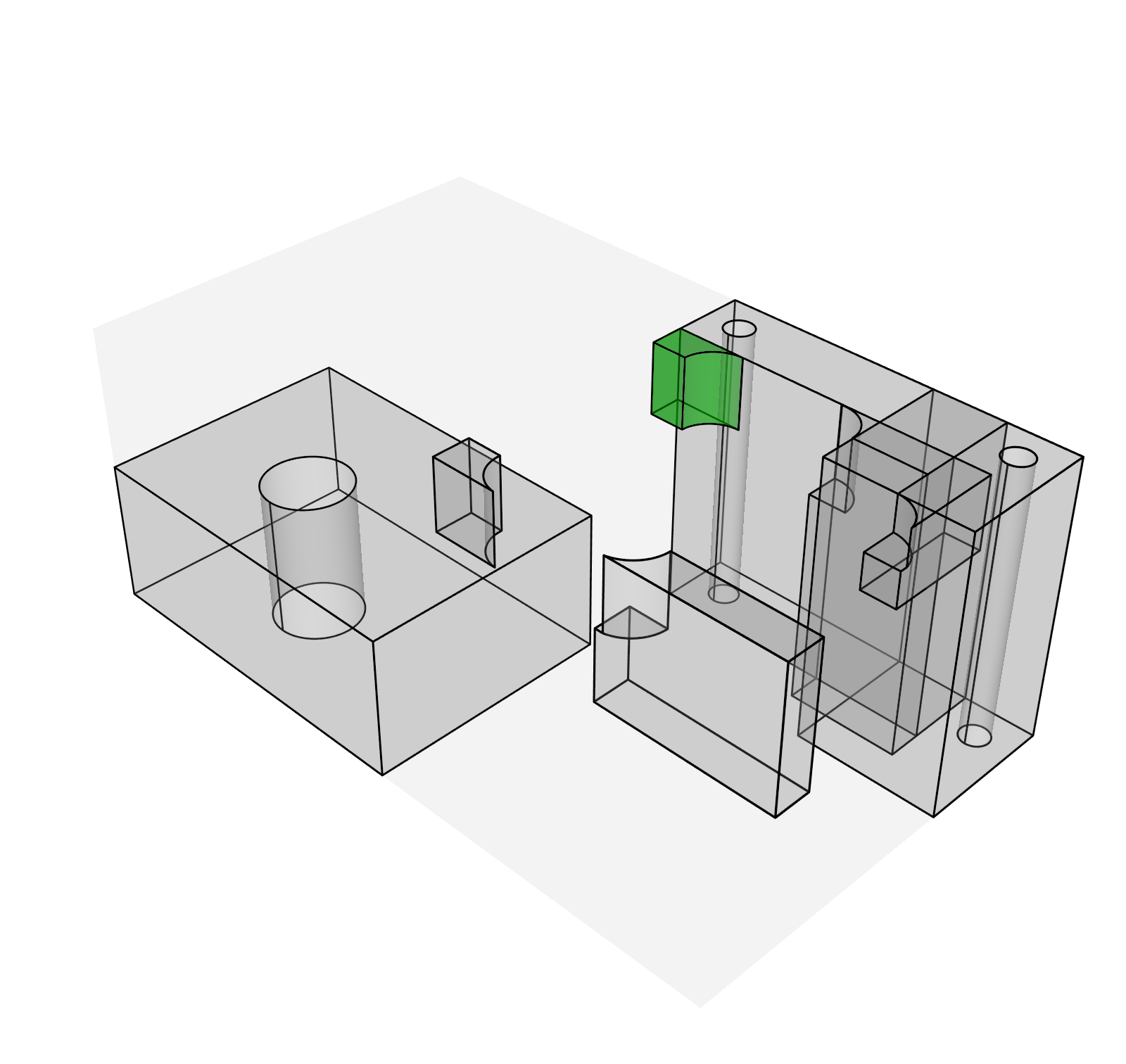} &
        \includegraphics[width=0.124\linewidth]{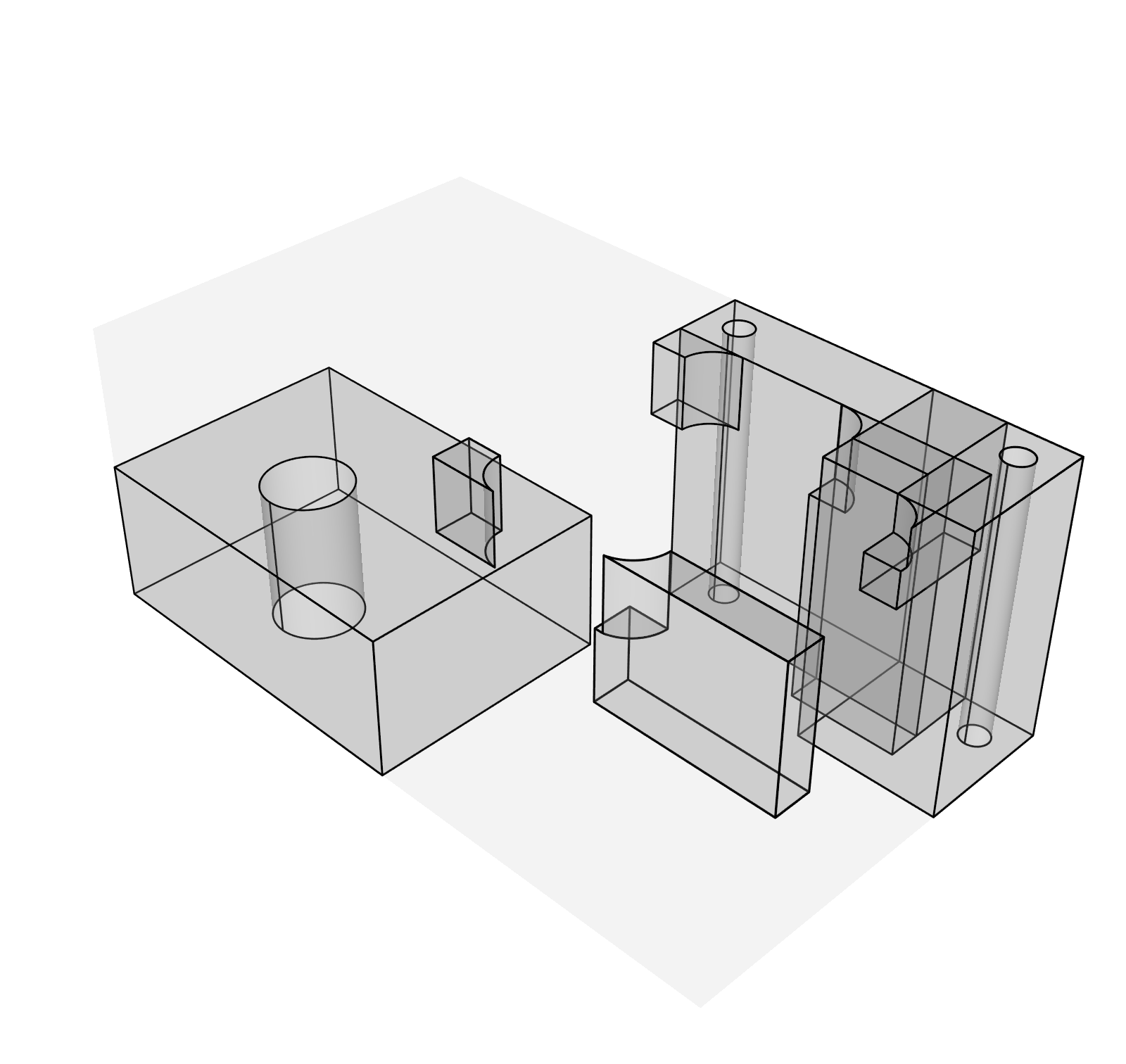}
        \\
        \includegraphics[width=0.124\linewidth]{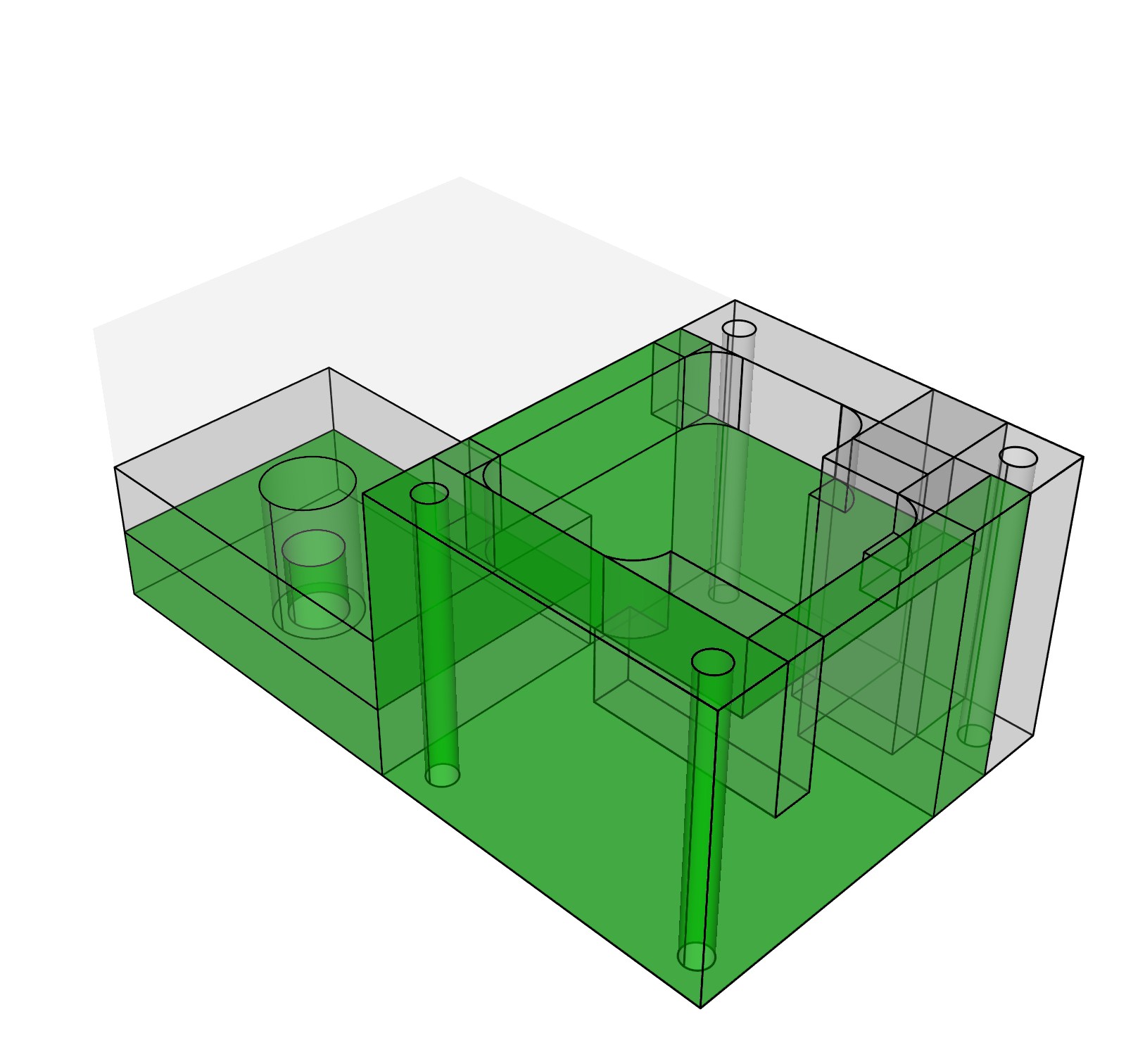} &
        \includegraphics[width=0.124\linewidth]{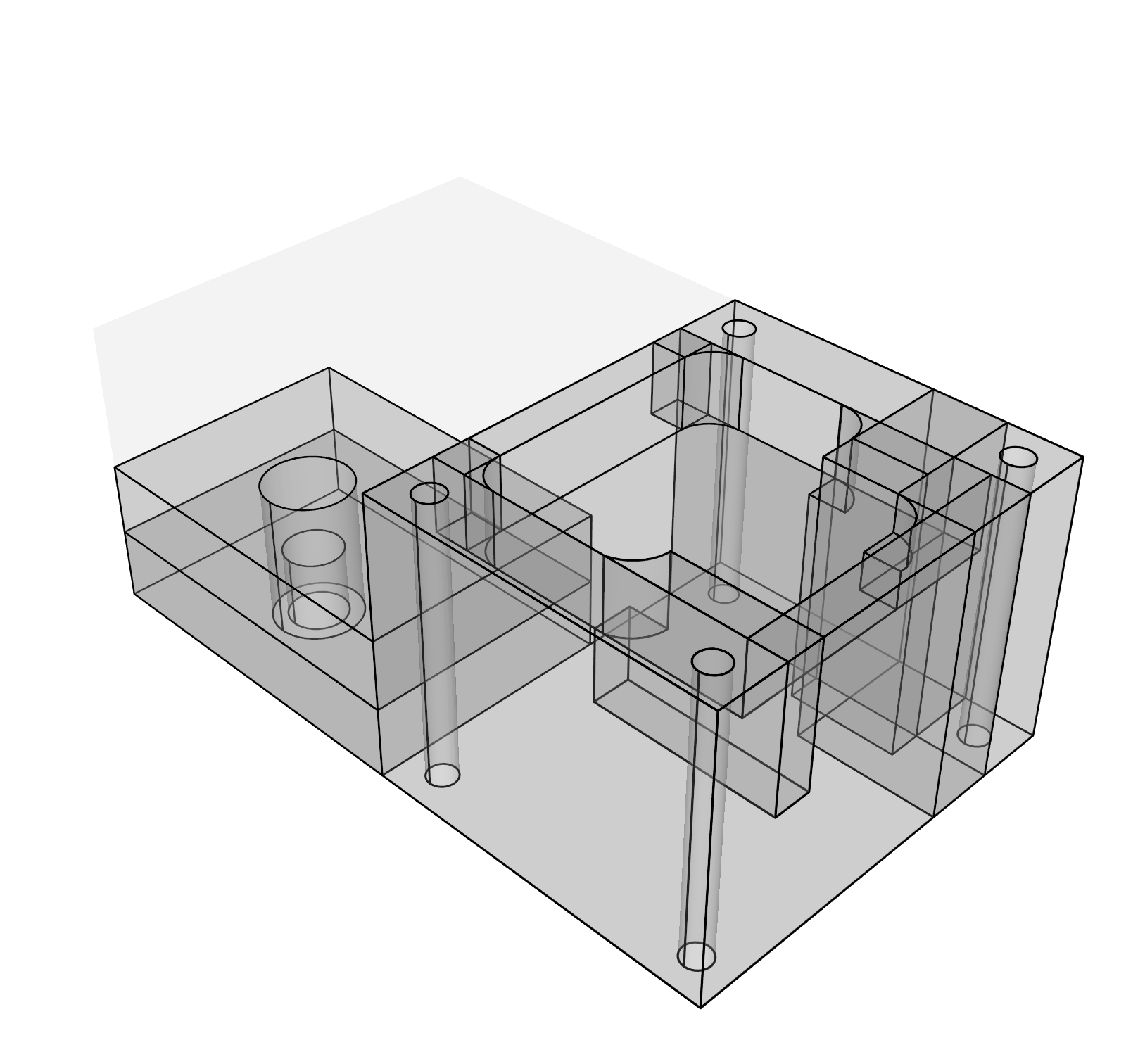} &
        \includegraphics[width=0.124\linewidth]{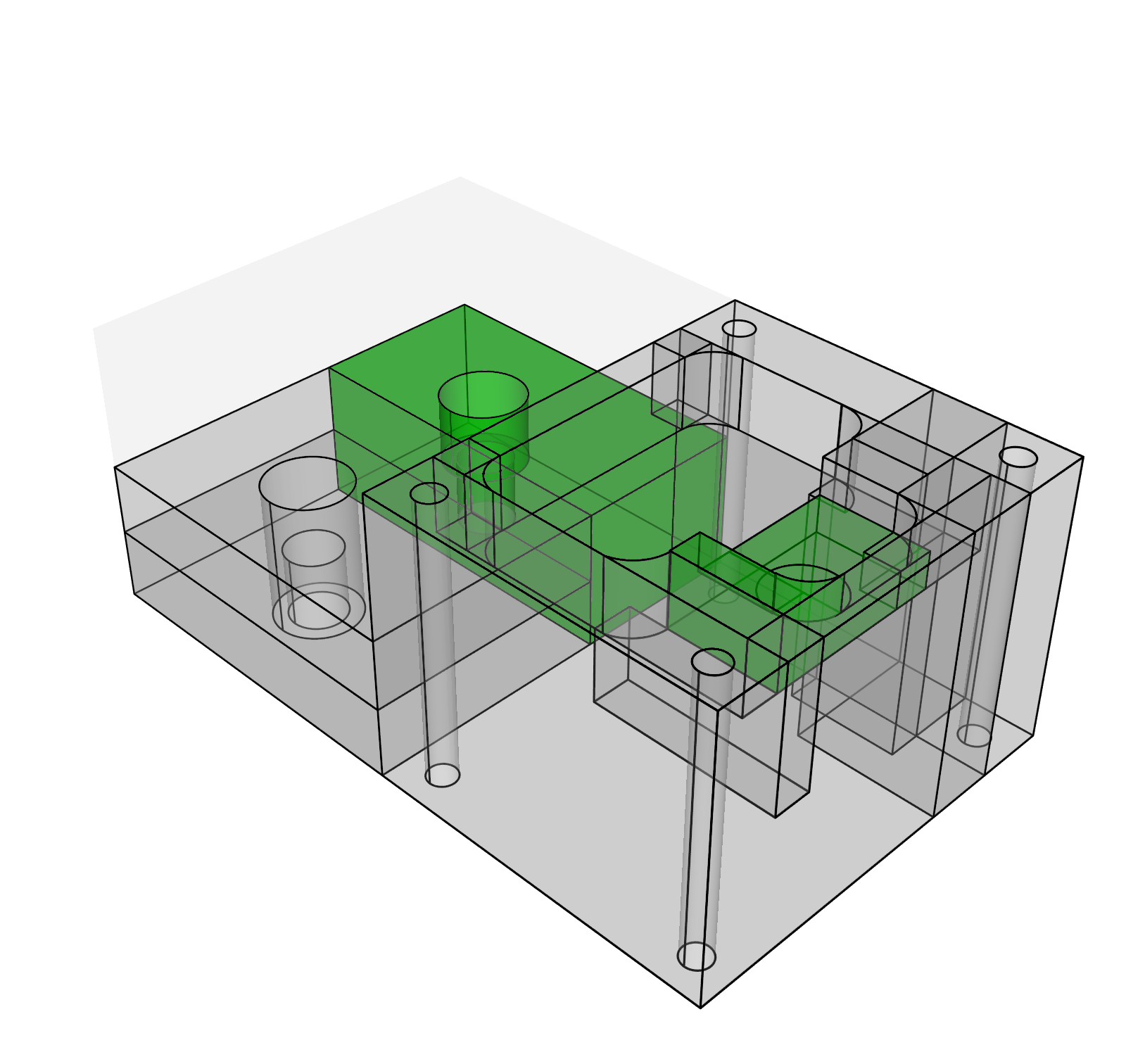} &
        \includegraphics[width=0.124\linewidth]{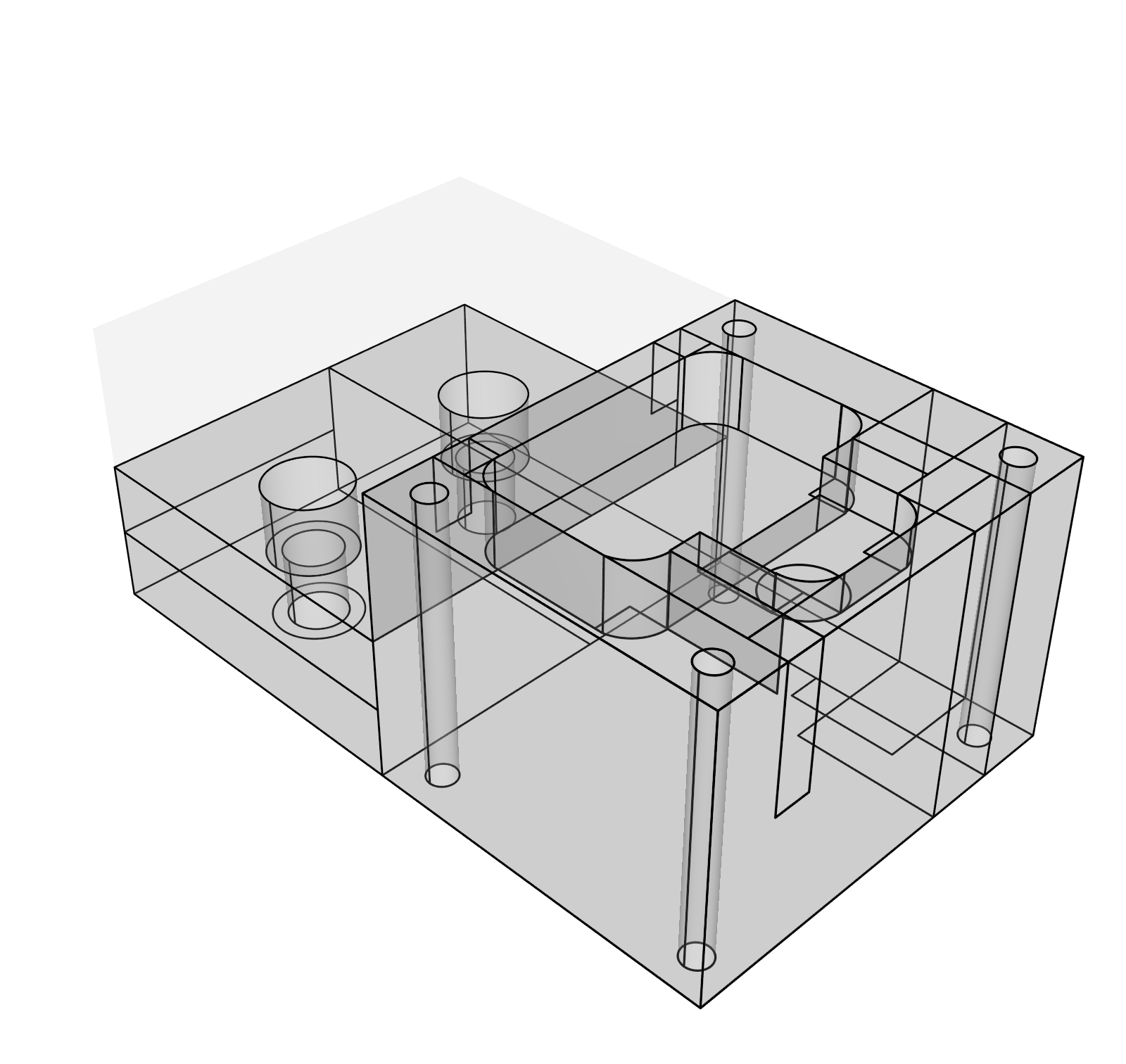}
        &
        &
        &
        \\

        \multicolumn{1}{l}{Ours Net} & & & & &
        \\
        \includegraphics[width=0.124\linewidth]{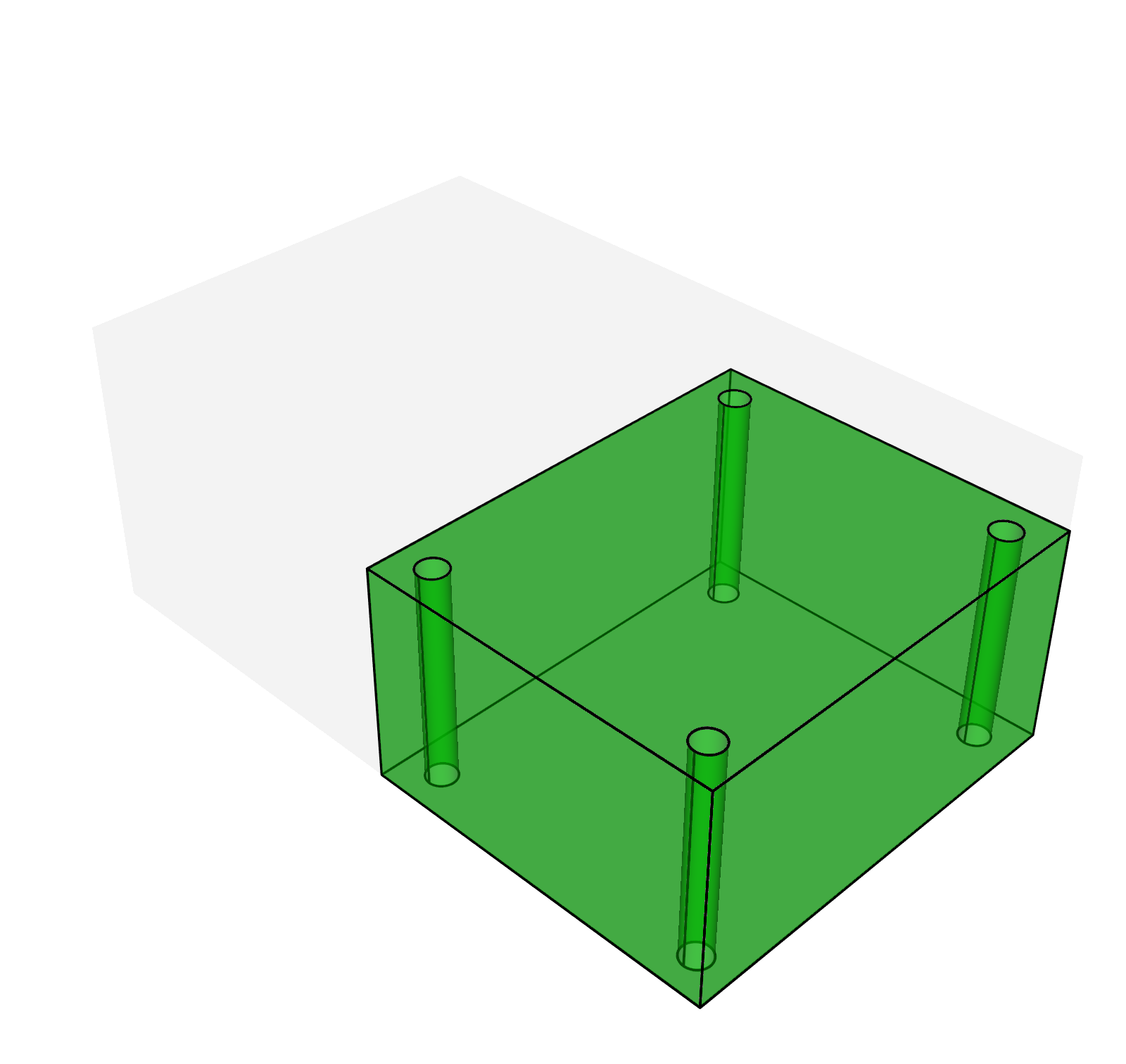} &
        \includegraphics[width=0.124\linewidth]{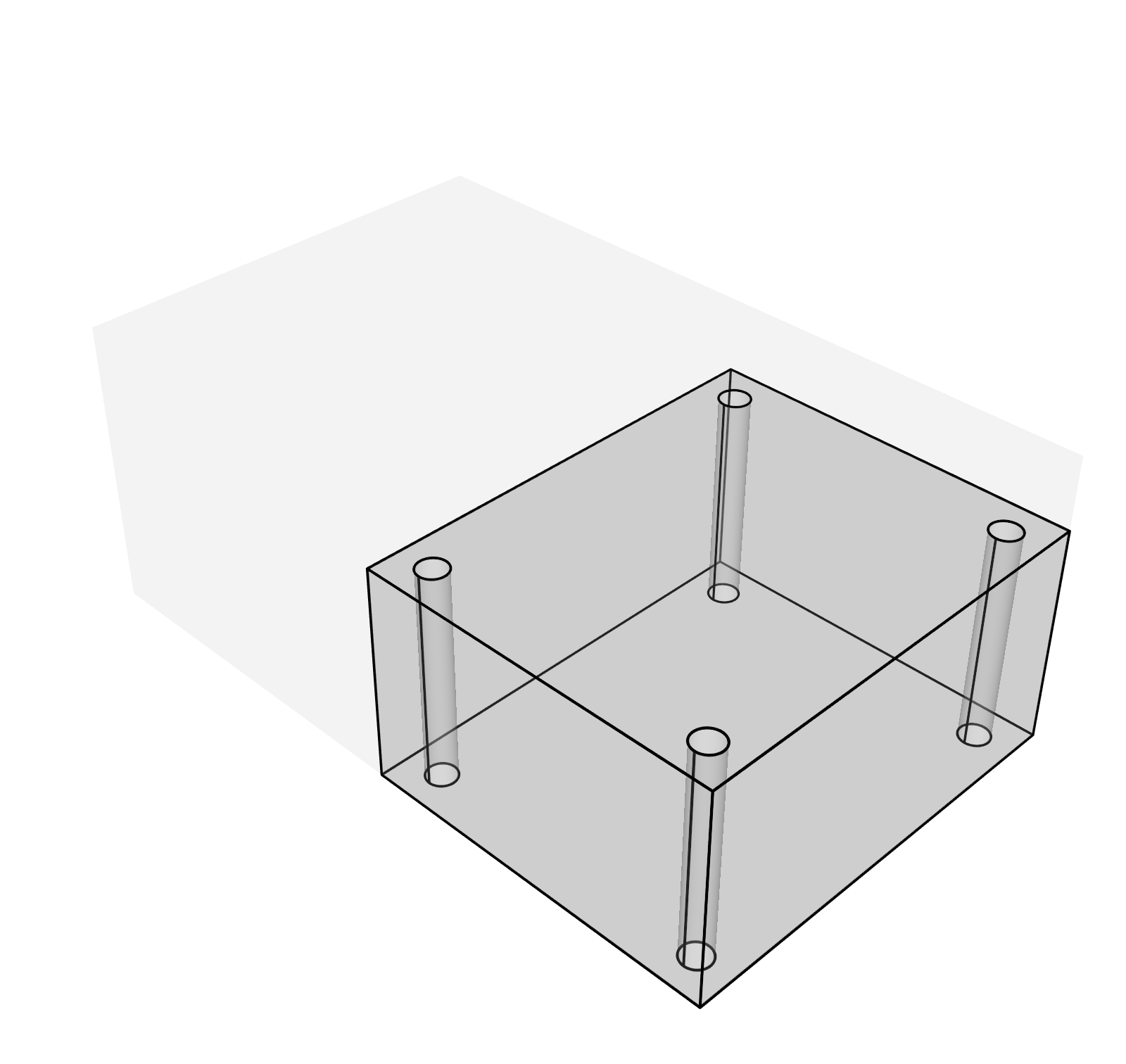} &
        \includegraphics[width=0.124\linewidth]{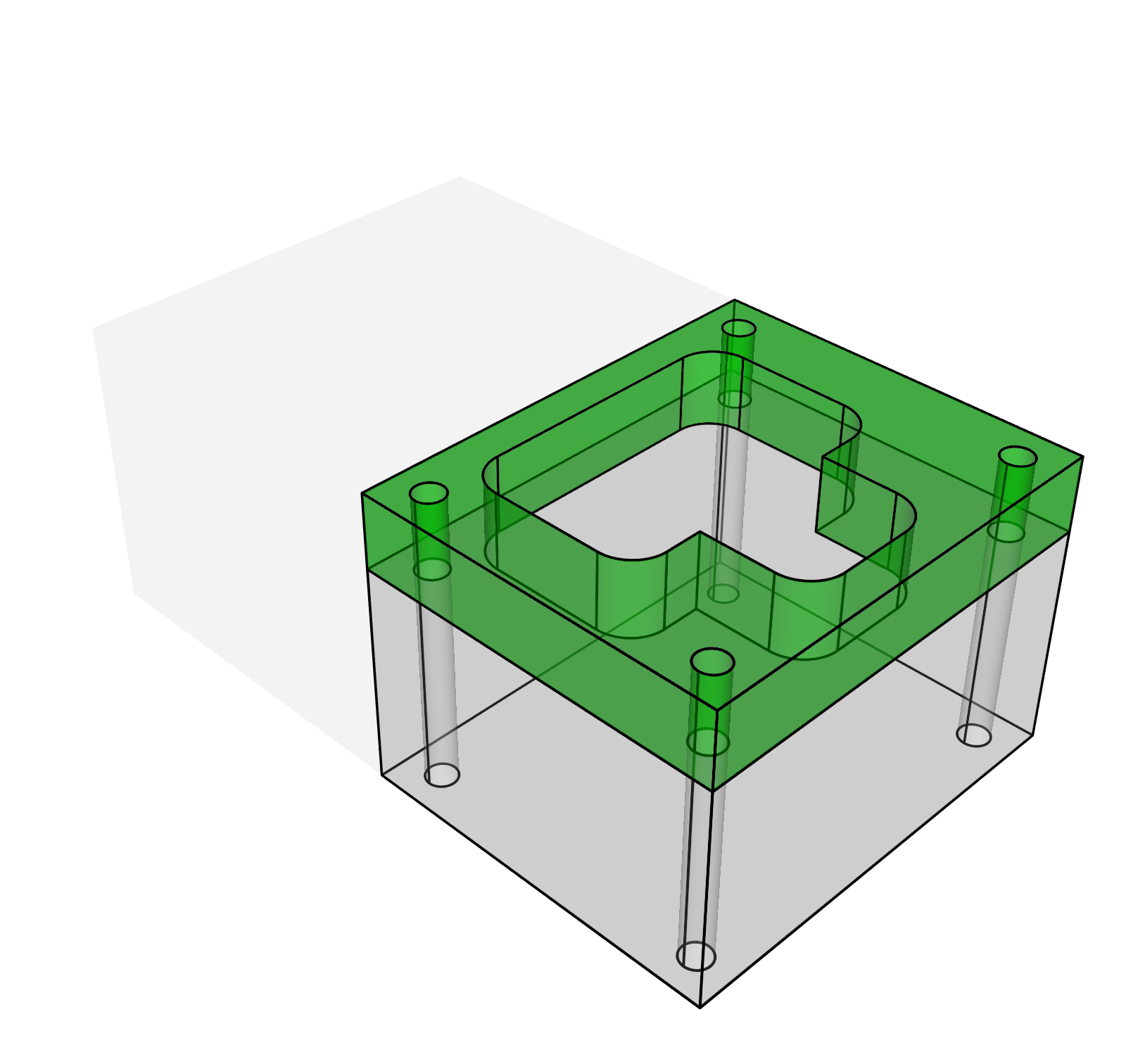} &
        \includegraphics[width=0.124\linewidth]{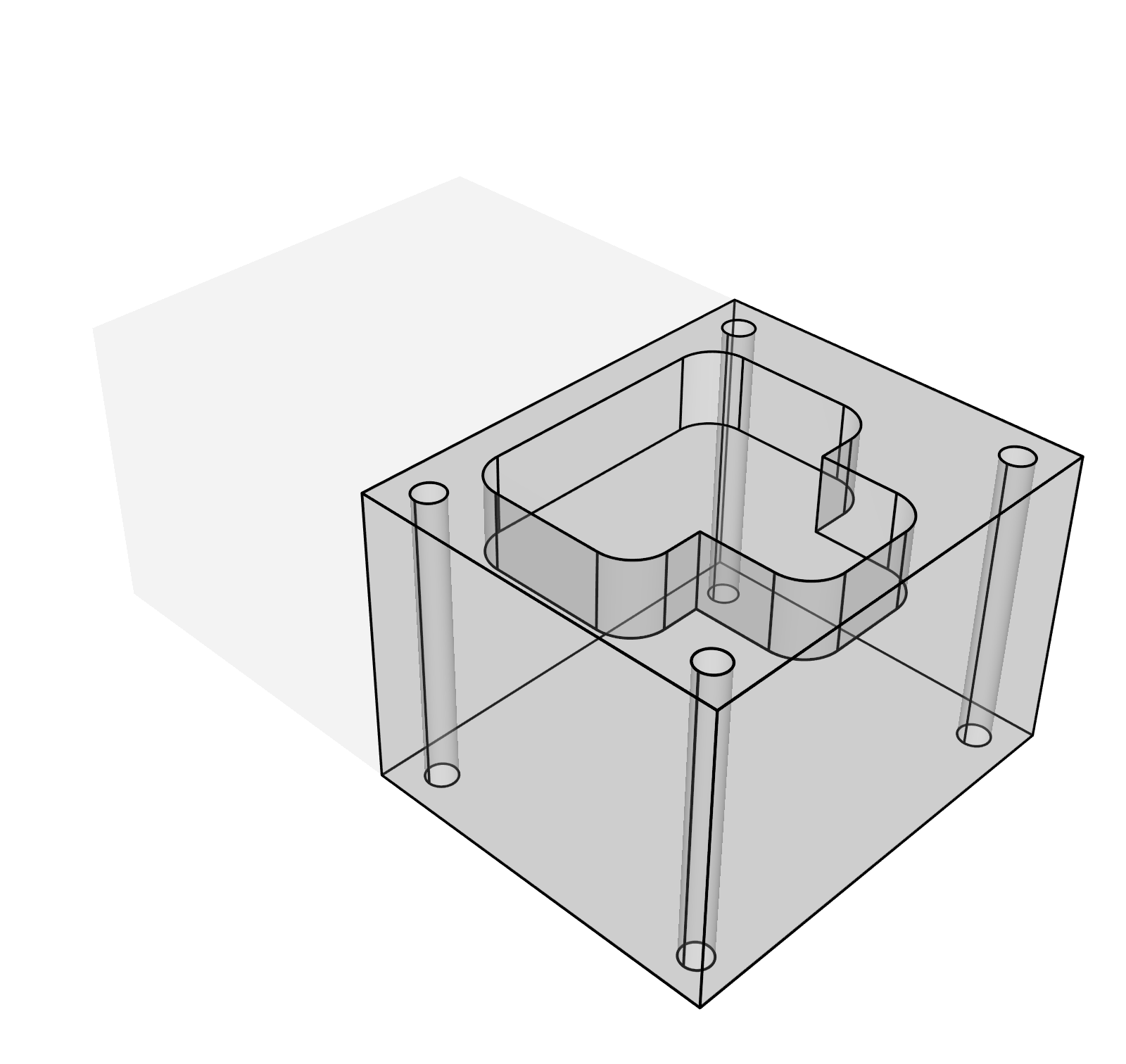} &
        \includegraphics[width=0.124\linewidth]{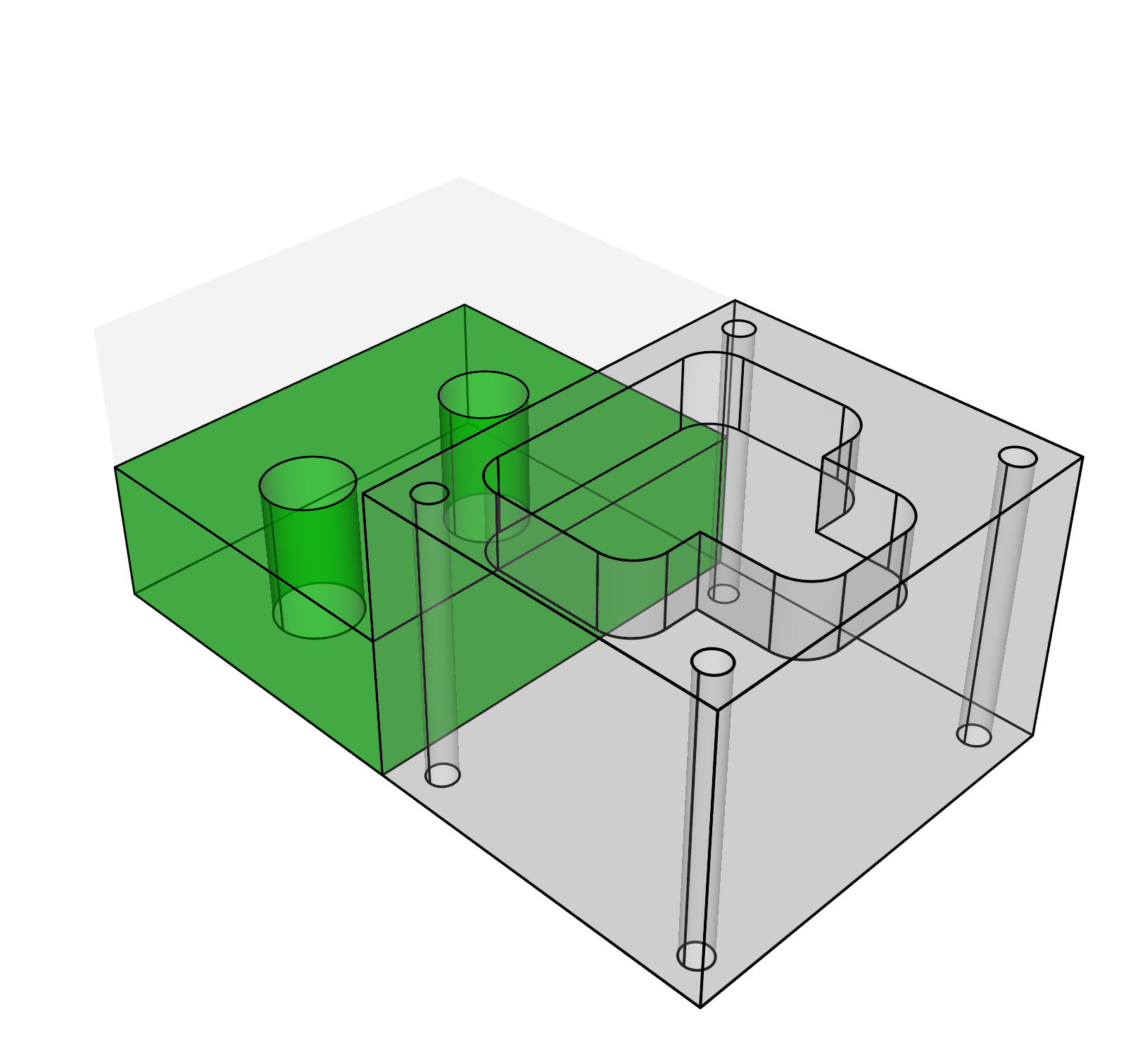} &
        \includegraphics[width=0.124\linewidth]{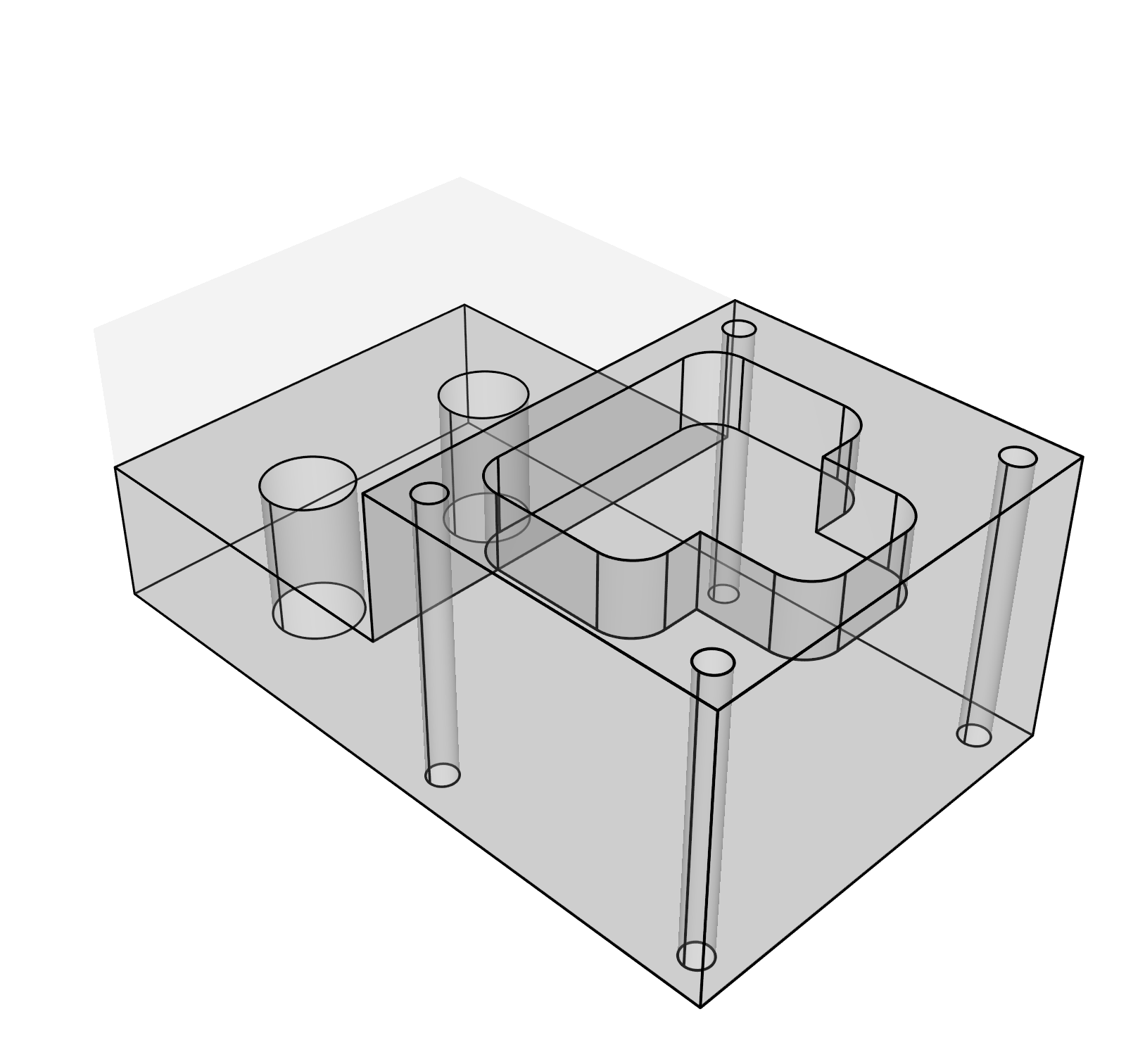} &
        \includegraphics[width=0.124\linewidth]{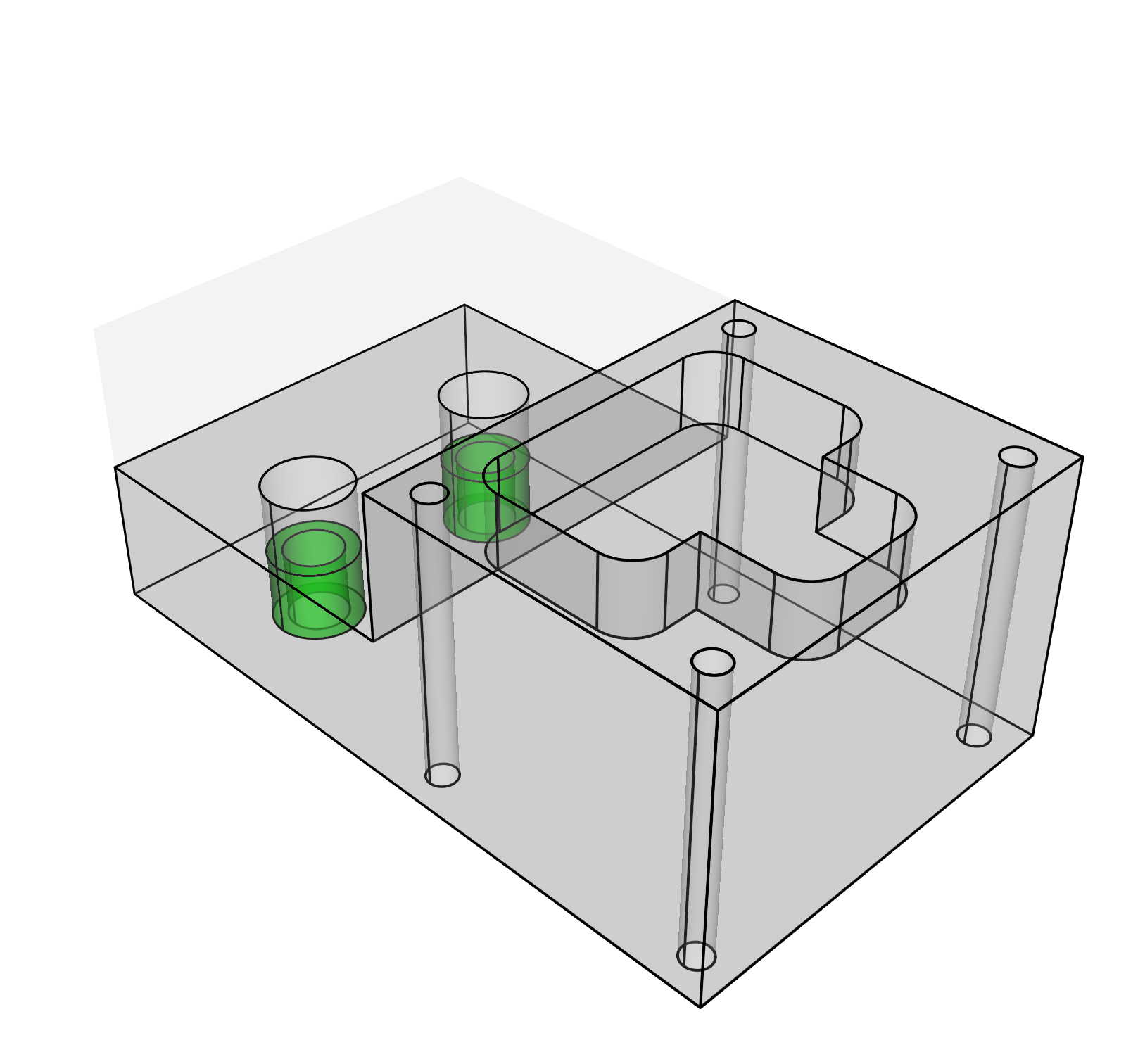} &
        \includegraphics[width=0.124\linewidth]{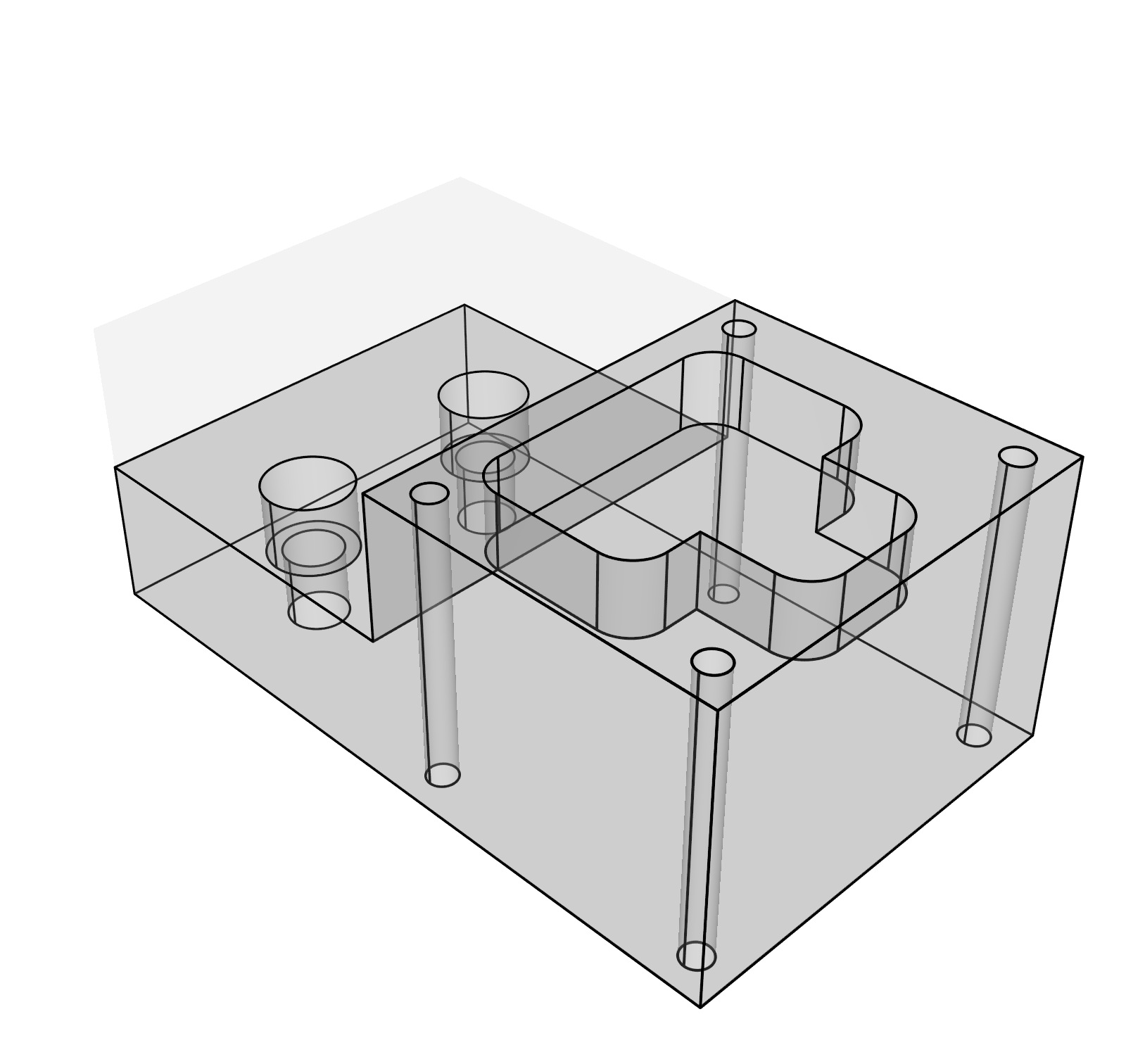}
        \\
        \includegraphics[width=0.124\linewidth]{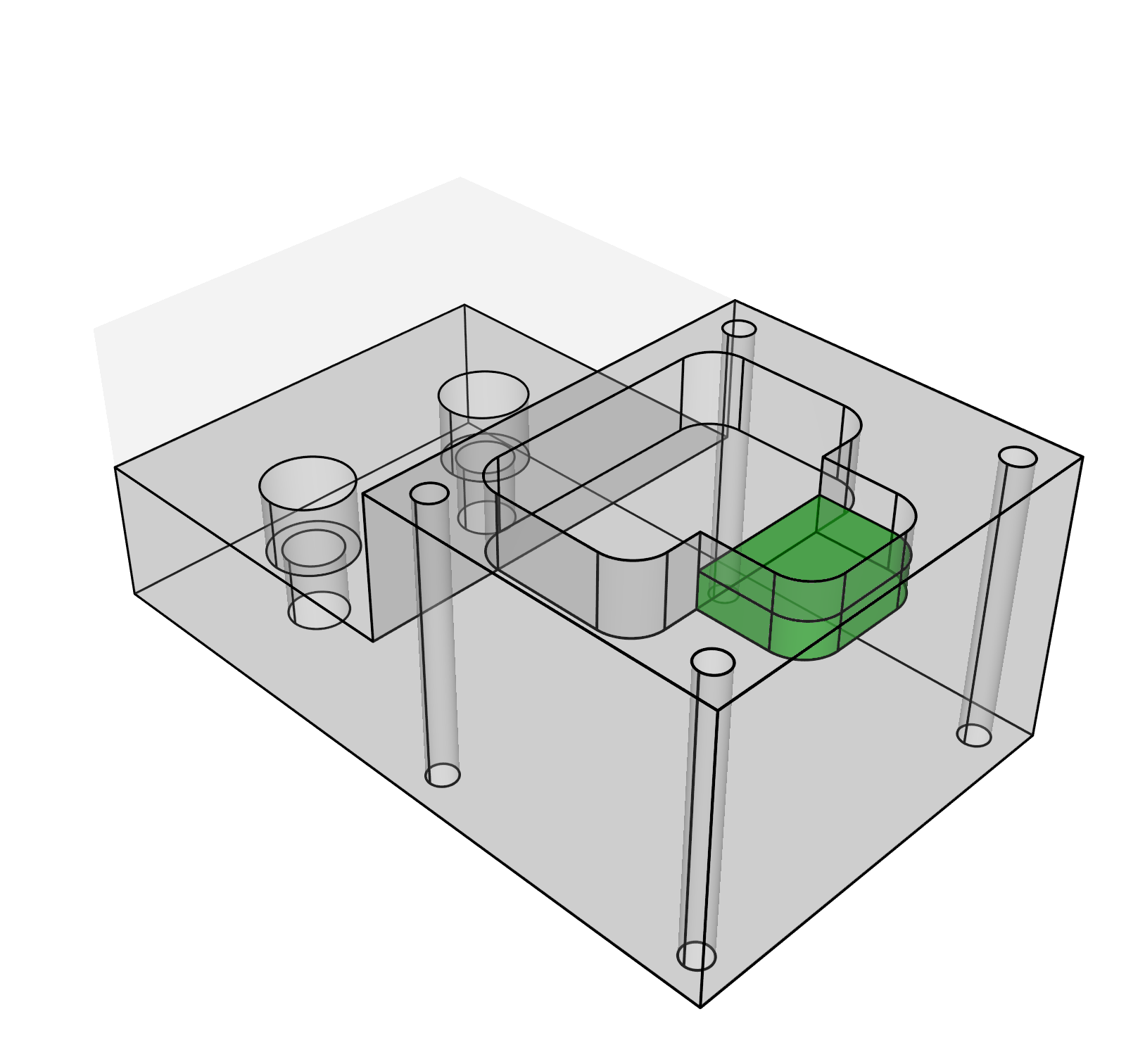} &
        \includegraphics[width=0.124\linewidth]{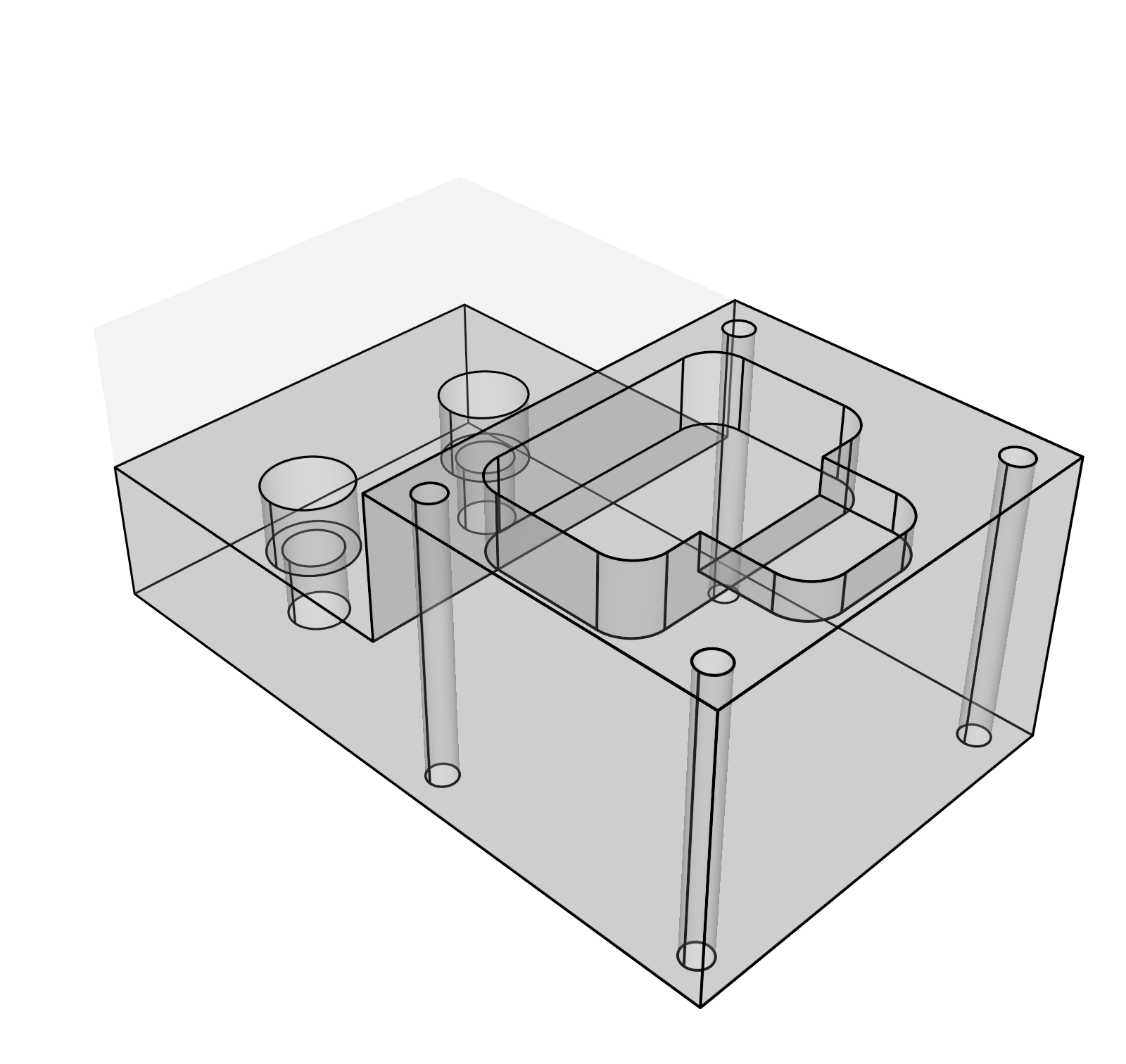}
        &
        &
        &
        &
        &
        \\
    \end{tabular}
    \caption{
    Qualitative comparison of the output of our model's inferred programs (Ours Net) vs. InverseCSG. Green: addition, Red: subtraction, Blue: intersection, Grey: current. (Case 2, Part 2)
    }
    \label{figure:qualitative_comparision_inv2}
\end{figure*}

